\documentclass{aa}  

\usepackage{graphicx}
\usepackage{xcolor}
\usepackage{txfonts}

\usepackage{subfig}
\usepackage{pdflscape}

\newcommand{\kms }{km s$^{-1}$}

\newcommand{ \Msun } {M$_{\odot}$}
\newcommand{ \Lsun } {$L_{\odot}$}

\newcommand{ \Herschel } {\textit{Herschel}}

\newcommand{ \OIII} {[O\,\textsc{iii}]}

\newcommand{ \micron}{$\mu$m }
\newcommand{ \vmin}{$v_{min}$}
\newcommand{ \vmax}{$v_{max}$}
\newcommand{\Nnuclei}{38}
\newcommand{ \vout}{$v_{out}$}
\newcommand{ \Rout}{$R_{out}$}
\newcommand{ \Mout}{$M_{out}$}
\newcommand{ \Mrate}{$\dot{M}_{out}$}

\usepackage[colorlinks=true, citecolor=blue, linkcolor=violet]{hyperref}
\begin{document} 

   \title{Physics of ULIRGs with MUSE and ALMA: The PUMA project \\ IV.  No tight relation between cold molecular outflow rates and AGN luminosities}
  \titlerunning{PUMA IV. Spatially resolved cold molecular outflows in ULIRGs}

   \author{I.  Lamperti \inst{1}\fnmsep\thanks{E-mail: isabellalamperti@gmail.com}   
   		 M.  Pereira-Santaella\inst{1}  \and
   		 M.  Perna\inst{1, 2}  \and
  		  L.  Colina\inst{1} \and
  		  S.  Arribas\inst{1} \and
  		  S.  García-Burillo\inst{3} \and
  		  E.  González-Alfonso\inst{4} \and 
  		  S.  Aalto \inst{5} \and
		 A.  Alonso-Herrero\inst{6} \and
		 F.  Combes\inst{7} \and
	     A.  Labiano\inst{8} \and
	    J.  Piqueras-López\inst{1} \and
		D.  Rigopoulou\inst{9} \and
		P.  van der Werf\inst{10} 
         }
\authorrunning{I.  Lamperti et al.}

\institute{
Centro de Astrobiología (CAB),  CSIC-INTA,  Ctra. de Ajalvir Km. 4, 28850 Torrejón de Ardoz, Madrid, Spain
   \and  INAF - Osservatorio Astrofisico di Arcetri, Largo Enrico Fermi 5, I-50125 Firenze, Italy 
   \and Observatorio Astronómico Nacional (OAN-IGN)-Observatorio de Madrid, Alfonso XII, 3, 28014, Madrid, Spain
   \and Universidad de Alcalá,  Departamento de Física y Matemáticas,  Campus Universitario, 28871 Alcalá de Henares,  Madrid, Spain
     \and  Department of Space, Earth and Environment, Division of Astronomy and Plasma Physics, SE-412 96 Gothenburg, Sweden
    \and  Centro de Astrobiología (CAB), CSIC-INTA, Camino Bajo del Castillo s/n, 28692, Villanueva de la Cañada, Madrid, Spain
\and  LERMA,  Obs. de Paris,  PSL Univ.,  Collége de France, CNRS, Sorbonne Univ., Paris, France
\and Telespazio UK for ESA, ESAC, E-28692 Villanueva de la Cañada, Madrid, Spain
\and Department of Physics,  University of Oxford,  Keble Road, Oxford OX1 3RH, UK
\and Leiden Observatory,  Leiden University,  PO Box 9513, 2300, RA Leiden, The Netherlands
   }


 
  \abstract
  { We study molecular outflows in a sample of 25 nearby ($z< 0.17$,  $d<750$~Mpc) ultra-luminous infrared galaxy (ULIRG) systems (\Nnuclei\ individual nuclei) as part of the  Physics of ULIRGs with  MUSE and ALMA (PUMA) survey,   using $\sim400$~pc (0.1-1.0" beam FWHM) resolution ALMA CO(2-1) observations.
  We used a spectro-astrometry analysis to identify high-velocity ($> 300$~\kms) molecular gas disconnected from the galaxy rotation, which we attribute to outflows. 
  In 77\% of the 26 nuclei with $\log L_{IR}/L_{\odot}>11.8$,  we identified molecular outflows with an average $v_{out}= 490$~\kms, outflow masses $1-35 \times 10^7$  \Msun, mass outflow rates \Mrate$=6-300$~\Msun~yr$^{-1}$,  mass-loading factors $\eta = \dot{M}_{out}/SFR = 0.1-1$, and an average outflow mass escape fraction of $45\pm6\%$.
 The majority of these outflows (18/20) are spatially resolved with radii of $0.2-0.9$~kpc and have short dynamical times ($t_{dyn}=R_{out}/v_{out}$)  in the range $0.5-2.8$~Myr. 
  The outflow detection rate is higher in nuclei dominated by starbursts (SBs,  $14/15=93\%$) than in active galactic nuclei (AGN,  $6/11=55\%$). 
Outflows perpendicular to the kinematic major axis are mainly found in interacting SBs.
   We also find that our sample does not follow the \Mrate\ versus  AGN luminosity relation reported in previous works. 
 In our analysis, we include a sample of nearby main-sequence galaxies  (SFR = $0.3-17$~\Msun~yr$^{-1}$) with detected molecular outflows from the PHANGS-ALMA survey to increase the $L_{IR}$ dynamic range.
 Using these two samples, we find a correlation between the outflow velocity and the star-formation rate (SFR),  as traced by $L_{IR}$ ($v_{out} \propto SFR^{0.25\pm0.01})$,  which is consistent with what was found for the atomic ionised and neutral phases.  
 Using this correlation,  and the relation between \Mout/\Rout\ and \vout, we conclude that these  outflows are  likely  momentum-driven.
Finally, we compare the CO outflow velocities with the ones derived from the OH 119\micron\ doublet.  
In 76\% of the targets,  the outflow is detected in both CO and OH,  while in three  targets (18\%)  the outflow is only detected in CO, and in one target  the outflow is detected in OH but not in CO.
 The difference between the OH and CO outflow velocities could be due to the far-IR background source required by the OH  absorption which makes these observations more dependent on the specific outflow geometry.
 }

   {}
   {}
   {}
   {}
   {}


	\maketitle
%

\section{Introduction}
Local ultraluminous infrared galaxies (ULIRGs) are extreme objects with infrared (IR, 8-1000~$\mu$m) luminosities $ L_{IR}>10^{12} L_{\odot}$.
ULIRGs are mostly gas-rich major mergers  \citep{Lonsdale2006} and represent an important stage in galaxy evolution.
The nuclei of ULIRGs host intense starbursts,  and active galactic nuclei (AGN) can also account for a significant fraction of their IR luminosity \citep{Farrah2003, Nardini2010}.
Nuclear outflows powered by starbursts and AGN  are thought to play a significant role in the evolution of galaxies.  They can influence star-formation by injecting energy into the interstellar medium (ISM),  heating  or expelling gas from the galaxy \citep[e.g.][]{Fabian2012, Veilleux2020}. 
In ULIRGs, outflows have been detected  in different gas phases: atomic neutral \citep[e.g.][]{Rupke2005a, Cazzoli2016},  atomic ionised \citep[e.g.][]{Westmoquette2012,  Bellocchi2013,  Arribas2014},  
hot molecular \citep[e.g.][]{Dasyra2011,   Dasyra2014,   Emonts2017}, 
  and cold molecular \citep[e.g.][]{Fischer2010, Feruglio2010, Sturm2011, Cicone2014, Sakamoto2014, Garcia-Burillo2015, Aalto2015,  Feruglio2015, Pereira-Santaella2018, Fluetsch2019, Lutz2020}.

In this work, we focus on the cold molecular phase, which is thought to account for most of the mass outflow rate \citep[e.g.][]{Rupke2013,  Carniani2015,  RamosAlmeida2019,   Fluetsch2021}.
The most common tracers of the cold molecular phase are the low-J transitions of CO \citep[][]{Feruglio2010,  Chung2011,  Cicone2014, Pereira-Santaella2018,  Lutz2020}, HCN    \citep[][]{Aalto2012, Aalto2015, Walter2017, Barcos-Munoz2018},   and FIR OH lines \citep[e.g.][]{Fischer2010, Sturm2011, Spoon2013, Veilleux2013, Gonzalez-Alfonso2017}.

To  accurately measure the outflow properties, such as the  mass outflow rate, outflow energy and momentum rate,  it is necessary to know the distribution of the outflowing gas.  Spatially resolved studies of molecular outflows in ULIRGs  have only been  performed on a few sources \citep[e.g.][]{Garcia-Burillo2015, Saito2018, Barcos-Munoz2018, Pereira-Santaella2018}. 
In most cases, the targeted  objects were selected based on the presence of an outflow,  which was detected through previous low-resolution observations. This could introduce a bias in the samples,  in which only ULIRGs with the most-extreme outflows are selected.

 In this work, we present  high spatial resolution (0.1-1.0",  $\sim400$~pc) ALMA observations of the CO(2-1) line for an unbiased sample of 25 ULIRGs,  selected only based on their IR luminosity.
This paper is part of the Physics of ULIRGs with MUSE and ALMA (PUMA) project. The main goals of PUMA are the following: (i) to study the prevalence of outflows in different gas phases (ionised, neutral, and molecular) as a function of the galaxy properties, (ii) to determine the driving mechanisms of the outflows (e.g. distinguish between starburst- and AGN-powered outflows), and (iii) to  characterise the effects of outflow feedback on the host galaxies.
 The PUMA survey combines VLT/MUSE optical integrated field spectroscopy and ALMA CO(2-1) and continuum observations to study the multi-phase (ionised,  neutral, and molecular) properties of outflows in ULIRGs. \citet{Perna2021} have presented the first results on the spatial distribution of the ionised gas and the resolved stellar kinematics derived from the MUSE data. A detailed analysis of the MUSE data for Arp~220 has been presented in \citet{Perna2020}.  \citet{Perna2022} studied the properties and incidence of the ionised gas disks in the PUMA ULIRGs,  the associated velocity dispersion and its relation with the offset from the main sequence.
 \citet{Pereira-Santaella2021}  analysed the ALMA 220~GHz continuum and  provided evidence for the ubiquitous presence of deeply obscured AGN in ULIRGs, which could substantially contribute to their IR emission.

This paper is organised as follows.
In Section~\ref{sec:sample}, we describe the sample selection criteria and the general properties of the PUMA targets.  
Section~\ref{sec:ALMA_data} describes the ALMA observations and the data reduction.  
In Section~\ref{sec:analysis},  we present the CO(2-1) moment maps (Sec.~\ref{sec:moment_maps}), the spectro-astrometry analysis  (Sec. ~\ref{sec:spectroastrometry}), and the method used to derive the properties of the molecular outflows  (Sec.~\ref{sec:outflow_def}- \ref{sec:comp_lit}),  as well as the analysis of the OH 119~\micron\ spectra (Sec. ~\ref{sec:OH_analysis}).
 The outflow properties,  detection rate,  energetics, and  launching mechanisms are discussed in Section~\ref{sec:mean_out}-\ref{sec:out_prop_energetics}.
In Section~\ref{sec:OH_comp}, we compare the outflow properties derived from the CO(2-1) data with the ones derived from the OH~119~\micron doublet.  
 The discussion is presented in Section~\ref{sec:discussion}.
In Section~\ref{sec:conclusions}, we summarise the main results and our conclusions.
 Throughout this work, we assume a cosmological model with $\Omega_{\lambda} = 0.7$, $\Omega_{\text{M}}= 0.3$ , and $H_0 = 70$ km s$^{-1}$ Mpc$^{-1}$.

\section{Sample}
\label{sec:sample}

The PUMA sample consists of 25 nearby ($z<0.165$,  $d<750$~Mpc) ULIRG systems (\Nnuclei\ individual nuclei) with IR luminosities ($8-1000~\mu$m) in the range $\log L_{IR}/L_{\odot} =11.9-12.7$.  The individual nuclei have luminosities from  $\log L_{IR}/L_{\odot} < 10.5$ to $\log L_{IR}/L_{\odot} =12.7$,  based on the relative contribution of the nuclei to the total ALMA continuum fluxes \citep[see ][]{Pereira-Santaella2021}.
The sample has been selected to cover a range of interacting phases:  12 systems are classified as advanced mergers (with distance between the nuclei $d_{nuclei} < 1$~kpc) and 13 systems are classifies as interacting (with  $d_{nuclei}=1.8-8.3$~kpc). 
The nuclei of the ULIRGs can be classified as AGN-dominated or starburst- (SB-) dominated based on the  AGN contribution in the MIR \citep[$\alpha_{AGN}$,  see][]{Perna2021}: eight  ULIRG systems are dominated by AGN ($\alpha_{AGN}\geq0.5$) and 17  by a SB ($\alpha_{AGN}<0.5$). 
Based on the optical classification,  our sample includes the following: nine Seyfert galaxies (two Seyfert~1 and seven Seyfert~2), eight HII and eight low ionisation nuclear emission regions
 \citep[LINERs, ][]{Perna2021}.
In this paper we adopt a combined classification,  in which we consider as AGN all nuclei classified as AGN either based on the MIR criterion ($\alpha_{AGN}\geq0.5$) or based on the optical classification (Seyfert~1 or Seyfert~2).  According to this combined classification,  11/25 systems  (14/\Nnuclei\ individual nuclei) are classified as AGN, while the others are classified as SBs.

\begin{figure}[t]
\centering
\includegraphics[width=0.45\textwidth]{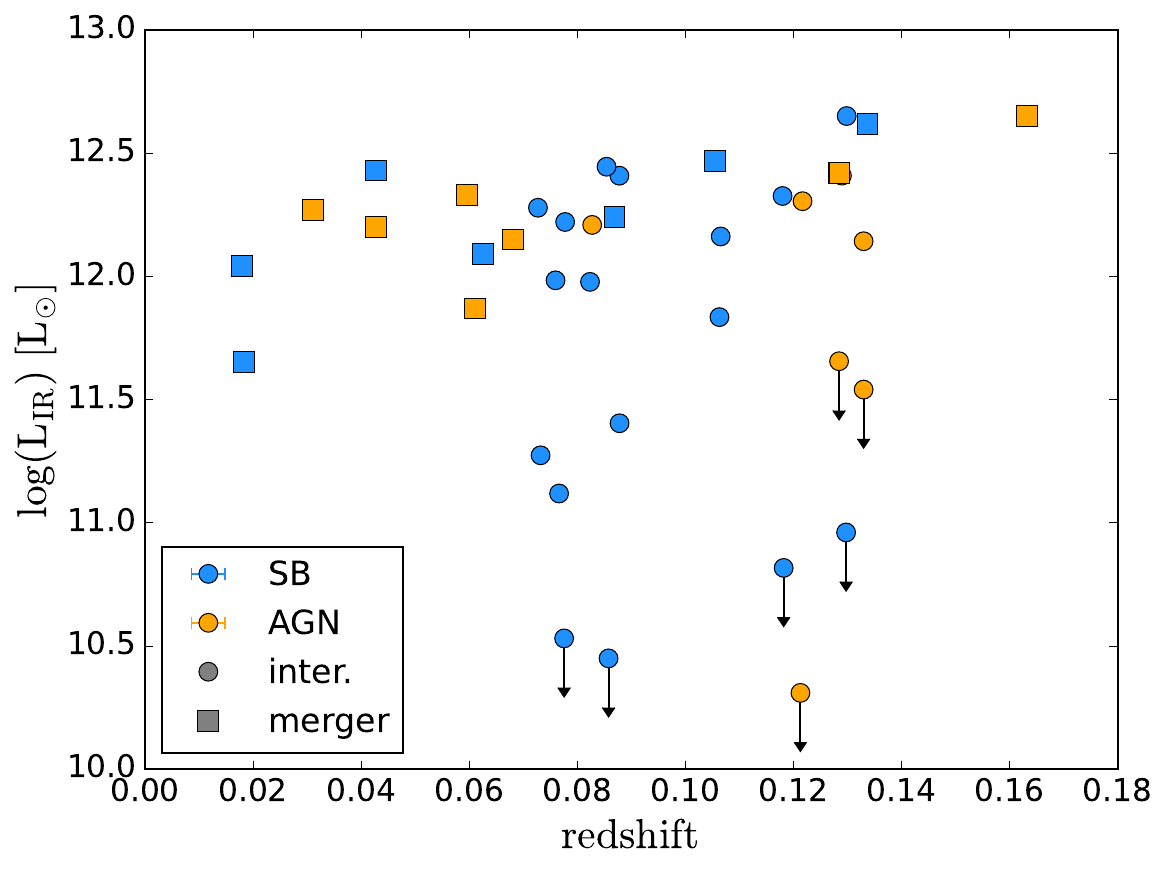}

\caption{Infrared luminosity (8-1000~$\mu$m) and redshift of the individual nuclei in the PUMA sample.  
AGN are shown in orange,  while starburst (SB) dominated nuclei are in lightblue. 
Circles indicate interacting systems and squares indicate advanced mergers. The nuclei with $\log L_{IR}/L_{\odot}<12$ are the fainter nuclei in interacting systems, where the IR luminosity is dominated by the other nucleus.}
\label{fig:sample}
\end{figure}

An overview of  the sample properties is shown in Table~\ref{tab:sample}.
Figure~\ref{fig:sample} shows the IR luminosities and redshift of the sample.

\begin{table*}
\centering
\caption{Properties of the sample.  }
\setlength{\tabcolsep}{3pt}
\begin{tabular}{llccccccccccc}
\hline

IRAS name & other name & nucleus & RA  & Dec  & log $L_{IR}$& morph.   & $d_{nuclei}$ & opt. class.  &  $\alpha_{AGN}$ & AGN/SB & CON \\ 
   &  & & [deg] & [deg] &   [log \Lsun] &   &  [kpc] &  \\
 (1) & (2)  & (3) & (4) &(5) & (6) & (7) & (8) & (9) & (10) & (11) & (12) \\ 
  \hline \hline

00091-0738 &  &S & 2.930301 & -7.368707 & 12.33 & I & 2.3 & HII & 0.46 & SB & Y\\
 & & N & 2.930424 & -7.368383 & < 10.82 &  & & & & & & \\
00188-0856 &  & - & 5.360472 & -8.657220 & 12.42 & M & < 0.3 & Sy 2 & 0.50 & AGN & N\\
00509+1225 & I Zw 1 & - & 13.395560 & 12.693316 & 11.87 & M & < 0.2 & Sy 1 & 0.92 & AGN & N\\
01572+0009 & Mrk 1014 & - & 29.959380 & 0.394688 & 12.65 & M & < 0.4 & Sy 1 & 0.65 & AGN & N\\
05189-2524 &  & - & 80.255834 & -25.362582 & 12.20 & M & < 0.1 & Sy 2 & 0.72 & AGN & N\\
07251-0248 &  &E & 111.906721 & -2.915070 & 12.41 & I & 1.8 & HII & 0.30 & SB & Y\\
 & & W & 111.906385 & -2.915106 & 11.40 &  & & & & & & \\
09022-3615 &  & - & 136.052940 & -36.450537 & 12.33 & M & < 0.1 & HII & 0.54 & AGN & N\\
10190+1322 &  &E & 155.428143 & 13.115447 & 11.98 & I & 7.2 & HII & 0.17 & SB & N\\
 & & W & 155.427056 & 13.114952 & 11.12 &  & & & & & & \\
11095-0238 &  &NE & 168.014095 & -2.906371 & 12.16 & I & 1.1 & LINER & 0.44 & SB & Y\\
 & & SW & 168.013994 & -2.906468 & 11.84 &  & & & & & & \\
12071-0444 &  &N & 182.438000 & -5.020377 & 12.41 & I & 2.3 & Sy 2 & 0.75 & AGN & N\\
 & & S & 182.437998 & -5.020812 & < 11.66 &  & & & & & & \\
12112+0305 &  &NE & 183.441903 & 2.811542 & 12.28 & I & 4.2 & LINER & 0.17 & SB & N\\
 & & SW & 183.441417 & 2.810866 & 11.27 &  & & & & & & \\
13120-5453 & WKK 2031 & - & 198.776347 & -55.156339 & 12.27 & M & < 0.1 & Sy 2 & 0.33 & AGN & N\\
13451+1232 & 4C+12.50 &W & 206.889004 & 12.290064 & 12.31 & I & 4.3 & Sy 2 & 0.82 & AGN & N\\
 & & E & 206.889583 & 12.289944 & < 10.31 &  & & & & & & \\
14348-1447 &  &SW & 219.409503 & -15.006729 & 12.21 & I & 5.5 & LINER & 0.17 & AGN$^{\star}$ & Y\\
 & & NE & 219.409988 & -15.005908 & 11.98 &  & & & & SB$^{\star}$& & \\
14378-3651 &  & - & 220.245888 & -37.075537 & 12.15 & M & < 0.1 & Sy 2 & 0.21 & AGN & N\\
15327+2340$^{\star \star}$ & Arp220 &E & 233.738722 & 23.503135 & 11.65 & M & 0.4 & LINER & 0.06 & SB & Y\\
 & & W & 233.738431 & 23.503177 & 12.04 &  & & & & & & \\
16090-0139 &  & - & 242.918411 & -1.785098 & 12.62 & M & < 0.7 & HII & 0.41 & SB & Y\\
16156+0146 &  &NW & 244.539016 & 1.656043 & 12.14 & I & 8.3 & Sy 2 & 0.70 & AGN & Y\\
 & & SE & 244.539754 & 1.655458 & < 11.54 &  & & & & & & \\
17208-0014 &  & - & 260.841481 & -0.283582 & 12.43 & M & < 0.1 & LINER & 0.00 & SB & Y\\
19297-0406 &  &N & 293.092955 & -4.000286 & 12.45 & I & 1.1 & HII & 0.23 & SB & Y\\
 & & S & 293.092904 & -4.000500 & < 10.45 &  & & & & & & \\
19542+1110 &  & - & 299.149103 & 11.318064 & 12.09 & M & < 0.1 & LINER & 0.26 & SB & N\\
20087-0308 &  & - & 302.849441 & -2.997422 & 12.47 & M & < 0.9 & LINER & 0.20 & SB & N\\
20100-4156 &  &SE & 303.373149 & -41.793113 & 12.65 & I & 6.5 & LINER & 0.26 & SB & Y\\
 & & NW & 303.372813 & -41.792383 & < 10.96 &  & & & & & & \\
20414-1651 &  & - & 311.075663 & -16.671340 & 12.24 & M & < 0.2 & HII & 0.00 & SB & N\\
22491-1808 &  &E & 342.955620 & -17.873368 & 12.22 & I & 2.7 & HII & 0.15 & SB & Y\\
 & & W & 342.955158 & -17.873239 & < 10.53 &  & & & & & & \\

\hline
\end{tabular} 
\label{tab:sample}
\tablefoot{
(1) IRAS name.
(2) Alternative name. 
(3) Name of the nucleus. 
(4) and (5) Coordinates of the nuclei derived from the ALMA 1.4~mm continuum. If the continuum is not detected, the coordinates are measured  from the NIR or optical HST  \citep[see][]{Perna2021}. 
(6) Infrared luminosity of the nuclei.  For the interacting systems, the proportion of IR luminosity in each nucleus is based on their relative ALMA continuum fluxes (see \citet{Pereira-Santaella2021}).
(7) Morphology of the system.  I: interacting system with nuclear separation $>1$~kpc; M: advanced merger with nuclear separation $<1$~kpc.  
(8) Nuclear separation.
(9) Nuclear activity classification based on optical spectroscopy \citep[see][]{Perna2021}.
(10) Fraction of AGN contribution to $L_{bol}$ derived from the 30\micron to 15\micron\ flux ratio \citep[see][]{Perna2021}.
(11) Combined nuclear activity classification used in this paper. Nuclei are classified as AGN either if $\alpha_{AGN}\geq 0.5$ or if their optical classification is Seyfert (Sy 1 or Sy 2).  
(12) Objects classified as compact obscured nuclei (CONs)  using ratios of the equivalent widths of different polycyclic aromatic hydrocarbons (PAHs) based on the method by \citet{Garcia-Bernete2022}.
$^{\star}$: For 14348-1447,  there is evidence that the AGN is located in the SW nucleus based on high-angular resolution mid-IR imaging \citep{Alonso-Herrero2016}.
$^{\star \star}$: Even though this target  (Arp 220) is classified as a merger ($d_{nuclei}=0.37$~ kpc),  we report the position of the two nuclei that have been identified thanks to the low redshift of this source.  For consistency,  we treat this target as a single nucleus in the rest of the paper.
}
\end{table*}

\section{Observations and data reduction}
\label{sec:ALMA_data}

In this work, we analyse ALMA 12-m array CO(2-1) 230.538~GHz observations of the 25 ULIRGs systems in the PUMA sample.
Most of these observations were carried out as part of our programmes 2015.1.00263.S, 2016.1.00170.S,  2018.1.00486.S, and 2018.1.00699.S  (PI: M. ~Pereira-Santaella).  Additionally, we use archival data for  13120-5453 and 15327+2340 (Arp~220),  from programmes 2016.1.00777.S (PI: K. ~Sliwa) and 2015.1.00113.S (PI: N. ~Scoville),  respectively.
 The observations were taken between 2015 and  2021.  
We note that the analysis of the CO(2-1) observations of three of the PUMA  ULIRGs (12112+0305,  14348-1447,  and 22491-1808) has already  been presented in \citet{Pereira-Santaella2018},  but are included in this paper for completeness.

Depending on the redshift of the targets, the CO(2-1) transition falls in Band 5 or Band 6.
The synthesised beam full-width at half-maximum (FWHM) has been selected for each target so that it corresponds to a similar physical spatial resolution for all targets ($\sim400-500$~pc).
As a result, the synthesised beam FWHM  are in the range $0.13-1.05$".
 The maximum recoverable scales are $\sim10$ times the beam FWHM. 
Table~\ref{tab:ALMA_obs} presents the details of the observations: the beam FWHM,  line sensitivity and total CO(2-1) fluxes.   
The data-reduction is described in detail in the PUMA II paper \citep{Pereira-Santaella2021}.  We note that two targets (12071-0444 and 13451+1232) are not presented in that work because the ALMA observations were not available at the time of publication,  but they are included in this work.
The ALMA observations of 12071-0444 have been reduced following  the same procedure used for the other sources.  
The second target, 13451+1232 (4C+12.50),  is a radio AGN with a strong 230~GHz continuum dominated by synchrotron radiation \citep{Pereira-Santaella2021}.   Given that for this source the continuum is strong enough,  we decide to apply self-calibration to these data in order to increase the signal-to-noise.  
We apply five rounds of phase self-calibration and one round of amplitude self-calibration.
Due to the steep slope of its continuum,  we  use a first-order polynomial  to subtract the continuum from the CO(2-1) spectral window in the $uv$ plane.  For the rest of the data-reduction, we follow the same procedure as presented in  \citet{Pereira-Santaella2021}.

\begin{table*}
\centering
\caption{Properties of the ALMA CO(2-1) observations.  }
\setlength{\tabcolsep}{3pt}
\begin{tabular}{lcccccccccccc}
\hline

IRAS name & nucleus & synthesised beam & sensitivity & int.  time & $z_{CO}$ & $S_\text{CO} $ & $\log L'_\text{CO}$ & $\log M(H_2)$ & FWHM & PA \\ 
   & & (arcsec$\times$arcsec)  & [$\mu$Jy beam$^{-1}$] & [min.] &  & [Jy~\kms] &   [K \kms pc$^2$] & [$M_{\odot}$]  & [\kms]  & [deg.]\\
 (1) & (2)  & (3) & (4) &(5) & (6) & (7) & (8) & (9) & (10) & (11) \\ 
  \hline \hline

00091-0738 & S & 0.29$\times$0.24  & 552 & 100 & 0.117984 & 14.4$\pm$0.2 & 9.33 & 9.26 & 294$\pm$5 & 303 \\
 & N &  & & &0.118188 & 12.1$\pm$ 0.2 & 9.25 & 9.18 & 203$\pm$5 & 300 \\
00188-0856 &  & 0.14$\times$0.12  & 265 & 69 & 0.128500 & 20.6$\pm$0.2 & 9.55 & 9.48 & 274$\pm$5 & 268 \\
00509+1225 &  & 0.35$\times$0.31  & 332 & 48 & 0.061112 & 56.9$\pm$0.7 & 9.29 & 9.22 & 376$\pm$5 & 136 \\
01572+0009 &  & 0.16$\times$0.13  & 279 & 69 & 0.163271 & 10.8$\pm$0.3 & 9.47 & 9.40 & 183$\pm$16 & 306 \\
05189-2524 &  & 0.59$\times$0.49  & 440 & 34 & 0.042731 & 101.1$\pm$0.4 & 9.33 & 9.26 & 183$\pm$5 & 93 \\
07251-0248 & E & 0.36$\times$0.31  & 332 & 41 & 0.087787 &  &  &  &  &55 \\
 & W &  & & &0.087817 &  &  &  & &271 \\
 & tot & & & & &  68.8$\pm$ 0.7 & 9.77 & 9.70 & 376$\pm$5 & \\
09022-3615 &  & 0.34$\times$0.30  & 348 & 37 & 0.059577 & 329.4$\pm$0.8 & 10.14 & 10.07 & 376$\pm$5 & 0 \\
10190+1322 & E & 0.36$\times$0.31  & 330 & 49 & 0.075970 & 61.2$\pm$0.2 & 9.61 & 9.54 & 416$\pm$5 & 250 \\
 & W &  & & &0.076626 & 53.2$\pm$ 0.8 & 9.55 & 9.48 & 325$\pm$5 & 113 \\
11095-0238 & NE & 0.31$\times$0.25  & 460 & 121 & 0.106535 &  &  &  &  &6 \\
 & SW &  & & &0.106302 &  &  &  & & - \\
 & tot & & & & &  40.7$\pm$ 0.1 & 9.71 & 9.64 & 264$\pm$5 & \\
12071-0444 & N & 0.22$\times$0.16  & 367 & 62 & 0.128969 & 40.8$\pm$0.6 & 9.86 & 9.79 & 183$\pm$5 & 79 \\
 & S &  & & &0.128441 & 4.6$\pm$ 0.2 & 8.92 & 8.85 & 335$\pm$15 & 172 \\
12112+0305 & NE & 0.34$\times$0.30  & 265 & 74 & 0.072717 & 122.3$\pm$0.5 & 9.88 & 9.81 & 335$\pm$5 & 80 \\
 & SW &  & & &0.073203 & 27.8$\pm$ 0.4 & 9.23 & 9.17 & 274$\pm$5 & 290 \\
13120-5453 &  & 0.65$\times$0.65  & 1065 & 19 & 0.031114 & 650.7$\pm$2.1 & 9.87 & 9.81 & 335$\pm$5 & 91 \\
13451+1232 & W & 0.23$\times$0.17  & 362 & 81 & 0.121680 & 41.5$\pm$0.6 & 9.82 & 9.76 & 386$\pm$15 & 242 \\
 & E &  & & &0.121291 & 2.8$\pm$ 0.4 & 8.65 & 8.58 & 223$\pm$36 &  - \\
14348-1447 & SW & 0.35$\times$0.30  & 345 & 79 & 0.082750 & 116.1$\pm$0.6 & 9.95 & 9.89 & 193$\pm$5 & 231 \\
 & NE &  & & &0.082351 & 64.9$\pm$ 1.0 & 9.70 & 9.63 & 213$\pm$5 & 196 \\
14378-3651 &  & 0.41$\times$0.28  & 554 & 47 & 0.068097 & 63.7$\pm$0.6 & 9.52 & 9.45 & 183$\pm$5 & 211 \\
15327+2340 &  & 1.28$\times$0.81  & 2145 & 2 & 0.018120 & 1693.7$\pm$3.5 & 9.84 & 9.77 & 477$\pm$5 & 41 \\
16090-0139 &  & 0.20$\times$0.17  & 436 & 49 & 0.133690 & 75.5$\pm$1.0 & 10.15 & 10.09 & 345$\pm$10 & 181 \\
16156+0146 & NW & 0.24$\times$0.16  & 498 & 76 & 0.132985 & 6.2$\pm$0.3 & 9.06 & 8.99 & 223$\pm$25 & 323 \\
 & SE &  & & & - & < 0.05$^\star$ & < 6.96 & < 6.89 & - &  - \\
17208-0014 &  & 0.47$\times$0.47  & 1043 & 7 & 0.042800 & 551.4$\pm$2.5 & 10.06 & 10.00 & 427$\pm$5 & 109 \\
19297-0406 & N & 0.34$\times$0.31  & 282 & 89 & 0.085407 &  &  &  &  &230 \\
 & S &  & & &0.085785 &  &  &  & & - \\
 & tot & & & & &  140.3$\pm$ 0.5 & 10.05 & 9.98 & 386$\pm$5 & \\
19542+1110 &  & 0.40$\times$0.34  & 368 & 44 & 0.062479 & 40.9$\pm$0.3 & 9.28 & 9.21 & 274$\pm$5 & 229 \\
20087-0308 &  & 0.31$\times$0.25  & 287 & 192 & 0.105461 & 94.6$\pm$0.3 & 10.05 & 9.98 & 528$\pm$5 & 250 \\
20100-4156 & SE & 0.20$\times$0.14  & 336 & 49 & 0.129848 & 31.3$\pm$0.3 & 9.74 & 9.68 & 376$\pm$5 & 228 \\
 & NW &  & & &0.129740 & 2.2$\pm$ 0.2 & 8.59 & 8.52 & 111$\pm$6 & 268 \\
20414-1651 &  & 0.24$\times$0.18  & 375 & 47 & 0.086874 & 51.4$\pm$0.6 & 9.62 & 9.56 & 427$\pm$5 & 59 \\
22491-1808 & E & 0.44$\times$0.33  & 342 & 58 & 0.077742 & 59.4$\pm$0.2 & 9.59 & 9.52 & 284$\pm$5 & 348 \\
 & W &  & & &0.077560 & 4.1$\pm$ 0.2 & 8.40 & 8.34 & 121$\pm$5 &  - \\

\hline

\end{tabular}
\label{tab:ALMA_obs}
\tablefoot{
(1) IRAS name.
(2) Name of the nucleus. 
(3) FWHM of the synthesised ALMA beam.
(4) 1$\sigma$ line rms sensitivity per 10~\kms channel.
(5) Integration time on source.
(6) CO(2-1) redshift, calculated based on the velocity  of the moment 1 map at the ALMA continuum peak position.  
(7) CO(2-1)  integrated flux.  For the interacting systems for which it is not possible to separate the flux belonging to the two nuclei (07251-0248,  11095-0238, and 19297-0406), we report the total flux of the system,  as well as the total CO(2-1) luminosity  (column 7) and FWHM (column 8).
$^\star$For the undetected nucleus 16156+014546 SE, we provide a 3$\sigma$ upper limit, calculated based on the rms and a typical line FWHM of 300~\kms.
(8) CO(2-1) luminosity, calculated as $L'_{CO}= 3.25\cdot 10^7 S_{CO} \nu_{rest}^{-2} D_L^2 (1+z)^{-1} $, where  $\nu_{rest}$ is the line rest-frequency, $D_L$ the luminosity distance and $z$ the redshift \citep{Solomon1997}.
(9) Molecular gas mass calculated as $M(H_2) = 1/r_{21} \cdot \alpha_{CO} \cdot  L'_{CO(2-1)}$, where $r_{21}= L'_{CO(2-1)}/L'_{CO(1-0)}=0.91$ \citep{Bolatto2013} and $\alpha_{CO}=0.78$ $M_\odot$/(K \kms\ pc$^2$) is the ULIRGs-like CO-to-H$_2$ conversion factor. 
(10) FWHM of the integrated CO(2-1) line, corrected for instrumental resolution.
(11) Position angle (PA) of the kinematic major axis measured from the spectro-astrometry of the low-velocity CO channels.  The PA  is measured east of north (anticlockwise) for the receding half of the galaxy.
}
\end{table*}

\section{Analysis}
\label{sec:analysis}

\subsection{CO(2-1)  moment maps}
\label{sec:moment_maps}
We produced the maps of the CO(2-1) intensity (moment 0),  velocity (moment 1),  and velocity dispersion (moment 2)  for our sample.  
 Before producing the moment maps,  we masked pixels with low signal-to-noise ratio (S/N) in each velocity channel, where the noise was estimated using the median absolute deviation (MAD).  Specifically,  for each velocity channel,  we masked spaxels with S/N $<3.5-5$  depending on the overall S/N of the observations.
 
 Moreover,  we masked individual spaxels which show spurious emission applying the following procedure.
For each spectral channel,  we applied a Gaussian kernel with a size of three pixels to smooth the channel image and then we masked individual pixels with S/N < 0.4  in the smoothed image.  In this way,  for each velocity channel, we remove isolated pixels  with spurious emission, which could bias the calculation of the moment maps.  
The moment 0, moment 1,  and moment 2 maps were produced using  
the cubes obtained with this masking process,  while the CO(2-1) peak map was obtained from the original data cubes.
 We define the zero velocity (and consequently the redshift) based on the moment~1 velocity at the position of the continuum peak. The CO(2-1) redshifts (reported in Table~\ref{tab:ALMA_obs})  are in good agreement with previously reported redshifts. 
 Throughout this paper,  we use the radio velocity definition. 
 
\begin{figure*}[!]
\centering

\includegraphics[width=0.98\textwidth]{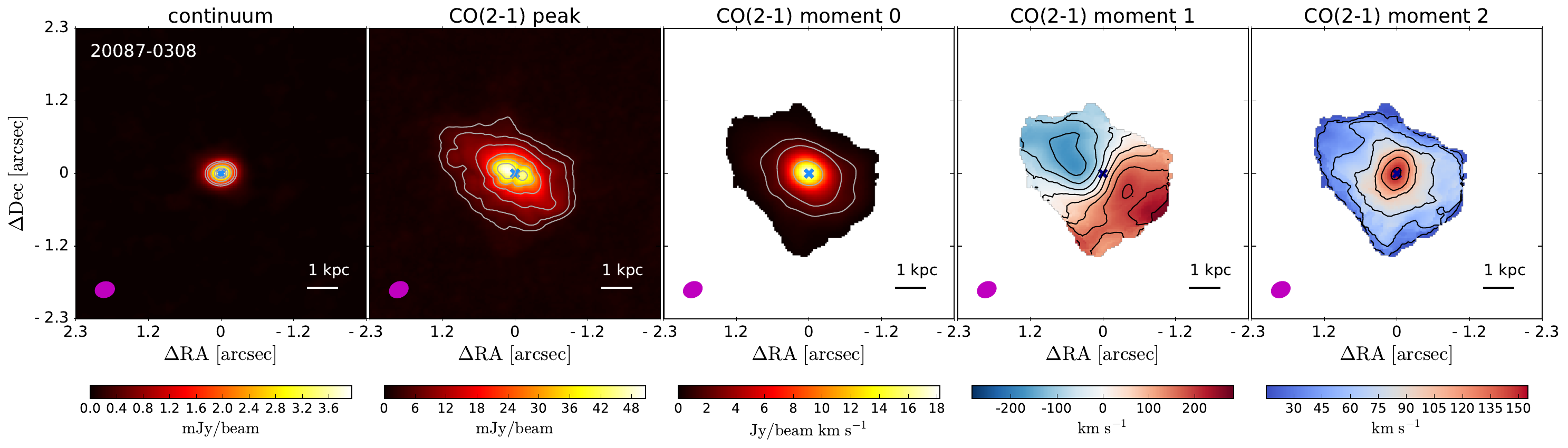}
\includegraphics[width=0.98\textwidth]{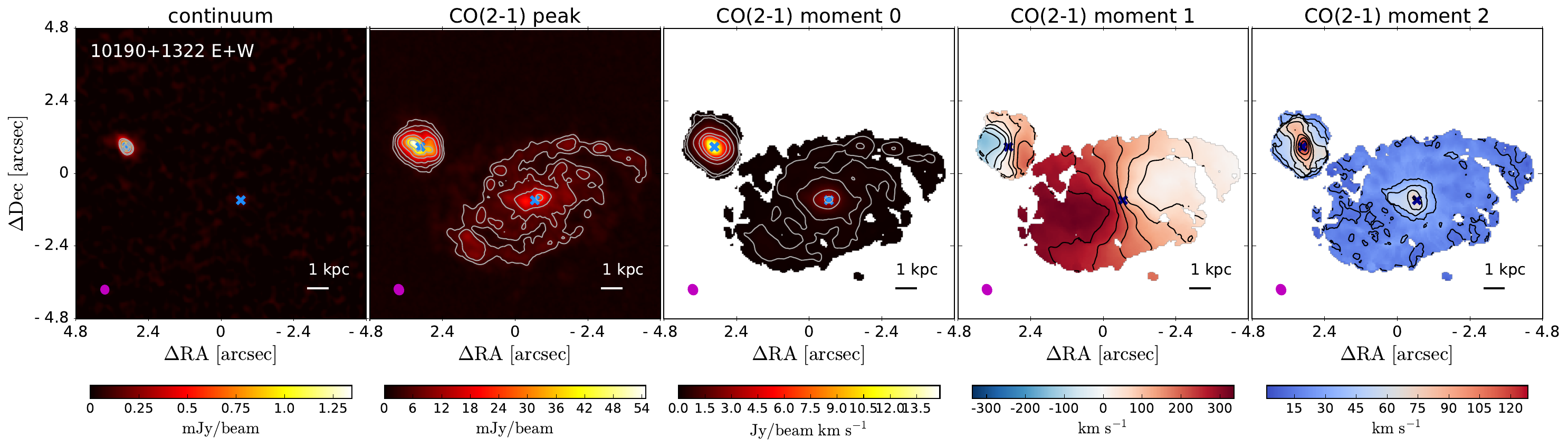}
\includegraphics[width=0.98\textwidth]{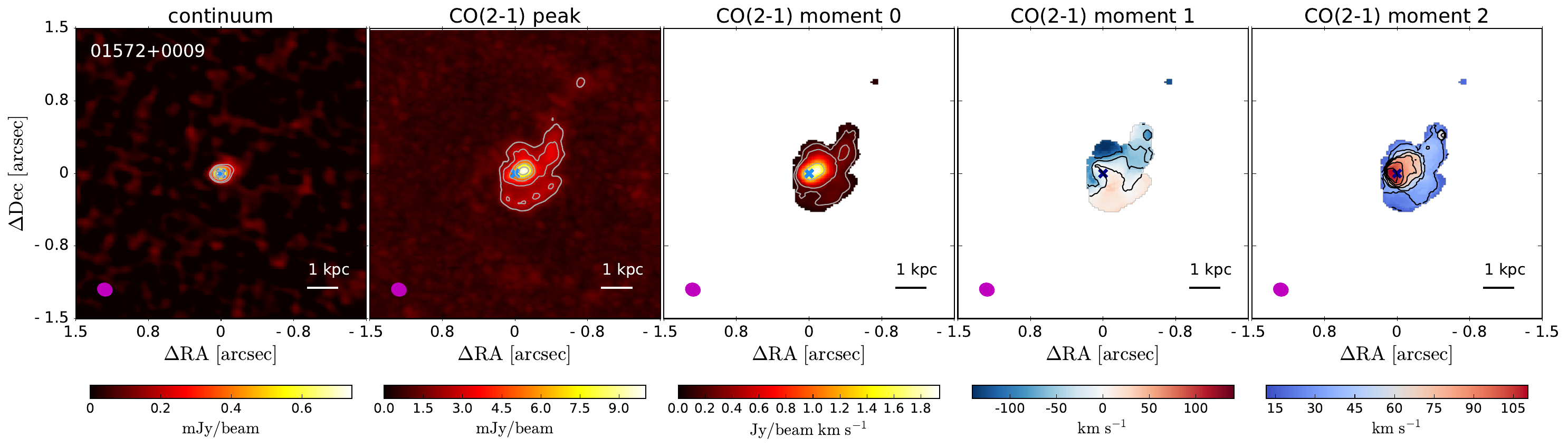}
\includegraphics[width=0.98\textwidth]{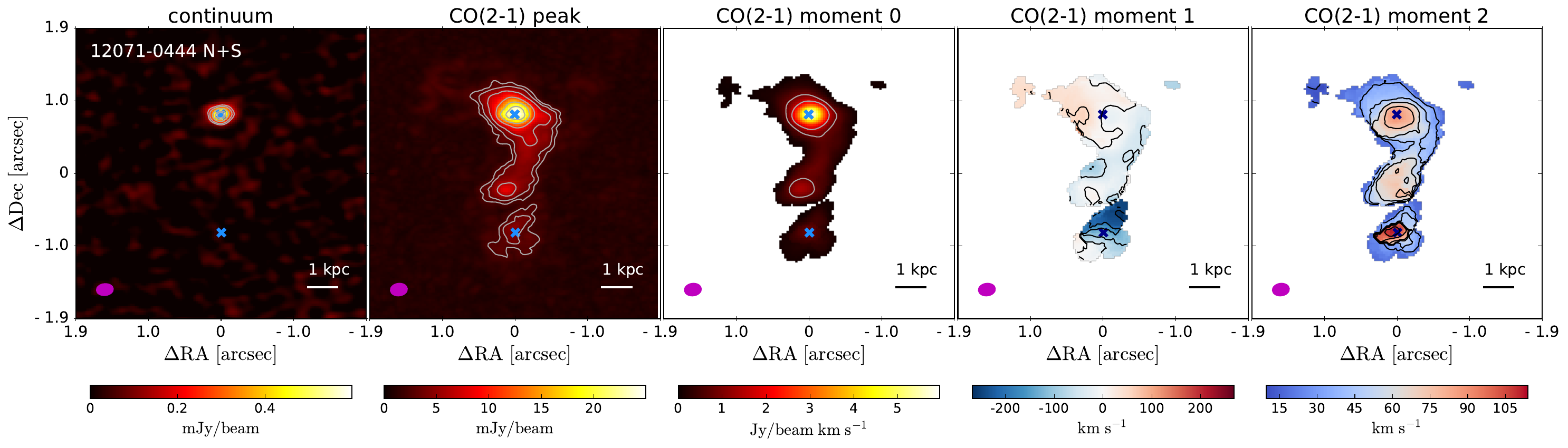}

\caption{Examples of the ALMA $\sim220-250$~GHz continuum and CO(2-1) moment maps (from left to right: peak map,  moment 0,  1,  and 2)  for a merger (20087-0308) and an interacting (10190+1322) starburst dominated system  and  for a merger (01572+0009) and an interacting (12071-0444) AGN dominated  system. The blue crosses mark the position of the nuclei (see Table~ \ref{tab:sample}). 
The magenta ellipse shows the FWHM and position angle of the ALMA beam .
The contours in the maps are:
\textit{continuum map}: [0.3,  0.4,  0.6, 0. 8,  0.9] of the maximum;
\textit{peak map}: 1.5$\times\sigma$ (where $\sigma$ is the rms) and [0.1, 0. 2,  0.4,  0.6,  0.8] of the maximum;
\textit{moment 0}: [3, 6, 25, 50, 75]$\times\sigma$ (where $\sigma$ is the rms),
\textit{moment 1:} every  50~\kms (every 25~\kms\ if the maximum value $< 100$~\kms),
\textit{moment 2:} every 25~ \kms (every 15~\kms\  if the maximum value $< 150$~\kms).
 }
\label{fig:moment_maps_ex}
\end{figure*}

 Figure~\ref{fig:moment_maps_ex} shows the CO(2-1) moment 0, 1, and 2 maps,  along with  the CO(2-1) peak map and continuum map  for four targets as an example.  The maps of the full sample are shown in the Appendix (Fig.~\ref{fig:moment_maps_appendix}). 
The CO(2-1) emission is detected with S/N~$ > 3$ 
 in all individual nuclei but 16156+0146 SE.
In the majority of the nuclei (32/37), the kinematic major axis is clearly visible in the moment 1 maps. The moment 1 maps of five nuclei show a less ordered motion (01572+0009, 05189-2524,  12112+0305 NE,   14348-1447 SW,  22491-1808 E+W).
 In 22/37 nuclei the moment 2 map shows an increase in the velocity dispersion close to the peak continuum position.

 In  09022-3615, we detect an increase in velocity dispersion south of the nucleus (distance $\sim 1$",  equivalent to $\sim 1$~kpc), which corresponds to the most blue-shifted velocities in the moment 1 map. The spectrum in this location shows two peaks. This could be due to a blue-shifted outflow (or inflow) in this location, or a cloud pushed by an outflow.  An alternative explanation is the presence of a second very obscured nucleus, which is not detected in the  ALMA millimetre continuum.
As we do not see evidence of rotation at the position of the putative second nucleus,  we consider more likely the former explanations.

The CO(2-1) peak maps of 10/\Nnuclei\  nuclei show a dip in the centre, corresponding roughly to the position of the continuum peak (e.g.  00188-0856,   12112+0305 NE,  13120-5453,  13451+1232 N).  
This drop in the centre is compatible with an extreme central optical depth.  
 In a future work,  we will present ongoing ALMA observations of the optically thin $^{13}$CO isotopologue to investigate this. 

In the appendix~\ref{sec:channel_maps}, we also show the maps of the CO(2-1) emission integrated in 50~\kms\ channels for all targets.


\subsection{Spectro-astrometry}
\label{sec:spectroastrometry}
We perform a spectro-astrometry analysis  to identify high-velocity molecular gas that is decoupled from the main rotation of the galaxy disk.
This analysis consists in determining  how the centroid  position of the CO(2-1) emission changes as a function of velocity.
We follow a similar methodology to the one used by \citet{Garcia-Burillo2015} and \citet{Pereira-Santaella2018}.

\begin{figure*}[h]
\centering
\includegraphics[width=0.8\textwidth]{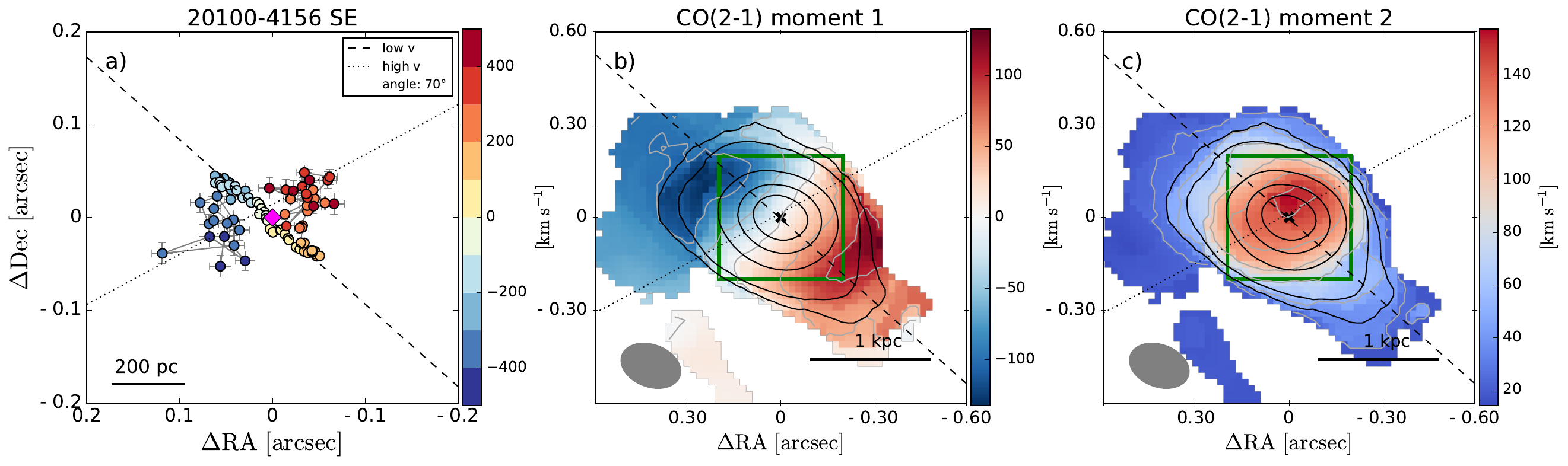}
\includegraphics[width=0.245\textwidth]{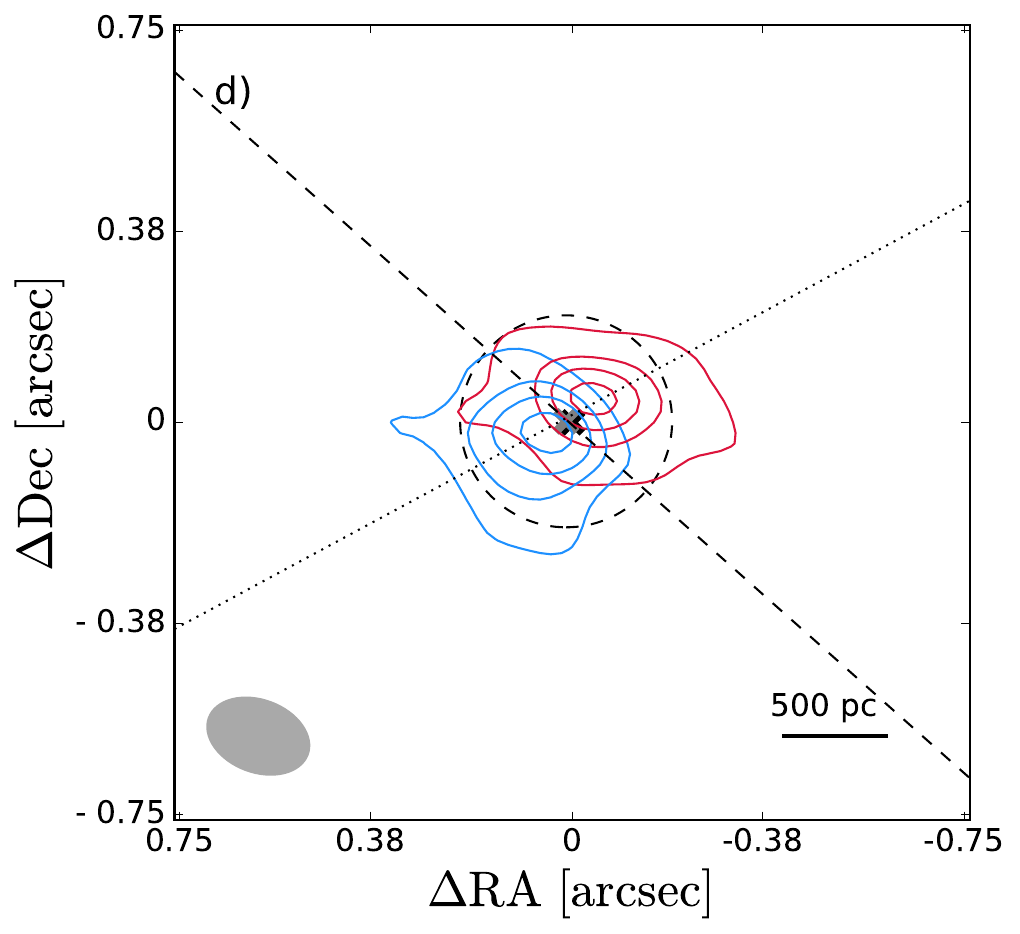}
\includegraphics[width=0.3\textwidth]{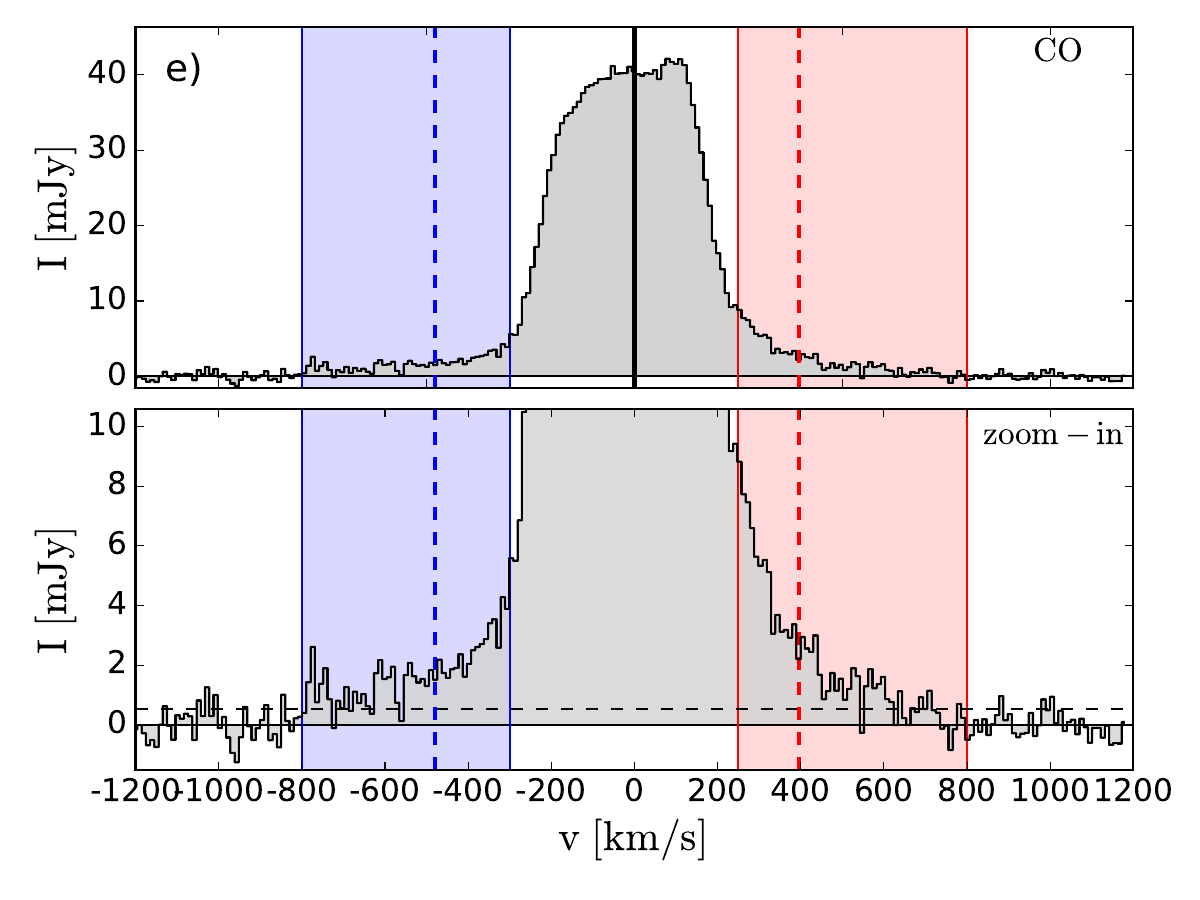}
\includegraphics[width=0.3\textwidth]{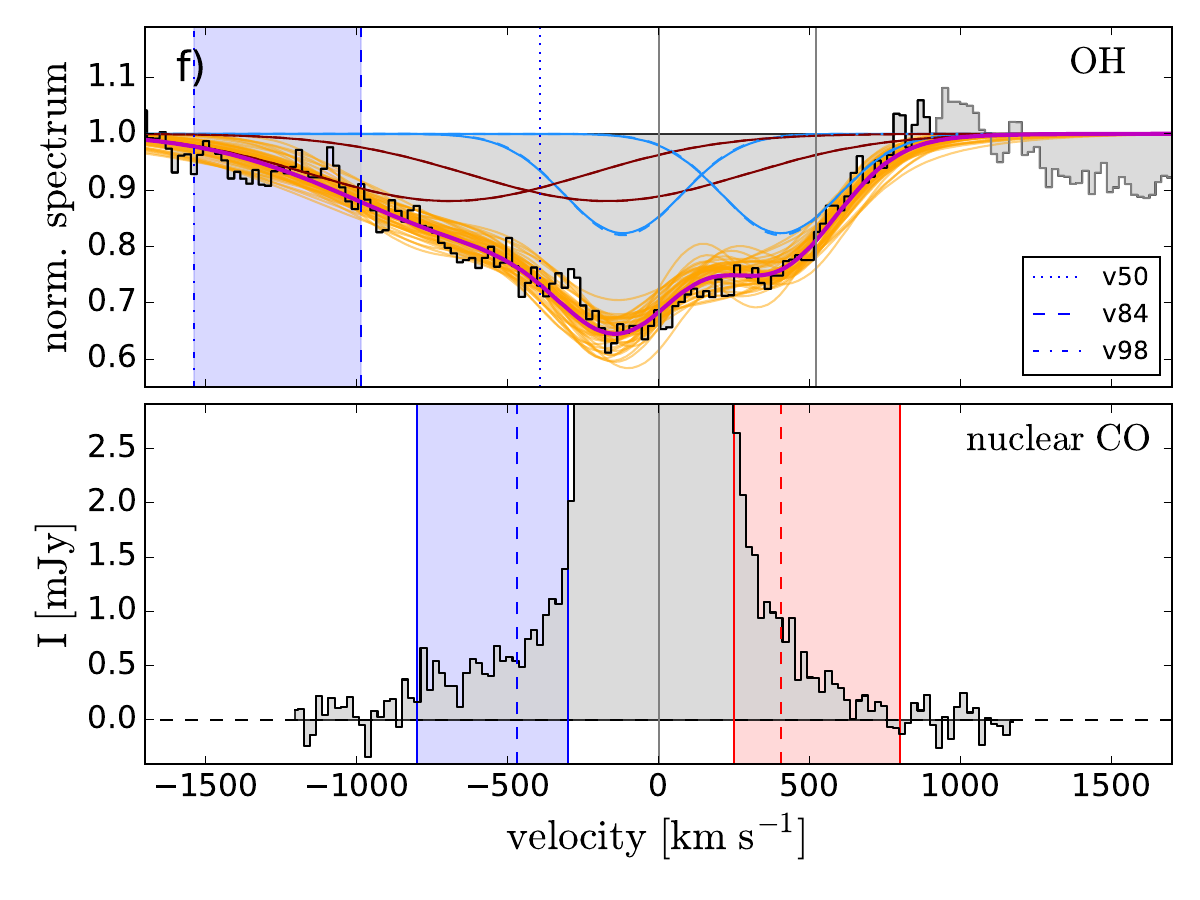}

\caption{Example of spectro-astrometry and outflow maps for one target (20100-4156 SE) with outflow direction perpendicular to the kinematic major axis.  
\textit{Panel a)} Spectro-astrometry of the CO(2-1) emission line,  i.e.  centroid position of the CO(2-1) emission in the different velocity channels. The points are colour-coded by the channel velocity. The pink diamond indicates the peak ALMA millimetre continuum position.  The dashed line is a linear fit to the low-velocity points (kinematic major axis)  and the dotted line is a fit to the high-velocity points (indicating the outflow direction, if present).
\textit{Panel b) and} \textit{c)}  show the moment 1 and moment 2 maps,  where the green square indicate the field of view of  panel \textit{(a)}. The grey ellipse illustrates the ALMA beam FWHM.   The grey contours on the moment 1 maps are every  50~\kms\ (every 25~\kms\ if the maximum value $< 100$~\kms),  and  every 25~\kms\ (every 15~\kms\  if the maximum value $< 150$~\kms) on the moment 2 map.  In black  are the CO(2-1) moment 0  contours ([3, 6, 25, 50, 75]$\times\sigma$).
\textit{Panel d)} Emission of the high-velocity channels,  integrated over the velocity ranges indicated on the CO(2-1) spectrum (shown in panel \textit{e}).  Blue- and red-shifted channels are shown with blue and red contours,  respectively (dashed lines indicate negative contour levels). The lowest contour corresponds to the 3$\sigma$ level. The next contour levels are (0.5, 0.7, 0.9) of the peak of the  emission,  if these are above the 3$\sigma$ level.  The dashed circle shows the size of the outflow (\Rout).
\textit{Panel e)} CO(2-1) continuum-subtracted spectrum  extracted from a circle with radius equivalent to the outflow size (\Rout). The lower panel shows an y-axis zoom-in to highlight the emission in the wings.  The horizontal dashed line shows the 1$\sigma$ noise level.
The vertical dashed lines indicated the `flux-weighted' velocity of the blue and red-shifted outflow ($v_{out}$).
\textit{Panel f)} OH 119~\micron spectrum (upper) compared with the nuclear CO(2-1) spectrum (bottom),  convolved to the resolution of the OH spectrum  (FWHM$\sim 270$~\kms).  The total fit to the OH lines is shown with a magenta line, while the Gaussian components of the fit are shown in lightblue and brown. The orange lines show the 50 Monte Carlo iterations used to estimate the uncertainties on the fit  (see Sec.~\ref{sec:OH_analysis}). The vertical dotted, dashed and dot-dashed lines show the $v_{50}$,   $v_{84}$,  and $v_{98}$ percentile velocities, respectively. The blue-shaded area in the upper panel shows the wavelength range between $v_{84}$ and $v_{98}$.  Figures for the rest of the sample are in the appendix.}
\label{fig:spectroastrometry}
\end{figure*}

\begin{figure*}[h]
\centering
\includegraphics[width=0.8\textwidth]{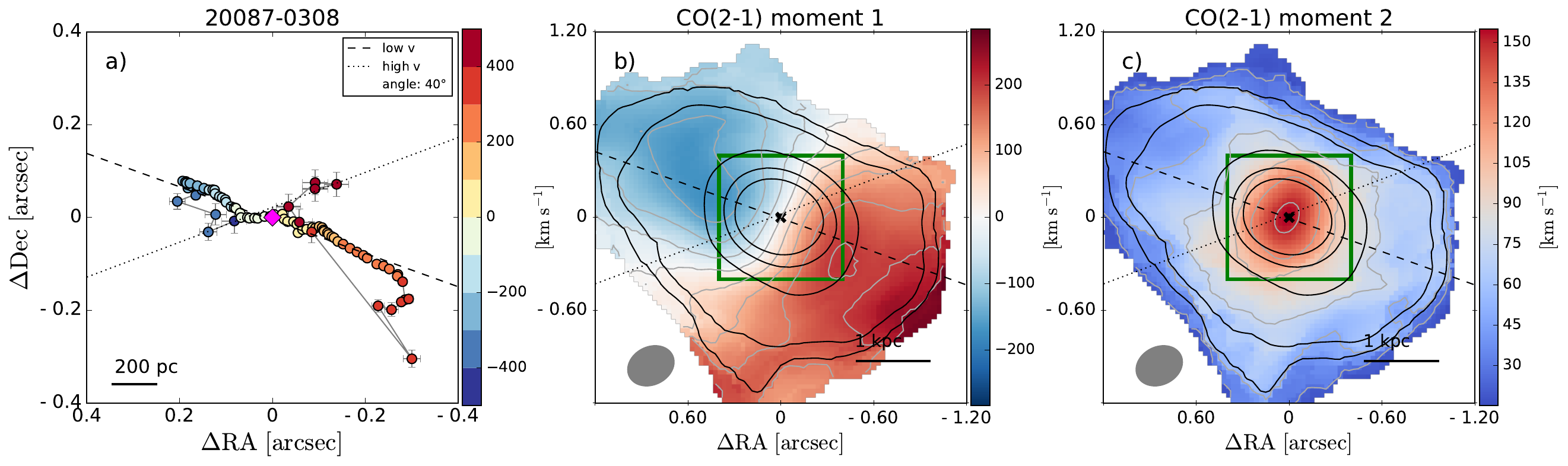}
\includegraphics[width=0.245\textwidth]{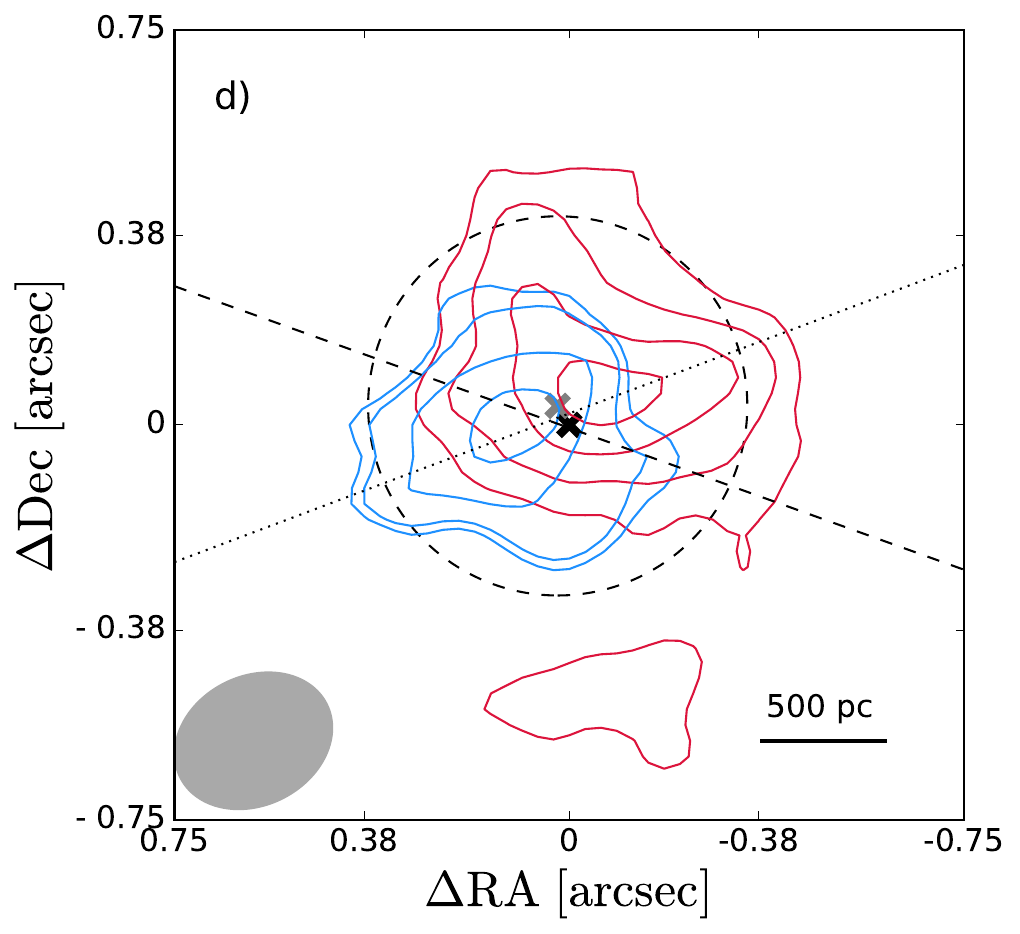}
\includegraphics[width=0.3\textwidth]{Figures/Spectra/IRAS_20100-4156_CO21_spectrum_zoom_s1.pdf}
\includegraphics[width=0.3\textwidth]{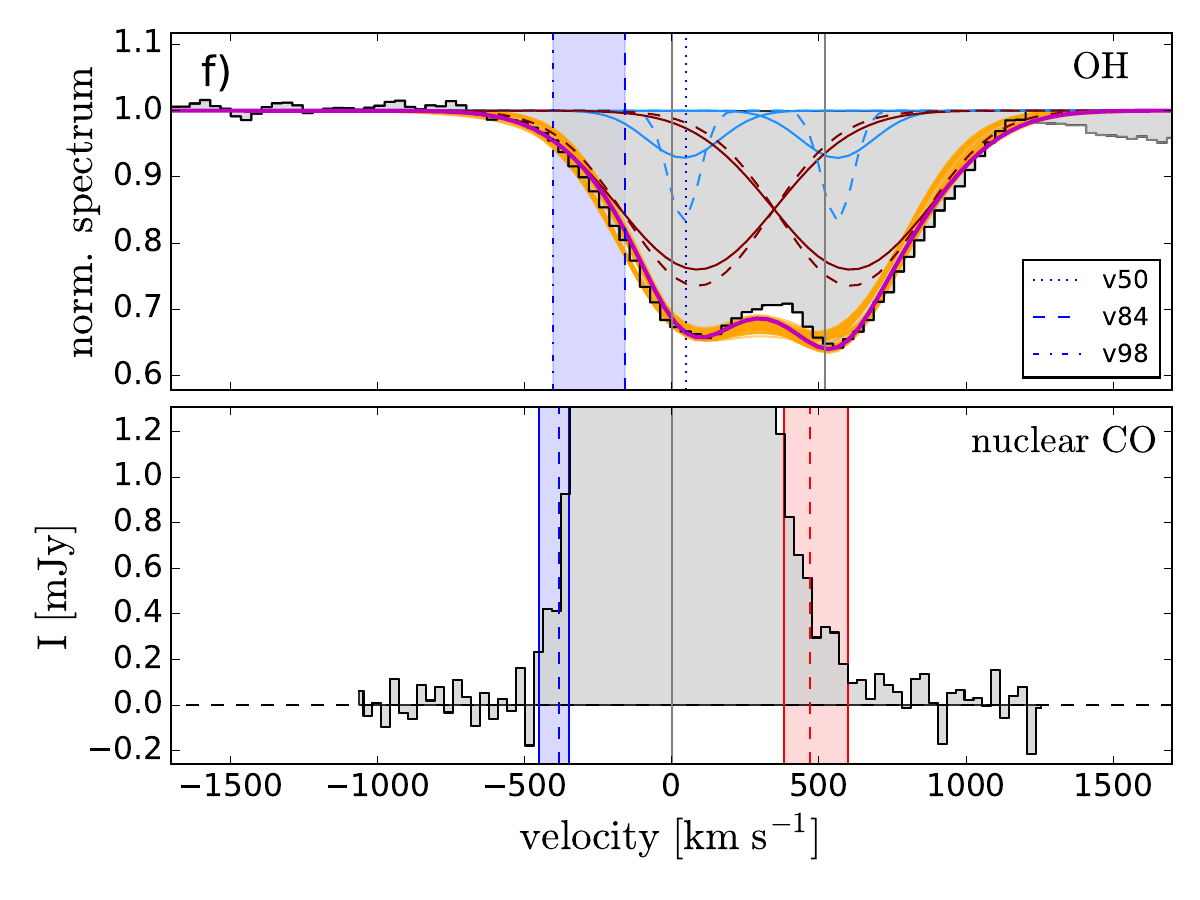}
\caption{Same as Figure~\ref{fig:spectroastrometry},  but for 20087-0308,  a target with outflow direction not perpendicular to the kinematic major axis. }
\label{fig:spectroastrometry2}
\end{figure*}

We binned together the velocity channels in order to achieve a minimum S/N of five,  necessary to reliably determine the position of the peak of the emission. 
To determine the peak position in each binned channel, we first identify the spaxel with the highest flux. Then, we consider a region of $5\times5$ spaxels centred on the maximum spaxel and we perform a 2D Gaussian fit. In this way, we can determine the position at sub-pixel scales. 
 In some targets,  the CO(2-1) peak map presents a dip in the centre (see previous section) in the central velocity-channels.  Thus,  the position determined from the pixel with the maximum flux is not representative of the centroid of the emission.  In these cases (07251-0248, 10190+1322, 13120-5453,  13451+1232,  19297-0406,  19542+1110,  and 20414-1651), we determine the centroid emission by fitting  a 2D Gaussian.  The uncertainties on the centroids are calculated as $\Delta x= FWHM_{beam}/(2\times S/N)$, where $FWHM_{beam}$ is the beam size and S/N is the signal-to-noise ratio of the binned channel \citep{Condon1997}. 
The spectro-astrometry analysis of  13120-5453 and 20100-4156 SE  are shown in Figure~\ref{fig:spectroastrometry} and  \ref{fig:spectroastrometry2}  as examples.

We perform a linear bisector fit of the centroids of the low-velocity channels,  to determine the orientation of the kinematic major axis (see dashed line in Fig.~\ref{fig:spectroastrometry}, \ref{fig:spectroastrometry2},  and \ref{fig:spectroastrometry_app}).  As low-velocity channels,  we consider absolute velocities $|v|< 300$~\kms. 
 If there are channels at $|v|< 300$~\kms\ whose position deviates significantly from the direction described by the other low-velocity channels  (e.g.  00091-0738 N), we do not include them in the fit of the kinematic major axis.
 To determine the direction of the high-velocity gas,   we perform a bisector fit to the high-velocity centroids  (see dotted line in Fig.~\ref{fig:spectroastrometry}, \ref{fig:spectroastrometry2},  and \ref{fig:spectroastrometry_app}).   In general,  we consider as high-velocity the channels with  $|v|> 300$~\kms, and we exclude channels that do not agree with the direction of the highest-velocity centroids or that follow the directions of the kinematic major axis. 
In some cases,  the centroids of the blue-shifted and red-shifted high-velocity channels  occupy a very similar region (e.g.   13451+1232~W or 14378-3651),  
 which prevents us from determining the direction of the outflow axis.  It is possible that the emission of the high-velocity gas is compact and unresolved,  or that the outflow is pointing  towards our line of sight,  so that the blue- and red-sides overlap. 
 In some cases, we observe some gas deviating from the kinematic major axis,  however, because of its low relative velocity,  $\sim200-300$~\kms, it is more likely due to a tidal tail than to an outflow  (e.g.  00091-0738~N, 01572+0009).

\subsection{Comparison of position angles of the molecular gas,  stellar and ionised gas disks}

\begin{figure}[h]
\centering
\includegraphics[width=0.5\textwidth]{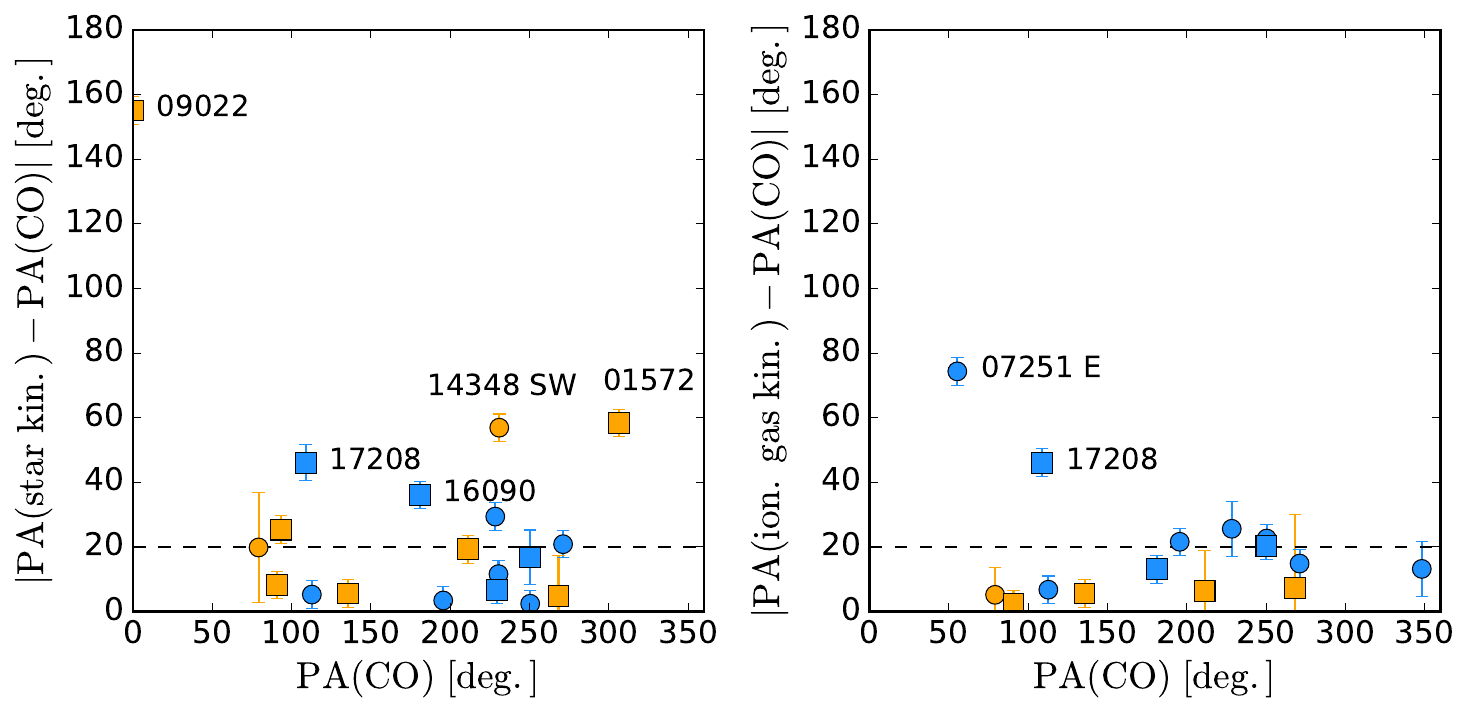}
\caption{Absolute difference between the position angles (PAs) of the kinematic major axis derived from CO and the PAs derived from the stellar kinematics (\textit{left}), and the ionised gas kinematics (\textit{right}).  The PAs of the stars and ionised gas are taken from \citet{Perna2022}. 
The dashed lines show a difference of  $20^{\circ}$. The names of galaxies with large PA differences ($\gg 20^{\circ}$) are shown on the figure.  The colours and shapes of the symbols are as in Fig.~\ref{fig:sample}.}
\label{fig:comp_PA} 
\end{figure}

From the fit of the spectro-astrometry low-velocity channels, we derive the position angles (PAs) of the kinematic major axis of the molecular gas disk  (listed in Table~\ref{tab:ALMA_obs}). We obtain the PA of the major axis on the receding half of the galaxy,  measured  east of north (anticlockwise). 
We compare these PAs with the ones of the stellar and ionised gas (traced by H$\alpha$) disks presented in \citet{Perna2022}. 
Figure~\ref{fig:comp_PA} shows the absolute differences between  PA(CO) and the  PA derived from the stellar and  ionised gas kinematics.

Overall, there is a general agreement between these measurements,  with differences within $\sim~20^{\circ}$,  hence it is  consistent with what observed in non-interacting galaxies  \citep[see ][]{Perna2022}.  
We identify only six outliers with PA differences $\gg 20^{\circ}$ (01572+0009, 07251-0248 E, 09022-3615,14348-1447 SW,  16090-0139, and 17208-0014).
The PAs of CO are measured on smaller scales  ($\sim$1~kpc),   compared to the scales used to derive the PAs of the stellar disk and ionised gas ($\sim$5-10~kpc). Thus,  we expect to see some differences between the PAs, especially considering the fact that our targets are mergers or interacting systems, many of which do not show regular rotating disk \citep[only 27\% and < 50\% of nuclei in the PUMA sample show  rotating disks in the  ionised gas and stars, respectively, ][]{Perna2022}.  
In summary,  we find that in most of the targets the molecular gas disk has an orientation (PA)  similar to the stellar and ionised gas disk.

\subsection{Outflow velocity range definition}
\label{sec:outflow_def}

One of the main goals of this work is to identify and characterise high-velocity ($> 300$~\kms) outflows of  cold molecular gas,  which produce broad wings in the line profile.
Moreover,  outflowing gas does not have to follow the disk rotation, thus it can be identified as high velocity gas that does not follow the main rotation pattern of the galaxy.
 For each nucleus, we define the velocity range where a potential outflow is located, using the spectro-astrometry analysis and the integrated CO(2-1) spectrum.
We define the minimum ($v_{min}$) and maximum ($v_{max}$) velocities of the outflow (separately for the blue and red part of the outflow) and consider the emitting gas in the [$v_{min}, v_{max}$] velocity range as part of the outflow.

To define $v_{min}$ and $v_{max}$ for each nucleus,  we perform the following procedure,  separately for the blue-shifted and red-shifted emission.
We used the spectro-astrometry plot to select the minimum velocity at which the centroid position of the gas starts to deviate from the direction of the major axis of rotation.  In particular, we look for velocity channels whose centroid position deviates significantly from the direction of the kinematic major axis, or that does not follow the rotation pattern (from blue to red). 
The $v_{min}$  values are in the range $|v_{min}|=180-450$~\kms.
To define $v_{max}$,  we started from the channel corresponding to $v_{min}$ and we continued adding velocity channels to create the map of the high-velocity emission,  until the peak S/N of the integrated map starts to decrease.  
 We checked by looking at the integrated CO(2-1) spectrum that we are not missing significant emission at $v>v_{max}$ due to a particular low S/N channel. The $v_{max}$  values are in the range $|v_{max}|=300-800$~\kms. 
Figure~\ref{fig:spectroastrometry},  \ref{fig:spectroastrometry2},  and \ref{fig:spectroastrometry_app}  show the CO(2-1) spectra with the blue-shifted and red-shifted outflow velocity ranges highlighted with blue and red shaded areas.  
The integrated maps of the blue- and red-shifted outflow channels are shown as contours in the bottom-left panel of the figures.

One caveat of our analysis is that since we are observing projected velocities,  we are not sensitive to high velocity gas in the plane of the sky. 
Additionally, with our method we are not considering outflowing gas with low projected velocities,  that is with velocities $v$ such that $v_{min}(blue)< v< v_{min}(red)$,  because it overlaps with the velocities of the rotating disk. In order to identify this gas, we would need to model the rotation of the system to identify non-rotating gas \citep[e.g.][]{Brusa2018, Gao2021, RamosAlmeida2022}.  We plan to investigate this in a future work.

\subsection{Measurements of the outflow properties and energetics}
\label{sec:outflow_Q}

In this section, we describe the method we use to measure the main outflow parameters: outflow radius ($R_{out}$),  outflow velocity (\vout),  and molecular gas mass in the outflow (\Mout).  We use these parameters  to derive the outflow energetics: mass outflow rate (\Mrate),  mass-loading factor ($\eta=\dot{M}_{out}/SFR$), outflow momentum rate ($\dot{P}_{out}$),  and kinetic luminosity ($L_{out}$).
 In the following, we explain how we measure the `projected' \Rout\ and  \vout.  We discuss the inclination corrections in Sec.~\ref{sec:inclination}.

Different methods have been used in the literature to separate the outflow and rotating disk emission.
A possible method \citep[used for example by][]{Pereira-Santaella2018} consists in  subtracting the flux belonging to the rotating disk,  by fitting the central velocity channels with one or two Gaussians and then considering only the flux in the residuals as part of the outflow.  However,  this method may underestimate the outflow mass as outflow flux with low (projected) velocities is generally assigned to the disk regardless of the actual position of the emission.
 An alternative method used in the literature consists in fitting the line profile using a systemic Gaussian component and a broader Gaussian component for the outflow \citep[e.g.][] {Fluetsch2019}. This method may be overestimating the flux of the outflowing gas, since it considers that a large portion of the outflowing gas is at low velocities.

Since the resolution of the observations allows us to determine the position of the gas and to identify the gas that is not following the galaxy rotation, we prefer to consider only the gas with high velocities as part of the outflow.
Moreover,  most of the line profiles of our targets  cannot be well fitted using a simple model with only one or two Gaussians (see Fig.~\ref{fig:spectroastrometry}, ~\ref{fig:spectroastrometry2} and ~\ref{fig:spectroastrometry_app}).  The line profiles are asymmetric and show multiple peaks, which could also include self-absorption (see Sec.~\ref{sec:moment_maps}). 
Thus, to measure the outflow gas mass,  we consider the total flux in the high-velocity channels,   highlighted  in the blue and red  shaded regions  on the spectra (see Fig. ~\ref{fig:spectroastrometry}, previous section), without subtracting the low-velocity Gaussian fit.

In Section~\ref{sec:caveats}),  we discuss the systematic effects affecting the derived outflow quantities depending on the different methods. 
A comparison of our outflow parameters with the ones reported by \citet{Fluetsch2019} and \citet{Lutz2020} is shown in Section~\ref{sec:comp_lit}.\\

\noindent \textbf{Outflow size}: 
  to measure the outflow radius, we fit a 2D Gaussian model to the high-velocity maps,  obtained integrating the flux over the high-velocity channels (see bottom left panel in Fig.~\ref{fig:spectroastrometry}),  separately for the blue and red part. 
 To take into account the beam size and obtain the `intrinsic radius',  we convolve our model with a 2D Gaussian with the shape of the ALMA beam, and we fit this `convolved model' to the maps.   The  radius of the blue (red) part of the outflow  is defined as:
 \begin{equation}
 R_{out}^b = d_c^b + \frac{FWHM^b}{2},
 \end{equation}
 where $d_c^b$ is the centroid distance from the nucleus and $FWHM^b$ is the average  size (deconvolved from the beam) of the two axes of the 2D Gaussian fit. 

 The outflow radius  $R_{out}$ is the mean of the  radii of the blue- and red-shifted wings:
  \begin{equation}
 R_{out} = 0.5 \cdot(R_{out}^{b}+R_{out}^{r}),
 \end{equation}
The outflow radii are shown in Fig.~\ref{fig:spectroastrometry}  as dashed circles (bottom left panel).
This method is analogous to the method used by \citet{Lutz2020},  although applied here to higher spatial resolution data (400~pc vs.  700~pc), which is enough to resolve the outflow structure.  
 Due to the limited S/N of the single channels, it is not possible to measure the radius in each channel, which would give a more accurate measurement  to derive the mass outflow rate.

 We also  measure the maximum  extent of the outflow,  by taking the maximum distance from the continuum position reached by the 3$\sigma$ contour. 
 We subtract the beam size in quadrature to obtain the `intrinsic' distance.  Then,  we take the mean between the radius of the red- and blue-shifted channels as the representative maximum radius of the outflow ($R_{3\sigma}$).

The outflow radii  $R_{out}$ measured from the 2D Gaussian fit are in the range $0.18-0.94$~kpc ($0.1-1.5$").
The maximum outflow radii  $R_{3\sigma}$  are in the range $ 0.1-2.1$~kpc ($0.05-2.5$").
The  ratio of the observed $R_{3\sigma}/R_{out}$ is in the range 1-3, with a mean of 1.45.  We note that in some cases the $R_{3\sigma}$ values reported in Table~\ref{tab:outflow_Q} are smaller than $R_{out}$. This is due to the different way used to deconvolve the beam and to the uncertainty on the Gaussian fit. 
 We calculate that $>50$\% of the outflow flux is within $R_{out}$, with a median of 76\%.  
Given that most of the outflow flux is located within $R_{out}$,  we decide to use  $R_{out}$ to compute the mass outflow rate and the energetics. 
If we were to use the outflow flux within $R_{3\sigma}$ instead of within $R_{out}$ to calculate the outflow mass,  \Mout\ would increase on average by a factor of 1.3 (0.11~dex).
The mass outflow rate is proportional to \Mout/\Rout, thus the larger molecular mass included within $R_{3\sigma}$ is counterbalanced by the larger radius.
If we were to use $R_{3\sigma}$ and the outflow flux within this radius, we would obtain very similar values (less than 8\% smaller,  $-0.04$~dex)  compared to the  \Mrate\ estimated using $R_{out}$.

We do not attempt to correct the radii of the single targets for inclination, since information about the inclination is not available for the full sample (see discussion in Sec.~\ref{sec:inclination}).\\

\noindent \textbf{Outflow mass}: To derive the outflow mass,  we extract the central  spectrum from a radius equal to the observed $R_{out}$ (not deconvolved from the beam) and  we integrate the flux in the high-velocity channels between \vmin\ and \vmax,  separately for the blue and the red part of the outflow.  The central spectra are shown in Fig.~\ref{fig:spectroastrometry}, ~\ref{fig:spectroastrometry2} and ~\ref{fig:spectroastrometry_app} (bottom row, middle panel).  Then, we sum the flux of the blue and red part of the outflow to obtain the total outflow flux ($F_{out}$).

We estimate the uncertainties on $F_{out}$ by extracting a spectrum from a region away from the source and measuring the standard deviation of the flux density in the high-velocity channels ($\sigma$ in units of mJy) .  The  uncertainty on $F_{out}$ is :
\begin{equation}
F_{out, err} = \sigma \cdot \Delta v_{ch} \cdot \sqrt{N_{ch}},
\label{eq:F_err}
\end{equation}
where $ \Delta v_{ch}$ is the width of a velocity channel in \kms, and $N_{ch}$ is the number of velocity channels in the high-velocity windows.
We transform the CO(2-1) flux into  luminosity (in units of K \kms\ pc$^2$)  using the formula:
\begin{equation}
 L'_{CO}= 3.25\cdot 10^7 S_{CO} \nu_{rest}^{-2} D_L^2 (1+z)^{-1},
 \end{equation}
  where $S_{CO}$  is the velocity-integrated CO line flux in Jy \kms,   $\nu_{rest}$ is the line rest-frequency in  GHz,  $D_L$ is the luminosity distance in Mpc,  and $z$ is the redshift \citep{Solomon1997}.
We convert the CO(2-1) luminosity to CO(1-0) luminosity using $r_{21} = L'_\text{CO(2-1)}/L'_\text{CO(1-0)} = 0.91$ \citep{Bolatto2013}. Then we multiply it by the ULIRGs-like CO-to-H$_2$ conversion factor $\alpha$= 0.78 M$_\odot$/(K \kms\ pc$^2$),  to obtain the outflow  molecular (H$_2$) gas mass \Mout.
Although, we note that the cold molecular gas conditions in the outflow likely differ from those in the disk and, therefore,  the CO-to-H$_2$ outflow conversion factor is uncertain \citep[see e.g.][]{Pereira-Santaella2020}.\\

\noindent  \textbf{Mean outflow velocity}: We calculate the mean velocity of the outflow separately for the blue- and red-shifted high-velocity wings,  by taking the flux-weighted mean of the velocity in the channels identified as part of the blue-shifted  (or red-shifted) emission (see Sec.~\ref{sec:outflow_def},  Fig.~\ref{fig:spectroastrometry},  middle panel of the bottom row):
\begin{equation}
v_{out} =  \frac{ \sum_{i} v_i \cdot F_{i}}{ \sum_{i}  F_{i}},
\label{eq:vout}
\end{equation} 
where $v_i$ is the velocity of  channel $i$ and $F_{i}$ is the CO(2-1) flux density in that channel.

{Different methods have been used in the literature to estimate the outflow velocity (see Sec.~\ref{sec:caveats}). 
We decide to use this `flux-weighted velocity' to calculate the mass outflow rate,  because it is independent from the modelling of the emission line profile and it gives more weight to the velocities at which most of the emission takes place.
We note that the outflow velocities measured with this method  are sensitive to the choice of the velocity window defined as `high-velocity gas'. 
In Sections~\ref{sec:caveats} and \ref{sec:comp_lit} we explore this possible bias.
\\

\noindent \textbf{Mass outflow rate}:
For the red and blue part of the outflow separately,  we calculate the mass outflow rate (in units of [\Msun\  yr$^{-1}$]) using the formula:
\begin{equation}
\label{eq:M_rate}
\dot{M}_{out} =  \frac{|v_{out}| \cdot M_{out}}{ R_{out}}= \frac{ \sum_{i} |v_i| \cdot M_{i}}{ R_{out}},
\end{equation} 
where $R_{out}$ is the outflow radius,  $M_{i}$ is the H$_2$ gas mass  in the channel with velocity  $v_i$, and the sum is over the velocity channels identified as part of the blue-shifted  (or red-shifted) emission (see Fig.~\ref{fig:spectroastrometry}). The total mass outflow rate is the sum of the blue and red $\dot{M}_{out}$. 
  We note that this formula corresponds to the assumption that the outflow has started at a point in the past  ($-t =-$\Rout/\vout) and has continued with a constant \Mrate\  \citep{Rupke2005b, Veilleux2005, Lutz2020}. 
 Under this assumption,  the average volume density  of the outflowing gas decreases with radius ($\propto R^{-2}$).
 Assuming that the outflowing gas fills a spherical or multi-conical volume with  a constant average volume density,  would increase $\dot{M}_{out}$ by a factor of three \citep[e.g. ][]{Maiolino2012, Cicone2014, Lutz2020}.

In the cases where the outflow is not detected,  we estimate 3$\sigma$ upper limits on $\dot{M}_{out}$ as:
\begin{equation}
\dot{M}_{out, UL} =  3\cdot\frac{\langle v_{out}\rangle \cdot M_{out, err}}{\langle R_{out}\rangle},
\end{equation} 
where $\langle v_{out}\rangle= 390$~\kms and $\langle R_{out}\rangle=0.52$~kpc are the median outflow velocity and radius of our sample.   
$M_{out, err}$ is calculated based on $F_{out, err}$ (eq.~\ref{eq:F_err}),  measured from the spectrum extracted from a radius $\langle R_{out}\rangle$. 
 The mass outflow rates are in the range $\sim5$ to $\sim300$ \Msun\ yr$^{-1}$.

 We compare the outflow properties (\vout, \Rout\ and \Mout) measured from the blue and red-shifted wings.  The \vout, \Rout\ and \Mout\ measured from the red and blue parts are similar within a factor of 1.6,  2.4 and 1.2 respectively. The mean and corresponding standard deviation of the ratio of the red- and blue-shifted outflow quantities are: $0.97\pm0.16$, $1.05\pm0.51$,  $0.99\pm0.04$, respectively.  

\subsubsection{Inclination corrections} 
\label{sec:inclination}
We do not attempt to correct the outflow radii and velocities of the single targets for the inclination, since this information is not available for all  objects. 
In order to apply an inclination correction to \vout\ and \Rout, we need to know the inclination of the outflow with respect to our line of sight.  Even if we knew the inclination of the molecular gas disk,  in order to use this information we would need to know the inclination of the outflow with respect to the disk.  In Sec.~\ref{sec:outflow_det} we discuss the projected angles between the outflow axis and the major kinematics axis.  For a handful of targets (6), there is evidence that the outflow may be perpendicular to the major kinematics axis.  However, only for one of them we have a measurement of the ionised gas (or stellar) disk inclination from \citet{Perna2022}.   Thus, we decide not to attempt to correct for inclination  \vout\ and \Rout\ of the single targets, and all the quantities reported here are the `projected' ones.

However, we derive an average inclination correction that we apply to the average outflow properties of the sample. 
To convert the observed (projected) mean outflow velocity to intrinsic velocity,  we  need to divide \vout\ by $\sin(i)$, where $i$ is the inclination.  Analogously, the observed outflow radius needs to be divided by $\cos(i)$ to recover the intrinsic value.  Following \citet{Law2009},  who considered the average for a collection of objects oriented isotropically in space, the average  correction for the velocities is $1/\left<\sin(i)\right>=1/0.79=1.27$.  Analogously,  we calculated the average correction for the radius: $1/\left<\cos(i)\right>=2$. 
We use these values to correct the mean outflow velocity and mean outflow radius reported in table~\ref{tab:outflow_Q}.

It is not possible to derive the average inclination correction for the mass outflow rate \Mrate\ in a similar way,  since the calculation of the integral over the entire solid angle gives infinity. However, the average inclination correction for the dynamical time ($t_{dyn}= R_{out}/v_{out} \propto M_{out}/\dot{M}_{out})$ is unity \citep[see][]{Cicone2015}.


\begin{table*}
\centering
\caption{CO(2-1) observed outflow properties.  }
\setlength{\tabcolsep}{3pt}
\begin{tabular}{llccccccccccc}
\hline

IRAS name  & n. &  [$v_{min}, v_{max}$] (b)  &  [$v_{min}, v_{max}$] (r) &  $S_{CO}$ (b) & $S_{CO}$ (r)& $R_{3\sigma}$ & $R_{out}$ &  $v_{out}$&  $\log M_{out}$  &  $\log \dot{M}_{out}$ & angle$_{out}$\\ 
  &  & [\kms] &  [\kms] & [Jy \kms] & [Jy \kms]  & [kpc] &  [kpc] & [\kms]  & [\Msun] &  [\Msun yr$^{-1}$] & [deg.] \\
 (1) & (2)  & (3) & (4) &(5) & (6) & (7) & (8) & (9) & (10) & (11) & (12) \\ 
  \hline \hline

00091-0738 & S & - & - &- & - & - & - & - &<7.26 & <1.15 & - & \\
 & N & - & - &- & - & - & - & - &<7.20 & <1.08 & - & \\
00188-0856 &   & - & - &- & - & - & - & - &<7.33 & <1.22 & - & \\
00509+1225 &   & - & - &- & - & - & - & - &<6.62 & <0.50 & - & \\
01572+0009 &   & - & - &- & - & - & - & - &<7.62 & <1.51 & - & \\
05189-2524 &   & [-200,-600] & [200,350] & 1.27$\pm$0.05 & 1.27$\pm$0.04 & 1.39 & 0.83$\pm$0.01 & 294$\pm$16 & 7.66$\pm$0.01 & 1.22$\pm$0.01 & - & \\
07251-0248 & E & [-320,-500] & [300,550] & - & 0.32$\pm$0.01 & 0.64 & 0.17$\pm$0.01 & 377$\pm$12 & 7.37$\pm$0.01 & 1.72$\pm$0.01 & - & \\
 & W & - & - &- & - & - & - & - &<6.87 & <0.75 & - & \\
09022-3615 &   & [-400,-700] & [300,500] & 6.87$\pm$0.05 & 2.05$\pm$0.04 & 1.68 & 0.94$\pm$0.01 & 423$\pm$8 & 8.50$\pm$0.01 & 2.17$\pm$0.01 & 12 & \\
10190+1322 & E & [-290,-410] & [350,420] & 0.06$\pm$0.02 & 0.09$\pm$0.02 & 0.60 & 0.62$\pm$0.02 & 366$\pm$143 & 6.96$\pm$0.08 & 0.74$\pm$0.08 & 75 & \\
 & W & - & - &- & - & - & - & - &<6.78 & <0.66 & - & \\
11095-0238 & NE & [-300,-500] & [300,500] & 0.25$\pm$0.02 & 0.22$\pm$0.02 & 0.46 & 0.37$\pm$0.01 & 373$\pm$37 & 7.70$\pm$0.02 & 1.71$\pm$0.02 & 32 & \\
 & SW & - & - &- & - & - & - & - &<7.05 & <0.94 & - & \\
12071-0444 & N & [-250,-430] & [260,400] & 0.22$\pm$0.03 & 0.22$\pm$0.03 & 0.36 & 0.40$\pm$0.01 & 385$\pm$65 & 7.83$\pm$0.04 & 1.82$\pm$0.04 & 37 & \\
 & S & - & - &- & - & - & - & - &<7.37 & <1.26 & - & \\
12112+0305 & NE & [-250,-700] & [250,700] & 2.43$\pm$0.03 & 4.09$\pm$0.03 & 2.07 & 0.59$\pm$0.01 & 357$\pm$6 & 8.54$\pm$0.01 & 2.33$\pm$0.01 & 90 & \\
 & SW & [-200,-600] & [200,400] & 0.51$\pm$0.04 & 0.29$\pm$0.03 & 0.73 & 0.63$\pm$0.01 & 292$\pm$32 & 7.63$\pm$0.03 & 1.31$\pm$0.03 & 84 & \\
13120-5453 &   & [-300,-800] & [300,600] & 5.38$\pm$0.15 & 2.48$\pm$0.13 & 0.70 & 0.37$\pm$0.01 & 432$\pm$25 & 7.89$\pm$0.01 & 1.96$\pm$0.01 & 29 & \\
13451+1232 & W & - & - &- & - & - & - & - &<7.62 & <1.50 & - & \\
 & E & - & - &- & - & - & - & - &<7.22 & <1.10 & - & \\
14348-1447 & SW & [-250,-700] & [250,700] & 1.46$\pm$0.04 & 1.59$\pm$0.04 & 1.78 & 0.67$\pm$0.01 & 380$\pm$13 & 8.31$\pm$0.01 & 2.07$\pm$0.01 & 56 & \\
 & NE & [-300,-550] & [300,600] & 0.57$\pm$0.03 & 0.46$\pm$0.03 & 1.36 & 0.69$\pm$0.01 & 393$\pm$35 & 7.83$\pm$0.02 & 1.60$\pm$0.02 & 82 & \\
14378-3651 &   & [-200,-600] & [200,300] & 1.16$\pm$0.05 & 0.32$\pm$0.02 & 0.69 & 0.51$\pm$0.01 & 311$\pm$24 & 7.82$\pm$0.02 & 1.61$\pm$0.02 & - & \\
15327+2340 &   & [-450,-600] & [400,600] & 4.06$\pm$0.15 & 2.08$\pm$0.18 & 0.96 & 0.56$\pm$0.01 & 474$\pm$43 & 7.33$\pm$0.02 & 1.26$\pm$0.02 & 64 & \\
16090-0139 &   & [-400,-600] & [350,600] & 0.33$\pm$0.04 & 0.70$\pm$0.04 & 0.69 & 0.50$\pm$0.01 & 454$\pm$58 & 8.22$\pm$0.02 & 2.19$\pm$0.02 & - & \\
16156+0146 & NW & - & - &- & - & - & - & - &<7.32 & <1.21 & - & \\
 & SE & - & - &- & - & - & - & - &<7.47 & <1.35 & - & \\
17208-0014 &   & [-400,-800] & [390,450] & 3.18$\pm$0.17 & 0.18$\pm$0.06 & 0.49 & 0.51$\pm$0.01 & 487$\pm$148 & 7.78$\pm$0.03 & 1.77$\pm$0.03 & 3 & \\
19297-0406 & N & [-350,-500] & [350,500] & 0.28$\pm$0.02 & 0.27$\pm$0.02 & 1.01 & 0.64$\pm$0.01 & 409$\pm$42 & 7.58$\pm$0.02 & 1.40$\pm$0.02 & 49 & \\
 & S & - & - &- & - & - & - & - &<6.76 & <0.64 & - & \\
19542+1110 &   & [-250,-400] & [300,500] & 0.19$\pm$0.02 & 0.16$\pm$0.02 & 0.48 & 0.38$\pm$0.01 & 346$\pm$52 & 7.14$\pm$0.03 & 1.11$\pm$0.03 & 72 & \\
20087-0308 &   & [-350,-450] & [380,600] & 0.22$\pm$0.01 & 0.39$\pm$0.02 & 1.12 & 0.67$\pm$0.01 & 425$\pm$32 & 7.79$\pm$0.01 & 1.60$\pm$0.01 & 40 & \\
20100-4156 & SE & [-300,-800] & [250,800] & 0.90$\pm$0.04 & 1.05$\pm$0.04 & 0.76 & 0.44$\pm$0.01 & 438$\pm$26 & 8.47$\pm$0.01 & 2.48$\pm$0.01 & 70 & \\
 & NW & - & - &- & - & - & - & - &<7.39 & <1.28 & - & \\
20414-1651 &   & [-350,-440] & [300,400] & 0.18$\pm$0.03 & 0.25$\pm$0.03 & 0.44 & 0.52$\pm$0.01 & 364$\pm$84 & 7.48$\pm$0.05 & 1.33$\pm$0.05 & 25 & \\
22491-1808 & E & [-200,-400] & [200,600] & 1.69$\pm$0.01 & 0.68$\pm$0.01 & 0.92 & 0.26$\pm$0.01 & 263$\pm$5 & 8.12$\pm$0.01 & 2.14$\pm$0.01 & 57 & \\
 & W & - & - &- & - & - & - & - &<6.60 & <0.49 & - & \\

\hline
average$^{*}$ & & & & &  &1.84 & 1.07 & 485 & 8.01 & 1.89 \\
\hline
\end{tabular}
\label{tab:outflow_Q}
\tablefoot{
(1) IRAS name.
(2) Name of the nucleus. 
(3) and (4) Velocity range considered to measure the blue- and red-shifted wings of the CO(2-1) profile with respect to the systemic velocity.
(5) and (6) CO(2-1) flux in the blue- and red-shifted wings.
(7) Maximum extent of the outflow estimated from the emission above 3$\sigma$ (see Sec.~\ref{sec:outflow_Q}).
(8) Outflow radius (of  the blue- and red-shifted wings) deconvolved from the beam, but not corrected for inclination.
(9) Flux-weighted outflow velocity (see eq.~\ref{eq:vout}),  not corrected for inclination.
(10) Outflow molecular gas mass,  calculated  assuming a ULIRG-like conversion factor $\alpha_{CO}$ of 0.78 \Msun/(K\kms\  pc$^{-2}$)$^{-1}$ and $r_{21}$ ratio of 0.91 \citep{Bolatto2013}.
(11) Mass outflow rate calculated using equation~\ref{eq:M_rate}.
(12) Angle between the outflow axis and the kinematic major axis, derived from the spectro-astrometry,  for the cases where it could be determined.
$^{*}$ Average outflow properties, excluding upper limits. The average of \Rout\ and \vout\  have been corrected for inclination assuming the average corrections: $ R_{out}= R_{out}(obs)/\left< \cos(i)\right> = 2\cdot R_{out}(obs)$,  $v_{out}= v_{out}(obs)/\left< \sin(i) \right> = 1.27\cdot v_{out}(obs)$ (see Section~\ref{sec:inclination}). 
}
\end{table*}

\subsubsection{Caveats: Methods to measure outflow parameters}
\label{sec:caveats}
In this section, we discuss how the outflow parameters (in particular  \vout, \Mout\ and \Mrate) change depending on the method adopted for the measurements. Readers who are less interested in the details about the methodology and the comparison with previous works may wish to go directly to Section~\ref{sec:OH_analysis}.\\

\noindent \textit{1) Low projected velocity ($ v < 300$~\kms) gas in the outflow}: since we are not considering low projected velocities ($|v|\lesssim 200-300$~\kms), it is possible that we are missing part of the outflow flux.  
 If we were to consider also this low velocities, the outflow mass \Mout\ would increase and the flux-weighted outflow velocity \vout\ would decrease.  
Overall, we expect \Mrate\ to increase, but by a lower factor than \Mout, because the increase in \Mout\ is counterbalanced by the decrease of \vout.

To test how much this effect could affect our measurements of the outflow properties,  we consider 3D models of biconical outflows based on the models presented by \citet{Bae2016}.
The outflow is a  bicone with a half opening angle of 40$^{\circ}$.
We set the maximum velocity of the outflow to 750~\kms. 
We consider several outflow radial velocity profiles, motivated by previous works in the literature \citep[e.g.][]{ForsterSchreiber2014,  Venturi2018, Meena2021}, and different outflow inclinations with respect to the line of sight.  More details on the simulations can be found in the appendix~\ref{sec:app_sim}.
We measure \vout, \Mout\ and \Mrate\ from the total simulated profile (outflow+systemic component) considering only $|v| > 300$~\kms, to mimic the method we are using with our data.  Then, we measure the outflow quantities (\vout, \Mout,  and \Mrate) from the simulated outflow emission profile (without the systemic component) considering the full velocity range.

The  \Mout\ measured from $|v| > 300$~\kms\ are underestimated compared to the values measured from the entire velocity range by a factor of $0.2-1$ (average 0.5), while the  \vout\ are overestimated by up to a factor of 2.2 (average of 1.6). 
 Consequently,  \Mrate\ would be underestimated by up to $\sim$0.45~dex (65\%) for outflow inclinations close to the plane of the sky (90$^{\circ}$). For an inclination of 10$^{\circ}$,  the measured \Mrate\  would be underestimated  by up to 0.1~dex (see Fig. ~\ref{fig:outflow_profiles}).

\begin{figure}
\centering
\includegraphics[width=0.48\textwidth]{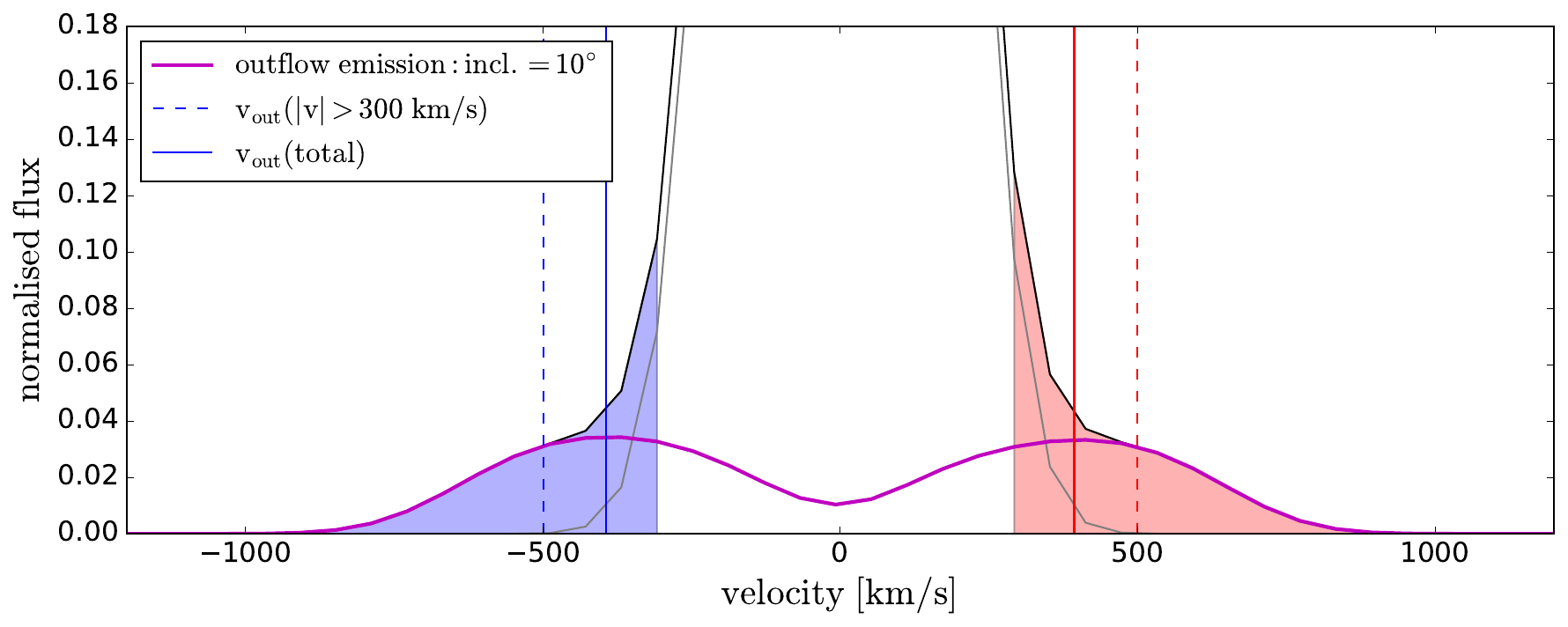}
\includegraphics[width=0.48\textwidth]{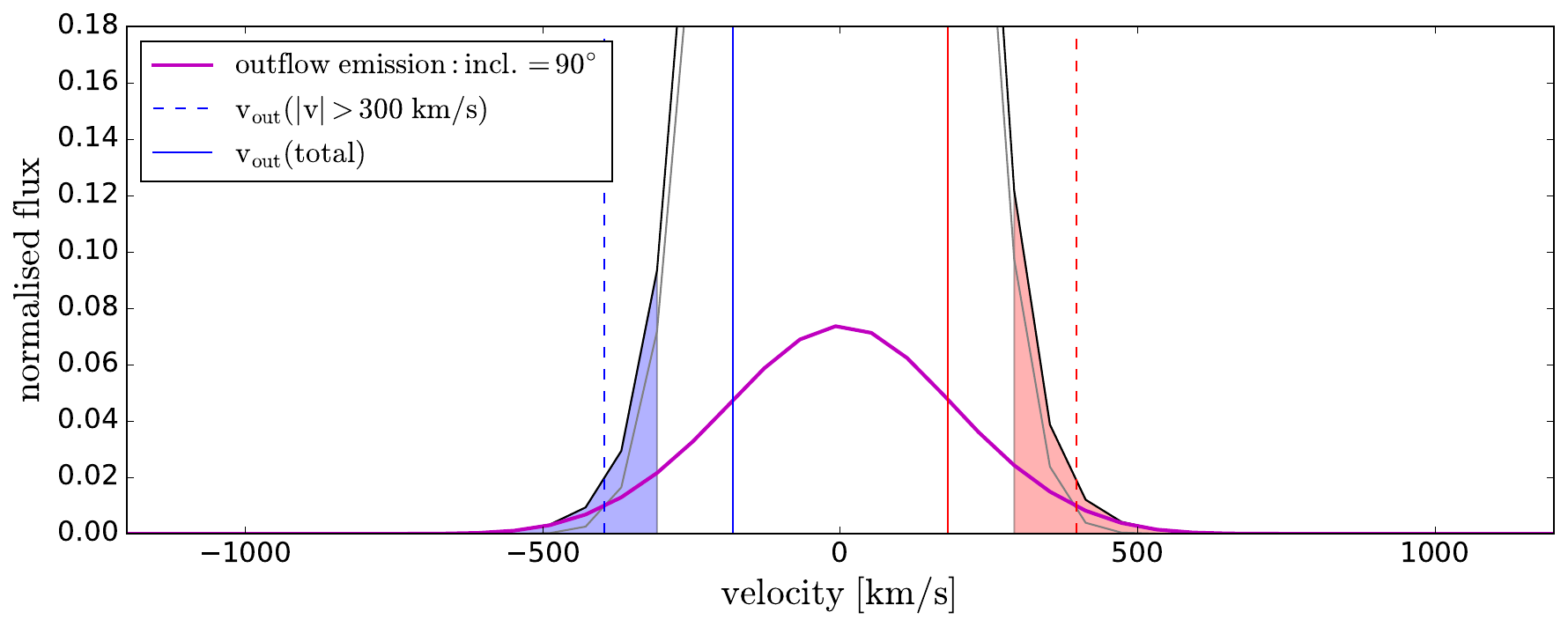}
\caption{Simulated outflow profiles for outflow  inclinations with respect to our line of sight of 10$^{\circ}$ (\textit{upper}) and  90$^{\circ}$ (\textit{bottom}). 
The outflow velocity in this particular simulation increases up to a turnover radius and then decreases. Up to four different velocity fields are considered in Appendix~\ref{sec:app_sim}. 
The magenta curve shows the outflow component,  the grey curve  the systemic component and the black curve the total profile. The vertical dashed lines show the flux-weighted \vout\ measured from the total profile at $|v|\geq300$~\kms\  (coloured in blue and red).
The vertical solid lines show the flux-weighted \vout\ measured from the outflow component over the full velocity range.
 For high outflow inclination (close to the plane of the sky),  the contribution to the outflow flux by the gas at low projected velocities ($|v|<300$~\kms)  increases.}
\label{fig:outflow_profiles}
\end{figure}

For targets with a wide CO(2-1) line  core (large FWHM),  the flux of the gas in the rotating disk can overlap with the outflow flux up to larger velocities. Indeed, there is a positive correlation between the velocity at which we start to consider the flux to be dominated by the outflow ($v_{min}$) and the FWHM of the line (Spearmann rank correlation coefficient $r=0.81$,  $p$-value$<0.1$).
We did a test to estimate how much flux we may be missing in our measurements of the outflow flux for the targets with large FWHM CO(2-1) line profiles.  In particular, we consider the 13 targets with $|v_{min}| > 300$~\kms (either in the blue or red side). 
 To estimate the amount of possible flux belonging to the outflow in the velocity range between  $|v|=300$~\kms\ and  $|v_{min}|$,  we assume the outflow flux density in this range to be equal to the value at $v_{min}$.   
 Using this assumption, we estimate the outflow parameters  (\Mout, \vout, and \Mrate) starting from $v_{min}=300$~\kms.
We find that  the value of \Mout\ increases by 0.28~dex on average (maximum 0.67~dex), while \vout\ decreases by a factor of $-0.07$~dex on average (minimum $-0.11$~dex).  So, the \Mrate\ estimates increase by $0.2$~dex on average (maximum 0.6~dex). \\

\noindent \textit{2) Possible overestimation of \Mrate\ due to the rotating disk contribution at $|v|=250-300$~\kms}: Since we do not model and  subtract the disk rotation,  it is possible that at low velocities ($250-350$~\kms) we are including in the outflow flux some flux emitted by the gas  in the rotating galaxy disk.
To test how large this contribution could be, we  subtract from the spectra the flux due to rotation estimated by modelling the core of the emission profile (with absolute velocities smaller than $\sim300$~\kms) with one,  two or three Gaussians \citep[e.g.][]{Pereira-Santaella2018}.  
Then, we compute the outflow parameters (\Mout, \vout, \Mrate) from the residuals,  considering the velocity range between $v_{max}$ and the velocity ($v_{min}$) at which the residuals approach zero.  
The outflow masses vary in the range  $-1.0$~dex to 0.5~dex ($-0.12$~dex on average).  In some cases,  the measured \Mout\ increases because we can extend the outflow velocity range to smaller velocities,  since there is no risk of including flux from the rotation. The outflow velocities vary by less than $\pm 50$~\kms.
The  \Mrate\ vary between $-0.82$~dex and $+0.45$~dex  ($-0.11$~dex on average). \\

\textit{3) Different methods to estimate the outflow velocity}:
In this work, we adopted the `flux-weighted' outflow velocity definition to compute the mass outflow rate. 
Other works instead have used different definitions of `maximum outflow velocity'  \citep[e.g.][]{Fluetsch2019, Lutz2020}.  
 If we assume a biconical outflow with constant gas velocity within the outflow, the range of velocities observed in the broad wings of the CO profile would be solely due to different orientation angles of the outflow gas clouds with respect to our line of sight. The part of the outflow closer to our line of sight would have the highest observed velocity. Thus, one could assume that this maximum velocity is the closest to the true/intrinsic velocity.  We do not know the  true radial profile of the velocity or the geometry of the outflowing gas for our targets.  However, we can estimate how much our `flux-weighted outflow velocity' differ from the `maximum outflow velocity' assumption in our sample.
We consider two definition of the maximum velocity.  Both definition require the fit of the line profile with multiple Gaussian components: a broad component for the outflow,  and other components for the systemic emission.  
The first definition that we consider is the prescription from \citet{Rupke2005a}, used for example by \citet{Fluetsch2019}:
\begin{equation}
v_{max, FWHM} = | \Delta v| + \frac{FWHM_{broad}}{2},
\label{eq:vout_Fluetsch2019}
\end{equation}
where $|\Delta v|$ is the shift of the broad Gaussian outflow component with respect to systemic velocity and $FWHM_{broad}$ its full width at a half maximum. The second definition, adopted for example by \citet{Lutz2020}, is:
\begin{equation}
v_{max, FW10} = | \Delta v| +\frac{FW_{10\%}}{2},
\label{eq:vout_Lutz2020}
\end{equation}
 where $FW_{10\%}$ is  the full width of the broad component at a tenth of its peak.
 We measure $v_{max, FWHM}$ and $v_{max, FW10}$ for our sample of galaxies with outflow detection.  We use up to three Gaussian components to model the systemic emission.  We stress that in most of the cases, the parameters of the Gaussian components are highly degenerate. Thus, the derived  maximum velocity can vary depending on the assumptions used in the fit. 
 In Fig.~\ref{fig:comp_vout_vmax} we compare the flux-weighted \vout\ with the maximum velocity  $v_{max, FWHM}$  and $v_{max, FW10}$ for our sample.  The ratio $v_{max, FWHM}$/\vout\ varies between  0.5 and 2.0, with a median 0.9.  The ratio $v_{max, FW10}$/\vout\ instead is almost always larger than one, with a maximum of 3.2 and a median of 1.6.  Assuming that    $v_{max}$ is the closest measure of the `true' outflow velocity, it does not need to be corrected for inclination, while the observed (projected) \vout\ would need to be corrected  by an average factor of 1.3 (see Sec.~\ref{sec:inclination}).  This factor can account for most of the average difference between $v_{max, FW10}$ and \vout.\\
 
\begin{figure}
\centering
\includegraphics[width=0.5\textwidth]{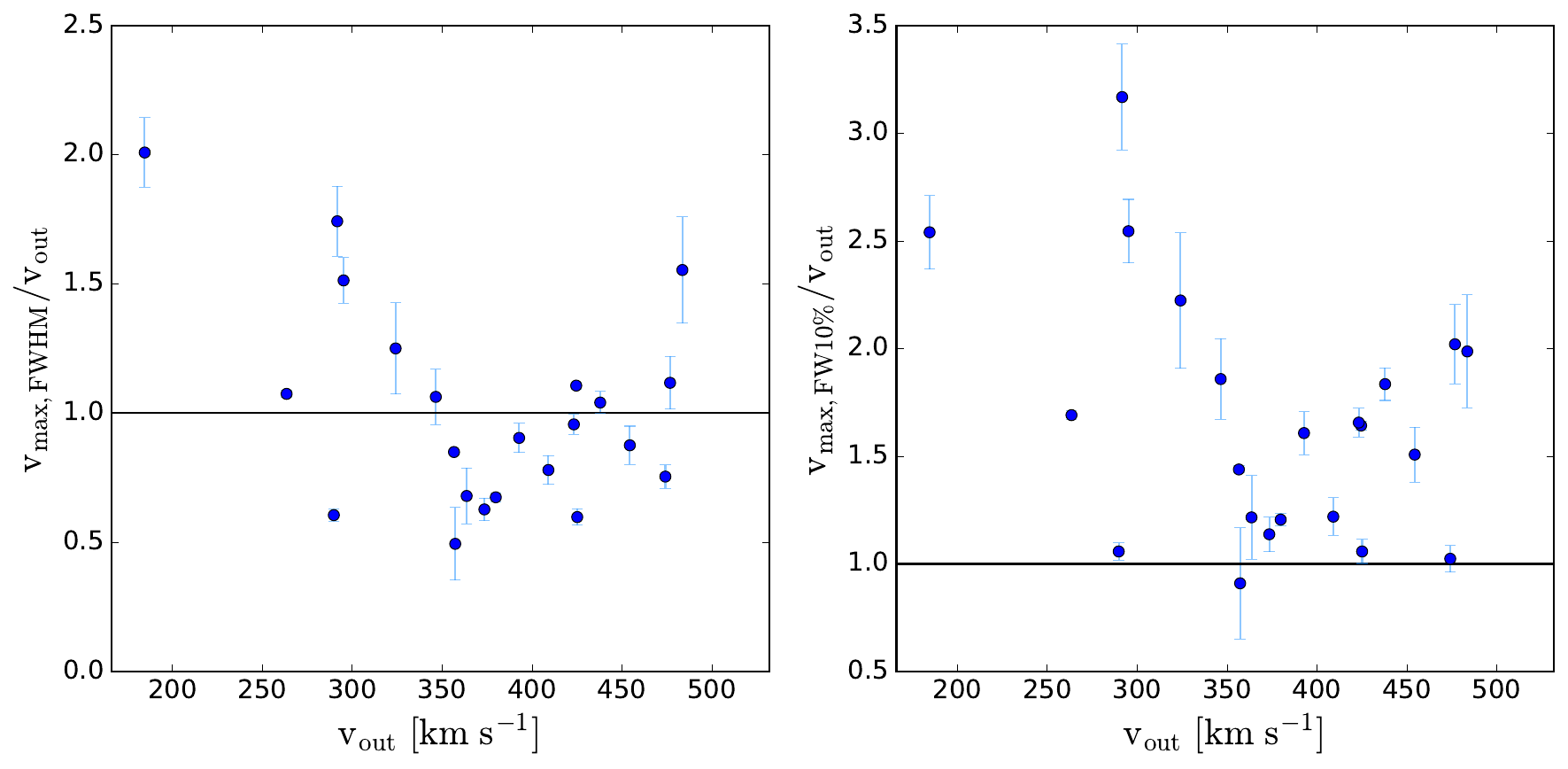}
\caption{Comparison of different methods for measuring the outflow velocity.  \textit{Left:} Ratio between the maximum outflow velocity $v_{max, FWHM}$ (eq. ~\ref{eq:vout_Fluetsch2019}) and the  flux-weighted outflow velocity ($v_{out}$) in our sample.   
\textit{Right:} Ratio between  $v_{max, FW10\%}$  (eq. ~\ref{eq:vout_Lutz2020}) and \vout. The horizontal lines show the 1-to-1 ratio.}
\label{fig:comp_vout_vmax}
\end{figure}
 
We decide to avoid the methods based on the fit of the emission line profile with multiple Gaussian components,  because the components are degenerate and this will introduce further uncertainties in the measurements.

An additional caveat is related to the definition of $v_{max}$, which will impact the $v_{out}$ estimates.  In selecting $v_{max}$ based on the S/N of the integrated high-velocity maps,  it is possible that we are excluding diffuse high-velocity flux which is below the 3$\sigma$ level.  In this way,  we may underestimate \vout, especially for cases in which the outflow velocity increases radially.  Unfortunately,  with our current data it is not possible to estimate the impact of this possible additional component.  Higher sensitivity observations are needed for this purpose.  

 In Table~\ref{tab:caveat}, we summarise the average biases  in the outflow properties (\vout, \Mout\ and \Mrate) due to the effects described in 1), 2) and 3).
Since the two effects described in 1) and 2) go in opposite directions, they tend to compensate each other.  
 Taking into account the two effects,  we may be underestimating \Mout\ $\sim 0.2$~dex and \Mrate\ by $\sim 0.04$~dex on average.

Based on the variations of outflow quantities from the  tests using different methods, we estimate the typical  uncertainties on the outflow quantities.  For the outflow velocity,  we estimate a typical uncertainty of 0.1~dex, while for \Mout\ and \Mrate\ we adopt a typical uncertainty of 0.3~dex. For \Rout, we estimate a typical uncertainty of 0.2~dex, based on the difference between $R_{out}$ and $R_{3\sigma}$.

The uncertainties on $\dot{M}_{out}$ are dominated by the uncertainties on the  $\alpha_{CO}$ conversion factor,  which can be up to 0.7~dex  \citep[][]{Papadopoulos2012, Bolatto2013, Pereira-Santaella2020}, 
 and on the outflow geometry.  
 If the $\alpha_{CO}$ is more similar to the Galactic value ($\alpha_{CO}$=4.3 M$_\odot$/(K \kms\ pc$^2$)), it would imply that all our \Mout\ and \Mrate\ measurements are underestimated by a factor of $\sim5$ (0.7~dex), while if the optically thin case applies ($\alpha_{CO}= 0.35$ M$_\odot$/(K \kms\ pc$^2$)), our measurements would be overestimated by a factor of $\sim2$ \citep{Bolatto2013}.

\begin{table}
\centering
\caption{Possible biases in the outflow properties measured with our method, due to different issues.  }
\setlength{\tabcolsep}{3pt}
\begin{tabular}{lccc}
\hline

 &  \multicolumn{3}{c}{ biases in outflow properties}  \\ 
   &   \vout & \Mout &   \Mrate    \\
  & [dex] &[dex]  & [dex]   \\ 
  \hline \hline
\\
 \multicolumn{4}{l}{ 1) Not considering outflow flux at $|v|<300$~\kms\ :} \\
average: & $0.20$  & $-0.30$ &  $-0.15$ \\
range: & $(0, 0.34)$  &  $(-0.70, 0)$ & $(-0.45, 0)$ \\ 
\hline
\\
 \multicolumn{4}{l}{2) Rotating disk contribution at $|v| = 250 -300$~\kms:} \\
average: & $-0.01$  & $0.12$ &  $0.11$  \\
range: & $(-0.08, 0, 08)$ & $(-1.0, 0.5)$ &  $(-0.82,  0.45)$ \\ 
 \hline
 \\
 \multicolumn{4}{l}{3) Flux-weighted velocity instead of $v_{max}$: }\\
average: & $-0.20$  & $-$ &  $-0.20$ \\
range: &  $(-0.50, 0.30)$ & $-$ &  $(-0.50, 0.30)$ \\
\hline
\end{tabular}
\label{tab:caveat}
\end{table}

\subsection{Comparison with previous works}
\label{sec:comp_lit}
In this Section we compare the derived outflow parameters with previous works that study properties of molecular outflow in nearby ULIRGs using CO observations.

\subsubsection{Comparison with \citet{Lutz2020}}
 \citet{Lutz2020} study the outflow properties in a sample of 54  nearby ($z<0.2$) galaxies with  CO(1-0),  CO(2-1) or CO(3-2) observations,  with a range of spatial  resolutions (30~pc to 5~kpc; $0.5-5$~arcsec beam FWHM).  
 They collect 41 nearby galaxies with molecular outflow detections or upper limits from the literature.
 They also present new NOEMA and ALMA observations,  with a  spatial resolution of $\sim$700~pc (0.5-3.7~arcsec beam FWHM),  for 13 compact far-infrared galaxies from the \citet{Lutz2016} sample.   
 To derive the outflow velocity and flux, they fitted the line profile with two Gaussian components (systemic and broad (outflow) component).
 They defined the outflow velocity using Eq.~\ref{eq:vout_Lutz2020}. 
To derive the outflow flux, they integrated the broad component only over the velocity ranges for which it contributes at least 50\% of the total flux density of the line profile. The outflow radius was defined similarly to our method: $R_{out} = |\Delta R| + FWHM/2$, where $|\Delta R|$ is the distance of the outflow emission centroid from the continuum position and $FWHM$ was derived from a Gaussian spatial fit in the $uv$-plane using a velocity range that is dominated by outflow,  although,  the spatial resolution is a factor of $\sim2$ lower than in our work. 
 For the other 41  targets,  they collect information about the outflow parameters (\vout, \Rout, \Mout) from the data published in the literature, trying to be consistent with their adopted definition of these parameters and their adopted methodology to separate the flux of the  line core and high velocity wings.

There are 12 objects in common with our sample,  with data published by  \citet{Cicone2014}  \citet{Barcos-Munoz2018}, \citet{Gowardhan2018},  \citet{Pereira-Santaella2018},  and \citet{Lutz2020}.
We compare our measurements of the outflow parameters (\vout, \Rout, \Mout, \Mrate) with these works in Figure~\ref{fig:comp_Lutz2020}.
There are some discrepancy between our measurements and the literature values. 
 In particular  for two galaxies (17208-0014 and 20100-4156) the literature  $v_{out} $ is larger than 900~\kms, while we measure  $v_{out}< 500$~\kms.  For 20100-4156, it is possible that the spectral range of our observations ($v$ = [-1200, 1200]~\kms) is not wide enough to detect the emission at very high velocities  ($|v_{out}|>1000$~\kms).  Additionally, this high-velocity outflow is detected in CO(1-0),  while in CO(3-2) no outflow is detected \citep{Gowardhan2018}.  Thus, it is possible that this difference also depends on the J-level observed.  For 17208-0014, the S/N of the outflow component presented in \citet{Lutz2020} is not very high, thus the uncertainty on $v_{out}$ should be large (even though it is not reported in the paper).
For four galaxies, \Rout\ values from the literature are considerably higher than our measurements ($\sim2-6$~times higher).   For three of these galaxies, the difference is due to the different definition of \Rout\ as the maximum radius at which the outflow is detected \citep[see $R_{max}$ definition in][]{Pereira-Santaella2018}.  For the other target (20100-4156),  the difference could be due to the different spatial resolution (1.5~arcsec vs.  0.2~arcsec). 
 For Arp~220 instead \Rout\ from the literature is smaller than our measurement; also in this case the difference could be due to the different spatial resolution (0.09~arcsec vs.  1.0~arcsec).
The \Mout\ values agree within a factor of 2.5, with some of our values being smaller and other larger than the values from \citet{Lutz2020}. 
Since  \Mrate\ is proportional to  \vout\ and \Rout$^{-1}$,  the differences in \Rout\ and \vout\ tend to balance each other and lead to similar  \Mrate (within a factor of 3).

\begin{figure*}
\centering
\includegraphics[width=0.9\textwidth]{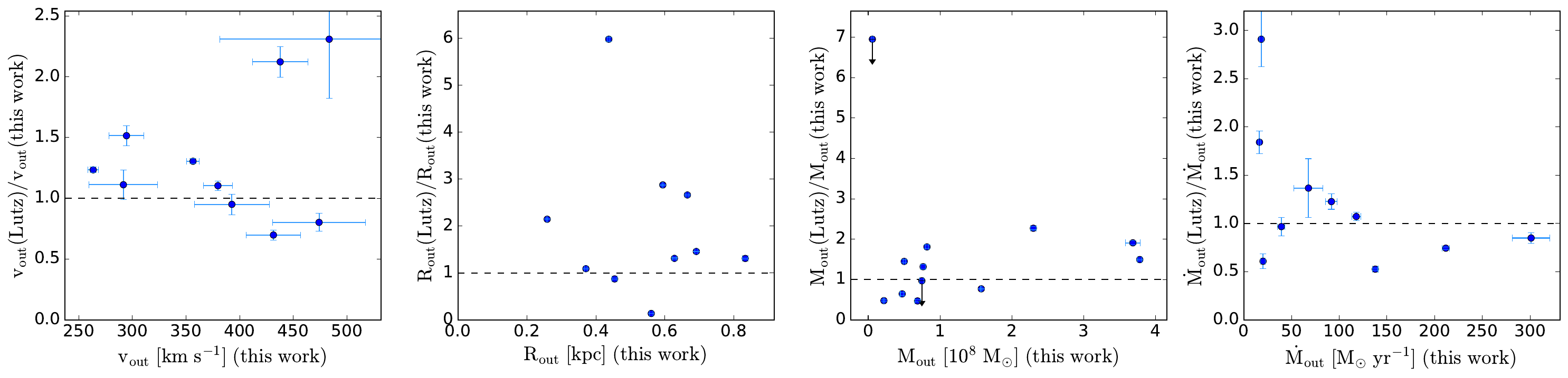}
\caption{Comparison of outflow parameters measured in this work with values reported in \citet{Lutz2020}. From left to right: outflow velocity, outflow radius, outflow mass and mass outflow rate.   
The dashed line marks the ratio of 1.
Even though there are differences in outflow velocities and radii, the mass outflow rates agree within a factor of 3.}
\label{fig:comp_Lutz2020}
\end{figure*}

\subsubsection{Comparison with \citet{Fluetsch2019}}
We also compared the \Mout\ and \Mrate\  of our sample with the values reported by \citet{Fluetsch2019} for a sample of 45 local ($z<0.2$) star-forming galaxies with CO(1-0), CO(2-1), or CO(3-2) ALMA observations.  
\citet{Fluetsch2019}  used a different method to estimate the outflow mass: they measured the outflow flux by fitting the line profile with two Gaussians, one for the core of the line and one broad component for the outflow, and they considered the total flux of the broad component as the outflow flux.  We expect that the outflow fluxes measured in this way will be higher than the ones measured with our method,  since  in addition to the flux in the wings, they are also considering  the low-velocity emission of the broad component as part of the outflow.  
 They measured the outflow velocity using Eq.~\ref{eq:vout_Fluetsch2019}.
 They fitted a 2D-Gaussian profile to the wing maps and used the beam-deconvolved major axis (FWHM) divided by two as the radius of the outflow.  Compared to our methodology, they did not include the distance between the centroids of the blue-shifted and red-shifted emission in the calculation of \Rout, thus their \Rout\ estimates are expected to be smaller.

Figure~\ref{fig:comp_Fluetsch2019_outflow_param} shows the comparison of the outflow parameters derived by \citet{Fluetsch2019}   with our results for the five sources in common between the two samples. Their $v_{out}$ tend to be higher than ours (maximum by factor of $1.7$) ,  while their \Rout\ tend to be smaller (maximum by a factor of  1/4) . The largest difference is in the outflow masses, that are larger by up to a factor of 16 (1.2~dex). This difference can be attributed to the different method used to estimate the flux belonging to the outflowing gas. This difference in \Mout\ propagates to the \Mrate.

Since the overlap between the two samples is small,   we decide to also compare the outflow properties of the two samples at the same IR luminosity. 
Figure~\ref{fig:comp_Fluetsch2019} shows  \Mout\ and \Mrate\ as a function of IR luminosity for the \citet{Fluetsch2019} sample and our sample. The diamond symbols with red borders show the average values of the two samples for different bins of IR luminosity  (with bin width of 0.5~dex). 
For the same IR luminosity  (in the range $\log L_{IR}/L_{\odot}= 12.0-12.5$),  their average \Mout\  and  \Mrate\ are $\sim0.6-0.8$~dex higher than our measurements (factor of $\times4-6$).  

Given the different methodology used to measure these parameters, the difference is not surprising. As discussed in Section~\ref{sec:caveats}, this comparison highlights the large effect that the choice of methodology can have on the measured outflow parameters.
 The approach used by \citet{Fluetsch2019} assumes that most of the outflow flux is at low projected velocity.   Based on simulations of biconical outflows \citep[e.g.][]{Bae2016},  this scenario is possible when the outflow is oriented in a direction close to the plane of the sky,  or if the outflow has low velocities.
 In general, this method would tend to over-estimate the outflow mass and it suffers from the degeneracy of the fit with multiple Guassian components.  With our method on the other hand, we could be missing the outflow contribution at the low projected velocities,  thus, we may be underestimating the outflow mass.

\begin{figure*}
\centering
\includegraphics[width=0.9\textwidth]{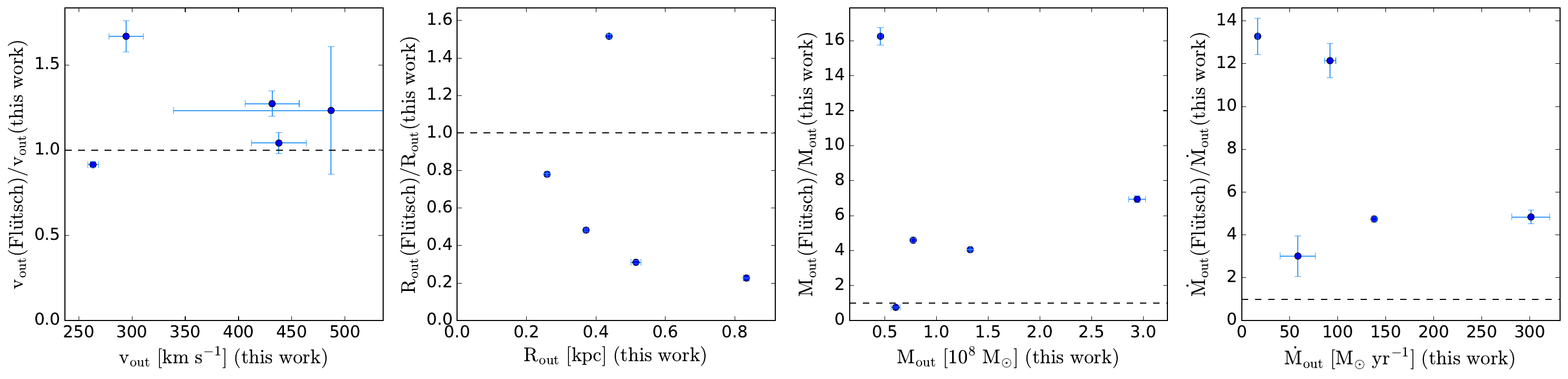}
\caption{Comparison of outflow parameters measured in this work with values reported in \citet{Fluetsch2019}.From left to right: outflow velocity, outflow radius, outflow mass and mass outflow rate.  Differences in \Mout\ and \Mrate\ can be up to a factor of 16.}
\label{fig:comp_Fluetsch2019_outflow_param} 
\end{figure*}
 
\begin{figure*}
\centering
\includegraphics[width=0.4\textwidth]{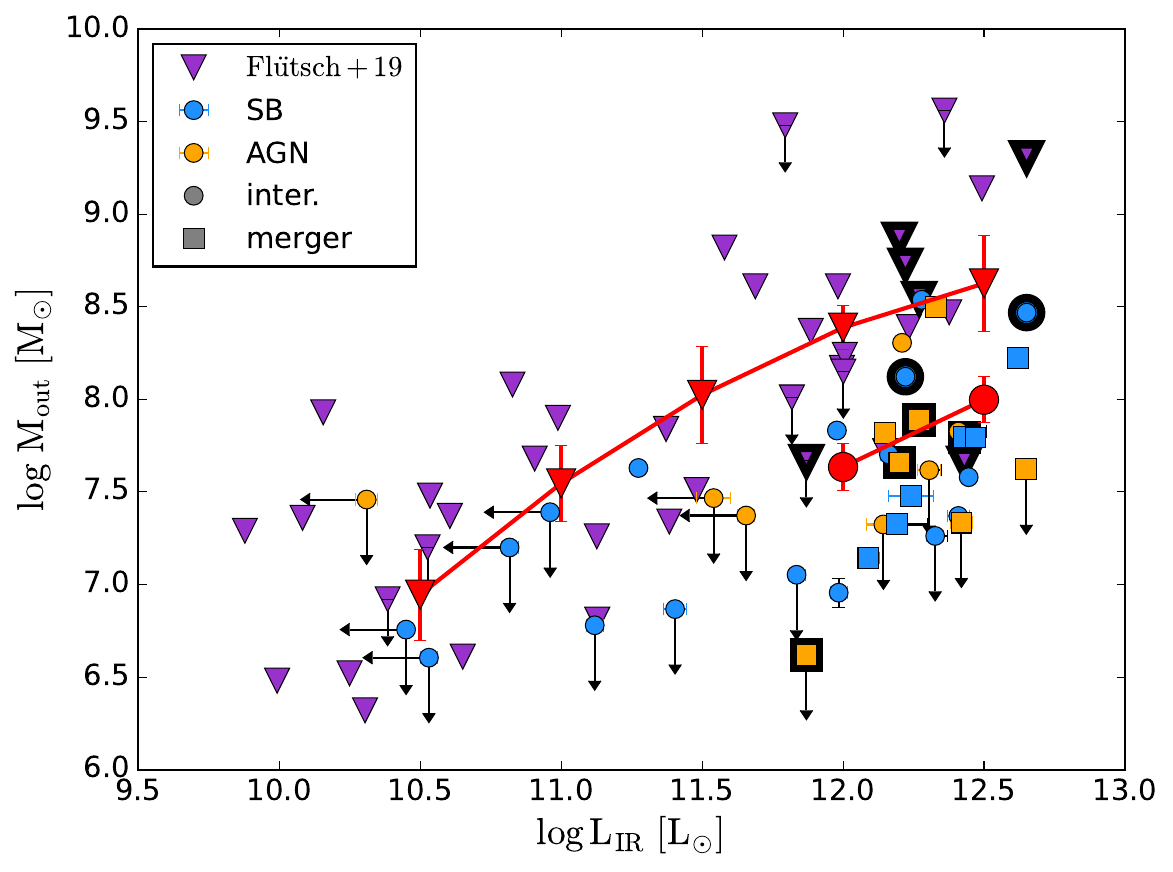}
\includegraphics[width=0.4\textwidth]{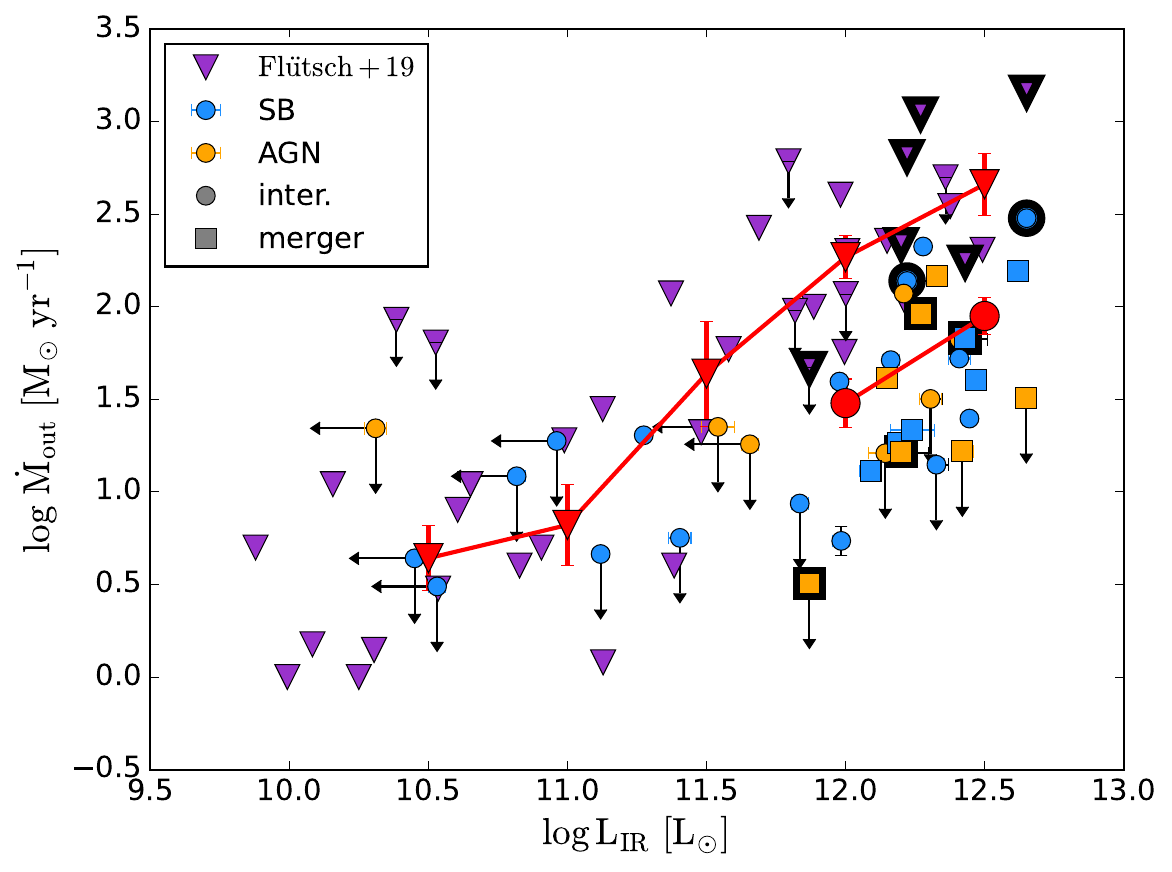}
\caption{Comparison of the outflow mass and mass outflow rate as a function of IR luminosity for our sample and the sample from \citet{Fluetsch2019} (purple triangles).  
The red symbols  show the average values (considering only the detections) in bins of $\log L_{IR}$ for the \citet{Fluetsch2019} sample (triangles) and the PUMA sample (circles). The sources in common between the two samples are highlighted with black contours. Due to the different methods,  the \Mout\ and \Mrate\ values estimated by  \citet{Fluetsch2019} are higher by a factor of $\times4-6$ compared with our measurements at the same infrared luminosity}.
\label{fig:comp_Fluetsch2019} 
\end{figure*}

\subsection{Analysis of the OH 119\micron\ spectra}
\label{sec:OH_analysis}
Other lines that are often used as tracer of molecular outflows are the OH (hydroxyl) FIR lines   \citep[e.g.][]{Fischer2010,  Sturm2011, Spoon2013, Veilleux2013,  Gonzalez-Alfonso2014b,  Stone2016, Gonzalez-Alfonso2017}. We compare the outflow parameters derived from CO(2-1) with the ones derived from the  OH 119\micron\ doublet, which is the strongest of the OH ground-state lines \citep[e.g.][]{Gonzalez-Alfonso2017}.
We collect archival \Herschel/PACS  spectra of the OH 119\micron\  doublet transition (hereafter OH) for 22/25 of our ULIRGs systems (no data available for 00091-0738, 10190+1322, and 16156+0146).
The majority of the spectra (20/22) were published in \citet{Veilleux2013}, \citet{Spoon2013}, and \citet{Gonzalez-Alfonso2017}; 11095-0238 and  13451+1232 are not presented in these works but were found in the \Herschel\ archive.

We extract the spectra from the central $9.4''\times 9.4''$ spaxel and apply the point source aperture correction. 
We fit the OH spectra to derive the velocity of the outflow.  In particular, we want to compare the velocity ranges where the OH outflow is detected with the ones of the CO outflow.
Outflow parameters were derived by \citet{Veilleux2013} for 14 of the ULIRGs in our sample,  but we repeat the analysis in order to obtain consistent parameters for the full sample. 
We set the zero velocity of the OH spectra based on the redshift of CO (see Table~\ref{tab:sample}), so that we can directly compare the outflow velocities of the two tracers.

We fit the line profile following the method used by \citet{Veilleux2013},   and we check that our derived parameters are consistent with their results. 
We first perform a linear fit of the continuum around the OH line and normalise the spectra to the continuum level. The OH doublet can appear in absorption (11/22), emission (4/22) or as P-Cygni profile (7/22).  
The P-Cygni profile is considered a clear indication of the presence of an outflow \citep[e.g. ][]{Prochaska2011, Fischer2010, Sturm2011,  Gonzalez-Alfonso2012,  Gonzalez-Alfonso2013, Veilleux2013}.
In our sample,  emission and P-Cygni profiles are more common in AGN (73\%) than in SB dominated nuclei, which is consistent with previous findings \citep{Veilleux2013, Stone2016,  Runco2020}.

 For the absorption profiles, we fit a model with two Gaussians (one for the systemic and one for the outflow) for each line of the OH doublet.  The separation between the two  lines of the $\Lambda$-doublet is fixed to 0.208~\micron (in rest-frame) and the amplitude and width of the two lines were tied to be the same in each component.  
  We convolve our model with the \Herschel/PACS  instrumental resolution \citep[FWHM$\sim 270$~\kms, ][]{Veilleux2013}, 
  to recover the intrinsic shape of the line.
For  profiles in emission, we fit a model with two Gaussians (one for the systemic and one for the outflow component) in emission for each OH line.
For the P-Cygni profiles, we consider one component in absorption and one in emission for each line.  Adding more components is not possible due to parameter degeneracy  \citep{Veilleux2013}. We add as an additional constraint that the model absorption can not be deeper than the observed absorption \citep[similar to][]{Veilleux2013}, to avoid a fitting result with unrealistically large amplitude of the emission and absorption components that  cancel each other out.

We use the  best-fit model to derive the characteristic velocity of the emission and absorption profiles, separately.  For the absorption components, $v_{50}$,  $v_{84}$ and $v_{98}$  are the velocities corresponding to the 50th, 84th and 98th percentile of the absorption line profile, i.e. the velocities above which 50\%,  84\%,  and 98\% of the absorption takes place.  Similarly,  for the emission components, $v_{50}$,  $v_{84}$ and $v_{98}$  are the velocities corresponding to the 50th, 84th and 98th percentile of the emission line profile.
\citet{Veilleux2013} consider $v_{84}$ to be a more robust estimate of the outflow velocity compared to the `maximum outflow velocity'.  We use $v_{98}$ as an estimate of the maximum outflow velocity,  keeping in mind that it may be more susceptible to noise variation than $v_{84}$.

We apply a Monte Carlo (MC)  approach to estimate the uncertainty on the derived parameters. We estimate the noise level on a region of the continuum away from the line (excluding also the region of the CH$^+$(3-2) and $^{18}$OH 120\micron\ lines), then we add random Gaussian noise, proportional to the noise level, to the best-fit model and we run the line fitting on this artificial spectrum.  We repeat this procedure 50 times and we use the 16th and 84th percentiles of the distribution of best-fit parameters to estimate the 1$\sigma$ uncertainties on the derived parameters. The 50 MC realisations are shown as orange curves on Figure~\ref{fig:OH_spectra}.
The velocities derived from the OH profiles are shown in Table~\ref{tab:OH_vel}.
 The properties of molecular outflows derived from OH will be discussed and compared with those derived from the CO tracer in Sec.~\ref{sec:OH_comp}.

\begin{figure}[h!]
\centering
\includegraphics[width=0.4\textwidth]{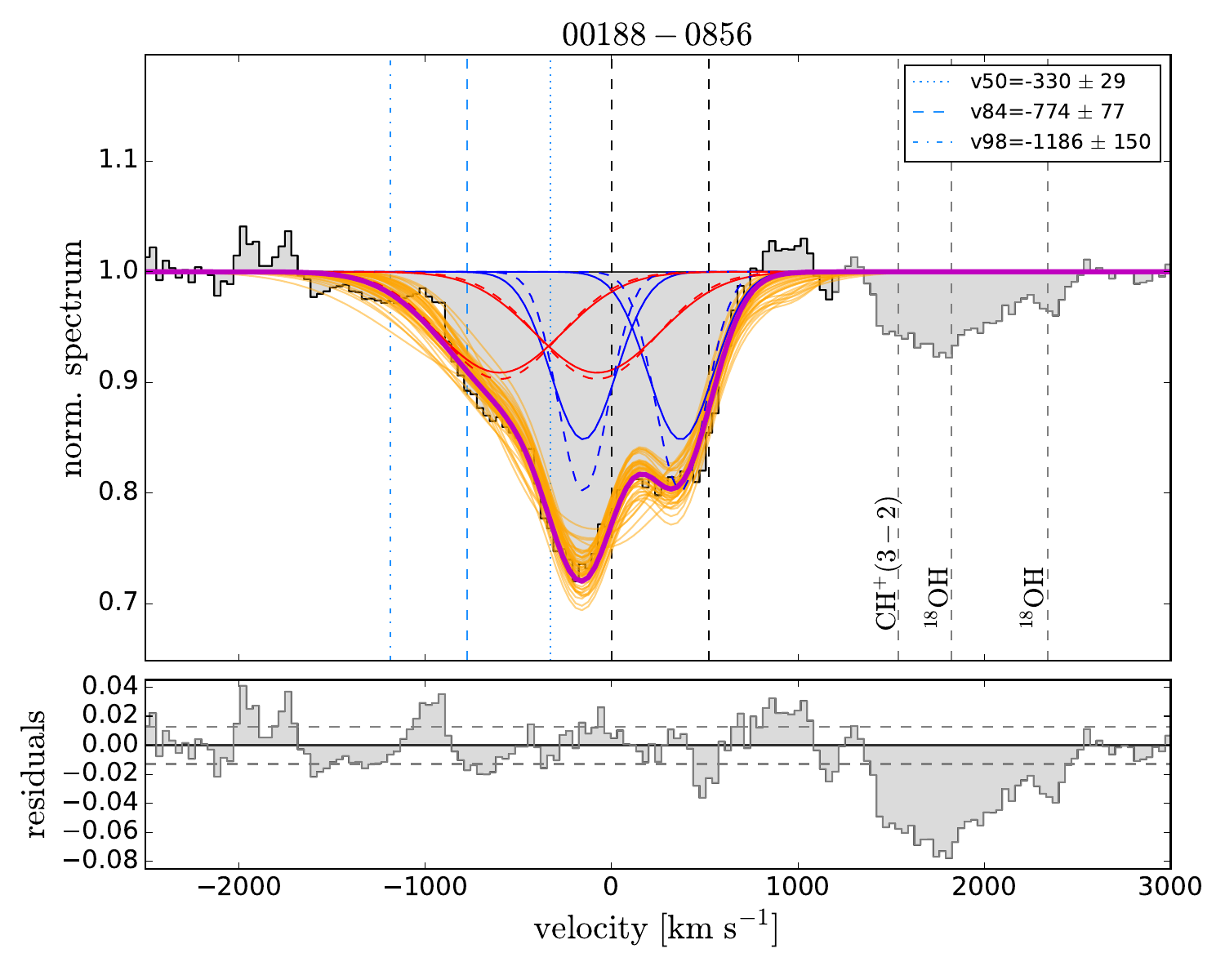}
\includegraphics[width=0.4\textwidth]{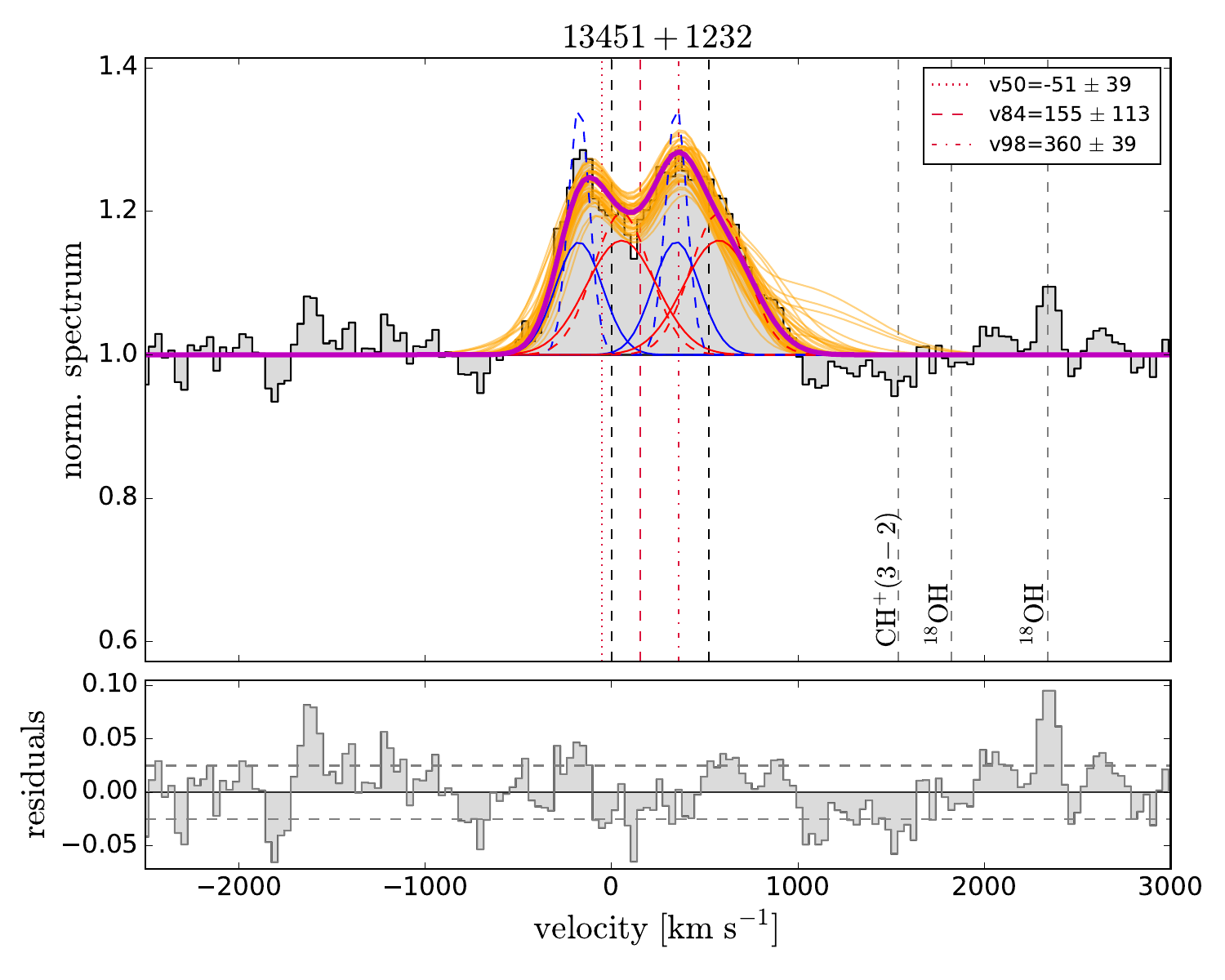}
\includegraphics[width=0.4\textwidth]{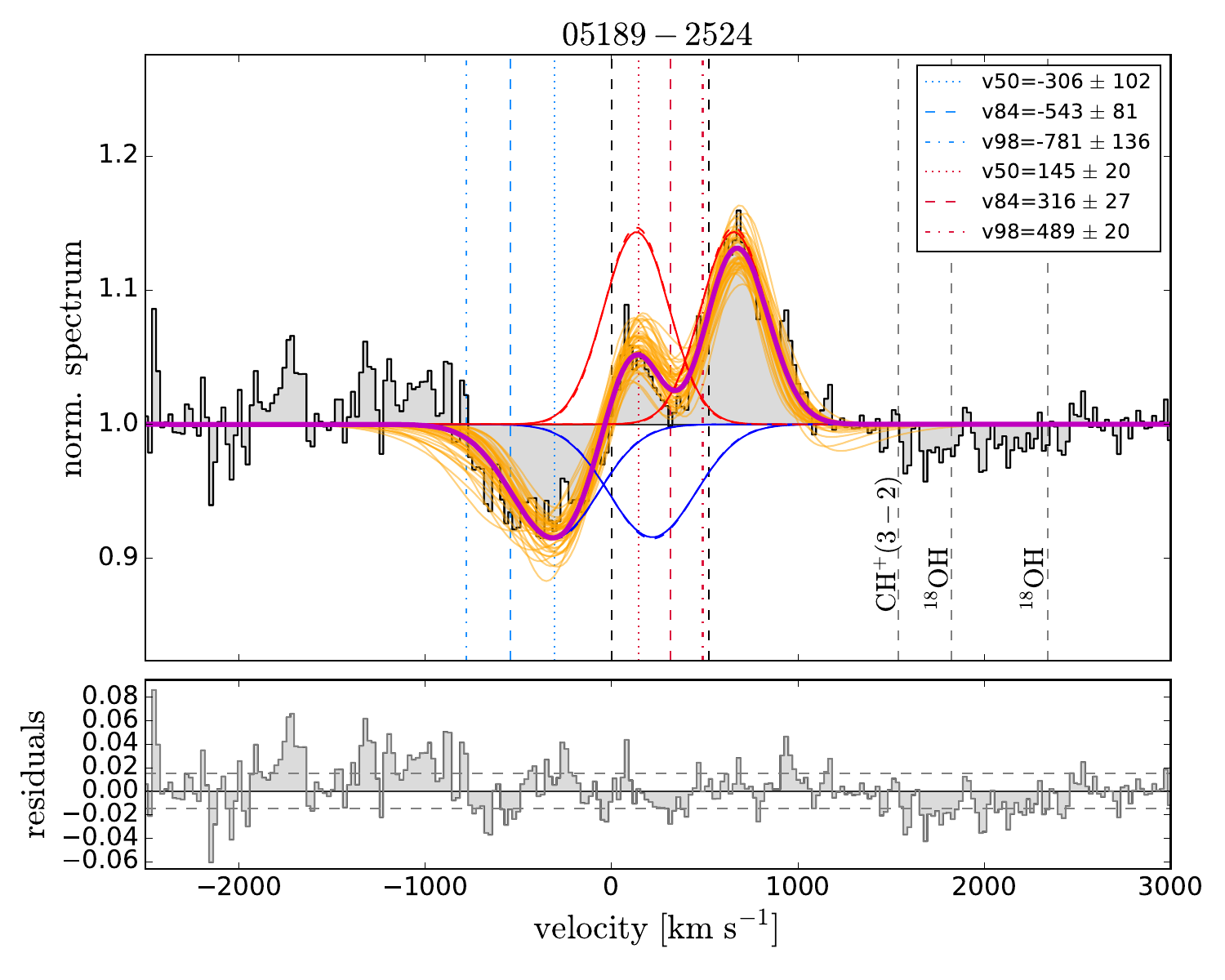}

\caption{Example of the fit of the OH 119\micron\ doublet  with absorption (\textit{upper}), emission  (\textit{middle}) and P-Cygni (\textit{bottom}) profile.   The velocity is relative to the blue component of the doublet (at 119.233~$\mu$m) at the systemic velocity inferred from the CO(2-1) line.
The best-fit model is shown in magenta,  while the orange models are the results of the 50 Monte Carlo realisations that illustrate the uncertainty of the fit.  Solid lines show the model convolved with the instrumental resolution, while dashed lines indicate the intrinsic (deconvolved) model.
Blue and red mark the two components of the fit. 
The vertical dotted, dashed and dot-dashed lines show the $v_{50}$,   $v_{84}$ and $v_{98}$ percentile velocities, respectively,  derived from the absorption (blue) and/or emission (red) profiles.  Grey dashed lines at $\sim2000$~\kms\ show the position of the $^{18}$OH 120~$\mu$m doublet and the CH$^{+}$(3-2) absorption line.}
\label{fig:OH_spectra}
\end{figure}

\begin{table*}
\centering
\caption{Velocities derived from the OH~119$\mu$m profiles. }
\setlength{\tabcolsep}{3pt}
\begin{tabular}{lccccccc}
\hline

IRAS name & $v_{50}$(abs) & $v_{84}$(abs) & $v_{98}$(abs)& $v_{50}$(em)  & $v_{84}$(em) & $v_{98}$(em)   \\ 
   &  [\kms] &[\kms]  &   [\kms] & [\kms]   &  [\kms]  & [\kms]   \\
 (1) & (2)  & (3) & (4) &(5) & (6) & (7)  \\ 
  \hline \hline

00091-0738$^{\star}$  &  - & - & - &  - & - & - & \\
00188-0856 & -330$\pm$29 & -774$\pm$69 & -1186$\pm$136 &  - & - & - & \\
00509+1225 &  - & - & - & 120$\pm$33 & 262$\pm$85 & 404$\pm$194 & \\
01572+0009 &  - & - & - & 88$\pm$139 & 155$\pm$209 & 188$\pm$318 & \\
05189-2524 & -306$\pm$84 & -543$\pm$72 & -781$\pm$105 & 145$\pm$23 & 316$\pm$22 & 489$\pm$33 & \\
07251-0248 & -127$\pm$20 & -347$\pm$22 & -475$\pm$126 &  - & - & - & \\
09022-3615 & -139$\pm$36 & -270$\pm$28 & -398$\pm$72 & 214$\pm$11 & 342$\pm$14 & 472$\pm$22 & \\
10190+1322$^{\star}$  &  - & - & - &  - & - & - & \\
11095-0238 &  - & - & - & 206$\pm$163 & 344$\pm$40 & 413$\pm$110 & \\
12071-0444 & -249$\pm$17 & -454$\pm$30 & -694$\pm$43 & 194$\pm$8 & 296$\pm$12 & 398$\pm$17 & \\
12112+0305 & 21$\pm$12 & -76$\pm$27 & -207$\pm$48 & 341$\pm$263 & 503$\pm$327 & 664$\pm$405 & \\
13120-5453 & -253$\pm$16 & -586$\pm$18 & -850$\pm$35 &  - & - & - & \\
13451+1232 &  - & - & - & -51$\pm$35 & 155$\pm$123 & 360$\pm$274 & \\
14348-1447 & -245$\pm$34 & -560$\pm$40 & -917$\pm$68 & 154$\pm$11 & 300$\pm$14 & 446$\pm$23 & \\
14378-3651 & -296$\pm$27 & -650$\pm$39 & -1007$\pm$58 & 189$\pm$15 & 319$\pm$14 & 478$\pm$20 & \\
15327+2340 & -28$\pm$0 & -194$\pm$0 & -359$\pm$13 &  - & - & - & \\
16090-0139 & -36$\pm$25 & -410$\pm$43 & -683$\pm$91 &  - & - & - & \\
16156+0146$^{\star}$   &  - & - & - &  - & - & - & \\
17208-0014 & 50$\pm$10 & -99$\pm$15 & -466$\pm$94 &  - & - & - & \\
19297-0406 & -121$\pm$39 & -500$\pm$84 & -879$\pm$199 &  - & - & - & \\
19542+1110 & -302$\pm$50 & -480$\pm$96 & -622$\pm$157 & 267$\pm$72 & 302$\pm$114 & 373$\pm$168 & \\
20087-0308 & 48$\pm$19 & -160$\pm$24 & -403$\pm$52 &  - & - & - & \\
20100-4156 & -393$\pm$39 & -985$\pm$73 & -1539$\pm$145 &  - & - & - & \\
20414-1651 & 17$\pm$0 & -53$\pm$16 & -88$\pm$85 &  - & - & - & \\
22491-1808 & 88$\pm$20 & 25$\pm$55 & -8$\pm$196 &  - & - & - & \\

\hline
\end{tabular}
\label{tab:OH_vel}
\tablefoot{
(1) IRAS name.
(2),  (3) and (4): 50th, 84th and 98th percentile velocities derived  from the absorption profile (when present).
(5),  (6) and (7): 50th, 84th and 98th percentile velocities derived  from the emission profile (when present).
$^{\star}$: OH 119$\mu$m spectra not available for 00091-0738, 10190+1322, and 16156+0146.
}
\end{table*}

\section{Results}
\label{sec:results}

\subsection{Mean outflow properties}
\label{sec:mean_out}
Here we summarise the ranges of outflow properties of our sample and their average values,  corrected for inclination as described in Sec.~\ref{sec:inclination}.
We measure projected outflow velocities of $\sim260-490$~\kms, with a mean outflow velocity (corrected for inclination) of $485\pm16$~\kms. Outflow radii are in the range 0.17-0.94~kpc and the mean inclination-corrected outflow radius is $1.1\pm0.1$~kpc.
The outflow masses are between $1-35  \times 10^7$~\Msun, with an average of $(10\pm2)  \times 10^7$~\Msun. These outflow masses corresponds to 0.2-6.5\% of the total molecular gas masses.  The mass outflow rates are in the range $6-302$~\Msun\ yr$^{-1}$, with an average outflow rate of $78\pm16$~\Msun\ yr$^{-1}$. The ranges and averages of the outflow parameters measured in this work are summarised in Table~\ref{tab:mean_out_prop}.

As a comparison,  \citet{RamosAlmeida2022} find molecular gas mass outflow rate \Mrate$=8-16$~\Msun\ yr$^{-1}$ in a sample of nearby type 2 quasars ($\log L_{AGN}/[\text{erg s}^{-1}] = 45.7-46.3$).  Lower   \Mrate$=0.3-5$~\Msun\ yr$^{-1}$ have been measured in lower luminosity ($\log L_{AGN}/[\text{erg s}^{-1}] = 43.2-44.2$) Seyfert galaxies \citep{Alonso-Herrero2019, Dominguez-Fernandez2020, Garcia-Bernete2021}.  Higher molecular gas mass outflow rates (\Mrate$=60-400$~\Msun\ yr$^{-1}$) have been measured in ULIRGs hosting an AGN \citep{Feruglio2010, Cicone2014}. We note that, as discussed in Sec.~\ref{sec:comp_lit}, the method used to derive the mass outflow rates can introduce systematic differences (up to factor of $\sim10$) between different samples.  

We also compare our  measurements  with the properties of the ionised outflows measured by \citet{Arribas2014} for a sample of nearby U/LIRGs. For ULIRGs, they find maximum outflow velocities in the range 100-1000~\kms, with a mean $393\pm 38$~\kms\ and mass outflow rates in the range  $1-100$~\Msun\ yr$^{-1}$ (from Fig. 13 in the paper), with an average 38~\Msun\ yr$^{-1}$.  The average \Mrate\ of the ionised gas is about a factor of 2 smaller than the one of the molecular phase.
   For U/LIRGs, they find outflow masses in the range  $0.14 -28  \times 10^7$~\Msun, with a average of  $6.7  \times 10^7$~\Msun,  similar to the masses of the molecular phase.

We find molecular outflow dynamical times ($t_{dyn} = R_{out}/v_{out} $) in the range $0.45-2.77$~Myr.
The mean $t_{dyn},$ based on the mean observed \Rout\ and \vout, is 1.37~Myr,  assuming a average inclination correction of unity \citep{Cicone2015}. 
This is similar to the outflow dynamical times 0.63-2.51~Myr reported in \citet{Pereira-Santaella2018}.  
The outflow depletion times ($ t_{dep} = M(H_2)_{tot}/\dot{M}_{out}$) in our sample are in the range $15-644$~Myr, with a median of 75~Myr. These are a bit longer than the values reported for ULIRGs in \citet{Pereira-Santaella2018} ($15-80$~Myr) and \citet{Cicone2015}  ($1.2-50$~Myr).
For comparison,  the star-formation depletion times  ($ t_{dep} = M(H_2)_{tot}/SFR$) for targets with detected outflows in our sample are in the range 9-77~Myr (median 27~Myr).

\begin{table}
\centering
\caption{Mean cold molecular outflow properties in ULIRGs. }
\setlength{\tabcolsep}{2pt}
\begin{tabular}{lccccccccc}
\hline
   &   \vout &   \Rout  &  \Mout &   \Mrate  &  \\
   &   [\kms] &  [kpc] &  [10$^7$ \Msun ] &   [\Msun\ yr$^{-1}$]  &   \\ 
  \hline \hline
mean: &  $485\pm16 $  &  $1.07\pm0.08$   &  $10\pm2$	&  $78\pm16$	\\ 
range:  &$ 260-490^{\star}$ & $0.26-0.94^{\star}$  & $ 1-35 $& $6-302$ \\  \hline
\end{tabular}
\label{tab:mean_out_prop}
\tablefoot{
$^{\star}$: Not corrected for inclination.
}
\end{table}

\subsection{Outflow characteristics for AGN/SB and interacting/mergers}
\label{sec:outflow_det}

\subsubsection{Outflow detection rate and direction}

In this Section, we present the statistics of the number of detected molecular outflows in our sample and we investigate whether there is any dependency of the outflow detection rate on the nuclear classification (AGN or SB) or on the merger stage (advanced mergers or interacting systems).

The top panel of Figure~\ref{fig:out_stats} shows the percentage of nuclei in the sample belonging to each category: AGN, SB, merger (M), interacting (I),  and the mixed categories (merger AGN, merger SB, interacting AGN, and interacting SB).  For the interacting systems, we consider only the nuclei with $\log L_{IR}/L_{\odot} > 11. 8$ in this statistical analysis. 
As shown in \citet{Pereira-Santaella2021},  in most of our interacting systems the IR luminosity is dominated by one nucleus, thus by applying this luminosity threshold,   we discard the fainter nuclei (with $\log L_{IR}/L_{\odot} = 10.3-11. 7$) which would not be classified as ULIRGs.
In this way,  we avoid that the outflow detection rate in interacting nuclei is artificially lower only because of the `secondary' faint nuclei.

The second panel of Fig.~\ref{fig:out_stats} presents the percentage of outflow detections in each category.  We detect an outflow, defined as high-velocity ($|v|>300$~\kms) CO(2-1) emission which deviates from the main rotation,  in 20/26 of the nuclei with $\log L_{IR}/L_{\odot} > 11. 8$ ($77\pm7\%$ \footnote{The uncertainties on the percentages represent the 90\% binomial confidence intervals. }).
The nuclei with outflow detections are  equally divided between mergers  and interacting systems.
The percentage of detections in SB ($93\pm4$\%,  14/15) is higher than in AGN ($55\pm14$\%,  6/11).  
 A possible explanation for this difference is related to the outflow inclination: if AGN outflows are in the plane of the disk (contrary to SB outflows that tend to be perpendicular to the disk), they are more difficult to detect with our method, but by modelling and subtracting the disk rotation, it may be possible to identify them.  Additionally,  outflows in the plane of the disk may be braked and prevented from reaching high velocities,   while outflows perpendicular to the disk can escape more freely,  reaching higher velocities and being more easily detectable.
 Another possibility is that AGN outflows could be more collimated,  and therefore more difficult to detect for some  orientations (i.e.  on the plane of the sky).
 A third possible explanation could be the different amount of molecular gas in the nucleus.  AGN in our sample have on average lower nuclear molecular gas masses ($\log M(H_2)/M_{\odot}= 9.2\pm 0.1$), than SB nuclei ($\log M(H_2)/M_{\odot}= 9.4\pm 0.1$).  The lower amount of material close to the nucleus may also be the reason that allows us to identify them as AGN,  contrary to deeply buried nuclei.  This is in agreement with an evolutionary scenario in which in a first merger phase the nuclei are more obscured and produced outflows, which expel gas and dust from the nuclear region; in a second phase,  after some of the  material has been removed, the nuclei are visible as AGN  \citep[e.g.][]{Hopkins2008}.
Similarly,   \citet{Stone2016} find a lower OH outflow detection fraction in X-ray selected AGN (24\%) than in ULIRGs ($\sim70\%$) and suggest that outflow detection in ULIRGs may be easier due to their higher gas fraction.

With the spectro-astrometry analysis (see Sec.~\ref{sec:spectroastrometry}),  we determine the outflow direction in 16/20 ($80\%$) of the  nuclei with outflow detection,  while in the remaining four nuclei the outflow direction is not clear. This could be due to the fact that the outflow is unresolved,  or to the fact that the outflow is pointing towards our line of sight, so that the blue and red-shifted sides of the outflow overlap.  For the nuclei with a well determined outflow direction, we measure the angle between the outflow and the major axis of rotation (see Sec.~\ref{sec:spectroastrometry}). 
We qualitatively compare the direction of the high-velocity gas with the integrated maps of the blue and red-shifted high-velocity channels (see lower left panel in Fig.~\ref{fig:spectroastrometry}, \ref{fig:spectroastrometry2} and \ref{fig:spectroastrometry_app}) to check that the direction is consistent with the position of the gas in the high-velocity channels.
 The bottom panel of Fig.~\ref{fig:out_stats} shows, for each category,  the fraction of outflows with 1)  direction perpendicular to the rotating disk (angle  $90\pm20^{\circ}$),  2) direction non-perpendicular,  or 3) unclear direction.
  The percentages of outflow with determined direction in mergers ($70\pm12 \%$,   7/10) is smaller than in interacting nuclei ($90\pm6\%$,  9/10).
 We could determine the outflow direction in  $86\pm7\%$ (12/14) of  SB nuclei with outflow detection,  but only in $67\pm15\%$ (4/6) of AGN.
If in AGN the outflow is oriented within the plane of the disk, it is possible that it can not travel very far, thus it appear very compact and unresolved. This could explain why we could not determine the outflow direction in many AGN.

We find that in six nuclei the outflow direction is nearly perpendicular (angle $90\pm20^{\circ}$) to the kinematic major axis. These nuclei are 12112+0305 NE,  12112+0305 SW,  14348-1447 NE  \citep[already presented in][]{Pereira-Santaella2018},   10190+1322 E,  19542+1110 and  20100-4156 SE. 
All these nuclei are SB dominated.  
 This supports the idea that outflow powered by SB tend to be perpendicular to the disk, while AGN outflow can have any orientation \citep[e.g.][]{Pjanka2017}. 
However,  there are also many SB nuclei ($50\pm13$\%) for which the outflow direction is not perpendicular. This could be due to a hidden (deeply buried) AGN that is powering the outflow,  even though it is not detected in the MIR or optical \citep[see][]{Pereira-Santaella2021}.  
 An alternative explanation is that in these SB  the molecular gas is still strongly disturbed and has not yet settled into a galactic plane, and consequently also the path of least resistance is not well defined.
We note that we measure the angle between outflow and major kinematic axis projected in the plane of the sky. Thus, there is the possibility that we measured a projected angle of $90^{\circ}$ for an outflow that is not perpendicular to the plane if we observe it from a particular orientation.  However,  for SB it is more likely that the outflow escape from the path of least resistance (perpendicular to the disk) than from any other direction, where the outflow will encounter more material.  
The majority of the nuclei with perpendicular outflow (5/6) are interacting systems.  Although the number statistics are small, this could be due to the poorly defined disk rotation axis in some of the more advanced mergers.

Our sample was selected to have a similar number of objects in each category (AGN/SB, interacting/mergers), thus it does not have the $L_{IR}$ distribution or AGN fraction of the general population of ULIRGs.
We estimate the outflow detection fraction  that would be measured in a sample of ULIRGs  with the same AGN fraction as the local ULIRGs population.
\citet{Veilleux2009} study the AGN contribution in a representative sample of  74 ULIRGs from the IRAS 1~Jy Survey \citep{Kim1998} and find that 24\% of their  sample has an AGN contribution $> 50$\% in the MIR. 
 We use this AGN fraction to correct our outflow detection statistics and we find that the expected outflow detection fraction in local ULIRGs is 84\%  (compared to 77\% measured in our sample). 

\begin{figure}[h!]
\centering
\includegraphics[width=0.46\textwidth]{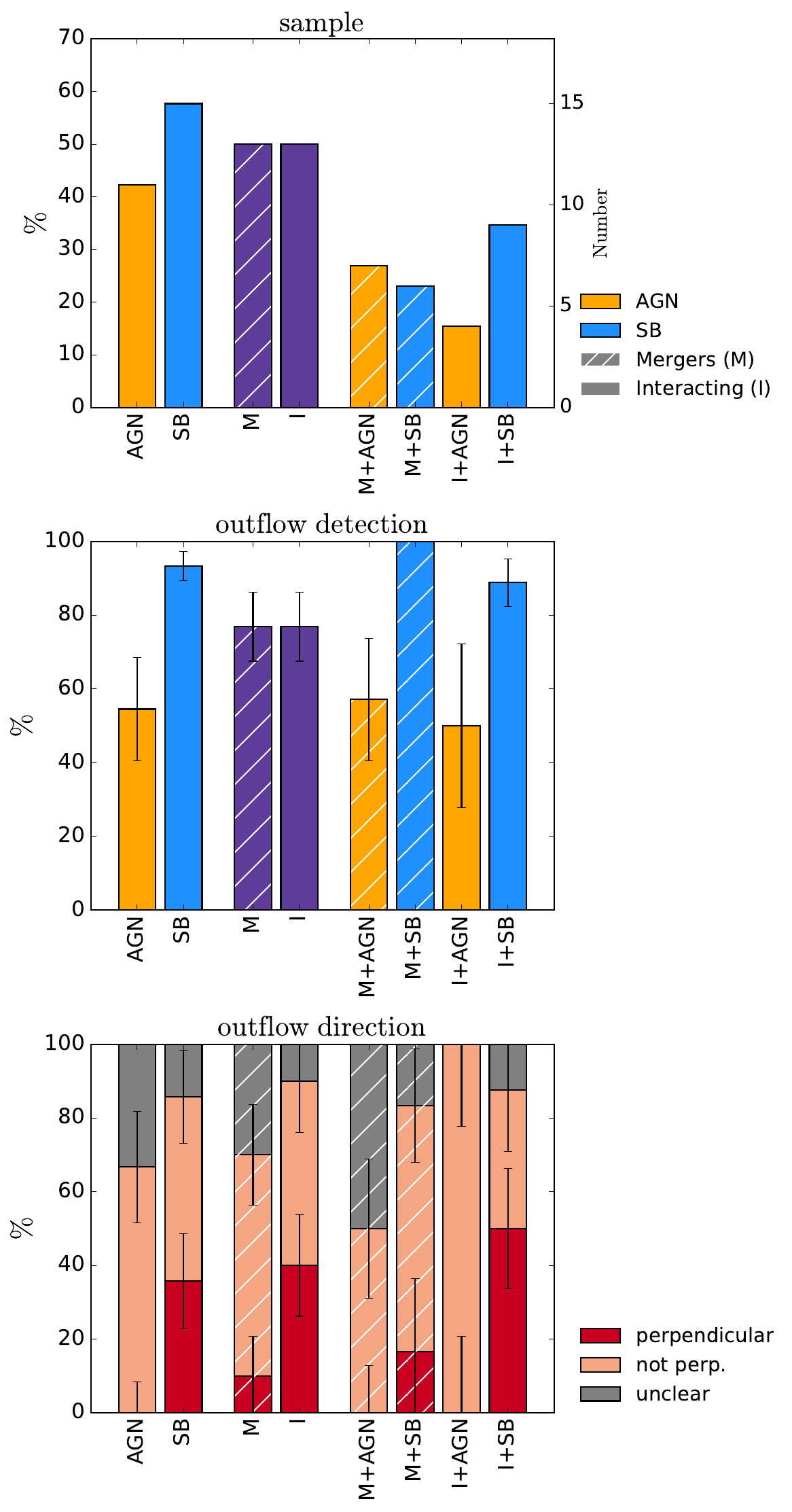}
\caption{Outflow detection statistics for different categories: AGN,  starburst (SB),  mergers (M, $d_{nuclei}<1$~kpc), interacting systems (I,  $d_{nuclei}>1$~kpc), and mixed categories (mergers AGN, mergers SB, interacting AGN, interacting SB).
\textit{Upper panel:} fraction of individual galaxy nuclei in each category with respect to the total sample.  The scale on the right axis shows the number of objects.
\textit{Middle panel:} outflow detections fractions. The percentages have been calculated with respect to the total number of nuclei in each category.  
\textit{Lower panel:} outflow orientation statistics divided in three groups:  outflow projected orientation perpendicular to the kinematic major axis of the disk (angle  $90\pm20^{\circ}$),  outflow projected orientation non-perpendicular, or orientation could not be determined. The percentages have been calculated with respect to the number of outflow detection in each category. 
Error bars  in middle and lower panel show the 90\% binomial confidence interval.}
\label{fig:out_stats}
\end{figure}

\subsubsection{Outflow quantities}

In this Section, we investigate whether there are differences in the outflow properties depending on these categories.
 We calculate the mean outflow parameters (\vout, \Rout, \Mout, \Mrate) for each category and we compare it with the mean of the total sample.  We do not include upper limits in this comparison. Even though there are small differences,  the mean quantities in all categories are consistent (within 2$\sigma$) with the  sample mean (see Fig.~\ref{fig:CDF_out_prop}).

Additionally, we compare the distribution of outflow properties for AGN vs. SB and mergers vs.  interacting systems, taking into account also the upper limits on the non-detections.
Figure~\ref{fig:CDF_out_prop} shows the cumulative distribution functions (CDF) of the outflow properties (\vout , \Rout, $\log M_{out}$,   $\log \dot{M}_{out}$,  and $\log \eta=\log (\dot{M}_{out}/SFR)$) for the AGN and SB categories (upper) and mergers and interacting  (bottom) categories. 
To test whether two samples have different distributions, we perform  a Two Sample test using the survival analysis package ASURV \citep{Feigelson1985} which allows us to take into account upper limits. We find that the distributions of the outflow properties of AGN and SB are not significantly different according to the Gehan's,  Logrank and Peto-Prentice's Two Sample Tests ($p$-value=0.2-0.7). 
Similarly, we do not find significant differences between mergers and interacting systems ($p$-value=0.2-0.9).

We also consider the mean outflow rate normalised by the total $L_{IR}$ (see Fig.~\ref{fig:out_stats_mean}), in order to remove the effect of the correlation between \Mrate\ and $L_{IR}$ (see Sec.~\ref{sec:out_prop_energetics}).  Also in this case there are no significant differences in the average \Mrate/$L_{IR}$  of the different categories. 
 The average for the total sample is \Mrate/$L_{IR}=(3.8\pm 0.6 )\times10^{-11}$ \Msun\ $yr^{-1}$ \Lsun$^{-1}$.  In the right panel, we also plot the average outflow rates normalised by AGN luminosity \Mrate/$L_{IR, AGN}$ for the AGN categories.  The average \Mrate/$L_{IR, AGN}$ for AGN are $\sim3-6$ times higher than the  \Mrate/$L_{IR}$ for starbursts.

\begin{figure*}
\centering
\includegraphics[width=0.98\textwidth]{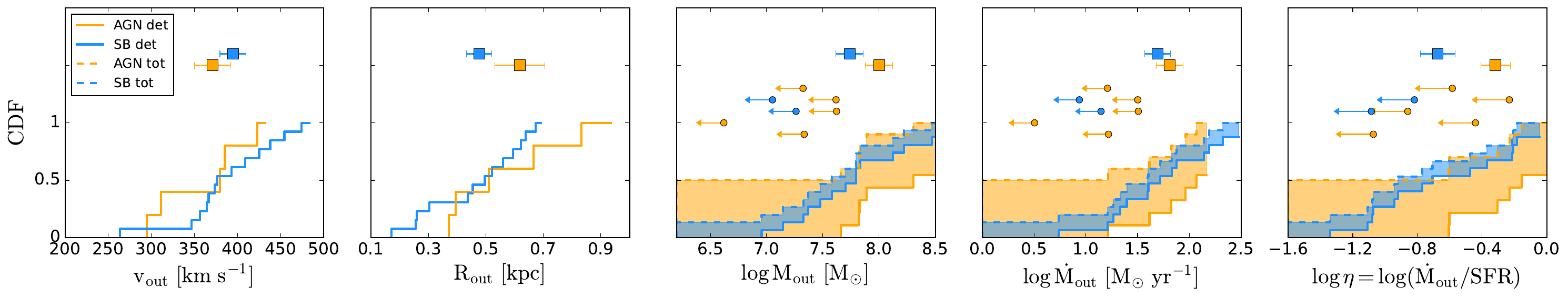}
\includegraphics[width=0.98\textwidth]{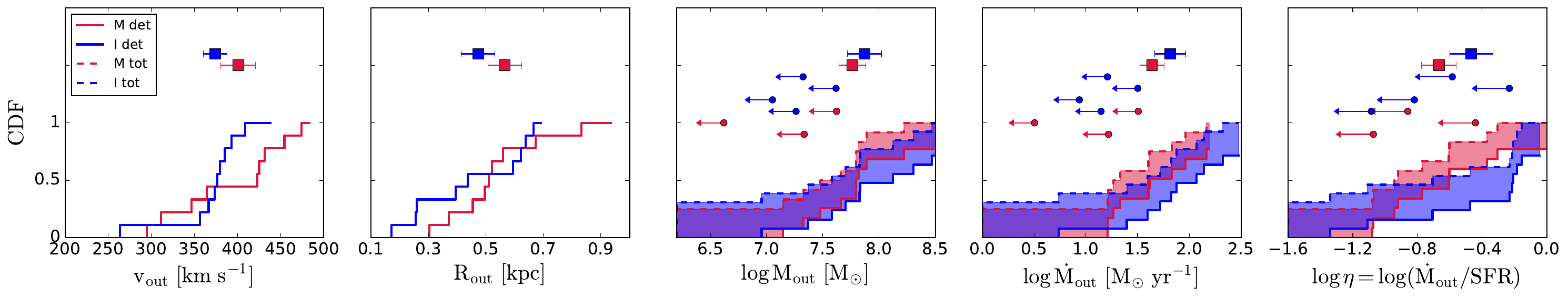}
\caption{Cumulative distribution function (CDF) of the outflow properties of AGN and starbursts (SB) (\textit{top}) and mergers (M) and interacting (I) (\textit{bottom}).  The solid lines show the CDFs for the outflow detections only,  while the dashed lines show the CDFs including upper limits  (only for  \Mout, \Mrate\ and $\eta$) . The shaded area mark the area between the CDFs with and without upper limits. The square symbols show the average values  for the outflow detections in the two categories,  the points with arrows show the upper limits (with arbitrary values on the y-axis).  According to a survival analysis Two Sample test,  the differences between AGN and SB and between interacting and mergers, are not statistically significant.}
\label{fig:CDF_out_prop}
\end{figure*}

\begin{figure}
\centering
\includegraphics[width=0.48\textwidth]{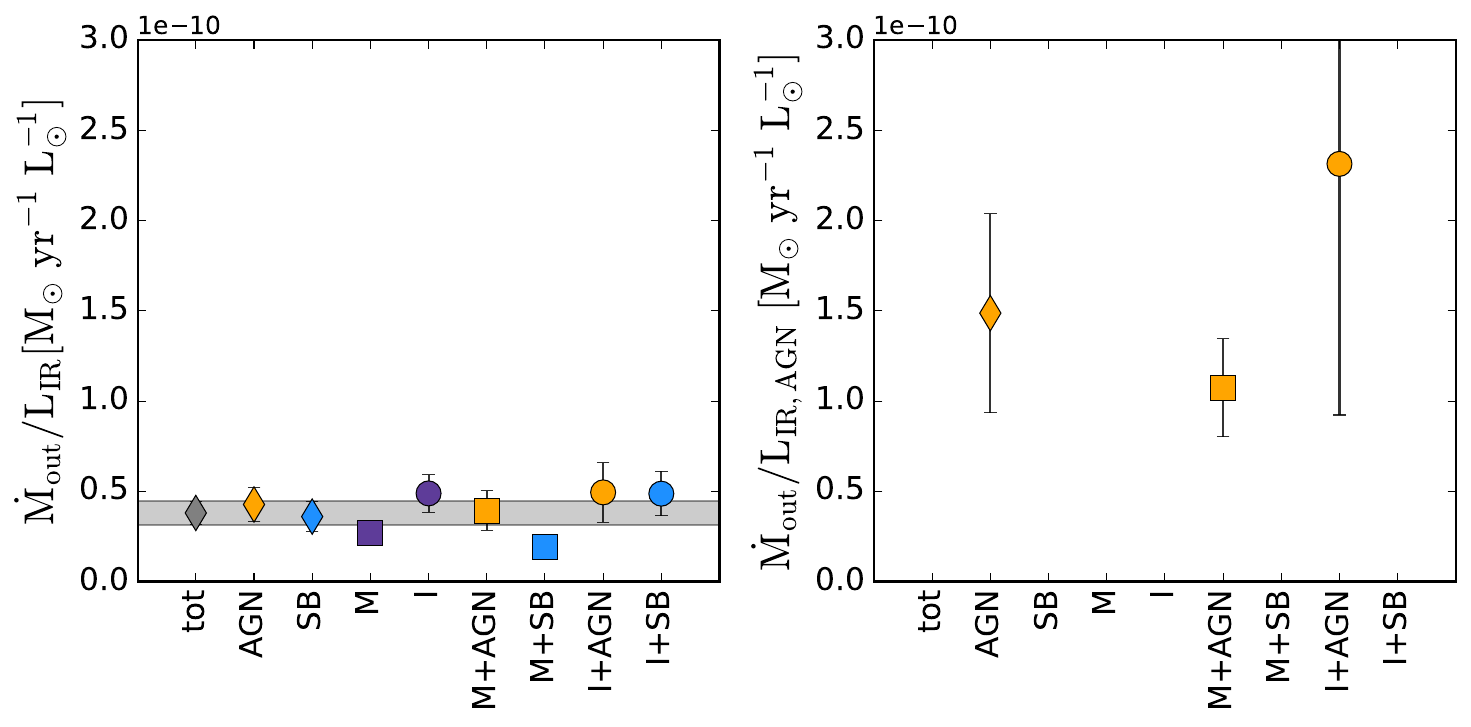}
\caption{\textit{Left:} Mean values of the ratio of \Mrate\ and total $L_{IR}$ for different categories: AGN,  starbursts (SB), mergers (M), interacting (I), and the mix categories mergers AGN, mergers SB, interacting AGN and interacting SB.  The errorbars show the uncertainty on the mean.
The grey diamond and horizontal band shows the mean of the total sample and the corresponding uncertainty. \textit{Right:}  For the AGN categories,  mean ratios of \Mrate\ and $L_{IR,AGN}$. }
\label{fig:out_stats_mean}
\end{figure}

\subsection{Outflow properties and energetics}
\label{sec:out_prop_energetics}

In this section,  we investigate the relation between the outflow properties and the total infrared luminosity,  as well as with the SFR and AGN luminosity.

\subsubsection{Outflow and AGN properties}
In this section,  we look at trends between the outflow properties and the AGN luminosity and AGN fraction.
 We calculate the AGN luminosity ($L_{IR,AGN}$)  as the fraction of the total infrared luminosity $L_{IR}$  due to AGN,  based on the MIR AGN fraction  \citep[$\alpha_{AGN}$, see][]{Perna2021}.   For the interacting systems,  $\alpha_{AGN}$ is determined only for the IR brightest nucleus, thus,  for the second faint nucleus we cannot estimate $L_{IR,AGN}$ and $L_{IR, SF}$.
 \citet{Cicone2014}  reported a trend for the mass-loading factor ($\eta=$\Mrate/SFR) to increase with AGN fraction ($\alpha_{AGN}=L_{AGN}/L_{bol}$),  while we do not find a correlation between these two quantities in our sample (Spearmann rank correlation coefficient for the detections $r=0.27$, $p$-value=0.28),  as it is shown in Figure~\ref{fig:alpha_vs_Mload}. The correlation found in \citet{Cicone2014} is driven mostly by the objects with high AGN fraction ($>0.8$), which have $\eta>10$ .  
 \citet{Fluetsch2019} expand the  \citet{Cicone2014} sample and  find that the correlation between $\eta$ and $\alpha_{AGN}$ is  only evident for $\alpha_{AGN}> 0.7$. They propose that the lack of correlation for $\alpha_{AGN}< 0.7$  could be due to i) the contribution from star-formation to the mass outflow rate for small AGN fractions  or ii) to the short timescale of AGN variability \citep[$10-10^5$~yr, ][]{Gilli2000, Schawinski2015}, compared to the outflow timescales ($10^6$~yr).
Since \citet{Cicone2014}  included in the study objects with known molecular outflows,  it is possible that they sample the upper end of the distribution of $\eta$ at large $\alpha_{AGN}$.  On the other hand,  our sample was selected without a prior knowledge of the presence of outflows.
The difference could be also due to the method used  to estimate the AGN fraction. We use $\alpha_{AGN}$ estimated  from the MIR 30~$\mu$m to 15~$\mu$m flux ratio.  \citet{Cicone2014} estimate the AGN fraction using different methods:  from the 5-8~$\mu m$ spectral range  using the method from \citet{Nardini2010},  from a combination of MIR and FIR diagnostics as described in \citet{Veilleux2009},  by taking the fraction of $L_{AGN}$ (estimated  from [OIII] or from the X-ray luminosity) and $L_{bol}$.   If we were to use the method from \citet{Nardini2010},  the $\alpha_{AGN}$ for our sample would be lower (< 0.7 for all sources).

 \citet{Cicone2014} also find a correlation between the molecular mass outflow rate and the AGN luminosity. 
 A similar correlation was also reported in \citet{Fiore2017},  who expanded the sample used by  \citet{Cicone2014}.
Figure~\ref{fig:LAGN_vs_Mrate} shows  \Mrate\  as a function of $L_{AGN}$ for our sample,  together with the sample from \citet{Cicone2014},  \citet{Lutz2020},  the nearby (non-ULIRGs) AGN from \citet{Stuber2021}, and the nearby ($z< 0.13$) type 2 AGN from \citet{RamosAlmeida2022}.  
 \citet{RamosAlmeida2022} find that their sample has \Mrate\ more than one order of magnitude below the relation derived by \citet{Cicone2014} (dashed line on the Fig.~\ref{fig:LAGN_vs_Mrate}).  
 Our sample occupies the parameter space between the  \citet{RamosAlmeida2022} objects and the \citet{Cicone2014} relation.    \citet{RamosAlmeida2022} suggest that this scaling relation represents the upper boundary of the \Mrate\ versus AGN luminosity relation.    Our sample confirms that for the same AGN luminosity,  ULIRGs can have a wide range of \Mrate.
 We note that most of the  objects in \citet{Cicone2014} were selected based on the presence of a molecular outflow (detected in CO or OH), while for our sample we do not apply this prior criterion.  This may  explain the fact that the sources in their sample have the highest \Mrate\ values for a given AGN luminosity.

\begin{figure}[th]
\centering
\includegraphics[width=0.48\textwidth]{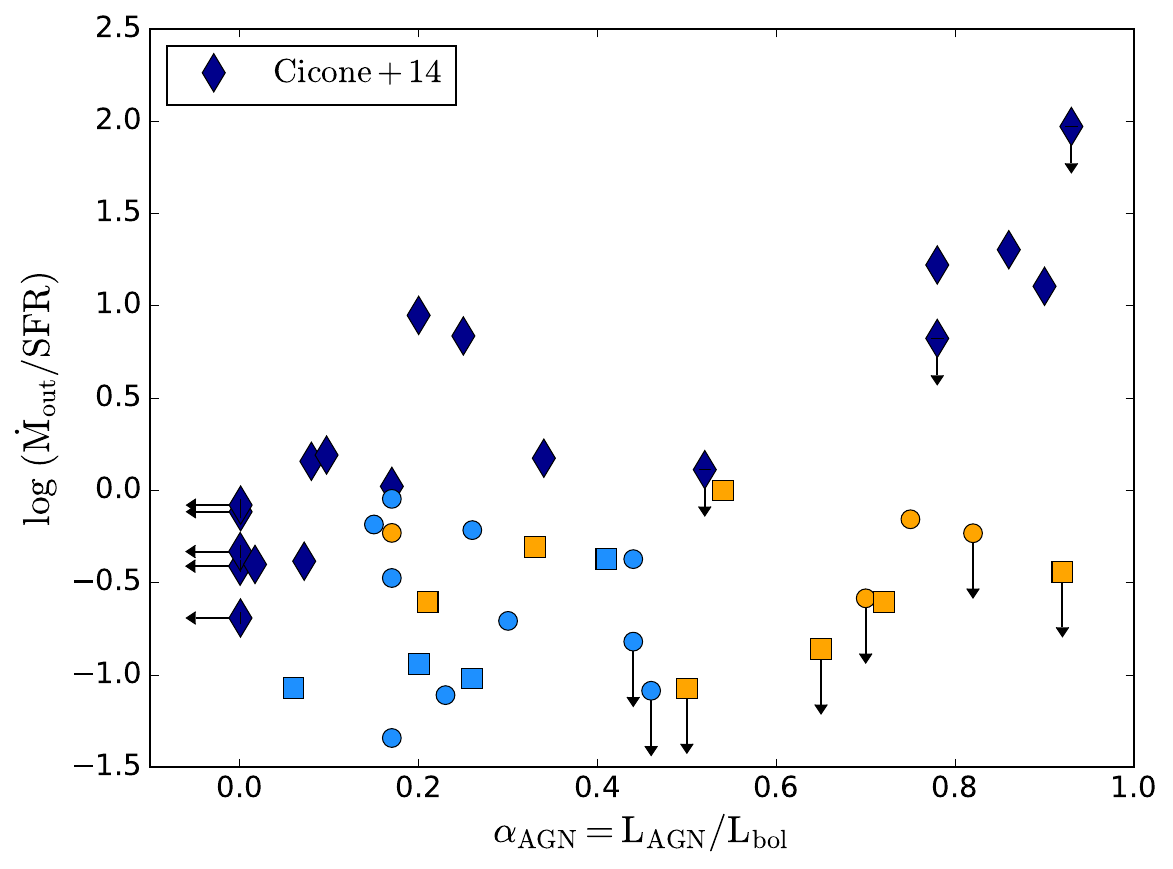}
\caption{Mass-loading factor ($\eta=\dot{M}_{out}/$SFR) versus AGN fraction ($\alpha_{AGN}$).  Star symbols show the \citet{Cicone2014} sample (scaled down by a factor of three to match our outflow geometry definition).  Symbols for the PUMA sample  are  as in Fig.~\ref{fig:sample}. }
\label{fig:alpha_vs_Mload}
\end{figure}

\begin{figure}[th]
\centering
\includegraphics[width=0.48\textwidth]{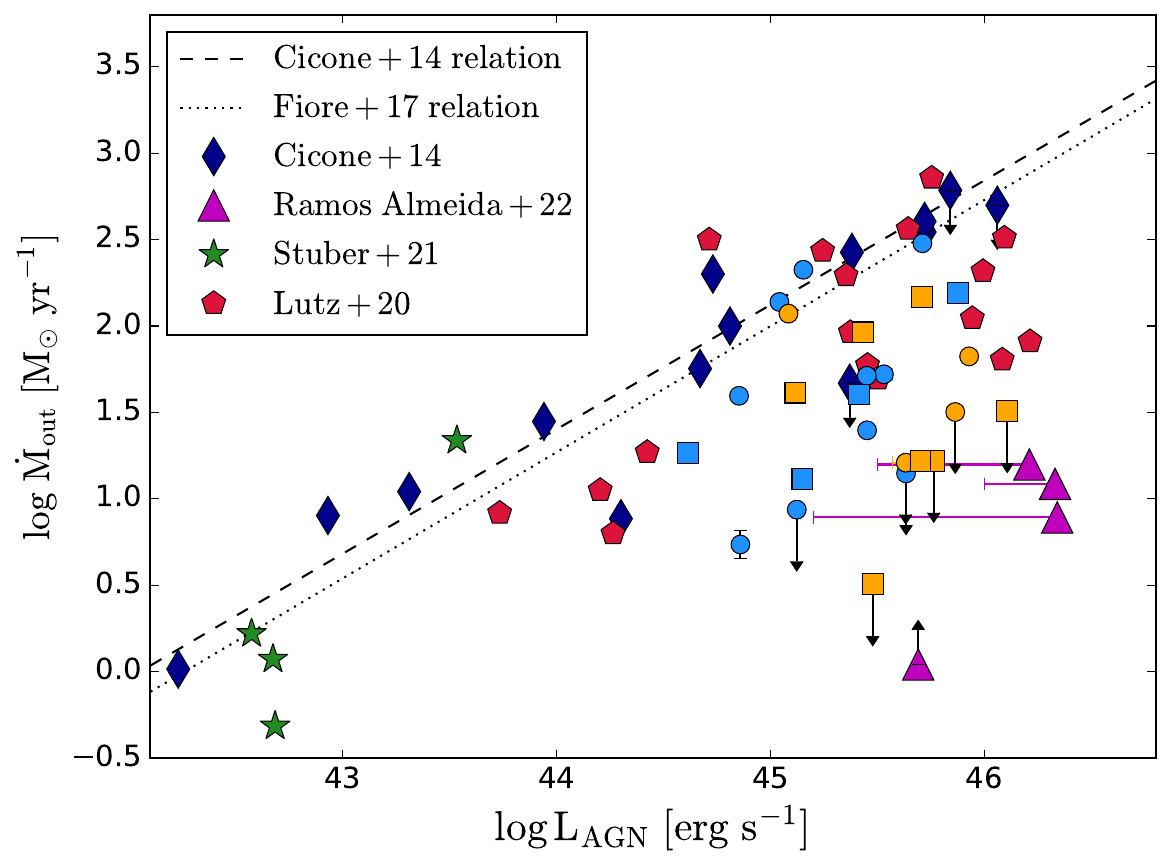}
\caption{Mass outflow rate versus AGN luminosity.  Diamonds symbols show the sample from \citet{Cicone2014}, for which \Mrate\ have been scaled down by a factor of three to match our outflow geometry definition.  The sample of \citet{RamosAlmeida2022} is shown with triangles.  Horizontal lines for the \citet{RamosAlmeida2022} sample indicate the position that they would occupy if we were to use the $L_{AGN}$ derived from SED fitting from \citet{Jarvis2019}, instead of the one derived from the \OIII\ luminosity \citep[see][for details]{RamosAlmeida2022}.    The \citet{Lutz2020} sample is shown with red pentagons.  For the \citet{Stuber2021} sample (stars),  the AGN luminosities have been derived from the 14-195~keV X-ray luminosity.  
 Symbols for the PUMA sample  are  as in Fig.~\ref{fig:sample}. 
The dashed and dotted lines show the relations presented by \citet{Cicone2014} and by \citet{Fiore2017}, respectively,  scaled down by a factor of three to match our outflow geometry definition).
}
\label{fig:LAGN_vs_Mrate}
\end{figure}

\subsubsection{Outflow energetics}
 
In Figure~\ref{fig:Q_out_vs_SFR} we show the outflow rate,  momentum rate ($\dot{P}_{out} =\dot{M}_{out} \cdot v_{out}$),  and kinetic luminosity ($L_{out} = \frac{1}{2}\dot{M}_{out} \cdot v_{out}^2$) as a function of  SFR.  The SFR has been estimated from the IR luminosity, using the \citet{Kennicutt2012} relation (with \citet{Kroupa2003} initial mass function (IMF)) after removing the fraction of luminosity associated with the AGN ($\alpha_{AGN}$).

To increase the dynamic range of SFR, we include in this figure also the 20 galaxies with candidate outflows from the Physics at High Angular resolution in Nearby GalaxieS (PHANGS\footnote{http://www.phangs.org}) ALMA survey  \citep{Leroy2021a,  Stuber2021}.The PHANGS-ALMA  sample consists of 90 nearby ($d < 24$~ Mpc) 
star-forming galaxies ($\log \text{sSFR/yr}^{-1} > -11$) with stellar masses in the range $9.3 \leq \log M_{\star}/M_\odot \leq 11.1$.
 \citet{Stuber2021} use  ALMA CO(2-1) observations  with a resolution of $\sim100$~pc to look for molecular outflows in their sample and they find 
20 (22\%) outflow candidates, of which 16 are classified as ‘secure’ candidates.
  Half of the candidates are classified as AGN hosts based on \citet{Veron-Cetty2010}.

The outflow parameters of this sample have been measured using a method comparable to ours.  They integrated the total flux within the defined outflow velocity range to derive the outflow mass and measure the flux-weighted outflow velocity. 
 They also measured flux-weighted outflow radii.  Because of the limited S/N,  we did not attempt to derive flux-weighted \Rout\ by measuring the outflow radii in each channel,  but the \Rout\ derived from the 2D Gaussian fit should be representative of the radius where most of the flux is located.
To convert the CO(2-1) outflow flux  to outflow mass, \citet{Stuber2021}  use the conversion factor $\alpha_{CO}=0.8$ M$_\odot$/(K \kms\ pc$^2$) \citep{Bolatto2013},  similar to our assumption of $\alpha_{CO}=0.78$ M$_\odot$/(K \kms\ pc$^2$),  and $r_{21} =0.65$ \citep{Bolatto2013a, Leroy2013,denBrok2021, Leroy2021b} while we use the value of  $r_{21} =0.91$ measured in ULIRGs  \citep{Papadopoulos2012}. 
 The SFR for this sample are derived by \citet{Leroy2021a}  from the WISE band 4 photometry, with a calibration consistent  with \citet{Kennicutt2012},  the one used for the PUMA targets.
 The molecular outflows detected in the PHANGS galaxies have weighted velocities $v_{out}= 65-238 $~\kms, $M_{out}= 0.35-102\times 10^6$~\Msun\ and $\dot{M}_{out} =0.15-21$~\Msun\ yr$^{-1}$.  These outflow parameters are lower than the ones we measure for PUMA,  as it is expected for galaxies with lower SFR compared to ULIRGs.

The left panel of  Figure~\ref{fig:Q_out_vs_SFR} shows the relation between mass outflow rate and SFR.  The dashed lines indicate constant mass-loading factors ($\eta = \dot{M}_{out}/$SFR). The PUMA objects have $\eta < 0.04-1$. The PHANGS-ALMA galaxies have slightly higher $\eta =0.15- 4$.    The dispersion in \Mrate\ for a given SFR is quite high ($\sim1$~dex).  
Mass-loading factors in local starburst galaxies are typically lower than $2-3$ \citep[e.g. ][]{Bolatto2013,  Cicone2014, Salak2016},  
thus, the outflows of the PUMA sample are consistent with being powered by starburst. 
We note that the nuclei classified as AGN through MIR or optical diagnostics (orange symbols) do not show higher mass-loading factors compared with the rest of the sample.
For the PUMA sample, the vertical errorbars are probably more extended towards higher values, given that with our method it is possible that we are underestimating \Mrate\ in some of the objects (see discussion in Sec.~\ref{sec:caveats}).

In Figure~\ref{fig:Q_out_vs_SFR}  we also show the outflow momentum rate ($\dot{P}_{out}$) as a function of SFR.
We follow \citet{Pereira-Santaella2018} to calculate the total momentum injected by supernova explosions. We assume that the momentum per supernova is $1.3\times 10^5$ \Msun\ \kms $\times(n_0/100 \text{ cm}^{-3})^{-0.17}$ \citep{Kim2015},  where $n_0=100 \text{ cm}^{-3}$  is the electron density; and that the supernova rate is $0.012\times$SFR (\Msun\ yr$^{-1}$) for the adopted IMF \citep{Leitherer1999}.
Since all the PUMA galaxies have $\dot{P}_{out}$ smaller than the total momentum injected by supernovae (SNe,  see dashed line in the middle panel of Fig.~\ref{fig:Q_out_vs_SFR}), their outflow momentum rate could be explained entirely by SNe.

The outflow kinetic luminosity is shown in the right panel of Figure~\ref{fig:Q_out_vs_SFR}. The dashed lines show  fractions of the energy produced by SNe (1\%, 10\%, 100\%).
The kinetic energy injected by supernova explosions is calculated as $L_{SNe}$ [erg s$^{-1}$]$= 9\times 10^{41}\times$SFR [\Msun yr$^{-1}$] \citep[][]{Leitherer1999}, adapted for a \citet{Kroupa2001} IMF.
For the PUMA ULIRGs, the outflow kinetic luminosity is $0.2-5.9$\% of the energy produced by SNe, for the PHANGS-ALMA sample it is slightly lower ($0.1-2.4$\%).
The linear fit to the two samples gives:
\begin{equation}
\log L_{out}\ \text{[erg s}^{-1}] = (39.4\pm0.08)+(1.3\pm0.05) \log SFR\ \text{[M}_{\odot}\ \text{yr}^{-1}].
\end{equation}
The slope is larger than one, meaning that,  assuming that these outflows are driven by SNe, the coupling efficiency between ISM and SNe increases slightly with SFR.
This slope is shallower than the value of $2\pm0.2$ found by \citet{Pereira-Santaella2018} using a smaller sample (15 objects).   The difference could be partly explained by the different method used to derive the outflow parameters in \citet{Pereira-Santaella2018}, which obtained higher mass outflow rates for ULIRGs (12-400~\Msun\ yr$^{-1}$).

\begin{figure*}
\centering
\includegraphics[width=0.33\textwidth]{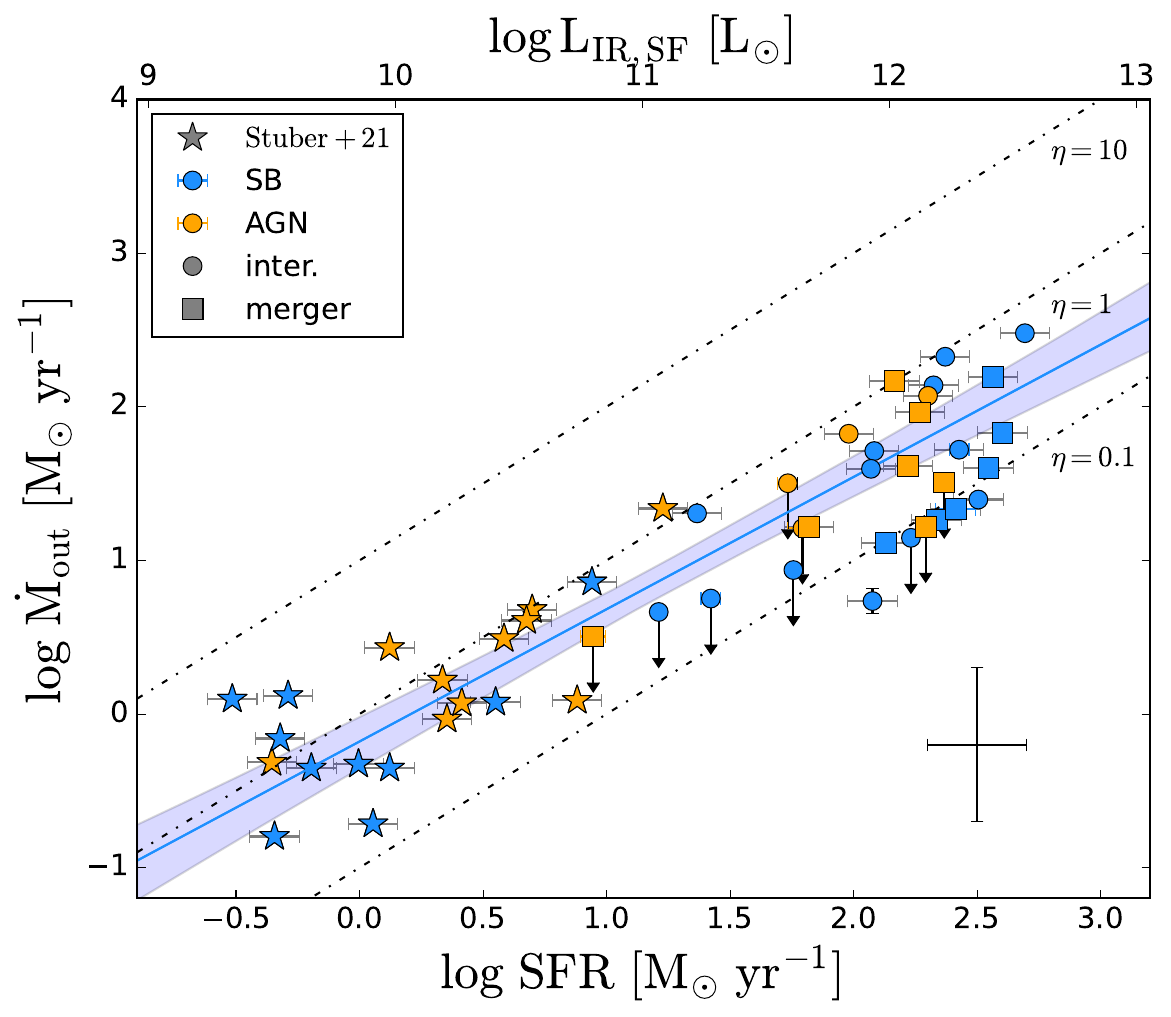}
\includegraphics[width=0.33\textwidth]{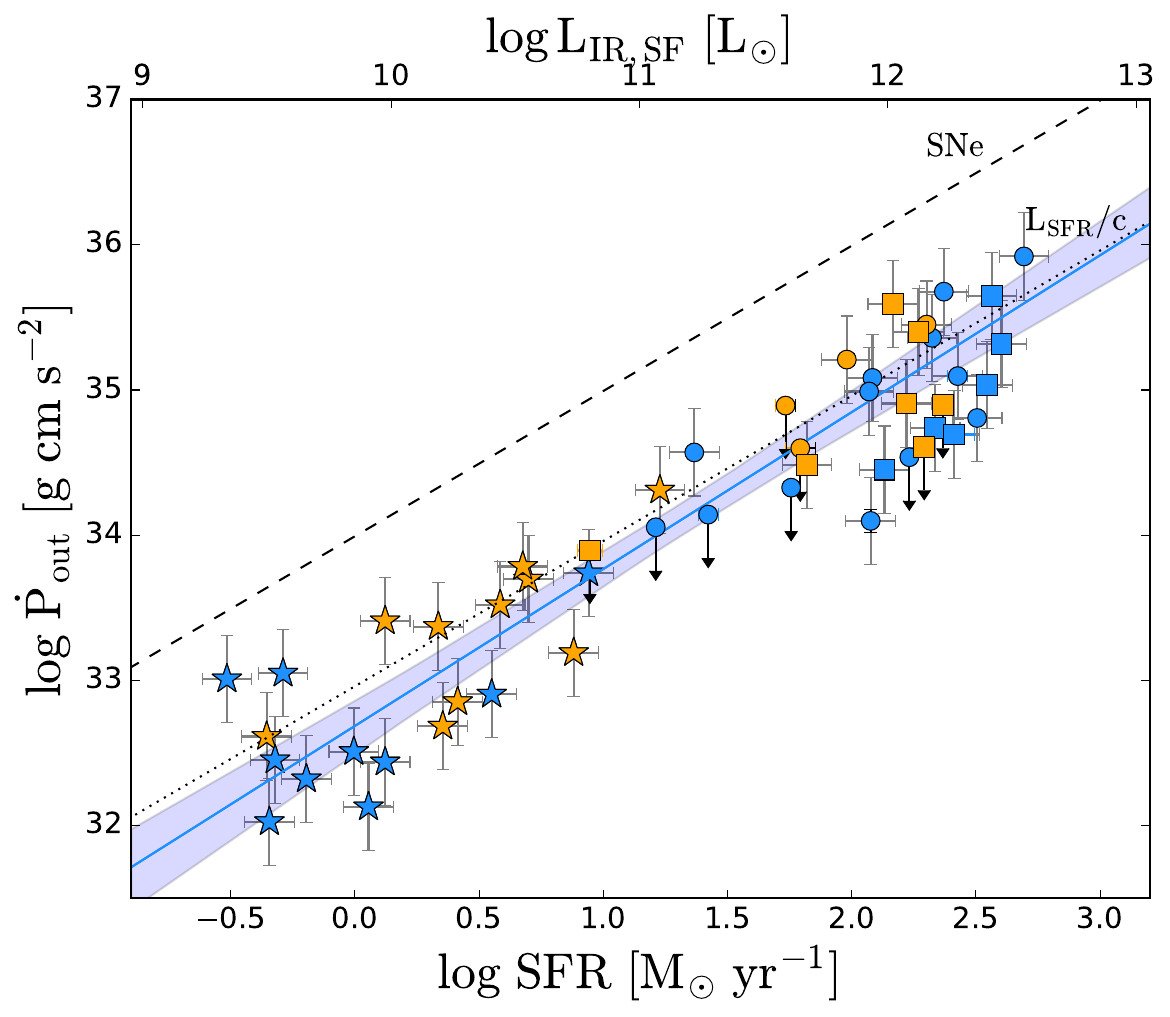}
\includegraphics[width=0.33\textwidth]{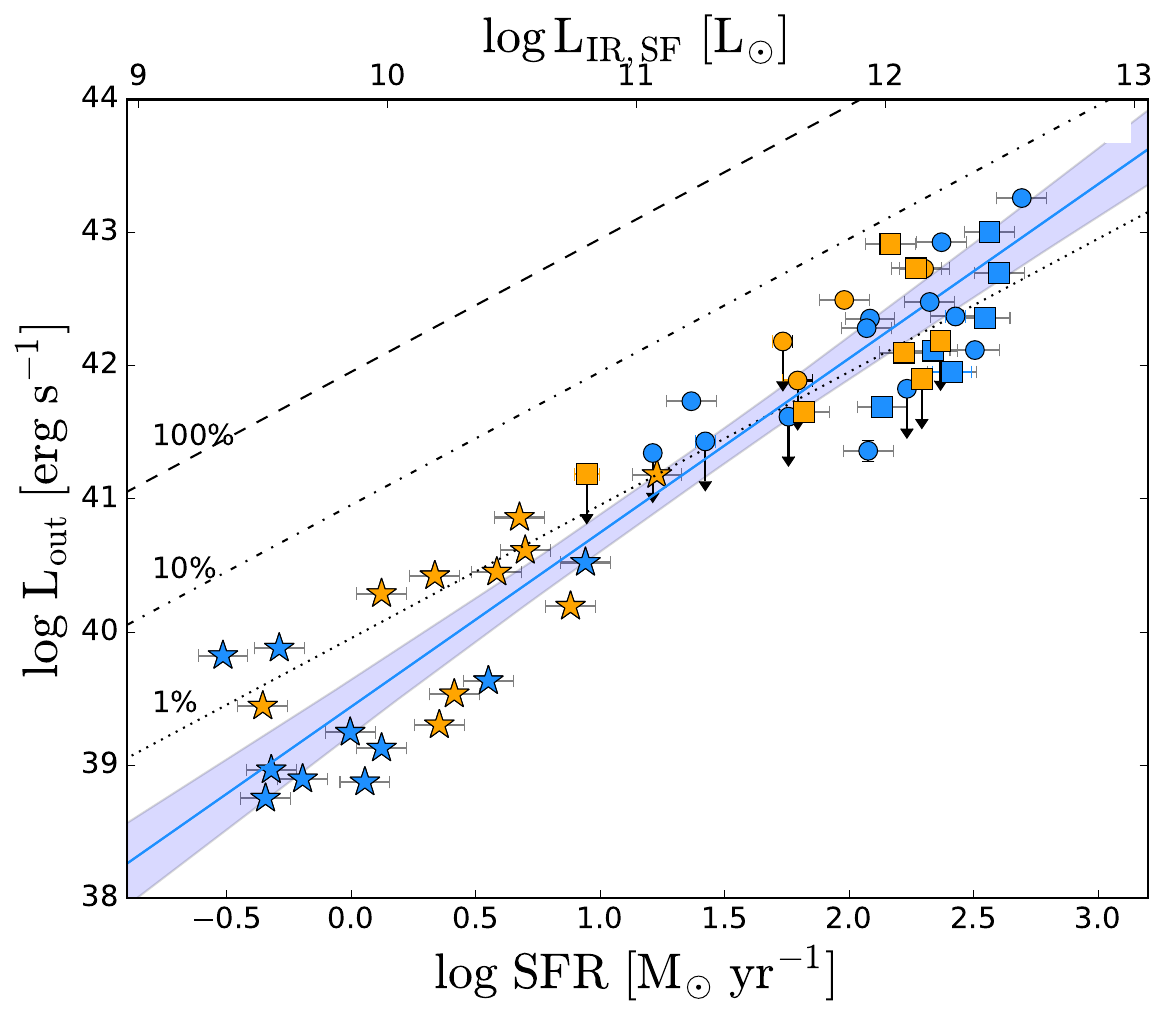}

\caption{ Mass outflow rate (\textit{left}), outflow momentum rate (\textit{middle}), and outflow kinetic luminosity (\textit{right}) as a function of SFR.  Lightblue and orange symbols indicate SB and AGN dominated nuclei, respectively.  Circles indicate interacting systems and squares indicate mergers from the PUMA sample (including  nuclei with $\log L_{IR}/L_{\odot}<11.8$).  
Stars show the PHANGS-ALMA targets from \citet{Stuber2021}.  A representative errorbar is shown on the lower right.
\textit{Left panel}: the black dotted-dashed lines  show lines of constant mass-loading factors ($\eta$=\Mrate/SFR) of 0.1, 1 and 10.  \textit{Middle panel}: the dashed line  indicates the total momentum injected by SNe as a function of SFR.  The dotted line shows the $L_{SFR}/c$ ratio, where $L_{SFR}$ is the IR luminosity for a given SFR derived using the \citep{Kennicutt2012} relation. 
 \textit{Right panel:}the lines show the  1\%, 10\% and  100\% of the energy produced by SNe. 
The lightblue lines in the three panels are the best linear fit to the data, with the shaded area indicating the 1$\sigma$ uncertainty.}
\label{fig:Q_out_vs_SFR}
\end{figure*}

\subsubsection{Outflow launching mechanism}
\label{sec:out_mechanism}
In this section we investigate the outflow launching mechanism.  In particular, we investigate whether the detected molecular outflows are  momentum-driven or energy-driven.
For momentum-driven outflows,  the mass-loading factor ($\eta = \dot{M}_{out}/SFR$) is proportional to $v_{out}^{-1}$, while for energy driven outflows $\eta \propto v_{out}^{-2}$ \citep[e.g.][]{Murray2005}.
It is possible to distinguish between the two scenarios by looking at the slope of the relation between $\log \eta$ and $\log v_{out}$:
\begin{equation}
\log \eta =\log \left( \frac{\dot{M}_{out}}{SFR} \right) = \alpha \cdot \log v_{out} + b,
\label{eq:mass_load}
\end{equation}
where $ \alpha$ and $b$ are the slope and intercept of the linear relation.  
In Figure~\ref{fig:Mload_vs_logv} we show $\log \eta$ as a function of $\log v_{out}$ for the PUMA and PHANGS-ALMA samples.  
We fit a linear relation between these two quantities using  the Monte Carlo Markov-Chain (MCMC) implementation PyStan\footnote{https://mc-stan.org/users/interfaces/pystan}.
The lightblue shaded region shows the 1~$\sigma$ uncertainty on the fit, obtained by sampling the posterior distribution. We assume a systematic error of 0.1~dex on $v_{out}$ and 0.3~dex on $\eta$. We also model the intrinsic scatter of the relation and we assume that the noise is normal distributed.  
 We derive the best fit values and the corresponding 1$\sigma$ uncertainties from the median and the 16th-84th percentiles, respectively,  of the marginalised posterior distributions of the parameters (more details on the fitting procedure can be found in appendix~\ref{sec:fitting_method}).
 
The best linear bisector fit gives a slope $\alpha=-1.43\pm0.28$ ($\sigma_{intr}$ = 0.35). This slope is  between the momentum-driven ($\alpha=-1$) and energy-driven ($\alpha=-2$) value.  From Fig.~\ref{fig:Mload_vs_logv}, we can see that both slopes are included within the 1$\sigma$ fit uncertainty (lightblue shaded area),  and therefore we cannot distinguish between the two options on the basis of this figure.

\begin{figure}
\centering
\includegraphics[width=0.48\textwidth]{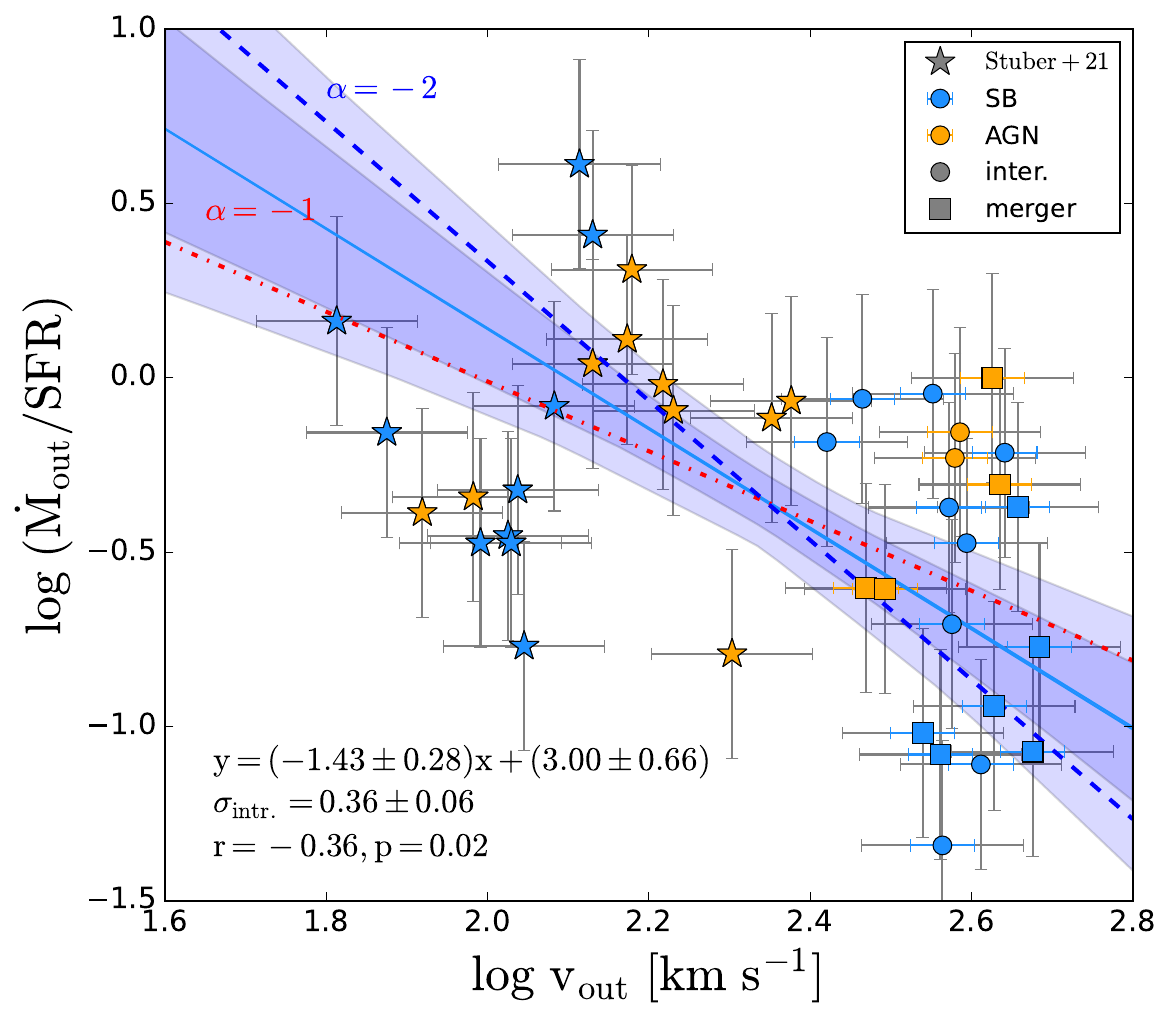}
\caption{Mass-loading factor versus outflow velocity for our sample (circles and square symbols) and the PHANGS-ALMA sample (stars). The lightblue line is the best linear bisector fit to the data  and the shaded areas indicates the 1$\sigma$ uncertainty on the fit with and without including the intrinsic scatter term (lighter and darker colour, respectively).  The blue dashed line  shows the predictions for a energy-driven outflow ($\alpha=-2$) and the red dotted-dashed line for a momentum-driven  outflow ($\alpha=-1$).}
\label{fig:Mload_vs_logv}
\end{figure}

\begin{figure}
\centering
\includegraphics[width=0.48\textwidth]{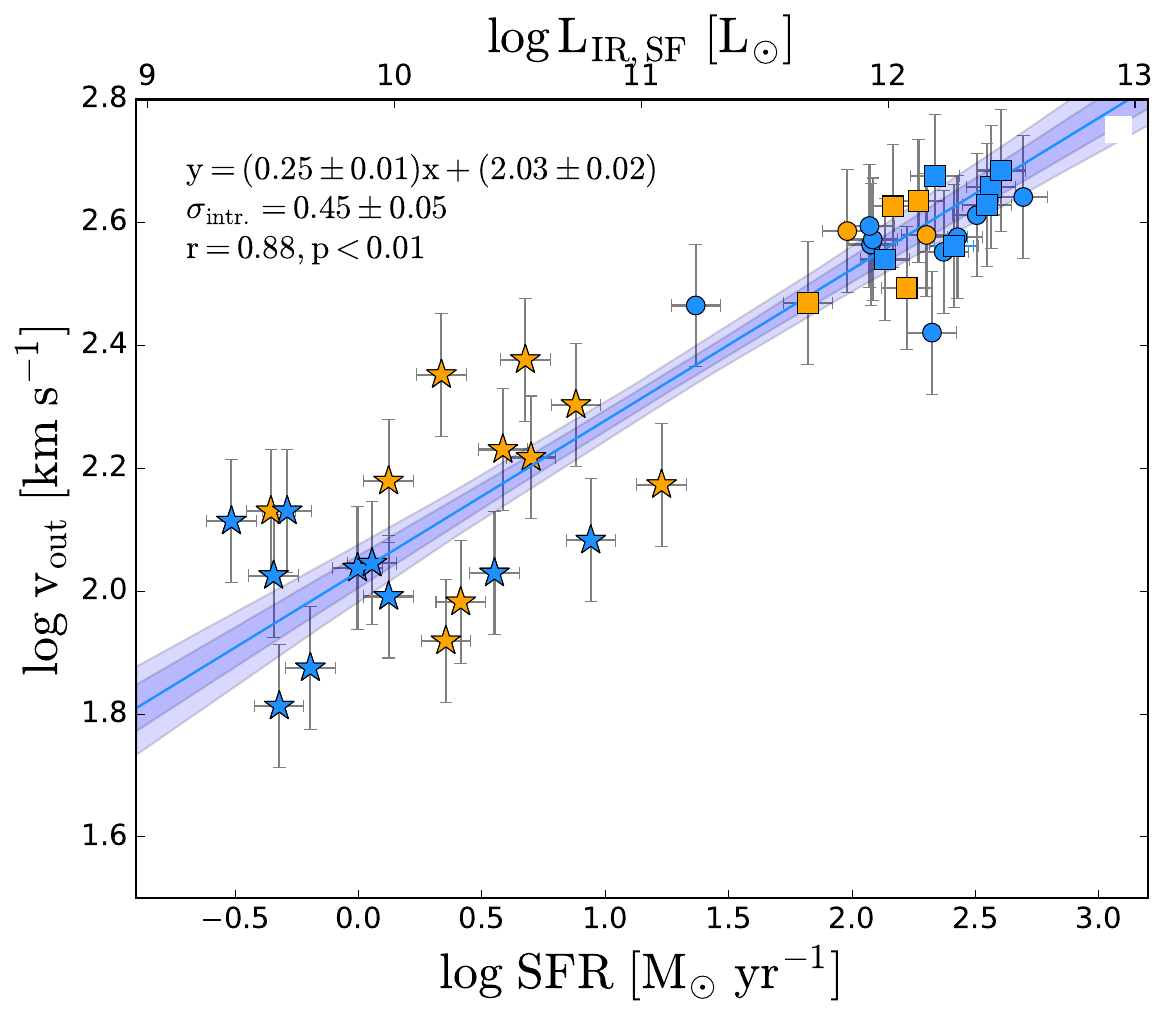}
\caption{Outflow velocity versus star-formation rate for our sample and the PHANGS-ALMA sample. The lightblue line is the best linear fit to the data (slope: $0.25\pm0.01$);  the 1$\sigma$ uncertainties on the fit  (with and without including the intrinsic scatter term) are indicated with shaded areas (lighter and darker colour, respectively).  Symbols as in Fig.~\ref{fig:Q_out_vs_SFR}.}
\label{fig:SFR_vs_logv}
\end{figure}

In  Figure~\ref{fig:SFR_vs_logv}, we show the SFR as a function of $v_{out}$.  There is a strong positive correlation between these two quantities ($r=0.84$,  $p$-value$<0.01$).The best linear fit gives:
\begin{equation}
\log v_{out} \propto (0.25\pm0.01) \cdot \log SFR.
\end{equation}
The slope of the relation is in agreement with the relation found for the ionised gas by \citet{Arribas2014} (slope $0.24\pm0.05$) and for neutral gas by \citet{Rupke2005b} (slope $0.21\pm0.04$).  It is also in agreement with the theoretical prediction from \citet{Heckman2000} of the relation between velocity and starburst luminosity $v_{max}\propto L_{bol}^{0.25}$.  
 This relation is derived under the assumption that starbursts have a maximum characteristic surface brightness \citep{Lehnert1996b, Meurer1997},  and therefore their bolometric luminosity is proportional to their radius squared.
 We note that these relations consider $v_{max}$,  not the mean outflow velocity.  We cannot use $v_{max}$ in our analysis, because it is not available for the PHANGS sample and we consider it  less robust than $v_{out}$.  Nonetheless, for the PUMA sample we find a good correlation between $v_{max}$ and $v_{out}$ ($r=0.58$,  $p$-value$<0.01$).

Using this relation, we can substitute $\log SFR$ with $4 \cdot \log v_{out}$ in the expression of the mass-loading factor:
\begin{equation}
\begin{split}
& \log \eta =\log \left( \frac{\dot{M}_{out}}{SFR} \right)
 = \log \dot{M}_{out}- \log SFR \\
& = \log \frac{v_{out}\cdot{M}_{out}}{R_{out}}- 4\cdot \log v_{out} \\
& = \log \frac{{M}_{out}}{R_{out}}- 3\cdot \log v_{out}.
 \end{split}
\end{equation}
Substituting this in equation \ref{eq:mass_load} gives:
\begin{equation}
\begin{split}
\log \eta = \log \frac{{M}_{out}}{R_{out}}- 3\cdot \log v_{out}= \alpha \cdot \log v_{out} +b
 \end{split}
\end{equation}
\begin{equation}
\begin{split}
 \log \frac{{M}_{out}}{R_{out}}= (\alpha+3) \cdot \log v_{out} +b.
 \end{split}
\end{equation}
Thus, by looking at the relation between ${M}_{out}/R_{out}$ and $v_{out}$, we can derive the value of $\alpha$. 
 For the fit, we assume a systematic error of 0.1~dex on $\log v_{out}$ and 0.3~dex on $\log ({M}_{out}/R_{out})$. 
Figure~\ref{fig:M_over_R_vs_logv} shows  $\log (M_{out}/R_{out})$ as a function of $\log v_{out}$.   

The best linear bisector fit relation has a slope of $2.61\pm0.25$, which implies $\alpha=-0.39$. The $\alpha=-1$ scenario (momentum-driven) is within the 1$\sigma$ uncertainty (see shaded area in Fig.~\ref{fig:M_over_R_vs_logv}), while the $\alpha=-2$ scenario does not agree with the fit. Therefore,  our analysis  favours the momentum-driven scenario as the primary launching mechanism for molecular outflows in ULIRGs.   Previous works investigating the molecular outflows in ULIRGs also reach this conclusion. 
\citet{Cicone2014} find $\alpha =-1.0\pm 0.5$ for a sample of five pure starburst galaxies, in agreement with the momentum-driven scenario.
\citet{Pereira-Santaella2018} find $\alpha=-0.3\pm0.2$ combining a sample U/LIRGs and lower luminosity starbursts.  
We note that for this analysis we adopt a different $r_{21}$ value for the \citet{Stuber2021} sample ($r_{21}=0.6$) and for the PUMA sample ($r_{21}=0.91$). If we were to use the same value for both samples, the fit would agree even more with  the momentum-driven scenario. 

\begin{figure}
\centering
\includegraphics[width=0.48\textwidth]{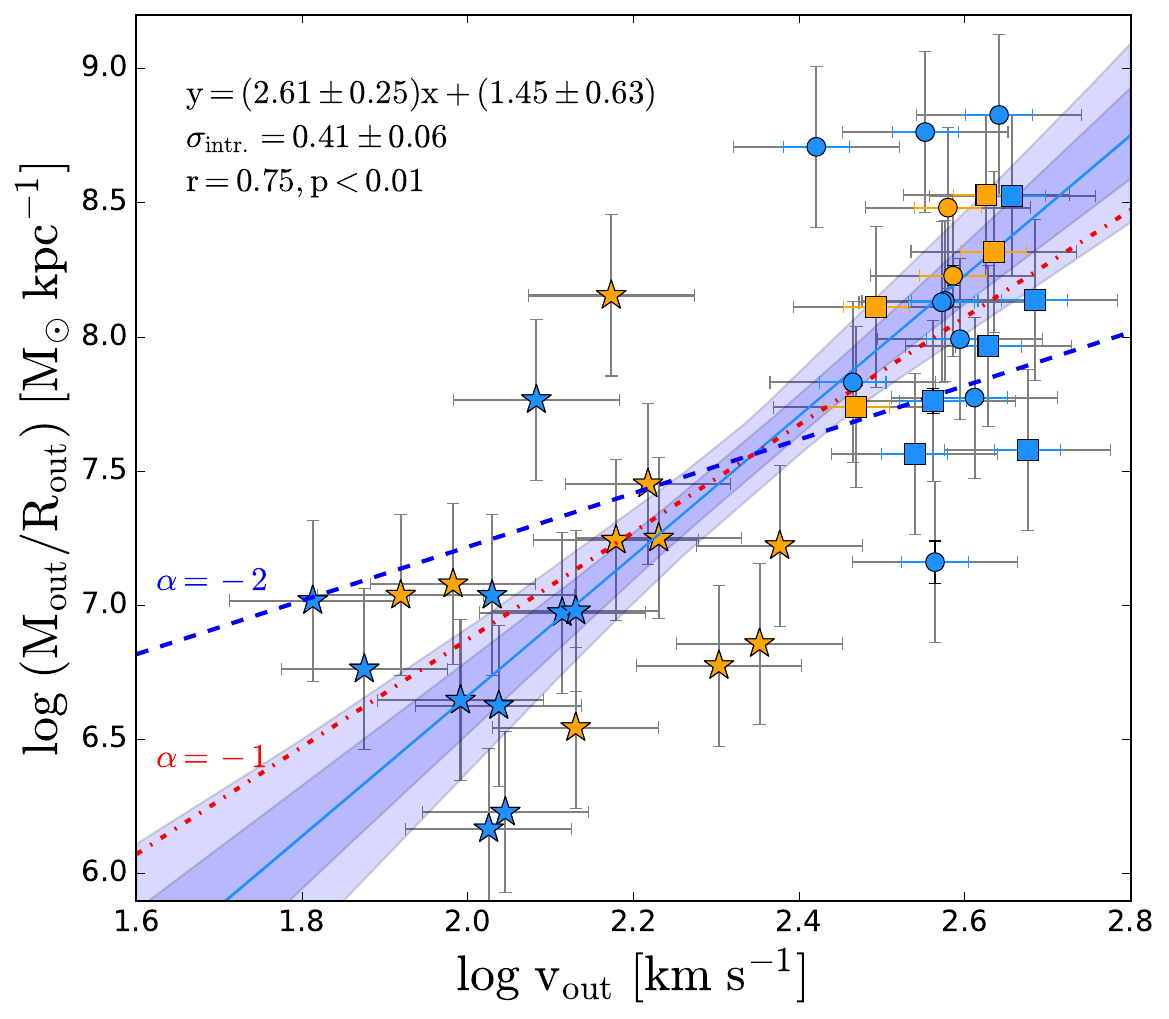}
\caption{Outflow mass divided by outflow radius (\Mout/\Rout) versus outflow velocity.  The lightblue line is the best linear fit to the data (slope=$2.61\pm0.25$). The shaded area indicates the 1$\sigma$ uncertainty   with and without including the intrinsic scatter term (lighter and darker colour, respectively). The blue dashed line shows the predictions for a energy-driven outflow ($\alpha=-2$) and the red dotted-dashed line for a momentum-driven  outflow ($\alpha=-1$).  Symbols as in Fig.~\ref{fig:Q_out_vs_SFR}.}
\label{fig:M_over_R_vs_logv}
\end{figure}

\subsubsection{Escape fractions}
\label{sec:escape}

We estimate the average fraction of the outflowing gas that can potentially escape from the gravitation potential of the galaxies.
To estimate the escape fraction, we compare the outflow velocity with the escape velocity derived from a gravitational model of the host
galaxy, which is assumed to be a truncated isothermal sphere \citep[e.g.][]{Veilleux2005}.
We use equation (7) in \citet{Arribas2014} to estimate the average escape velocity for our sample at  the average outflow radius $r=1$~kpc. 
 Since estimates of the dynamical masses  have been obtained by \citet{Perna2022} only for a subset of our targets (eight objects),  we use the mean dynamical mass of the PUMA sample for our calculation $\langle M_{dyn} \rangle = 3.9\cdot10^{10}$~\Msun. The dynamical masses have been obtained from  the modelling of the rotation of the ionised gas  \citep[see][ for more details]{Perna2022}.
As truncating radius we assume two times the average effective radius $\langle R_e \rangle = 2$~kpc \citep{Perna2022}.  We assume that the outflow velocity only has a radial component.With these assumptions,  we estimate an average escape velocity of $v_{esc}=486\pm40$~\kms.  We apply an inclination correction to determine the average `observed' $v_{esc, obs}=v_{esc}/1.27= 382$~\kms (see Sec.~\ref{sec:outflow_Q}). 

Integrating the CO(2-1) emission at velocities higher than $v_{esc, obs}$, we find  that  between $4-100$\% of the high-velocity gas will escape to the circumgalactic medium, with mean escape fraction $\langle f_{esc} \rangle= 45\pm6\%$.  
 This calculation depends on the assumption on the truncating radius. If we were to assume a larger truncating radius, the escape velocity would increase, and the escape fraction would decrease.  For example,  assuming a truncating radius larger by a factor of two ($r_{max}=4\times\langle R_e \rangle=8$~kpc),  $v_{esc}$ would increase by 13\%. 
The escape outflow rates are in the range 1-173~\Msun\ yr$^{-1}$,  with an average $\langle \dot{M}_{esc} \rangle = 40\pm10$~\Msun\ yr$^{-1}$.  
We find that interacting systems tend to have lower  $f_{esc}$ than mergers (mean $f_{esc}=33\pm6$\% vs.  $f_{esc}=60\pm10$\%).

We also compare the escape molecular gas mass ($M(H_2)_{esc}$) with the total $M(H_2)$  of the galaxy (i.e.  systemic and outflow). The escape $M(H_2)_{esc}$ is  $<5\%$ of the total  $M(H_2$), with a mean of 1\%.
 The mean depletion time based on the escape outflow rate ($t_{dep} = M(H_2)/\dot{M}_{esc}$) is 158~Myr (range $23-3715$~Myr).

\subsection{OH vs.  CO outflow properties}
\label{sec:OH_comp}

In this section we compare the molecular outflow properties derived from the  CO(2-1) emission line with the ones derived from the OH 119~\micron doublet, which is an alternative tracer used to identify molecular outflows.

We compare the OH and CO profile of each PUMA target in Figure~\ref{fig:spectroastrometry},  \ref{fig:spectroastrometry2},  and \ref{fig:spectroastrometry_app}. The OH absorption is produced in front of the continuum emission, thus,  we can assume that it is located in a region of the size of the continuum, which in most cases is equal or smaller than the beam size \citep{Pereira-Santaella2021}. Therefore,  to have a fair comparison,  we extract the nuclear CO spectrum  from a region equal to the beam size. We also smooth the CO spectrum to the same spectral resolution of the OH spectrum.
Qualitatively, it is evident from this comparison that there are some cases where a clear blue-shifted outflow is detected in OH but not in CO (e.g. 00188-0856), and vice-versa (e.g. 09022-3615, 17208-0014).
In Figure ~\ref{fig:comp_OH_CO_profile} we show four example cases:  i) broad OH absorption reaching  high negative velocities ($v< -1000$~\kms), but no CO blue-shifted emission at $v < -300$~\kms (00188-0856);
ii) CO outflow, but no sign of outflow in OH (22491-1808);
iii) outflow detected in both OH and CO,  but OH outflow maximum velocity ($v_{98}= -1540$~\kms) much larger than CO outflow maximum velocity $v_{max}=-800 $~\kms  (20100-4156);
iv) OH outflow maximum velocity ($v_{98}=-360$~\kms) smaller than CO outflow maximum velocity $v_{max}= -600$~\kms (15327+2340).

\begin{figure*}
\centering
\includegraphics[width=0.45\textwidth]{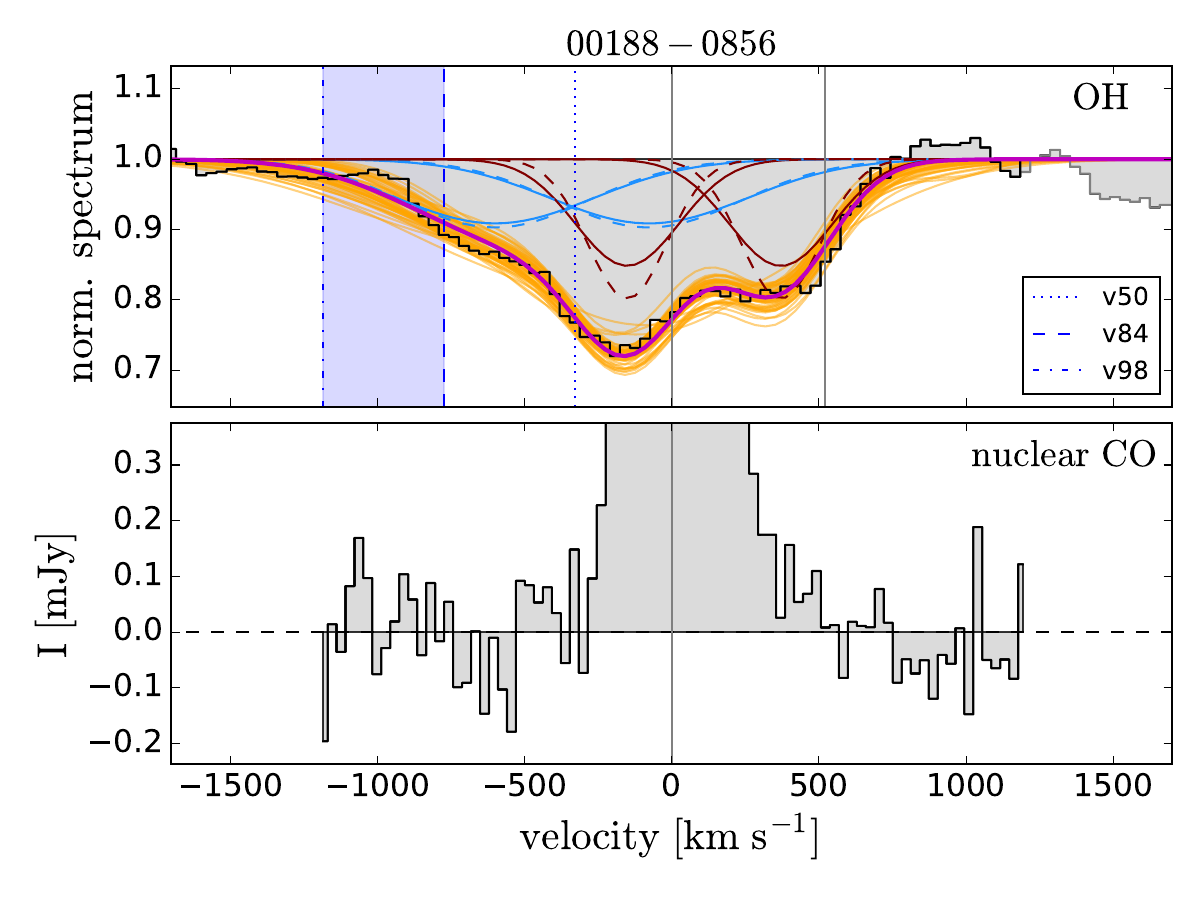}
\includegraphics[width=0.45\textwidth]{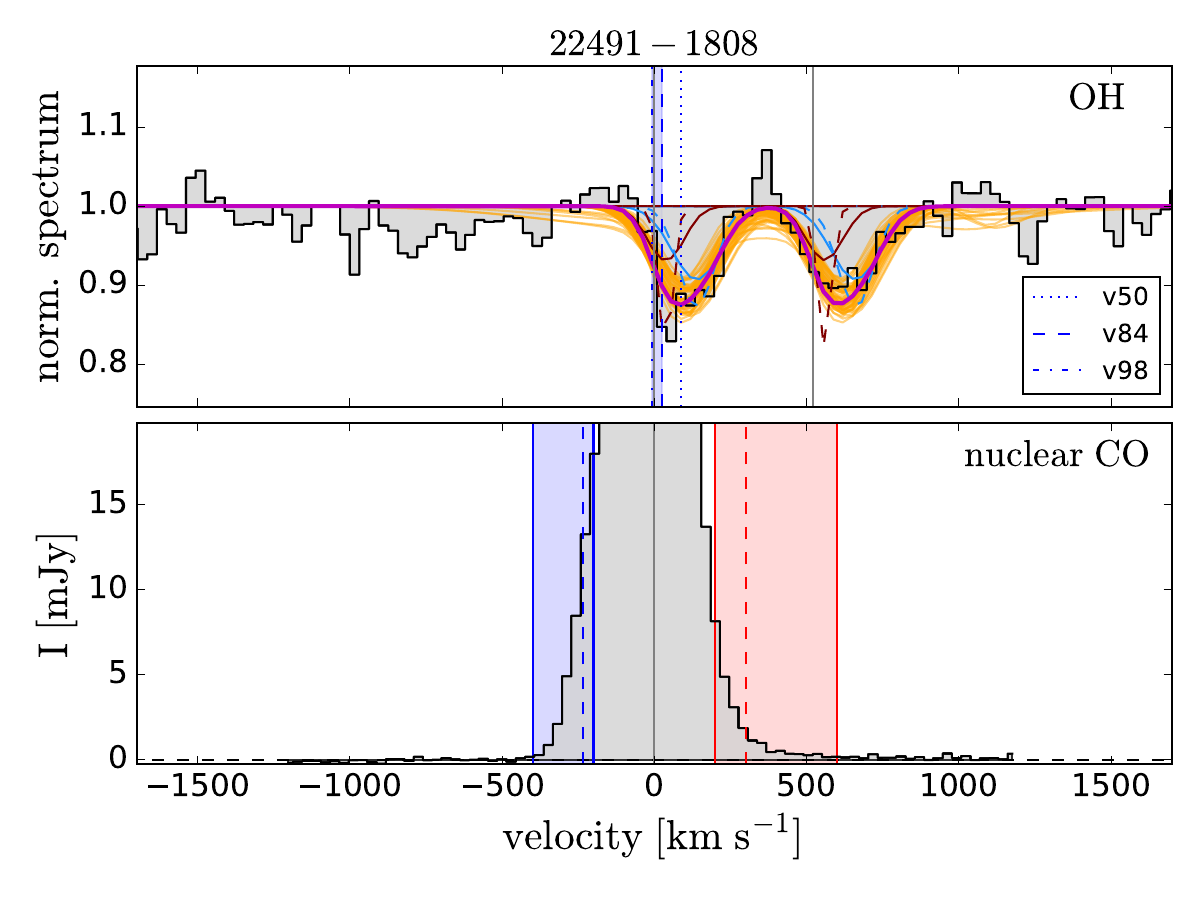}

\includegraphics[width=0.45\textwidth]{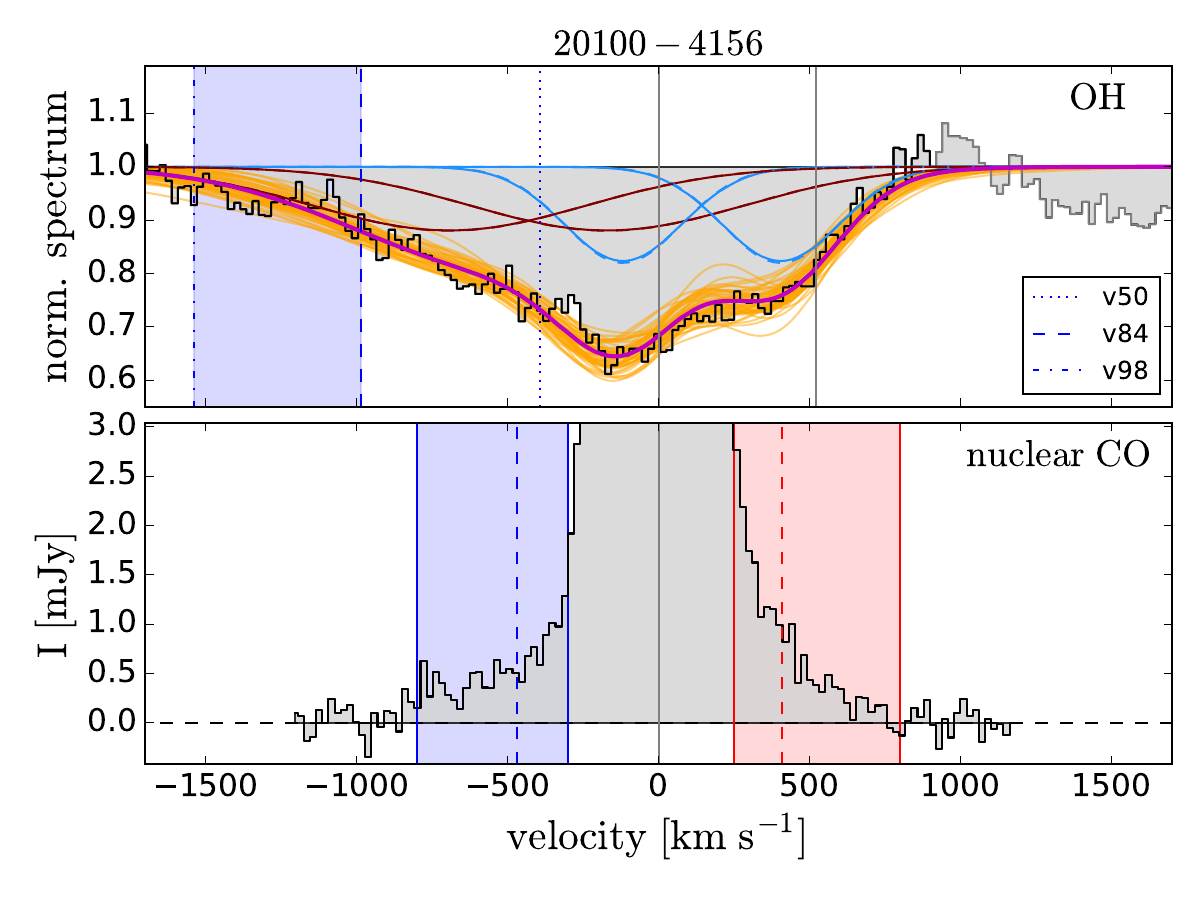}
\includegraphics[width=0.45\textwidth]{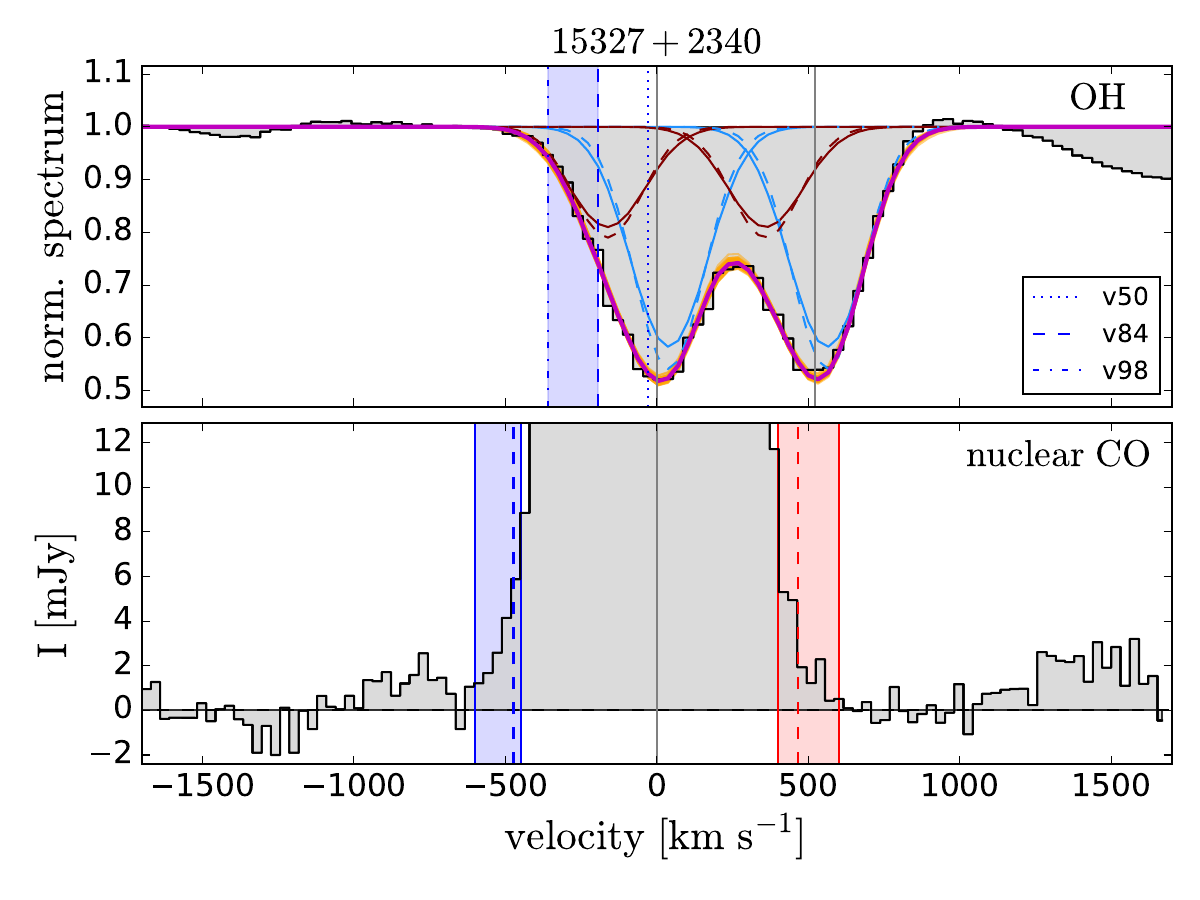}

\caption{Comparison of the OH 119~\micron and CO(2-1) line profile.  The CO(2-1) spectrum is extracted from a nuclear region corresponding to the ALMA beam size  and  is convolved to the resolution of the OH spectrum (FWHM$\sim270$~\kms). We identified four different cases: i) OH absorption reaching high velocity ($|v|> 1000$~\kms),  but no CO blue-shifted emission  (00188-0856);
ii) CO outflow, but no sign of OH outflow (22491-1808);
iii) outflow detected in both OH and CO,  but OH outflow maximum velocity ($v_{98}$) larger than CO outflow maximum velocity (20100-4156);
iv) OH outflow maximum velocity ($v_{98}$) smaller than CO outflow maximum velocity (15327+2340).
}
\label{fig:comp_OH_CO_profile}
\end{figure*}

In Figure~\ref{fig:comp_OH_CO}, we compare the OH and CO outflow velocities.  In the upper row, we compare the OH $v_{98}$ velocity of the absorption profile with the `weighted' outflow velocity (\vout, \textit{left}) and  maximum outflow velocity ($v_{max}$,  \textit{right}) of the blue-shifted wing of the CO profile.  We show the objects with OH absorption or P-Cygni profile,  for a total of 20 nuclei.  For 12112+0305 and 14348-1447, we show both nuclei,  since a CO outflow is detected around each nucleus and they have comparable continuum luminosity. 
For the other interacting systems, we consider only the nucleus with the brightest continuum,  since the second nucleus does not have a CO outflow detection. 

We identify three groups: i) targets with comparable OH and CO outflow velocity, ii) targets with OH outflow velocity larger than the ones from CO, and iii) targets with CO outflow velocity larger than the OH outflow velocity.
In 13 targets, both  |$v_{98}(abs)$(OH)| and $|v_{max}$(CO)| are $> 300$~\kms. The |$v_{98}$(abs)(OH)| values  tend to be higher (on average by 170~\kms ) than the $|v_{max}$(CO)| values,  but this is not surprising given that $v_{98}$ and $v_{max}$ are measured using different methods.

For 3/20 targets,   $|v_{98}(abs)\text{(OH)}|< 300$~\kms, } thus, there is no evidence of an outflow in the OH absorption profile, while a CO outflow is detected.  
This difference may be due to a collimated outflow in CO.  In this case,  the associated OH will absorb only a tiny fraction of the continuum behind the outflow and thus it will be undetectable.
 In Figure~\ref{fig:vdiff_Rout} we compare the velocity difference $|v_{98}(abs)\text{(OH)}-v_{max}\text{(CO)}|$ with the CO outflow radius \Rout. We find a weak anti-correlation  between the \Rout\ and the OH-CO  velocity difference ($r=-0.4$,  or $r=-0.3$ if we exclude the point with $v_{max}$(CO)=0~\kms), meaning that for more extended CO outflow, the CO velocity tend to be larger than the OH velocity. This is consistent with the scenario explained before,   where for an outflow extending to a larger distance from the nucleus,  a lower fraction of the outflowing gas may overlap with the background continuum, making the OH absorption weaker \citep{Veilleux2020}. 
 Additionally, if the velocity of the outflow increases with distance from the nucleus, the part with the highest velocity is more likely to not overlap with the continuum and thus, will not be detected in the OH absorption.   Indeed,  a radial velocity gradient of $\sim1$~km~s$^{-1}$~pc$^{-1}$ has been detected in the molecular outflows of two galaxies: in the nearby starburst galaxy NGC 253 by \citet{Walter2017} and in the LIRG ESO 320-G030 by \citet{Pereira-Santaella2020}.

For one target (00188-0856), the OH spectrum shows a clear outflow signature with $v_{98}(abs)\text{(OH)}\sim -1200$~\kms, but there is no outflow signature in the CO spectrum.  
The case of 00188-0856 can potentially be explained with extreme environments: if the  ionisation fraction of the molecular gas is high, the abundance of CO would decrease while the abundance of OH remains high \citep{Gonzalez-Alfonso2018}.  
This could be mostly associated with the highest-velocity gas.
 Alternatively,  this difference may be explained by  the geometry of the outflow. 

Another factor that can explain part of the differences in outflow velocities is the difference in sensitivities of the CO and OH observations.
 Thus,  the observation of one tracer may be more sensitive to weak outflow signatures at high velocities than the other.

We note that the AGN nuclei tend to have faster OH outflows than SB dominated nuclei.  Indeed,  all AGN nuclei have $|v_{98}\text{(OH)}|> 300$~\kms. There is a positive correlation ($r=0.51$) between the outflow velocity difference $|v_{98}\text{(OH)}$-$v\text{(CO)}|$ and the AGN fraction (see Fig.~\ref{fig:vdiff_Rout}).  
 \citet{Veilleux2013} find that AGN-dominated nuclei ($\alpha_{AGN} >0.5$) tend to have more negative OH outflow velocities than SB-dominated nuclei.  Their interpretation of this result is that once a significant fraction of the material has been pushed away from the nucleus by the outflow,  the AGN is more likely to be identifiable.  At the same time, the central high-velocity part of the outflow can be seen more easily.

To investigate the differences between the OH and CO outflow velocities, we consider the possibility that the presence of a compact obscured nucleus (CON)  could influence the OH outflow velocities. 
\citet{Falstad2019} find that CONs with the most luminous HCN-vib line lack signatures of  high velocity outflows in the OH 119~\micron\ absorption lines, even though some of them have molecular outflows detected in the CO or HCN emission lines. 
 We apply the method described in \citet{Garcia-Bernete2022} to identify CONs in our sample,  based on the ratios of the equivalent widths of different polycyclic aromatic hydrocarbons (PAHs) features from the literature (see Table~\ref{tab:sample}).  We find that CONs  do not occupy a specific region of the  $v_{98}(abs)$(OH) vs.  $v_{max}$(CO) diagram (see black circles in Fig.~\ref{fig:comp_OH_CO}).  We find that 6/7 CONs have |$v_{98}$(abs)(OH)|> 300~\kms, a sign of molecular outflows.

In Figure~\ref{fig:comp_OH_CO},  we also compared the velocities derived from the OH emission  ($v_{98}(em)$) with the velocities of the red-shifted CO outflow ($v_{out}(r)$ and $v_{max}(r)$). 
Most targets have $v_{98}(em)>300$~\kms.  For these targets,  $v_{98}(em)$ generally agree,  within the uncertainties,  with the CO outflow velocities.  The uncertainties on OH velocities derived from the emission profiles  tend to be larger than the ones derived from  the absorption,  because the emission features are weaker.

\begin{figure*}
\centering
\includegraphics[width=0.7\textwidth]{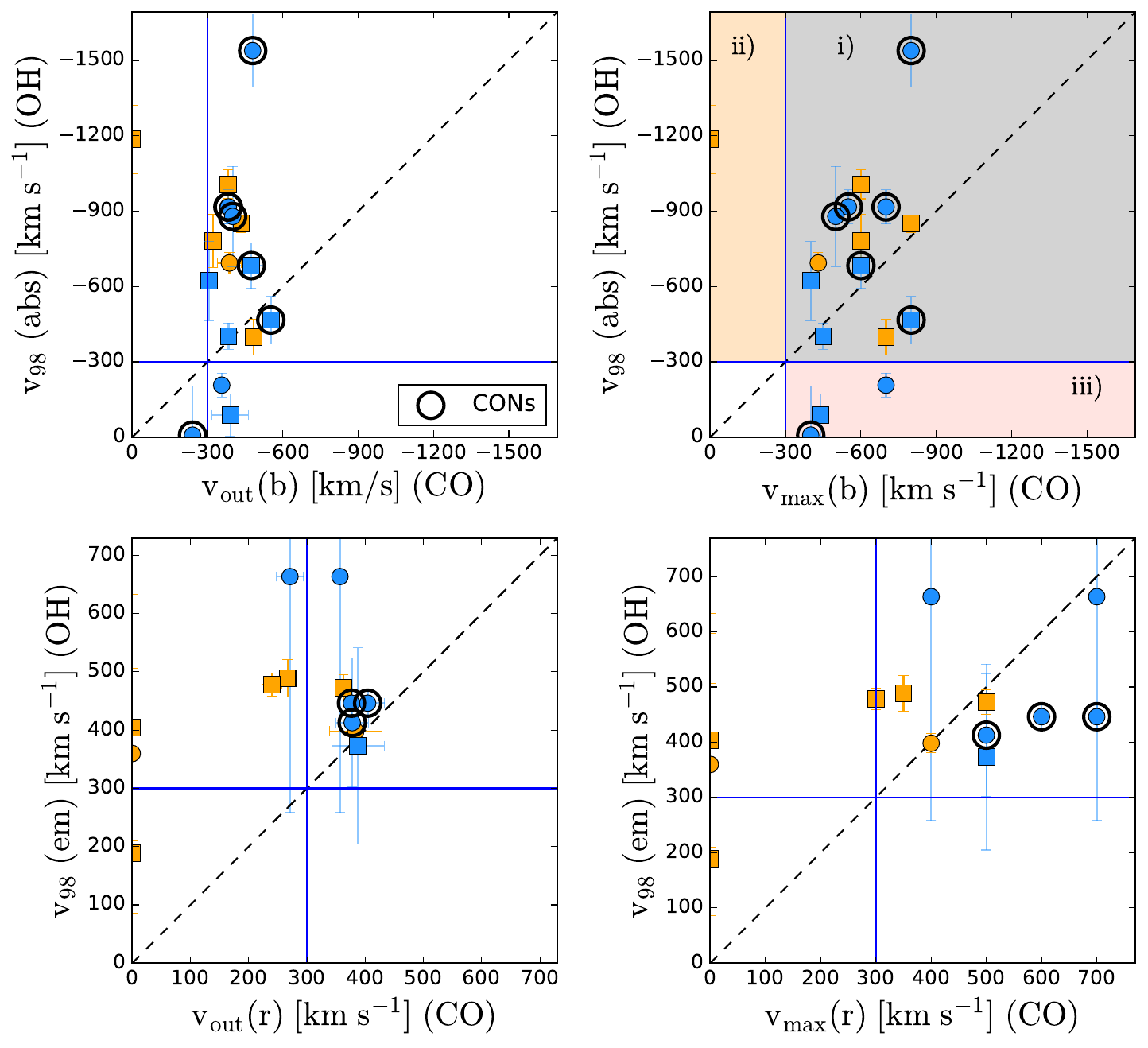}
\caption{Comparison of the outflow velocities of OH and CO. \textit{Upper row}: $v_{98}$ derived from the OH absorption profile compared with the mean ($v_{out}(b)$,  \textit{left}) and maximum ($v_{max}(b)$, \textit{right}) velocities of the CO  blue-shifted wing.  The background colours highlight the three regions described in section~\ref{sec:OH_comp}:  i) targets with $v_{98}$ similar to $v_{max}(b)$,  ii) targets with $v_{98} > v_{max}(b)$, and iii) targets $v_{84} < v_{max}(b)$. 
 \textit{Lower row}: $v_{98}$ derived from the OH emission profile compared with the mean ($v_{out}(r)$,  \textit{left}) and maximum ($v_{max}(r)$, \textit{right}) velocities of the CO  red-shifted wing.  The blue lines mark the outflow detection threshold of $|v|>300$~\kms. 
Black circles mark targets classified as compact obscured nuclei (CONs).  Colour-code as in Fig.~\ref{fig:sample}.}
\label{fig:comp_OH_CO}
\end{figure*}

\begin{figure}
\centering
\includegraphics[width=0.24\textwidth]{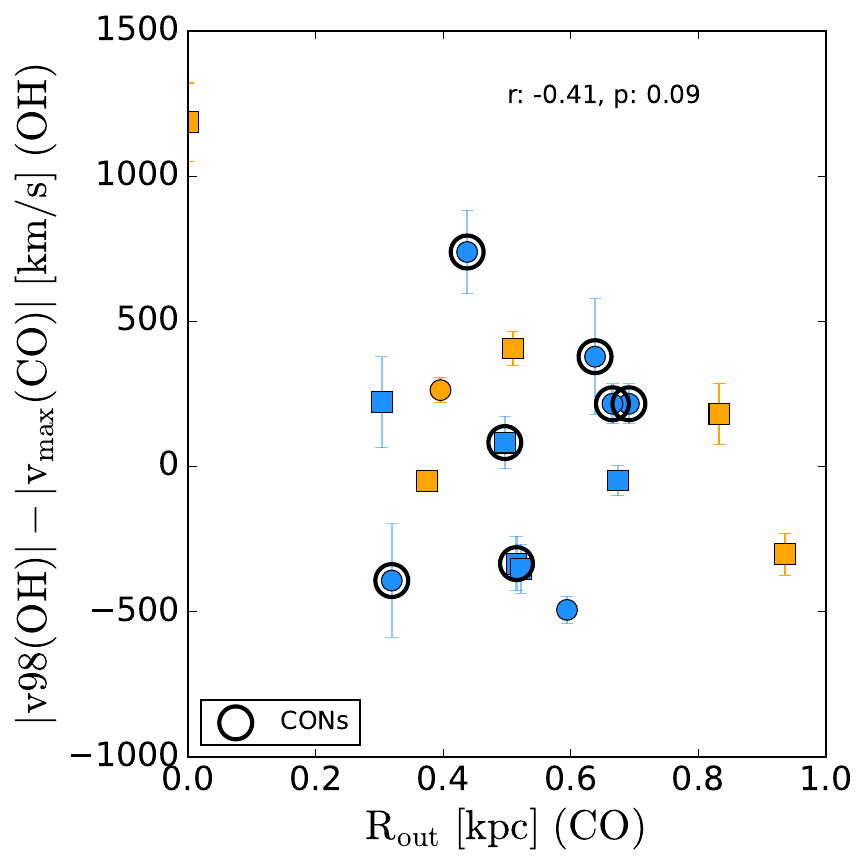}
\includegraphics[width=0.24\textwidth]{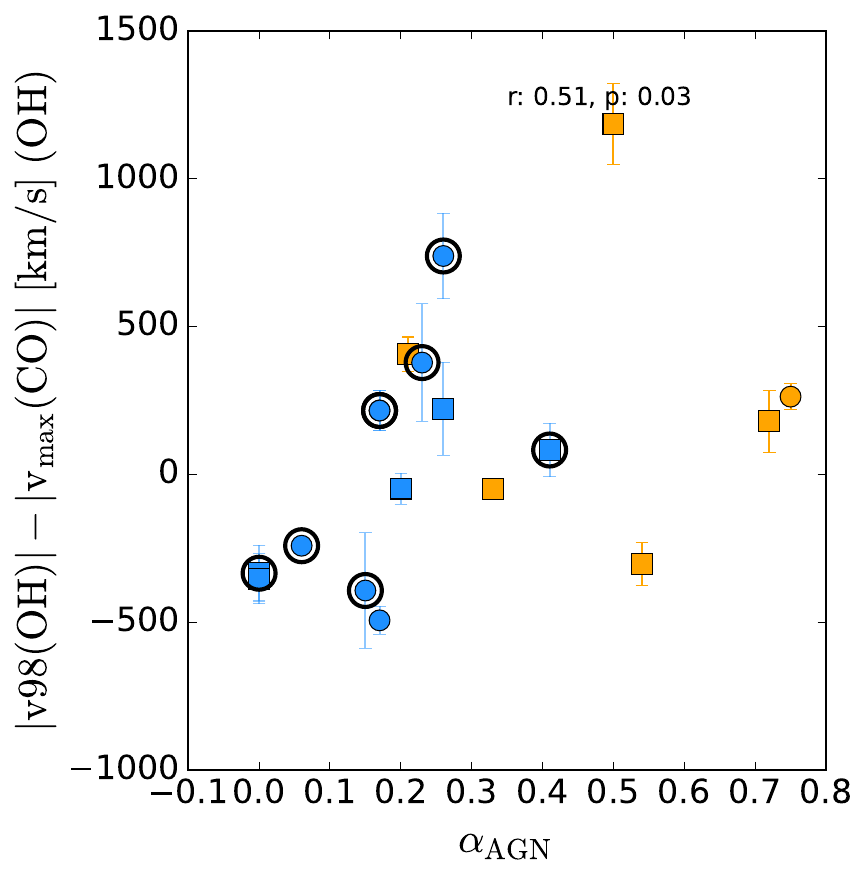}
\caption{Comparison of the difference between the outflow velocities of OH ($v_{98}$) and CO ($v_{max}$) with the CO outflow radius \Rout\ (\textit{left}) and  the AGN fraction $\alpha_{AGN}$ (\textit{right}).  Black circles mark targets classified as compact obscured nuclei (CONs).  Colour-code as in Fig.~\ref{fig:sample}.}
\label{fig:vdiff_Rout} 
\end{figure}

One  possible caveat of this analysis is the fact  that the OH spectrum includes the nuclear region with a beam size $\sim 20$ arcsec \footnote{http://herschel.esac.esa.int/Docs/PACS/html/pacs\_om.html}.  For the interacting systems, these include the two nuclei.
To detect the OH absorption, it is necessary to have the continuum emission in the background. In the majority of interacting cases, the ALMA continuum emission is dominated by one nucleus \citep[ > 80\% of the total emission for all but two cases, 11095-0238 (65\%) and 14348-1447 (60\%),  see Table 7 in][]{Pereira-Santaella2021},  thus we can associate the OH absorption to the nucleus with the brightest continuum emission.

\section{Discussion}
\label{sec:discussion}

In our sample, we find a lower molecular outflow detection rate in AGN ($55\pm14\%$) than in SB ($93\pm4\%$),  for the nuclei with $\log L_{IR}> 11.8$ \Lsun\ (see Sec. ~\ref{sec:outflow_det}).  This finding is in contrast with the result of \citet{Stuber2021}, who find higher molecular outflow detection rate in AGN (53\%) compared to non-AGN (17\%) in  main-sequence galaxies from the PHANGS sample.  A possible explanation is that the situation in ULIRGs is different than in main-sequence galaxies: in ULIRGs, SB-driven outflows are stronger and may be easier to detect than AGN-driven outflows.  In main-sequence galaxies on the other hand,   outflows driven by star-formation are weaker, and the AGN-outflows are more important.

Other than the detection rate,  we do not find significant differences in the outflow properties (\vout, \Rout, \Mout, \Mrate) in AGN and SB in the PUMA sample. 
Thus,  molecular outflow in AGN-dominated nuclei seem to be as powerful as SB-driven outflows, contrary to what was found in previous works \citep{Cicone2014,   Fluetsch2019}. 
\citet{Cicone2014} and \citet{Fiore2017} reported  a  positive correlation  between \Mrate\ and AGN luminosity,  while in the PUMA sample we do not see this correlation, in agreement with the findings of \citet{RamosAlmeida2022}.  
 The previous relations maybe trace the  most efficient coupling between AGN and molecular gas, so that the objects with the most extreme outflows follow those relations. 
This implies that there are other factors that set the mass outflow rate other than the AGN luminosity,  as for example the geometry of the outflow and the coupling between the outflow and the CO disk,  the efficiency of the energy/momentum transfer between the AGN and the molecular gas,  or the distribution of the gas around the AGN.  
Additionally,  starbursts could contribute to powering the molecular outflows even in AGN-dominated nuclei.

In our sample,  we find outflow dynamical times 0.5-2.8~Myr (see Sec.~\ref{sec:mean_out}).  These dynamical times are significantly shorter than the expected age of the star-formation burst  in ULIRGs \citep[$\sim60-100$~Myr, ][]{RodriguezZaurin2010}, the depletion times of star-formation ($10-100$~Myr for our sample)  or  the outflow depletion times  \citep[$\sim10-700$~Myr, ][]{Cicone2015, Pereira-Santaella2018}. 
These short dynamical times of the molecular outflows may be due  to the survival time of the molecular gas in the hot outflow environment \citep[e.g.][]{ Decataldo2017}.
An alternative explanation may be related to the geometry of the outflow.  If the outflow has a bi-conical geometry,  the area of the outflow increases with the squared of the radial distance from the nucleus ($r^2$). Thus, the column density of the outflow  decreases with radius,  making the emission fainter and more difficult to detect at large radii \citep{Pereira-Santaella2018}.

In our sample, we find that between 4-100\% of the outflow gas could escape from the gravitational potential of the galaxies into the circumgalactic medium (see Sec.~\ref{sec:escape}).  However,  this represents only 1-5\% of the total molecular gas reservoir of the galaxies.  Thus, it is unlikely that these outflows are able to impact star-formation by removing gas from the molecular gas reservoir.

\section{Summary and conclusions}
\label{sec:conclusions}
In this work, we present ALMA CO(2-1) observations of 25 nearby  ($z < 0.165)$ ULIRG systems (\Nnuclei\ individual nuclei) with a resolution of $\sim 400$~pc.  We have used these observations to study their molecular outflow  properties.
The main results of this work are as follows:
\begin{enumerate}

\item  We detect molecular outflows in 20/26 (77\%) nuclei with $\log L_{IR}/L_{\odot} > 11.8$.  The molecular outflows have an average outflow velocity $485\pm16$~\kms, outflow masses $1-35 \times 10^7$  \Msun, mass outflow rates \Mrate$=6-300$~\Msun~yr$^{-1}$,   and mass-loading factors $\eta = \dot{M}_{out}/SFR = 0.1-1$.
 The majority of the outflows (18/20) are spatially resolved with radii of $0.2-0.9$~kpc and have short dynamical times ($t_{dyn}=R_{out}/v_{out}$)  in the range $0.5-2.8$~Myr.  
We estimate the average escape velocity for our sample $v_{esc} = 486\pm40$~\kms. We find that, on average, $45\pm6\%$ of the high-velocity gas will escape to the circumgalactic medium,  
 which represents less than 5\% of the total molecular gas mass of the systems (see Sec.~\ref{sec:outflow_Q}.)\\

\item We find that the outflow detection rate is higher in SBs ($93\pm4$\%,  14/15) than in AGN ($55\pm14\%$, 6/11) (see Sec.~\ref{sec:outflow_det}, Fig.~\ref{fig:out_stats}). 
A possible interpretation is that SB outflows tend to be perpendicular to the disk, and thus they are easier to detect, while AGN outflow can have any inclination and/or can be more collimated.
Indeed,  we find that in 43\% (6/14) of the SB nuclei with outflow detection, the projected position angle  of the outflow is along the kinematic minor axis, which suggests that these outflows are perpendicular to the disk.
This fraction is higher in early interacting SBs (50\%) than in advanced merger SBs (17\%). Outflows powered by AGN do not have a preferred orientation with respect to the disk.
We do not find any significant difference in the mean outflow properties depending on the nuclear engine (AGN versus SBs) or on the merger stage (advanced mergers versus interacting systems) (see Fig.~\ref{fig:out_stats_mean}).\\

\item We find that our sample does not follow a tight  \Mrate\ versus AGN luminosity relation,  as  reported in previous works (see Fig.~\ref{fig:LAGN_vs_Mrate}).  For the same AGN luminosity,  we find \Mrate\ spanning up to 1.2~dex. 
This suggests that other factors may contribute in determining \Mrate\ in AGN ULIRGs,  such as the efficiency of the energy or momentum transfer between the AGN and the molecular gas, the outflow geometry,  or the distribution of the gas around the AGN.\\

\item We investigate whether the molecular outflows are momentum-driven or energy-driven, using also the PHANGS-ALMA sample \citep{Stuber2021} to extend the analysis to galaxies with lower SFR.  Based on the slope ($\alpha=-1.43\pm028$) of the mass-loading factor versus outflow-velocity relation (Fig.~\ref{fig:Mload_vs_logv}), we cannot distinguish between the momentum-driven ($\alpha=-1$) and energy-driven ($\alpha=-1$) scenarios. However,  using the relation between the SFR and the outflow velocity ($\log v_{out} \propto 0.25\cdot \log SFR $), we could derive the value of $\alpha$ by fitting the relation between \Mout/\Rout\ and \vout. From this analysis, we derived a slope $\alpha=-0.39\pm0.25$,  which is more consistent with the momentum-driven than with the energy-driven scenario (see Sec~\ref{sec:out_mechanism}, Fig.~\ref{fig:M_over_R_vs_logv}).\\

\item We compare the outflow velocities derived from CO  ($v_{out}$) with the ones derived from OH outflow ($v_{98}$) (see Sec.~\ref{sec:OH_comp}, Fig.~\ref{fig:comp_OH_CO}).  We find that for 13 targets, both the CO and OH velocities are above 300~\kms, which is a sign of outflow, while in three targets the OH velocities are considerably lower than $v_{out}$,  and therefore there is no evidence of outflows. This could be explained if the outflow is highly collimated,  and thus the outflowing OH covers a small region of the background continuum.  In one case (00188-0856), the OH outflow velocity is  very high, but no outflow is detected in CO.  This situation could be explained by an extreme environment, where the high ionisation would decrease the CO abundance more than the OH abundance,  by the outflow geometry or by  the different sensitivities of the OH and CO observations.\\

\end{enumerate}

\begin{acknowledgements}


 We thank the referee for the useful comments and suggestions which helped improve this manuscript.
 We thank Cristina Ramos Almeida for helpful discussion.
 We thank the staff of the European ALMA ARC for the support with the self-calibration of the data.
 This research has been funded by MDM-2017-0737 Unidad de Excelencia "María de Maeztu"-Centro de Astrobiología (INTA-CSIC) by the Spanish Ministry of Science and Innovation/State Agency of Research MCIN/AEI/ 10.13039/501100011033 and by "ERDF A way of making Europe".
IL and MPS  acknowledge support from the Comunidad de Madrid through the Atracci\'on de Talento  Investigador Grant 2018-T1/TIC-11035 and PID2019-105423GA-I00 funded by MCIN/AEI/10.13039/501100011033. 
 MP is supported by the Programa Atracci\'on de Talento de la Comunidad de Madrid via grant 2018-T2/TIC-11715. 
 MP and SA acknowledge support from the Spanish Ministerio de Economía y Competitividad through the grant ESP2017-83197-Py
  and PID2019-106280GB-I00.  
 EG-A is a Research Associate at the Harvard-Smithsonian Center for Astrophysics, and thanks the Spanish Ministerio de Econom\'ia y Competitividad for support under projects ESP2017-86582-C4-1-R and PID2019-105552RB-C41.
AAH and SGB acknowledge grant PGC2018-094671-B-I00 funded by MCIN/AEI/ 10.13039/501100011033 and by ERDF A way of making Europe.
SGB acknowledges support from the research project PID2019-106027GA-C44 of the Spanish Ministerio de Ciencia e Innovaci\'on.
SAa gratefully acknowledges support from an ERC Advanced
Grant 789410, from the Swedish Research Council and from the Knut and Alice Wallenberg Foundation (KAW).
 DR acknowledges support from STFC through grant ST/S000488/1.

This paper makes use of the following ALMA data: 
ADS/JAO.ALMA$\#$2015.1.00113.S, 
ADS/JAO.ALMA$\#$2015.1.00263.S,   
ADS/JAO.ALMA$\#$2016.1.00170.S,  
ADS/JAO.ALMA$\#$2016.1.00777.S, 
ADS/JAO.ALMA$\#$2018.1.00486.S, 
and ADS/JAO.ALMA$\#$2018.1.00699.S.  
ALMA is a partnership of ESO (representing its member states), NSF (USA) and NINS (Japan), together with NRC (Canada), MOST and ASIAA (Taiwan), and KASI (Republic of Korea), in cooperation with the Republic of Chile. The Joint ALMA Observatory is operated by ESO, AUI/NRAO and NAOJ.


This research has made use of the NASA/IPAC Extragalactic Database (NED) which is operated by the Jet Propulsion Laboratory, California Institute of Technology, under contract with the National Aeronautics and Space Administration.
This research made use of Astropy,  a community-developed core Python package for Astronomy \citep{astropy},  {\tt Matplotlib} \citep{Hunter2007}, {\tt NumPy} \citep{VanDerWalt2011}. 
This research used the {\tt TOPCAT} tool for catalogue cross-matching \citep{Taylor2005} and the {\tt Stan} interface for Python {\tt PyStan} \citep{pystan}. 

\end{acknowledgements}

%
%

\bibliographystyle{aa} 
\bibliography{Biblio/PUMA_paper_biblio_full.bib}

\begin{thebibliography}{107}
\expandafter\ifx\csname natexlab\endcsname\relax\def\natexlab#1{#1}\fi

\bibitem[{{Aalto} {et~al.}(2015){Aalto}, {Garcia-Burillo}, {Muller}, {Winters},
  {Gonzalez-Alfonso}, {van der Werf}, {Henkel}, {Costagliola}, \&
  {Neri}}]{Aalto2015}
{Aalto}, S., {Garcia-Burillo}, S., {Muller}, S., {et~al.} 2015, \aap, 574, A85

\bibitem[{{Aalto} {et~al.}(2012){Aalto}, {Garcia-Burillo}, {Muller}, {Winters},
  {van der Werf}, {Henkel}, {Costagliola}, \& {Neri}}]{Aalto2012}
{Aalto}, S., {Garcia-Burillo}, S., {Muller}, S., {et~al.} 2012, \aap, 537, A44

\bibitem[{{Alonso-Herrero} {et~al.}(2016){Alonso-Herrero}, {Esquej}, {Roche},
  {Ramos Almeida}, {Gonz{\'a}lez-Mart{\'\i}n}, {Packham}, {Levenson}, {Mason},
  {Hern{\'a}n-Caballero}, {Pereira-Santaella}, {Alvarez}, {Aretxaga},
  {L{\'o}pez-Rodr{\'\i}guez}, {Colina}, {D{\'\i}az-Santos}, {Imanishi},
  {Rodr{\'\i}guez Espinosa}, \& {Perlman}}]{Alonso-Herrero2016}
{Alonso-Herrero}, A., {Esquej}, P., {Roche}, P.~F., {et~al.} 2016, \mnras, 455,
  563

\bibitem[{Alonso-Herrero {et~al.}(2019)Alonso-Herrero, Garc{\'\i}a-Burillo,
  Pereira-Santaella, Davies, Combes, Vestergaard, Raimundo, Bunker,
  D{\'\i}az-Santos, Gandhi, Garc{\'\i}a-Bernete, Hicks, H{\"o}nig, Hunt,
  Imanishi, Izumi, Levenson, Maciejewski, Packham, Ramos~Almeida, Ricci,
  Rigopoulou, Roche, Rosario, Schartmann, Usero, \& Ward}]{Alonso-Herrero2019}
Alonso-Herrero, A., Garc{\'\i}a-Burillo, S., Pereira-Santaella, M., {et~al.}
  2019, A{\&}A, 628, A65

\bibitem[{{Arribas} {et~al.}(2014){Arribas}, {Colina}, {Bellocchi}, {Maiolino},
  \& {Villar-Mart{\'\i}n}}]{Arribas2014}
{Arribas}, S., {Colina}, L., {Bellocchi}, E., {Maiolino}, R., \&
  {Villar-Mart{\'\i}n}, M. 2014, \aap, 568, A14

\bibitem[{{Bae} \& {Woo}(2016)}]{Bae2016}
{Bae}, H.-J. \& {Woo}, J.-H. 2016, \apj, 828, 97

\bibitem[{{Barcos-Mu{\~n}oz} {et~al.}(2018){Barcos-Mu{\~n}oz}, {Aalto},
  {Thompson}, {Sakamoto}, {Mart{\'\i}n}, {Leroy}, {Privon}, {Evans}, \&
  {Kepley}}]{Barcos-Munoz2018}
{Barcos-Mu{\~n}oz}, L., {Aalto}, S., {Thompson}, T.~A., {et~al.} 2018, \apjl,
  853, L28

\bibitem[{{Bellocchi} {et~al.}(2013){Bellocchi}, {Arribas}, {Colina}, \&
  {Miralles-Caballero}}]{Bellocchi2013}
{Bellocchi}, E., {Arribas}, S., {Colina}, L., \& {Miralles-Caballero}, D. 2013,
  \aap, 557, A59

\bibitem[{{Bolatto} {et~al.}(2013{\natexlab{a}}){Bolatto}, {Warren}, {Leroy},
  {Walter}, {Veilleux}, {Ostriker}, {Ott}, {Zwaan}, {Fisher}, {Weiss},
  {Rosolowsky}, \& {Hodge}}]{Bolatto2013a}
{Bolatto}, A.~D., {Warren}, S.~R., {Leroy}, A.~K., {et~al.} 2013{\natexlab{a}},
  \nat, 499, 450

\bibitem[{{Bolatto} {et~al.}(2013{\natexlab{b}}){Bolatto}, {Wolfire}, \&
  {Leroy}}]{Bolatto2013}
{Bolatto}, A.~D., {Wolfire}, M., \& {Leroy}, A.~K. 2013{\natexlab{b}}, \araa,
  51, 207

\bibitem[{Brusa {et~al.}(2018)Brusa, Cresci, Daddi, Paladino, Perna, Bongiorno,
  Lusso, Sargent, Casasola, Feruglio, Fraternali, Georgiev, Mainieri, Carniani,
  Comastri, Duras, Fiore, Mannucci, Marconi, Piconcelli, Zamorani, Gilli,
  La~Franca, Lanzuisi, Lutz, Santini, Scoville, Vignali, Vito, Rabien, Busoni,
  \& Bonaglia}]{Brusa2018}
Brusa, M., Cresci, G., Daddi, E., {et~al.} 2018, A{\&}A, 612, A29

\bibitem[{Carniani {et~al.}(2015)Carniani, Marconi, Maiolino, Balmaverde,
  Brusa, Cano-D{\'\i}az, Cicone, Comastri, Cresci, Fiore, Feruglio, La~Franca,
  Mainieri, Mannucci, Nagao, Netzer, Piconcelli, Risaliti, Schneider, \&
  Shemmer}]{Carniani2015}
Carniani, S., Marconi, A., Maiolino, R., {et~al.} 2015, A{\&}A, 580, A102

\bibitem[{{Cazzoli} {et~al.}(2016){Cazzoli}, {Arribas}, {Maiolino}, \&
  {Colina}}]{Cazzoli2016}
{Cazzoli}, S., {Arribas}, S., {Maiolino}, R., \& {Colina}, L. 2016, \aap, 590,
  A125

\bibitem[{{Chung} {et~al.}(2011){Chung}, {Yun}, {Naraynan}, {Heyer}, \&
  {Erickson}}]{Chung2011}
{Chung}, A., {Yun}, M.~S., {Naraynan}, G., {Heyer}, M., \& {Erickson}, N.~R.
  2011, \apjl, 732, L15

\bibitem[{Cicone {et~al.}(2015)Cicone, Maiolino, Gallerani, Neri, Ferrara,
  Sturm, Fiore, Piconcelli, \& Feruglio}]{Cicone2015}
Cicone, C., Maiolino, R., Gallerani, S., {et~al.} 2015, A{\&}A, 574, A14

\bibitem[{Cicone {et~al.}(2014)Cicone, Maiolino, Sturm, Gracia-Carpio,
  Feruglio, Neri, Aalto, Davies, Fiore, Fischer, Garc{\'\i}a-Burillo,
  Gonz{\'a}lez-Alfonso, Hailey-Dunsheath, Piconcelli, \& Veilleux}]{Cicone2014}
Cicone, C., Maiolino, R., Sturm, E., {et~al.} 2014, A{\&}A, 562, A21

\bibitem[{Condon(1997)}]{Condon1997}
Condon, J.~J. 1997, Publications of the Astronomical Society of the Pacific,
  109, 166

\bibitem[{{Dasyra} \& {Combes}(2011)}]{Dasyra2011}
{Dasyra}, K.~M. \& {Combes}, F. 2011, \aap, 533, L10

\bibitem[{{Dasyra} \& {Combes}(2012)}]{Dasyra2012}
{Dasyra}, K.~M. \& {Combes}, F. 2012, \aap, 541, L7

\bibitem[{{Dasyra} {et~al.}(2014){Dasyra}, {Combes}, {Novak}, {Bremer},
  {Spinoglio}, {Pereira Santaella}, {Salom{\'e}}, \& {Falgarone}}]{Dasyra2014}
{Dasyra}, K.~M., {Combes}, F., {Novak}, G.~S., {et~al.} 2014, \aap, 565, A46

\bibitem[{{Decataldo} {et~al.}(2017){Decataldo}, {Ferrara}, {Pallottini},
  {Gallerani}, \& {Vallini}}]{Decataldo2017}
{Decataldo}, D., {Ferrara}, A., {Pallottini}, A., {Gallerani}, S., \&
  {Vallini}, L. 2017, \mnras, 471, 4476

\bibitem[{{den Brok} {et~al.}(2021){den Brok}, {Chatzigiannakis}, {Bigiel},
  {Puschnig}, {Barnes}, {Leroy}, {Jim{\'e}nez-Donaire}, {Usero}, {Schinnerer},
  {Rosolowsky}, {Faesi}, {Grasha}, {Hughes}, {Kruijssen}, {Liu}, {Neumann},
  {Pety}, {Querejeta}, {Saito}, {Schruba}, \& {Stuber}}]{denBrok2021}
{den Brok}, J.~S., {Chatzigiannakis}, D., {Bigiel}, F., {et~al.} 2021, \mnras,
  504, 3221

\bibitem[{{Dom{\'\i}nguez-Fern{\'a}ndez}
  {et~al.}(2020){Dom{\'\i}nguez-Fern{\'a}ndez}, {Alonso-Herrero},
  {Garc{\'\i}a-Burillo}, {Davies}, {Usero}, {Labiano}, {Levenson},
  {Pereira-Santaella}, {Imanishi}, {Ramos Almeida}, \&
  {Rigopoulou}}]{Dominguez-Fernandez2020}
{Dom{\'\i}nguez-Fern{\'a}ndez}, A.~J., {Alonso-Herrero}, A.,
  {Garc{\'\i}a-Burillo}, S., {et~al.} 2020, \aap, 643, A127

\bibitem[{{Emonts} {et~al.}(2017){Emonts}, {Colina}, {Piqueras-L{\'o}pez},
  {Garcia-Burillo}, {Pereira-Santaella}, {Arribas}, {Labiano}, \&
  {Alonso-Herrero}}]{Emonts2017}
{Emonts}, B.~H.~C., {Colina}, L., {Piqueras-L{\'o}pez}, J., {et~al.} 2017,
  \aap, 607, A116

\bibitem[{Fabian(2012)}]{Fabian2012}
Fabian, A.~C. 2012, Annual Review of Astronomy and Astrophysics, 50, 455

\bibitem[{{Falstad} {et~al.}(2019){Falstad}, {Hallqvist}, {Aalto}, {K{\"o}nig},
  {Muller}, {Aladro}, {Combes}, {Evans}, {Fuller}, {Gallagher},
  {Garc{\'\i}a-Burillo}, {Gonz{\'a}lez-Alfonso}, {Greve}, {Henkel}, {Imanishi},
  {Izumi}, {Mangum}, {Mart{\'\i}n}, {Privon}, {Sakamoto}, {Veilleux}, \& {van
  der Werf}}]{Falstad2019}
{Falstad}, N., {Hallqvist}, F., {Aalto}, S., {et~al.} 2019, \aap, 623, A29

\bibitem[{{Farrah} {et~al.}(2003){Farrah}, {Afonso}, {Efstathiou},
  {Rowan-Robinson}, {Fox}, \& {Clements}}]{Farrah2003}
{Farrah}, D., {Afonso}, J., {Efstathiou}, A., {et~al.} 2003, \mnras, 343, 585

\bibitem[{Feigelson \& Nelson(1985)}]{Feigelson1985}
Feigelson, E.~D. \& Nelson, P.~I. 1985, Astrophysical Journal, 293, 192

\bibitem[{Feruglio {et~al.}(2015)Feruglio, Fiore, Carniani, Piconcelli,
  Zappacosta, Bongiorno, Cicone, Maiolino, Marconi, Menci, Puccetti, \&
  Veilleux}]{Feruglio2015}
Feruglio, C., Fiore, F., Carniani, S., {et~al.} 2015, A{\&}A, 583, A99

\bibitem[{Feruglio {et~al.}(2010)Feruglio, Maiolino, Piconcelli, Menci, Aussel,
  Lamastra, \& Fiore}]{Feruglio2010}
Feruglio, C., Maiolino, R., Piconcelli, E., {et~al.} 2010, A{\&}A, 518, L155

\bibitem[{Fiore {et~al.}(2017)Fiore, Feruglio, Shankar, Bischetti, Bongiorno,
  Brusa, Carniani, Cicone, Duras, Lamastra, Mainieri, Marconi, Menci, Maiolino,
  Piconcelli, Vietri, \& Zappacosta}]{Fiore2017}
Fiore, F., Feruglio, C., Shankar, F., {et~al.} 2017, A{\&}A, 601, A143

\bibitem[{{Fischer} {et~al.}(2010){Fischer}, {Sturm}, {Gonz{\'a}lez-Alfonso},
  {Graci{\'a}-Carpio}, {Hailey-Dunsheath}, {Poglitsch}, {Contursi}, {Lutz},
  {Genzel}, {Sternberg}, {Verma}, \& {Tacconi}}]{Fischer2010}
{Fischer}, J., {Sturm}, E., {Gonz{\'a}lez-Alfonso}, E., {et~al.} 2010, \aap,
  518, L41

\bibitem[{{Fluetsch} {et~al.}(2021){Fluetsch}, {Maiolino}, {Carniani},
  {Arribas}, {Belfiore}, {Bellocchi}, {Cazzoli}, {Cicone}, {Cresci}, {Fabian},
  {Gallagher}, {Ishibashi}, {Mannucci}, {Marconi}, {Perna}, {Sturm}, \&
  {Venturi}}]{Fluetsch2021}
{Fluetsch}, A., {Maiolino}, R., {Carniani}, S., {et~al.} 2021, \mnras, 505,
  5753

\bibitem[{Fluetsch {et~al.}(2019)Fluetsch, Maiolino, Carniani, Marconi, Cicone,
  Bourne, Costa, Fabian, Ishibashi, \& Venturi}]{Fluetsch2019}
Fluetsch, A., Maiolino, R., Carniani, S., {et~al.} 2019, Monthly Notices of the
  Royal Astronomical Society, 483, 4586

\bibitem[{{F{\"o}rster Schreiber} {et~al.}(2014){F{\"o}rster Schreiber},
  {Genzel}, {Newman}, {Kurk}, {Lutz}, {Tacconi}, {Wuyts}, {Bandara}, {Burkert},
  {Buschkamp}, {Carollo}, {Cresci}, {Daddi}, {Davies}, {Eisenhauer}, {Hicks},
  {Lang}, {Lilly}, {Mainieri}, {Mancini}, {Naab}, {Peng}, {Renzini}, {Rosario},
  {Shapiro Griffin}, {Shapley}, {Sternberg}, {Tacchella}, {Vergani},
  {Wisnioski}, {Wuyts}, \& {Zamorani}}]{ForsterSchreiber2014}
{F{\"o}rster Schreiber}, N.~M., {Genzel}, R., {Newman}, S.~F., {et~al.} 2014,
  \apj, 787, 38

\bibitem[{{Gao} {et~al.}(2021){Gao}, {Egusa}, {Liu}, {Kohno}, {Bao},
  {Morokuma-Matsui}, {Kong}, \& {Chen}}]{Gao2021}
{Gao}, Y., {Egusa}, F., {Liu}, G., {et~al.} 2021, \apj, 913, 139

\bibitem[{{Garc{\'\i}a-Bernete} {et~al.}(2021){Garc{\'\i}a-Bernete},
  {Alonso-Herrero}, {Garc{\'\i}a-Burillo}, {Pereira-Santaella},
  {Garc{\'\i}a-Lorenzo}, {Carrera}, {Rigopoulou}, {Ramos Almeida}, {Villar
  Mart{\'\i}n}, {Gonz{\'a}lez-Mart{\'\i}n}, {Hicks}, {Labiano}, {Ricci}, \&
  {Mateos}}]{Garcia-Bernete2021}
{Garc{\'\i}a-Bernete}, I., {Alonso-Herrero}, A., {Garc{\'\i}a-Burillo}, S.,
  {et~al.} 2021, \aap, 645, A21

\bibitem[{{Garc{\'\i}a-Bernete} {et~al.}(2022){Garc{\'\i}a-Bernete},
  {Rigopoulou}, {Aalto}, {Spoon}, {Hern{\'a}n-Caballero}, {Efstathiou},
  {Roche}, \& {K{\"o}nig}}]{Garcia-Bernete2022}
{Garc{\'\i}a-Bernete}, I., {Rigopoulou}, D., {Aalto}, S., {et~al.} 2022, arXiv
  e-prints, arXiv:2203.14795

\bibitem[{{Garc{\'\i}a-Burillo} {et~al.}(2015){Garc{\'\i}a-Burillo}, {Combes},
  {Usero}, {Aalto}, {Colina}, {Alonso-Herrero}, {Hunt}, {Arribas},
  {Costagliola}, {Labiano}, {Neri}, {Pereira-Santaella}, {Tacconi}, \& {van der
  Werf}}]{Garcia-Burillo2015}
{Garc{\'\i}a-Burillo}, S., {Combes}, F., {Usero}, A., {et~al.} 2015, \aap, 580,
  A35

\bibitem[{Gelman {et~al.}(2004)Gelman, Carlin, Ster, Dunson, Vehtari, \&
  Rubin}]{Gelman2004}
Gelman, A., Carlin, J.~B., Ster, H.~S., {et~al.} 2004, Chapman and Hall/CRC

\bibitem[{{Gilli} {et~al.}(2000){Gilli}, {Maiolino}, {Marconi}, {Risaliti},
  {Dadina}, {Weaver}, \& {Colbert}}]{Gilli2000}
{Gilli}, R., {Maiolino}, R., {Marconi}, A., {et~al.} 2000, \aap, 355, 485

\bibitem[{{Gonz{\'a}lez-Alfonso} {et~al.}(2018){Gonz{\'a}lez-Alfonso},
  {Fischer}, {Bruderer}, {Ashby}, {Smith}, {Veilleux}, {M{\"u}ller}, {Stewart},
  \& {Sturm}}]{Gonzalez-Alfonso2018}
{Gonz{\'a}lez-Alfonso}, E., {Fischer}, J., {Bruderer}, S., {et~al.} 2018, \apj,
  857, 66

\bibitem[{{Gonz{\'a}lez-Alfonso} {et~al.}(2013){Gonz{\'a}lez-Alfonso},
  {Fischer}, {Bruderer}, {M{\"u}ller}, {Graci{\'a}-Carpio}, {Sturm}, {Lutz},
  {Poglitsch}, {Feuchtgruber}, {Veilleux}, {Contursi}, {Sternberg},
  {Hailey-Dunsheath}, {Verma}, {Christopher}, {Davies}, {Genzel}, \&
  {Tacconi}}]{Gonzalez-Alfonso2013}
{Gonz{\'a}lez-Alfonso}, E., {Fischer}, J., {Bruderer}, S., {et~al.} 2013, \aap,
  550, A25

\bibitem[{{Gonz{\'a}lez-Alfonso} {et~al.}(2014){Gonz{\'a}lez-Alfonso},
  {Fischer}, {Graci{\'a}-Carpio}, {Falstad}, {Sturm}, {Mel{\'e}ndez}, {Spoon},
  {Verma}, {Davies}, {Lutz}, {Aalto}, {Polisensky}, {Poglitsch}, {Veilleux}, \&
  {Contursi}}]{Gonzalez-Alfonso2014b}
{Gonz{\'a}lez-Alfonso}, E., {Fischer}, J., {Graci{\'a}-Carpio}, J., {et~al.}
  2014, \aap, 561, A27

\bibitem[{{Gonz{\'a}lez-Alfonso} {et~al.}(2012){Gonz{\'a}lez-Alfonso},
  {Fischer}, {Graci{\'a}-Carpio}, {Sturm}, {Hailey-Dunsheath}, {Lutz},
  {Poglitsch}, {Contursi}, {Feuchtgruber}, {Veilleux}, {Spoon}, {Verma},
  {Christopher}, {Davies}, {Sternberg}, {Genzel}, \&
  {Tacconi}}]{Gonzalez-Alfonso2012}
{Gonz{\'a}lez-Alfonso}, E., {Fischer}, J., {Graci{\'a}-Carpio}, J., {et~al.}
  2012, \aap, 541, A4

\bibitem[{Gonz{\'a}lez-Alfonso {et~al.}(2017)Gonz{\'a}lez-Alfonso, Fischer,
  Spoon, Stewart, Ashby, Veilleux, Smith, Sturm, Farrah, Falstad, Mel{\'e}ndez,
  Gracia-Carpio, Janssen, \& Lebouteiller}]{Gonzalez-Alfonso2017}
Gonz{\'a}lez-Alfonso, E., Fischer, J., Spoon, H. W.~W., {et~al.} 2017, The
  Astrophysical Journal, 836, 11

\bibitem[{{Gowardhan} {et~al.}(2018){Gowardhan}, {Spoon}, {Riechers},
  {Gonz{\'a}lez-Alfonso}, {Farrah}, {Fischer}, {Darling}, {Fergulio}, {Afonso},
  \& {Bizzocchi}}]{Gowardhan2018}
{Gowardhan}, A., {Spoon}, H., {Riechers}, D.~A., {et~al.} 2018, \apj, 859, 35

\bibitem[{{Heckman} {et~al.}(2000){Heckman}, {Lehnert}, {Strickland}, \&
  {Armus}}]{Heckman2000}
{Heckman}, T.~M., {Lehnert}, M.~D., {Strickland}, D.~K., \& {Armus}, L. 2000,
  \apjs, 129, 493

\bibitem[{{Hopkins} {et~al.}(2008){Hopkins}, {Hernquist}, {Cox}, \&
  {Kere{\v{s}}}}]{Hopkins2008}
{Hopkins}, P.~F., {Hernquist}, L., {Cox}, T.~J., \& {Kere{\v{s}}}, D. 2008,
  \apjs, 175, 356

\bibitem[{Hunter(2007)}]{Hunter2007}
Hunter, J.~D. 2007, Computing in Science and Engineering, 9, 90

\bibitem[{Jarvis {et~al.}(2019)Jarvis, Harrison, Thomson, Circosta, Mainieri,
  Alexander, Edge, Lansbury, Molyneux, \& Mullaney}]{Jarvis2019}
Jarvis, M.~E., Harrison, C.~M., Thomson, A.~P., {et~al.} 2019, Monthly Notices
  of the Royal Astronomical Society, 485, 2710

\bibitem[{Kennicutt \& Evans(2012)}]{Kennicutt2012}
Kennicutt, R.~C. \& Evans, N.~J. 2012, Annual Review of Astronomy and
  Astrophysics, 50, 531

\bibitem[{{Kim} \& {Ostriker}(2015)}]{Kim2015}
{Kim}, C.-G. \& {Ostriker}, E.~C. 2015, \apj, 802, 99

\bibitem[{{Kim} \& {Sanders}(1998)}]{Kim1998}
{Kim}, D.~C. \& {Sanders}, D.~B. 1998, \apjs, 119, 41

\bibitem[{{Kroupa}(2001)}]{Kroupa2001}
{Kroupa}, P. 2001, \mnras, 322, 231

\bibitem[{{Kroupa} \& {Weidner}(2003)}]{Kroupa2003}
{Kroupa}, P. \& {Weidner}, C. 2003, \apj, 598, 1076

\bibitem[{{Law} {et~al.}(2009){Law}, {Steidel}, {Erb}, {Larkin}, {Pettini},
  {Shapley}, \& {Wright}}]{Law2009}
{Law}, D.~R., {Steidel}, C.~C., {Erb}, D.~K., {et~al.} 2009, \apj, 697, 2057

\bibitem[{{Lehnert} \& {Heckman}(1996)}]{Lehnert1996b}
{Lehnert}, M.~D. \& {Heckman}, T.~M. 1996, \apj, 472, 546

\bibitem[{{Leitherer} {et~al.}(1999){Leitherer}, {Schaerer}, {Goldader},
  {Delgado}, {Robert}, {Kune}, {de Mello}, {Devost}, \&
  {Heckman}}]{Leitherer1999}
{Leitherer}, C., {Schaerer}, D., {Goldader}, J.~D., {et~al.} 1999, \apjs, 123,
  3

\bibitem[{{Leroy} {et~al.}(2021{\natexlab{a}}){Leroy}, {Schinnerer}, {Hughes},
  {Rosolowsky}, {Pety}, {Schruba}, {Usero}, {Blanc}, {Chevance}, {Emsellem},
  {Faesi}, {Herrera}, {Liu}, {Meidt}, {Querejeta}, {Saito}, {Sandstrom}, {Sun},
  {Williams}, {Anand}, {Barnes}, {Behrens}, {Belfiore}, {Benincasa},
  {Be{\v{s}}li{\'c}}, {Bigiel}, {Bolatto}, {den Brok}, {Cao}, {Chandar},
  {Chastenet}, {Chiang}, {Congiu}, {Dale}, {Deger}, {Eibensteiner}, {Egorov},
  {Garc{\'\i}a-Rodr{\'\i}guez}, {Glover}, {Grasha}, {Henshaw}, {Ho}, {Kepley},
  {Kim}, {Klessen}, {Kreckel}, {Koch}, {Kruijssen}, {Larson}, {Lee}, {Lopez},
  {Machado}, {Mayker}, {McElroy}, {Murphy}, {Ostriker}, {Pan}, {Pessa},
  {Puschnig}, {Razza}, {S{\'a}nchez-Bl{\'a}zquez}, {Santoro}, {Sardone},
  {Scheuermann}, {Sliwa}, {Sormani}, {Stuber}, {Thilker}, {Turner}, {Utomo},
  {Watkins}, \& {Whitmore}}]{Leroy2021a}
{Leroy}, A.~K., {Schinnerer}, E., {Hughes}, A., {et~al.} 2021{\natexlab{a}},
  arXiv e-prints, arXiv:2104.07739

\bibitem[{{Leroy} {et~al.}(2021{\natexlab{b}}){Leroy}, {Schinnerer}, {Hughes},
  {Rosolowsky}, {Pety}, {Schruba}, {Usero}, {Blanc}, {Chevance}, {Emsellem},
  {Faesi}, {Herrera}, {Liu}, {Meidt}, {Querejeta}, {Saito}, {Sandstrom}, {Sun},
  {Williams}, {Anand}, {Barnes}, {Behrens}, {Belfiore}, {Benincasa},
  {Be{\v{s}}li{\'c}}, {Bigiel}, {Bolatto}, {den Brok}, {Cao}, {Chandar},
  {Chastenet}, {Chiang}, {Congiu}, {Dale}, {Deger}, {Eibensteiner}, {Egorov},
  {Garc{\'\i}a-Rodr{\'\i}guez}, {Glover}, {Grasha}, {Henshaw}, {Ho}, {Kepley},
  {Kim}, {Klessen}, {Kreckel}, {Koch}, {Kruijssen}, {Larson}, {Lee}, {Lopez},
  {Machado}, {Mayker}, {McElroy}, {Murphy}, {Ostriker}, {Pan}, {Pessa},
  {Puschnig}, {Razza}, {S{\'a}nchez-Bl{\'a}zquez}, {Santoro}, {Sardone},
  {Scheuermann}, {Sliwa}, {Sormani}, {Stuber}, {Thilker}, {Turner}, {Utomo},
  {Watkins}, \& {Whitmore}}]{Leroy2021b}
{Leroy}, A.~K., {Schinnerer}, E., {Hughes}, A., {et~al.} 2021{\natexlab{b}},
  \apjs, 257, 43

\bibitem[{{Leroy} {et~al.}(2013){Leroy}, {Walter}, {Sandstrom}, {Schruba},
  {Munoz-Mateos}, {Bigiel}, {Bolatto}, {Brinks}, {de Blok}, {Meidt}, {Rix},
  {Rosolowsky}, {Schinnerer}, {Schuster}, \& {Usero}}]{Leroy2013}
{Leroy}, A.~K., {Walter}, F., {Sandstrom}, K., {et~al.} 2013, \aj, 146, 19

\bibitem[{{Lonsdale} {et~al.}(2006){Lonsdale}, {Farrah}, \&
  {Smith}}]{Lonsdale2006}
{Lonsdale}, C.~J., {Farrah}, D., \& {Smith}, H.~E. 2006, in Astrophysics Update
  2, ed. J.~W. {Mason}, 285

\bibitem[{{Lutz} {et~al.}(2016){Lutz}, {Berta}, {Contursi}, {F{\"o}rster
  Schreiber}, {Genzel}, {Graci{\'a}-Carpio}, {Herrera-Camus}, {Netzer},
  {Sturm}, {Tacconi}, {Tadaki}, \& {Veilleux}}]{Lutz2016}
{Lutz}, D., {Berta}, S., {Contursi}, A., {et~al.} 2016, \aap, 591, A136

\bibitem[{{Lutz} {et~al.}(2020){Lutz}, {Sturm}, {Janssen}, {Veilleux}, {Aalto},
  {Cicone}, {Contursi}, {Davies}, {Feruglio}, {Fischer}, {Fluetsch},
  {Garcia-Burillo}, {Genzel}, {Gonz{\'a}lez-Alfonso}, {Graci{\'a}-Carpio},
  {Herrera-Camus}, {Maiolino}, {Schruba}, {Shimizu}, {Sternberg}, {Tacconi}, \&
  {Wei{\ss}}}]{Lutz2020}
{Lutz}, D., {Sturm}, E., {Janssen}, A., {et~al.} 2020, \aap, 633, A134

\bibitem[{{Maiolino} {et~al.}(2012){Maiolino}, {Gallerani}, {Neri}, {Cicone},
  {Ferrara}, {Genzel}, {Lutz}, {Sturm}, {Tacconi}, {Walter}, {Feruglio},
  {Fiore}, \& {Piconcelli}}]{Maiolino2012}
{Maiolino}, R., {Gallerani}, S., {Neri}, R., {et~al.} 2012, \mnras, 425, L66

\bibitem[{{Meena} {et~al.}(2021){Meena}, {Crenshaw}, {Schmitt}, {Revalski},
  {Fischer}, {Polack}, {Kraemer}, \& {Dashtamirova}}]{Meena2021}
{Meena}, B., {Crenshaw}, D.~M., {Schmitt}, H.~R., {et~al.} 2021, \apj, 916, 31

\bibitem[{{Meurer} {et~al.}(1997){Meurer}, {Heckman}, {Lehnert}, {Leitherer},
  \& {Lowenthal}}]{Meurer1997}
{Meurer}, G.~R., {Heckman}, T.~M., {Lehnert}, M.~D., {Leitherer}, C., \&
  {Lowenthal}, J. 1997, \aj, 114, 54

\bibitem[{{Murray} {et~al.}(2005){Murray}, {Quataert}, \&
  {Thompson}}]{Murray2005}
{Murray}, N., {Quataert}, E., \& {Thompson}, T.~A. 2005, \apj, 618, 569

\bibitem[{{Nardini} {et~al.}(2010){Nardini}, {Risaliti}, {Watabe}, {Salvati},
  \& {Sani}}]{Nardini2010}
{Nardini}, E., {Risaliti}, G., {Watabe}, Y., {Salvati}, M., \& {Sani}, E. 2010,
  \mnras, 405, 2505

\bibitem[{Papadopoulos {et~al.}(2012)Papadopoulos, van~der Werf, Xilouris,
  Isaak, Gao, \& M{\"u}hle}]{Papadopoulos2012}
Papadopoulos, P.~P., van~der Werf, P.~P., Xilouris, E.~M., {et~al.} 2012,
  Monthly Notices of the Royal Astronomical Society, 426, 2601

\bibitem[{{Pereira-Santaella} {et~al.}(2018){Pereira-Santaella}, {Colina},
  {Garc{\'\i}a-Burillo}, {Combes}, {Emonts}, {Aalto}, {Alonso-Herrero},
  {Arribas}, {Henkel}, {Labiano}, {Muller}, {Piqueras L{\'o}pez}, {Rigopoulou},
  \& {van der Werf}}]{Pereira-Santaella2018}
{Pereira-Santaella}, M., {Colina}, L., {Garc{\'\i}a-Burillo}, S., {et~al.}
  2018, \aap, 616, A171

\bibitem[{{Pereira-Santaella} {et~al.}(2020){Pereira-Santaella}, {Colina},
  {Garc{\'\i}a-Burillo}, {Gonz{\'a}lez-Alfonso}, {Alonso-Herrero}, {Arribas},
  {Cazzoli}, {Piqueras-L{\'o}pez}, {Rigopoulou}, \&
  {Usero}}]{Pereira-Santaella2020}
{Pereira-Santaella}, M., {Colina}, L., {Garc{\'\i}a-Burillo}, S., {et~al.}
  2020, \aap, 643, A89

\bibitem[{{Pereira-Santaella} {et~al.}(2021){Pereira-Santaella}, {Colina},
  {Garc{\'\i}a-Burillo}, {Lamperti}, {Gonz{\'a}lez-Alfonso}, {Perna},
  {Arribas}, {Alonso-Herrero}, {Aalto}, {Combes}, {Labiano},
  {Piqueras-L{\'o}pez}, {Rigopoulou}, \& {van der
  Werf}}]{Pereira-Santaella2021}
{Pereira-Santaella}, M., {Colina}, L., {Garc{\'\i}a-Burillo}, S., {et~al.}
  2021, \aap, 651, A42

\bibitem[{Perna {et~al.}(2020)Perna, Arribas, Catalan-Torrecilla, Colina,
  Bellocchi, Fluetsch, Maiolino, Cazzoli, Hernan~Caballero, Pereira-Santaella,
  Piqueras~Lopez, \& Rodriguez~del Pino}]{Perna2020}
Perna, M., Arribas, S., Catalan-Torrecilla, C., {et~al.} 2020, A{\&}A, 643,
  A139

\bibitem[{{Perna} {et~al.}(2022){Perna}, {Arribas}, {Colina}, {Pereira
  Santaella}, {Lamperti}, {Di Teodoro}, {{\"U}bler}, {Costantin}, {Maiolino},
  {Cresci}, {Bellocchi}, {Catal{\'a}n-Torrecilla}, {Cazzoli}, \& {Piqueras
  L{\'o}pez}}]{Perna2022}
{Perna}, M., {Arribas}, S., {Colina}, L., {et~al.} 2022, \aap, 662, A94

\bibitem[{{Perna} {et~al.}(2021){Perna}, {Arribas}, {Pereira Santaella},
  {Colina}, {Bellocchi}, {Catal{\'a}n-Torrecilla}, {Cazzoli}, {Crespo
  G{\'o}mez}, {Maiolino}, {Piqueras L{\'o}pez}, \& {Rodr{\'\i}guez del
  Pino}}]{Perna2021}
{Perna}, M., {Arribas}, S., {Pereira Santaella}, M., {et~al.} 2021, \aap, 646,
  A101

\bibitem[{{Pjanka} {et~al.}(2017){Pjanka}, {Greene}, {Seth}, {Braatz},
  {Henkel}, {Lo}, \& {L{\"a}sker}}]{Pjanka2017}
{Pjanka}, P., {Greene}, J.~E., {Seth}, A.~C., {et~al.} 2017, \apj, 844, 165

\bibitem[{{Prochaska} {et~al.}(2011){Prochaska}, {Kasen}, \&
  {Rubin}}]{Prochaska2011}
{Prochaska}, J.~X., {Kasen}, D., \& {Rubin}, K. 2011, \apj, 734, 24

\bibitem[{{Ramos Almeida} {et~al.}(2019){Ramos Almeida}, {Acosta-Pulido},
  {Tadhunter}, {Gonz{\'a}lez-Fern{\'a}ndez}, {Cicone}, \&
  {Fern{\'a}ndez-Torreiro}}]{RamosAlmeida2019}
{Ramos Almeida}, C., {Acosta-Pulido}, J.~A., {Tadhunter}, C.~N., {et~al.} 2019,
  \mnras, 487, L18

\bibitem[{{Ramos Almeida} {et~al.}(2022){Ramos Almeida}, {Bischetti},
  {Garc{\'\i}a-Burillo}, {Alonso-Herrero}, {Audibert}, {Cicone}, {Feruglio},
  {Tadhunter}, {Pierce}, {Pereira-Santaella}, \& {Bessiere}}]{RamosAlmeida2022}
{Ramos Almeida}, C., {Bischetti}, M., {Garc{\'\i}a-Burillo}, S., {et~al.} 2022,
  \aap, 658, A155

\bibitem[{{Rodr{\'\i}guez Zaur{\'\i}n} {et~al.}(2010){Rodr{\'\i}guez
  Zaur{\'\i}n}, {Tadhunter}, \& {Gonz{\'a}lez Delgado}}]{RodriguezZaurin2010}
{Rodr{\'\i}guez Zaur{\'\i}n}, J., {Tadhunter}, C.~N., \& {Gonz{\'a}lez
  Delgado}, R.~M. 2010, \mnras, 403, 1317

\bibitem[{{Runco} {et~al.}(2020){Runco}, {Malkan}, {Fern{\'a}ndez-Ontiveros},
  {Spinoglio}, \& {Pereira-Santaella}}]{Runco2020}
{Runco}, J.~N., {Malkan}, M.~A., {Fern{\'a}ndez-Ontiveros}, J.~A., {Spinoglio},
  L., \& {Pereira-Santaella}, M. 2020, \apj, 905, 57

\bibitem[{{Rupke} {et~al.}(2005{\natexlab{a}}){Rupke}, {Veilleux}, \&
  {Sanders}}]{Rupke2005a}
{Rupke}, D.~S., {Veilleux}, S., \& {Sanders}, D.~B. 2005{\natexlab{a}}, \apj,
  632, 751

\bibitem[{{Rupke} {et~al.}(2005{\natexlab{b}}){Rupke}, {Veilleux}, \&
  {Sanders}}]{Rupke2005b}
{Rupke}, D.~S., {Veilleux}, S., \& {Sanders}, D.~B. 2005{\natexlab{b}}, \apjs,
  160, 115

\bibitem[{{Rupke} \& {Veilleux}(2013)}]{Rupke2013}
{Rupke}, D. S.~N. \& {Veilleux}, S. 2013, \apj, 768, 75

\bibitem[{{Saito} {et~al.}(2018){Saito}, {Iono}, {Ueda}, {Espada}, {Sliwa},
  {Nakanishi}, {Lu}, {Xu}, {Michiyama}, {Kaneko}, {Yamashita}, {Ando}, {Yun},
  {Motohara}, \& {Kawabe}}]{Saito2018}
{Saito}, T., {Iono}, D., {Ueda}, J., {et~al.} 2018, \mnras, 475, L52

\bibitem[{{Sakamoto} {et~al.}(2014){Sakamoto}, {Aalto}, {Combes}, {Evans}, \&
  {Peck}}]{Sakamoto2014}
{Sakamoto}, K., {Aalto}, S., {Combes}, F., {Evans}, A., \& {Peck}, A. 2014,
  \apj, 797, 90

\bibitem[{{Salak} {et~al.}(2016){Salak}, {Nakai}, {Hatakeyama}, \&
  {Miyamoto}}]{Salak2016}
{Salak}, D., {Nakai}, N., {Hatakeyama}, T., \& {Miyamoto}, Y. 2016, \apj, 823,
  68

\bibitem[{Schawinski {et~al.}(2015)Schawinski, Koss, Berney, \&
  Sartori}]{Schawinski2015}
Schawinski, K., Koss, M., Berney, S., \& Sartori, L.~F. 2015, Monthly Notices
  of the Royal Astronomical Society, 451, 2517

\bibitem[{Solomon {et~al.}(1997)Solomon, Downes, Radford, \&
  Barrett}]{Solomon1997}
Solomon, P.~M., Downes, D., Radford, S. J.~E., \& Barrett, J.~W. 1997, The
  Astrophysical Journal, 478, 144

\bibitem[{{Spoon} {et~al.}(2013){Spoon}, {Farrah}, {Lebouteiller},
  {Gonz{\'a}lez-Alfonso}, {Bernard-Salas}, {Urrutia}, {Rigopoulou},
  {Westmoquette}, {Smith}, {Afonso}, {Pearson}, {Cormier}, {Efstathiou},
  {Borys}, {Verma}, {Etxaluze}, \& {Clements}}]{Spoon2013}
{Spoon}, H.~W.~W., {Farrah}, D., {Lebouteiller}, V., {et~al.} 2013, \apj, 775,
  127

\bibitem[{{Stan Development Team}(2018)}]{pystan}
{Stan Development Team}. 2018, PyStan: the Python interface to Stan, Version
  2.17.1.0

\bibitem[{{Stone} {et~al.}(2016){Stone}, {Veilleux}, {Mel{\'e}ndez}, {Sturm},
  {Graci{\'a}-Carpio}, \& {Gonz{\'a}lez-Alfonso}}]{Stone2016}
{Stone}, M., {Veilleux}, S., {Mel{\'e}ndez}, M., {et~al.} 2016, \apj, 826, 111

\bibitem[{{Stuber} {et~al.}(2021){Stuber}, {Saito}, {Schinnerer}, {Emsellem},
  {Querejeta}, {Williams}, {Barnes}, {Bigiel}, {Blanc}, {Dale}, {Grasha},
  {Klessen}, {Kruijssen}, {Leroy}, {Meidt}, {Pan}, {Rosolowsky}, {Schruba},
  {Sun}, \& {Usero}}]{Stuber2021}
{Stuber}, S.~K., {Saito}, T., {Schinnerer}, E., {et~al.} 2021, \aap, 653, A172

\bibitem[{Sturm {et~al.}(2011)Sturm, Gonz{\'a}lez-Alfonso, Veilleux, Fischer,
  Gracia-Carpio, Hailey-Dunsheath, Contursi, Poglitsch, Sternberg, Davies,
  Genzel, Lutz, Tacconi, Verma, Maiolino, \& de~Jong}]{Sturm2011}
Sturm, E., Gonz{\'a}lez-Alfonso, E., Veilleux, S., {et~al.} 2011, The
  Astrophysical Journal Letters, 733, L16

\bibitem[{Taylor(2005)}]{Taylor2005}
Taylor, M.~B. 2005, Astronomical Data Analysis Software and Systems XIV ASP
  Conference Series, 347, 29

\bibitem[{{The Astropy Collaboration} {et~al.}(2013){The Astropy
  Collaboration}, {Robitaille, Thomas P.}, {Tollerud, Erik J.}, {Greenfield,
  Perry}, {Droettboom, Michael}, {Bray, Erik}, {Aldcroft, Tom}, {Davis, Matt},
  {Ginsburg, Adam}, {Price-Whelan, Adrian M.}, {Kerzendorf, Wolfgang E.},
  {Conley, Alexander}, {Crighton, Neil}, {Barbary, Kyle}, {Muna, Demitri},
  {Ferguson, Henry}, {Grollier, Fr\'ed\'eric}, {Parikh, Madhura M.}, {Nair,
  Prasanth H.}, {G\"unther, Hans M.}, {Deil, Christoph}, {Woillez, Julien},
  {Conseil, Simon}, {Kramer, Roban}, {Turner, James E. H.}, {Singer, Leo},
  {Fox, Ryan}, {Weaver, Benjamin A.}, {Zabalza, Victor}, {Edwards, Zachary I.},
  {Azalee Bostroem, K.}, {Burke, D. J.}, {Casey, Andrew R.}, {Crawford, Steven
  M.}, {Dencheva, Nadia}, {Ely, Justin}, {Jenness, Tim}, {Labrie, Kathleen},
  {Lim, Pey Lian}, {Pierfederici, Francesco}, {Pontzen, Andrew}, {Ptak, Andy},
  {Refsdal, Brian}, {Servillat, Mathieu}, \& {Streicher, Ole}}]{astropy}
{The Astropy Collaboration}, {Robitaille, Thomas P.}, {Tollerud, Erik J.},
  {et~al.} 2013, A\&A, 558, A33

\bibitem[{Van Der~Walt {et~al.}(2011)Van Der~Walt, Colbert, \&
  Varoquaux}]{VanDerWalt2011}
Van Der~Walt, S., Colbert, S.~C., \& Varoquaux, G. 2011, Computing in Science
  and Engineering, 13, 22

\bibitem[{{Veilleux} {et~al.}(2005){Veilleux}, {Cecil}, \&
  {Bland-Hawthorn}}]{Veilleux2005}
{Veilleux}, S., {Cecil}, G., \& {Bland-Hawthorn}, J. 2005, \araa, 43, 769

\bibitem[{{Veilleux} {et~al.}(2020){Veilleux}, {Maiolino}, {Bolatto}, \&
  {Aalto}}]{Veilleux2020}
{Veilleux}, S., {Maiolino}, R., {Bolatto}, A.~D., \& {Aalto}, S. 2020, \aapr,
  28, 2

\bibitem[{Veilleux {et~al.}(2013)Veilleux, Mel{\'e}ndez, Sturm, Gracia-Carpio,
  Fischer, Gonz{\'a}lez-Alfonso, Contursi, Lutz, Poglitsch, Davies, Genzel,
  Tacconi, de~Jong, Sternberg, Netzer, Hailey-Dunsheath, Verma, Rupke,
  Maiolino, Teng, \& Polisensky}]{Veilleux2013}
Veilleux, S., Mel{\'e}ndez, M., Sturm, E., {et~al.} 2013, The Astrophysical
  Journal, 776, 27

\bibitem[{{Veilleux} {et~al.}(2009){Veilleux}, {Rupke}, {Kim}, {Genzel},
  {Sturm}, {Lutz}, {Contursi}, {Schweitzer}, {Tacconi}, {Netzer}, {Sternberg},
  {Mihos}, {Baker}, {Mazzarella}, {Lord}, {Sanders}, {Stockton}, {Joseph}, \&
  {Barnes}}]{Veilleux2009}
{Veilleux}, S., {Rupke}, D.~S.~N., {Kim}, D.~C., {et~al.} 2009, \apjs, 182, 628

\bibitem[{Venturi {et~al.}(2018)Venturi, Nardini, Marconi, Carniani, Mingozzi,
  Cresci, Mannucci, Risaliti, Maiolino, Balmaverde, Bongiorno, Brusa, Capetti,
  Cicone, Ciroi, Feruglio, Fiore, Gallazzi, La~Franca, Mainieri, Matsuoka,
  Nagao, Perna, Piconcelli, Sani, Tozzi, \& Zibetti}]{Venturi2018}
Venturi, G., Nardini, E., Marconi, A., {et~al.} 2018, A{\&}A, 619, A74

\bibitem[{{V{\'e}ron-Cetty} \& {V{\'e}ron}(2010)}]{Veron-Cetty2010}
{V{\'e}ron-Cetty}, M.~P. \& {V{\'e}ron}, P. 2010, \aap, 518, A10

\bibitem[{{Walter} {et~al.}(2017){Walter}, {Bolatto}, {Leroy}, {Veilleux},
  {Warren}, {Hodge}, {Levy}, {Meier}, {Ostriker}, {Ott}, {Rosolowsky},
  {Scoville}, {Weiss}, {Zschaechner}, \& {Zwaan}}]{Walter2017}
{Walter}, F., {Bolatto}, A.~D., {Leroy}, A.~K., {et~al.} 2017, \apj, 835, 265

\bibitem[{{Westmoquette} {et~al.}(2012){Westmoquette}, {Clements}, {Bendo}, \&
  {Khan}}]{Westmoquette2012}
{Westmoquette}, M.~S., {Clements}, D.~L., {Bendo}, G.~J., \& {Khan}, S.~A.
  2012, \mnras, 424, 416

\end{thebibliography}


\begin{appendix} 

\section{Simulations of biconical outflows}
\label{sec:app_sim}
In this section we provide some additional information about the simulations used to estimate the biases in the outflow properties measured with our method (see Sec.~\ref{sec:caveats}).
We simulate a biconcal  outflow  with a half opening angle of 40$^{\circ}$.
The maximum velocity of the outflow is 750~\kms\ and the maximum outflow extension is $R_{out}=2$~kpc.  The outflow  flux decreases with increasing radius (as it is expected in a scenario with constant mass outflow rate).
We consider four  distributions of the outflow velocity as a function of radius: 1) constant velocity,  2)  velocity linearly decreasing with radius,  3) linear velocity law with an increasing velocity going from zero to $v_{max}$ at a given turnover radius $R_{to}=0.5\times R_{out}$, followed by a linearly decreasing trend with $v(R_{out})=0$~\kms,
 4) a radial velocity profile with turnover radius as in the previous case, but accounting for an initial velocity $v(r=0)=400$~\kms.
We consider four values for the inclination of the outflow with respect to our line of sight: 10$^{\circ}$, 40$^{\circ}$, 70$^{\circ}$, and 90$^{\circ}$  (where an inclination $i=90^{\circ}$ corresponds to the plane of the sky). 
 The systemic component has a FWHM of 360~\kms and a flux $\sim10$ times the outflow flux.
We measure \vout, \Mout\ and \Mrate\ from the total simulated profile (outflow+systemic component) considering only $|v| > 300$~\kms, to mimic the method we are using with our data.  Then, we measure the outflow quantities (\vout,\Mout\ and \Mrate) from the simulated outflow emission profile (without the systemic component) considering the full velocity range.  

\begin{figure*}[h!]
\centering
\includegraphics[width=0.9\textwidth]{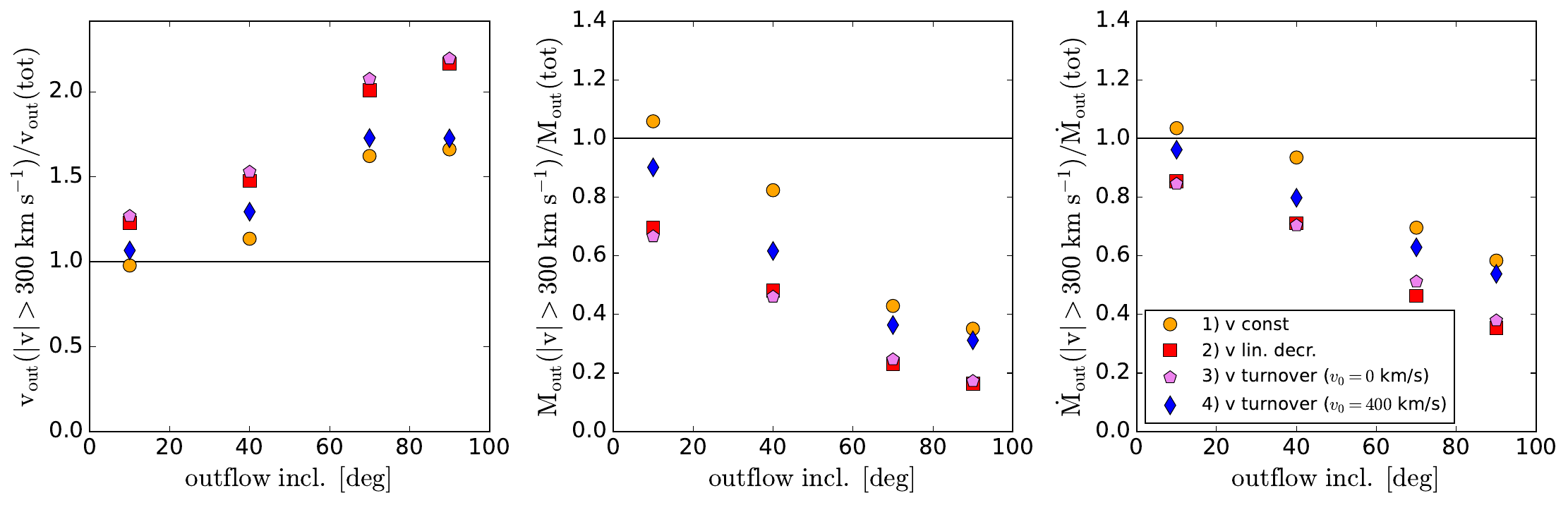}
\caption{Ratios of the outflow properties (\vout, \Mout\ and \Mrate) measured with our method from the total profile (systemic+outflow) considering only the $|v|> 300$~\kms\ range and from only the outflow profile considering the full velocity range.  The ratios are shown as a function of the outflow inclination with respect to our line of sight (where $i= 90^{\circ}$ is in the plane of the sky).  The points are colour-coded according to outflow velocity radial distribution used in the simulation.
The black lines show the one-to-one ratio.  
}
\label{fig:comp_sim}
\end{figure*}

Figure~\ref{fig:comp_sim} shows the ratios of the outflow quantities measured with our method ($|v| > 300$~\kms) and from the total outflow profile.  The differences between the two methods increase with increasing outflow inclination.

\section{Bisector fit methodology}
\label{sec:fitting_method}
Here we explain in detail the methodology used for the bisector fit used in Figures~\ref{fig:Q_out_vs_SFR}, \ref{fig:Mload_vs_logv}, \ref{fig:SFR_vs_logv},  and  \ref{fig:M_over_R_vs_logv}.
Considering two generic quantities $x $ and$ y$,  we fit a linear relation between them using  the Monte Carlo Markov-Chain (MCMC) implementation PyStan\footnote{https://mc-stan.org/users/interfaces/pystan}.
 We assume that the noise is normal distributed.  
 We fit for the parameter $\alpha$ and  $\beta$ that minimises the quantity:
\begin{equation}
\epsilon_y = \sum_i \left( \frac{(f(x_i, \alpha, \beta)-y_i)^2}{y_{err, i}^2+\sigma_{intr, y}^2}\right),
\end{equation}
where $f(x_i, \alpha, \beta)= \alpha \cdot x_i +\beta$,   $y_{err, i}$ is the error on the quantity $y_i$,  $\sigma_{intr}$ is the intrinsic scatter, and the sum is over the data points.  The analogous expression can be defined with respect to the quantity $x_i$ and the error $x_{err, i}$:
\begin{equation}
\epsilon_x = \sum_i \left( \frac{(f(y_i, \gamma, \delta)-x_i)^2}{x_{err, i}^2+\sigma_{intr, x}^2}\right),
\end{equation}
 We run the fitting algorithm to find the best parameters that minimise $\epsilon_y$ and $\epsilon_x$,  and then we perform a bisector fit.  We run the algorithm for 2000 steps with 4 chains  as `burn-in' phase.  
 We monitor the convergence by looking at the effective sample size
($N_{eff}$), which is defined as the number of iterations divided by the
integrated autocorrelation time $N_{eff} = N_{iter}/\tau_{int}$. 
We check that the number of effective samples is $ > 10$, which indicate that the algorithm has converged \citep{Gelman2004}.

We derive the best fit values and the corresponding 1$\sigma$ uncertainties from the median and the 16th-84th percentiles, respectively,  of the marginalised posterior distributions of the parameters.  The intrisic scatter is obtained by adding in quadrature $\sigma_{intr, x}$ and $\sigma_{intr, y}$.  
The lightblue shaded regions in the Figures shows the 1~$\sigma$ uncertainty on the fit, obtained by sampling the posterior distribution. 

\section{CO outflow detection in absorption}
\begin{figure*}
\centering

\includegraphics[width=0.32\textwidth]{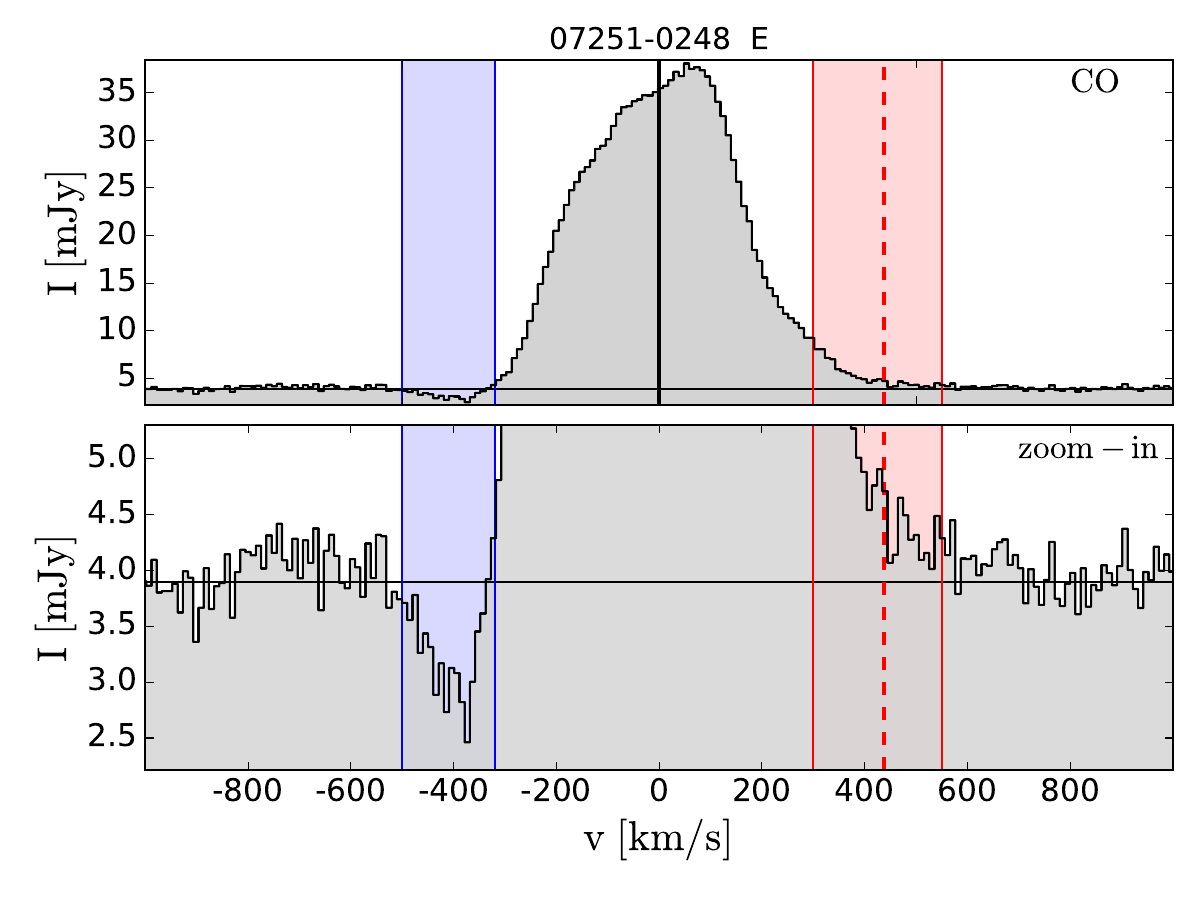}
\includegraphics[width=0.26\textwidth]{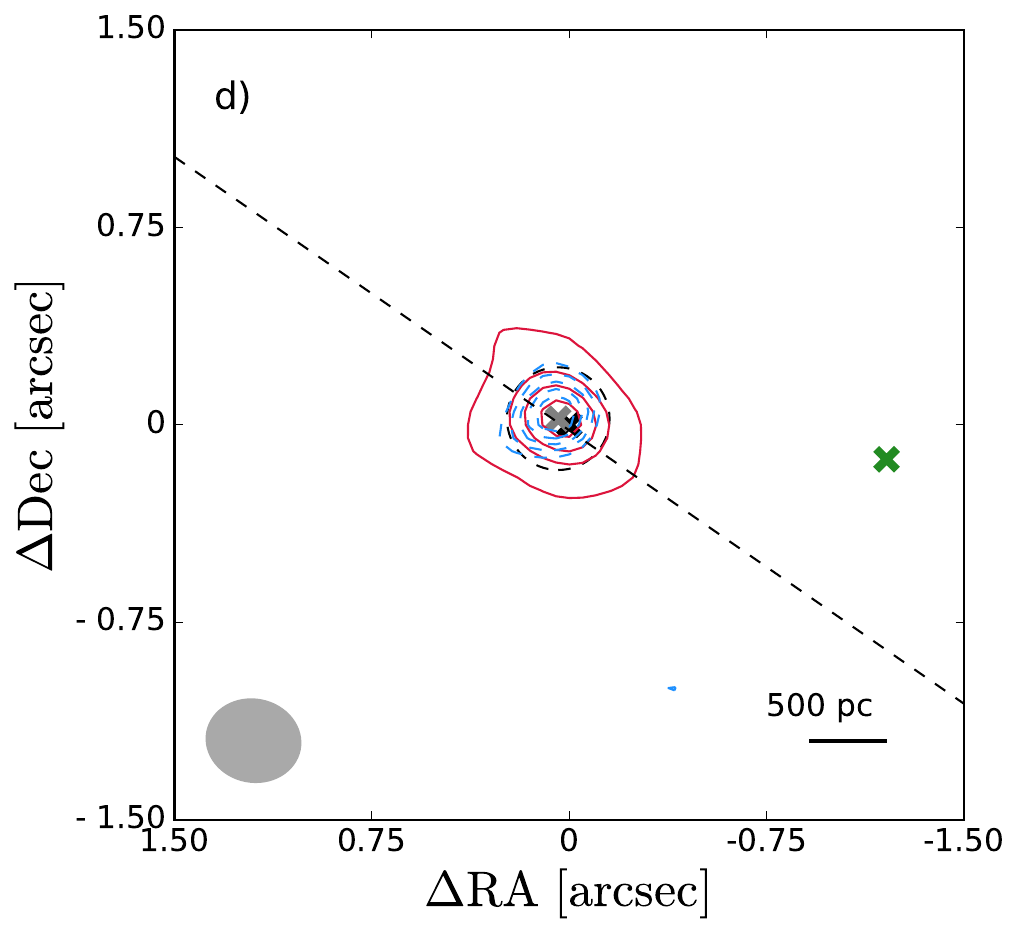}
\includegraphics[width=0.32\textwidth]{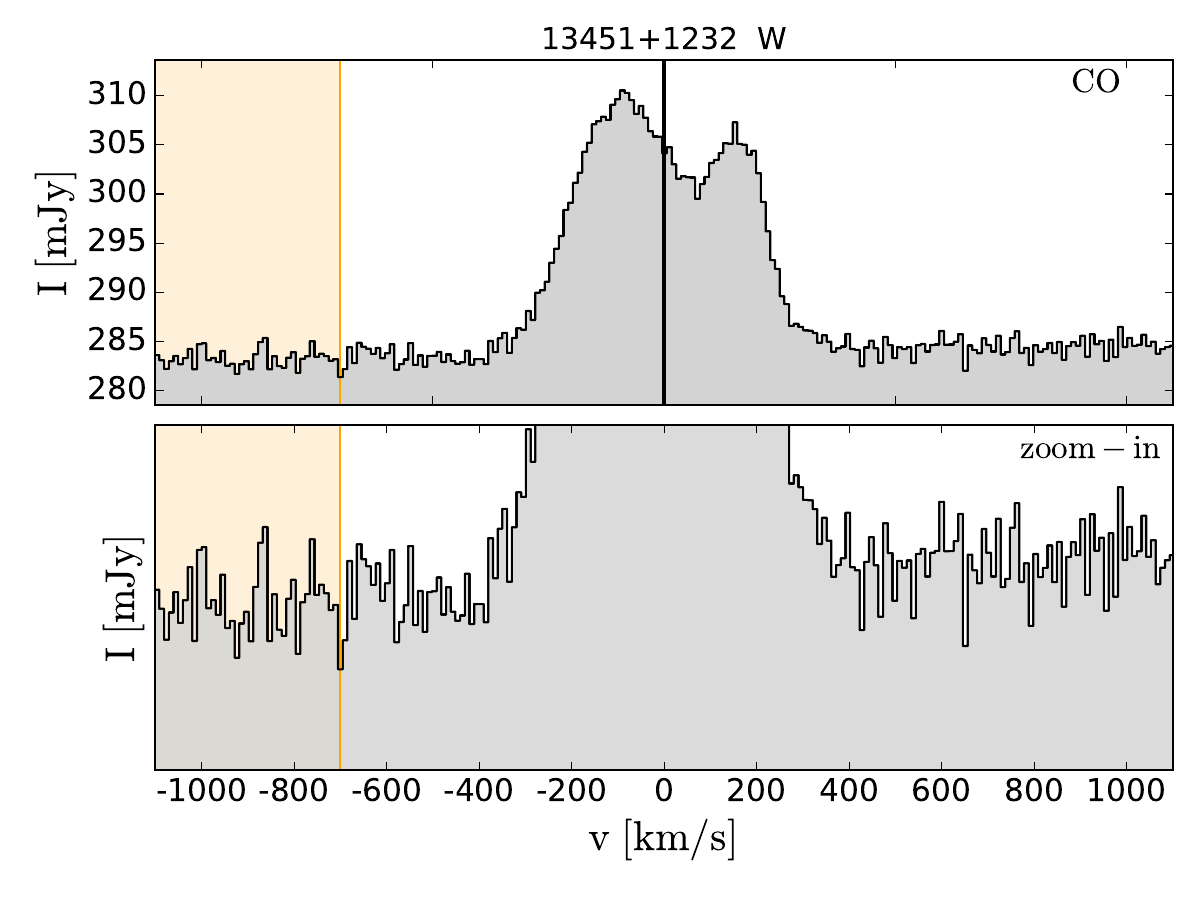}
\caption{\textit{Left:} CO(2-1) spectrum (with continuum) of 07251-0248 E , showing an absorption  at velocities  $v=[ -320, -500]$~\kms.
 \textit{Middle:} Emission of the blue- and red-shifted  high-velocity channels of 07251-0248 E,  shown with blue and red contours,  respectively. The lowest contour corresponds to the 3$\sigma$ level. The next contour levels are (0.5, 0.7, 0.9) of the peak of the  emission.  Dashed lines indicate negative contour levels ([-3, -4, -5, -6, -7]$\times \sigma$).   The dashed black circle shows the size of the outflow (\Rout) and the grey cross the central position of the outflow. The black and green crosses shows the position of the continuum peaks of the E and W nuclei,  respectively.  
 \textit{Right:} CO(2-1) spectrum (with continuum) of 13451+1232 W.  No clear absorption is detected in this spectrum at the velocities of the an absorption  detected in the CO(3-2) spectrum by \citet{Dasyra2012} at $v= [ -700, -1200]$~\kms\ (see orange shaded region). }
\label{fig:absorption}
\end{figure*}

We detect a blue-shifted absorption in the CO(2-1) spectrum of 07251-0248 E at velocities $v=[ -320, -500]$~\kms\  (see Fig.~\ref{fig:absorption}). The absorption is compact, centred at the position of the continuum peak and nearly coincident with the position of the red-shifted outflow emission at $v=[ 300, 550]$~\kms.
This absorption can be interpreted as an outflow located in front of the continuum source and moving towards us.  Alternatively, it could be due to an extremely compact outflow. The redshifted part of the outflow, identified in emission,  appears compact and moves away from us along a direction compatible with our line of sight.

This is the only target for which we could identify an absorption.  \citet{Dasyra2012} detected a blue-shifted absorption at  $-950$~\kms\ in the CO(3-2) spectrum of 13451+1232	(4C+12.50),  obtained with the IRAM 30m telescope.  No absorption was detected in the CO(1-0) data observed with the IRAM Plateau de Bure Interferometer (PdBI)  by \citet{Dasyra2014}.
We inspect the  ALMA CO(2-1) spectrum of the west nucleus of 13451+1232 (shown in Fig.~\ref{fig:absorption}),  which is the nucleus dominating the FIR continuum emission and therefore,  the nucleus where an absorption could be produced.
We do not observe any significant absorption in the spectral channels on the blue-side of the CO(2-1) line in this target.  
Although, we note that the CO(3-2) absorption feature would be near the edge of the spectral window of our CO(2-1) spectrum.

\section{CO(2-1) moment maps}

\begin{figure*} 
\centering 
\includegraphics[width=0.9\textwidth]{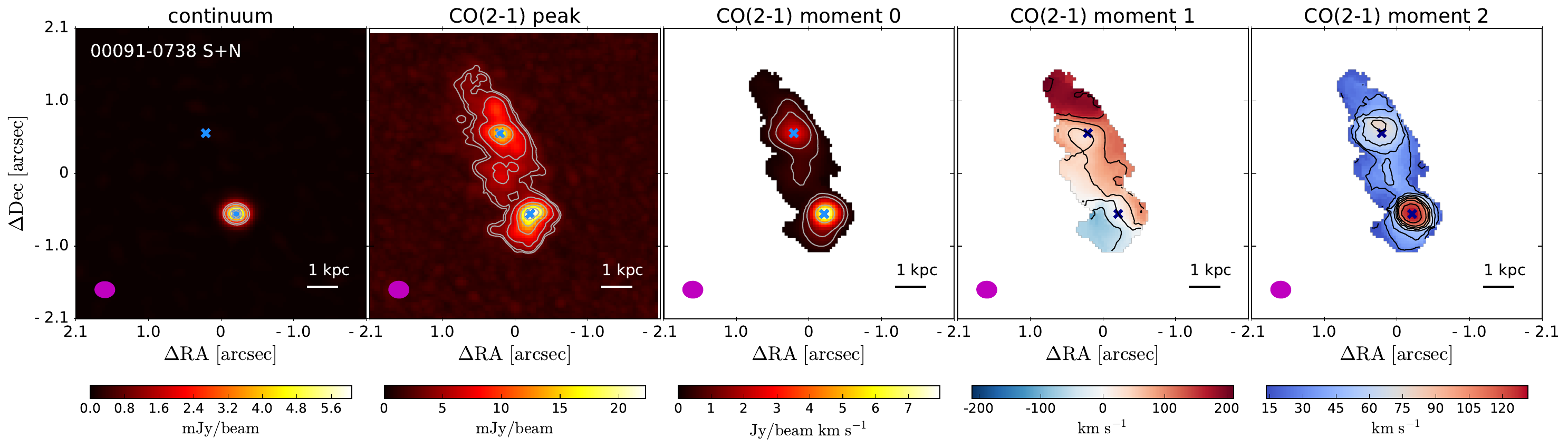}\\ 
\includegraphics[width=0.9\textwidth]{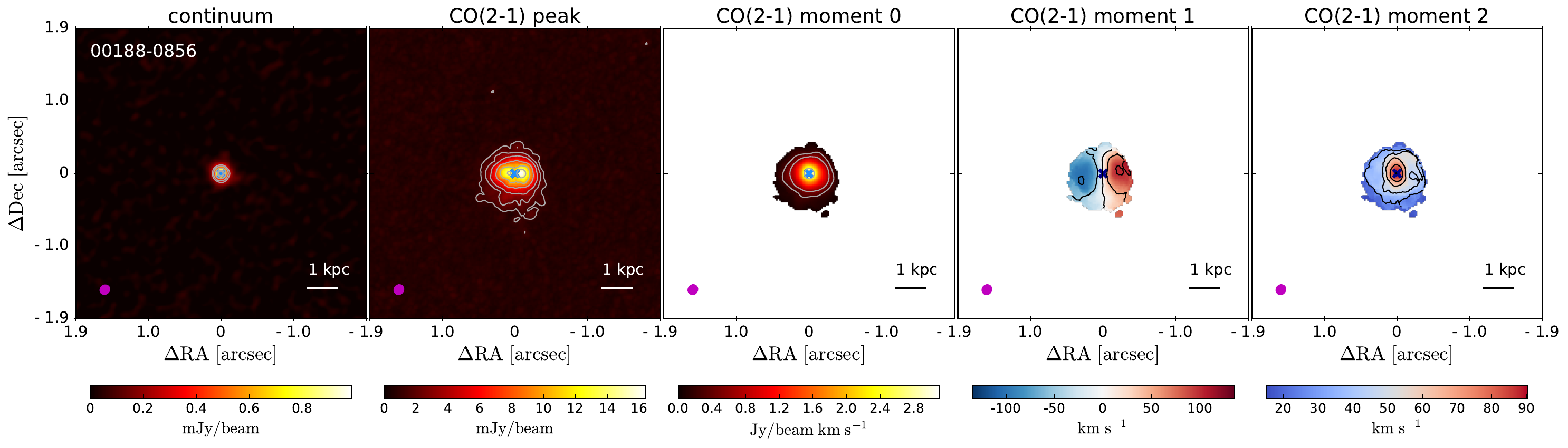}\\ 
\includegraphics[width=0.9\textwidth]{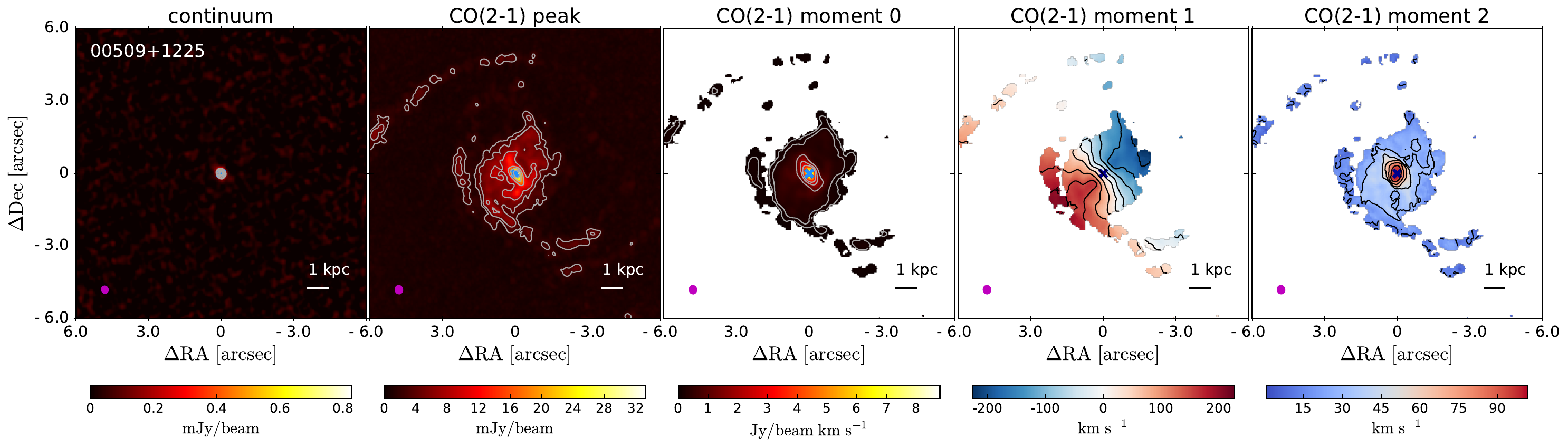}\\ 
\includegraphics[width=0.9\textwidth]{Figures/Moment_maps/IRAS_01572+0009_moment_maps.pdf}\\ 
\includegraphics[width=0.9\textwidth]{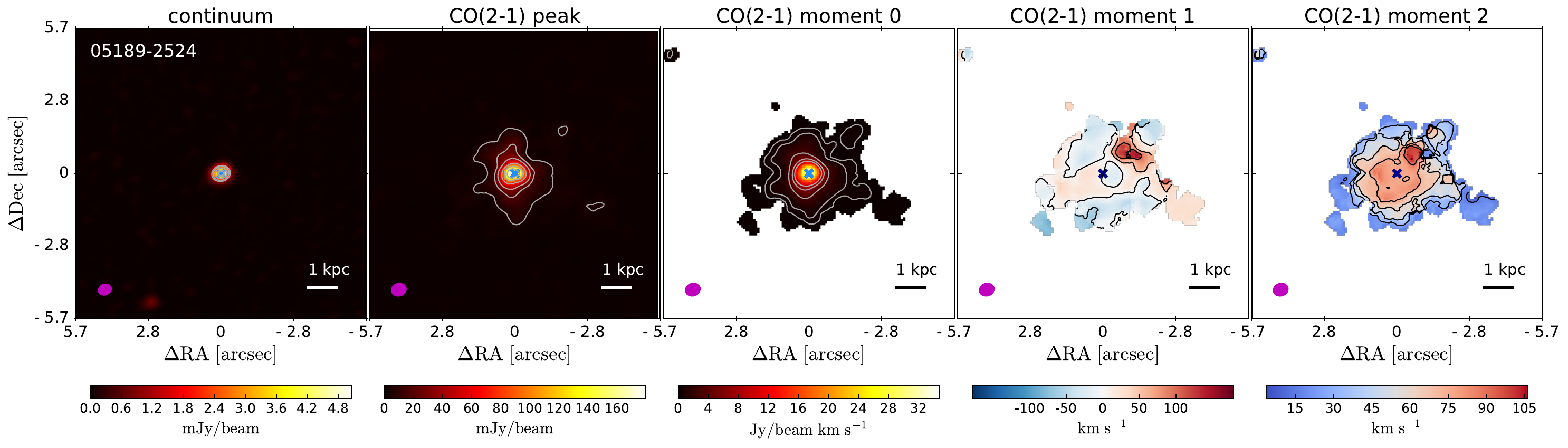}\\ 
 \caption{See caption of Figure~\ref{fig:moment_maps_ex}}
\label{fig:moment_maps_appendix}
\end{figure*} 

\begin{figure*}\ContinuedFloat 
\centering 
\includegraphics[width=0.9\textwidth]{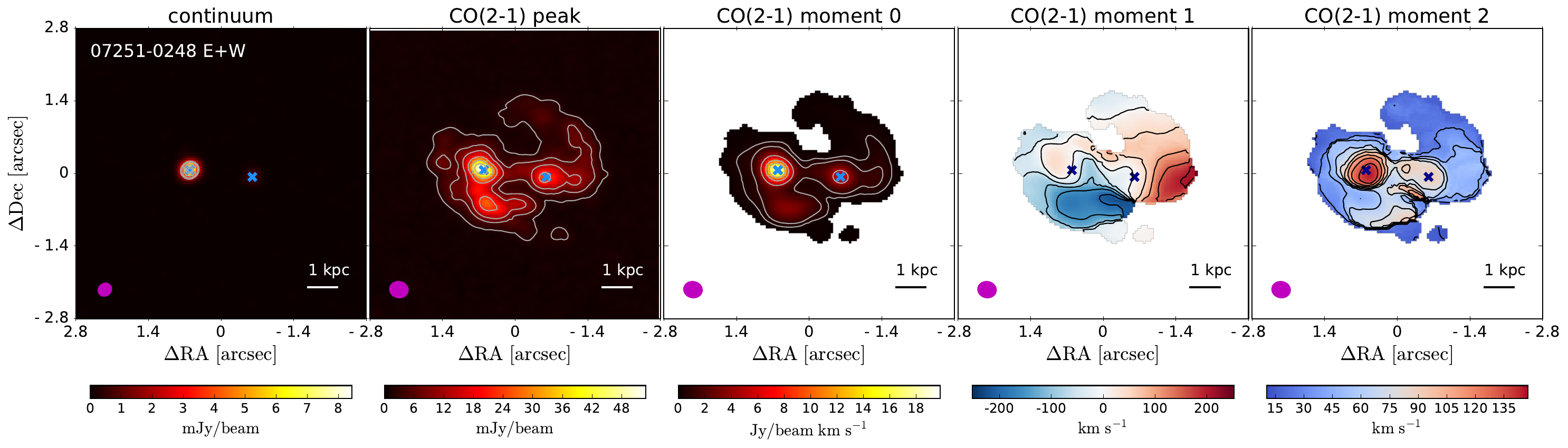}\\ 
\includegraphics[width=0.9\textwidth]{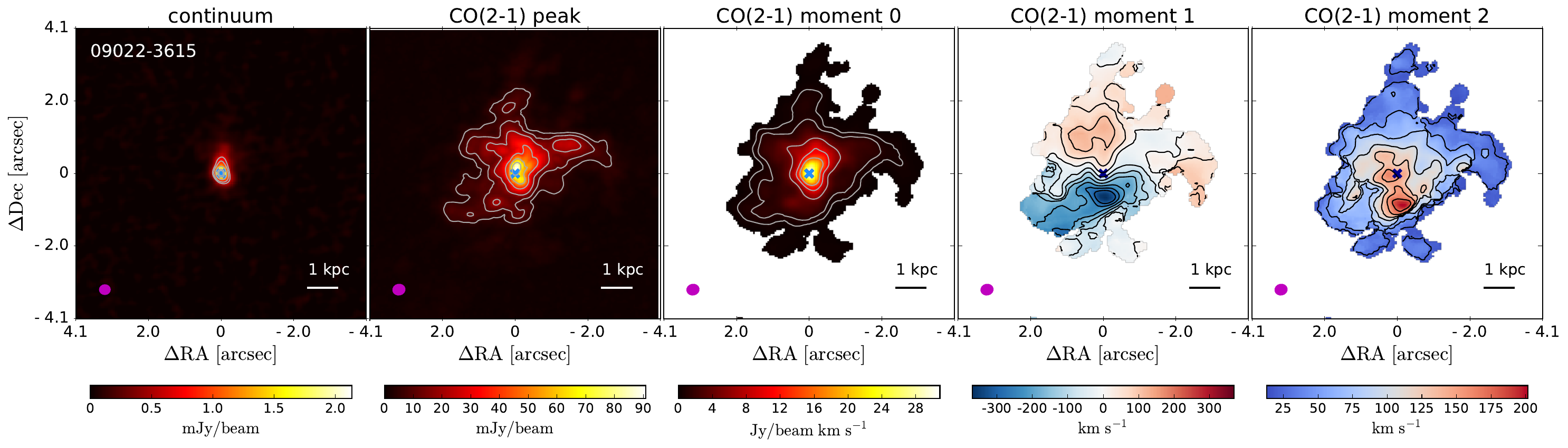}\\ 
\includegraphics[width=0.9\textwidth]{Figures/Moment_maps/IRAS_10190+1322_moment_maps.pdf}\\ 
\includegraphics[width=0.9\textwidth]{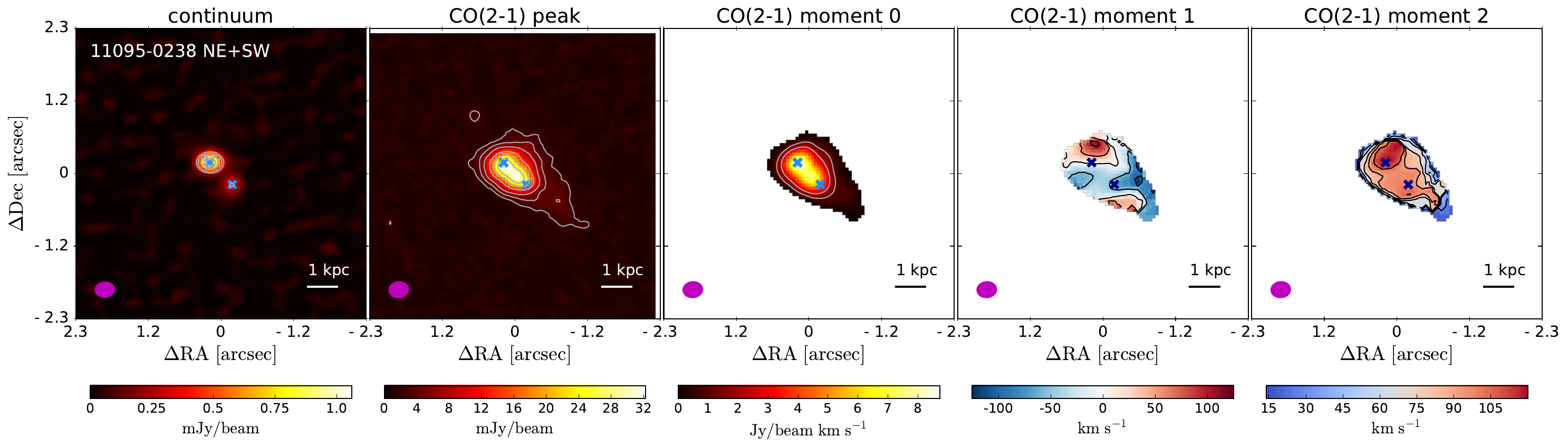}\\ 
\includegraphics[width=0.9\textwidth]{Figures/Moment_maps/IRAS_12071-0444_moment_maps.pdf}\\ 
\caption{continued.} 
 \end{figure*}

\begin{figure*}\ContinuedFloat 
\centering 
\includegraphics[width=0.9\textwidth]{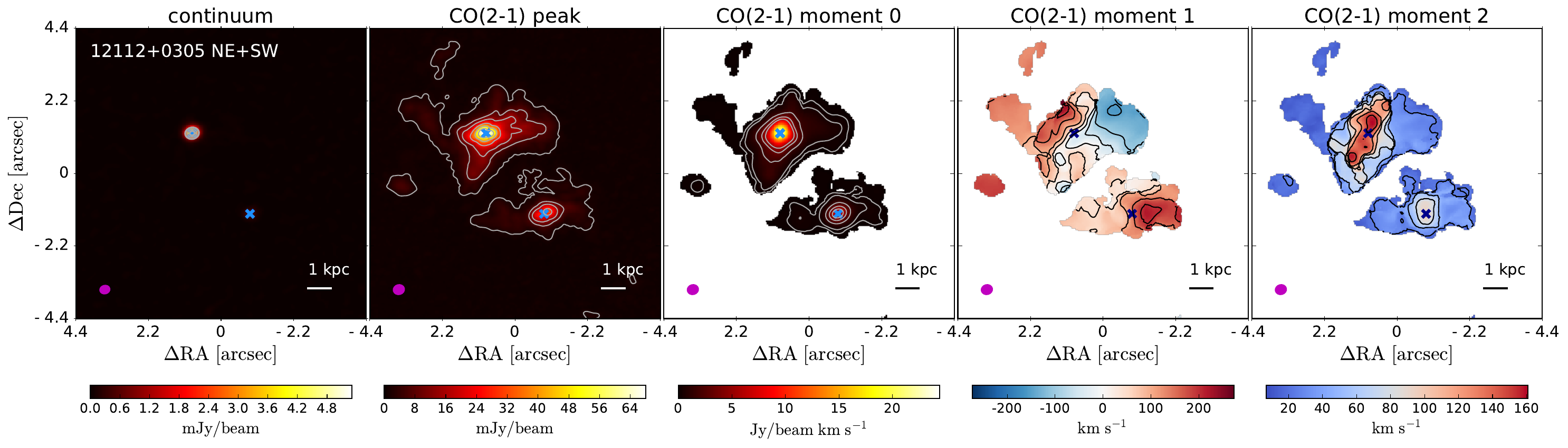}\\ 
\includegraphics[width=0.9\textwidth]{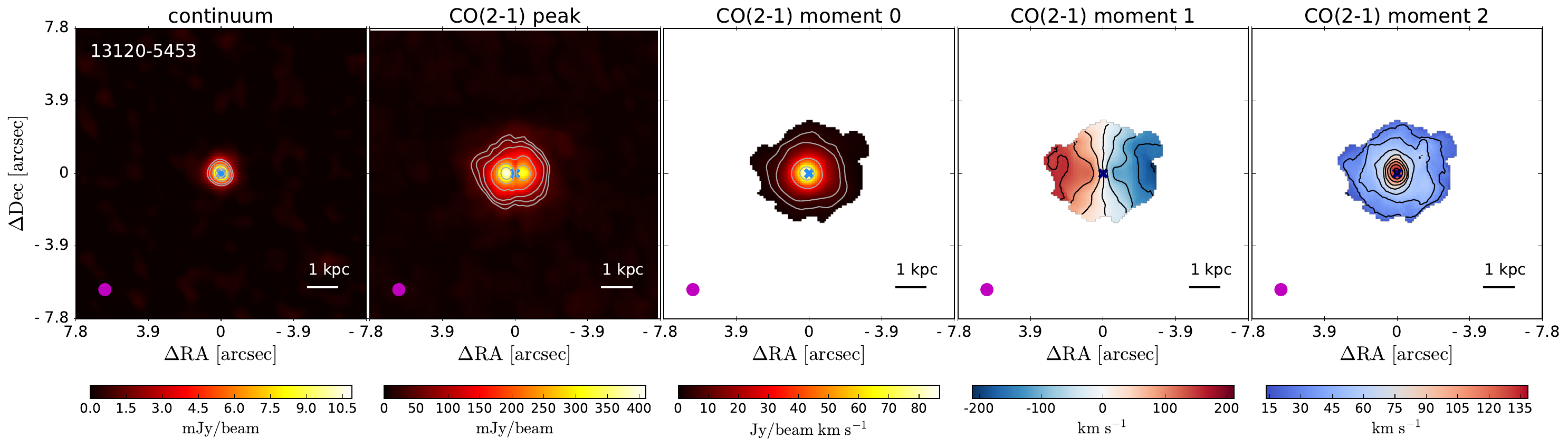}\\ 
\includegraphics[width=0.9\textwidth]{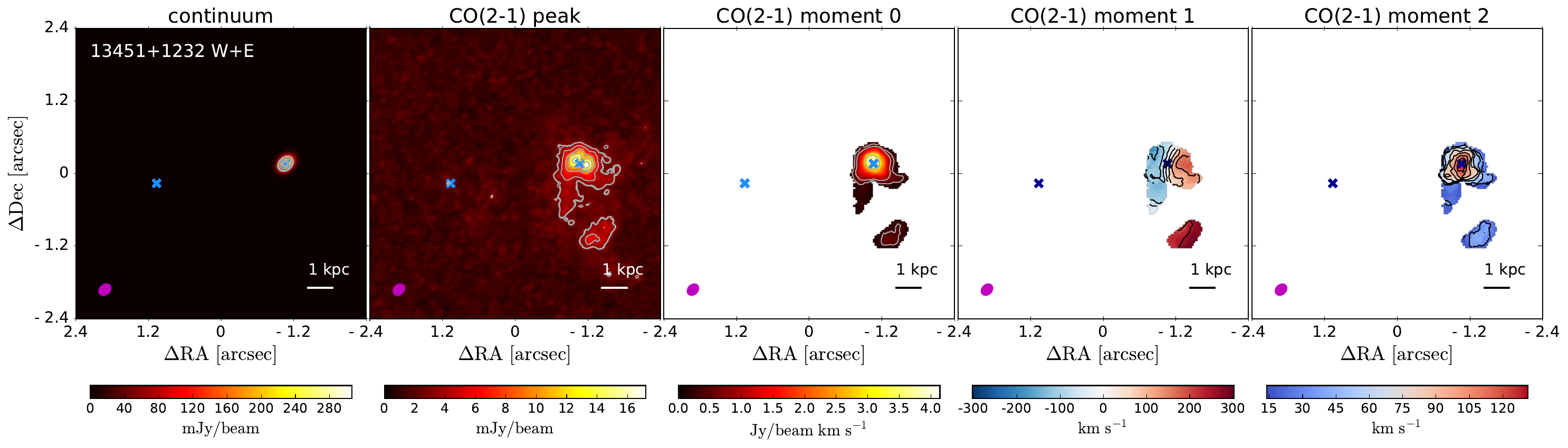}\\ 
\includegraphics[width=0.9\textwidth]{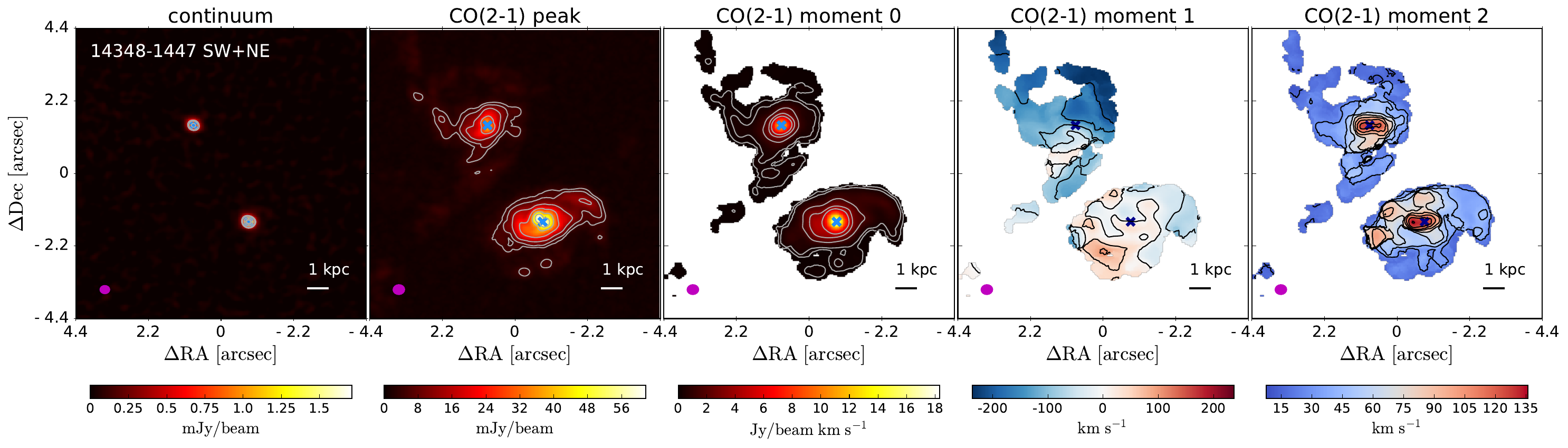}\\ 
\includegraphics[width=0.9\textwidth]{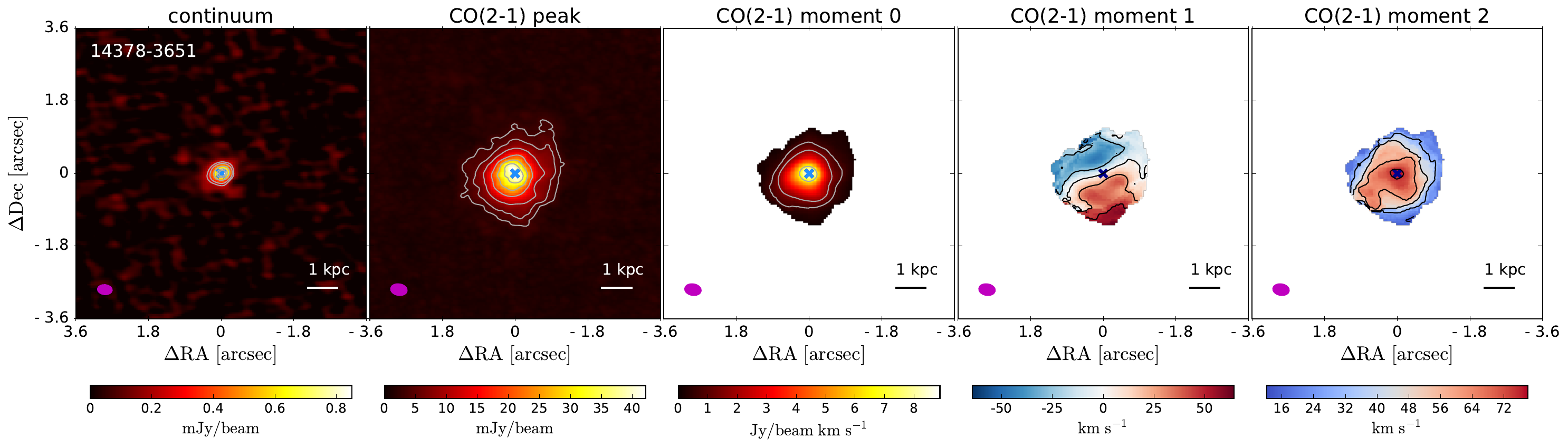}\\ 
\caption{continued.} 
 \end{figure*} 
 
 \begin{figure*}\ContinuedFloat 
\centering 
\includegraphics[width=0.9\textwidth]{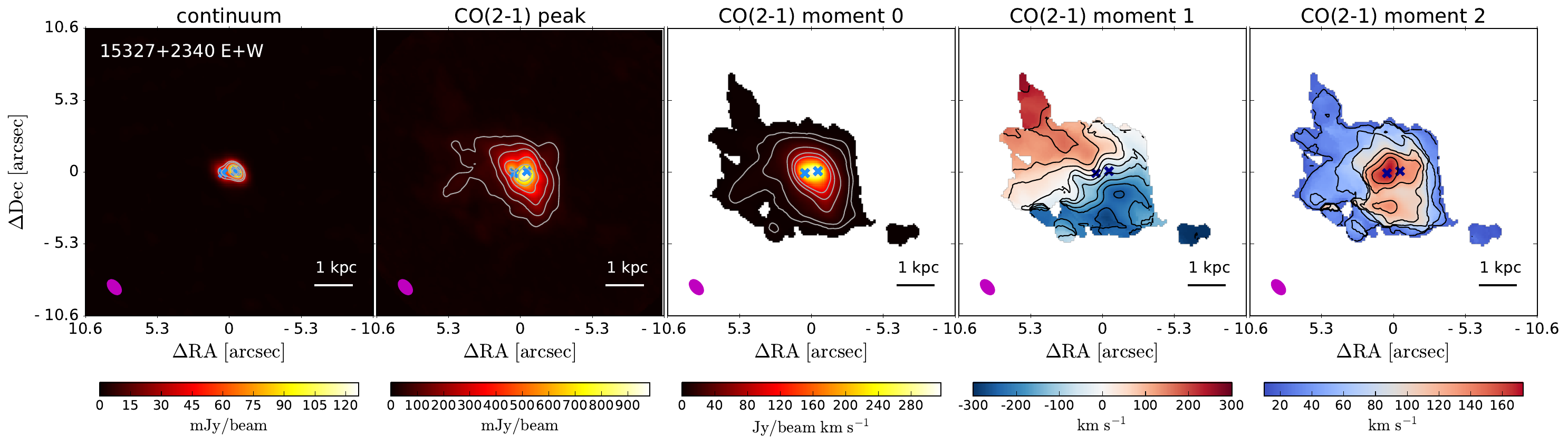}\\ 
\includegraphics[width=0.9\textwidth]{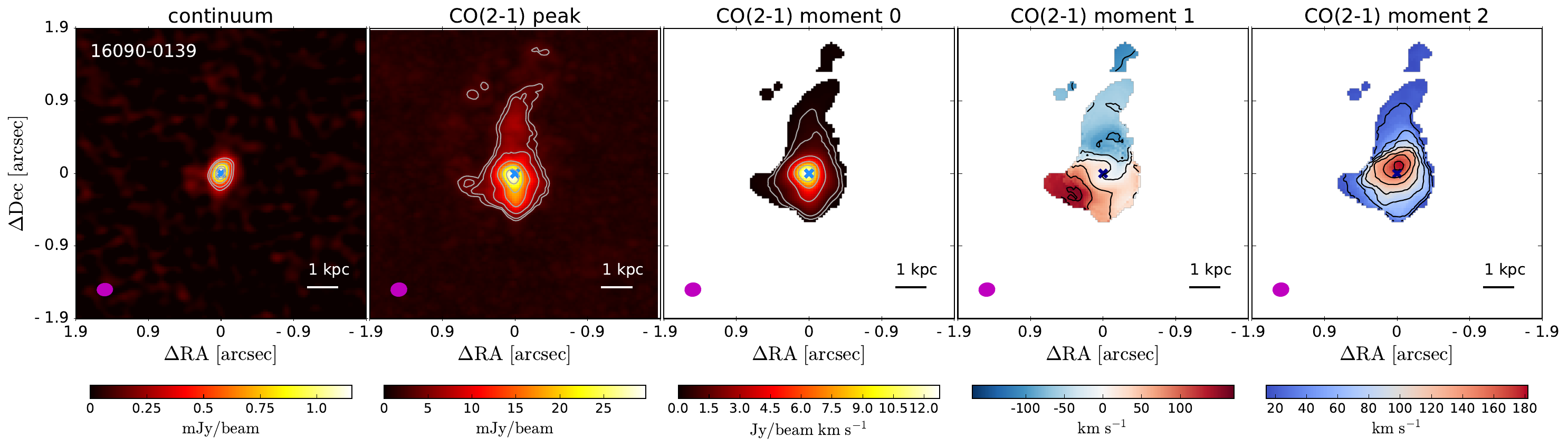}\\ 
\includegraphics[width=0.9\textwidth]{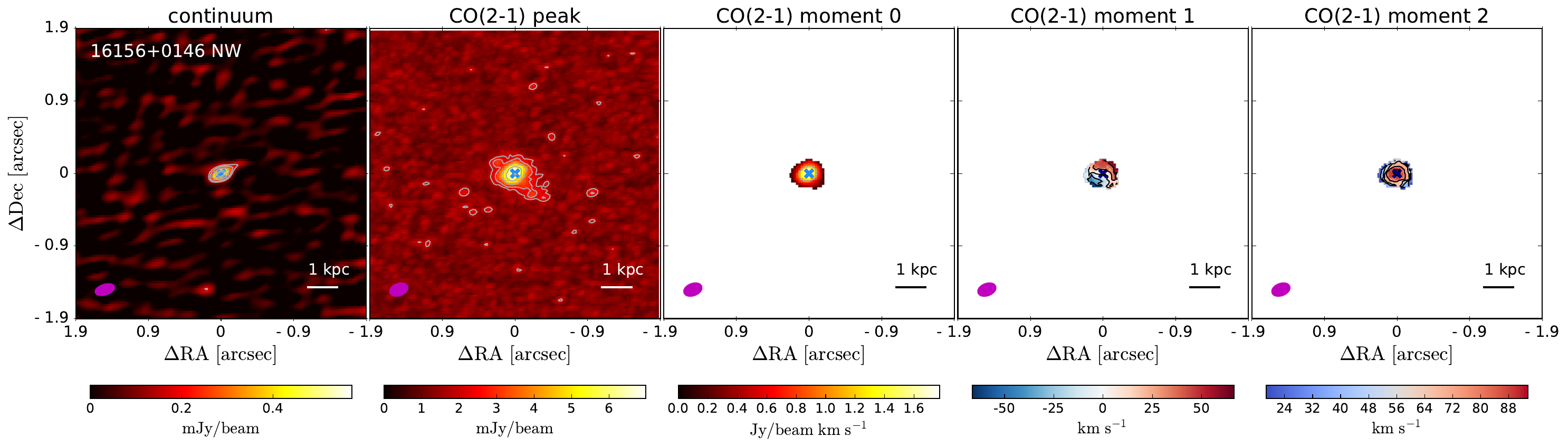}\\ 
\includegraphics[width=0.9\textwidth]{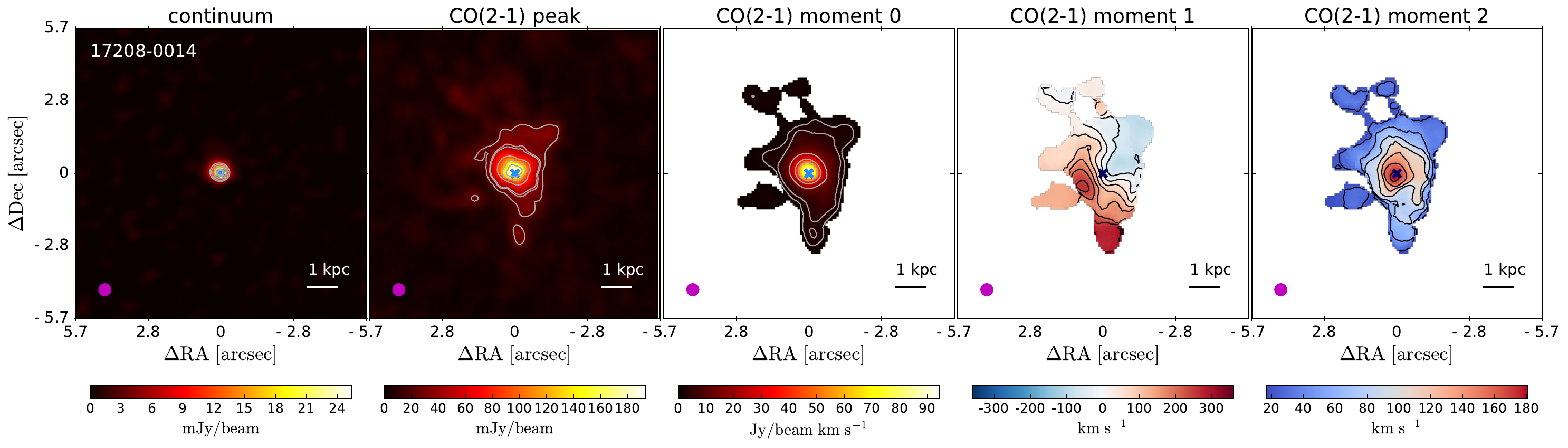}\\ 
\includegraphics[width=0.9\textwidth]{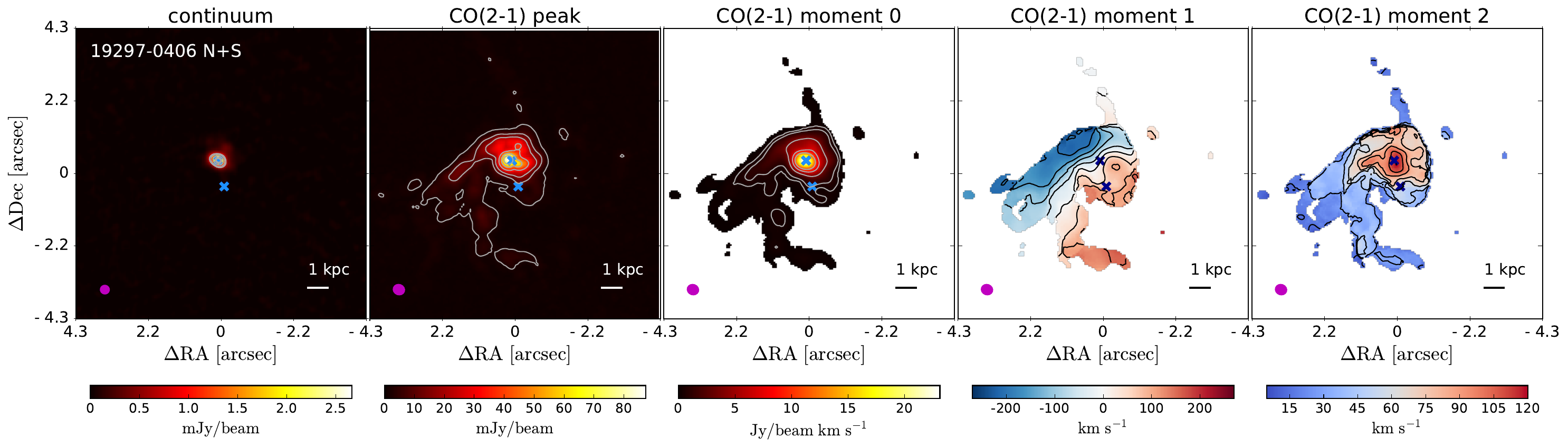}\\ 
\caption{continued.} 
 \end{figure*}

\begin{figure*}\ContinuedFloat 
\centering 
\includegraphics[width=0.9\textwidth]{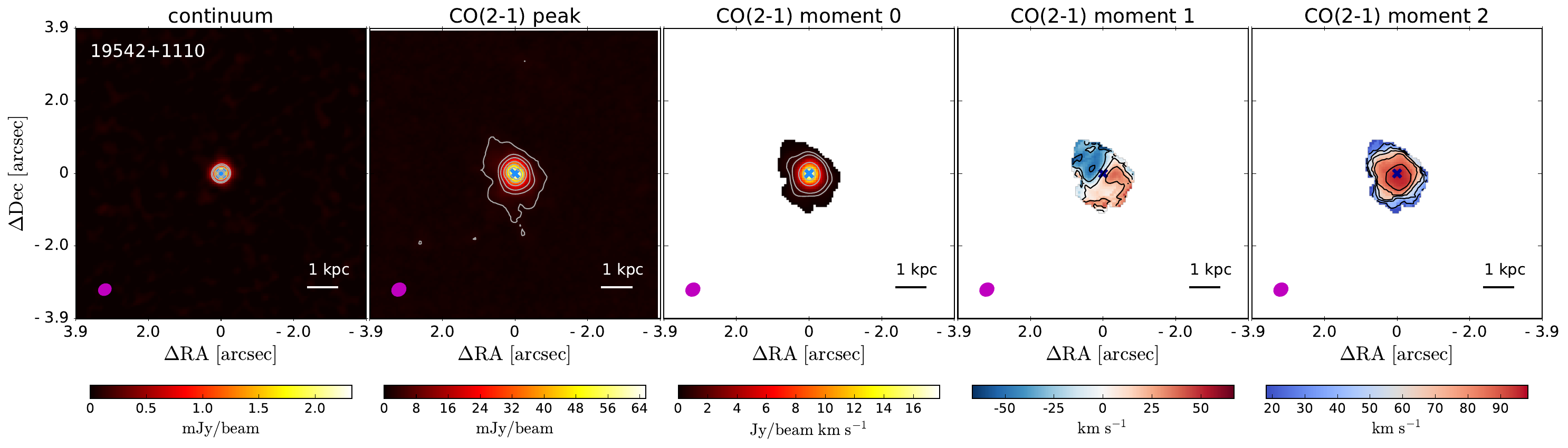}\\ 
\includegraphics[width=0.9\textwidth]{Figures/Moment_maps/IRAS_20087-0308_moment_maps.pdf}\\ 
\includegraphics[width=0.9\textwidth]{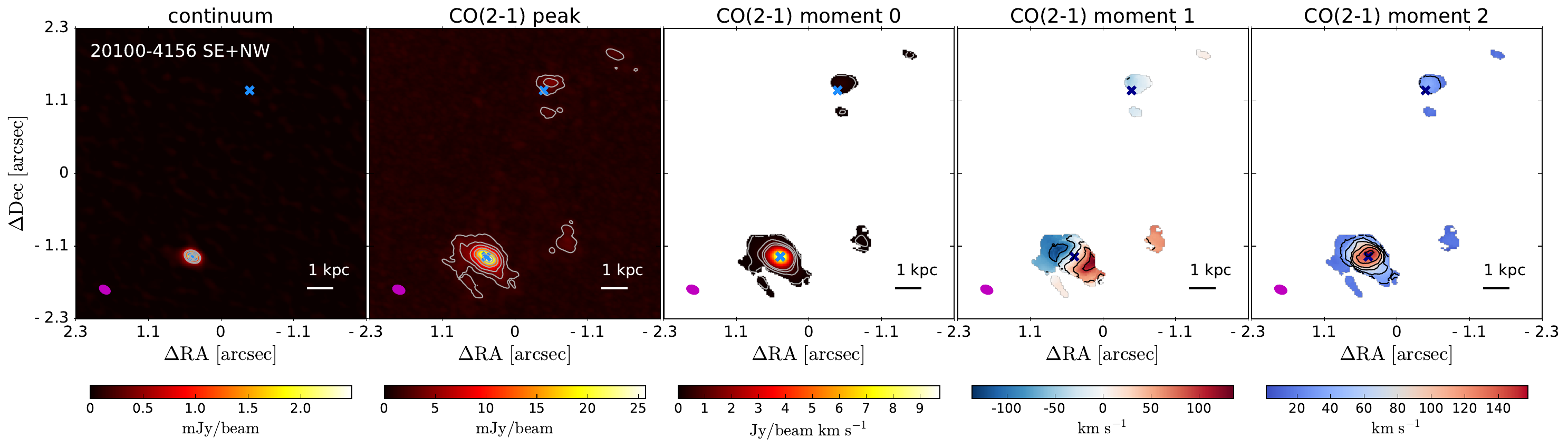}\\ 
\includegraphics[width=0.9\textwidth]{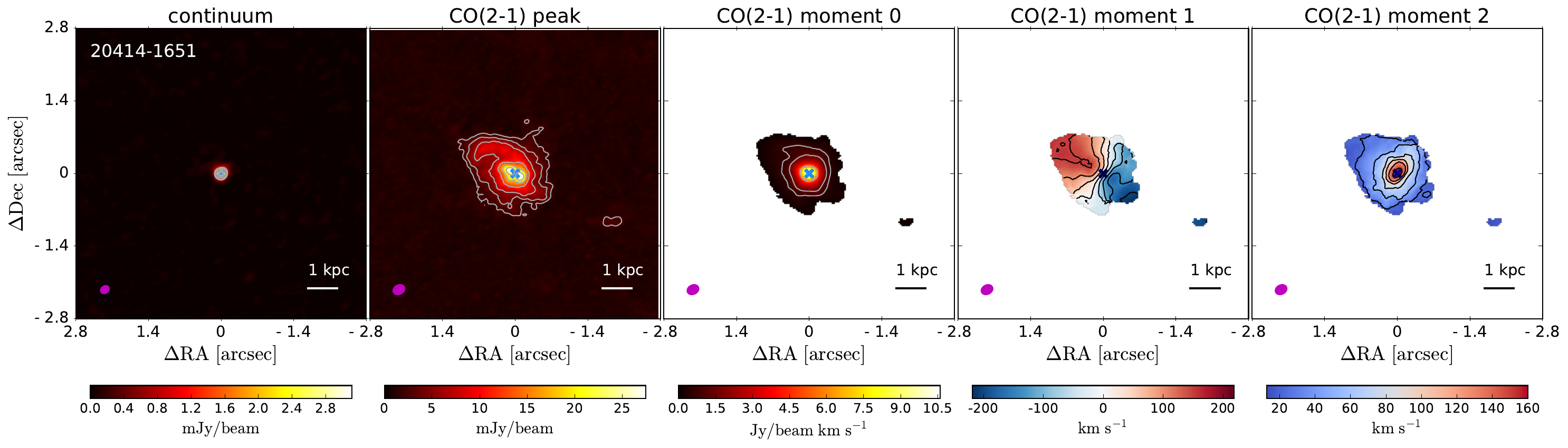}\\ 
\includegraphics[width=0.9\textwidth]{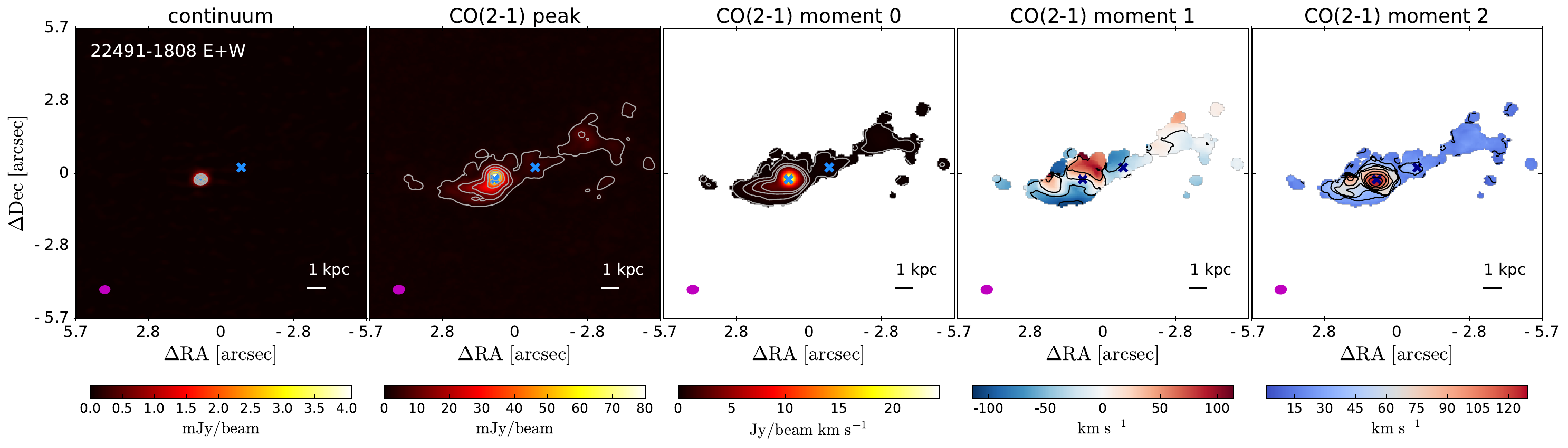}\\ 
\caption{continued.} 
 \end{figure*}

\section{CO(2-1) spectra and spectro-astrometry }

\begin{figure*} 
\centering 
\includegraphics[width=0.75\textwidth]{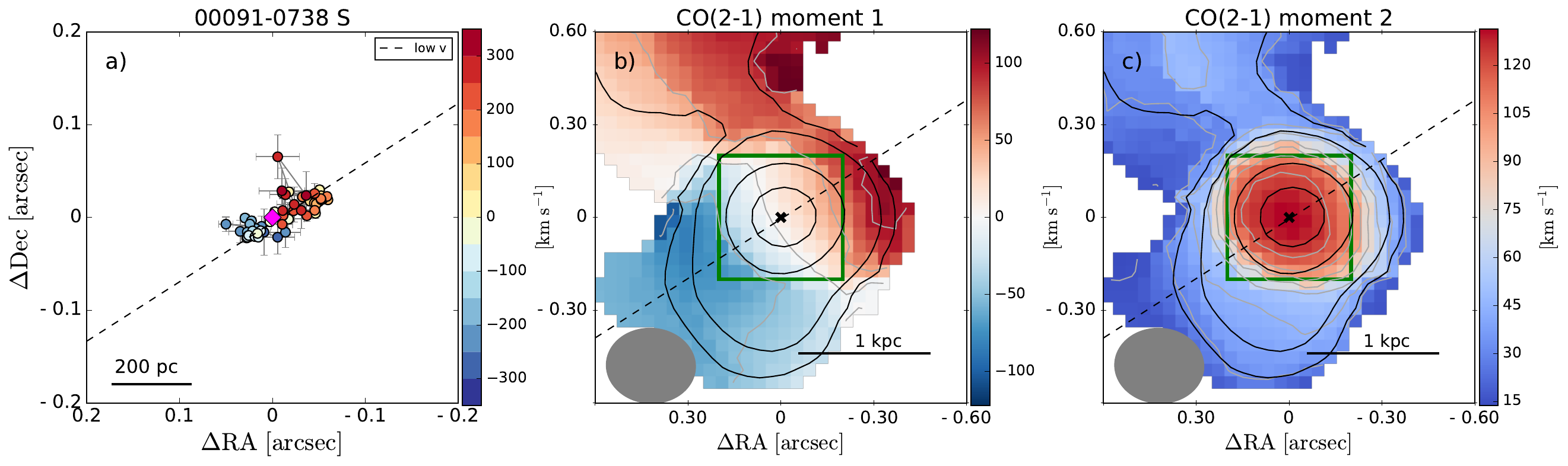}\\ 
\includegraphics[width=0.27\textwidth]{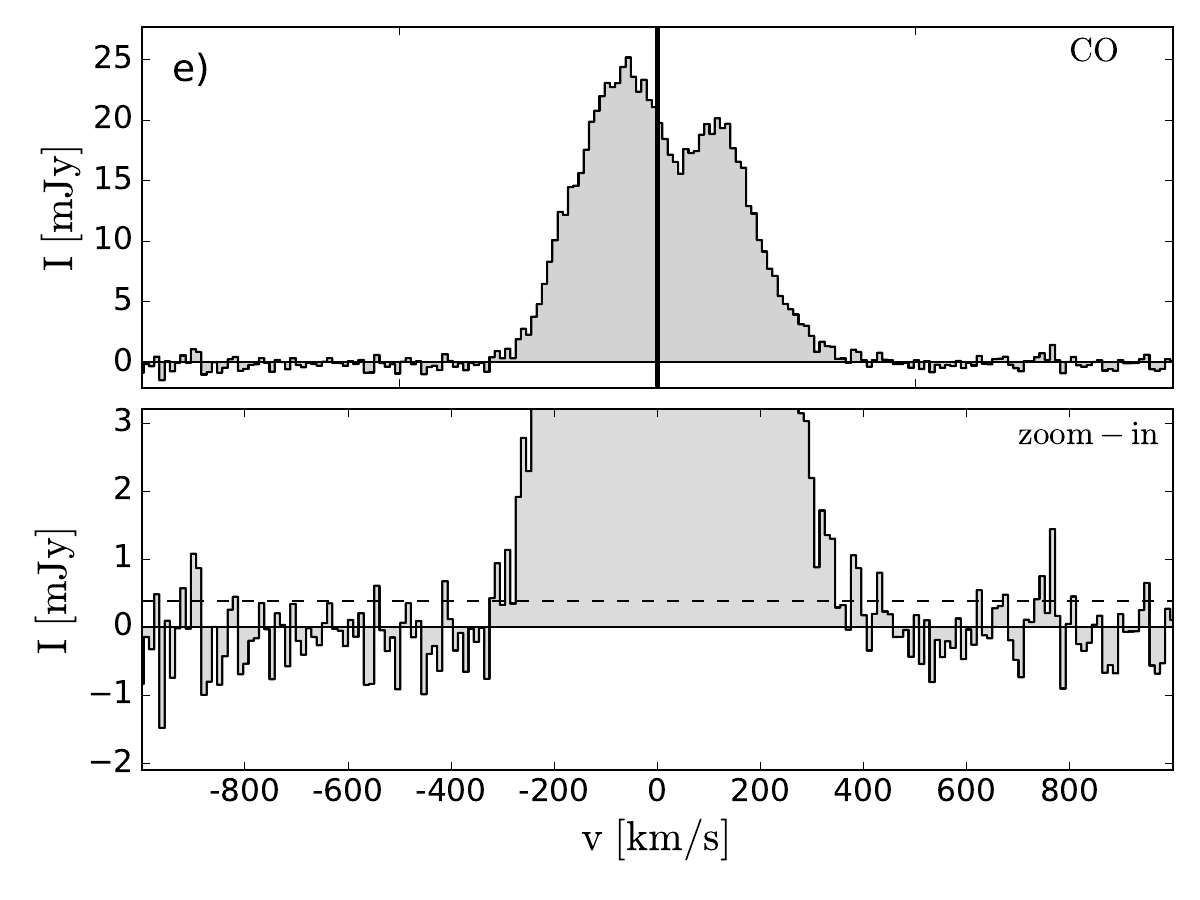} 

\includegraphics[width=0.75\textwidth]{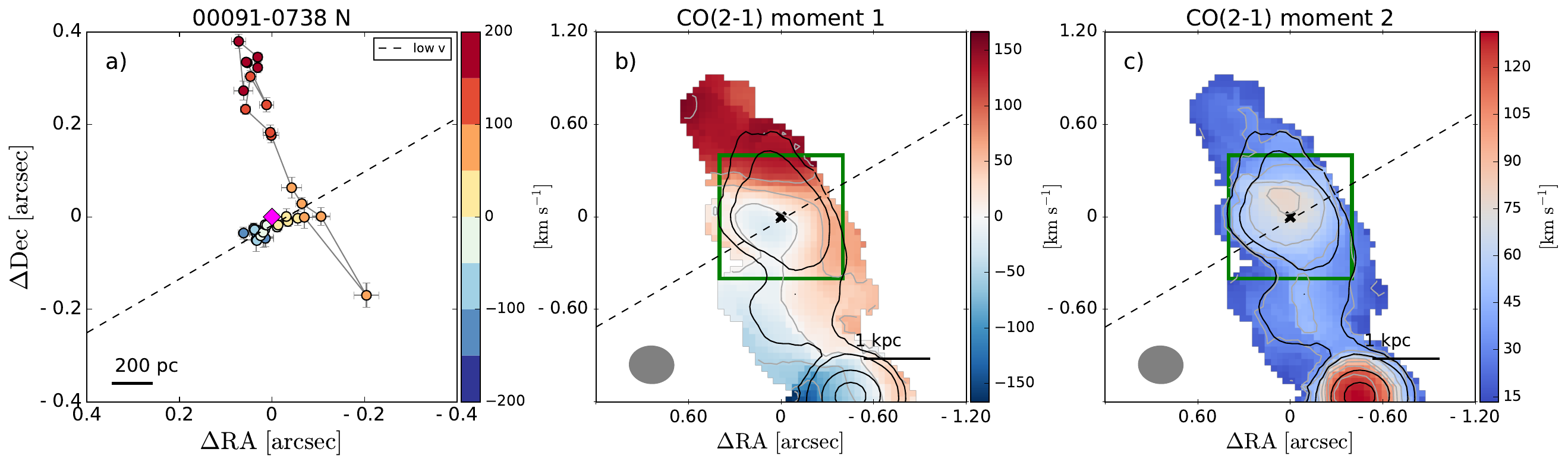}\\ 
\includegraphics[width=0.27\textwidth]{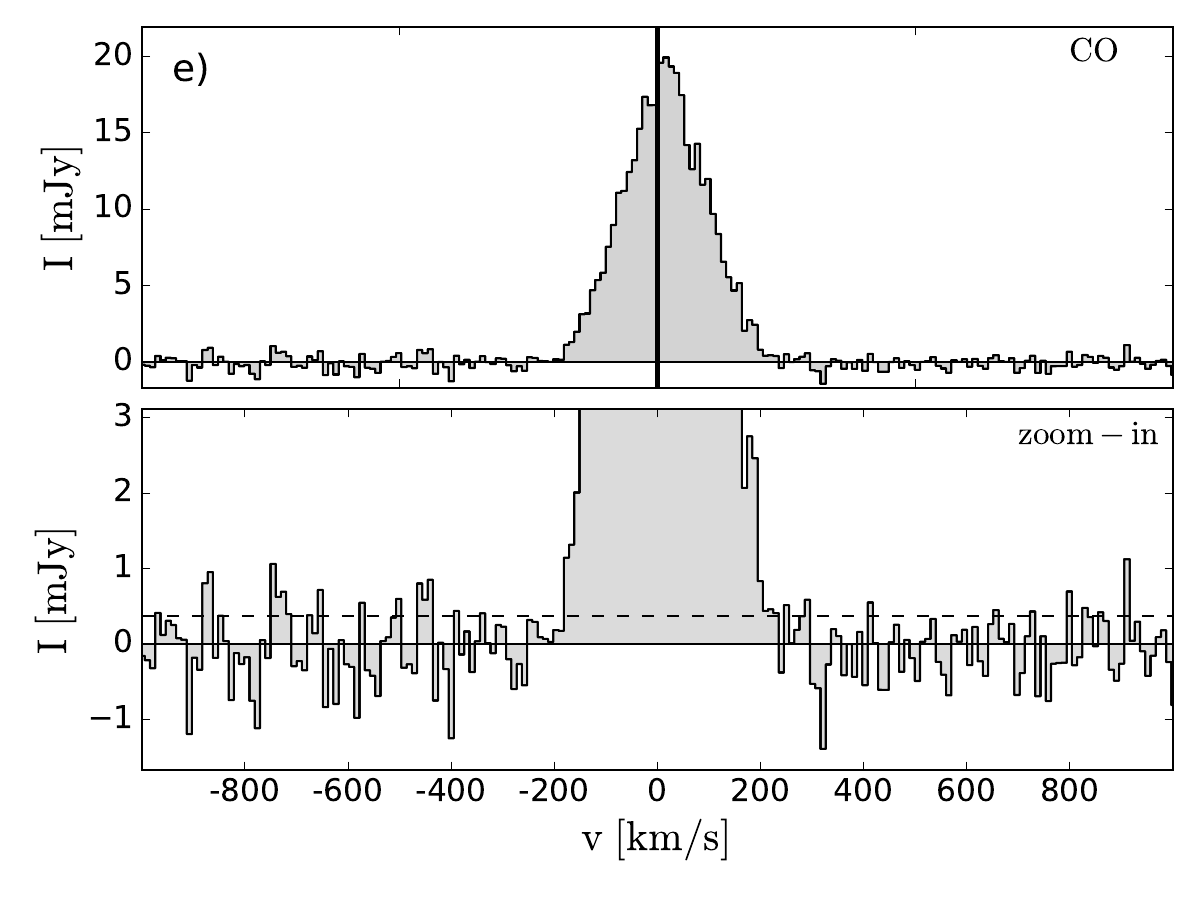}

\caption{ Spectro-astrometry and outflow maps for 00091-0738 S and 00091-0738 N. 
\textit{a)} Spectro-astrometry of the CO(2-1) emission line,  i.e.  centroid position of the CO(2-1) emission in the different velocity channels. The points are colour-coded by the channel velocity. The dotted line is a linear fit to the low-velocity points (kinematic major axis)  and the dashed line is a fit to the high-velocity points (indicating the outflow direction, if present).
Panel \textit{b)} and \textit{c)}  show the moment 1 and moment 2 maps,  where the green square indicate the field of view of  panel \textit{(a)}. The grey ellipse illustrates the ALMA beam FWHM.   The grey contours on the moment 1 maps are every  50~\kms (every 25~\kms\ if the maximum value $< 100$~\kms),  and  every 25~ \kms (every 15~\kms\  if the maximum value $< 150$~\kms) on the moment 2 map.  In black  are the CO(2-1) moment 0  contours ([3, 6, 25, 50, 75]$\times\sigma$).
\textit{d)} Emission of the high-velocity channels(if detected),  integrated over the velocity ranges indicated on the CO(2-1) spectrum (shown in panel \textit{e}).  Blue- and red-shifted channels are shown with blue and red contours,  respectively (dashed lines indicate negative contour levels). The lowest contour corresponds to the 3$\sigma$ level. The next contour levels are (0.5, 0.7, 0.9) of the peak of the  emission,  if these are above the 3$\sigma$ level.  The dashed circle shows the size of the outflow (\Rout).
\textit{e)} CO(2-1) continuum-subtracted spectrum  extracted from a circle with radius equivalent to the outflow size (\Rout). The lower panel shows an y-axis zoom-in to highlight the emission in the wings.  The horizontal dashed line show the 1$\sigma$ noise level.
The vertical dashed lines indicated the `flux-weighted' velocity of the blue and red-shifted outflow ($v_{out}$).
\textit{f)} OH 119~\micron spectrum (\textit{upper},  if available) compared with the nuclear CO spectrum (\textit{bottom}),  convolved to the resolution of the OH spectrum  (FWHM$\sim 270$~\kms).  The total fit to the OH lines is shown with a magenta line, while the Gaussian components of the fit are shown in lightblue and brown. The orange lines show the 50 Monte Carlo iterations used to estimate the uncertainties on the fit  (see Sec.~\ref{sec:OH_analysis}). The vertical dotted, dashed and dot-dashed lines show the $v_{50}$,   $v_{84}$,  and $v_{98}$ percentile velocities, respectively. The blue-shaded area in the upper panel shows the wavelength range between $v_{84}$ and $v_{98}$.}
\label{fig:spectroastrometry_app}
 \end{figure*}

\begin{figure*}\ContinuedFloat 
\centering 
\includegraphics[width=0.75\textwidth]{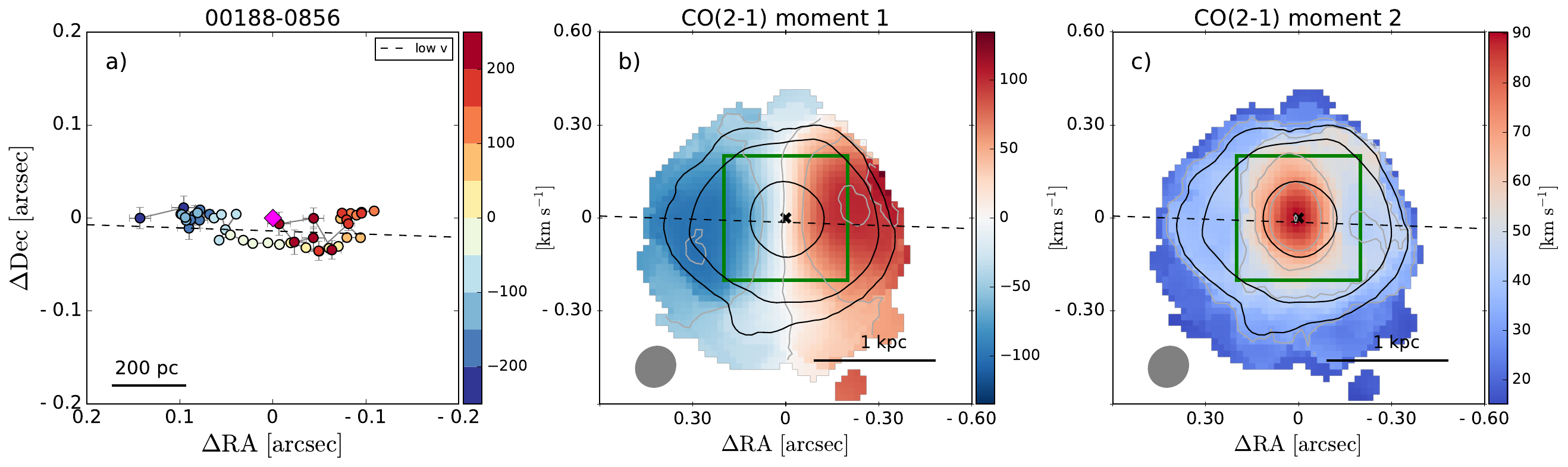}\\ 
\includegraphics[width=0.27\textwidth]{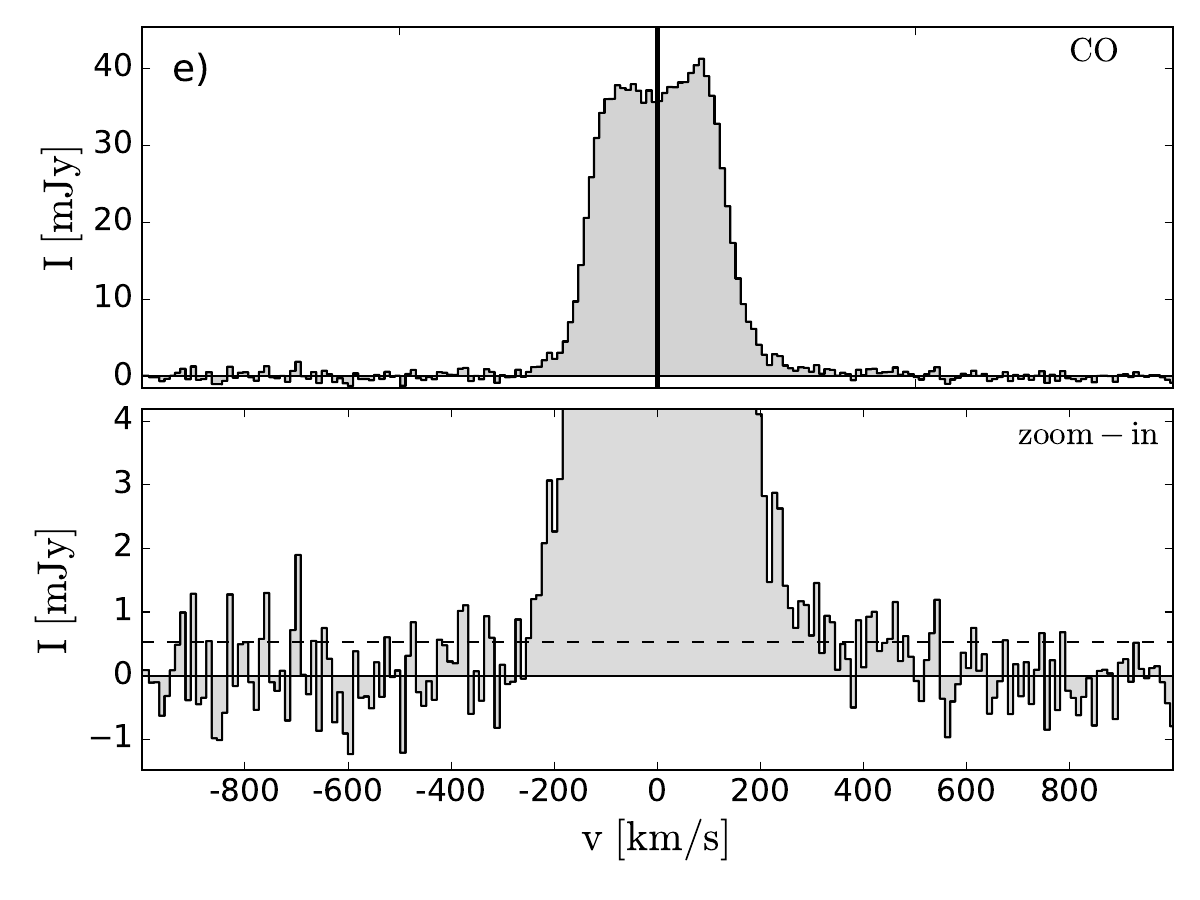} 
\includegraphics[width=0.27\textwidth]{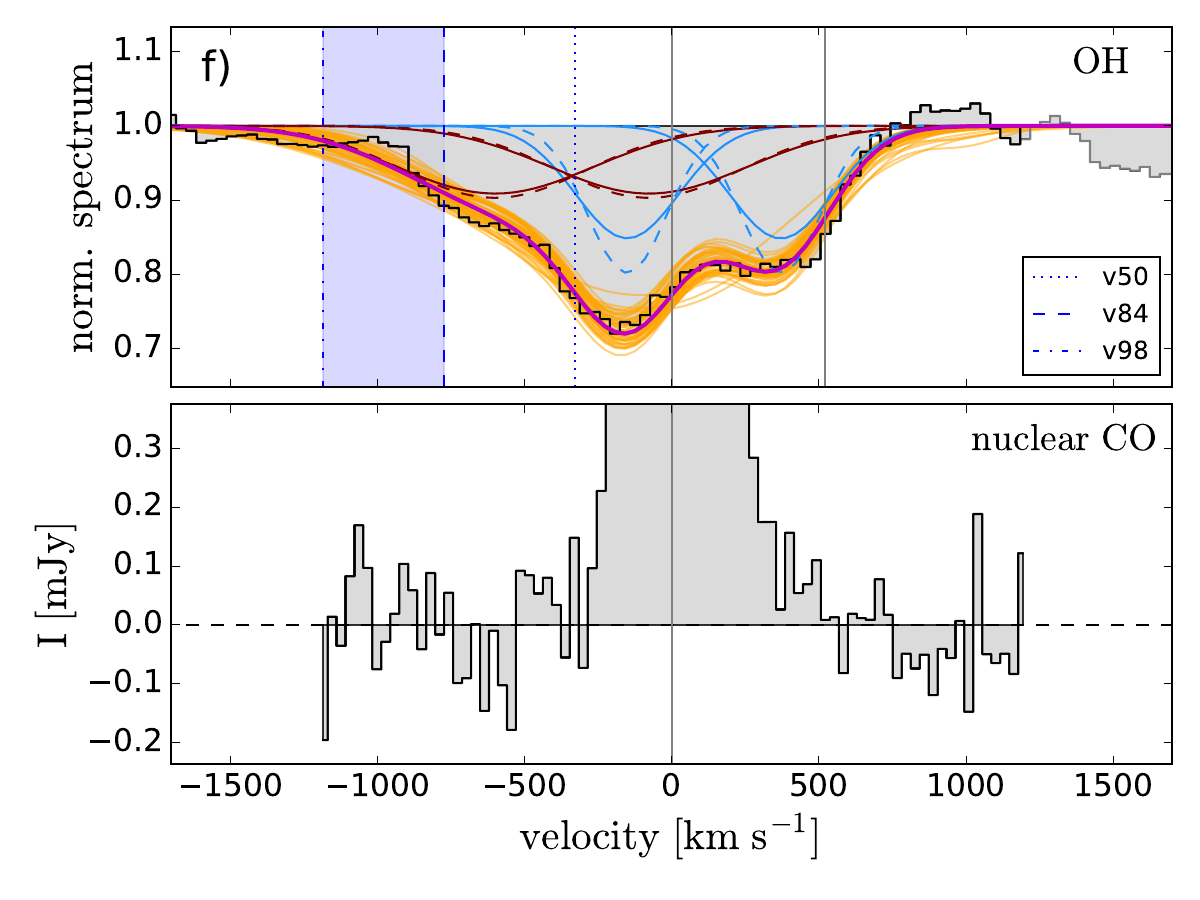}\\ 

\includegraphics[width=0.75\textwidth]{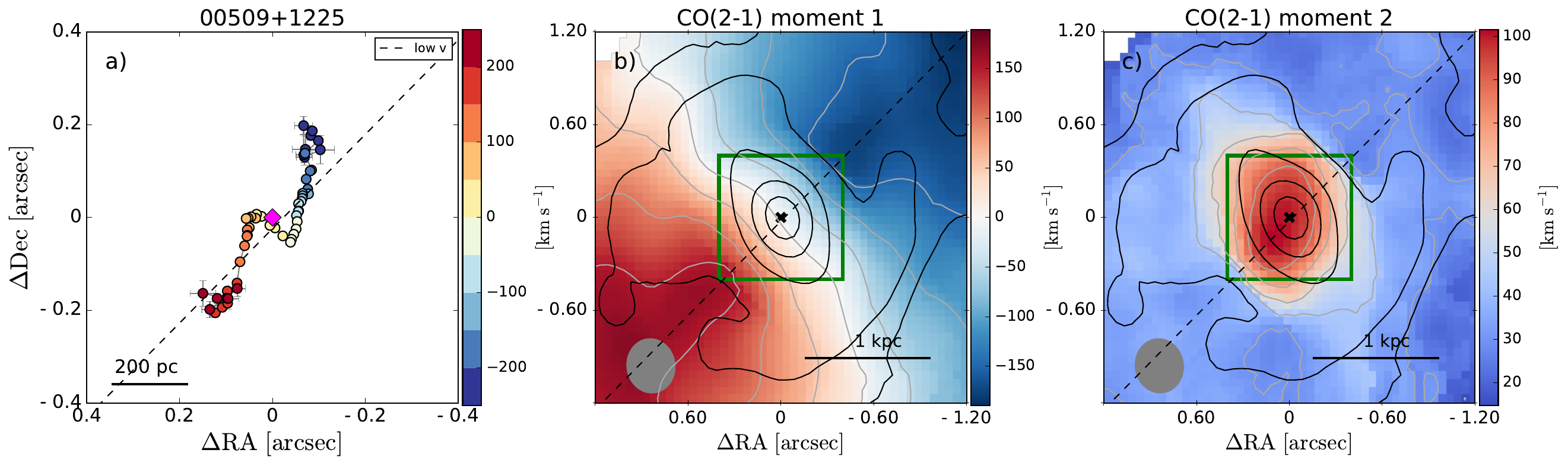}\\ 
\includegraphics[width=0.27\textwidth]{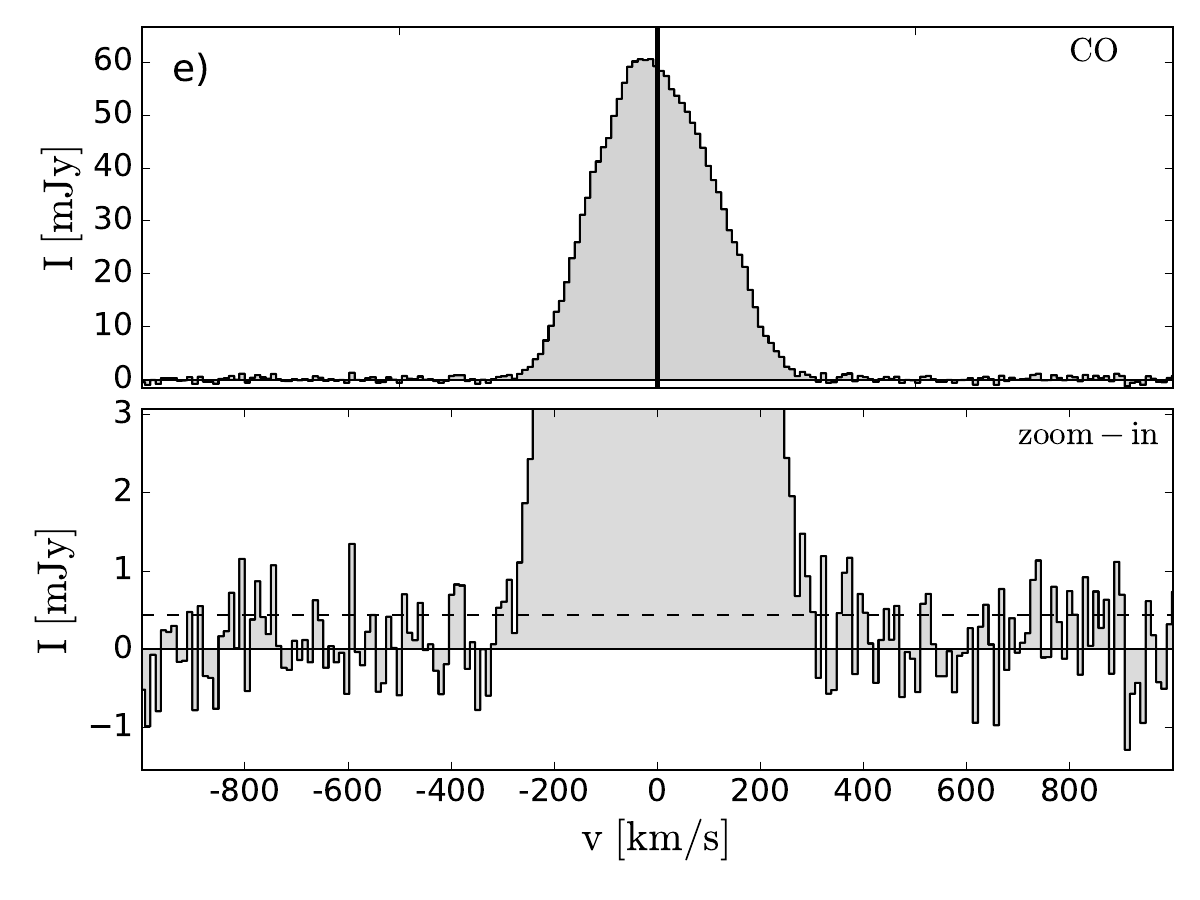} 
\includegraphics[width=0.27\textwidth]{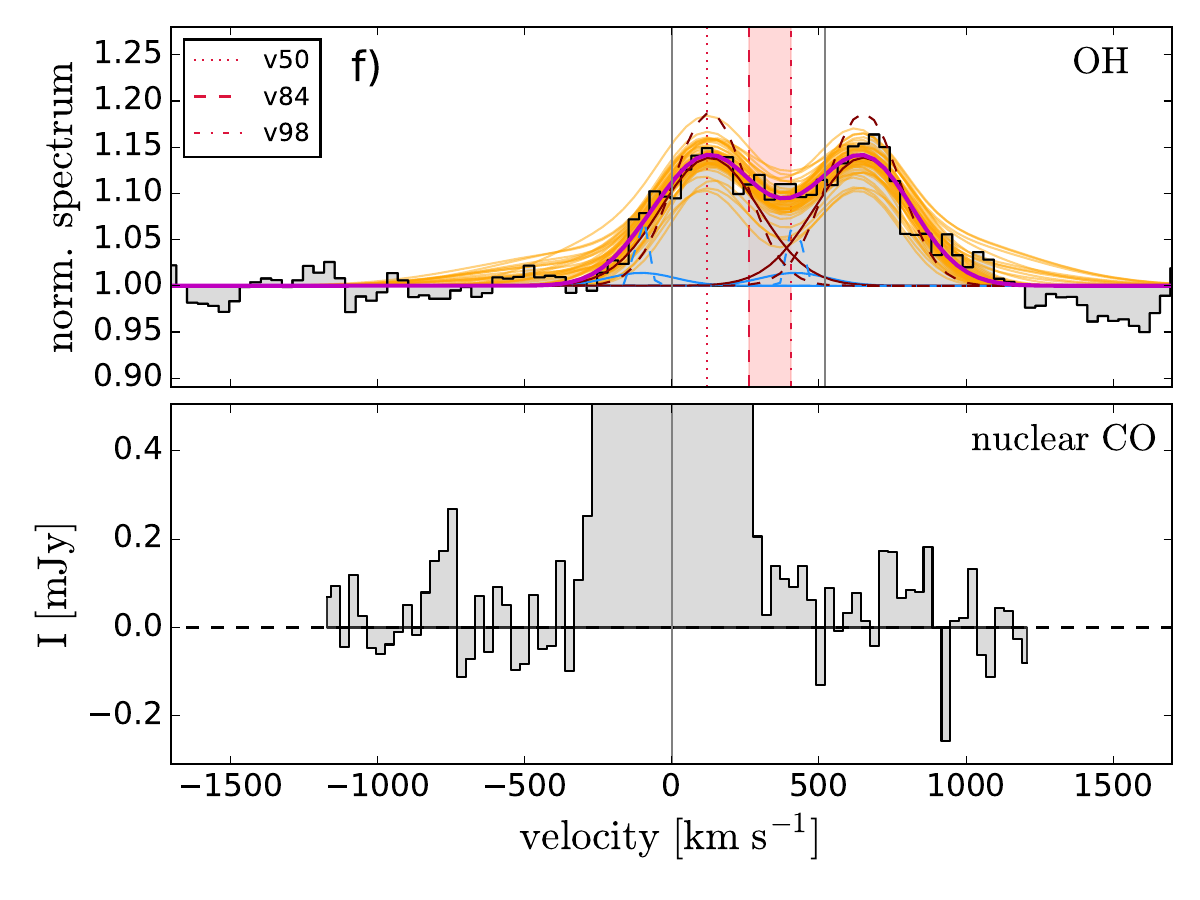}\\ 

\includegraphics[width=0.75\textwidth]{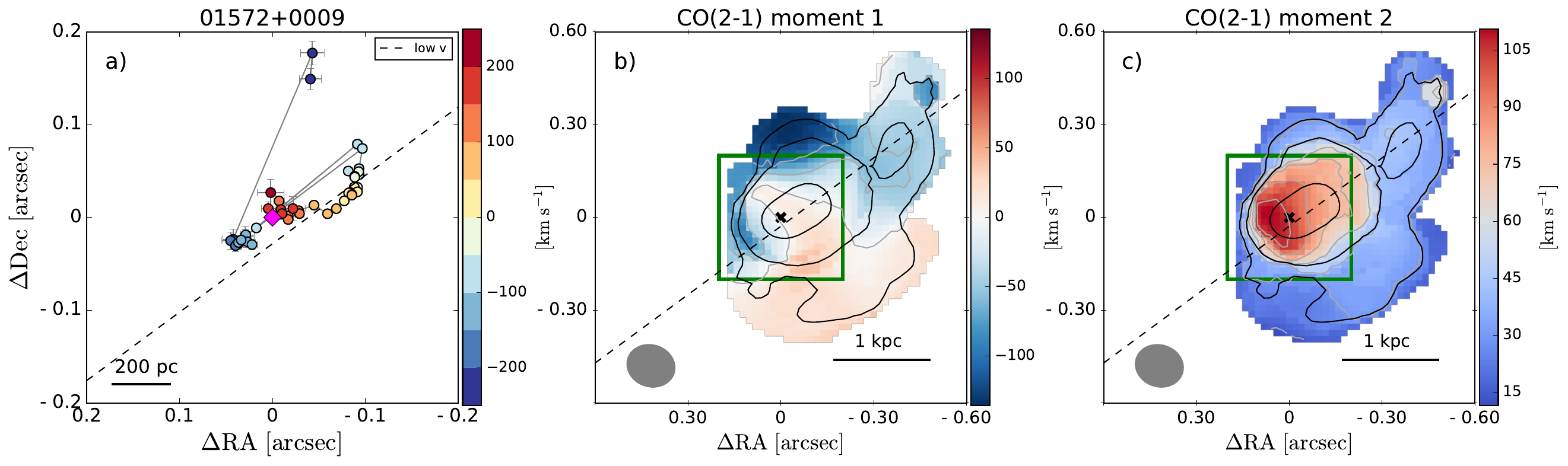}\\ 
\includegraphics[width=0.27\textwidth]{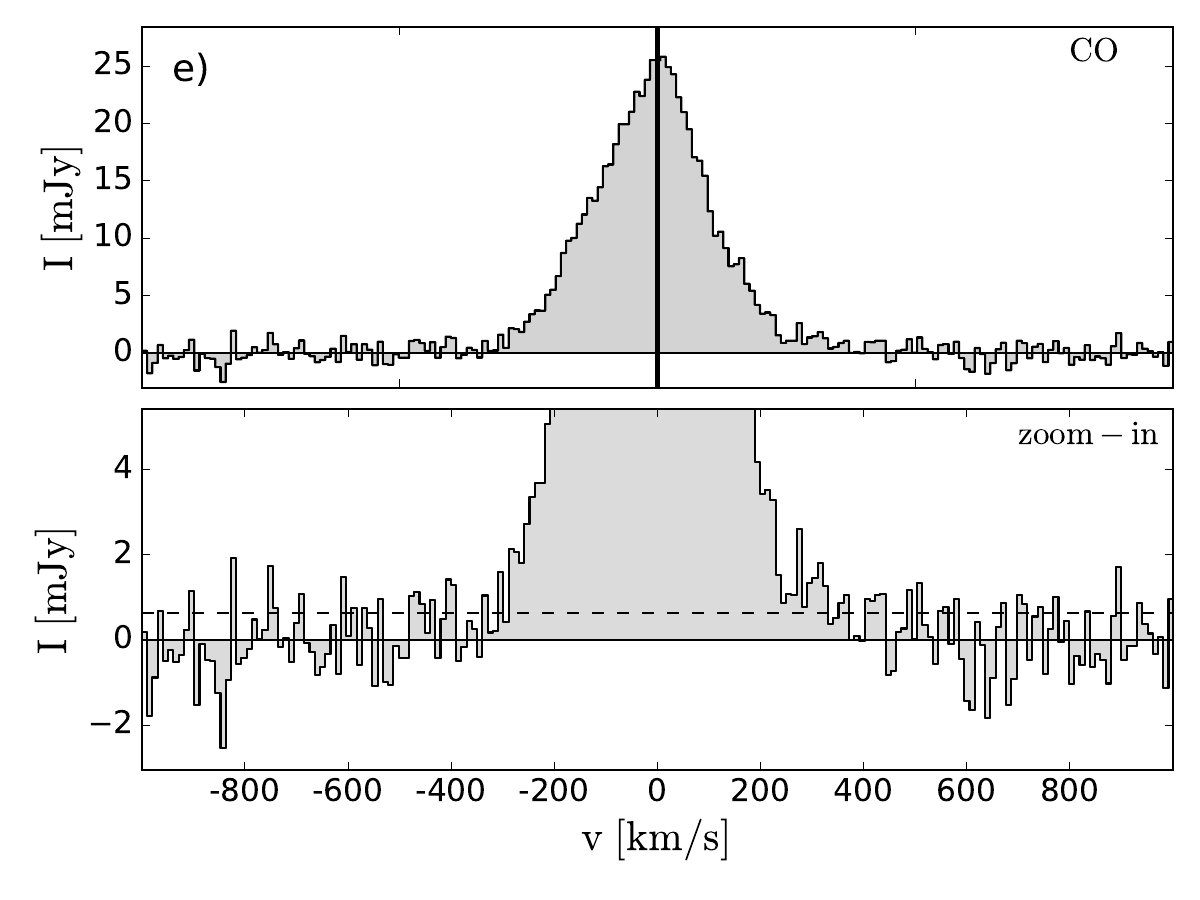} 
\includegraphics[width=0.27\textwidth]{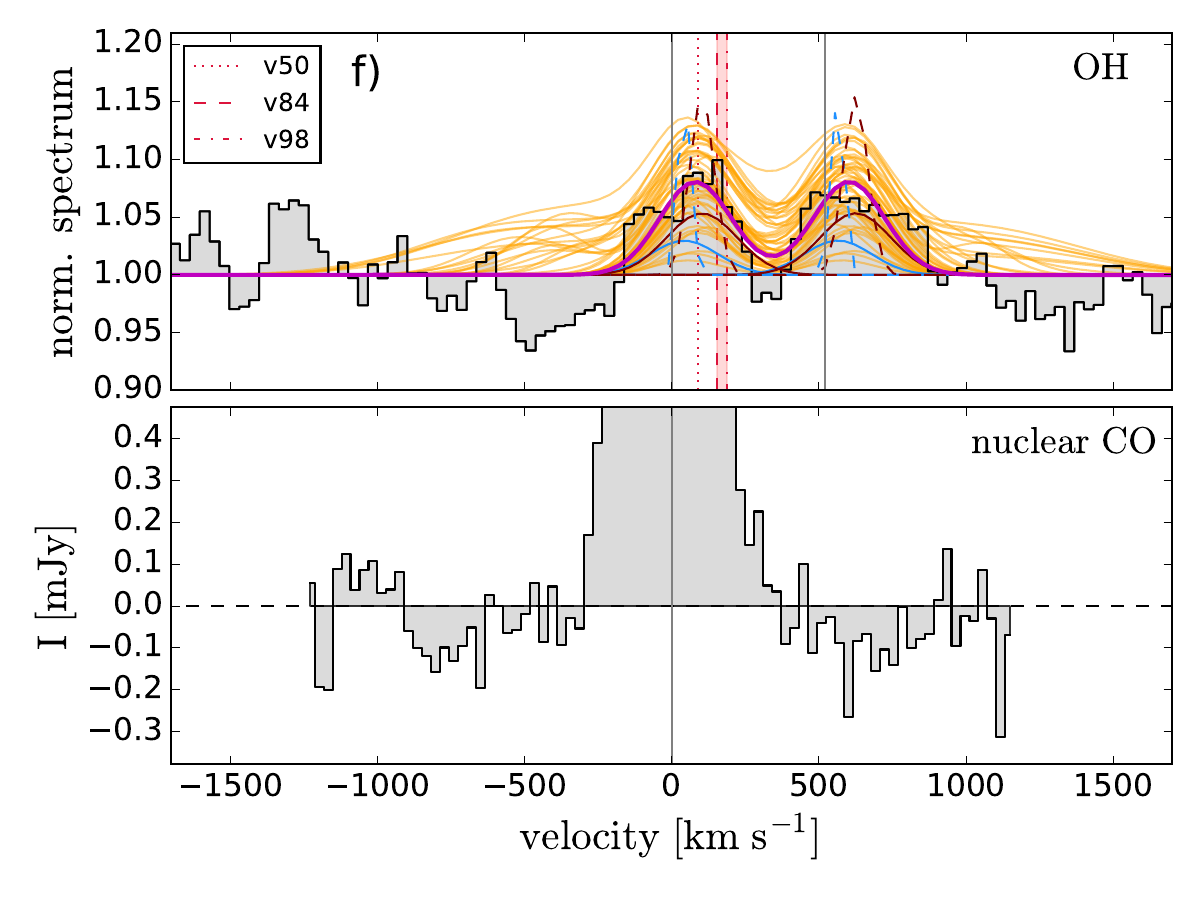}\\

\caption{continued.} 
 \end{figure*}

\begin{figure*}\ContinuedFloat 
\centering 
\includegraphics[width=0.75\textwidth]{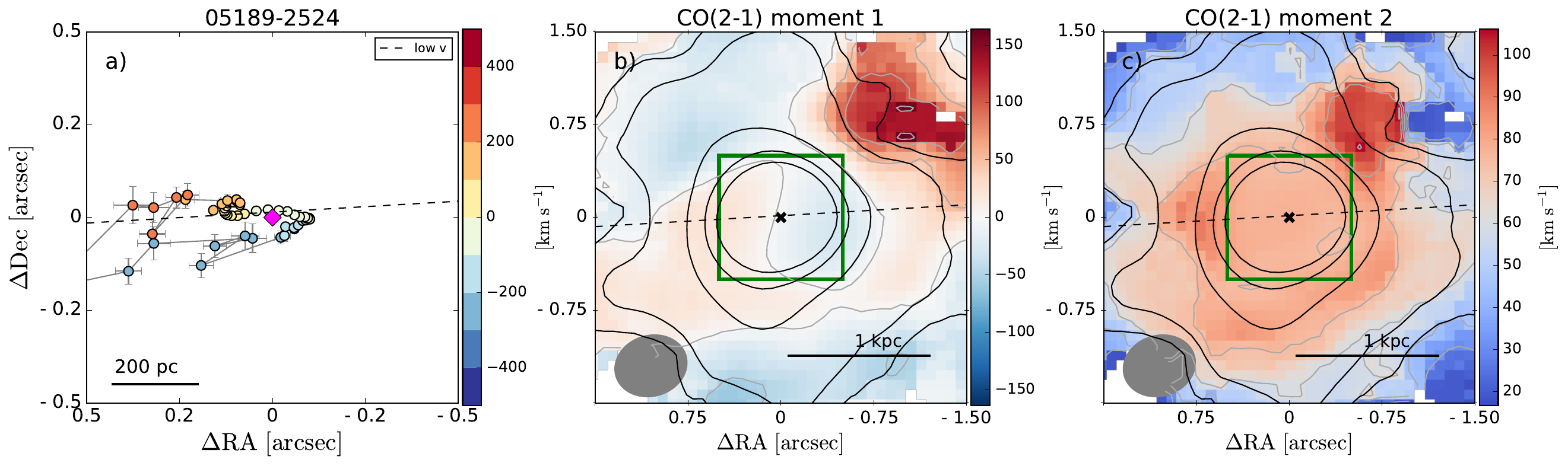}\\ 
\includegraphics[width=0.23\textwidth]{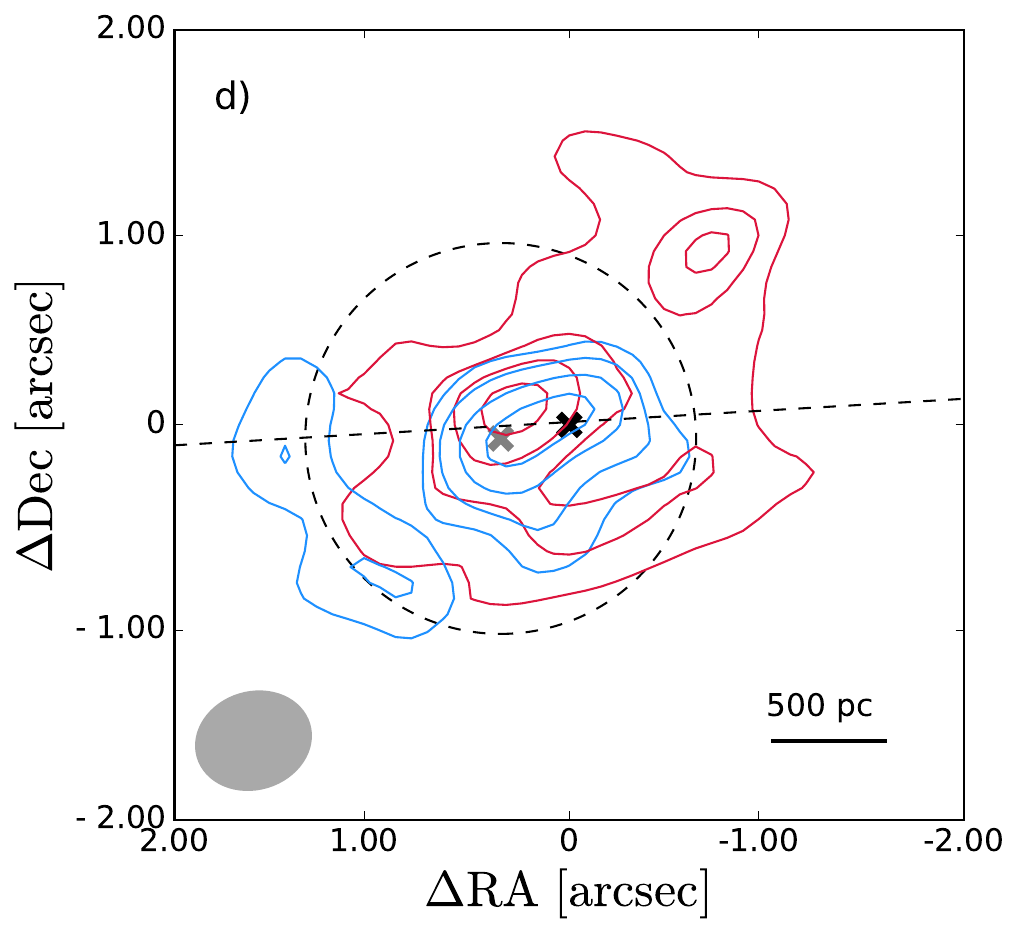} 
\includegraphics[width=0.27\textwidth]{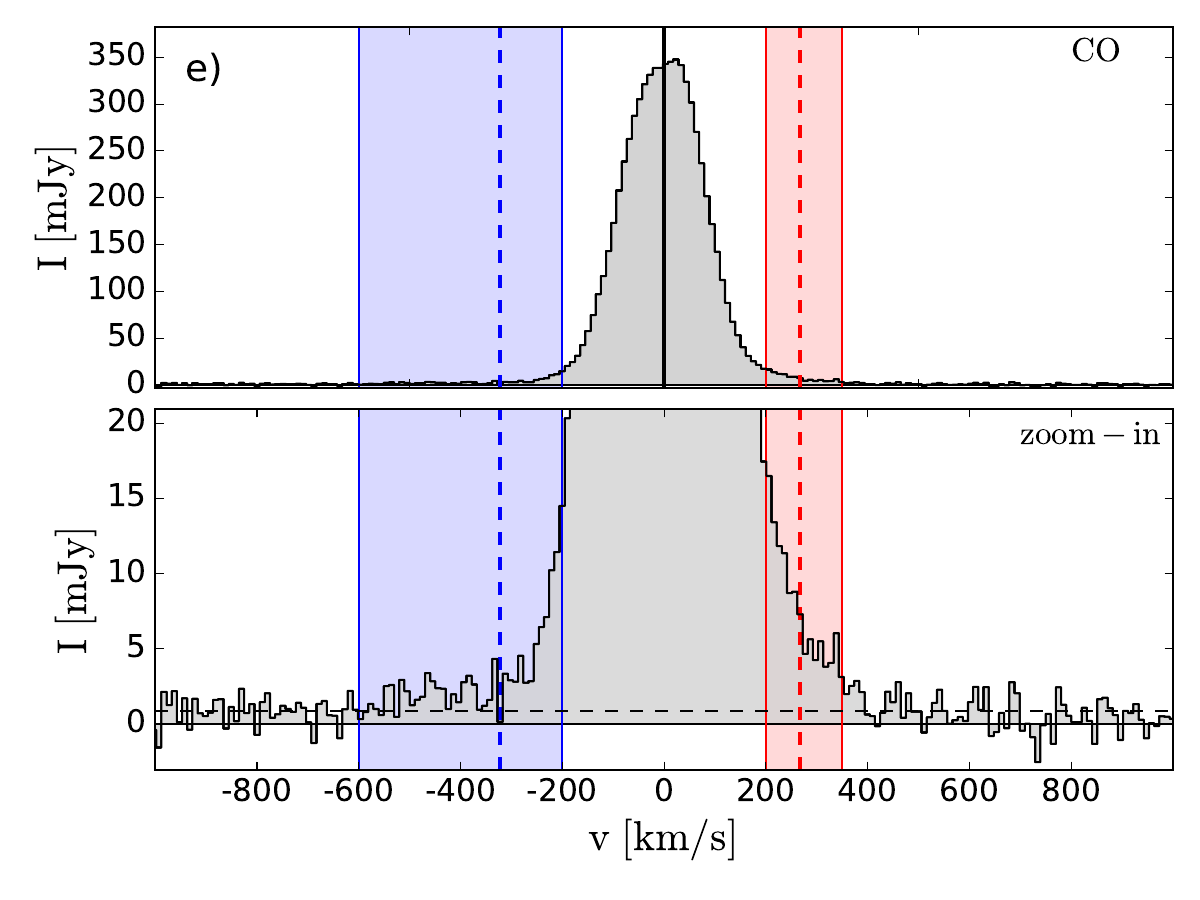} 
\includegraphics[width=0.27\textwidth]{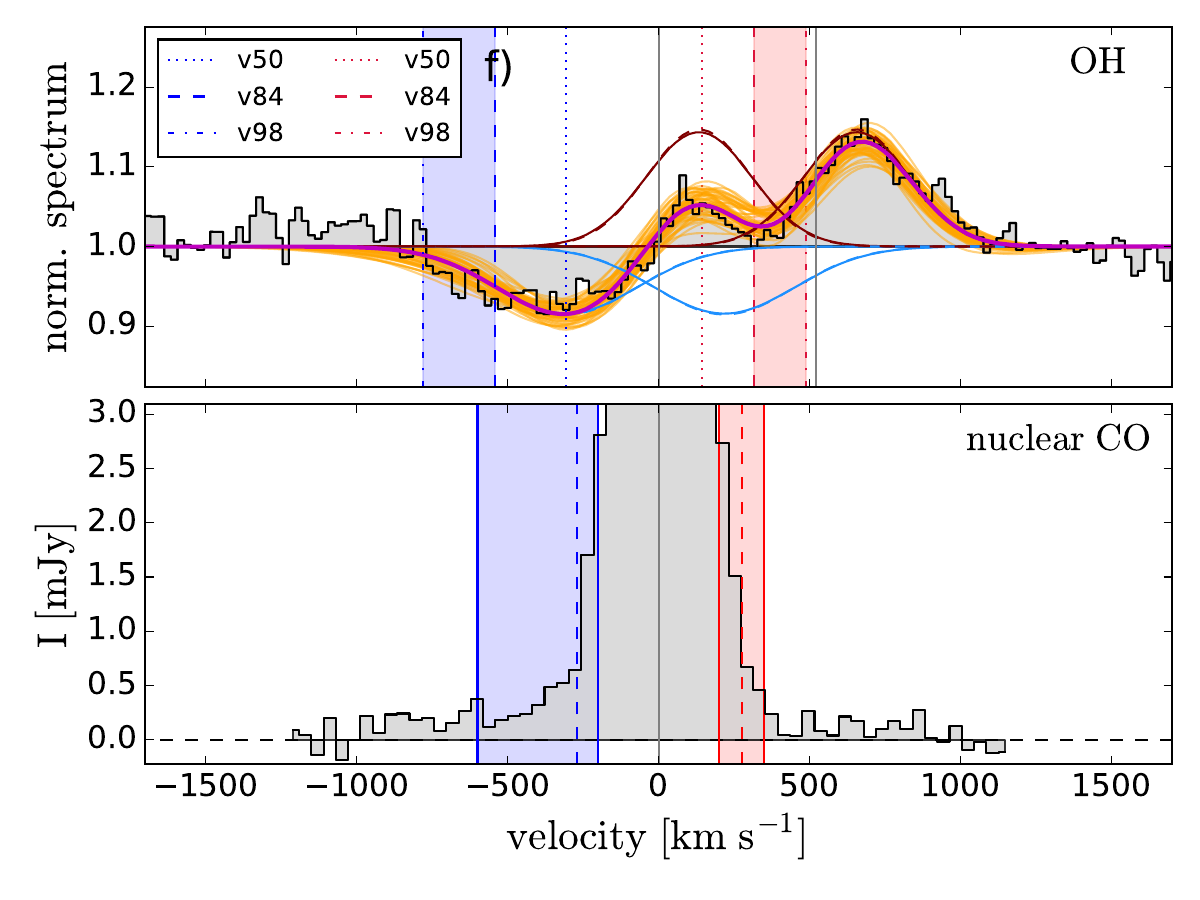}\\ 

\includegraphics[width=0.75\textwidth]{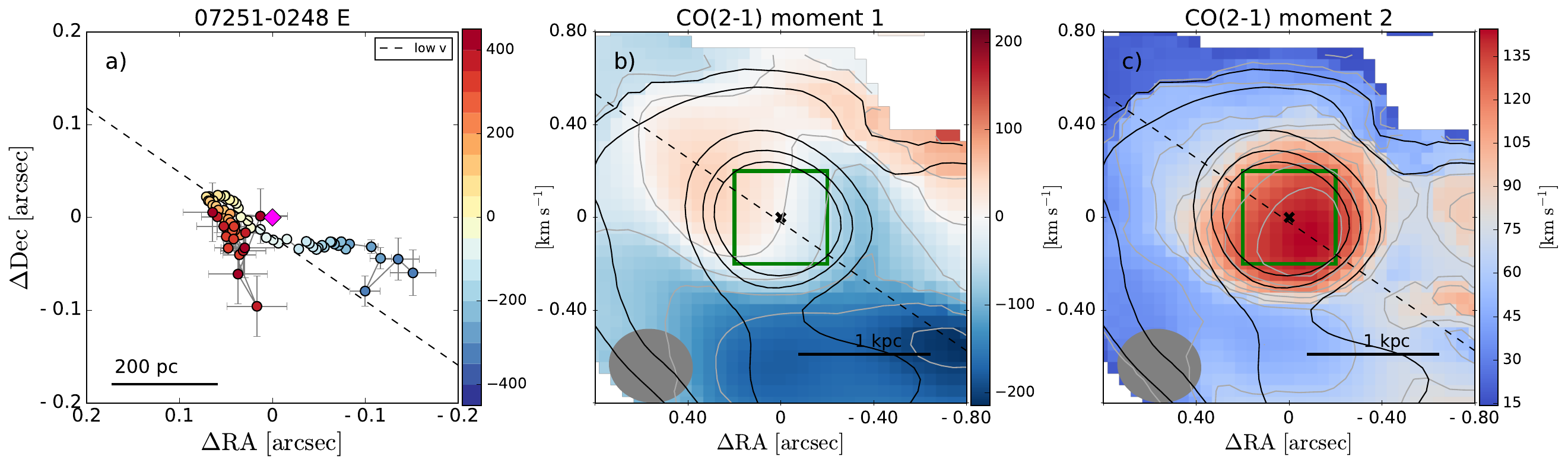}\\ 
\includegraphics[width=0.23\textwidth]{Figures/Outflow_maps/IRAS_07251-0248_b_r_channels_radius_fit.pdf} 
\includegraphics[width=0.27\textwidth]{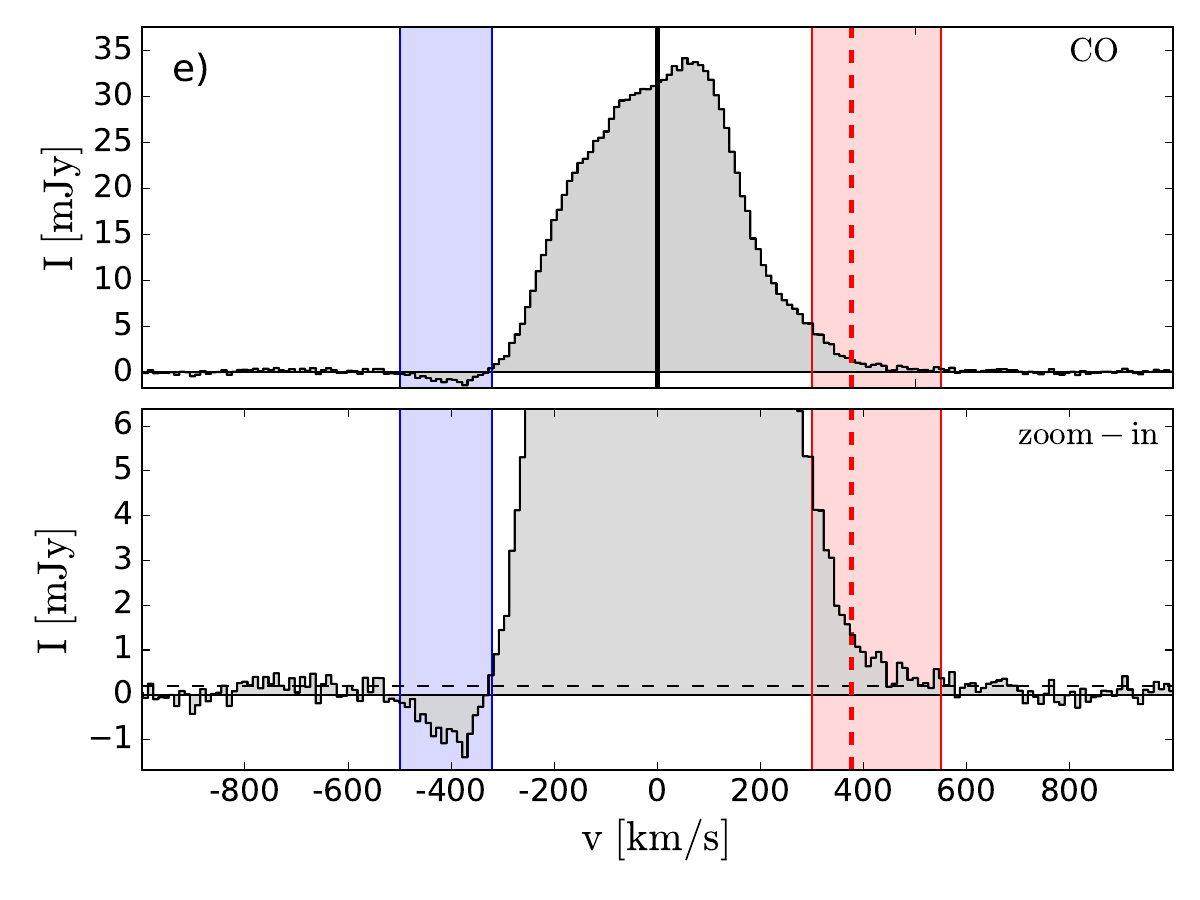} 
\includegraphics[width=0.27\textwidth]{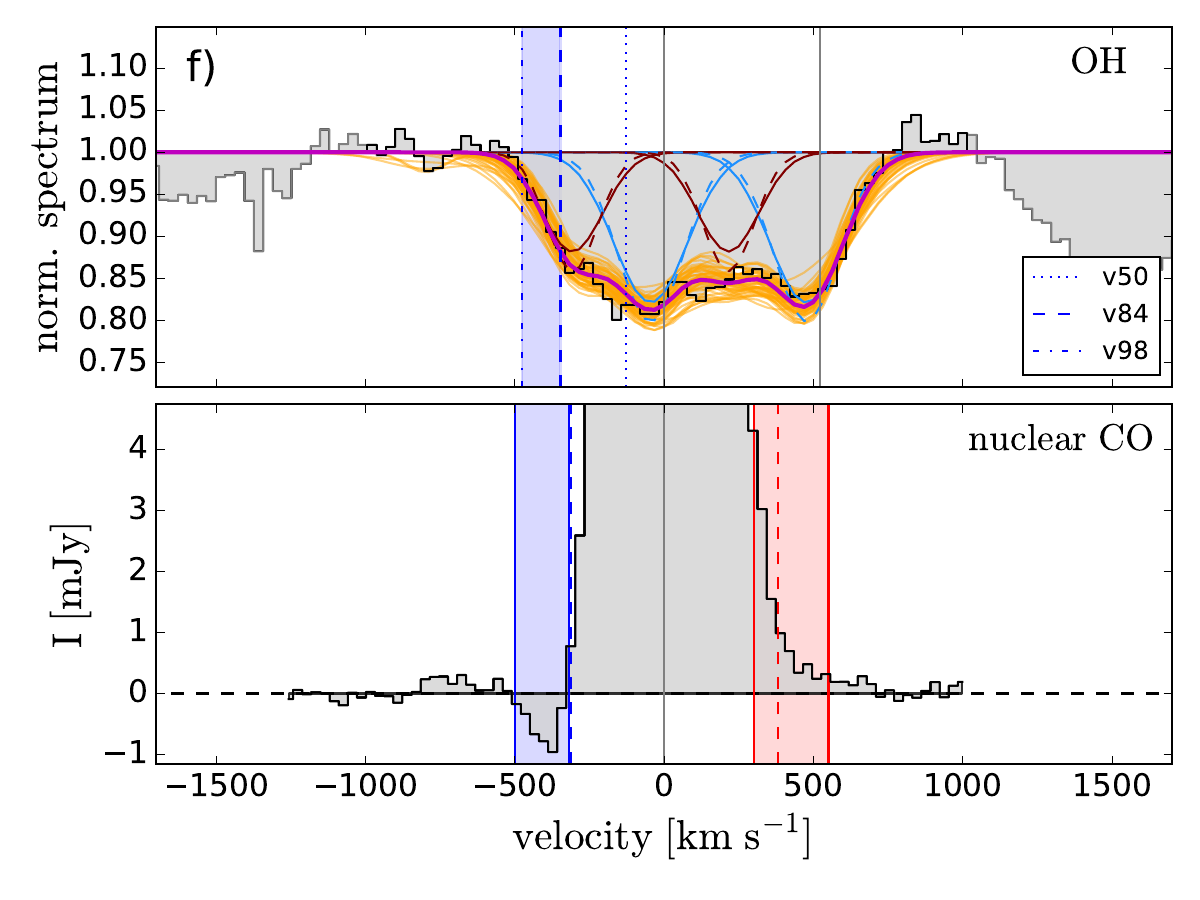}\\ 

\includegraphics[width=0.75\textwidth]{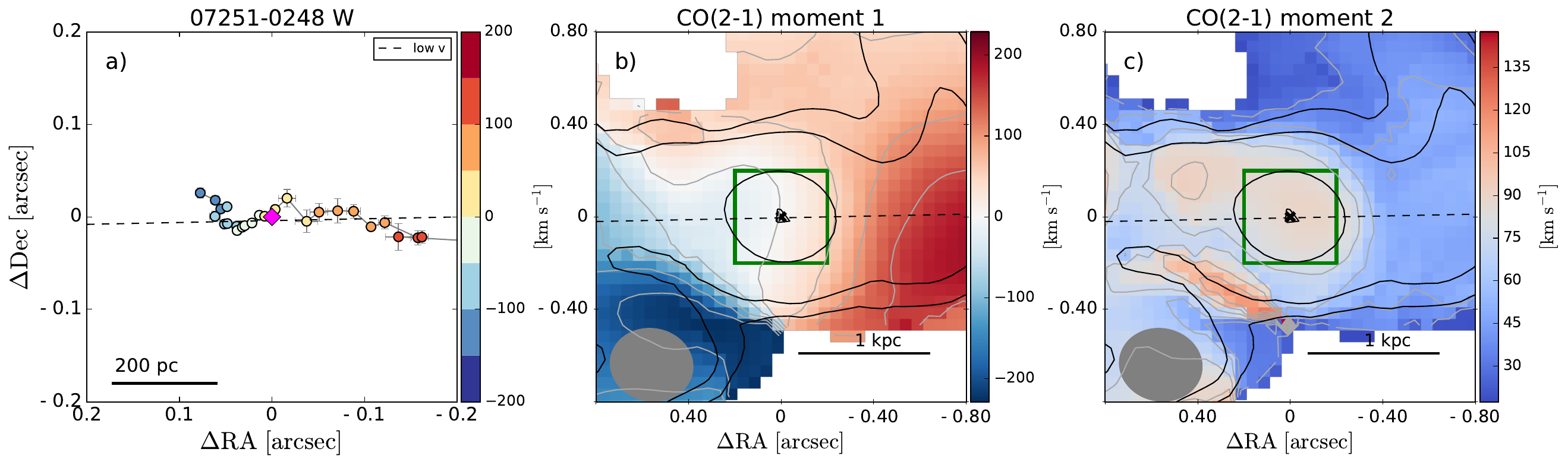}\\ 
\includegraphics[width=0.27\textwidth]{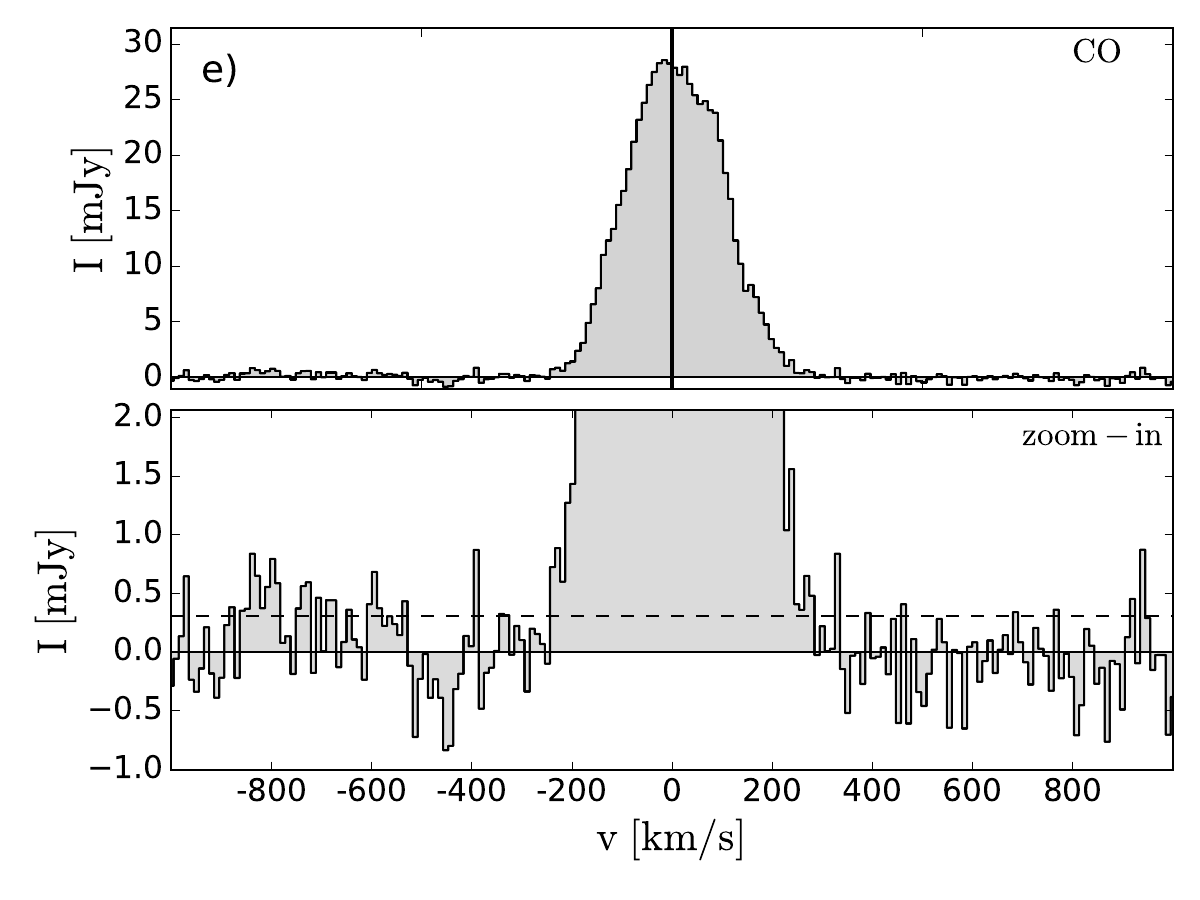} 
\includegraphics[width=0.27\textwidth]{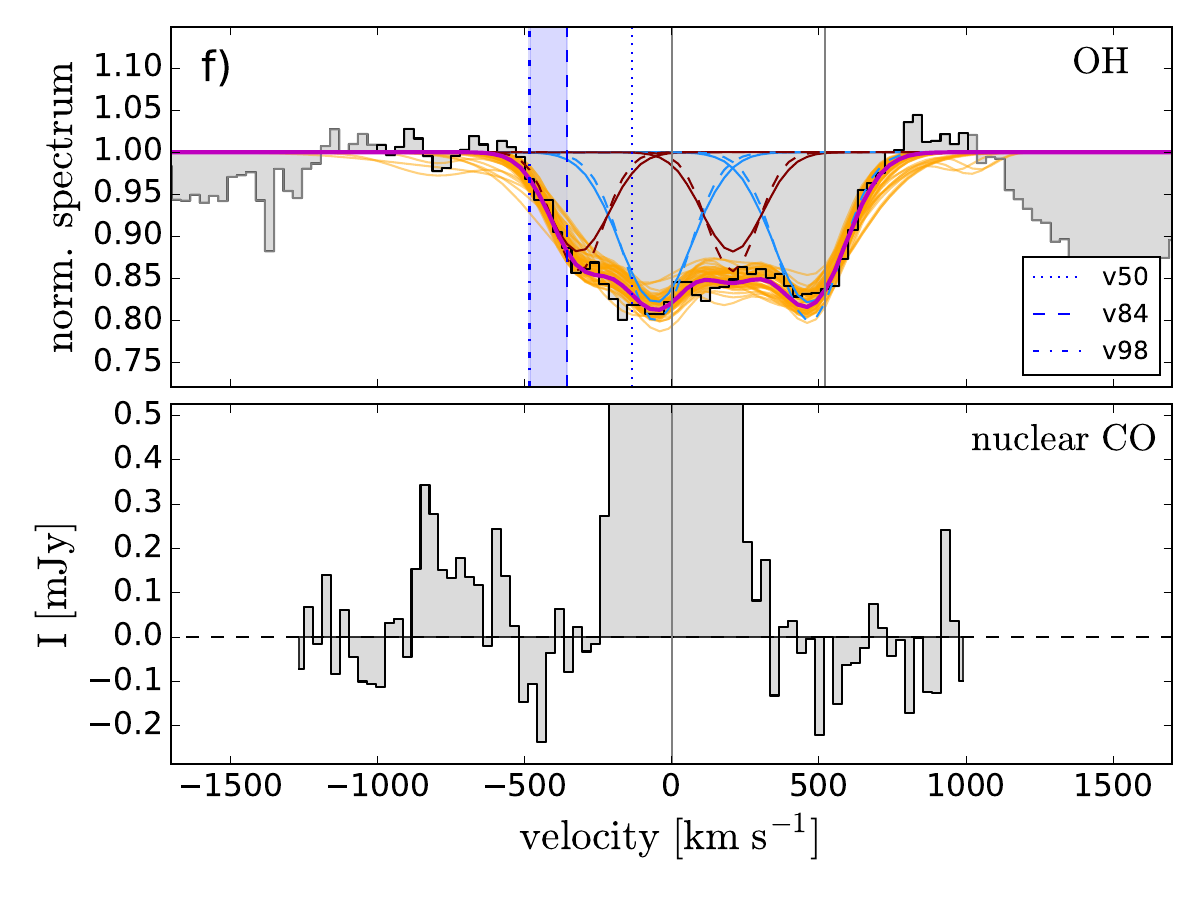}\\

\caption{continued.} 
 \end{figure*}

\begin{figure*}\ContinuedFloat 
\centering 
\includegraphics[width=0.75\textwidth]{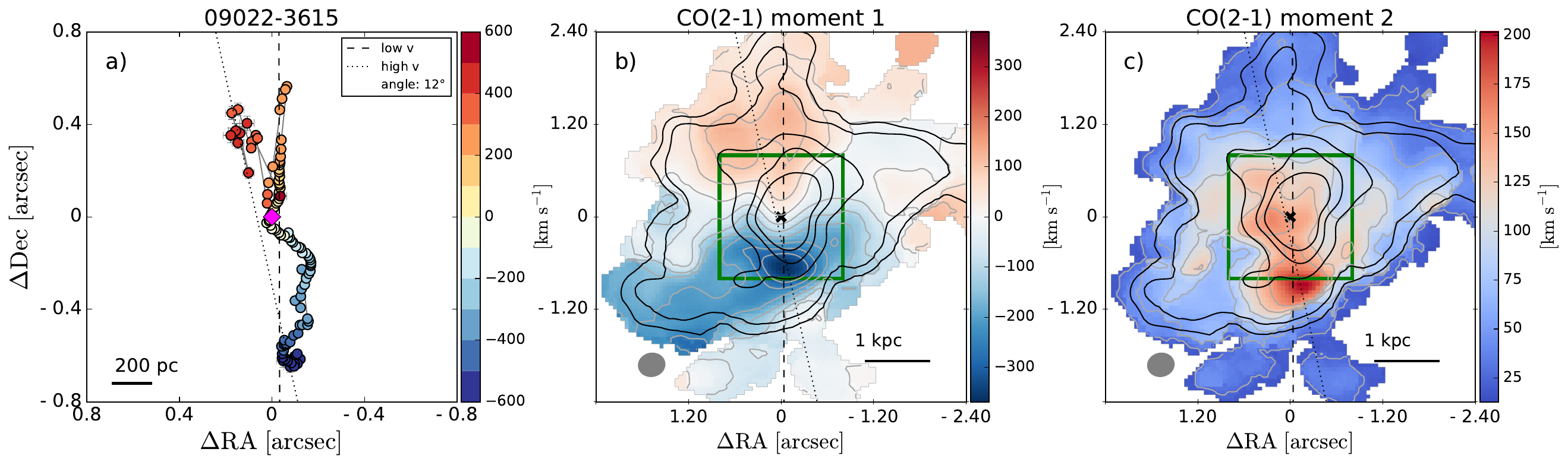}\\ 
\includegraphics[width=0.23\textwidth]{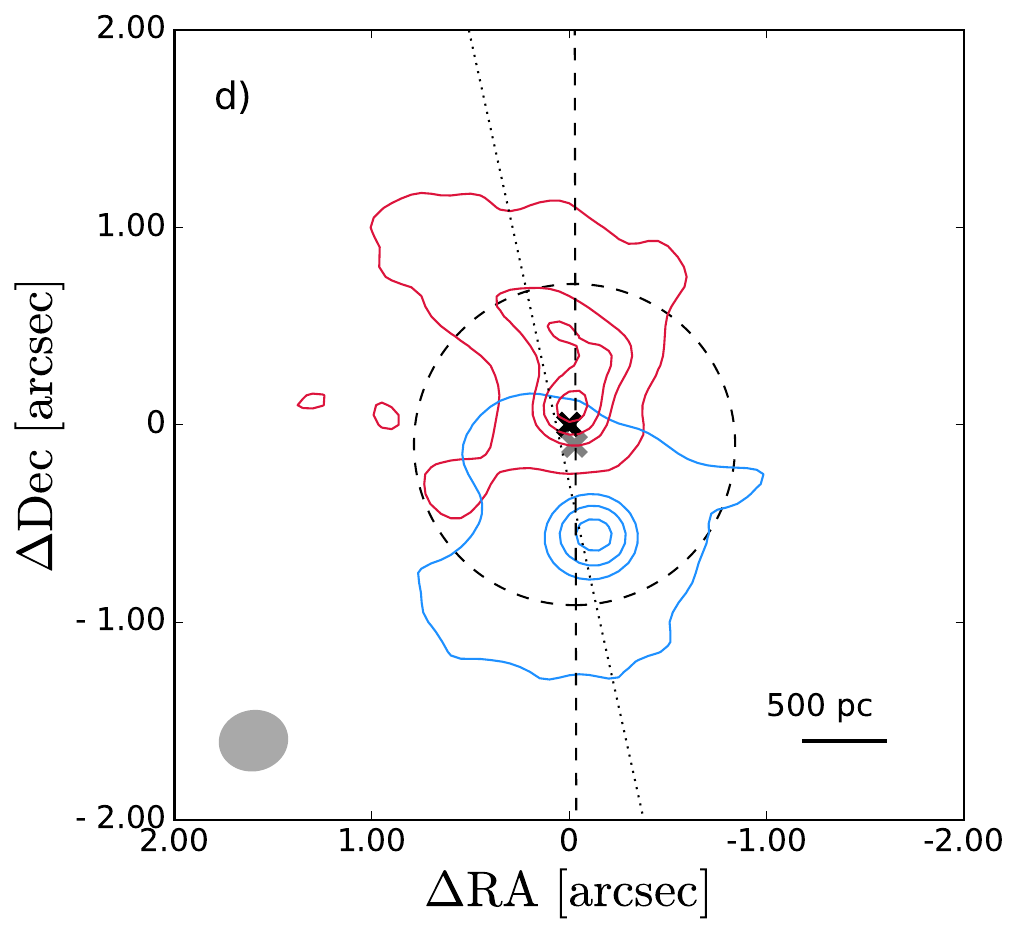} 
\includegraphics[width=0.27\textwidth]{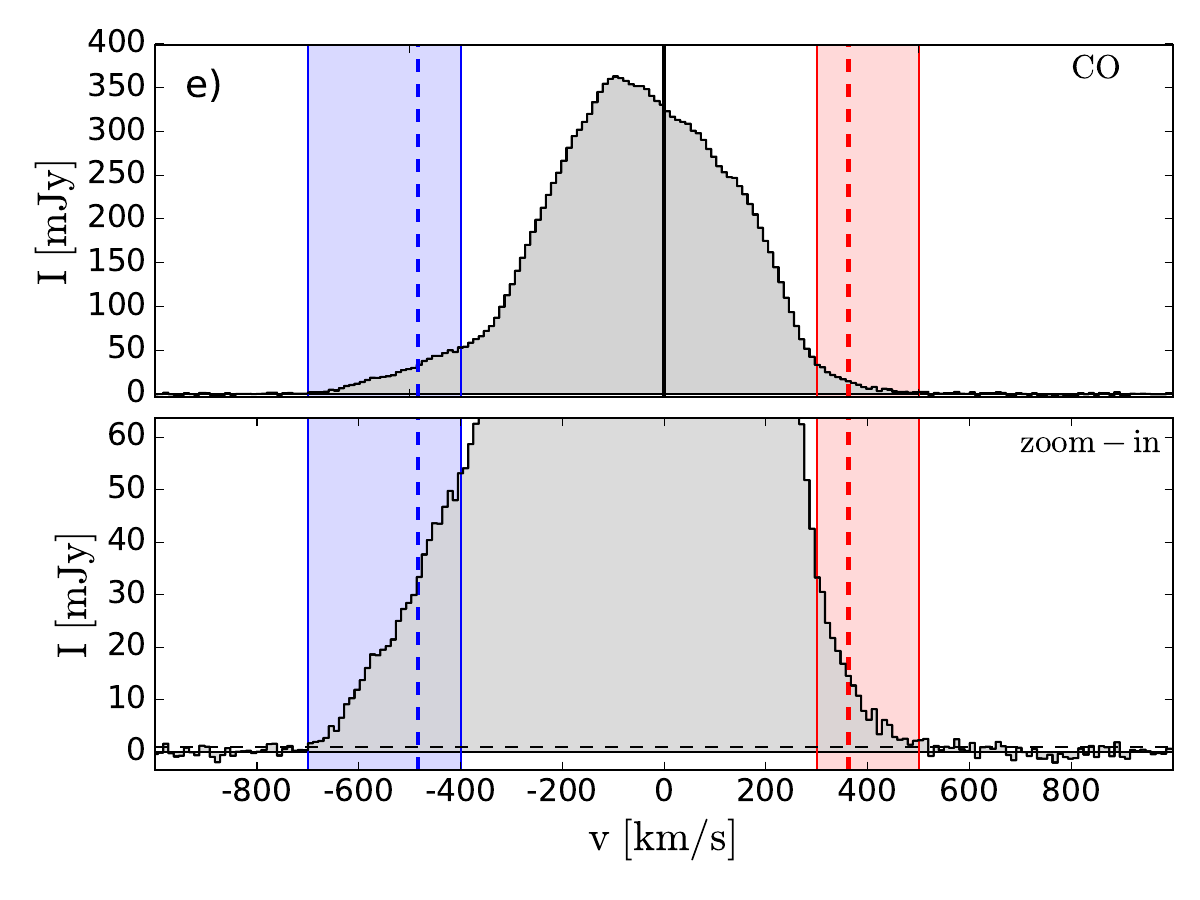} 
\includegraphics[width=0.27\textwidth]{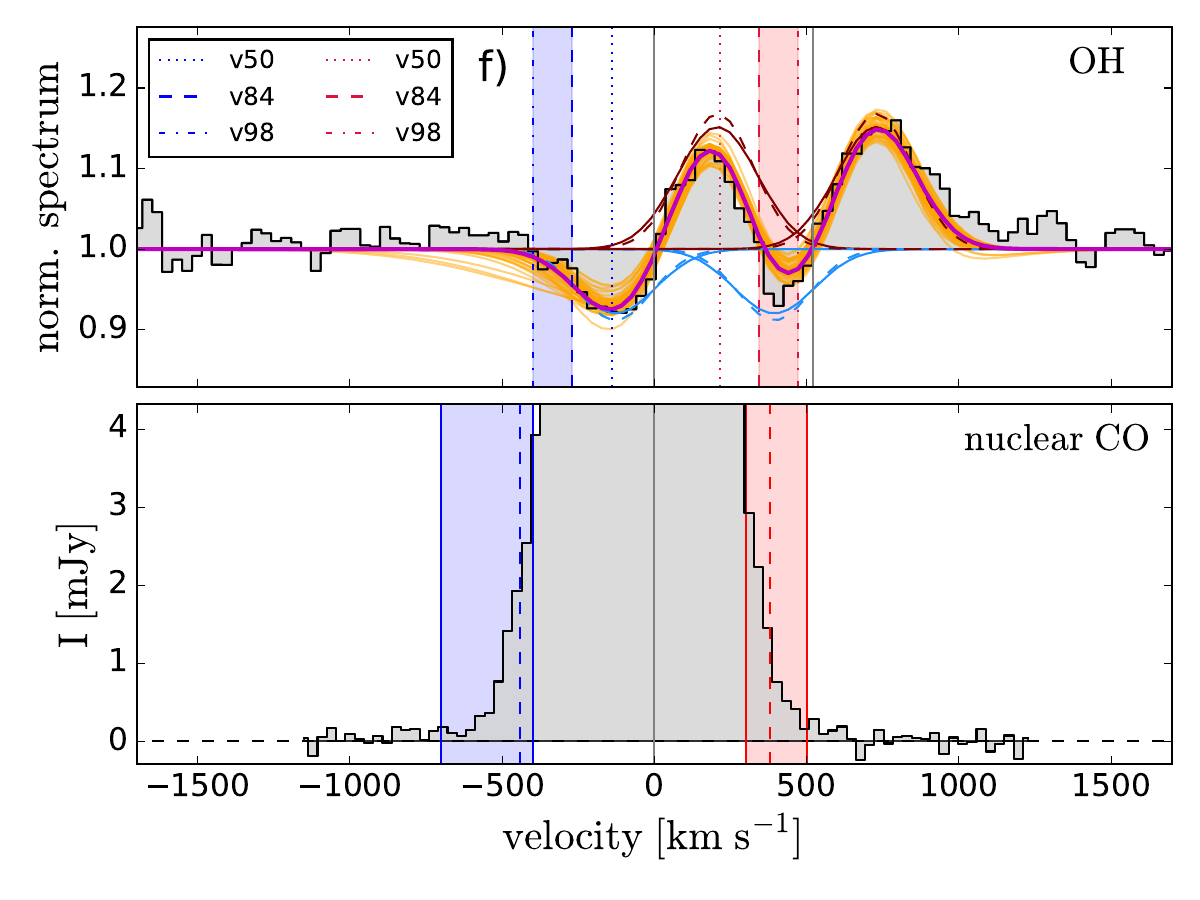}\\ 

\includegraphics[width=0.75\textwidth]{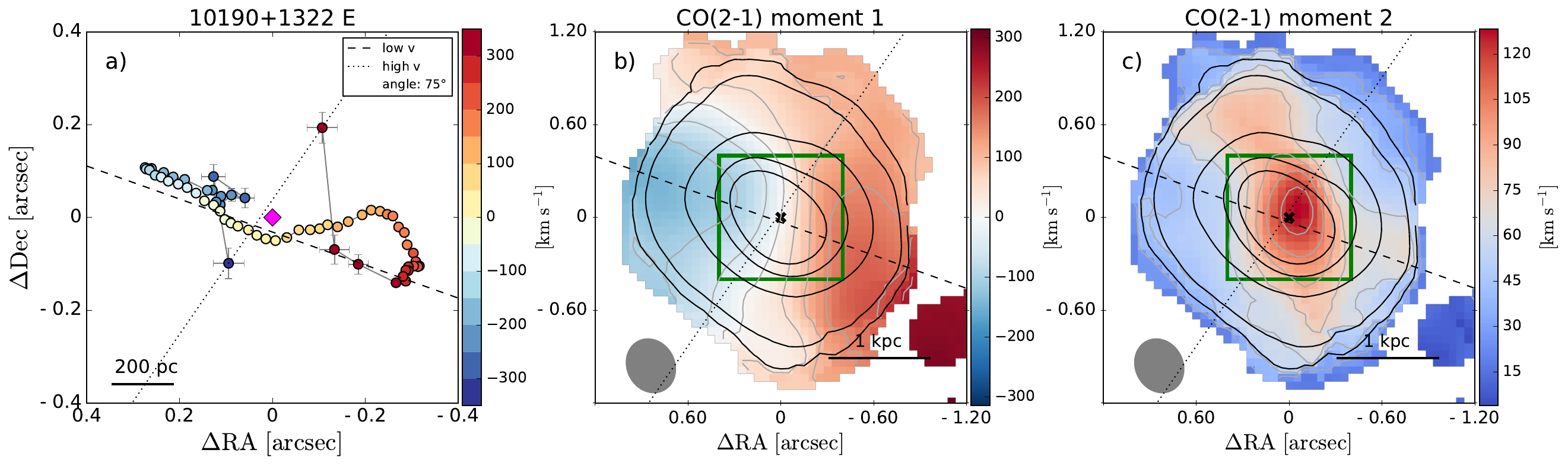}\\ 
\includegraphics[width=0.23\textwidth]{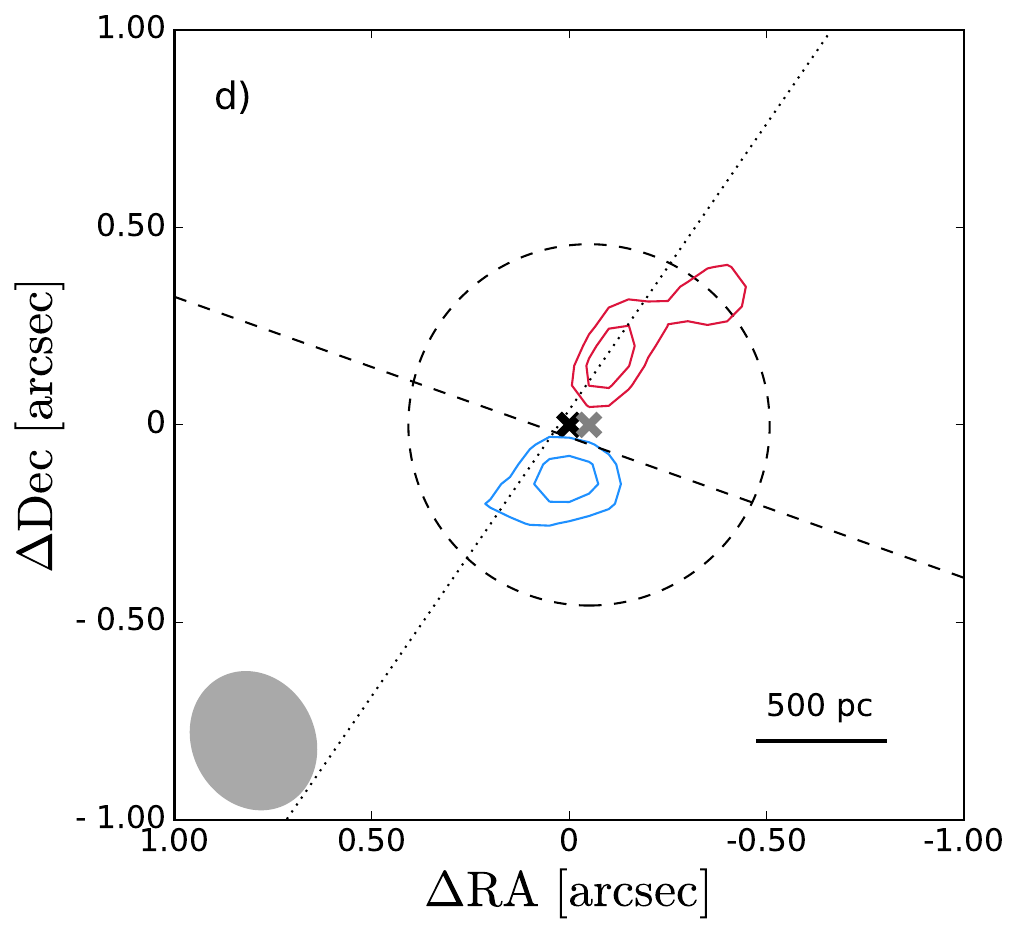} 
\includegraphics[width=0.27\textwidth]{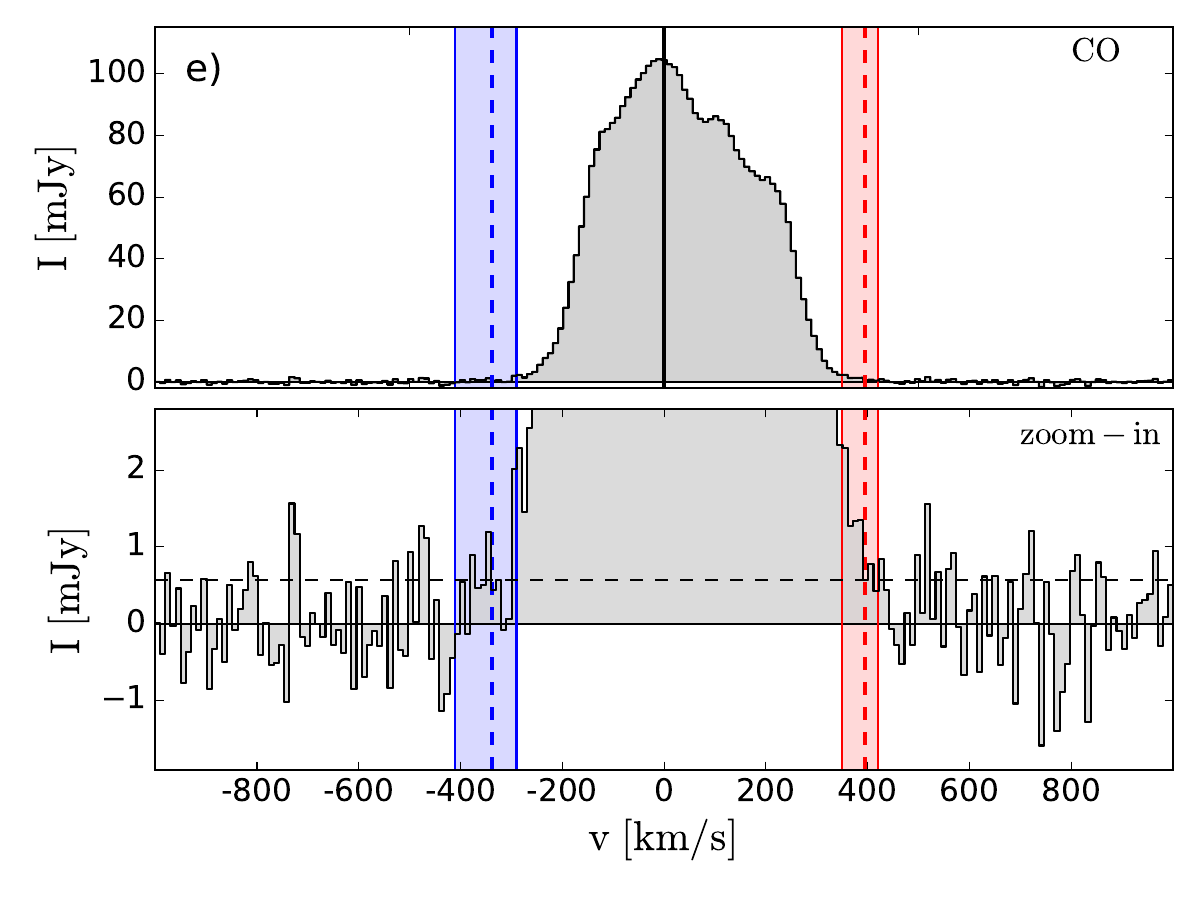} \\ 

\includegraphics[width=0.75\textwidth]{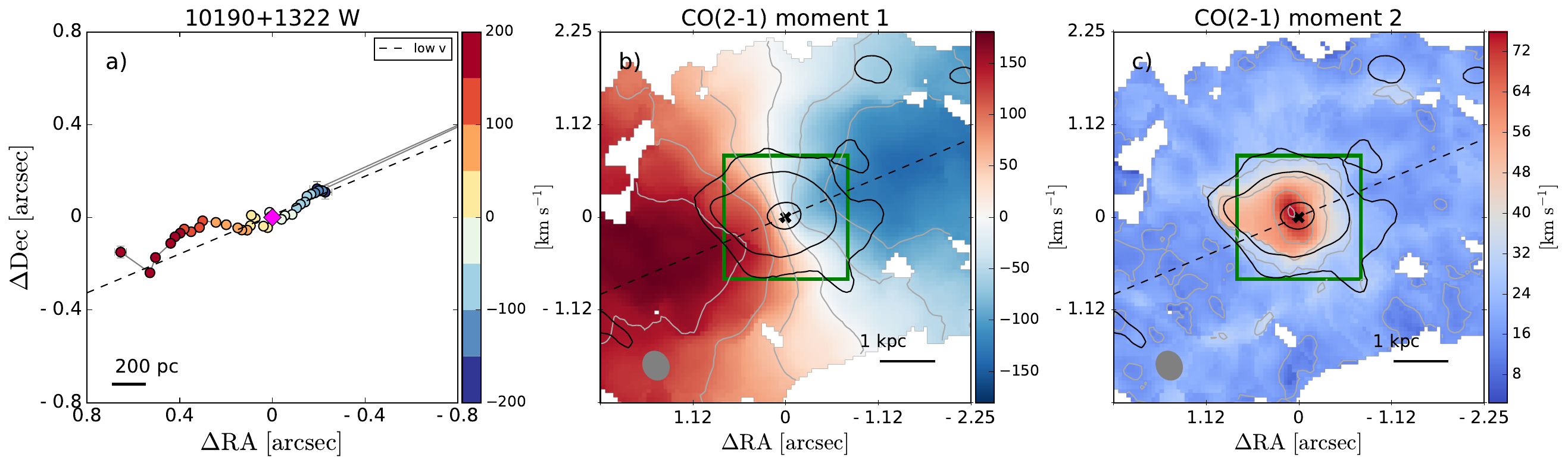}\\ 
\includegraphics[width=0.27\textwidth]{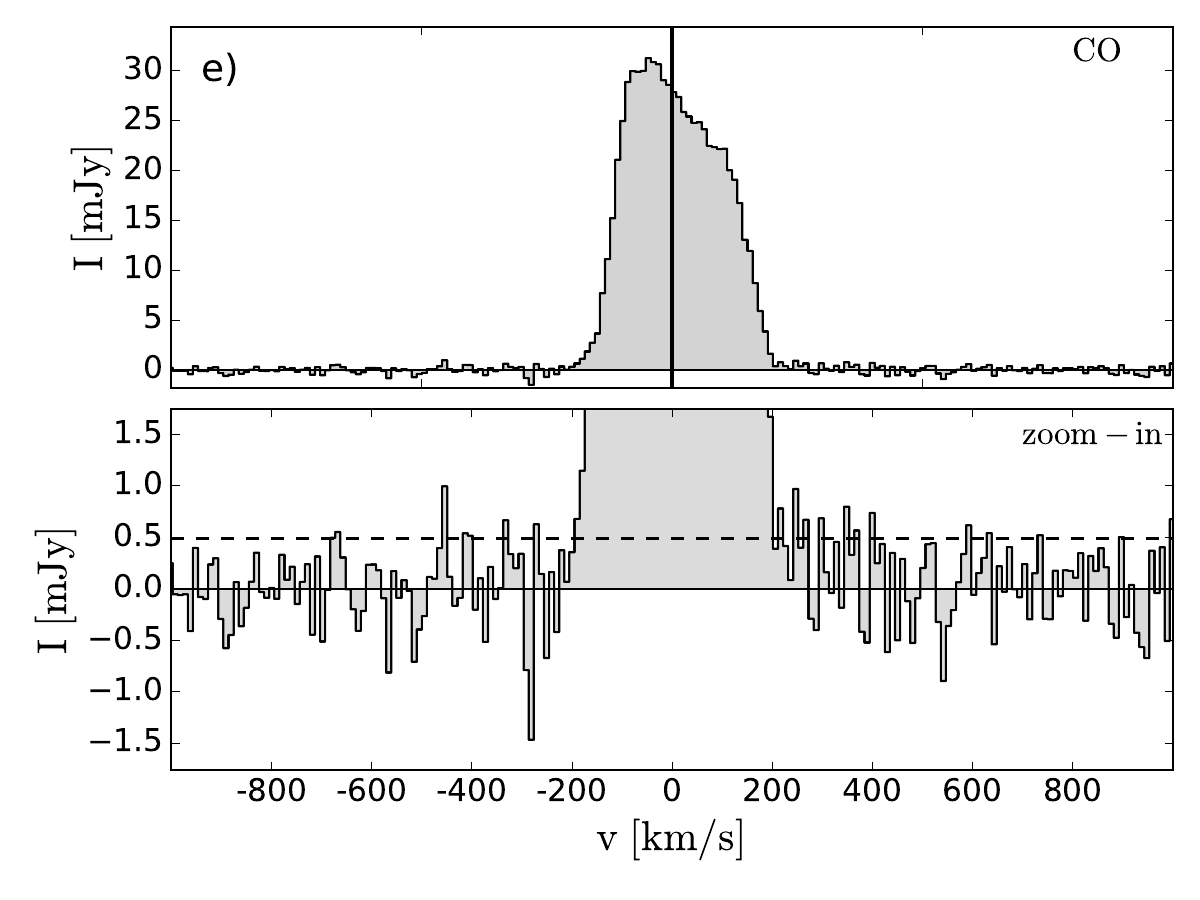} \\

\caption{continued.} 
 \end{figure*}

\begin{figure*}\ContinuedFloat 
\centering 
\includegraphics[width=0.75\textwidth]{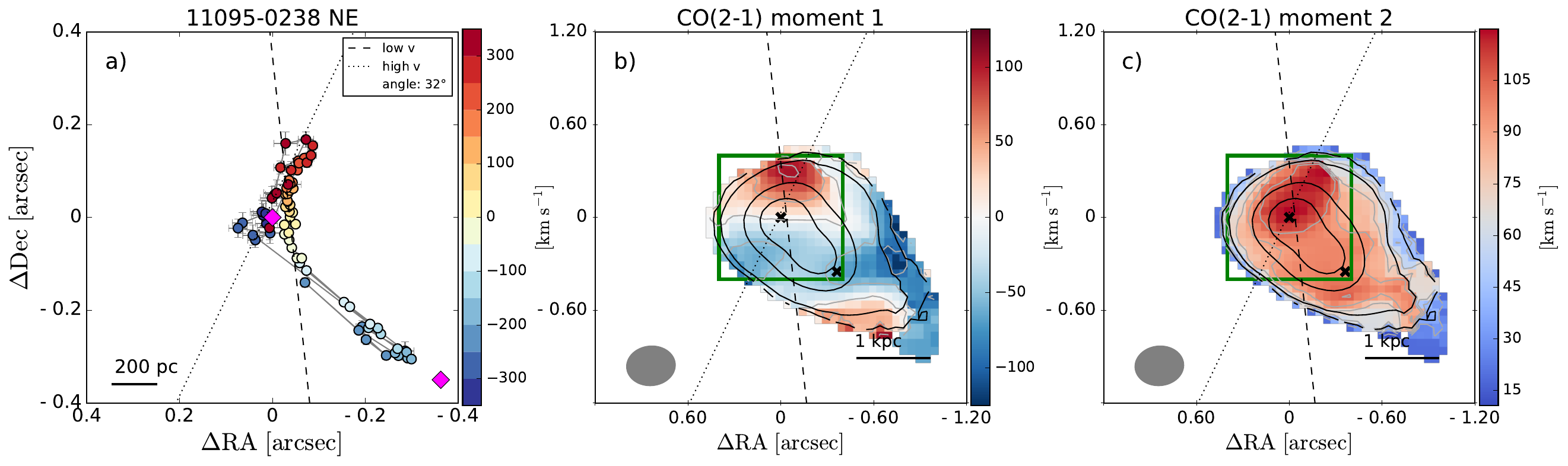}\\ 
\includegraphics[width=0.23\textwidth]{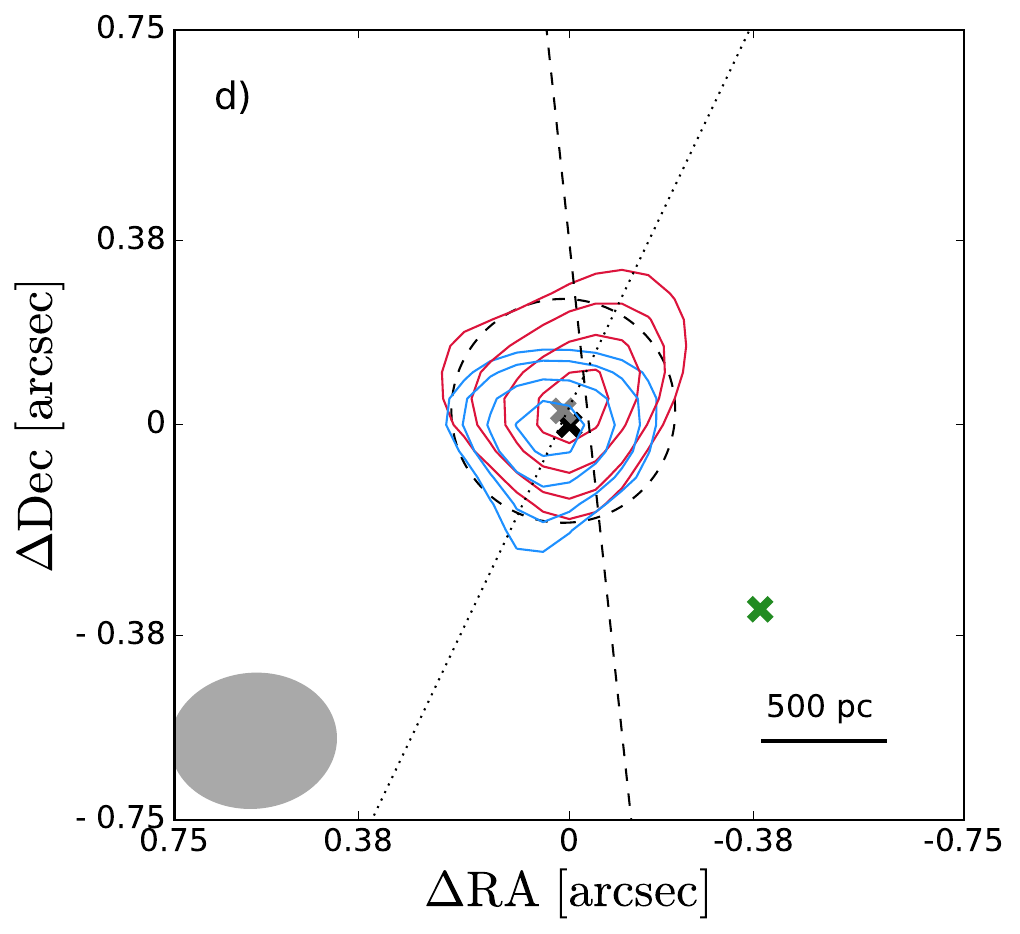} 
\includegraphics[width=0.27\textwidth]{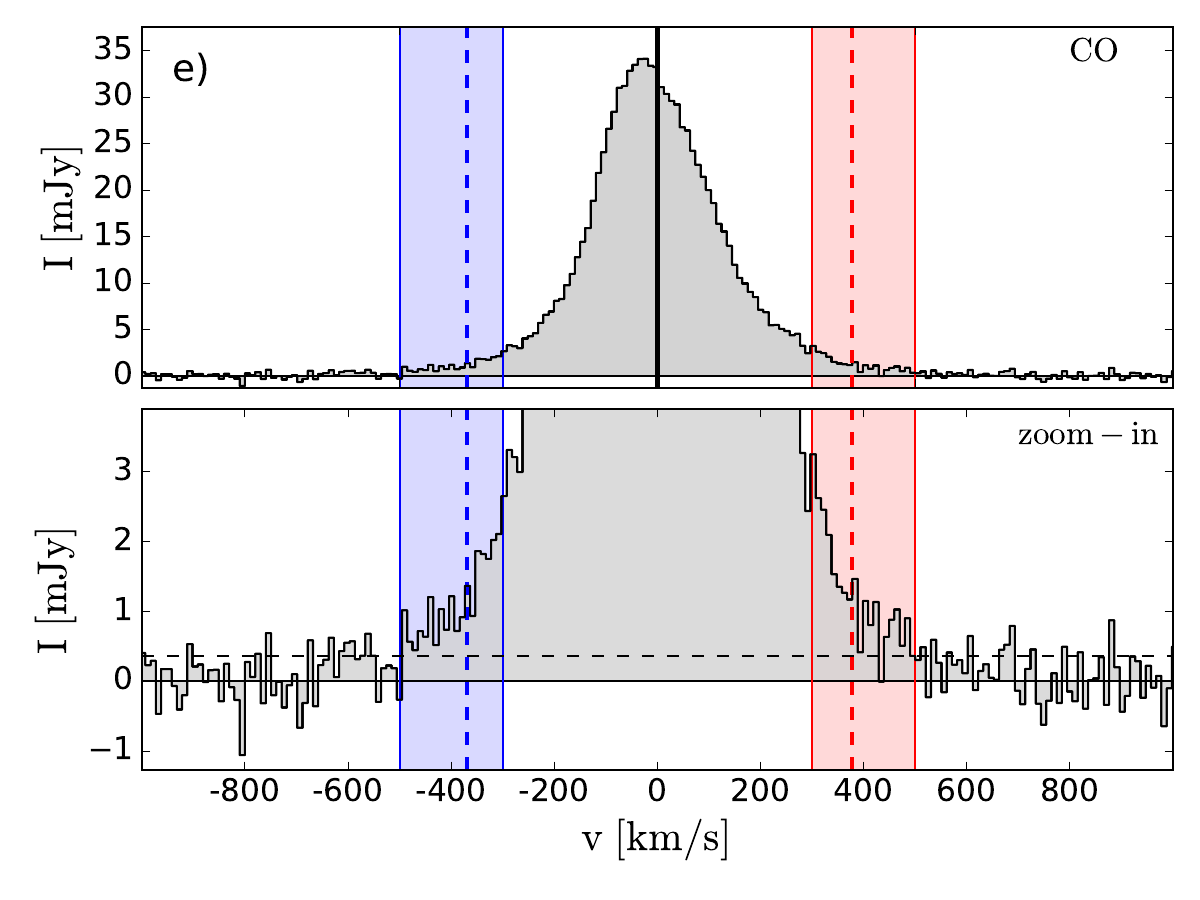} 
\includegraphics[width=0.27\textwidth]{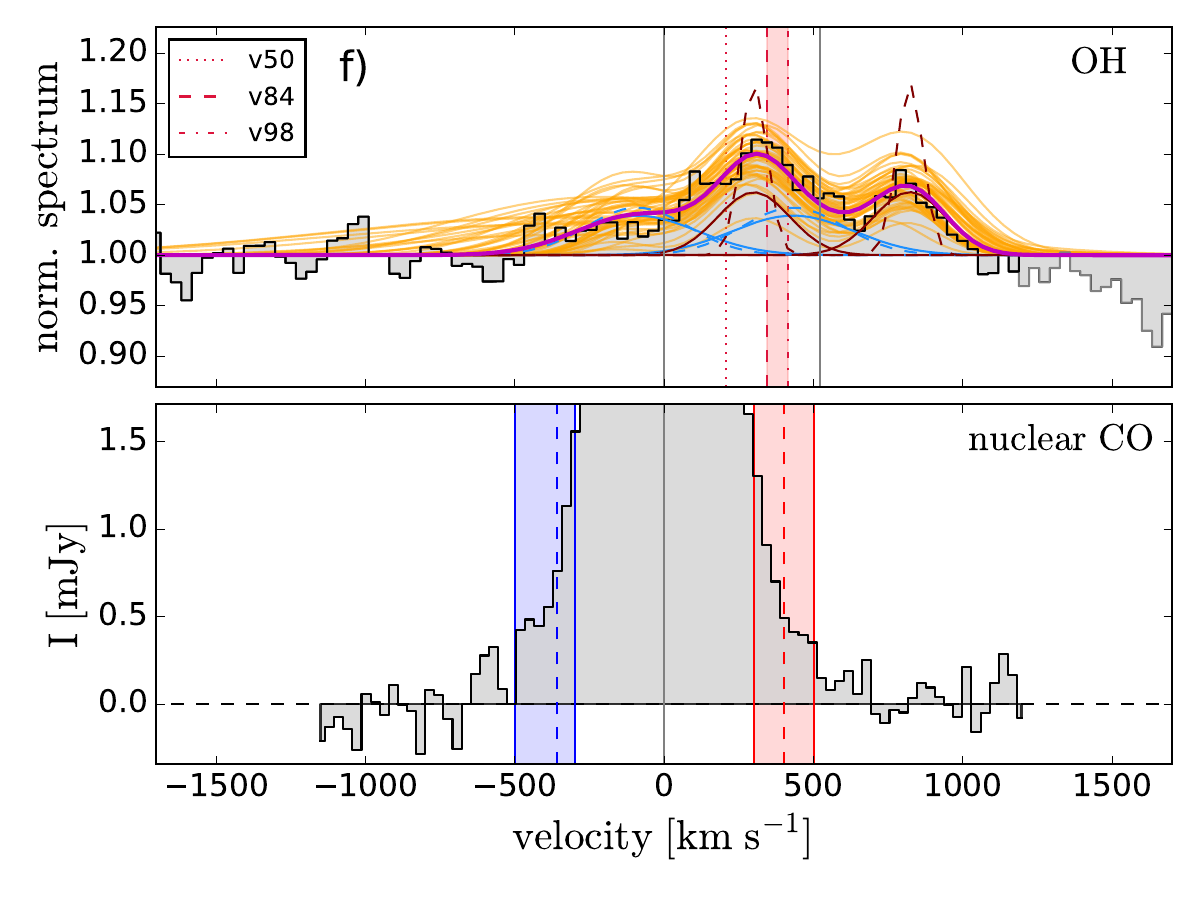}\\ 

\includegraphics[width=0.75\textwidth]{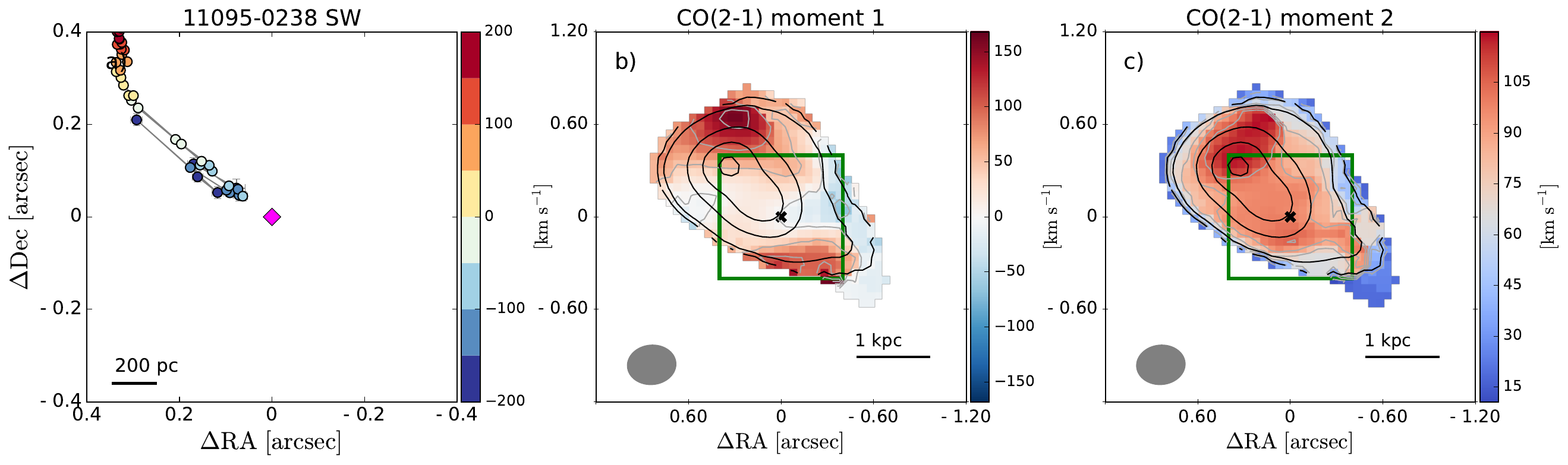}\\ 
\includegraphics[width=0.27\textwidth]{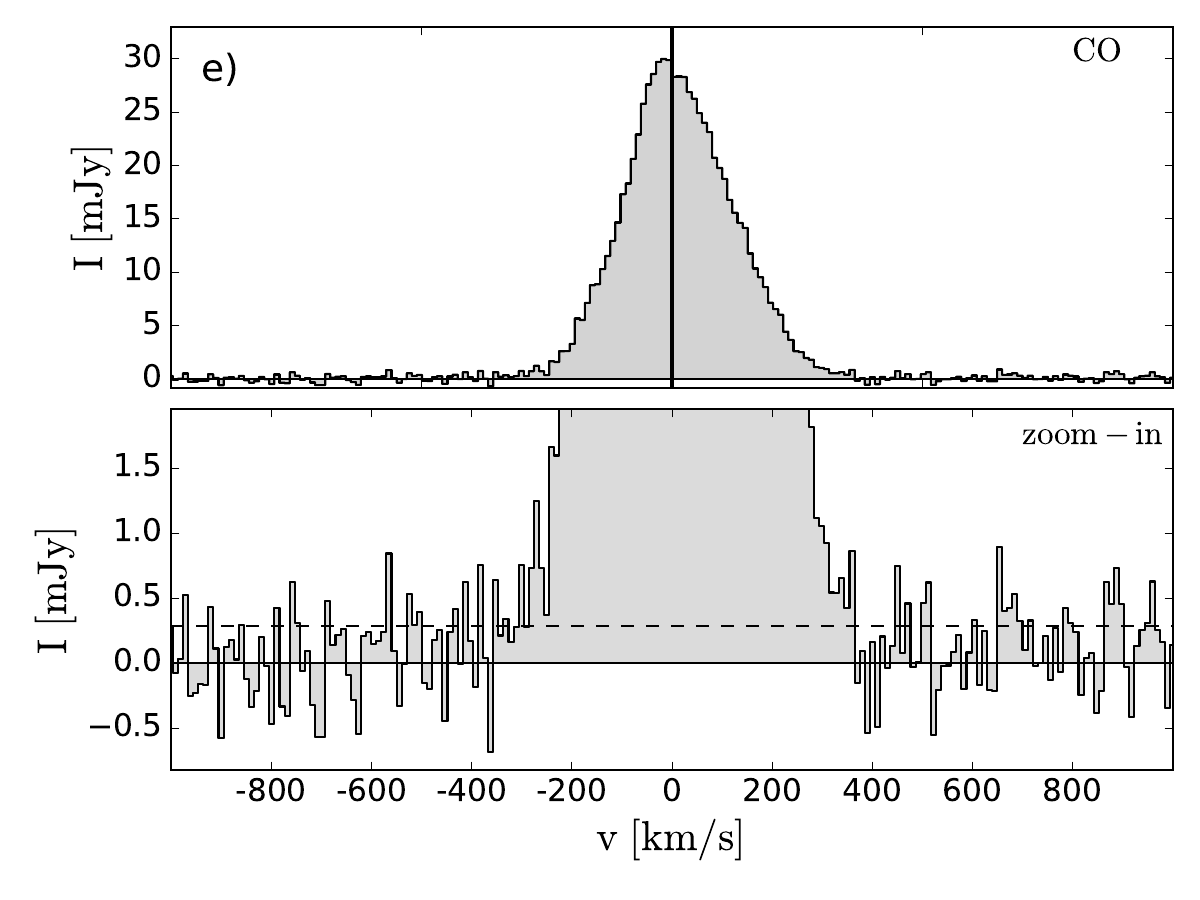} 
\includegraphics[width=0.27\textwidth]{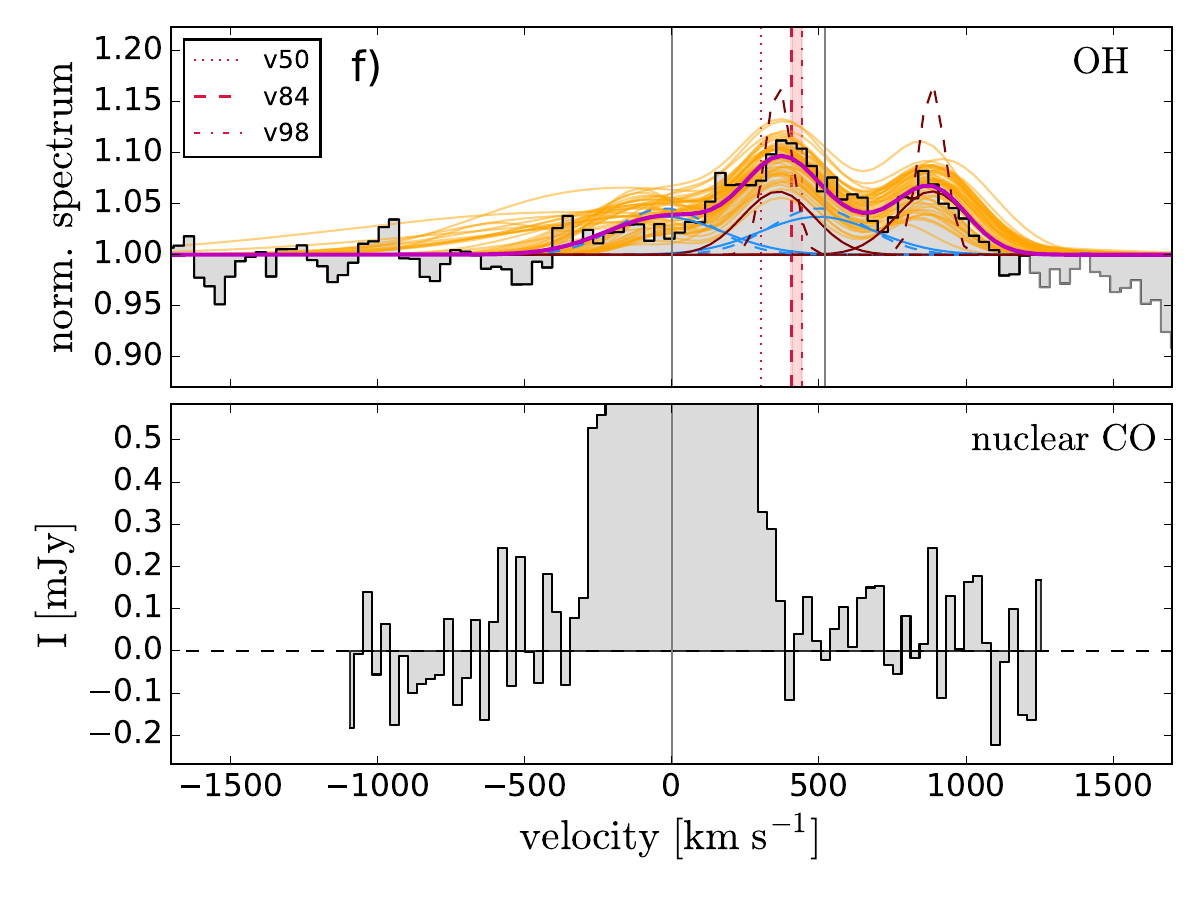}\\ 

\includegraphics[width=0.75\textwidth]{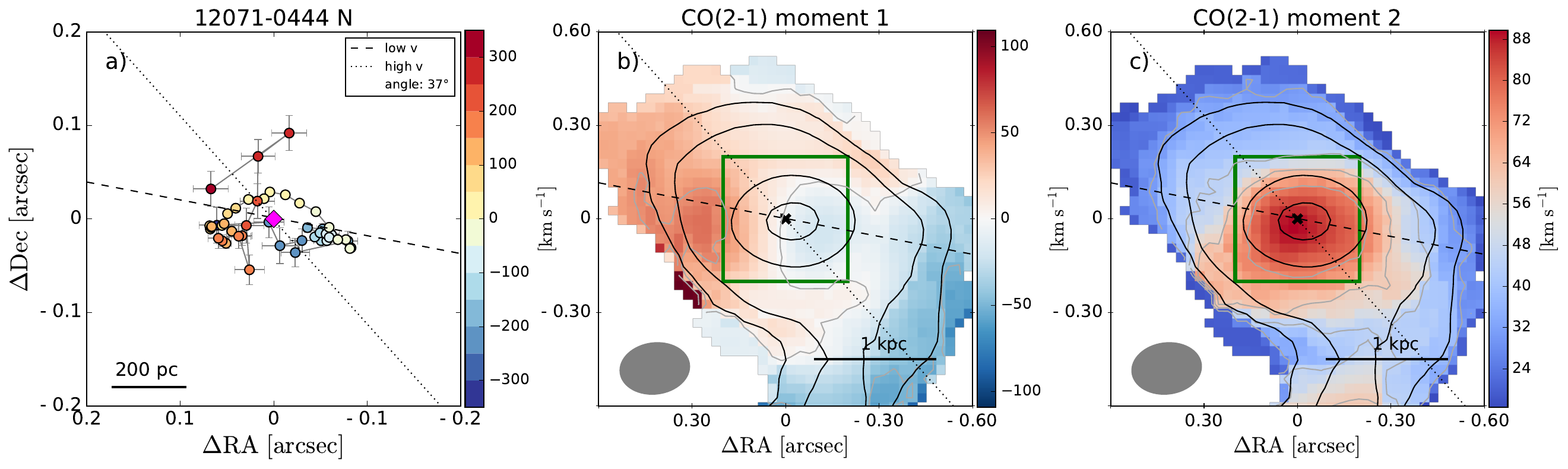}\\ 
\includegraphics[width=0.23\textwidth]{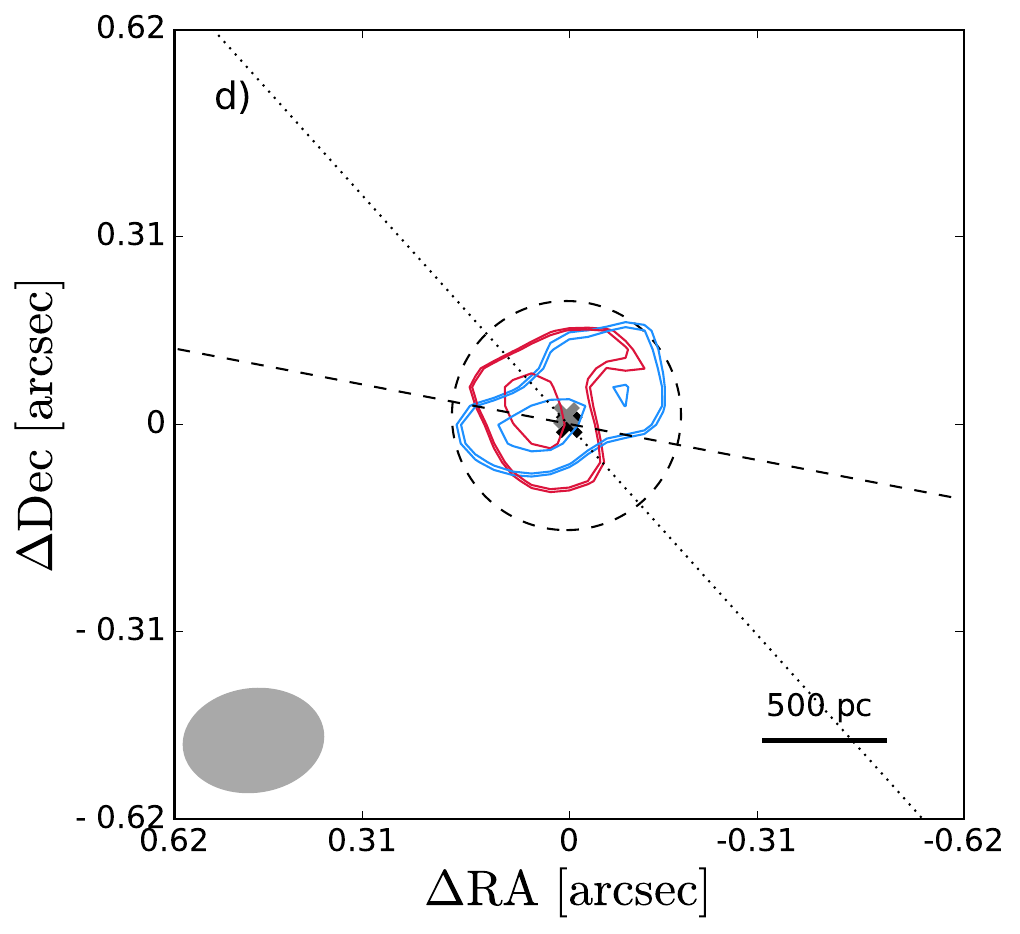} 
\includegraphics[width=0.27\textwidth]{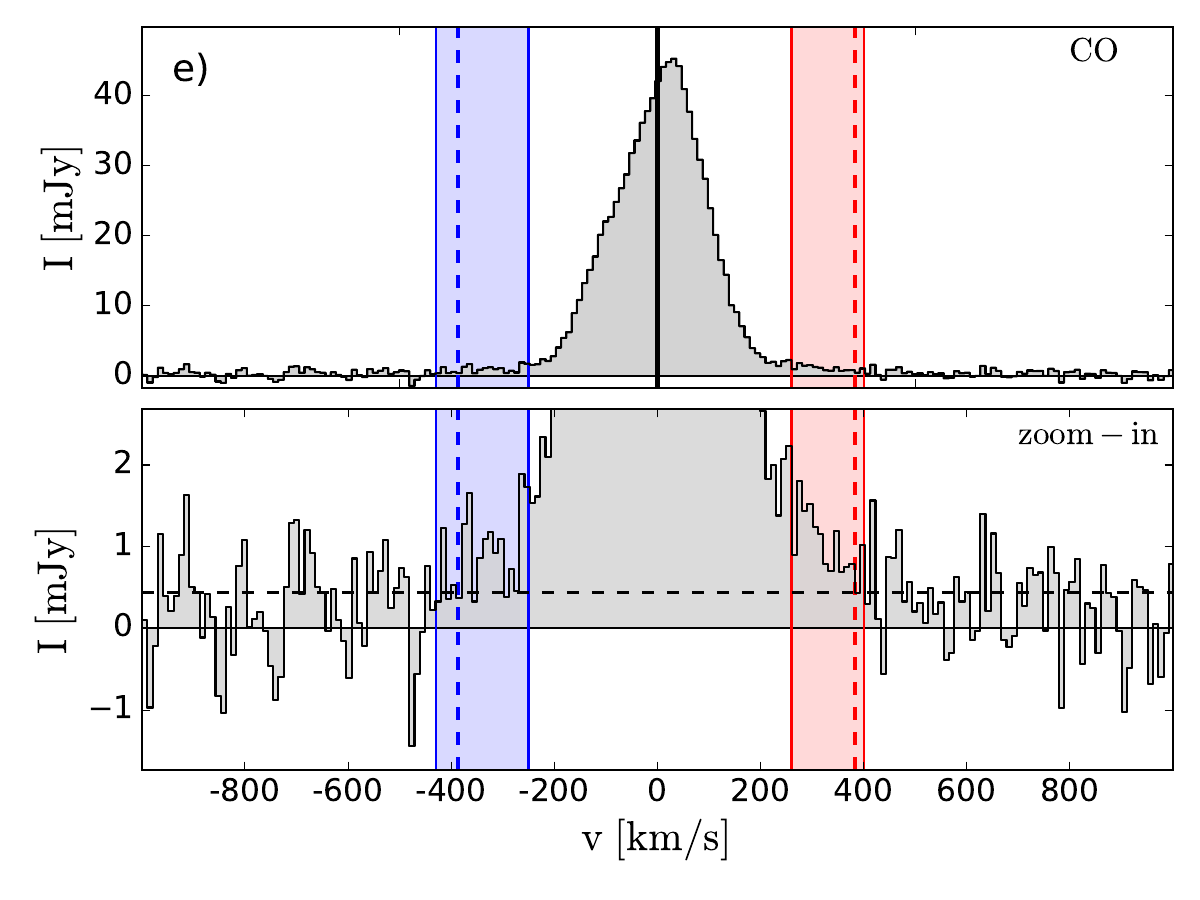} 
\includegraphics[width=0.27\textwidth]{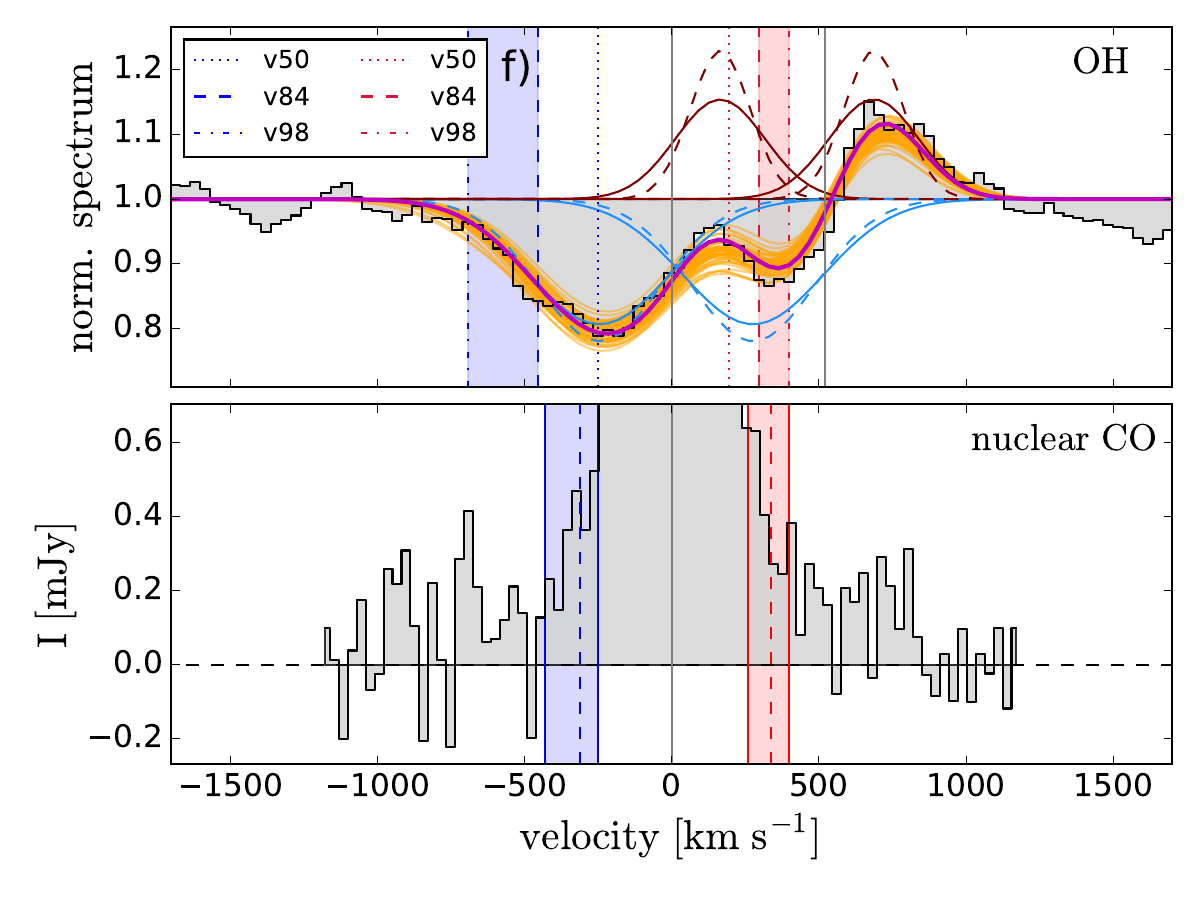}\\ 

\caption{continued.} 
 \end{figure*}

\begin{figure*}\ContinuedFloat 
\centering 

\includegraphics[width=0.75\textwidth]{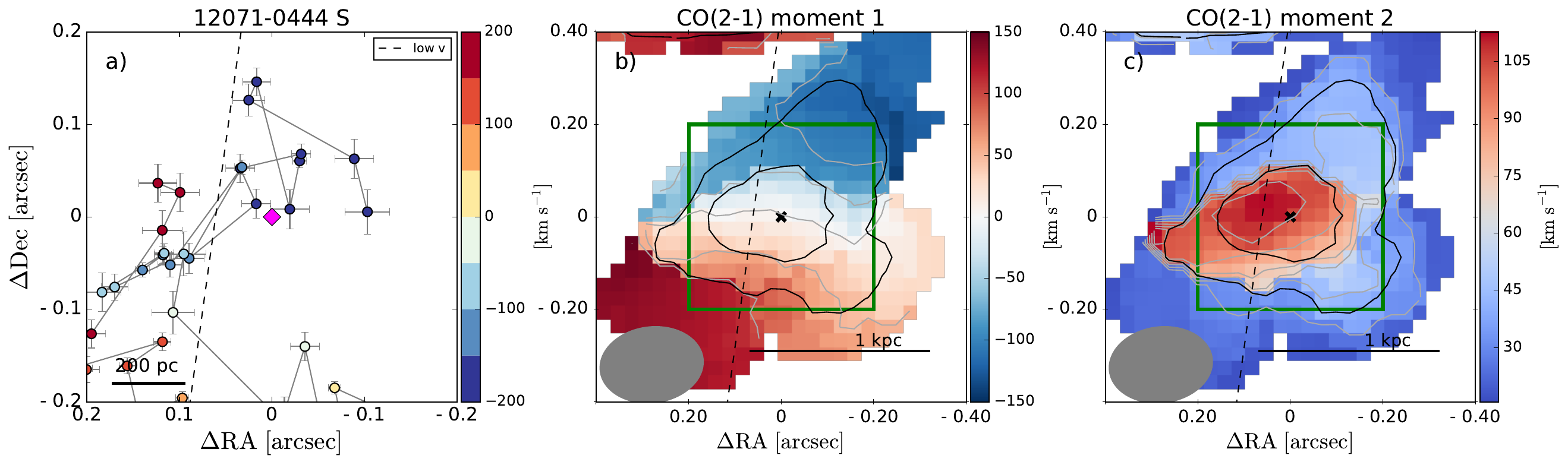}\\ 
\includegraphics[width=0.27\textwidth]{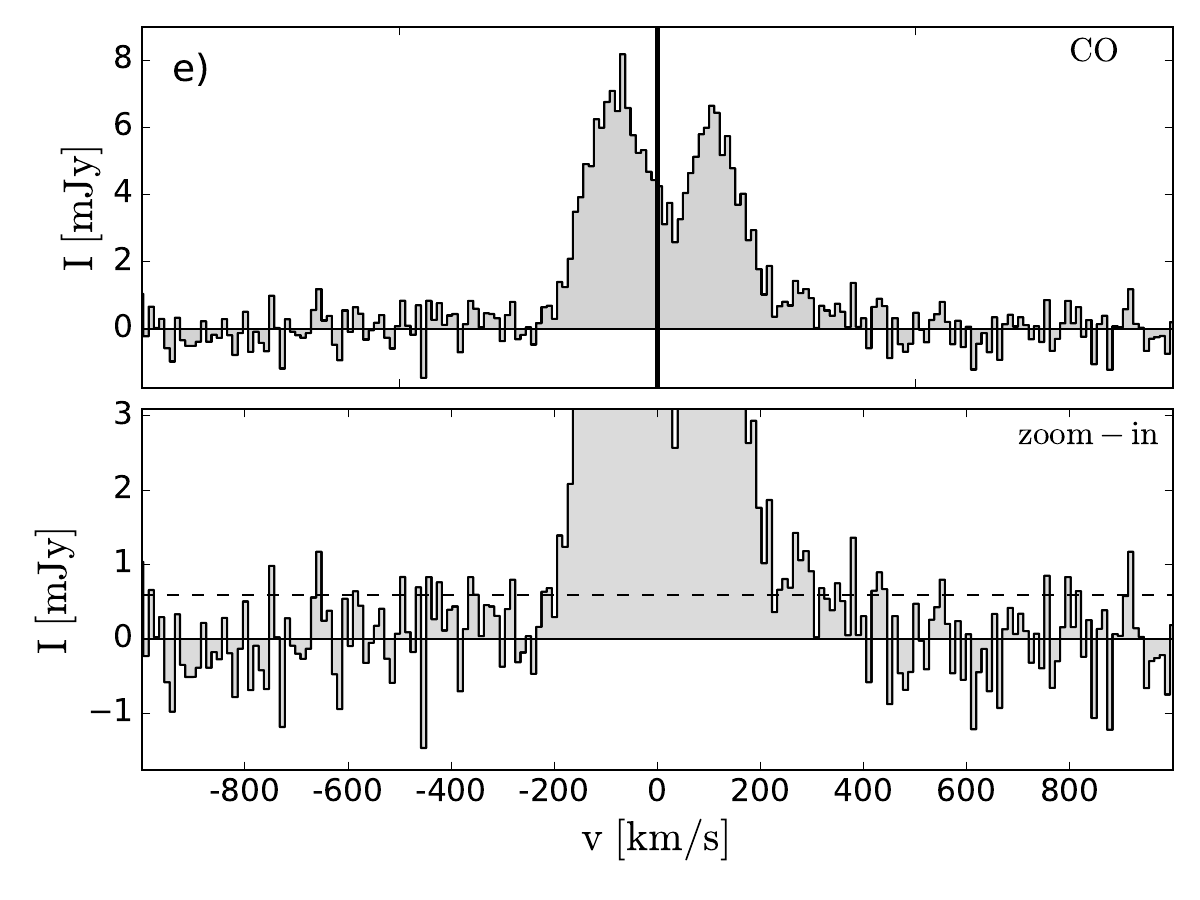} 

\includegraphics[width=0.75\textwidth]{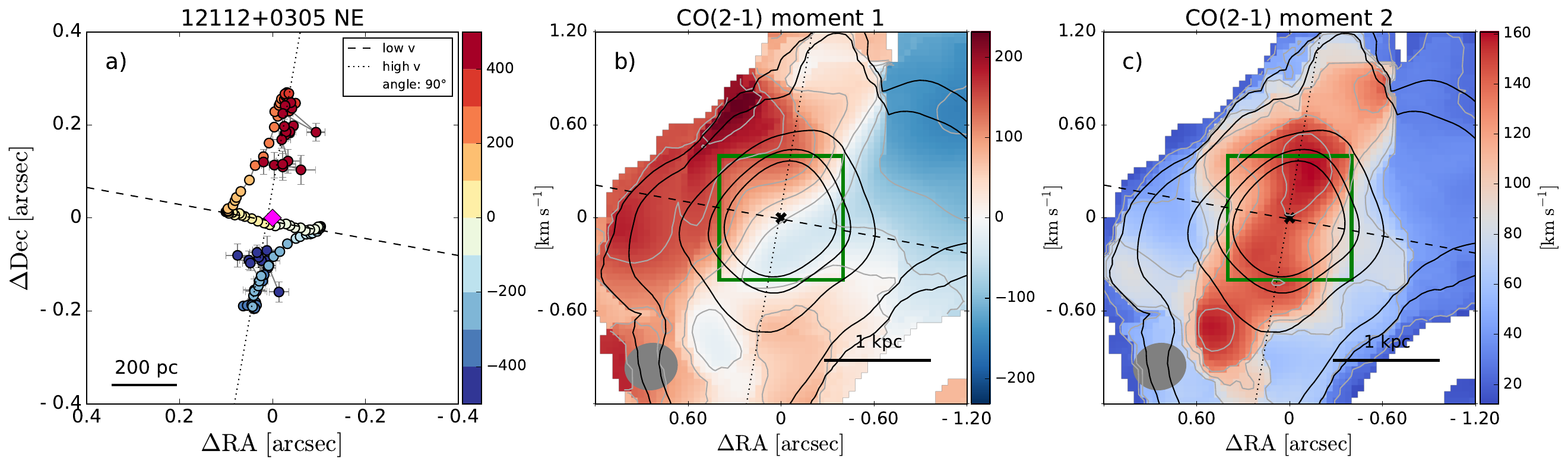}\\ 
\includegraphics[width=0.23\textwidth]{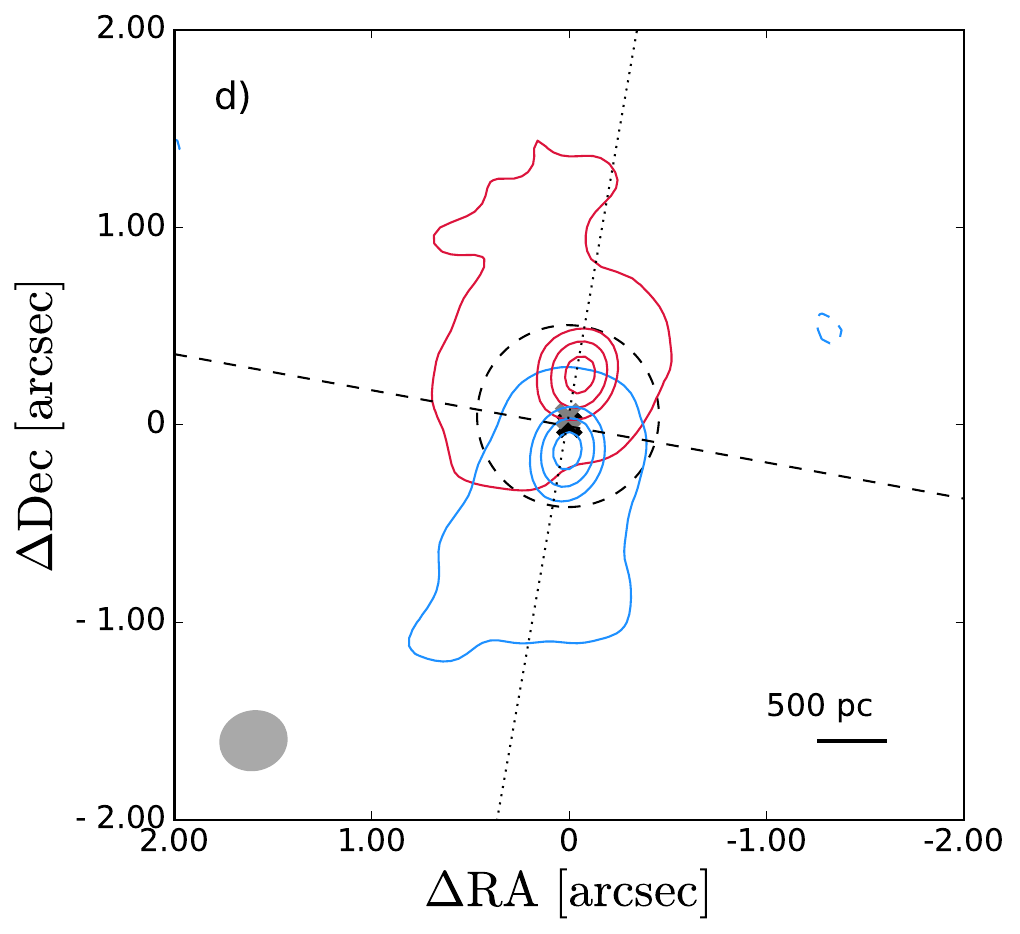} 
\includegraphics[width=0.27\textwidth]{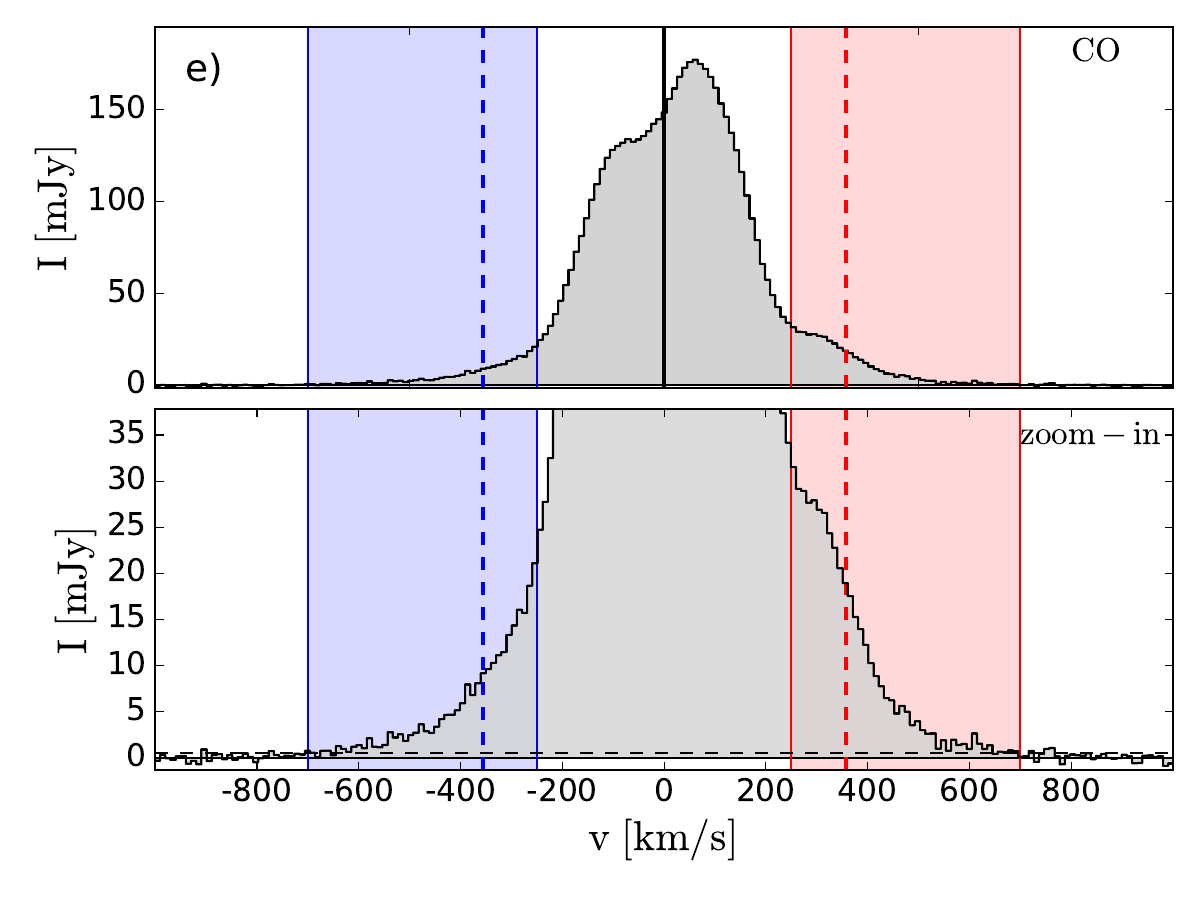} 
\includegraphics[width=0.27\textwidth]{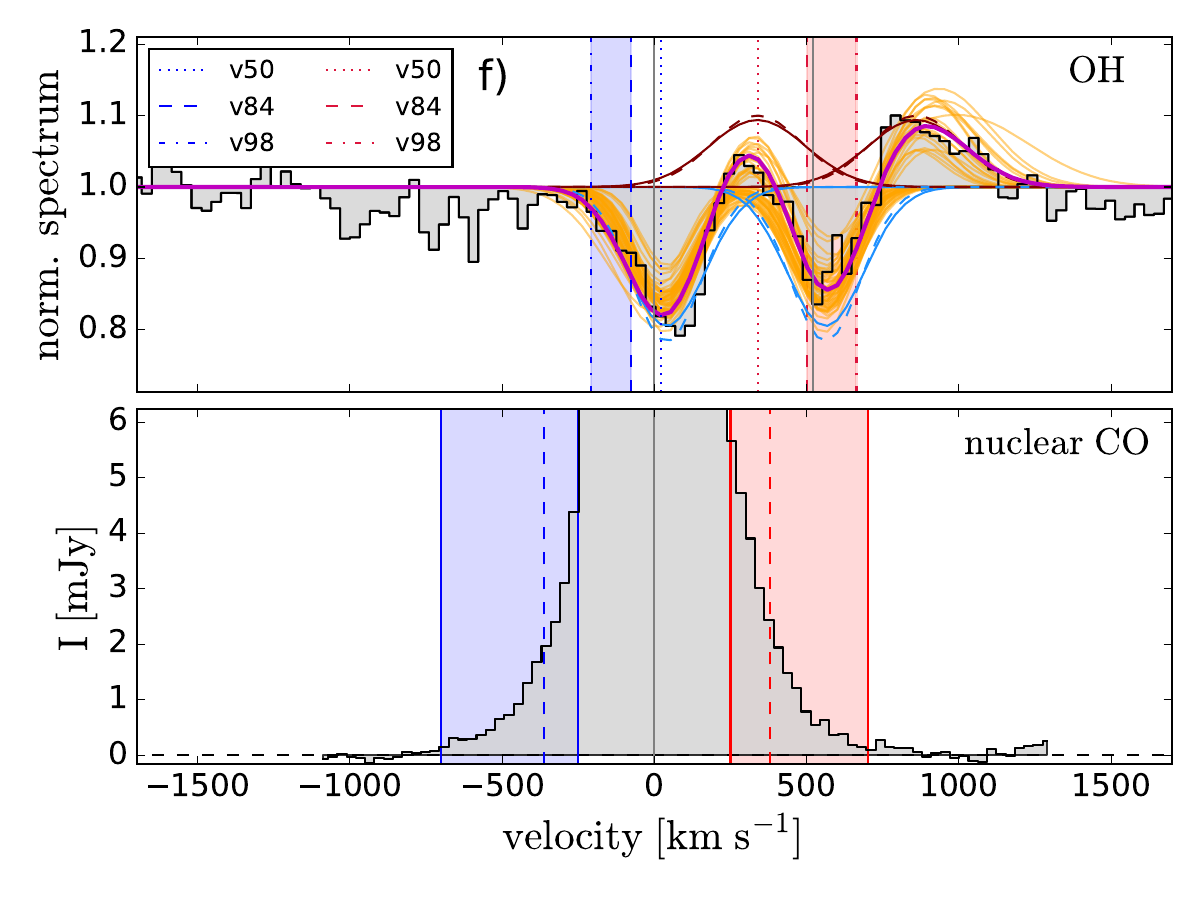}\\ 

\includegraphics[width=0.75\textwidth]{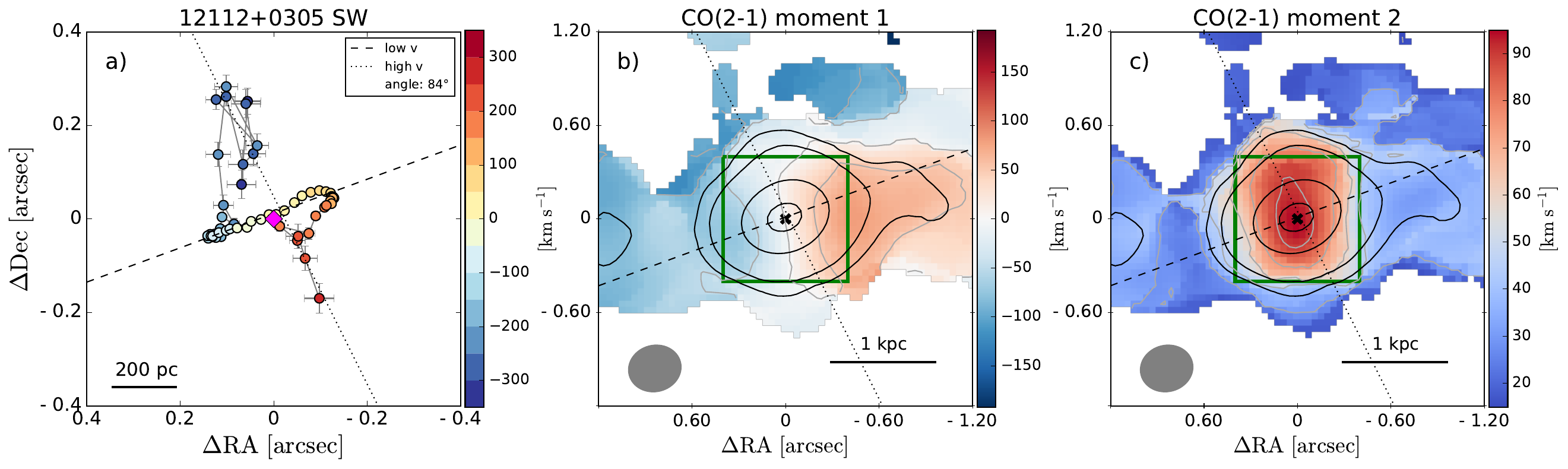}\\ 
\includegraphics[width=0.23\textwidth]{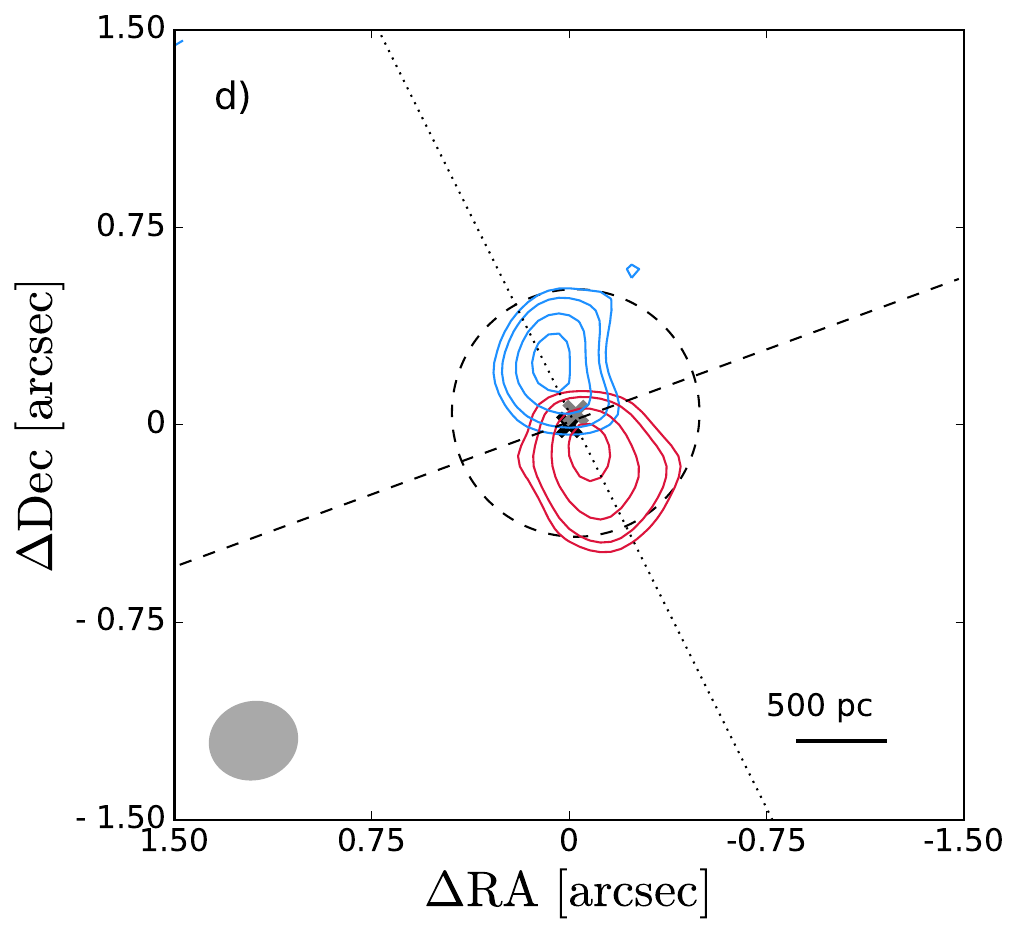} 
\includegraphics[width=0.27\textwidth]{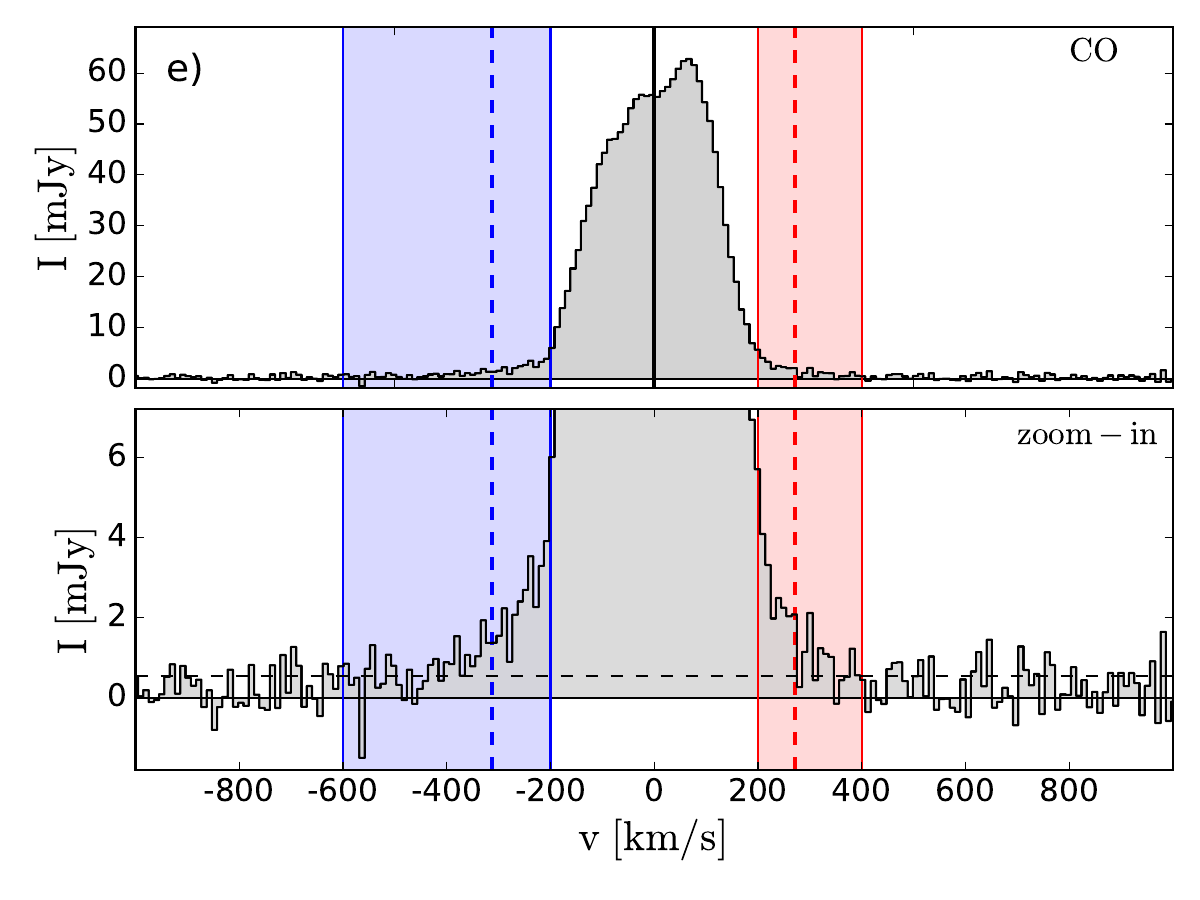} 
\includegraphics[width=0.27\textwidth]{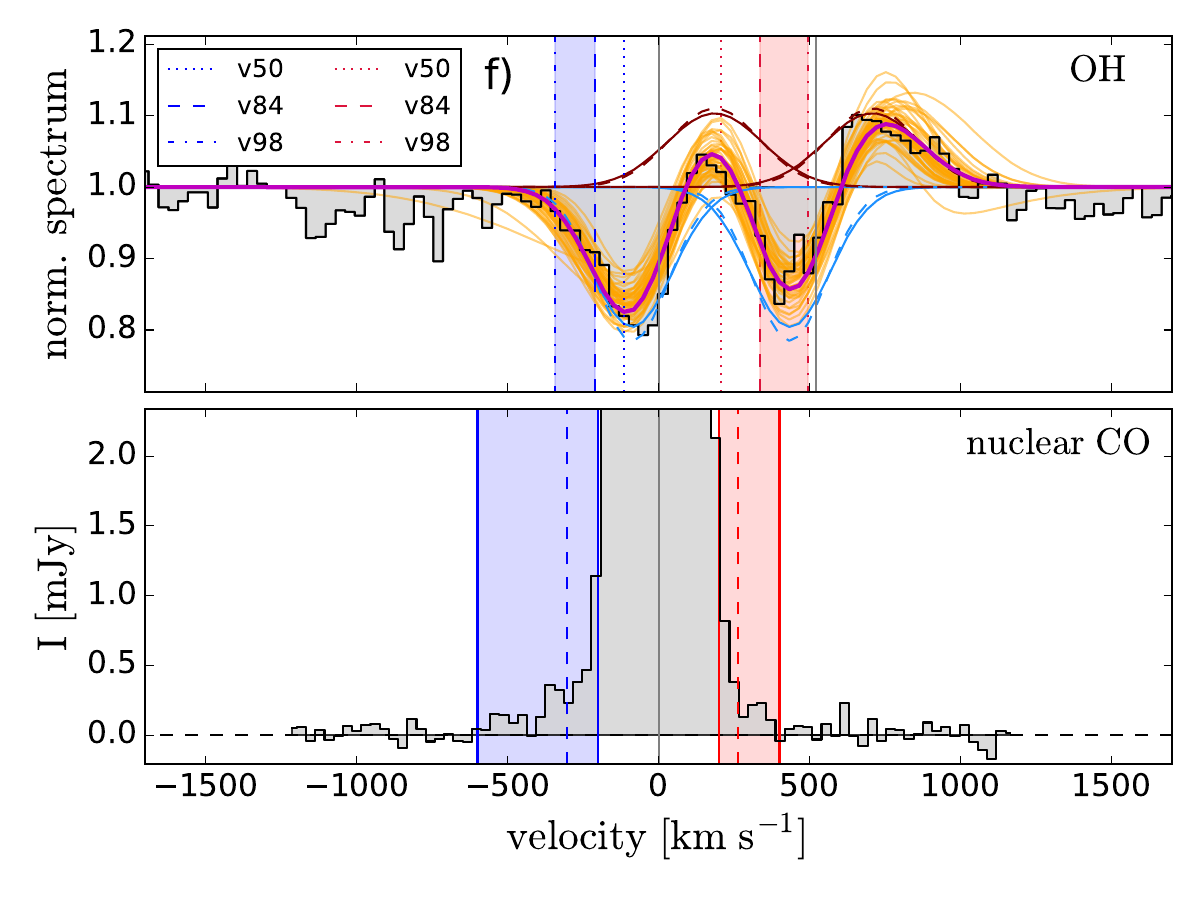}\\

\caption{continued.} 
 \end{figure*}

\begin{figure*}\ContinuedFloat 
\centering 

\includegraphics[width=0.75\textwidth]{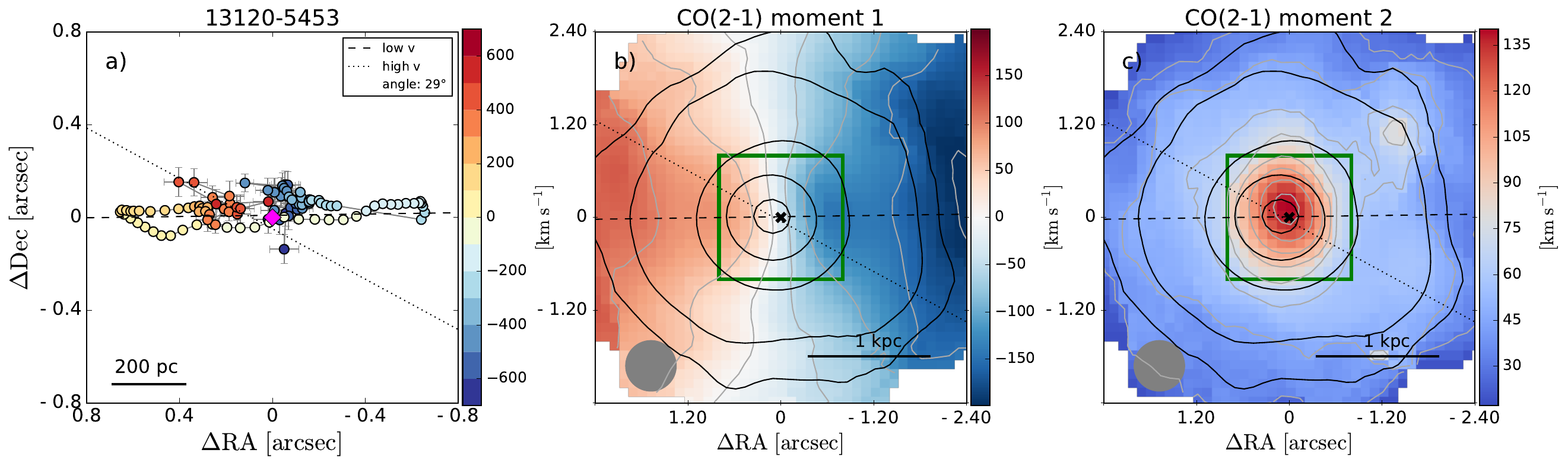}\\ 
\includegraphics[width=0.23\textwidth]{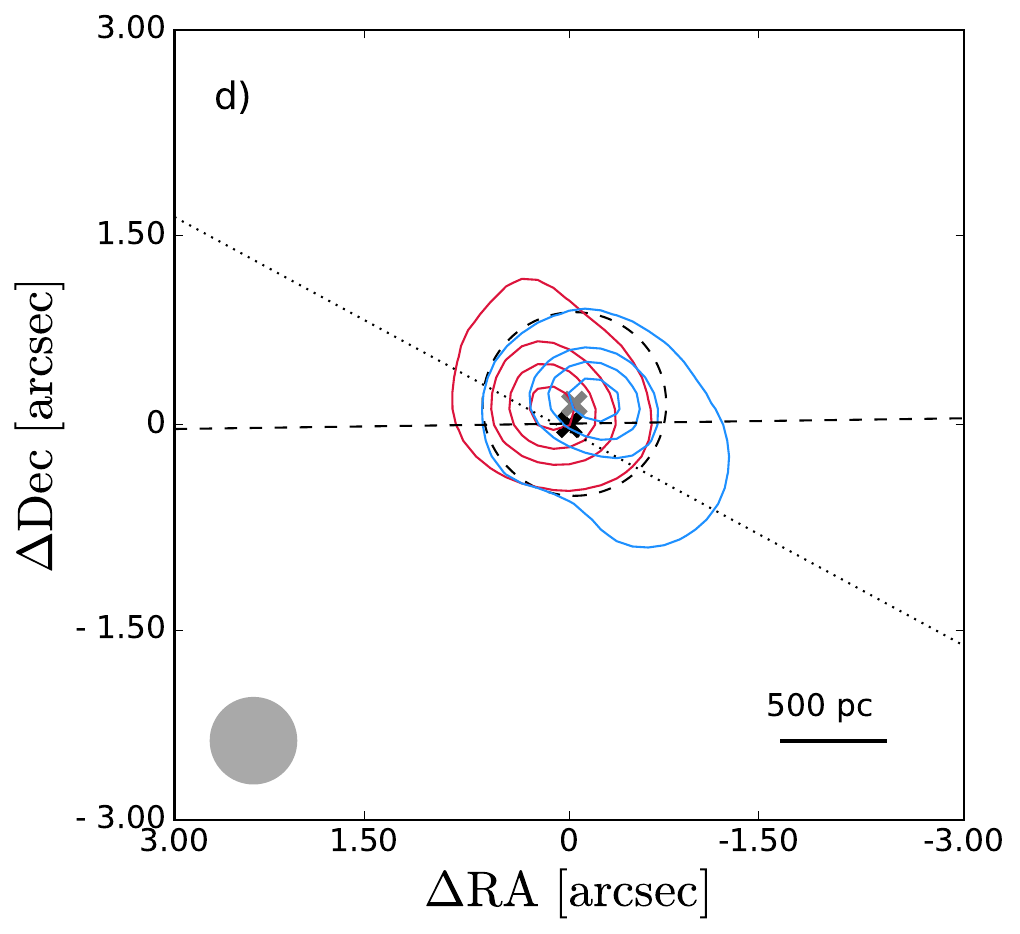} 
\includegraphics[width=0.27\textwidth]{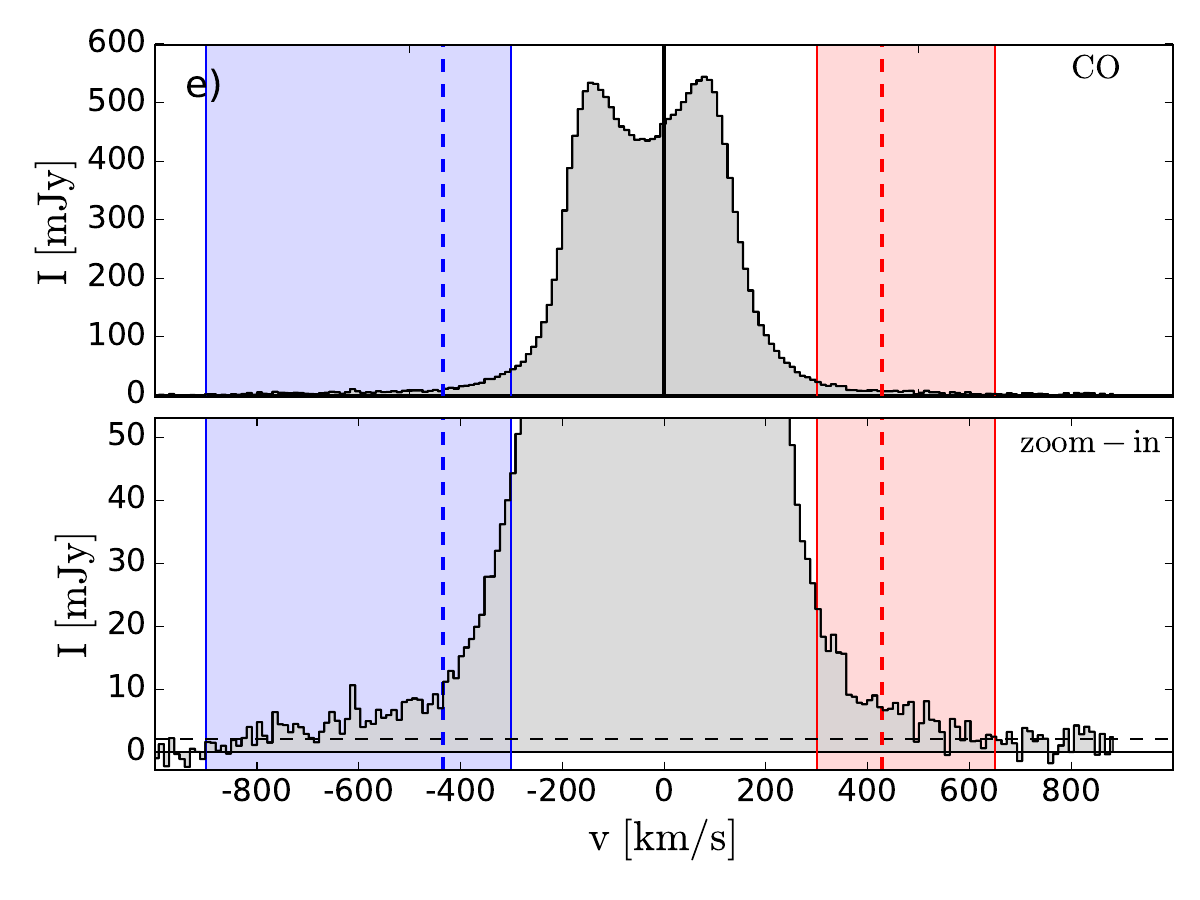} 
\includegraphics[width=0.27\textwidth]{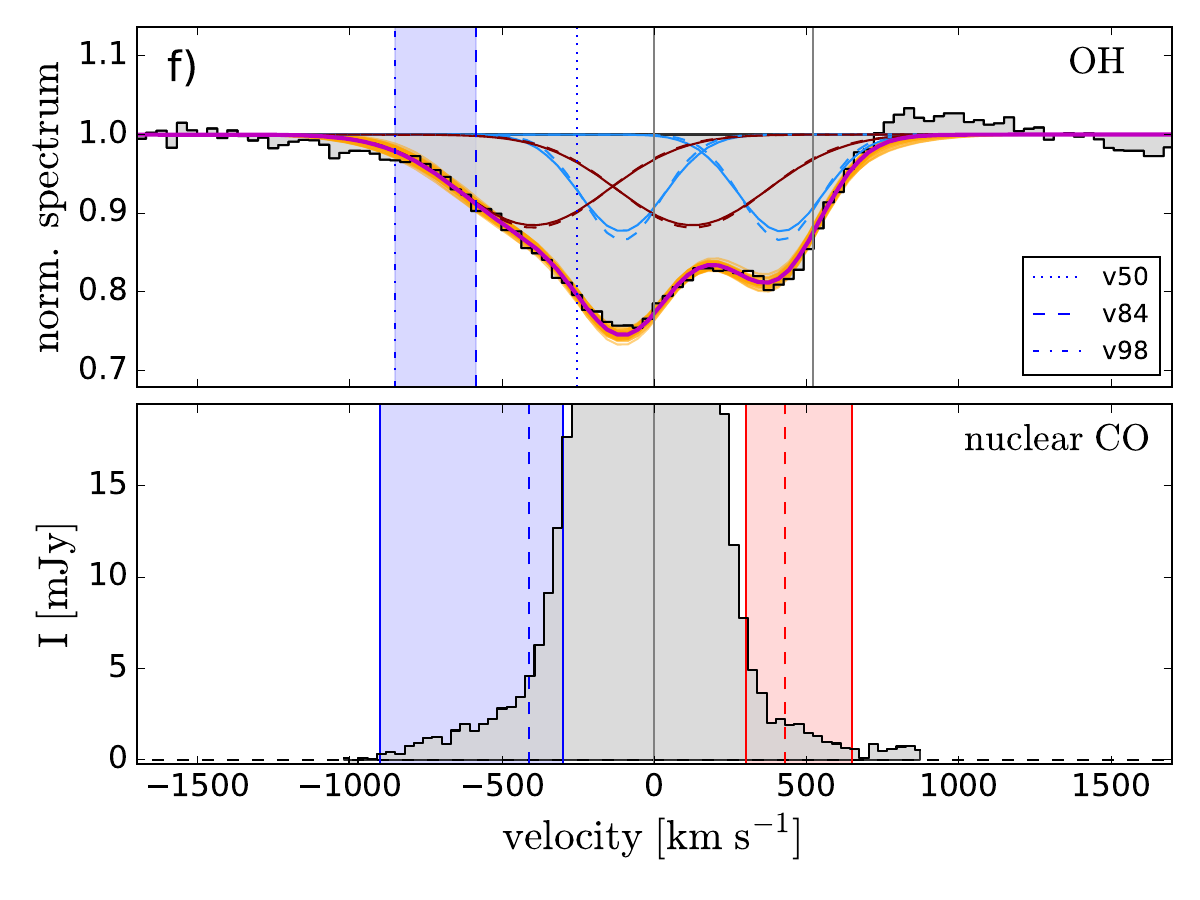}\\ 

\includegraphics[width=0.75\textwidth]{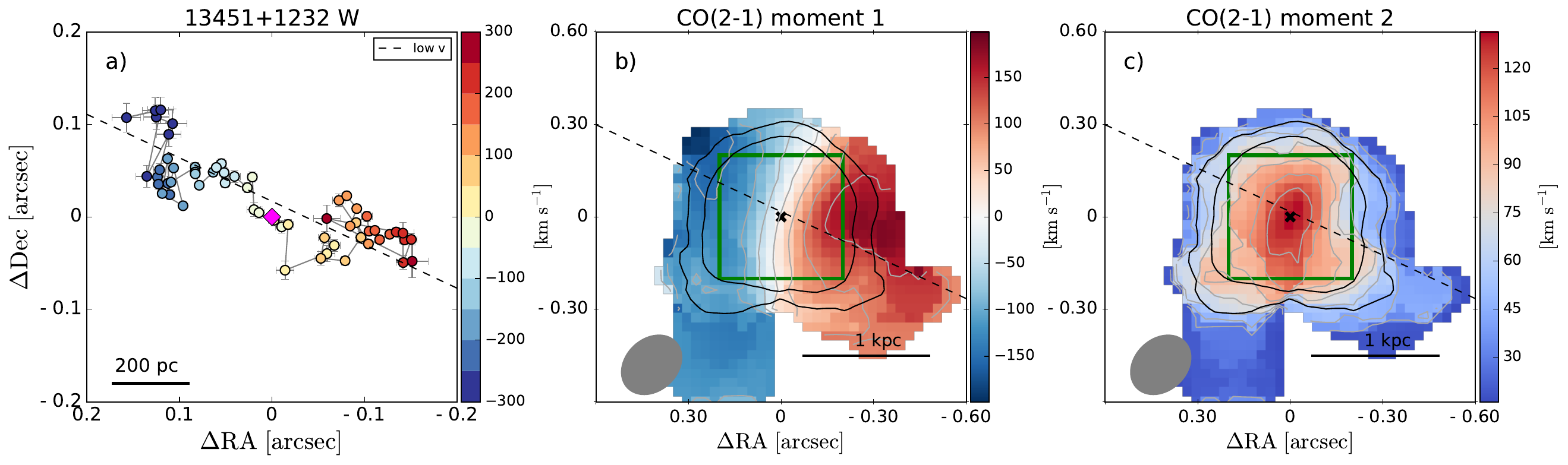}\\ 
\includegraphics[width=0.27\textwidth]{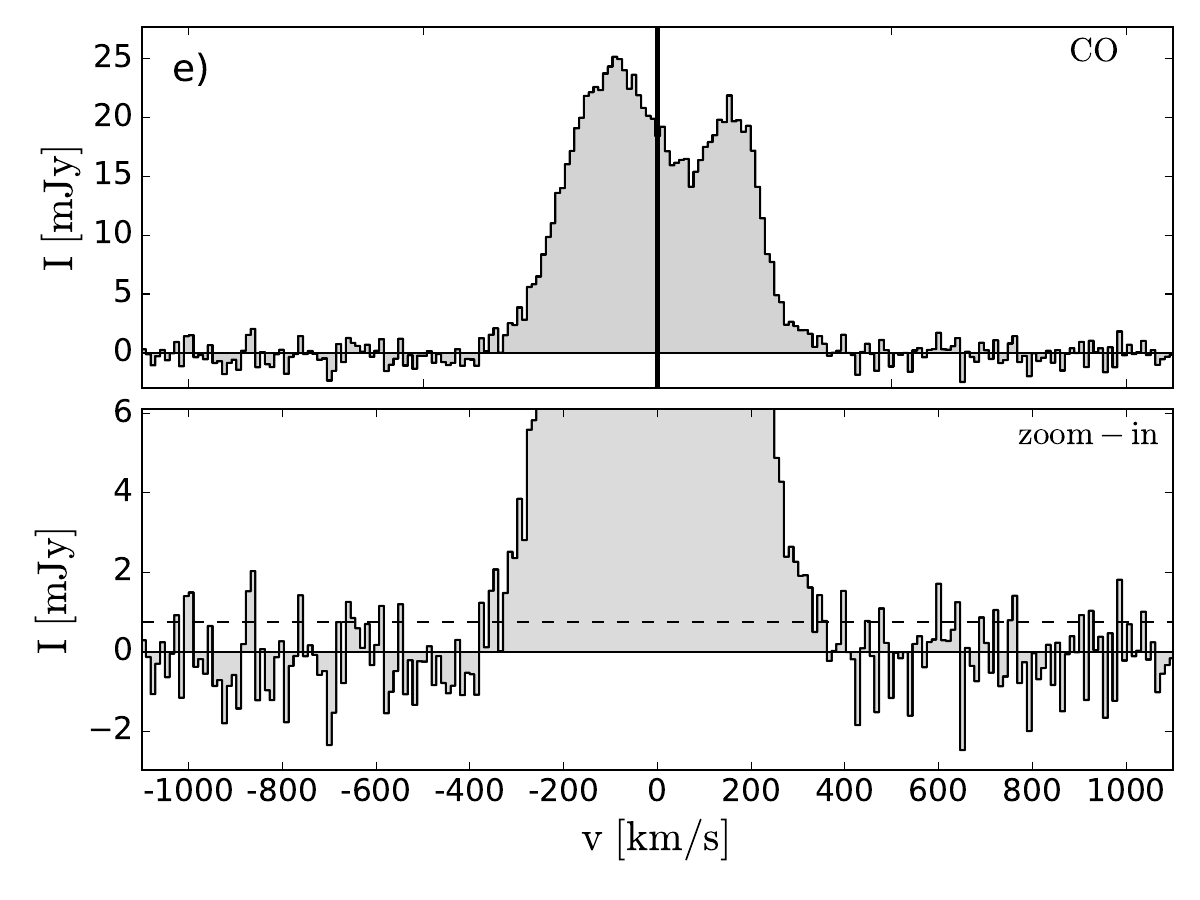} 
\includegraphics[width=0.27\textwidth]{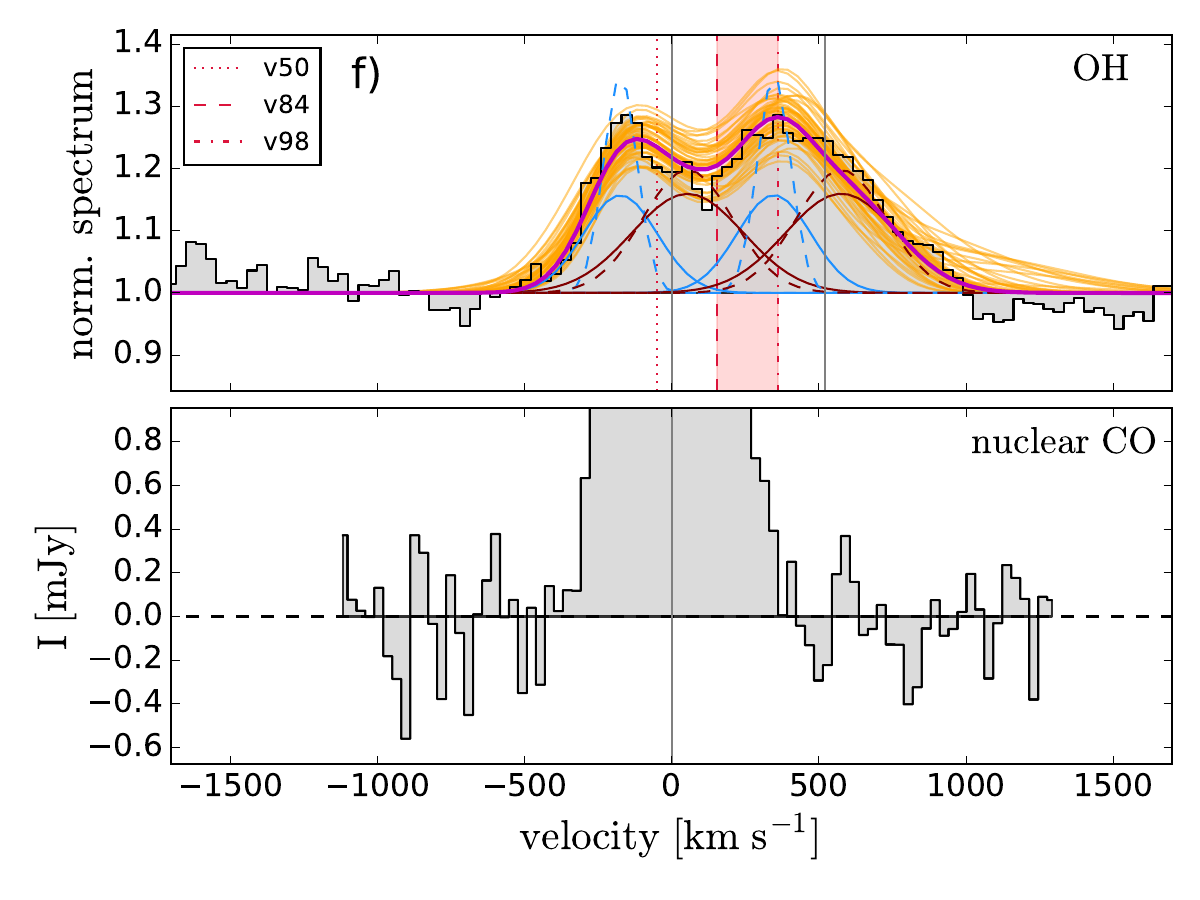}\\ 

\includegraphics[width=0.75\textwidth]{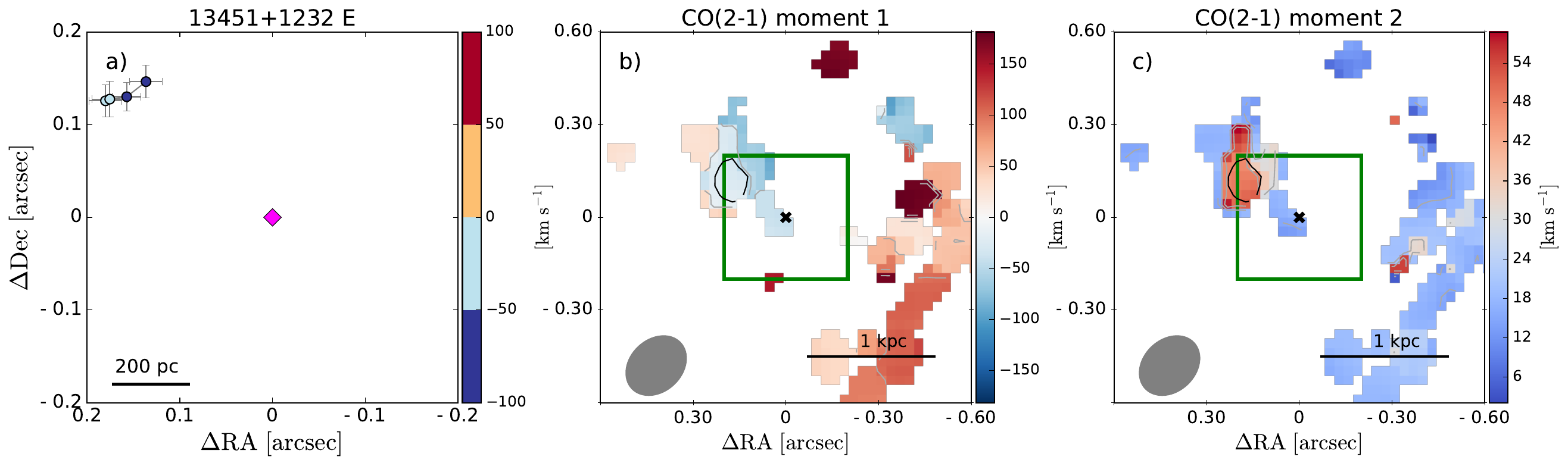}\\ 
\includegraphics[width=0.27\textwidth]{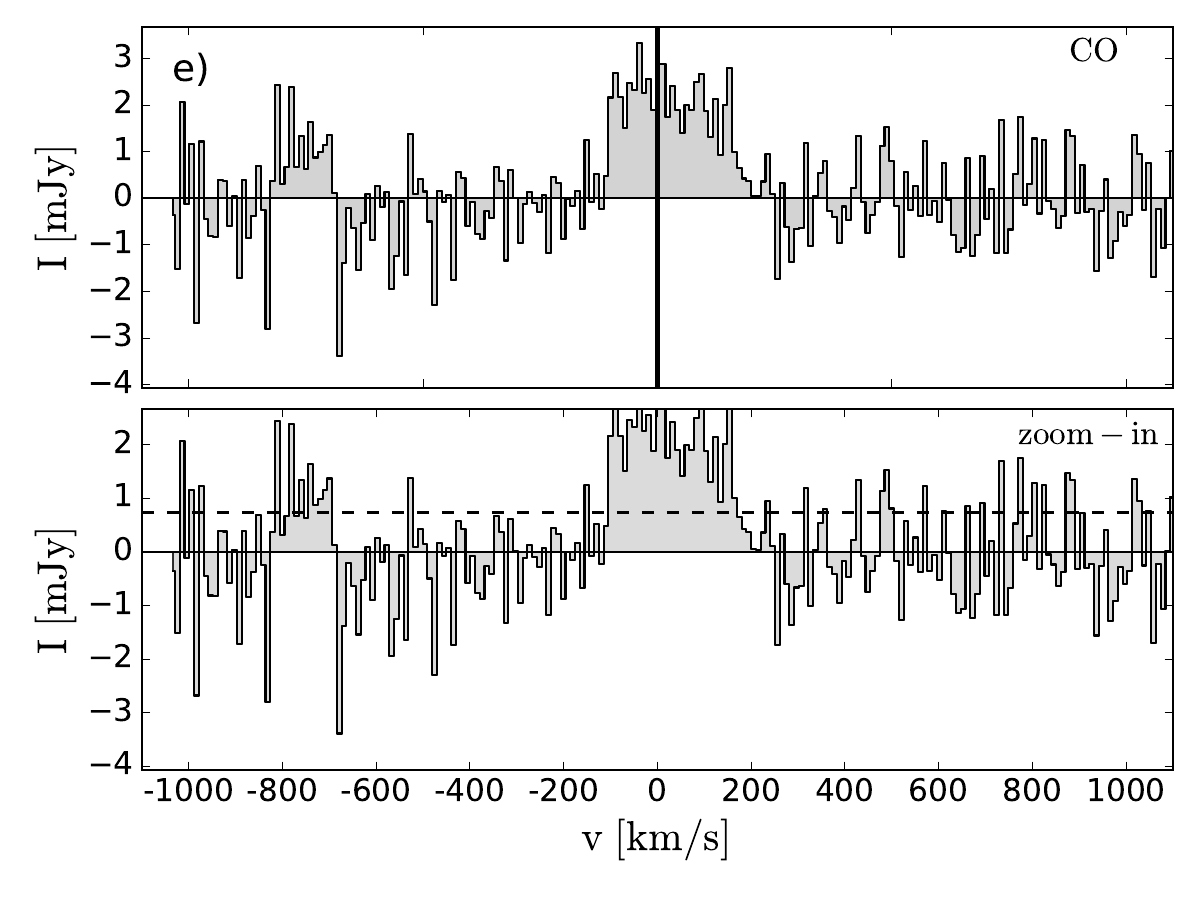}

\caption{continued.} 
 \end{figure*}

\begin{figure*}\ContinuedFloat 
\centering 
\includegraphics[width=0.75\textwidth]{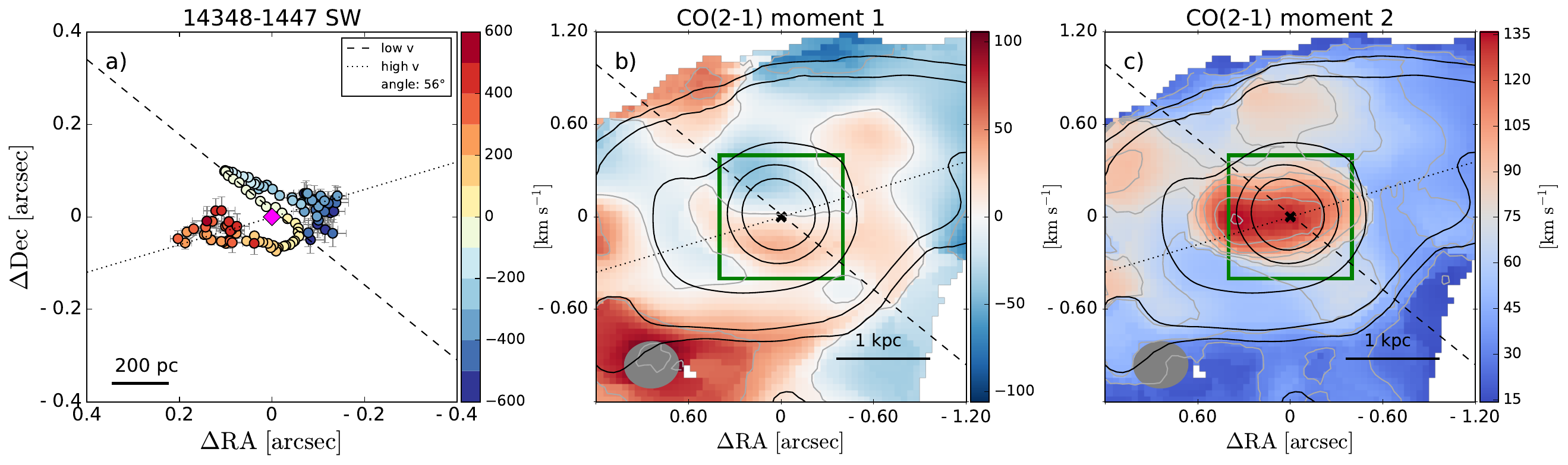}\\ 
\includegraphics[width=0.23\textwidth]{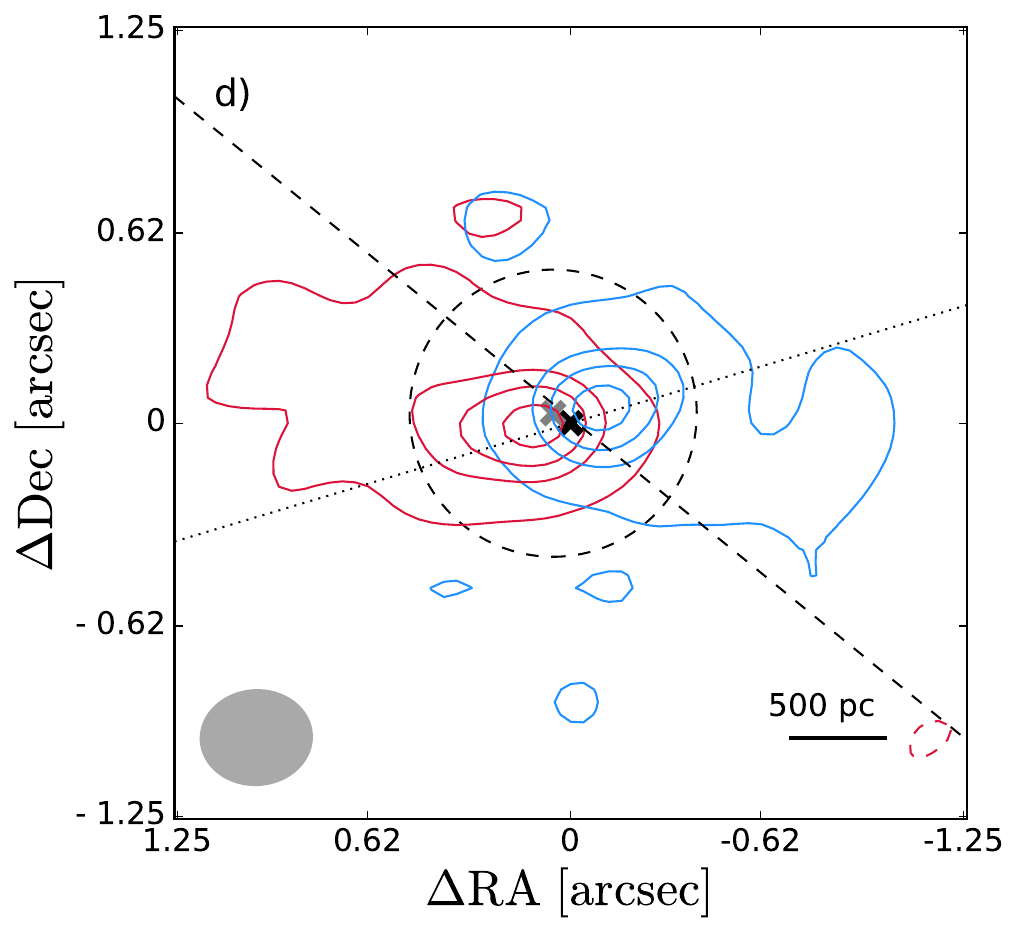} 
\includegraphics[width=0.27\textwidth]{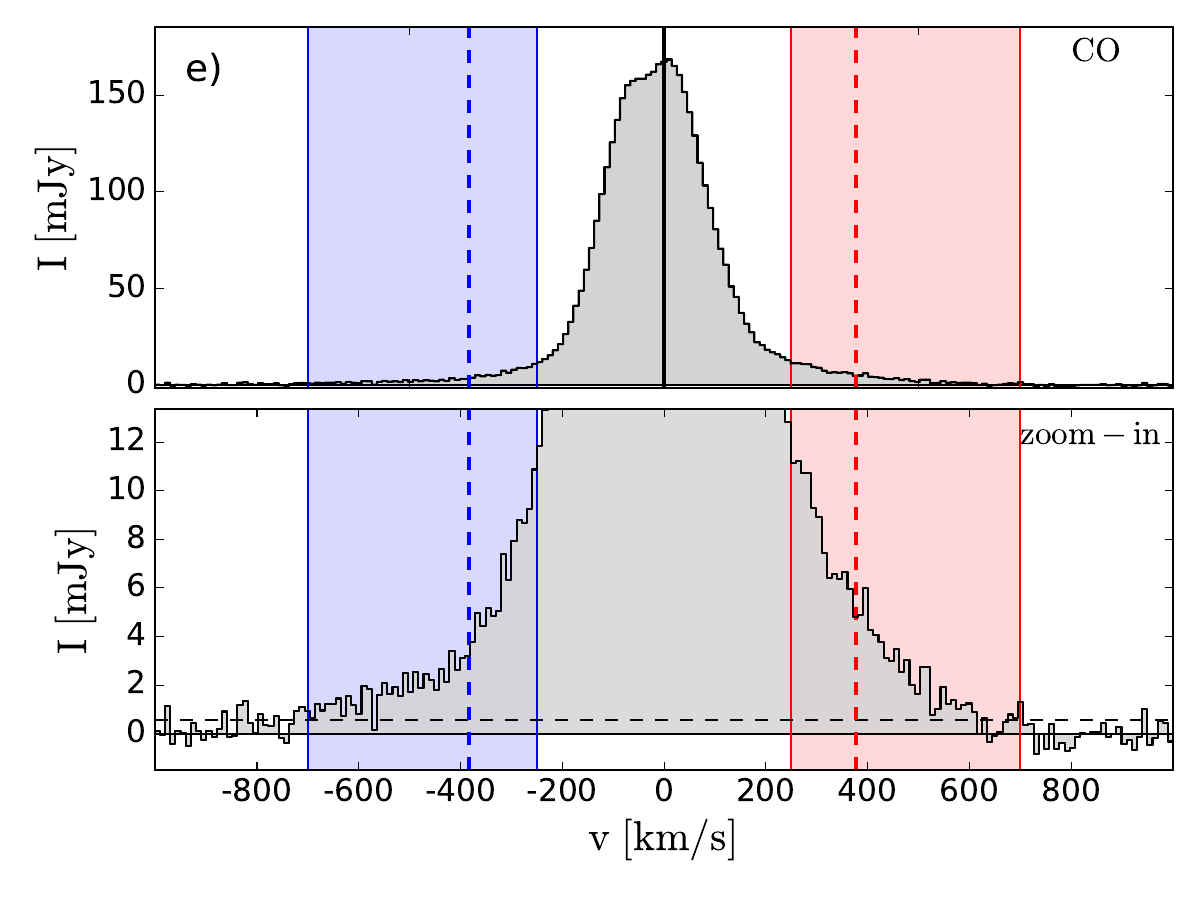} 
\includegraphics[width=0.27\textwidth]{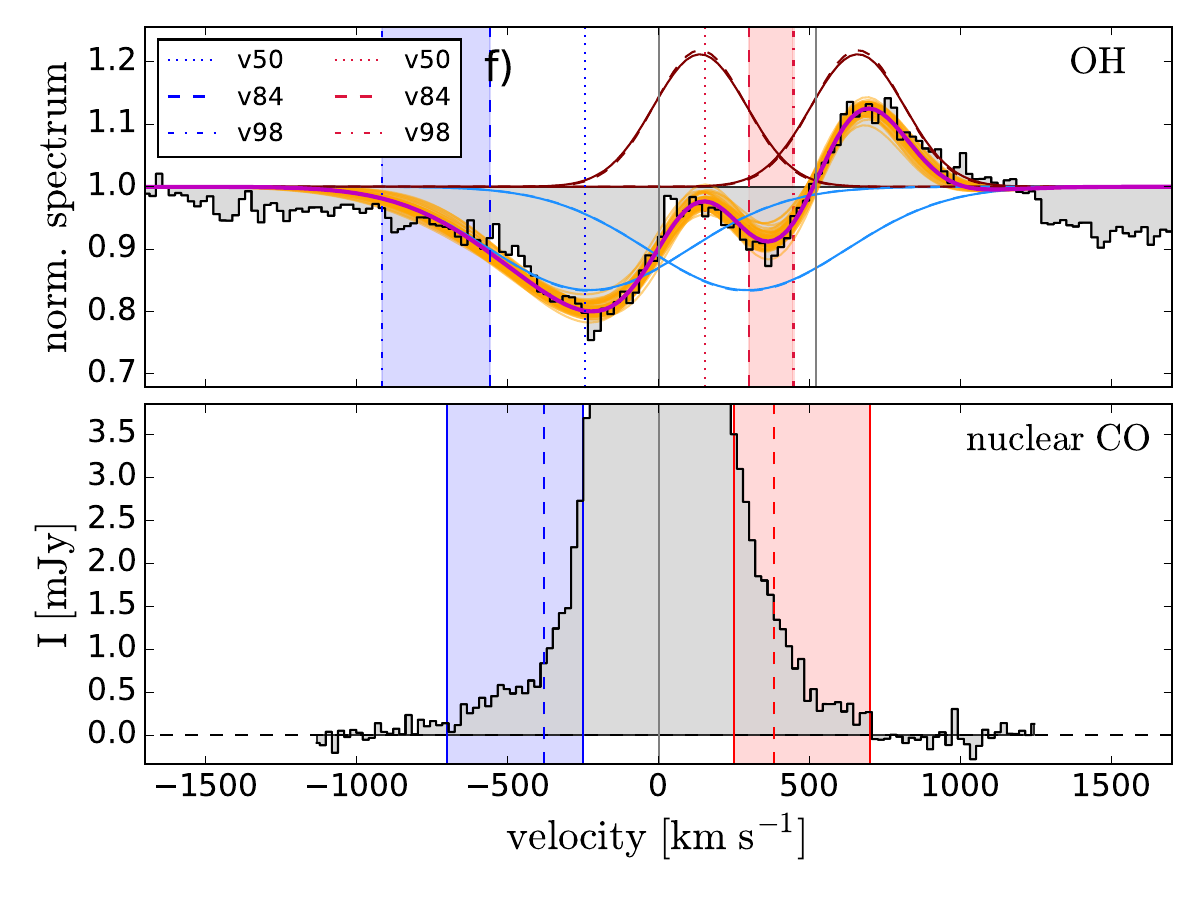}\\ 

\includegraphics[width=0.75\textwidth]{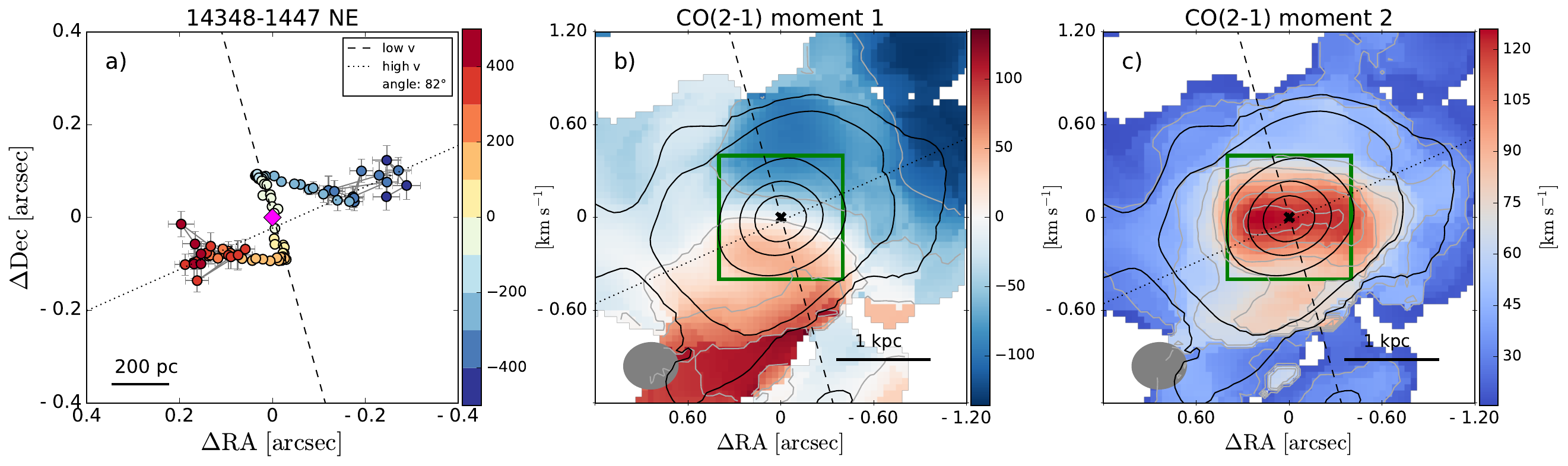}\\ 
\includegraphics[width=0.23\textwidth]{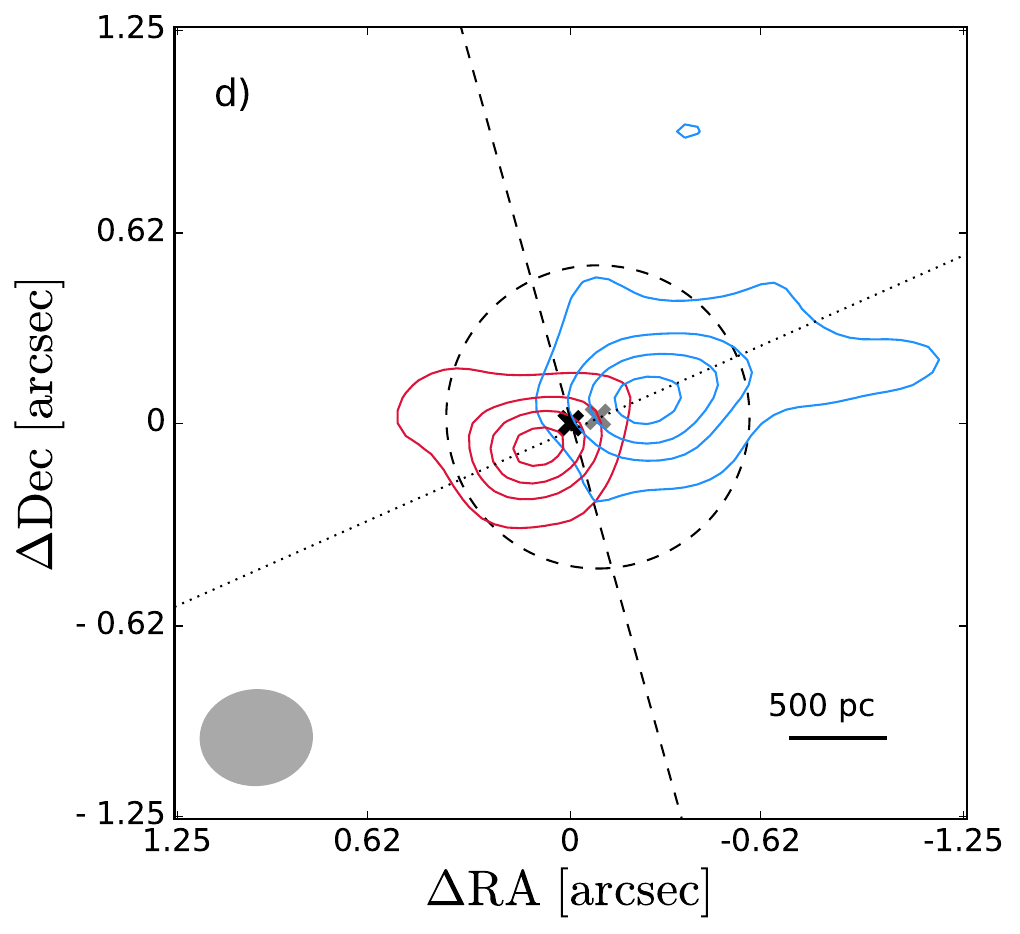} 
\includegraphics[width=0.27\textwidth]{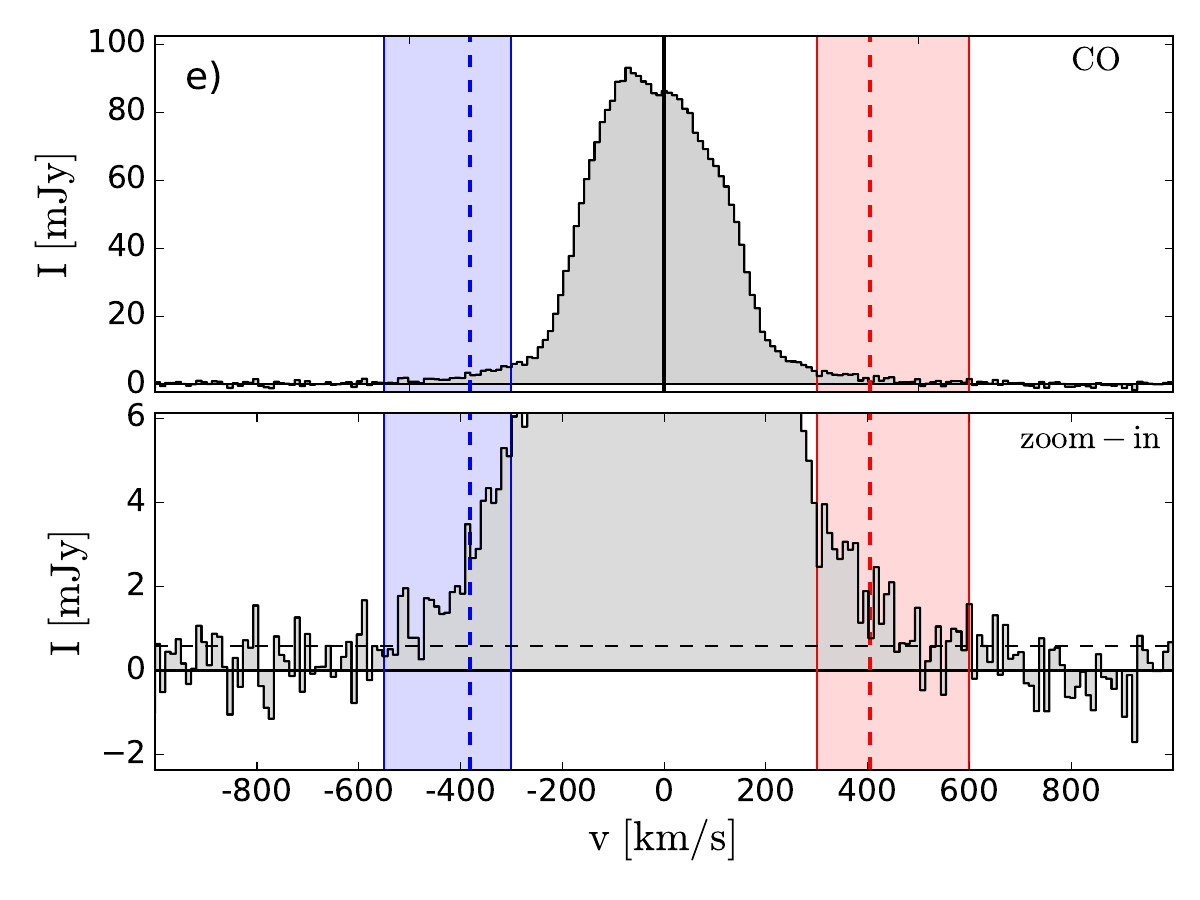} 
\includegraphics[width=0.27\textwidth]{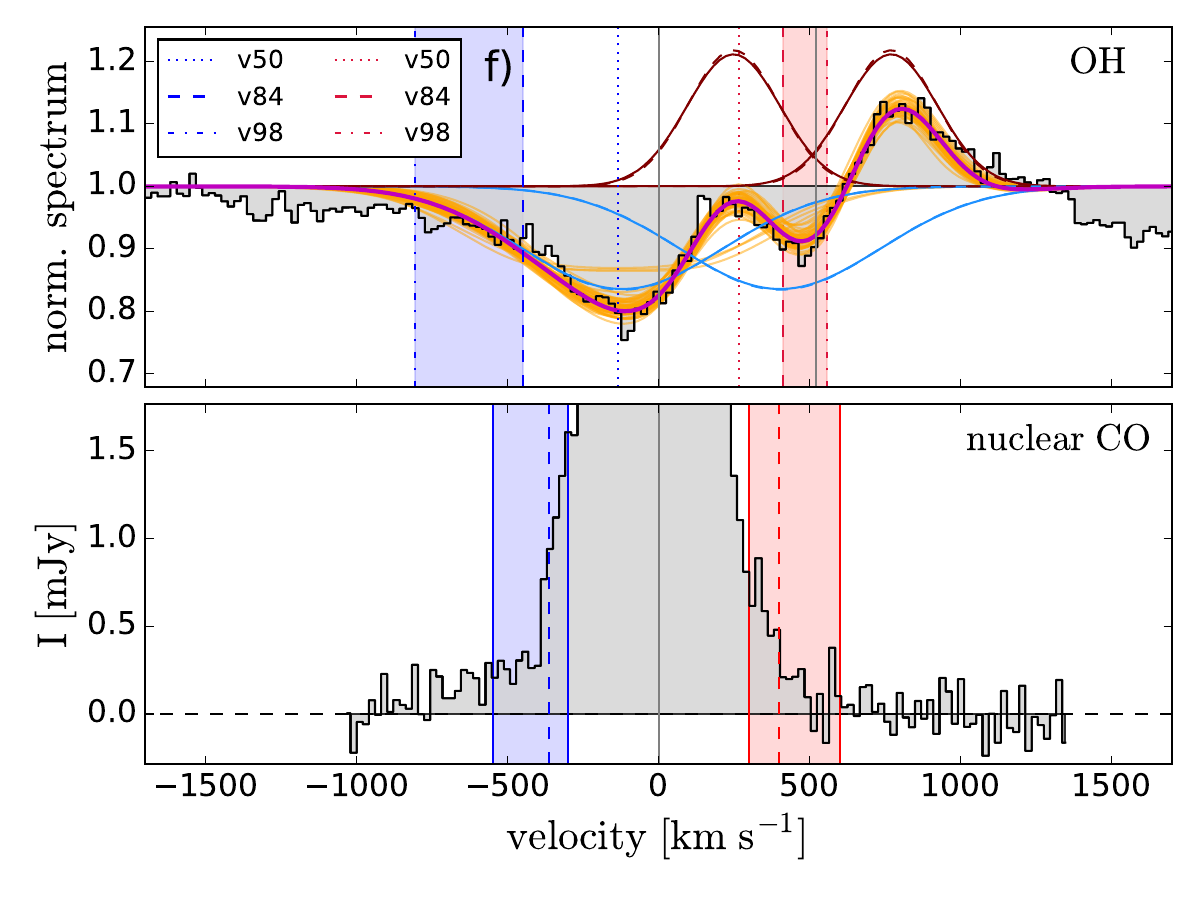}\\ 

\includegraphics[width=0.75\textwidth]{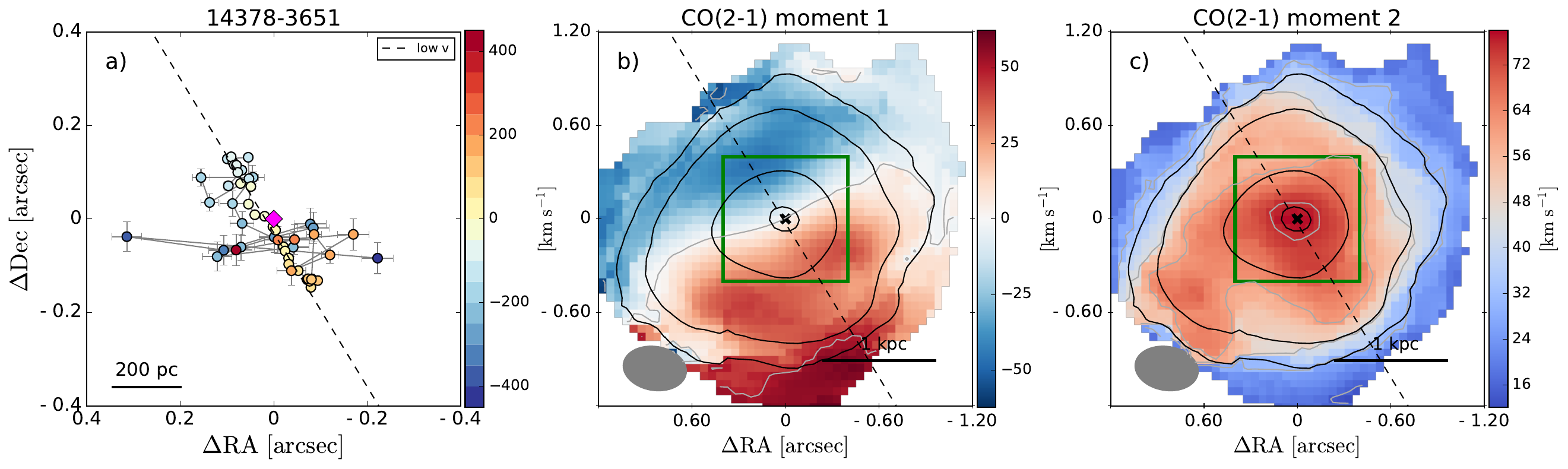}\\ 
\includegraphics[width=0.23\textwidth]{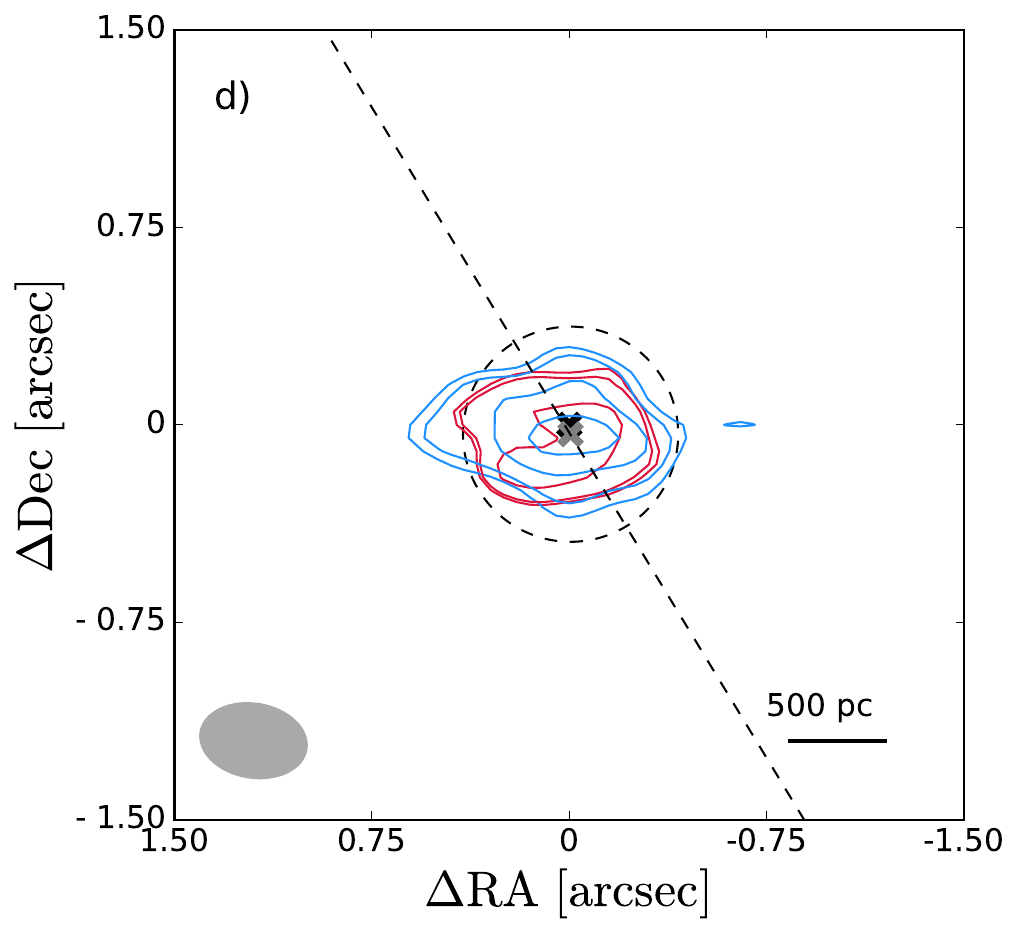} 
\includegraphics[width=0.27\textwidth]{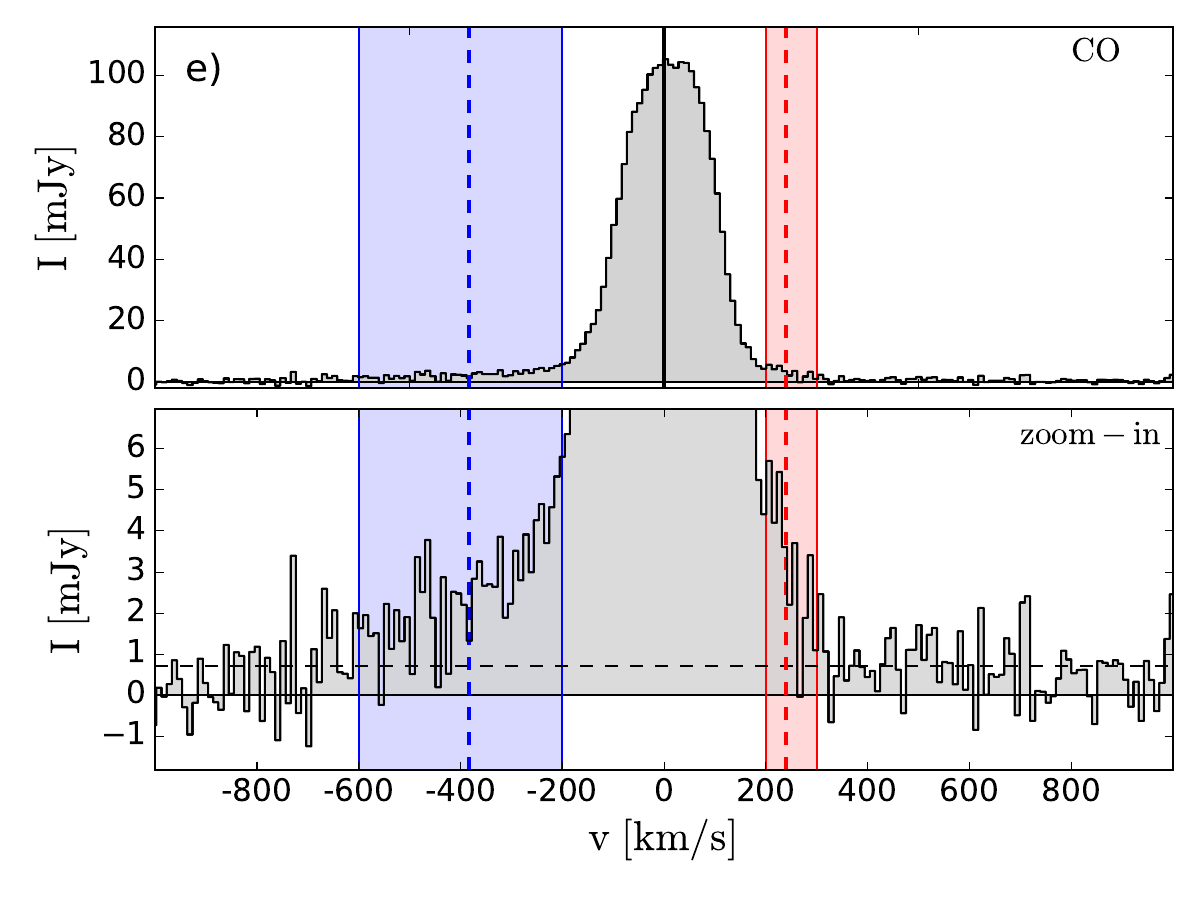} 
\includegraphics[width=0.27\textwidth]{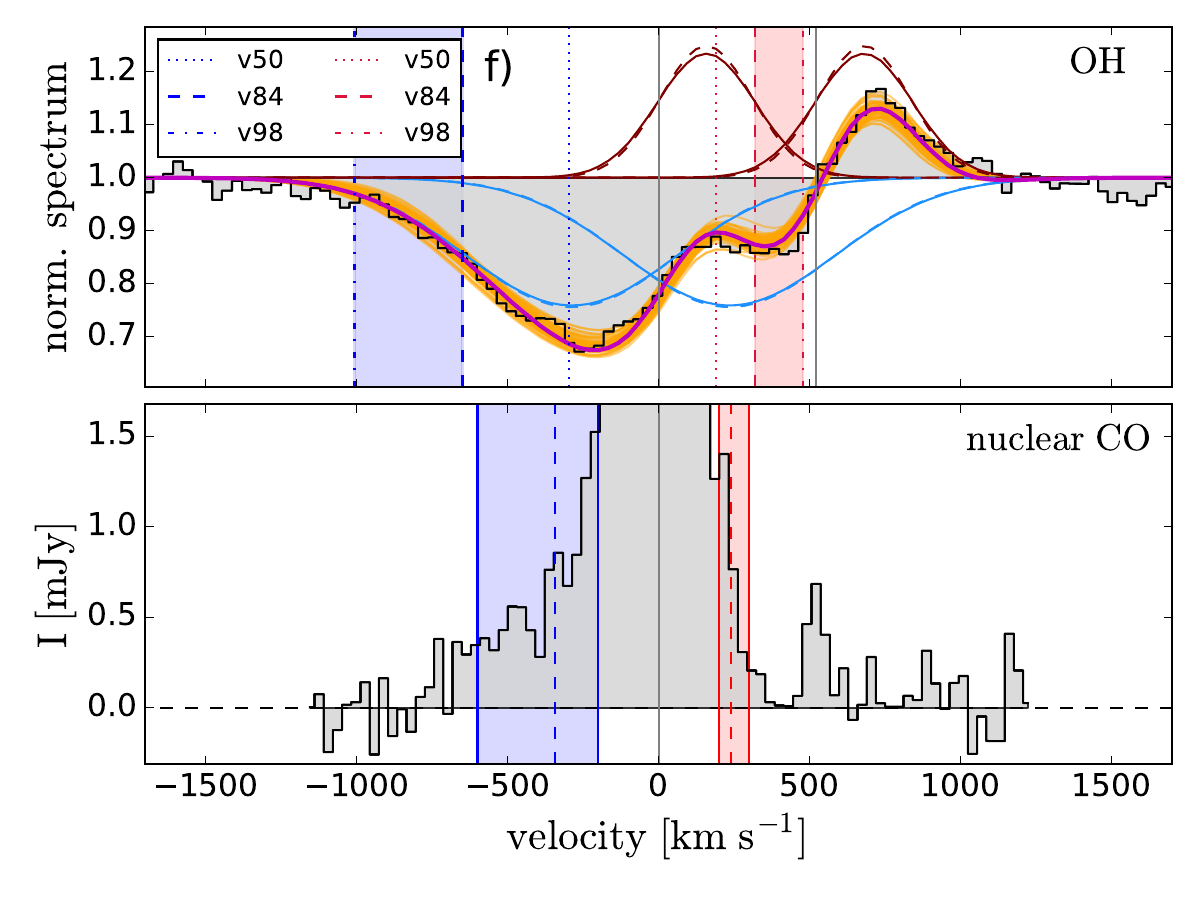}\\

\caption{continued.} 
 \end{figure*}

\begin{figure*}\ContinuedFloat 
\centering 
\includegraphics[width=0.75\textwidth]{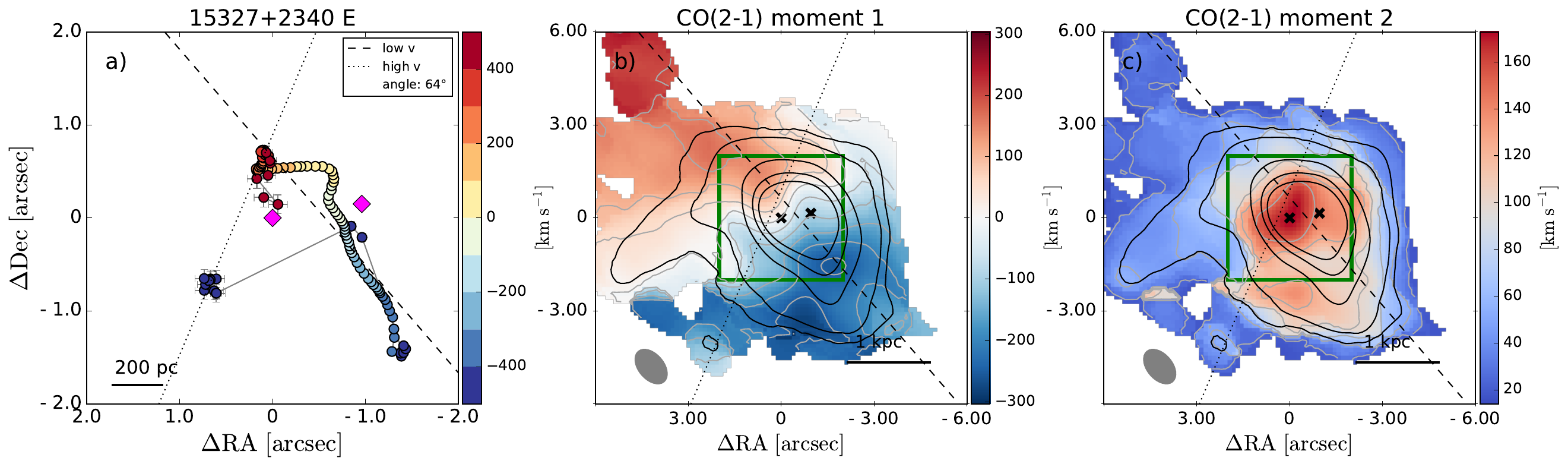}\\ 
\includegraphics[width=0.23\textwidth]{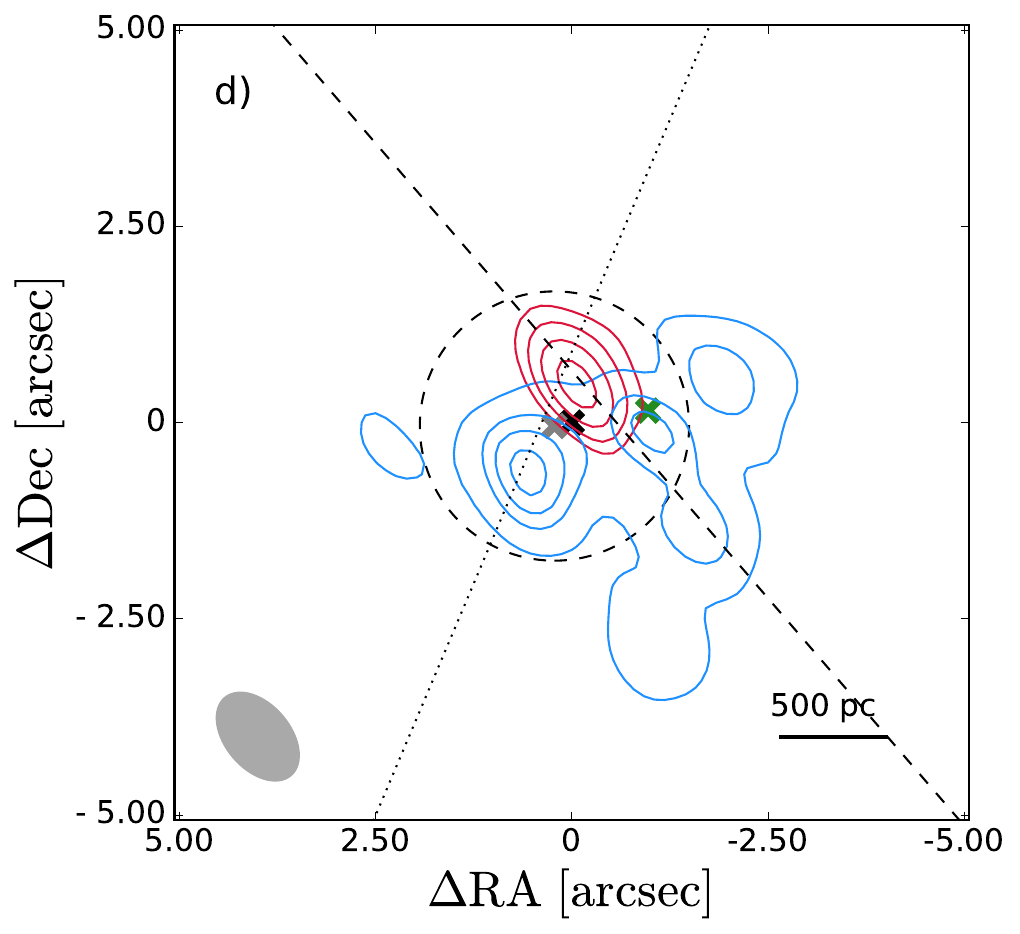} 
\includegraphics[width=0.27\textwidth]{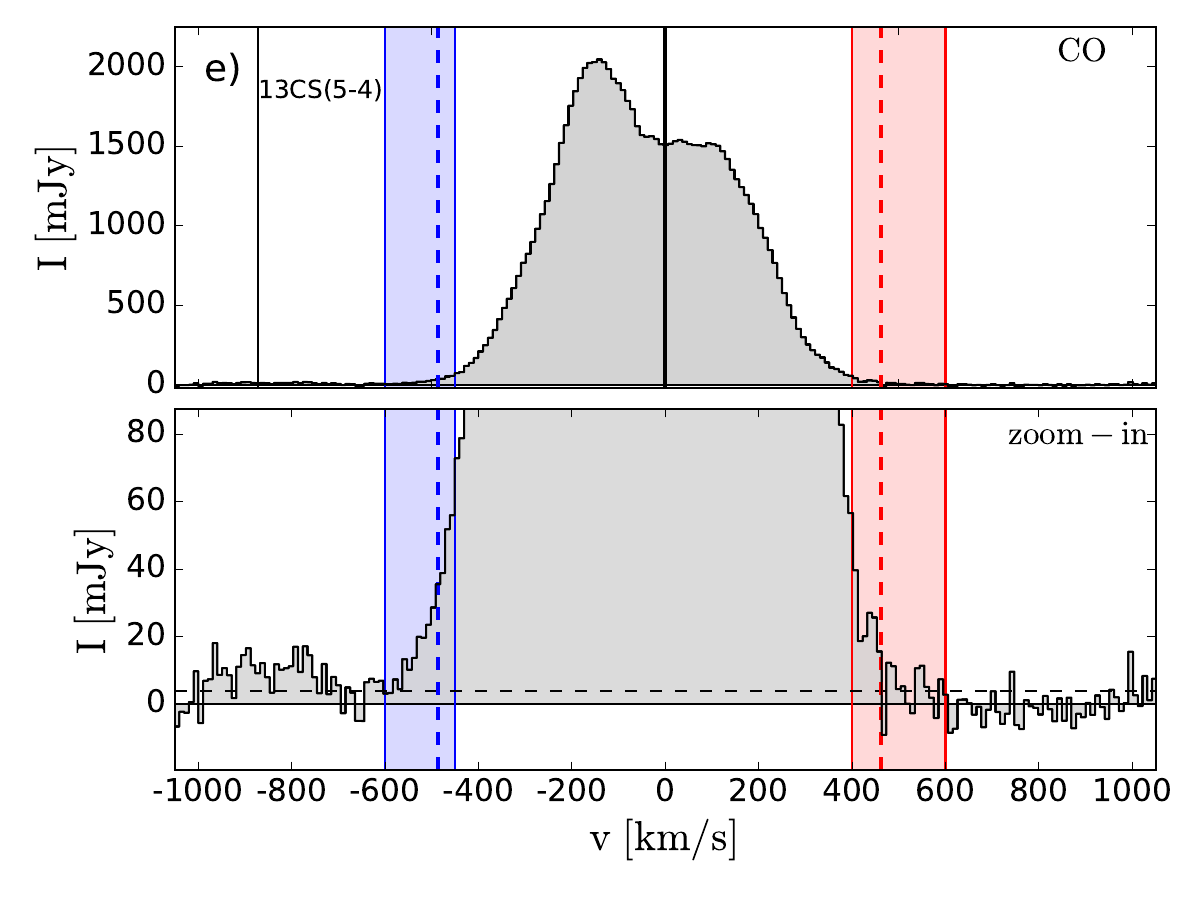} 
\includegraphics[width=0.27\textwidth]{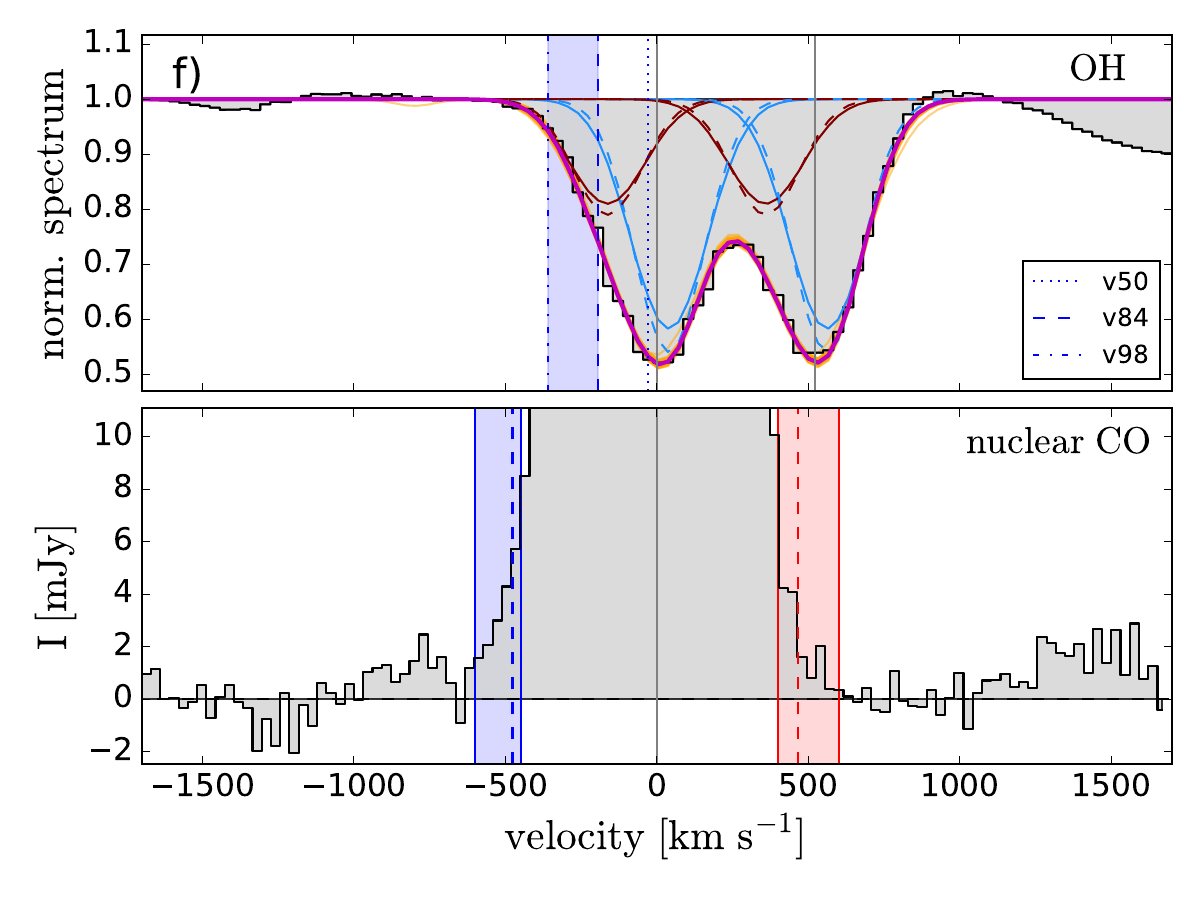}\\ 

\includegraphics[width=0.75\textwidth]{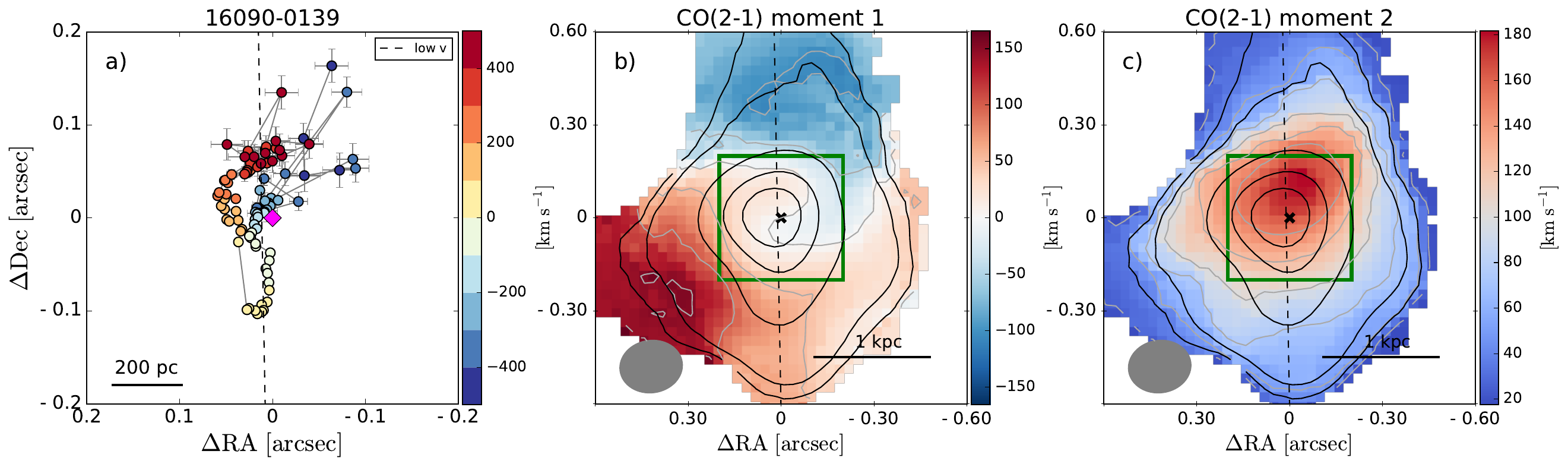}\\ 
\includegraphics[width=0.23\textwidth]{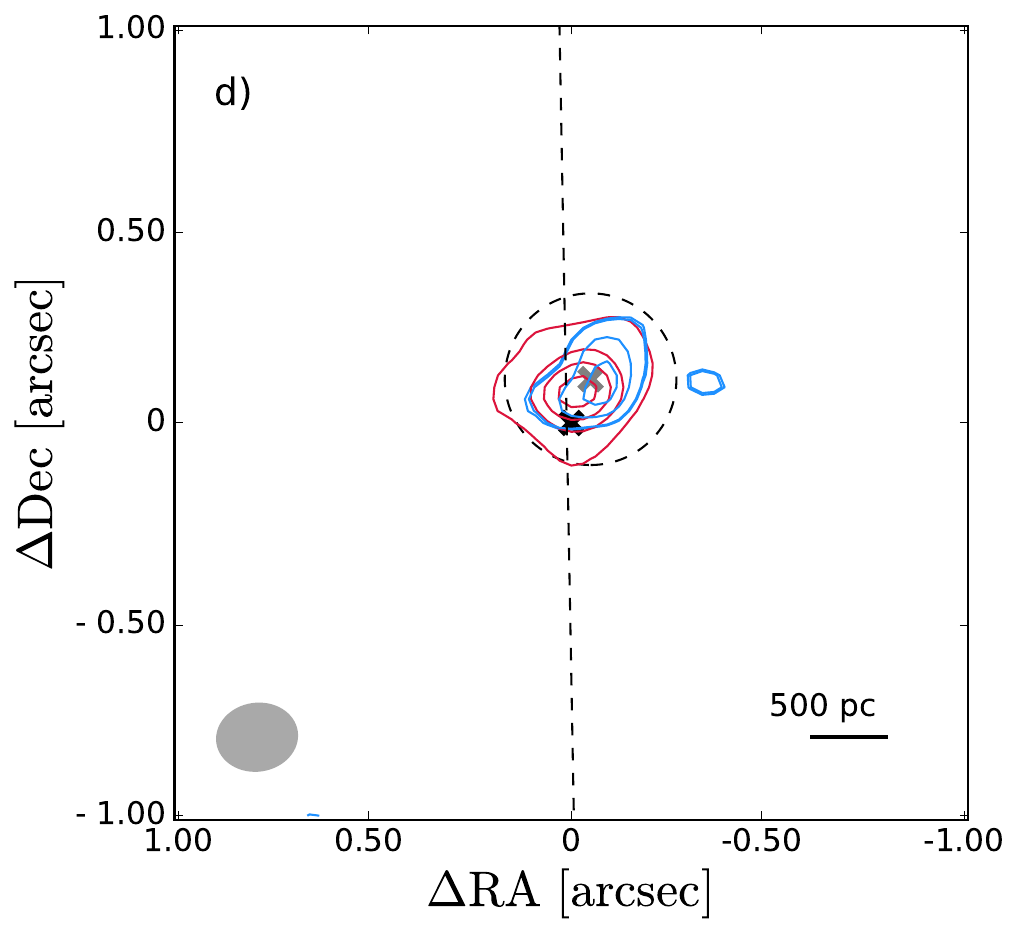} 
\includegraphics[width=0.27\textwidth]{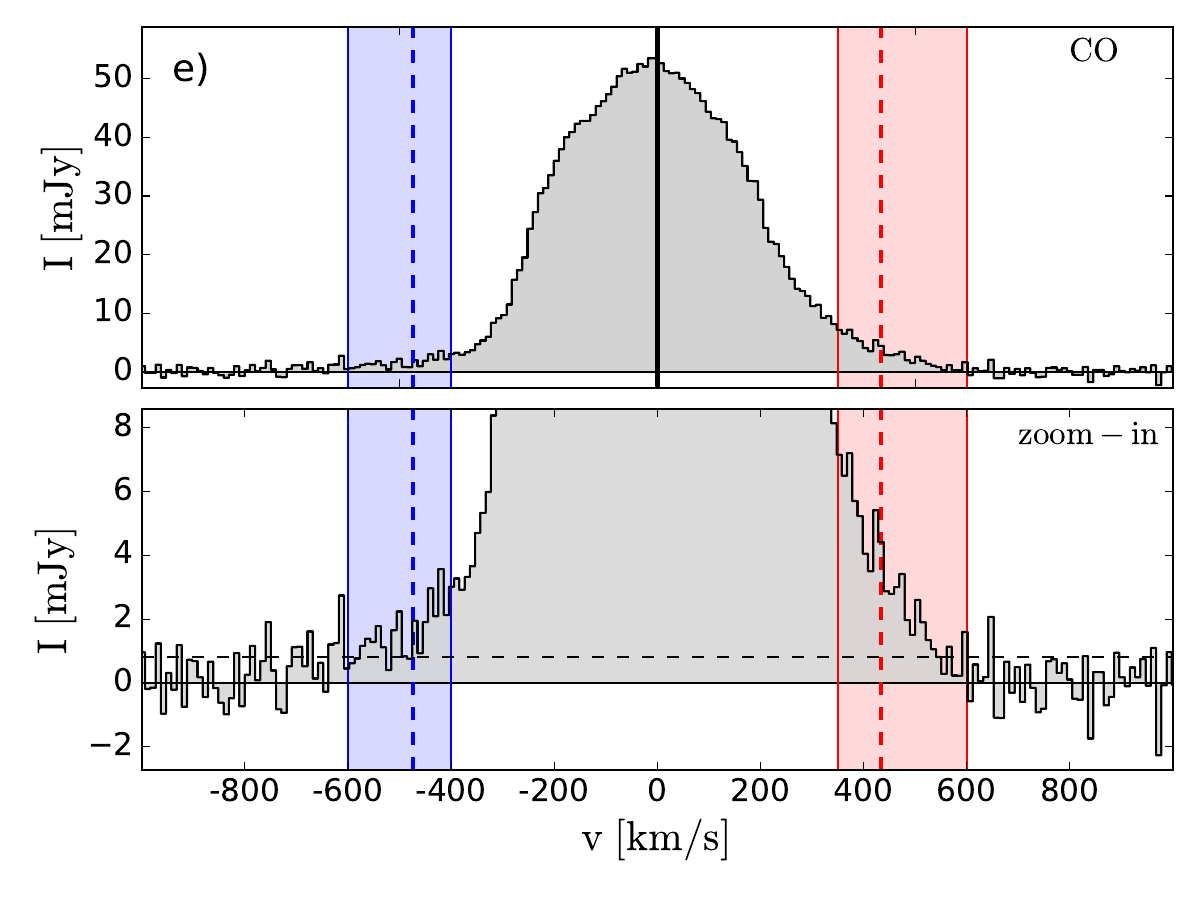} 
\includegraphics[width=0.27\textwidth]{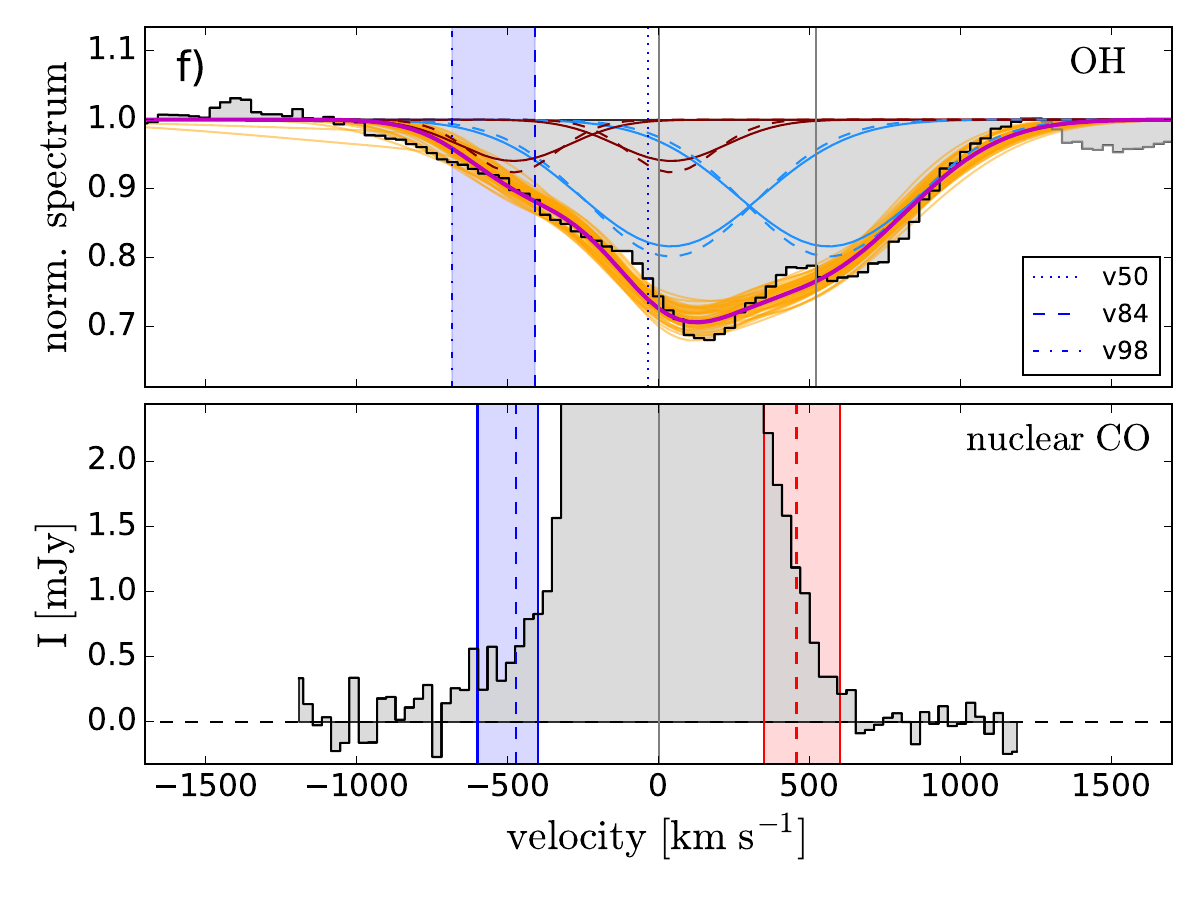}\\ 

\includegraphics[width=0.75\textwidth]{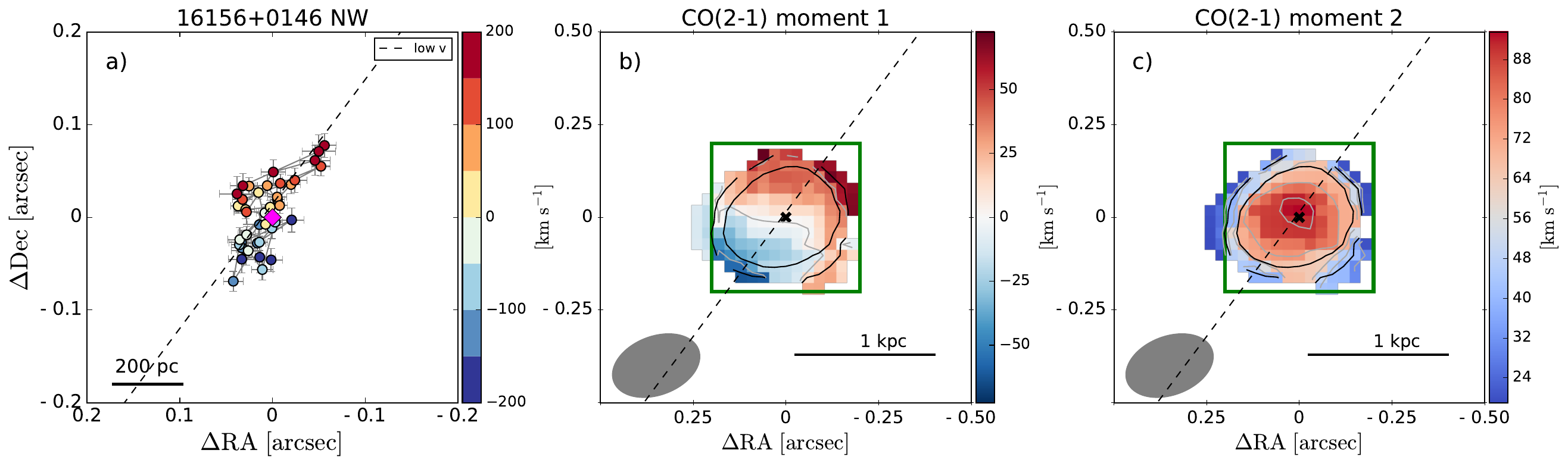}\\ 
\includegraphics[width=0.27\textwidth]{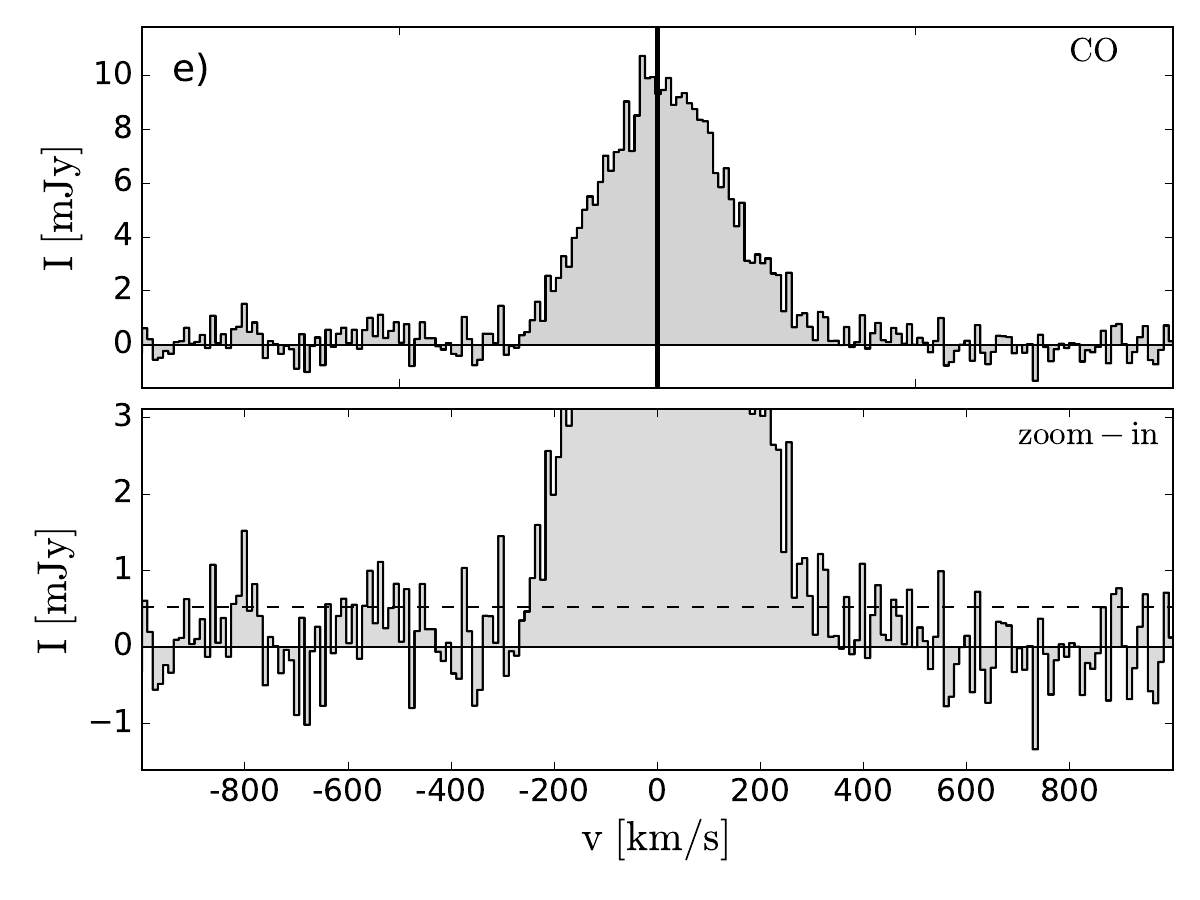}

\caption{continued.} 
 \end{figure*}

\begin{figure*}\ContinuedFloat 
\centering 
\includegraphics[width=0.75\textwidth]{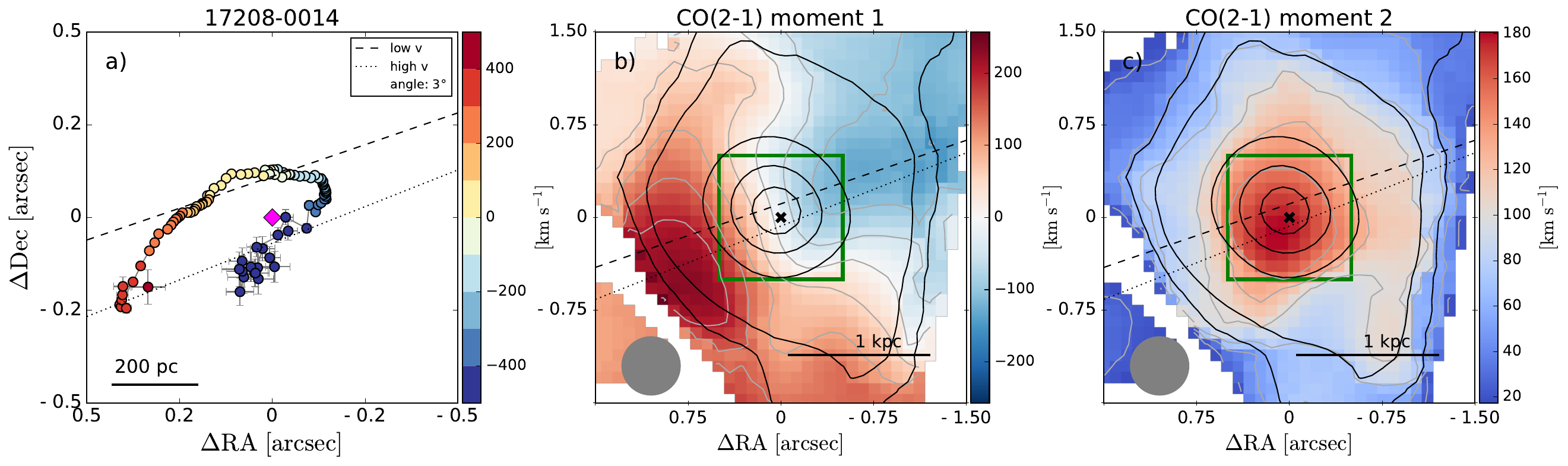}\\ 
\includegraphics[width=0.23\textwidth]{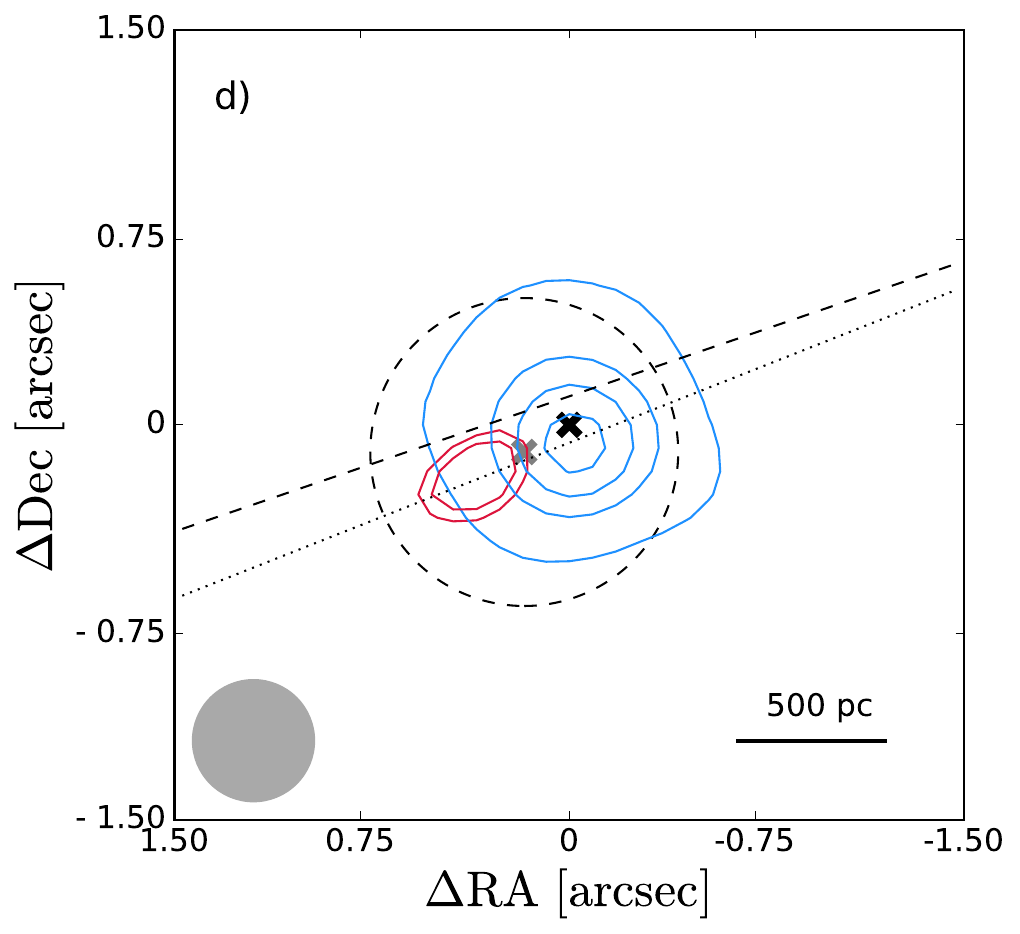} 
\includegraphics[width=0.27\textwidth]{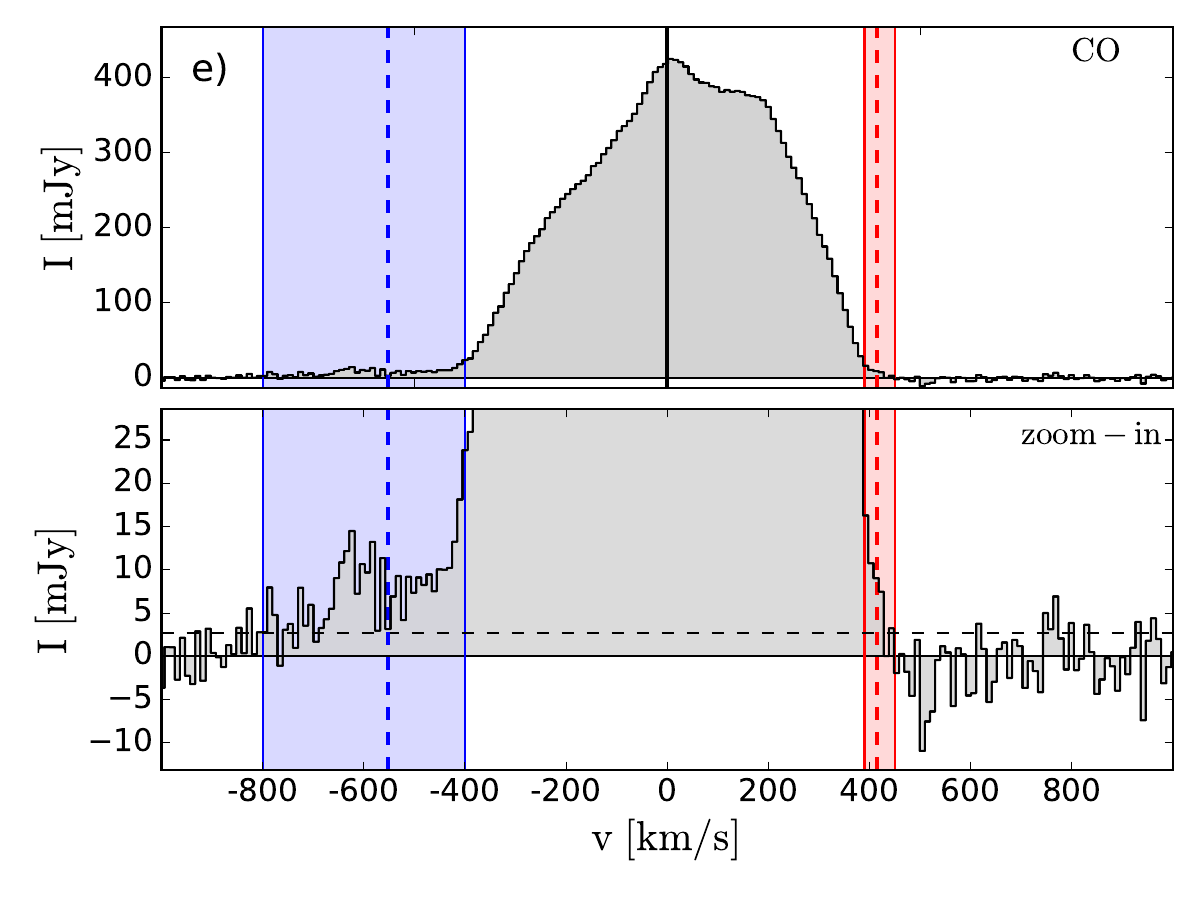} 
\includegraphics[width=0.27\textwidth]{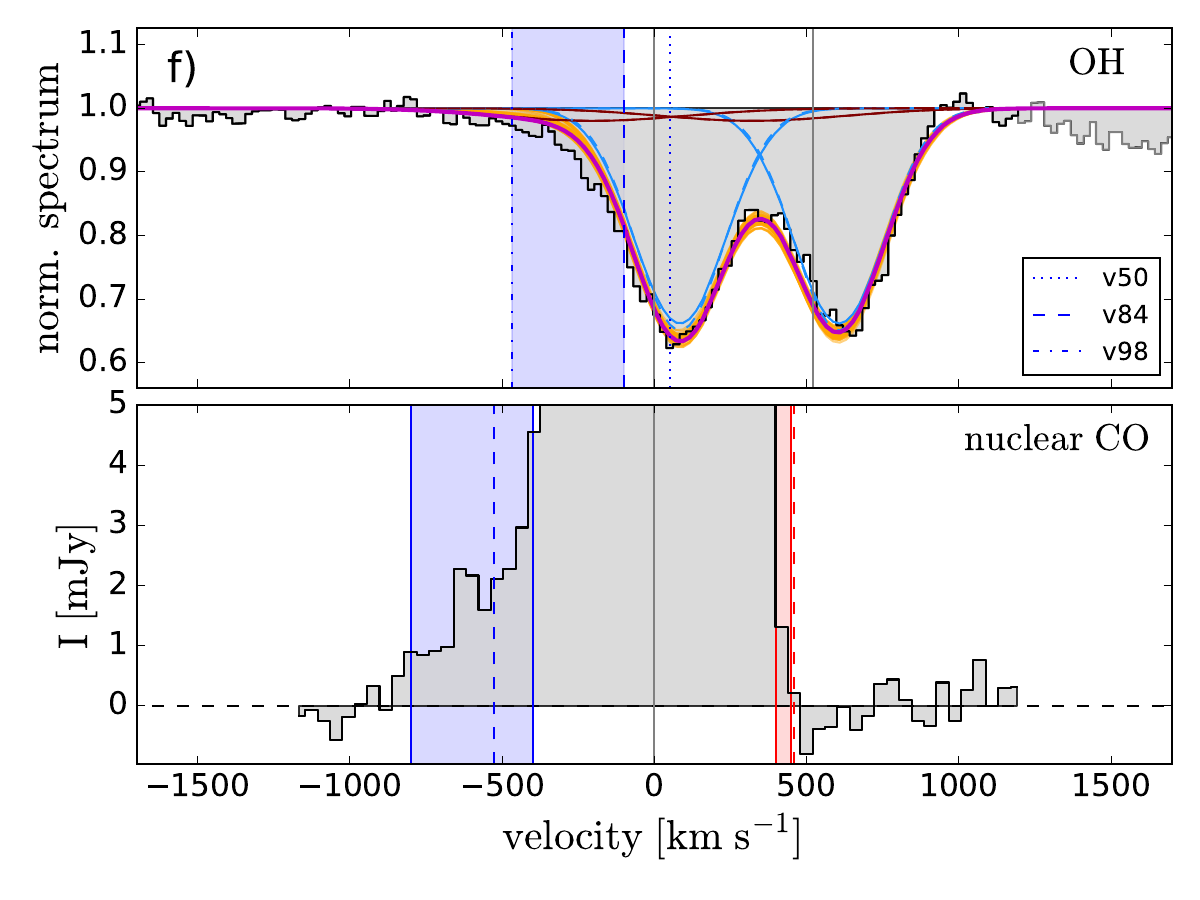}\\ 

\includegraphics[width=0.75\textwidth]{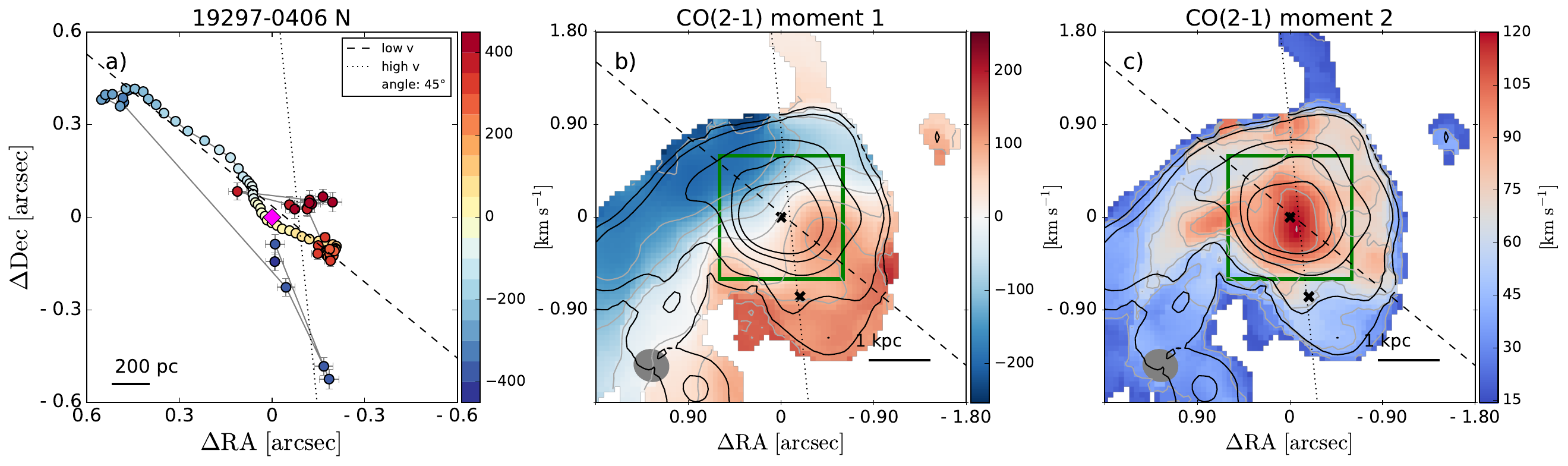}\\ 
\includegraphics[width=0.23\textwidth]{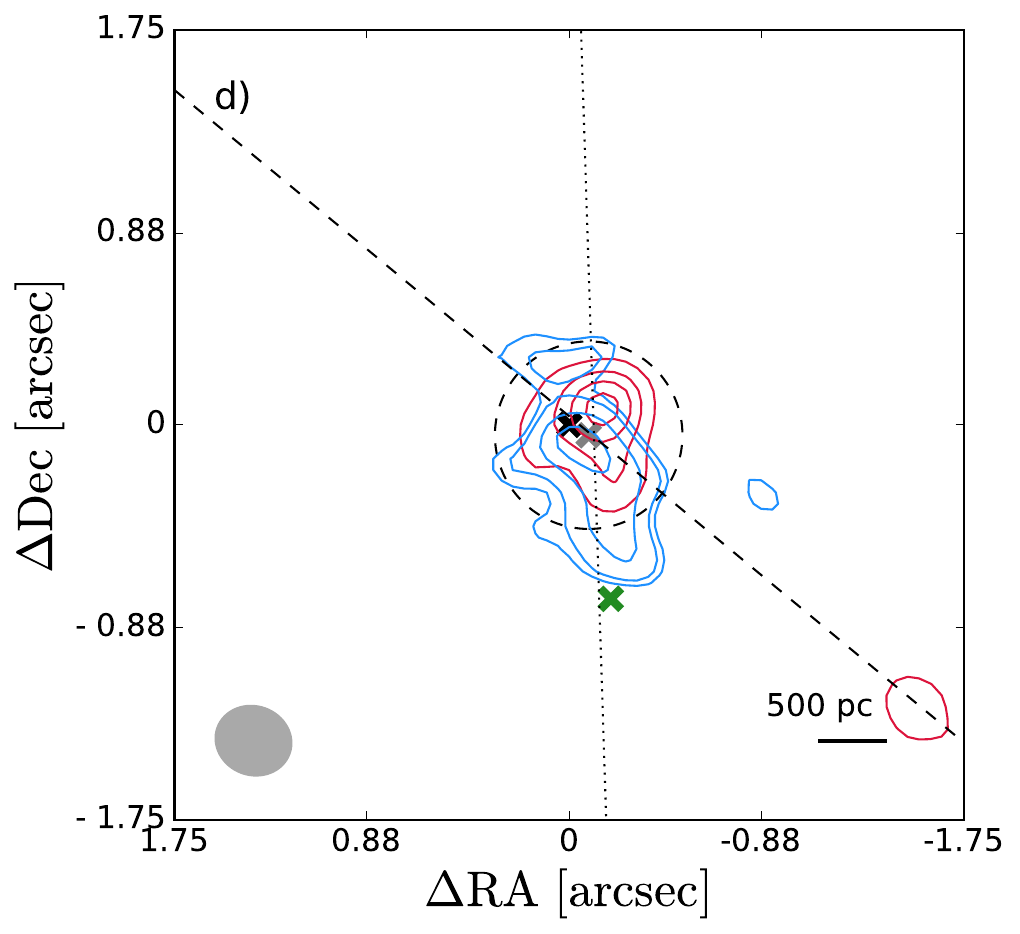} 
\includegraphics[width=0.27\textwidth]{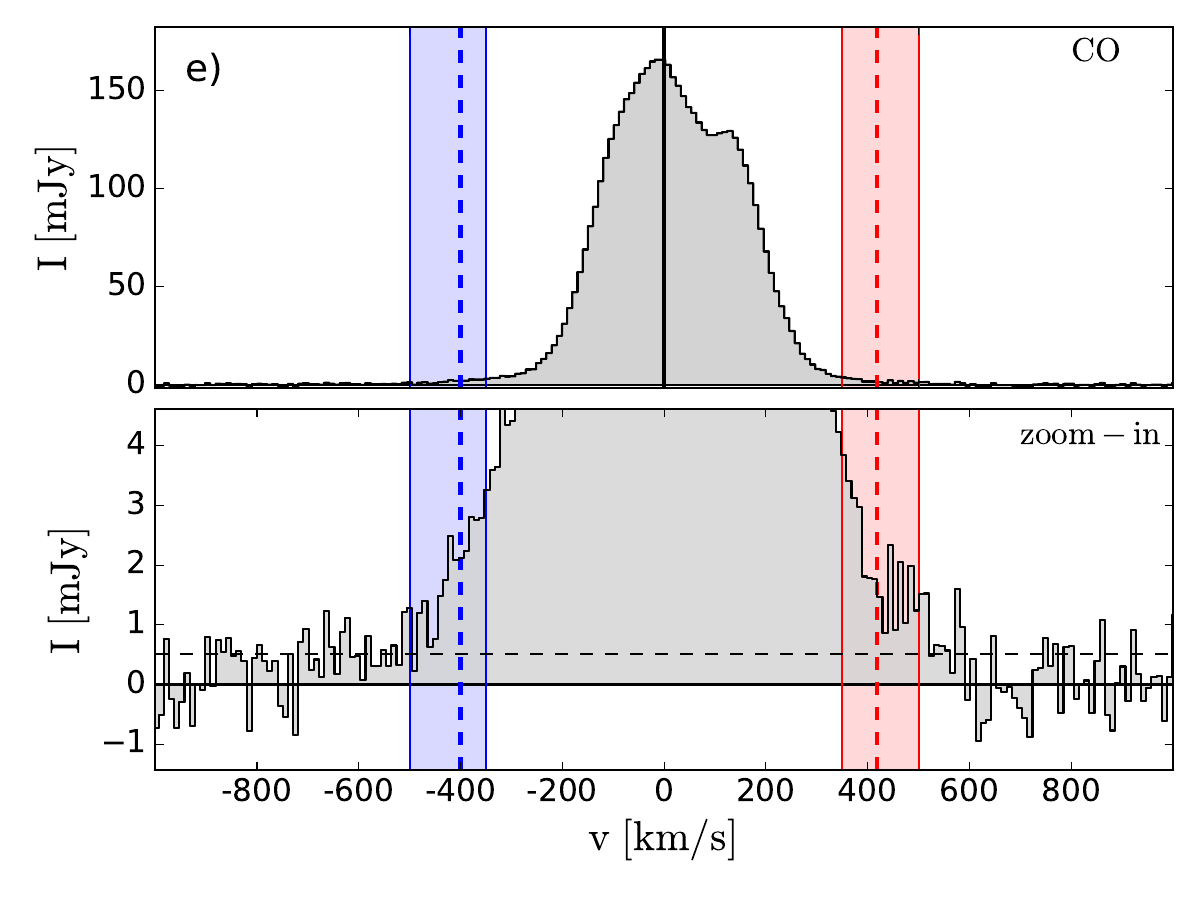} 
\includegraphics[width=0.27\textwidth]{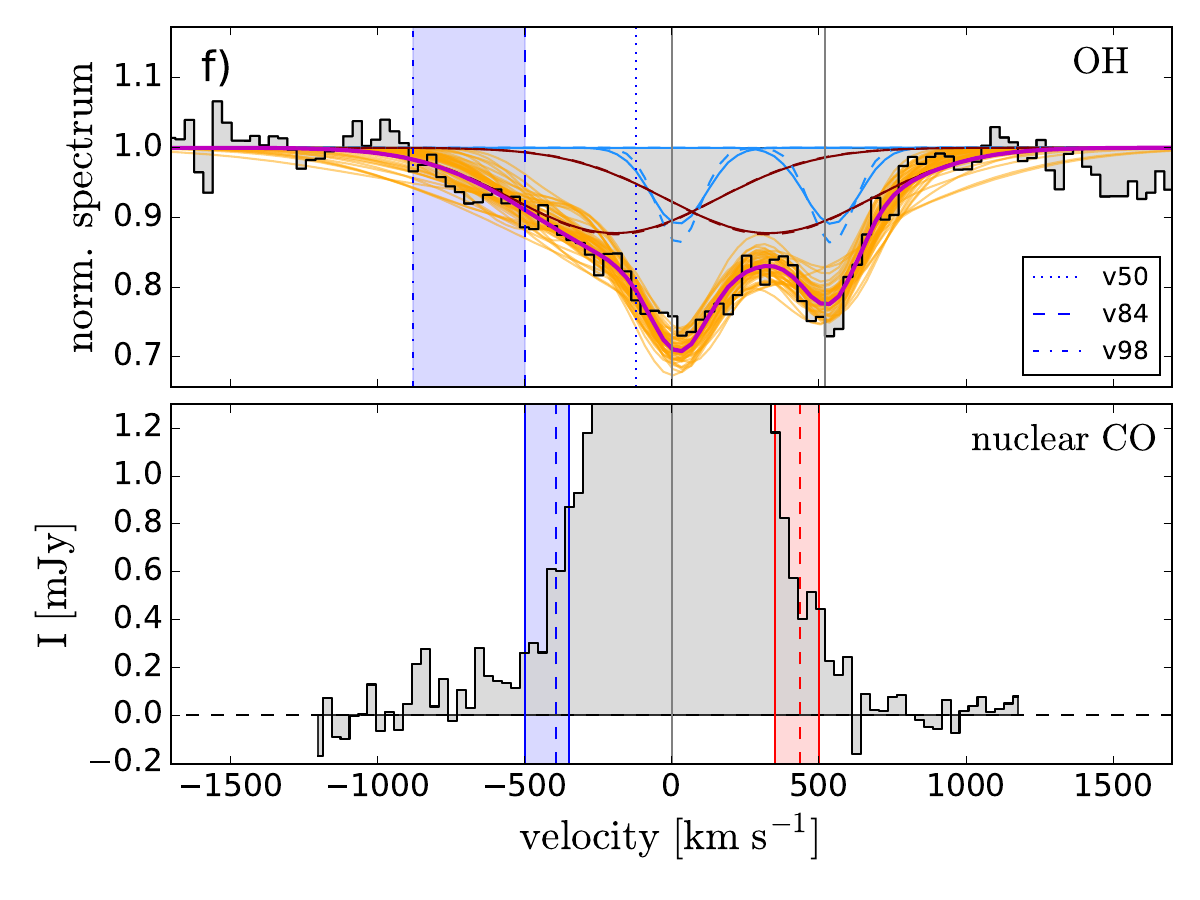}\\ 

\includegraphics[width=0.75\textwidth]{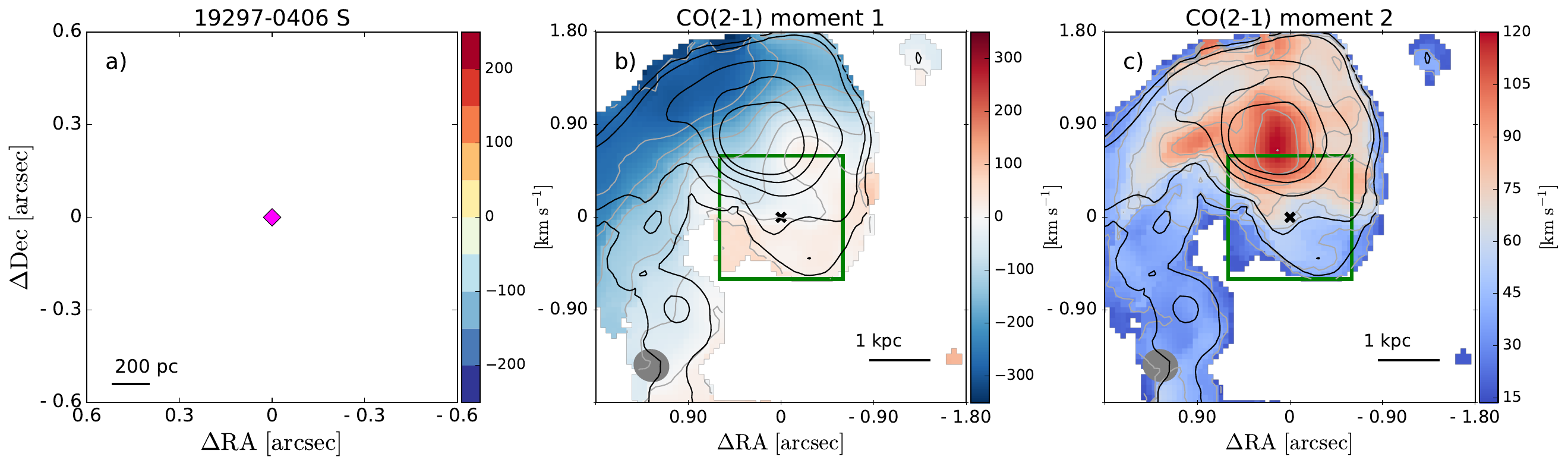}\\ 
\includegraphics[width=0.27\textwidth]{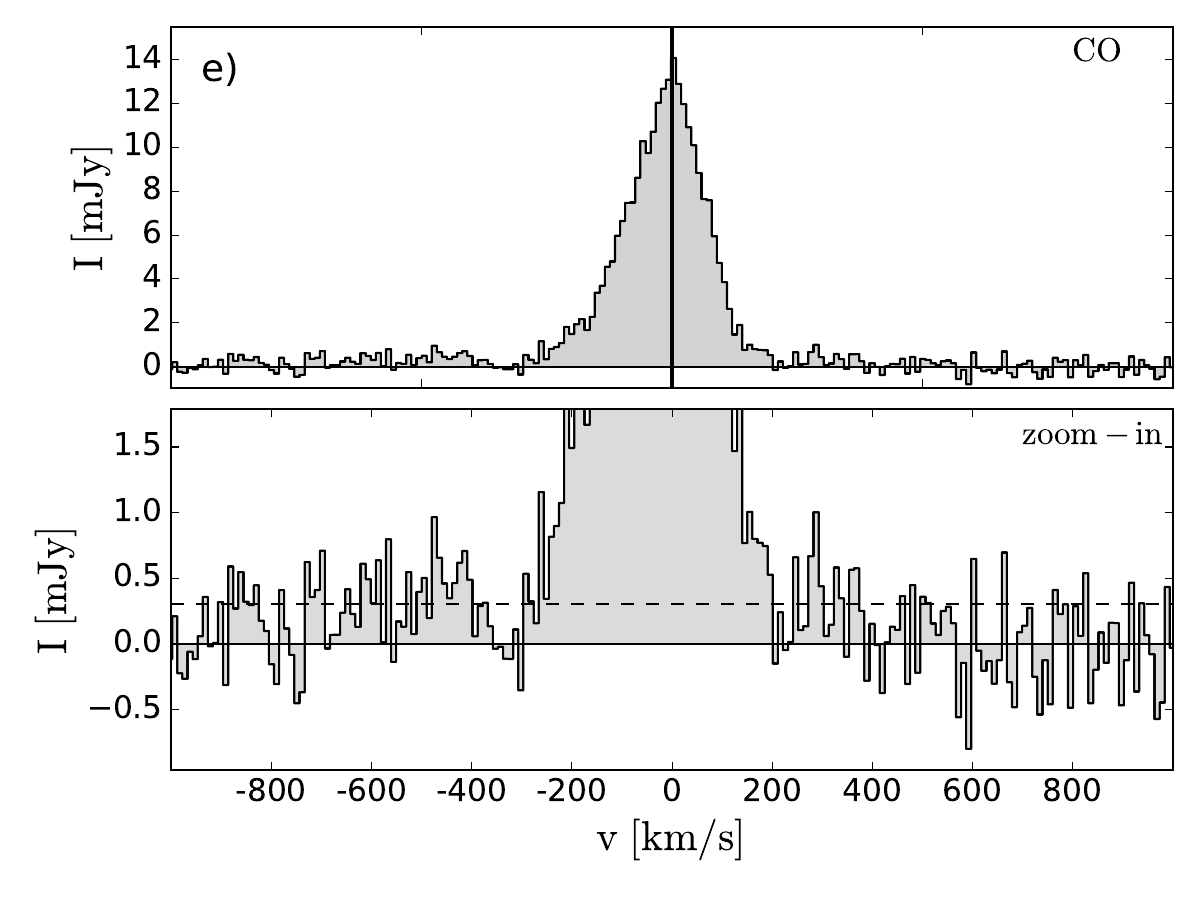} 

\caption{continued.} 
 \end{figure*}

\begin{figure*}\ContinuedFloat 
\centering 
\includegraphics[width=0.75\textwidth]{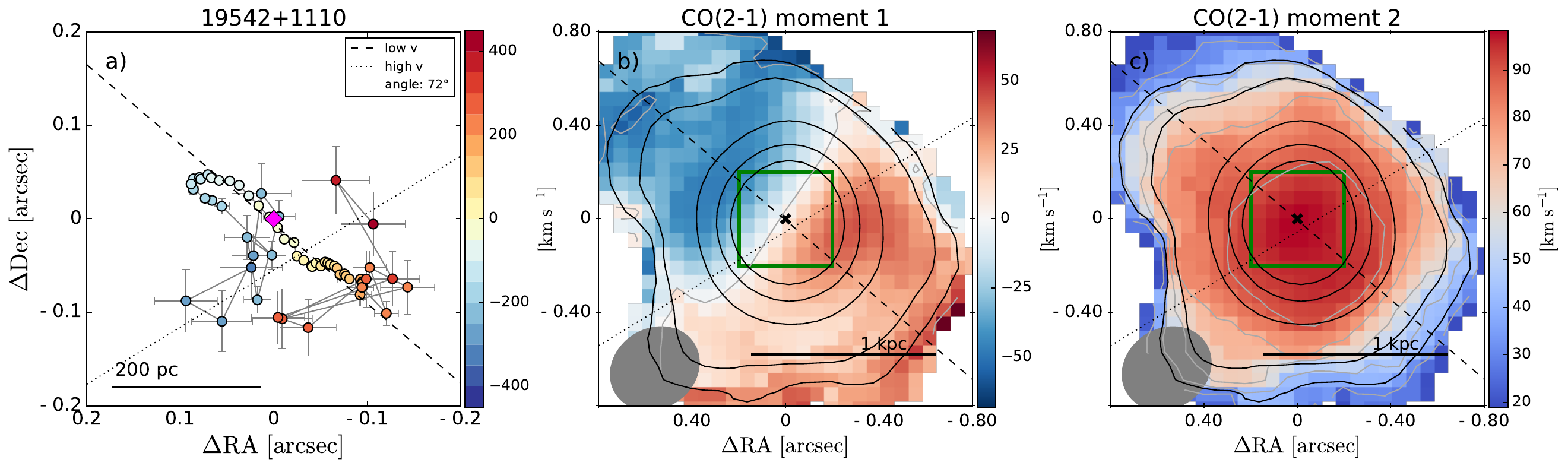}\\ 
\includegraphics[width=0.23\textwidth]{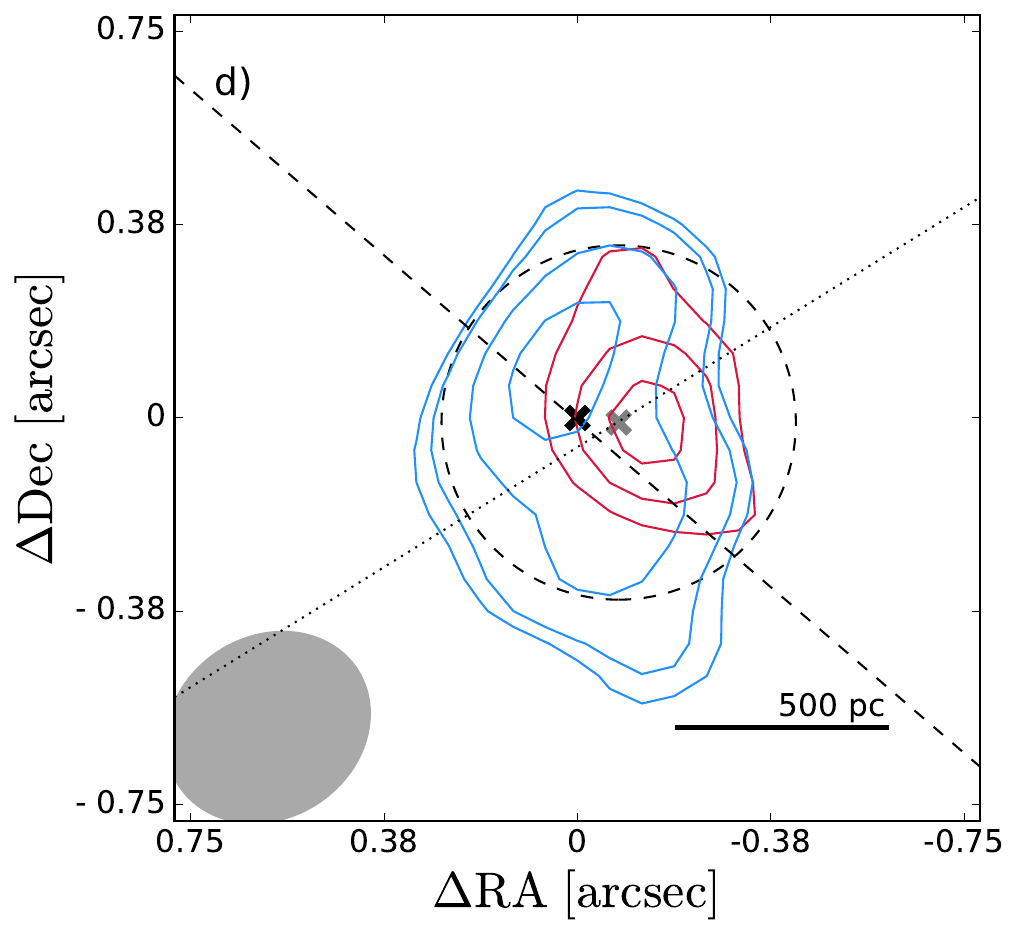} 
\includegraphics[width=0.27\textwidth]{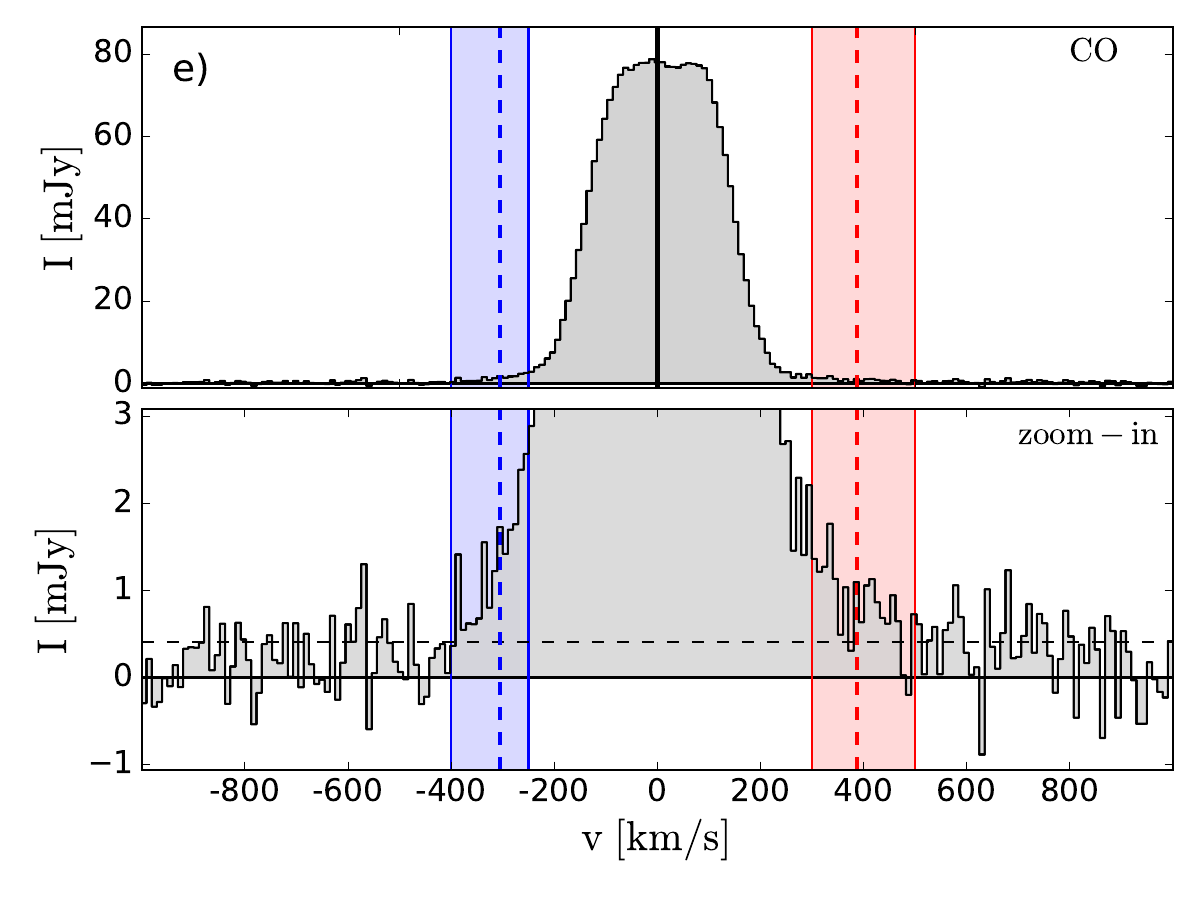} 
\includegraphics[width=0.27\textwidth]{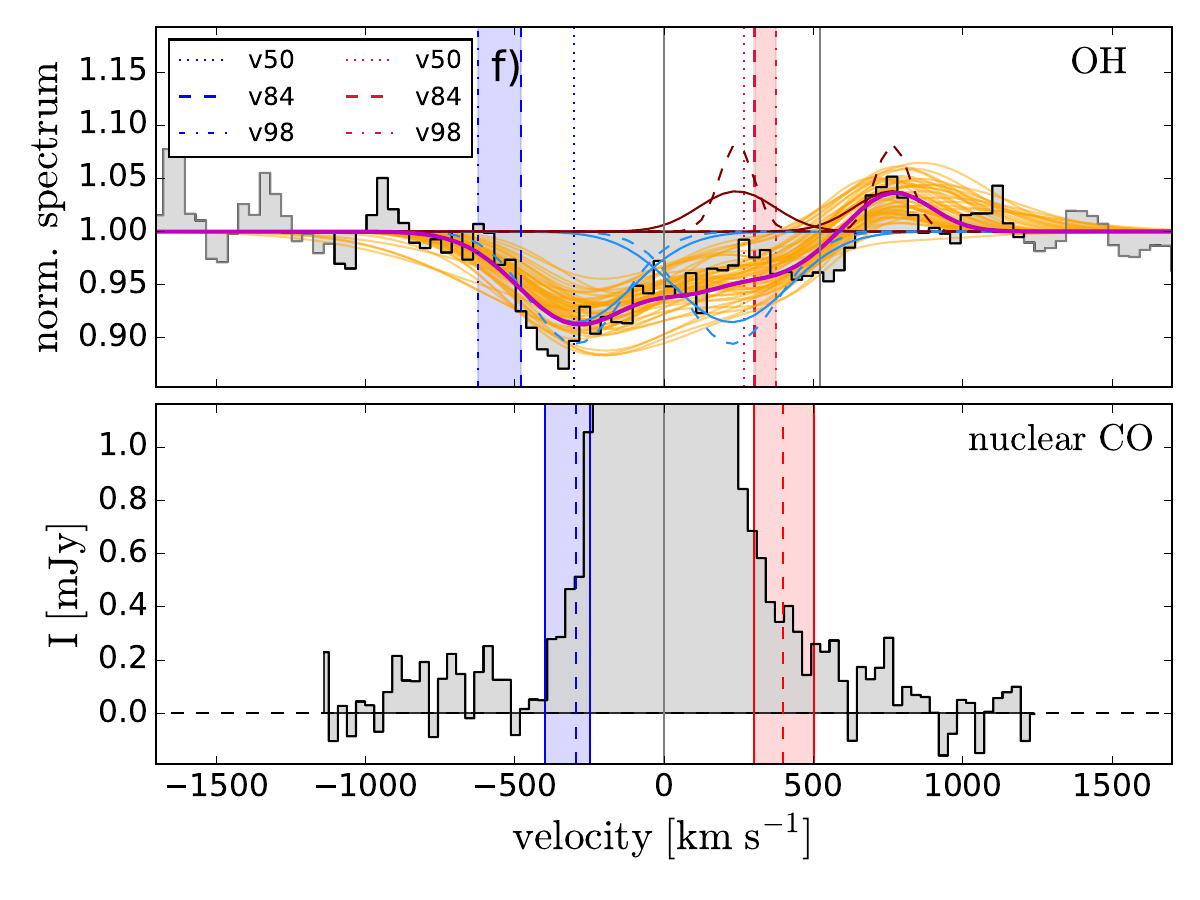}\\ 

\includegraphics[width=0.75\textwidth]{Figures/Spectroastrometry/IRAS_20087-0308_CO21_ch_pos_moment_maps_s1.pdf}\\ 
\includegraphics[width=0.23\textwidth]{Figures/Outflow_maps/IRAS_20087-0308_b_r_channels_radius_fit.pdf} 
\includegraphics[width=0.27\textwidth]{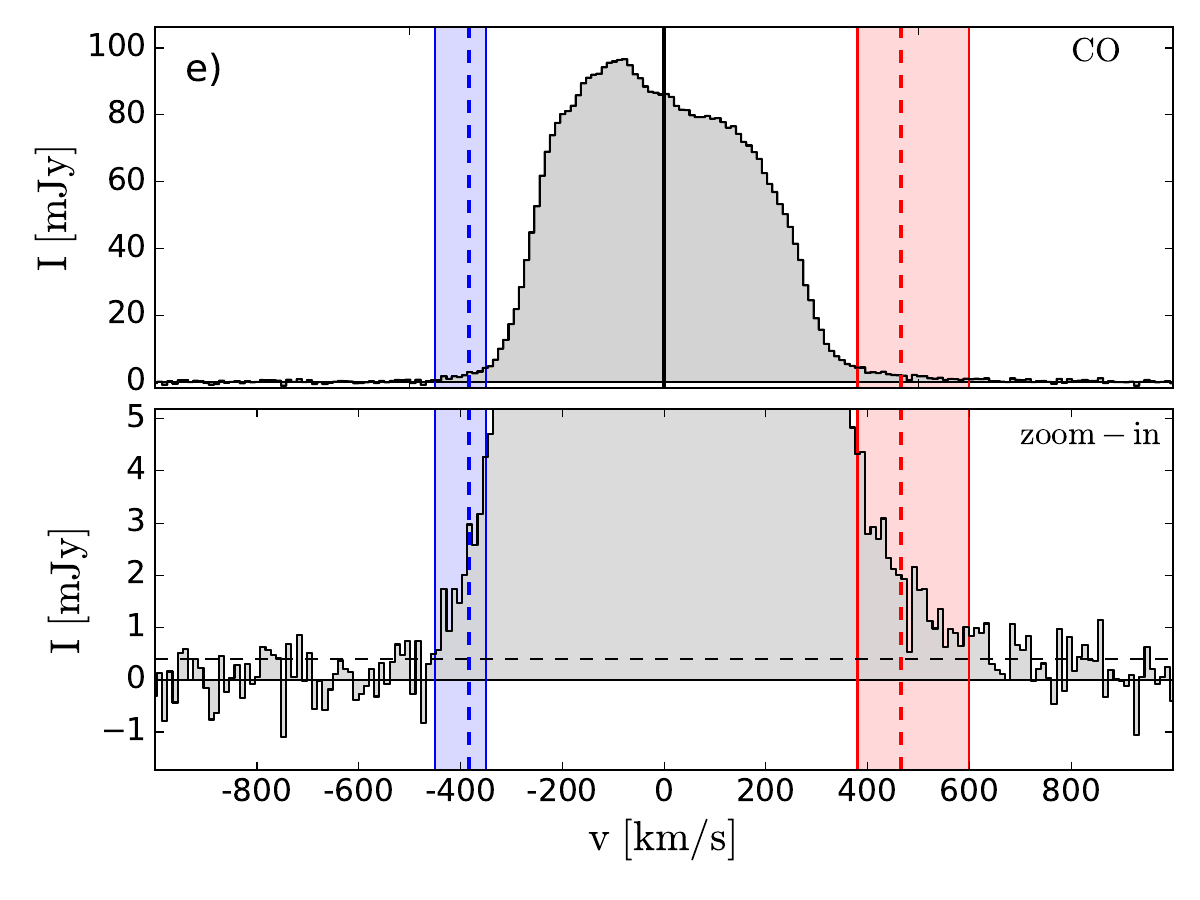} 
\includegraphics[width=0.27\textwidth]{Figures/Spectra_OH/CO_OH_spectra/20087-0308_fit_OH_spectrum_CO_paper.pdf}\\ 

\includegraphics[width=0.75\textwidth]{Figures/Spectroastrometry/IRAS_20100-4156_CO21_ch_pos_moment_maps_s1.pdf}\\ 
\includegraphics[width=0.23\textwidth]{Figures/Outflow_maps/IRAS_20100-4156_b_r_channels_radius_fit.pdf} 
\includegraphics[width=0.27\textwidth]{Figures/Spectra/IRAS_20100-4156_CO21_spectrum_zoom_s1.pdf} 
\includegraphics[width=0.27\textwidth]{Figures/Spectra_OH/CO_OH_spectra/20100-4156_fit_OH_spectrum_CO_paper.pdf}\\ 

\caption{continued.} 
 \end{figure*}

\begin{figure*}\ContinuedFloat 
\centering 
\includegraphics[width=0.75\textwidth]{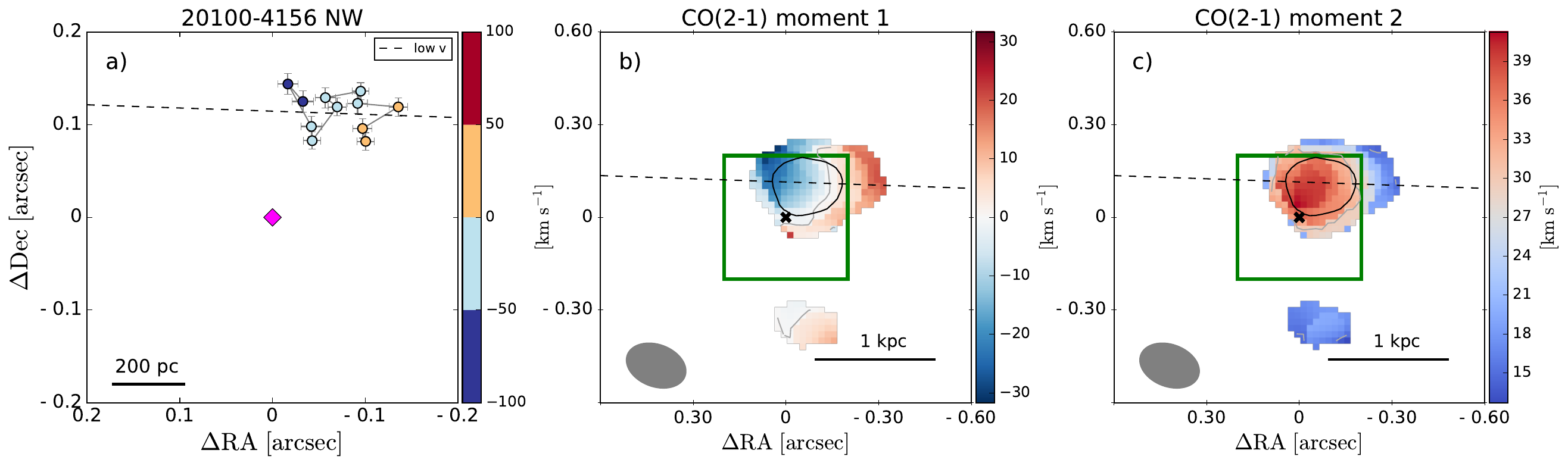}\\ 
\includegraphics[width=0.27\textwidth]{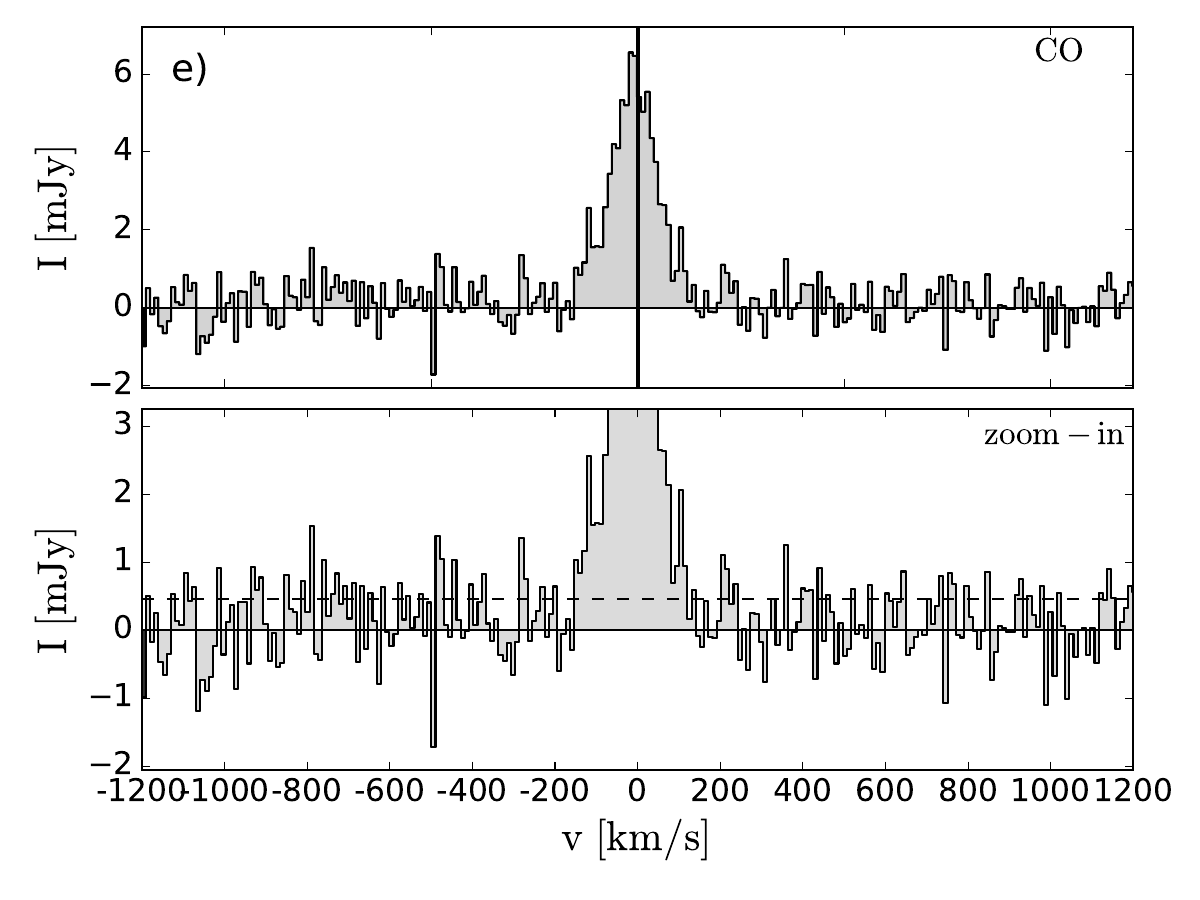} 

\includegraphics[width=0.75\textwidth]{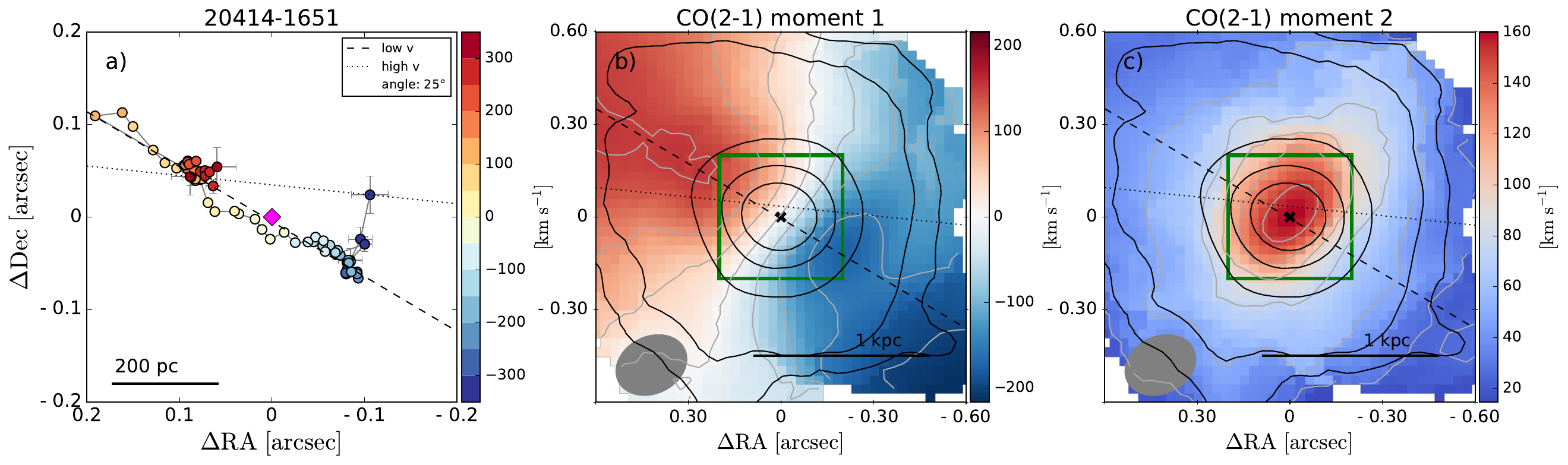}\\ 
\includegraphics[width=0.23\textwidth]{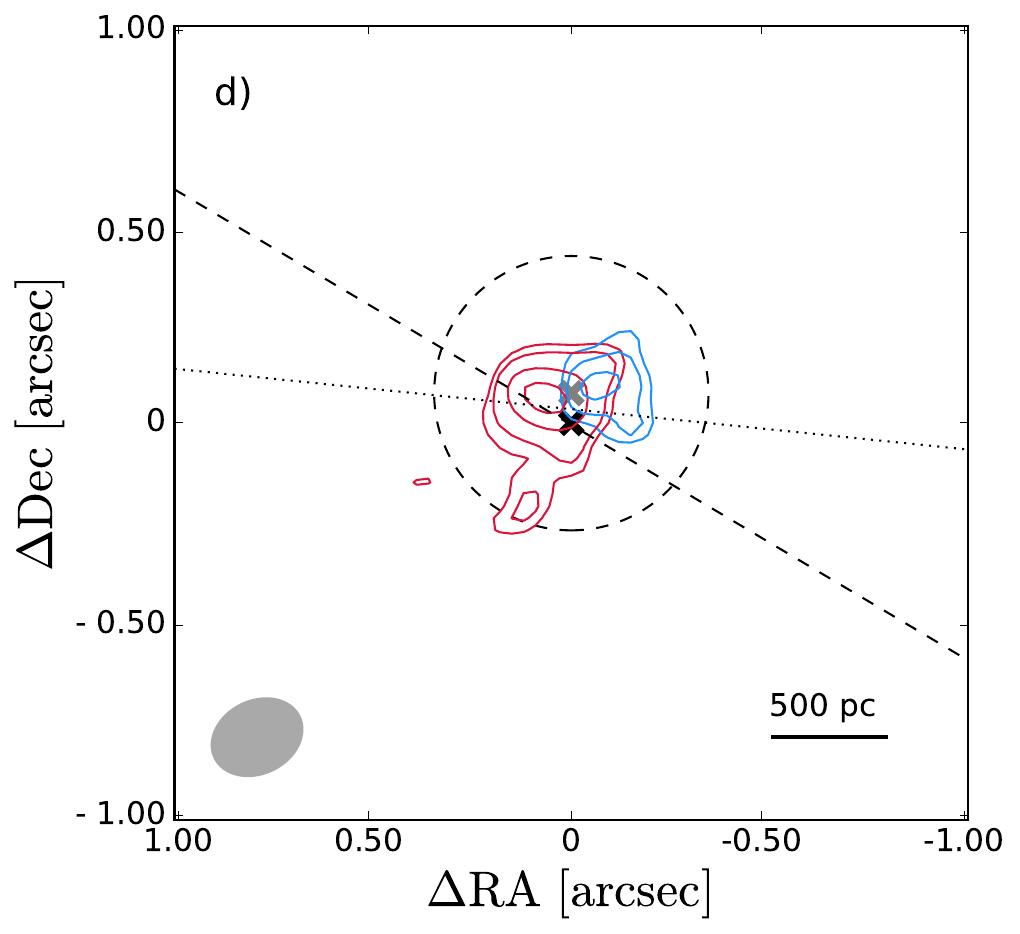} 
\includegraphics[width=0.27\textwidth]{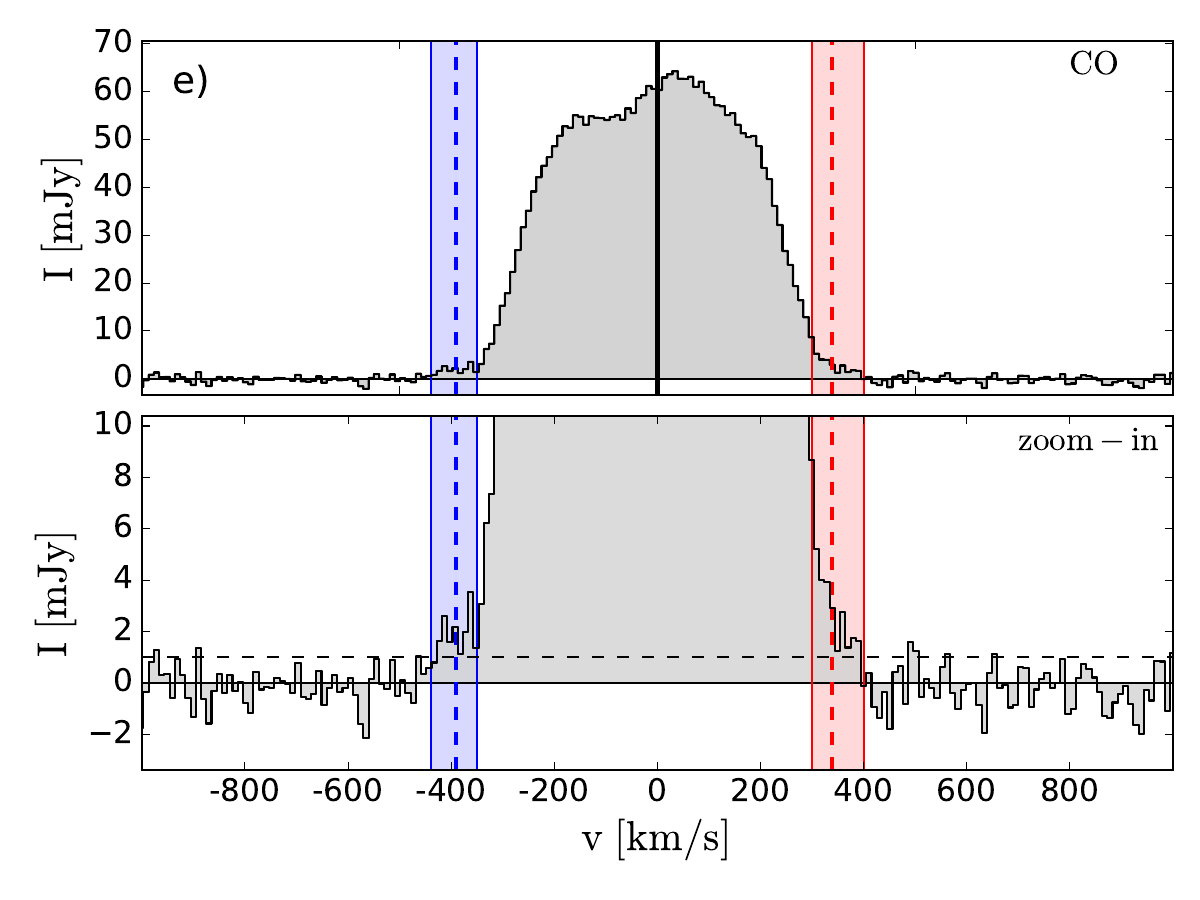} 
\includegraphics[width=0.27\textwidth]{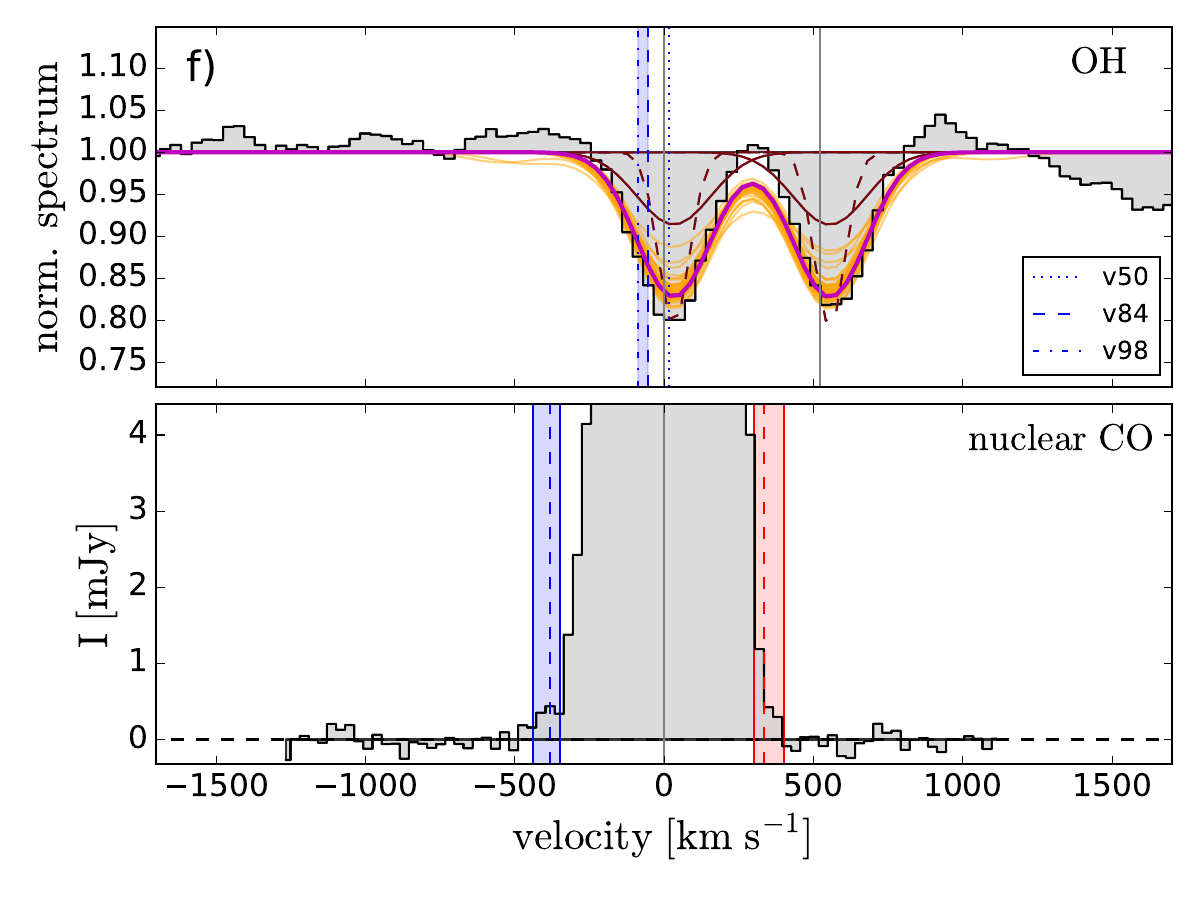}\\ 

\includegraphics[width=0.75\textwidth]{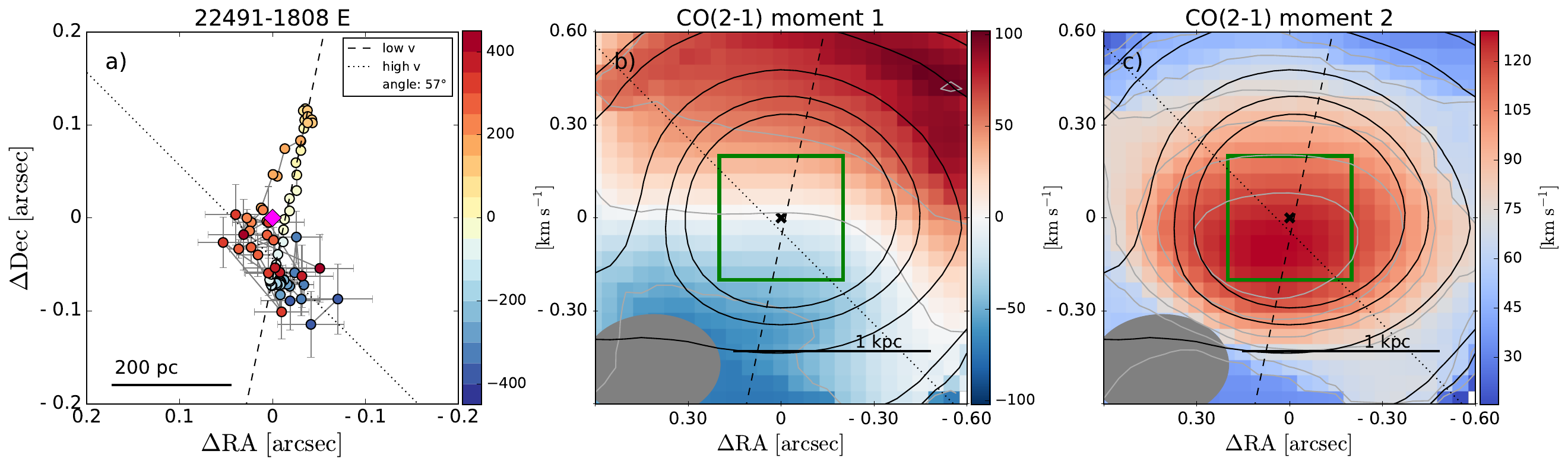}\\ 
\includegraphics[width=0.23\textwidth]{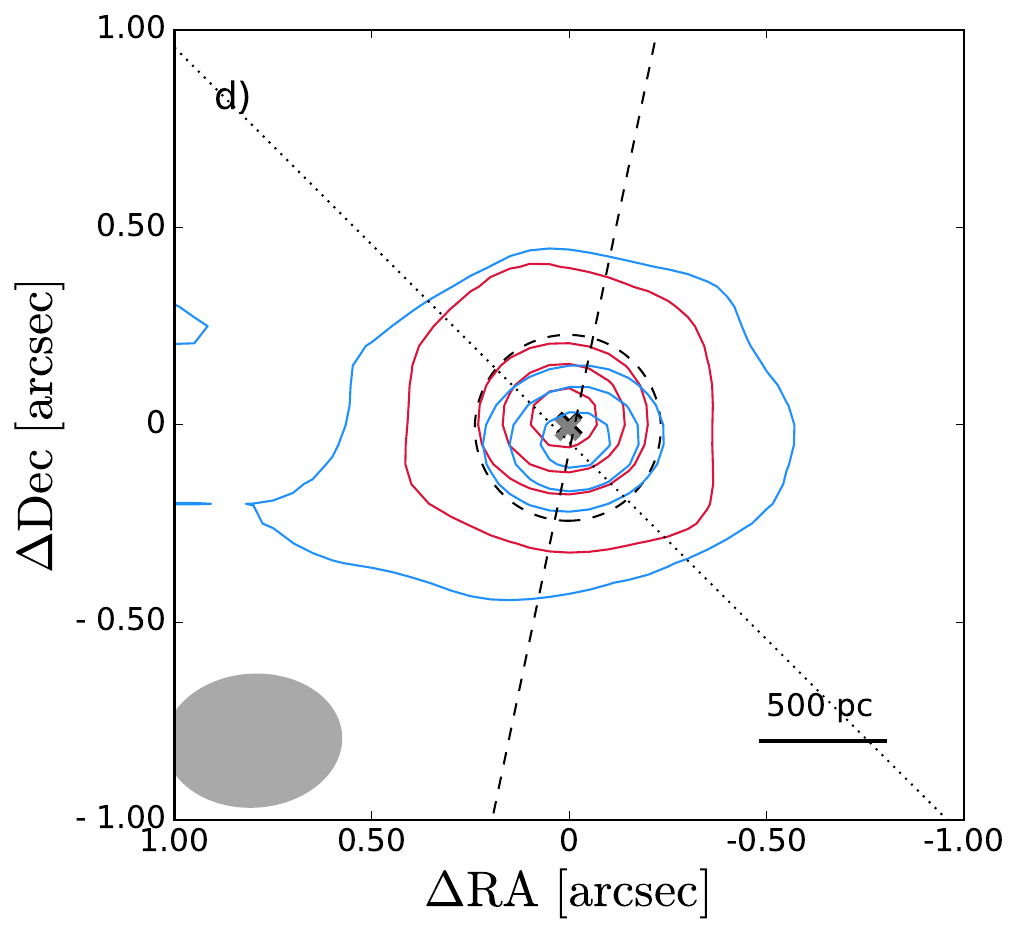} 
\includegraphics[width=0.27\textwidth]{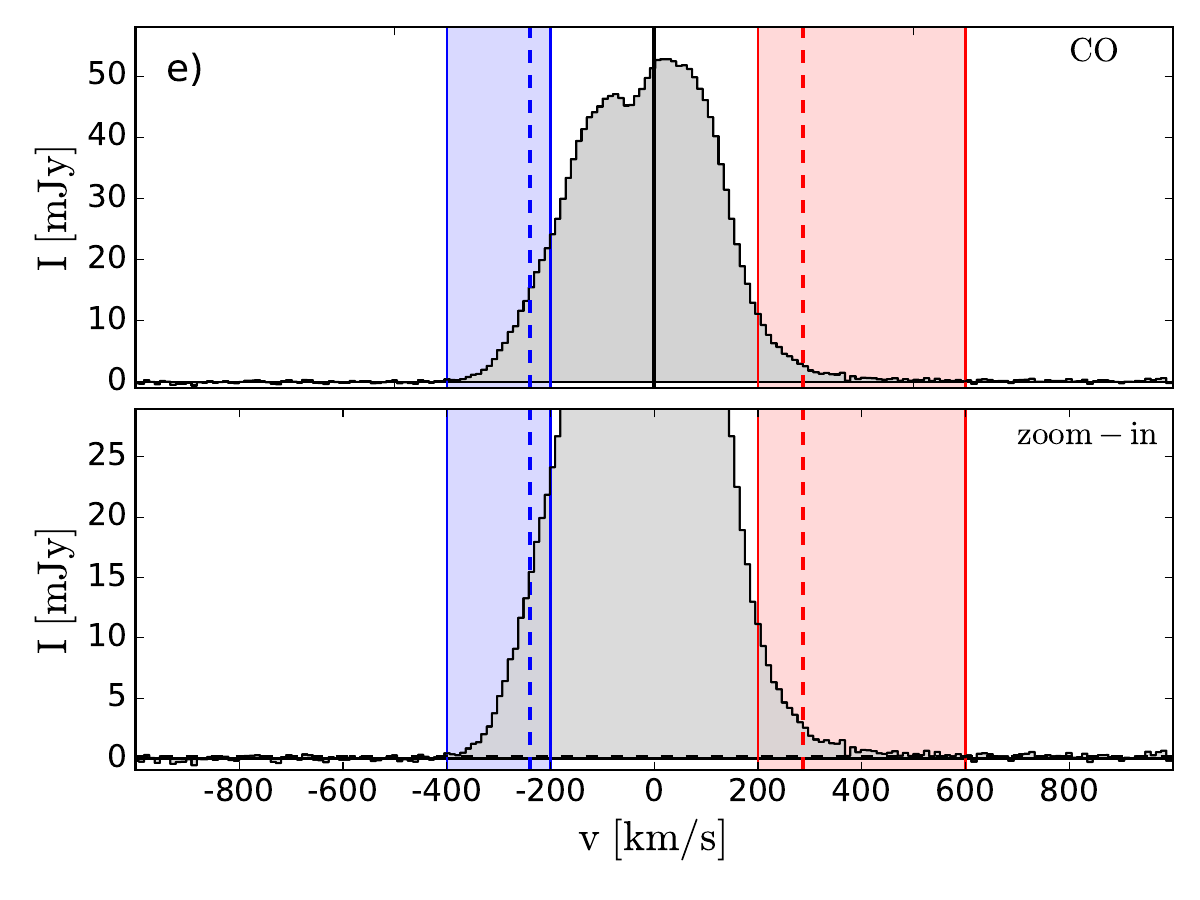} 
\includegraphics[width=0.27\textwidth]{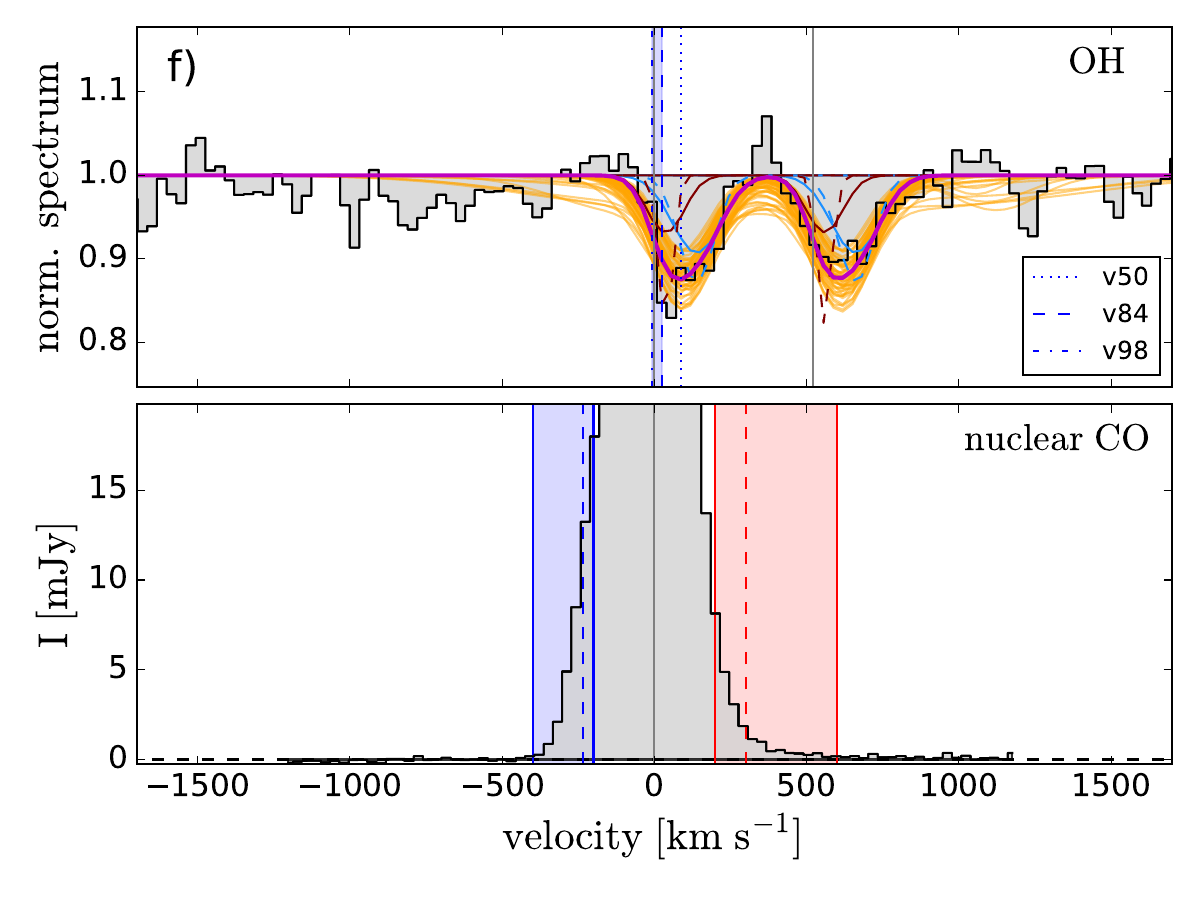}\\ 

\caption{continued.} 
 \end{figure*} 
 
 \begin{figure*}\ContinuedFloat 
\centering 
 \includegraphics[width=0.75\textwidth]{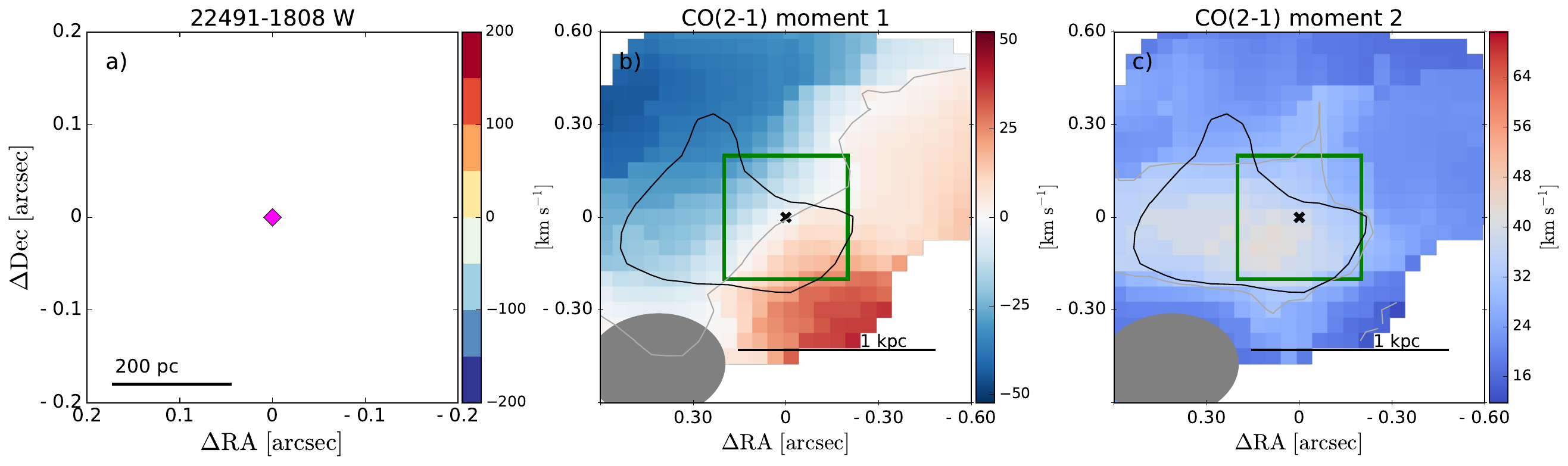}\\ 
\includegraphics[width=0.27\textwidth]{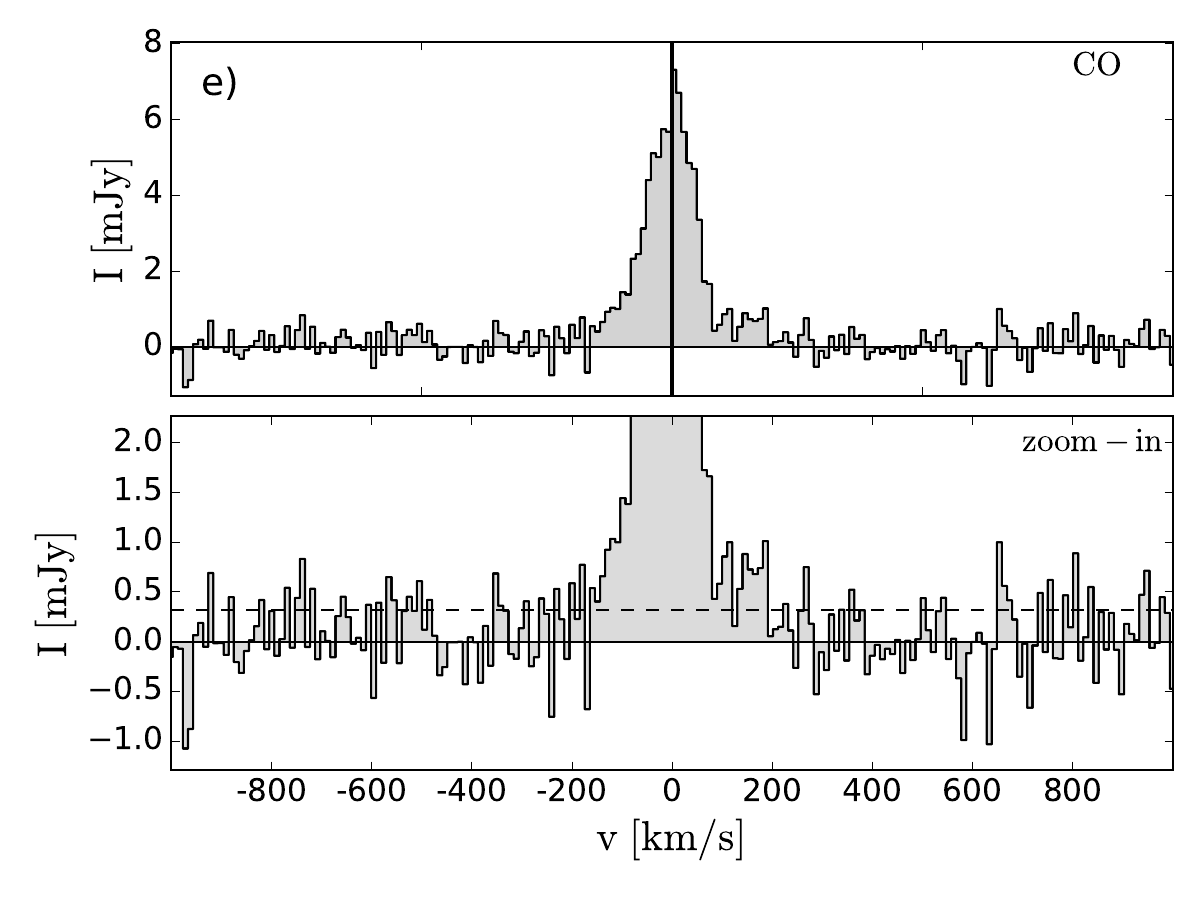} 
\caption{continued.} 
 \end{figure*} 

 \section{Channel maps}
  \label{sec:channel_maps}
 
 \begin{figure*} 
\centering 
 \caption{CO channel maps of 00091-0738 S showing the emission in channels of 50~\kms\ on the blue and red-shifted side with respect to the CO redshift ($z_\text{CO}$).  The lowest contour corresponds to 3$\times \sigma$, where $\sigma$ is the rms measured in each 50~\kms\ channel (in mJy), and the other contours correspond to 0.1, 0.2, 0.3, 0.5, 0.7, 0.9$\times$ maximum flux. Contours are shown only if they are above the 3$\times \sigma$ level.  Dashed contours show negative -3, -4, -5 ...$\times\sigma$  levels. The  $\sigma$ values (in mJy) for the blue and red channels are indicated in the top-right corner. The dotted line shows the kinematic major axis,  the dashed line shows the direction of the outflow (if present).  The black cross shows the peak position of the ALMA continuum.  The green cross shows the continuum position of the second nucleus in interacting systems. 
}
\includegraphics[width=0.99\textwidth]{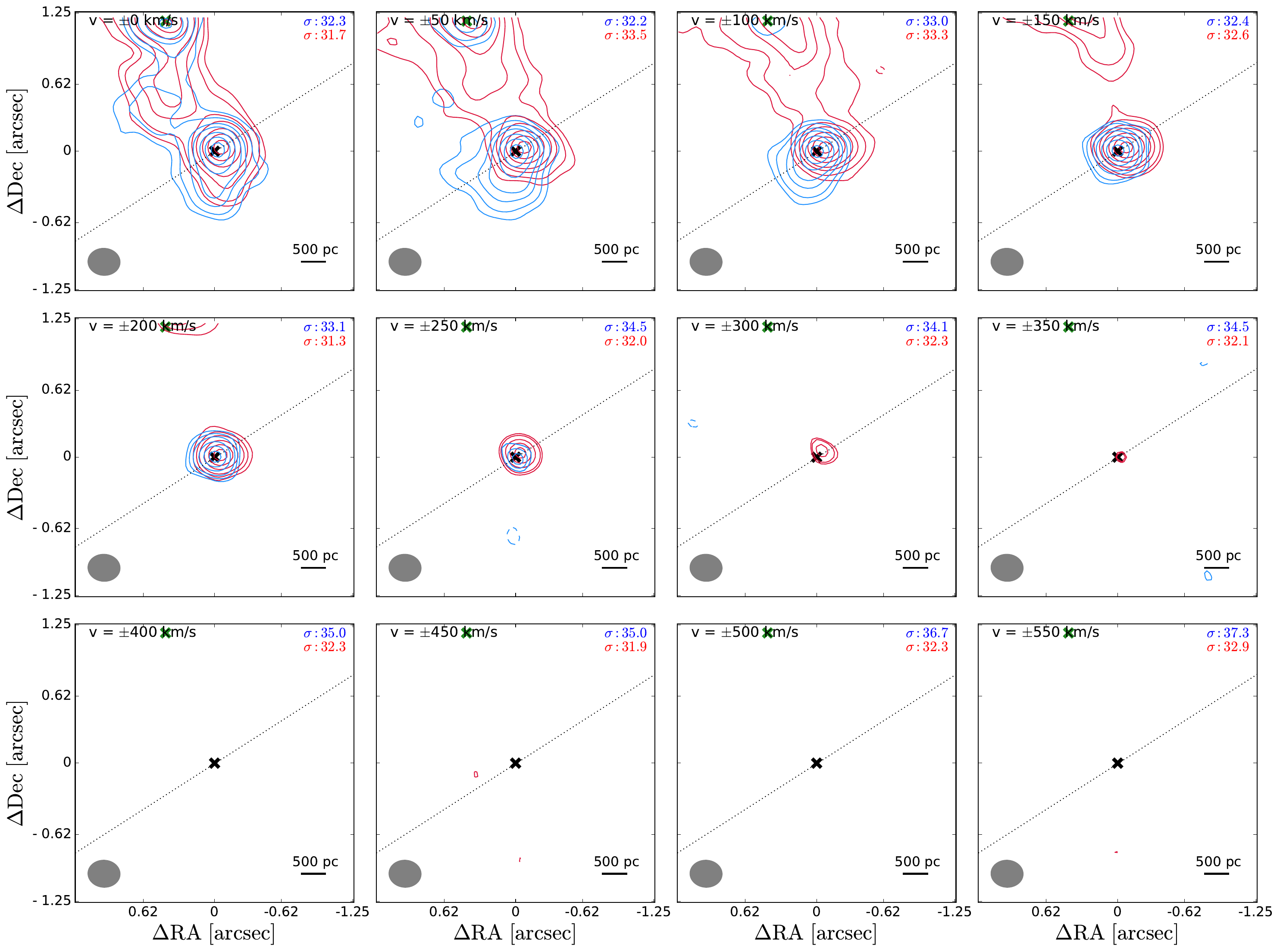} 
\label{fig:channel_maps}
 \end{figure*} 
 
  \begin{figure*} 
\centering 
\includegraphics[width=0.99\textwidth]{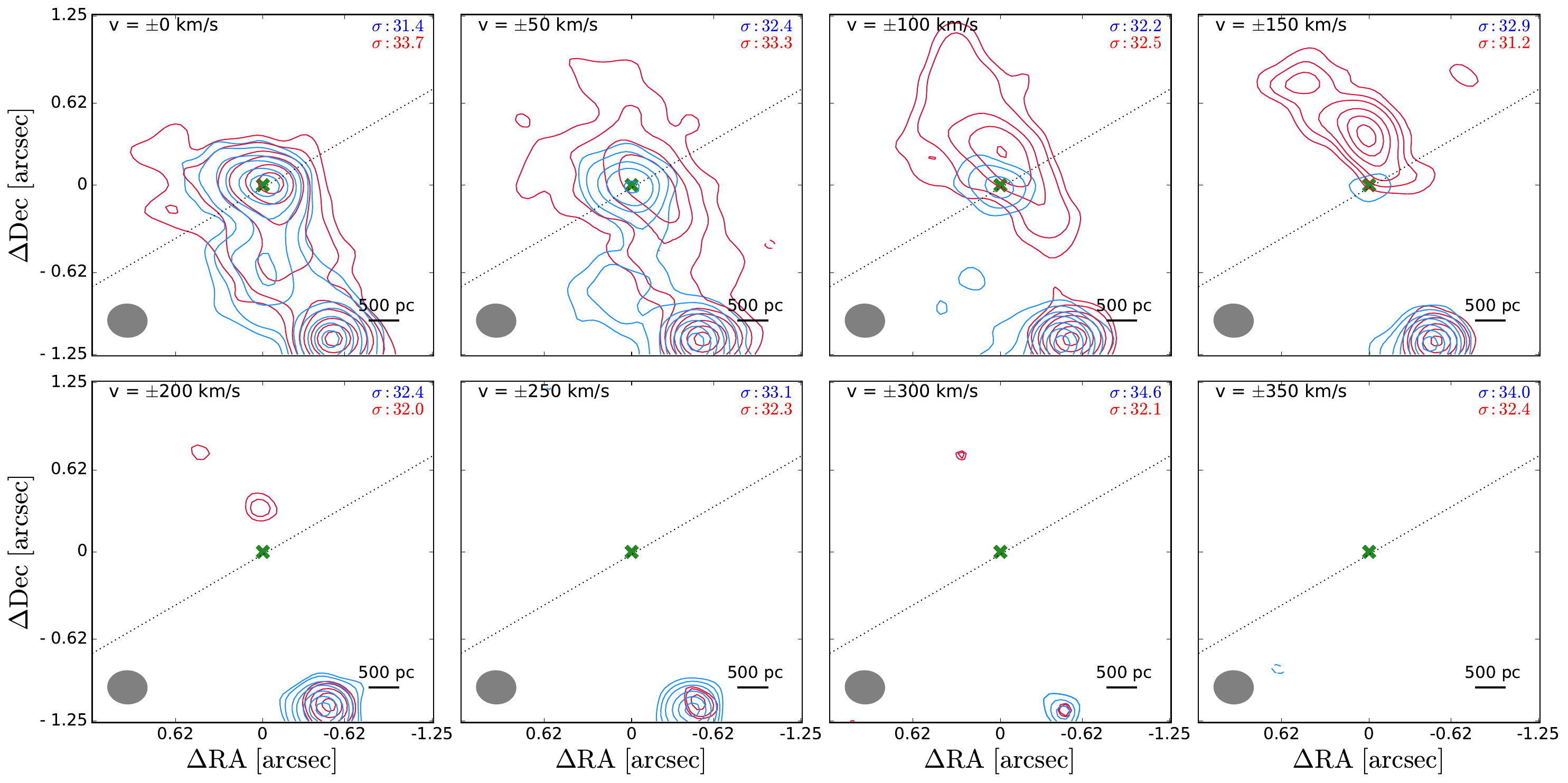} 
\caption{00091-0738 N.} 
\label{fig:channel_maps2}
 \end{figure*} 
 
  \begin{figure*} 
\centering 
\includegraphics[width=0.99\textwidth]{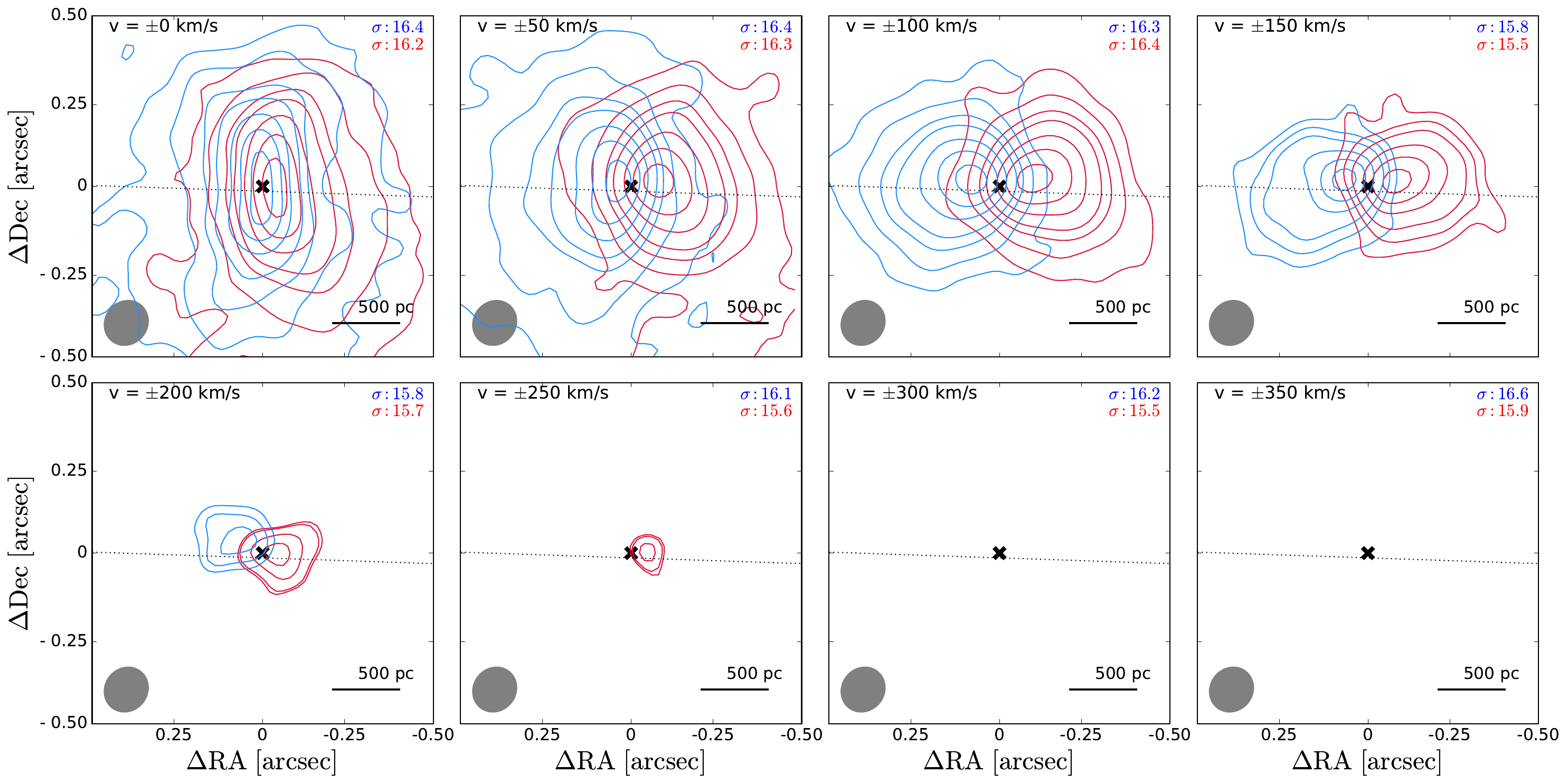} 
\caption{00188-0856.} 
\end{figure*}

\begin{figure*} 
\centering 
\includegraphics[width=0.99\textwidth]{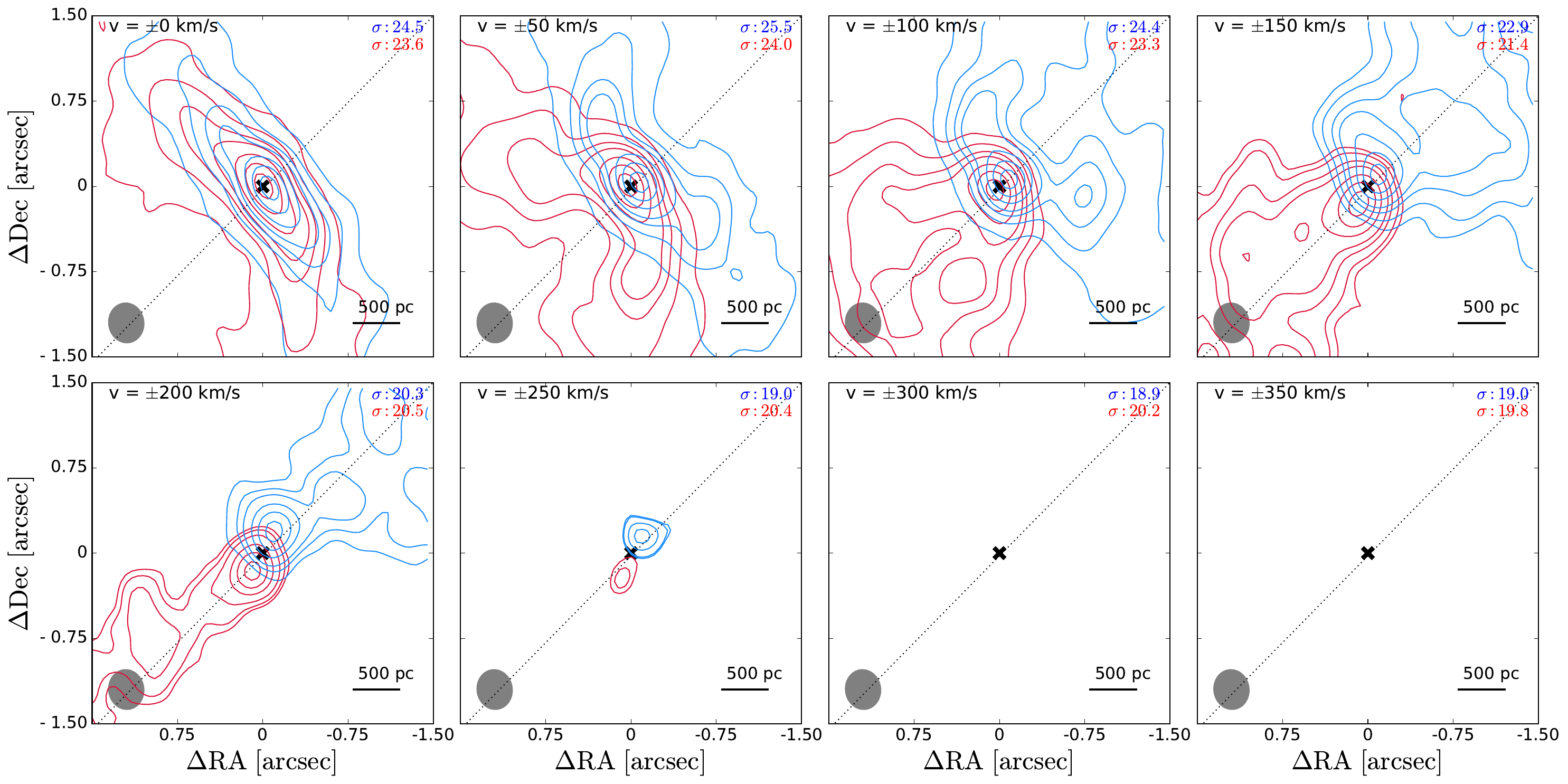} 
\caption{00509+1225.} 
\end{figure*} 
 
 \begin{figure*} 
\centering 
\includegraphics[width=0.99\textwidth]{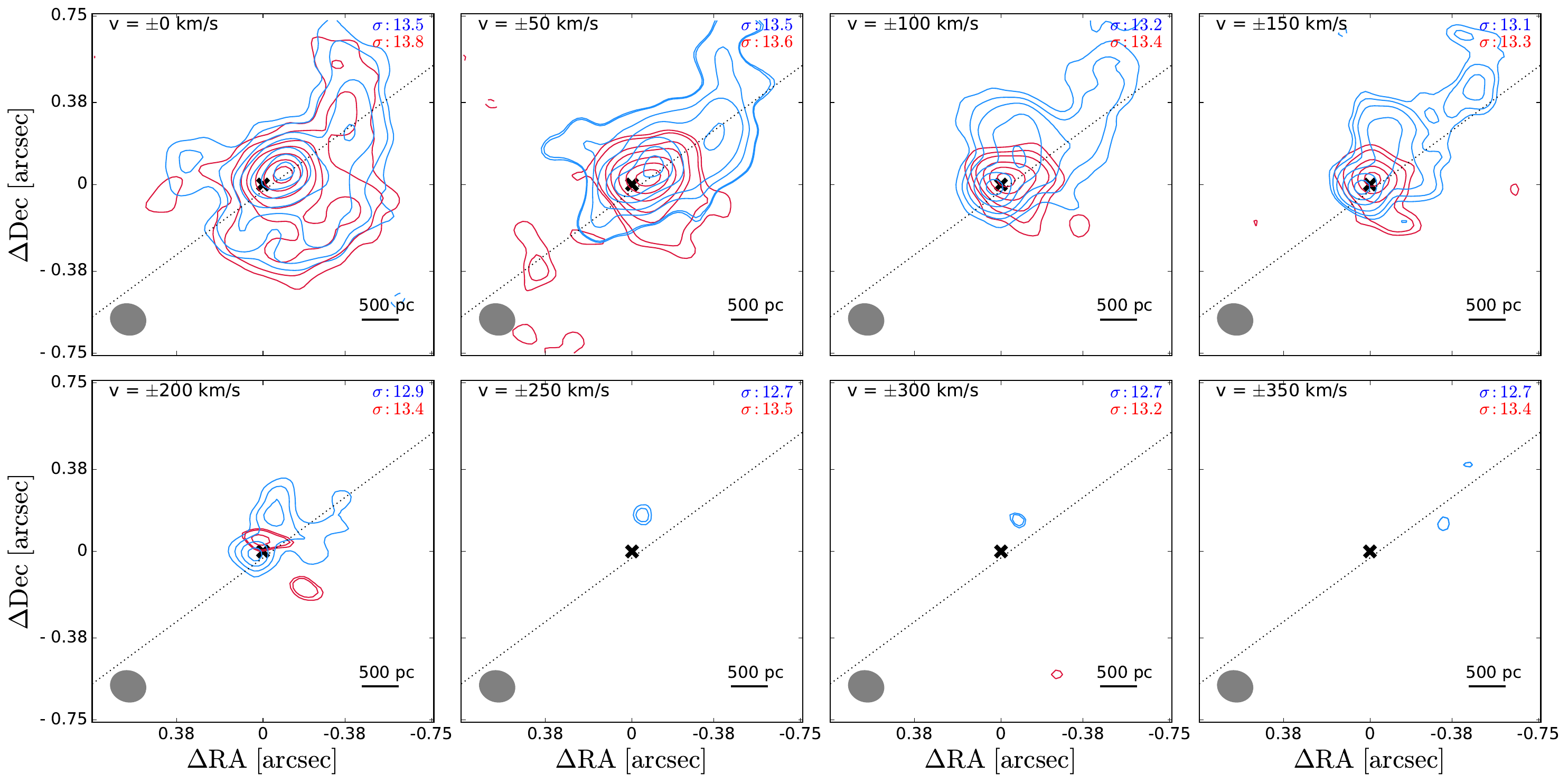} 
\caption{01572+0009.} 
\end{figure*}

\begin{figure*} 
\centering 
\includegraphics[width=0.99\textwidth]{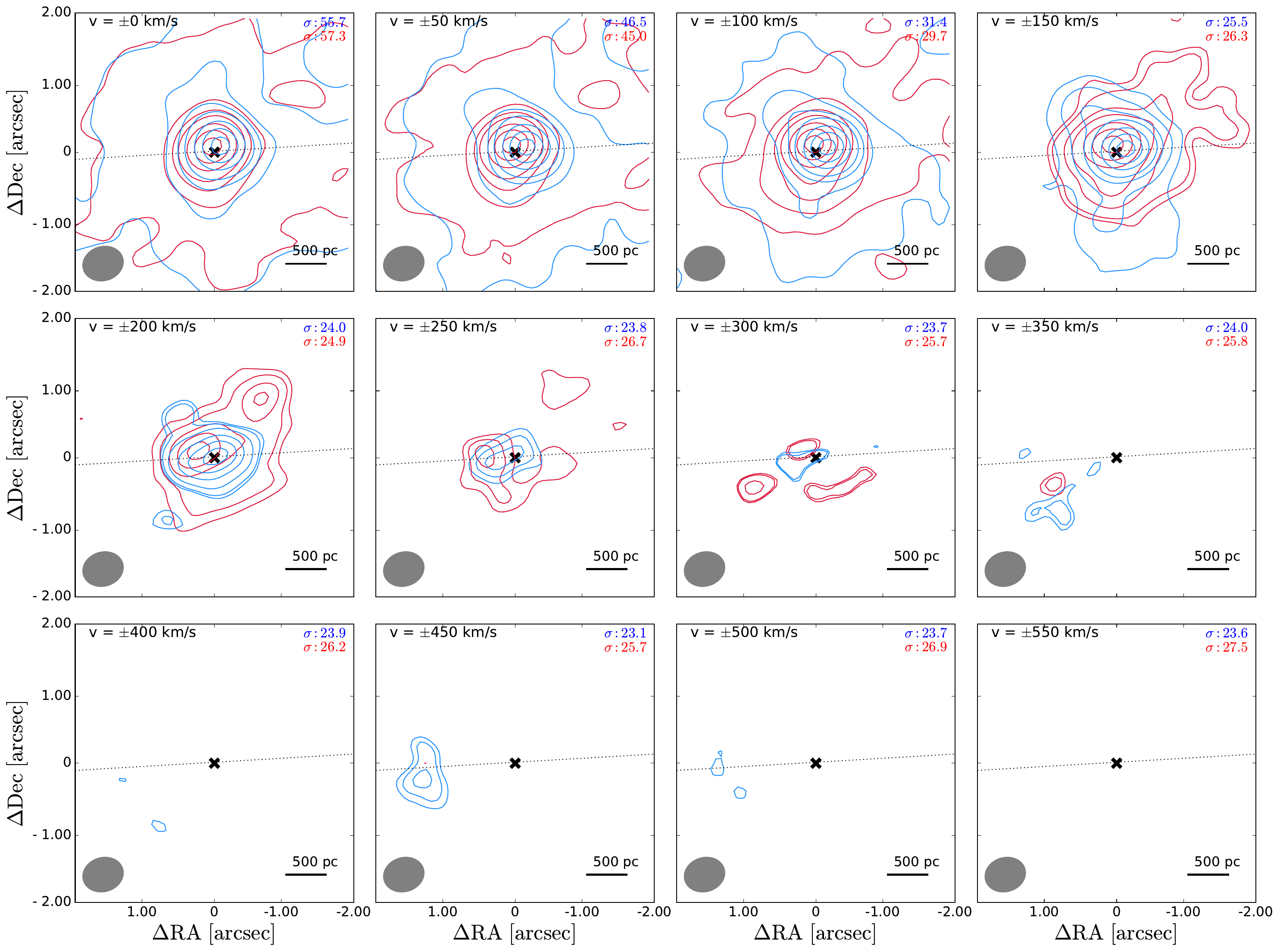} 
\caption{05189-2524.} 
\end{figure*}

\begin{figure*} 
\centering 
\includegraphics[width=0.99\textwidth]{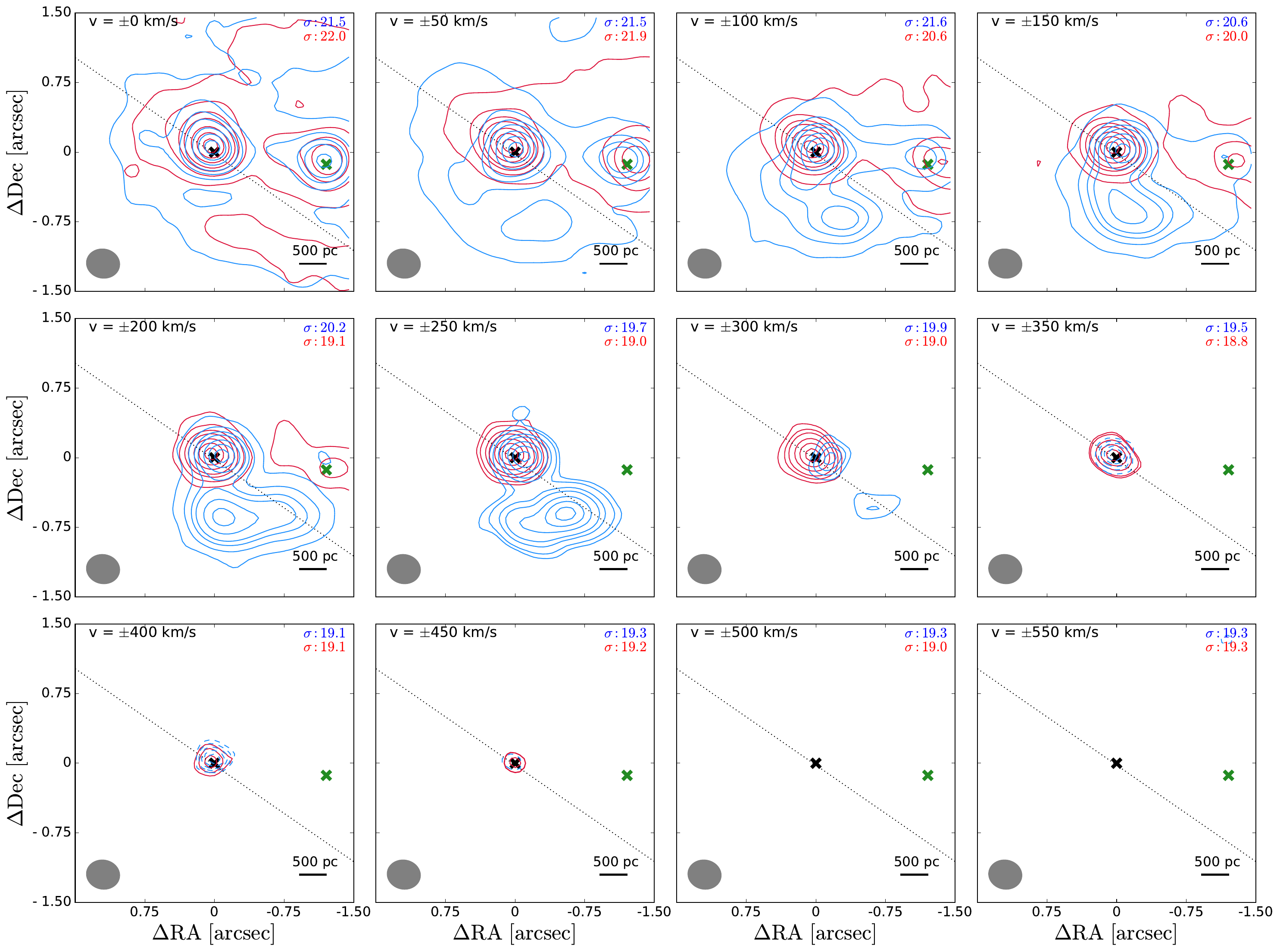} 
\caption{07251-0248 E.} 
\end{figure*}

\begin{figure*} 
\centering 
\includegraphics[width=0.99\textwidth]{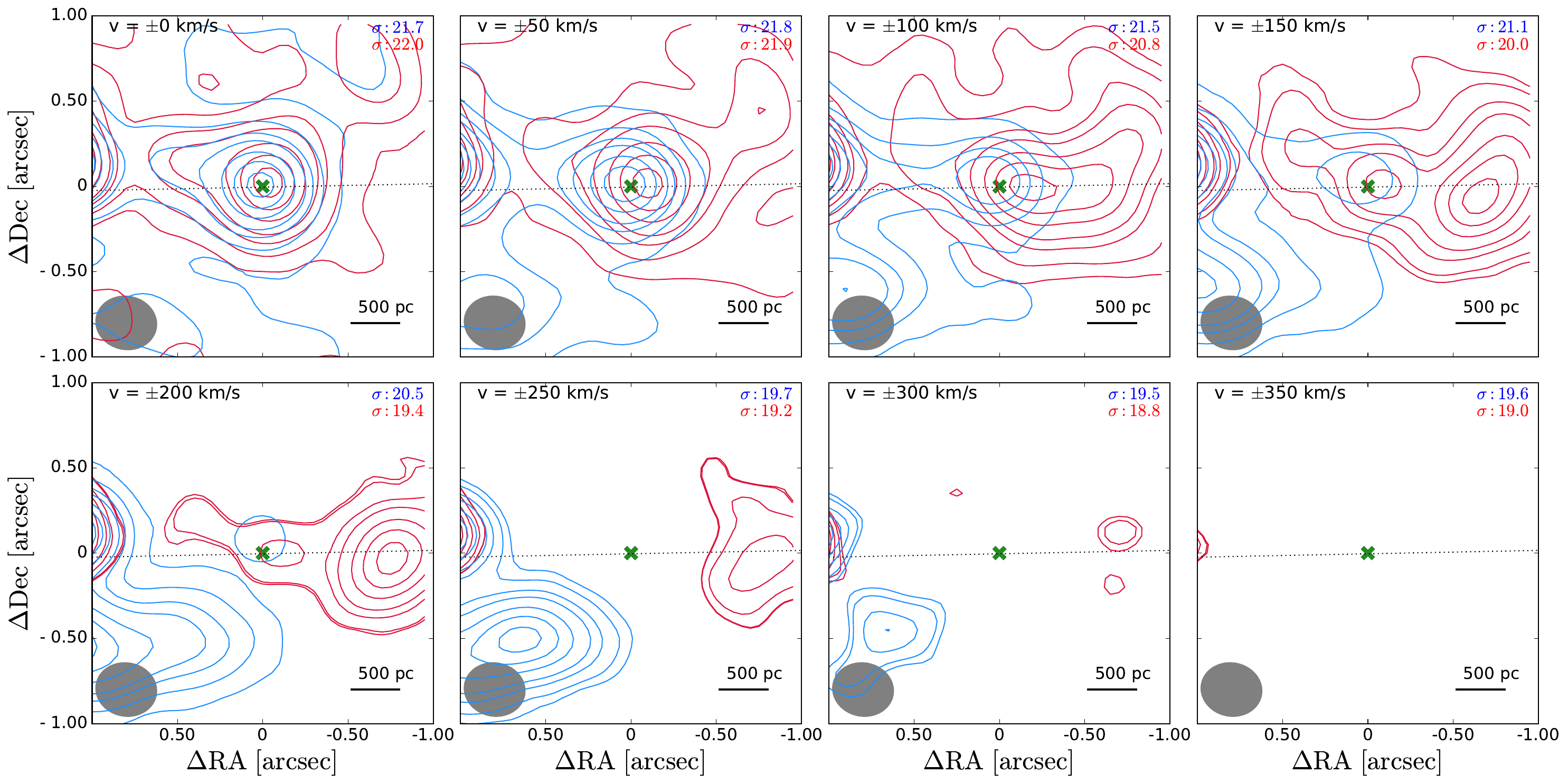} 
\caption{07251-0248 W.} 
 \end{figure*}

\begin{figure*} 
\centering 
\includegraphics[width=0.99\textwidth]{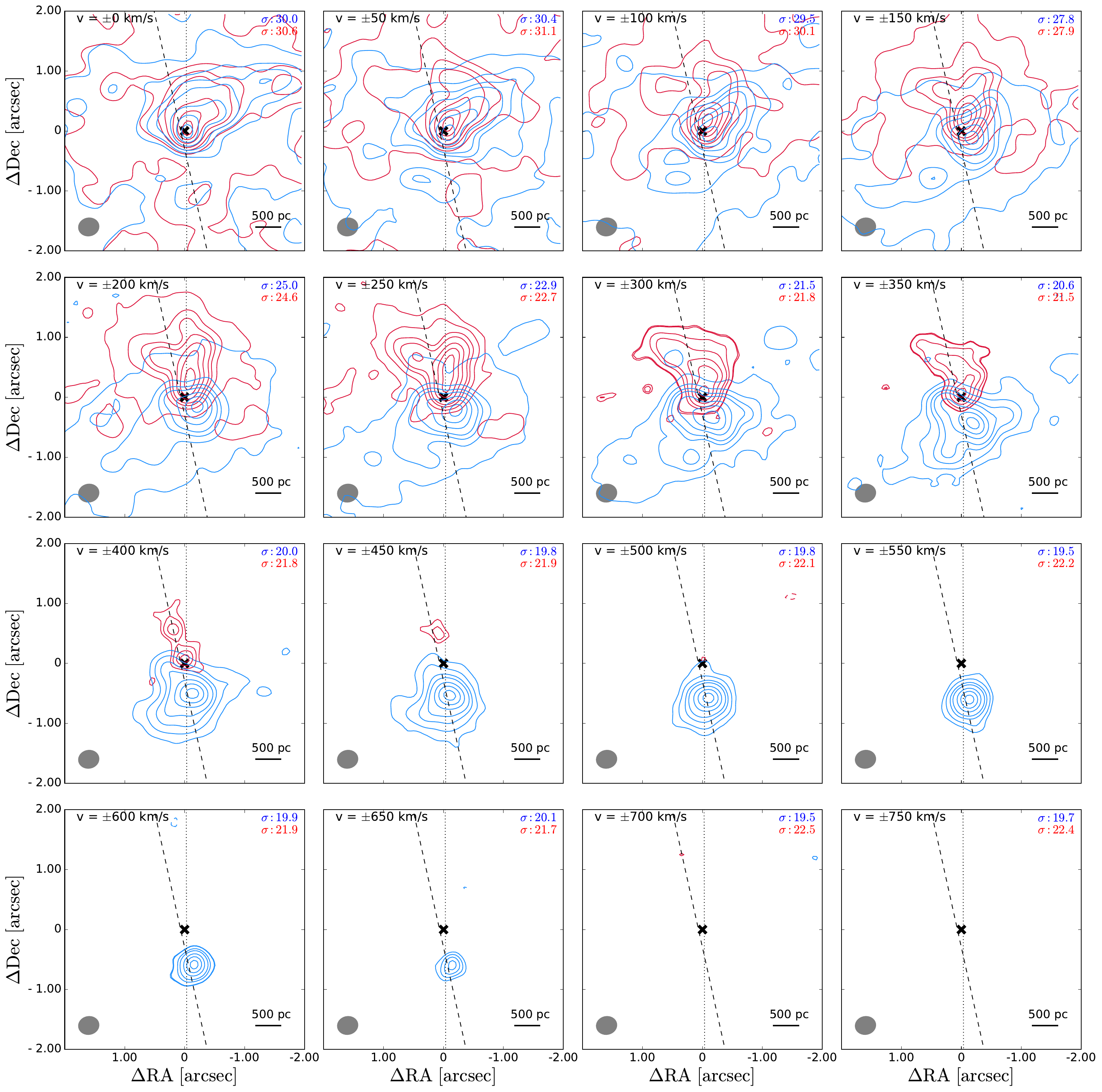} 
\caption{09022-3615.} 
\end{figure*} 

\begin{figure*} 
\centering 
\includegraphics[width=0.99\textwidth]{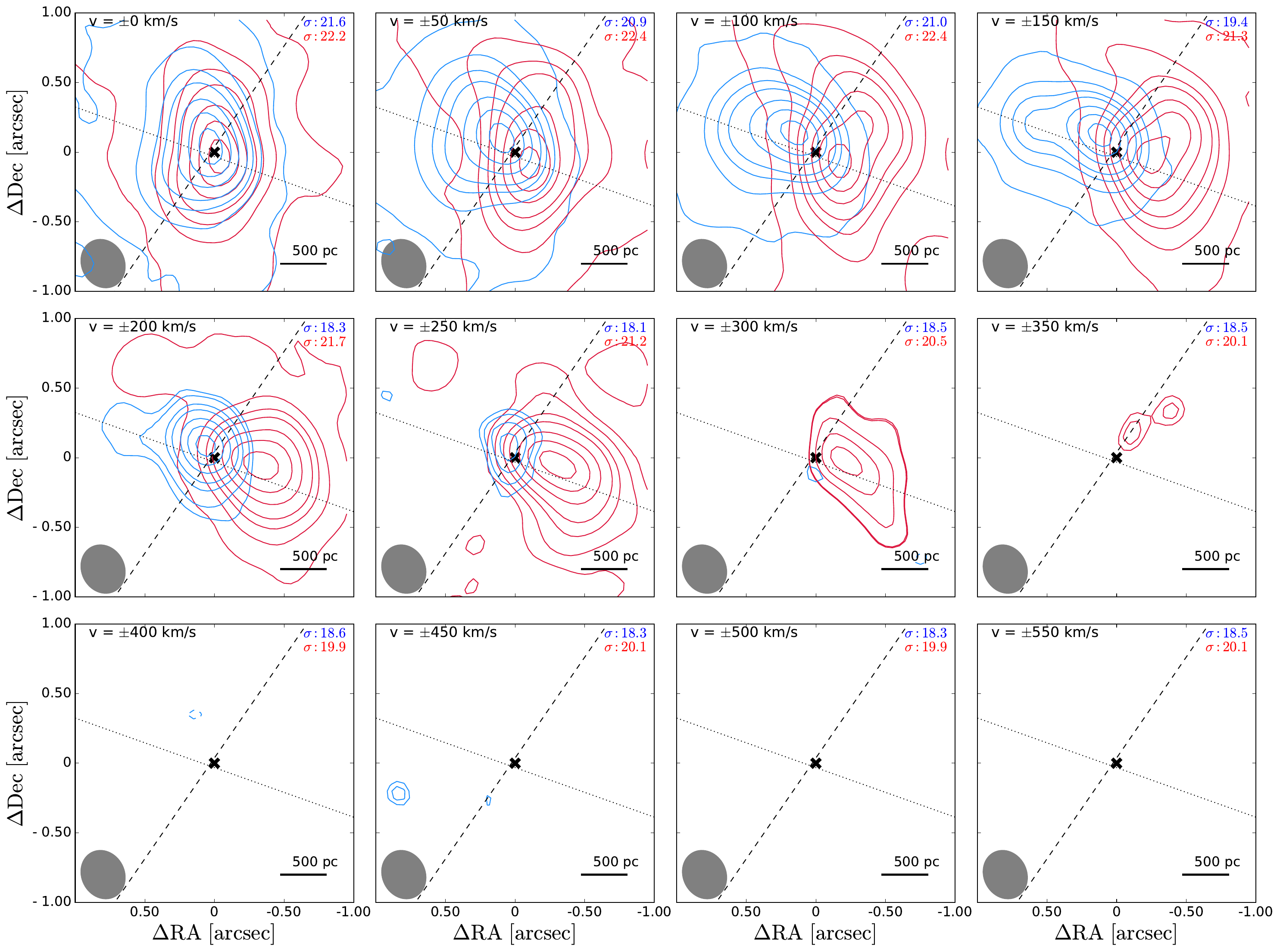} 
\caption{10190+1322 E.} 
\end{figure*} 
 
\begin{figure*} 
\centering 
\includegraphics[width=0.99\textwidth]{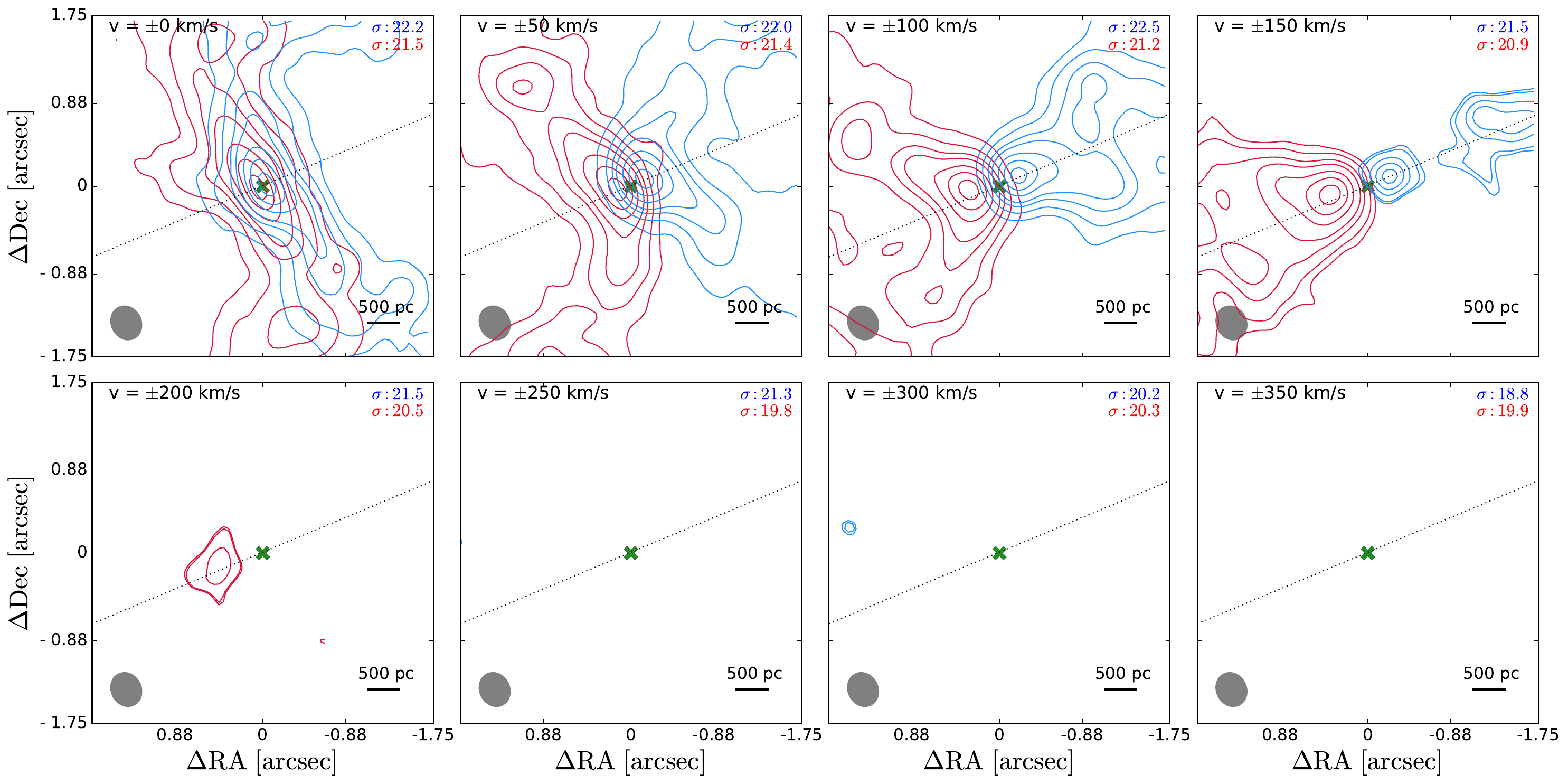} 
\caption{10190+1322 W.} 
 \end{figure*} 
 
\begin{figure*} 
\centering 
\includegraphics[width=0.99\textwidth]{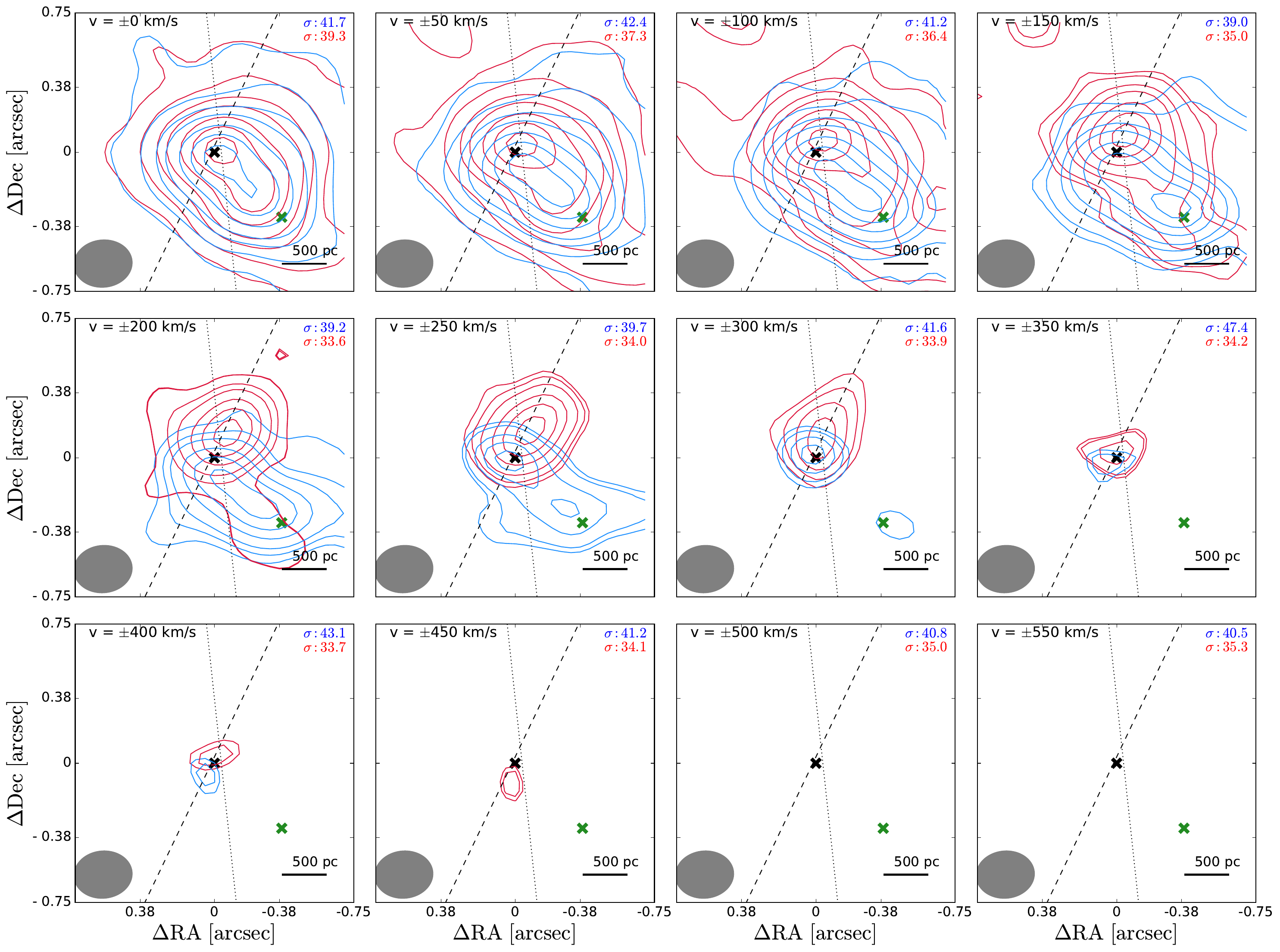} 
\caption{11095-0238 NE.} 
\end{figure*}

\begin{figure*} 
\centering 
\includegraphics[width=0.99\textwidth]{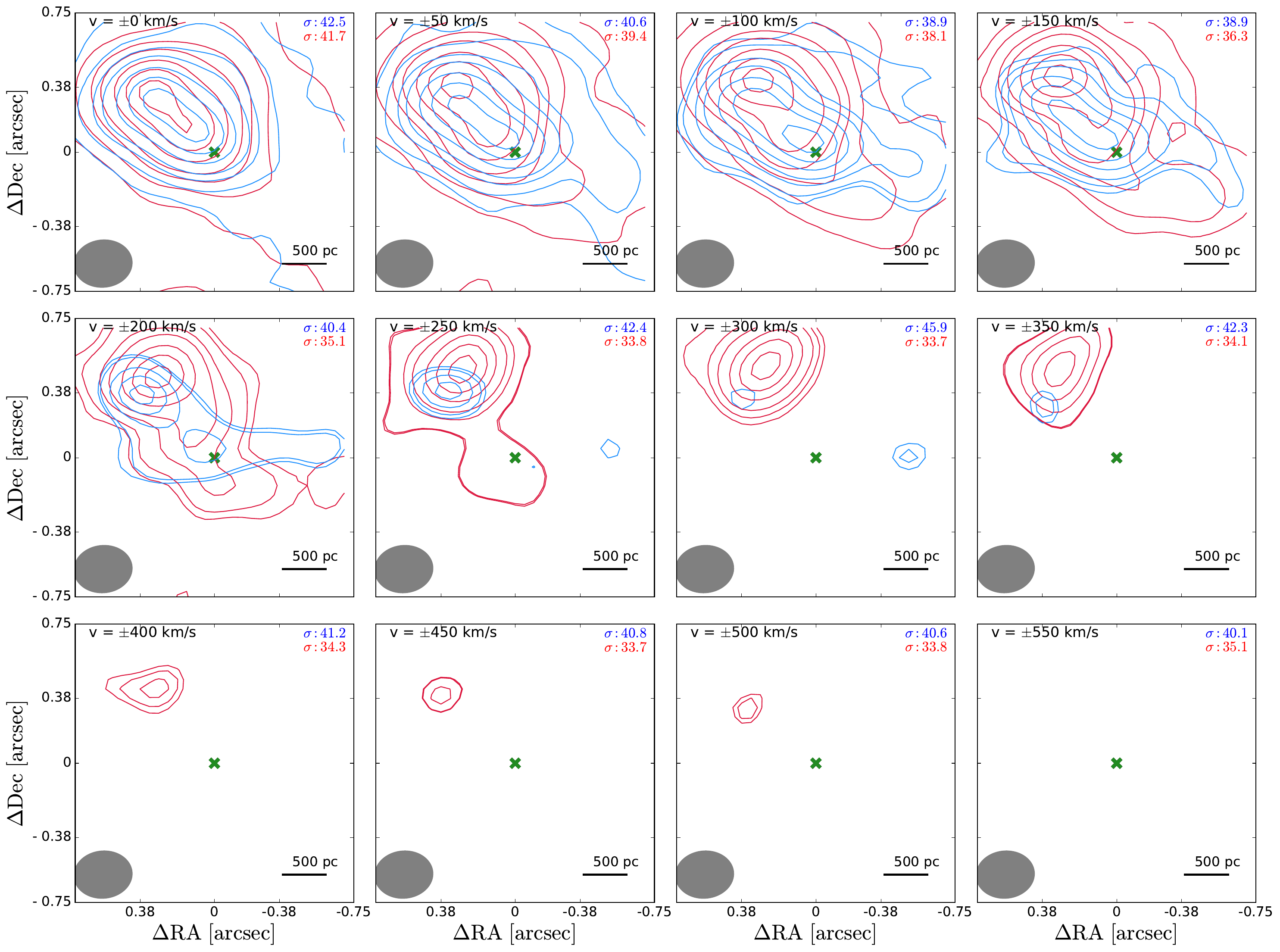} 
\caption{11095-0238 SW.} 
 \end{figure*}

\begin{figure*} 
\centering 
\includegraphics[width=0.99\textwidth]{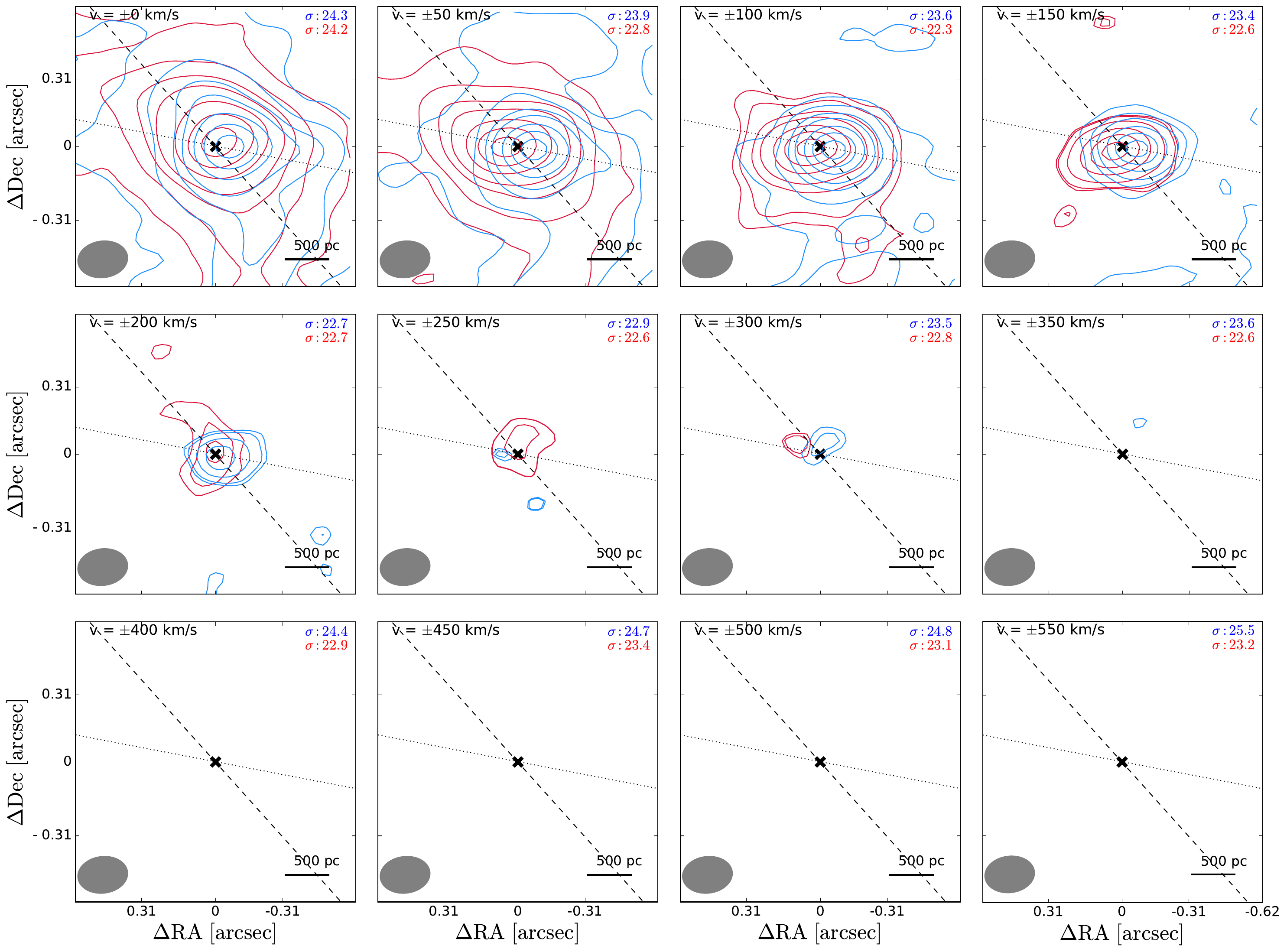} 
\caption{12071-0444 N.} 
\end{figure*}

\begin{figure*} 
\centering 
\includegraphics[width=0.99\textwidth]{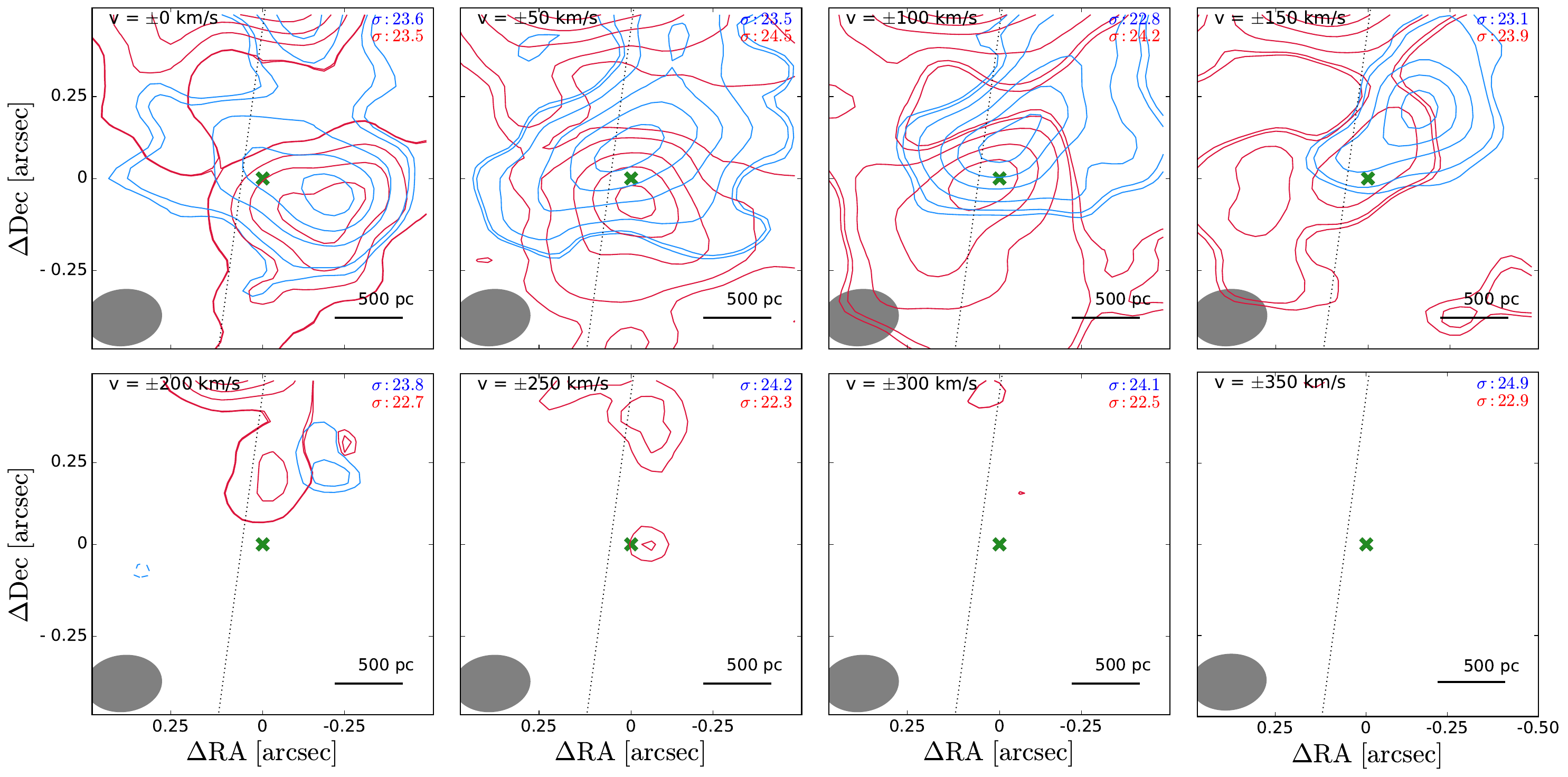} 
\caption{12071-0444 S.} 
 \end{figure*}

\begin{figure*} 
\centering 
\includegraphics[width=0.99\textwidth]{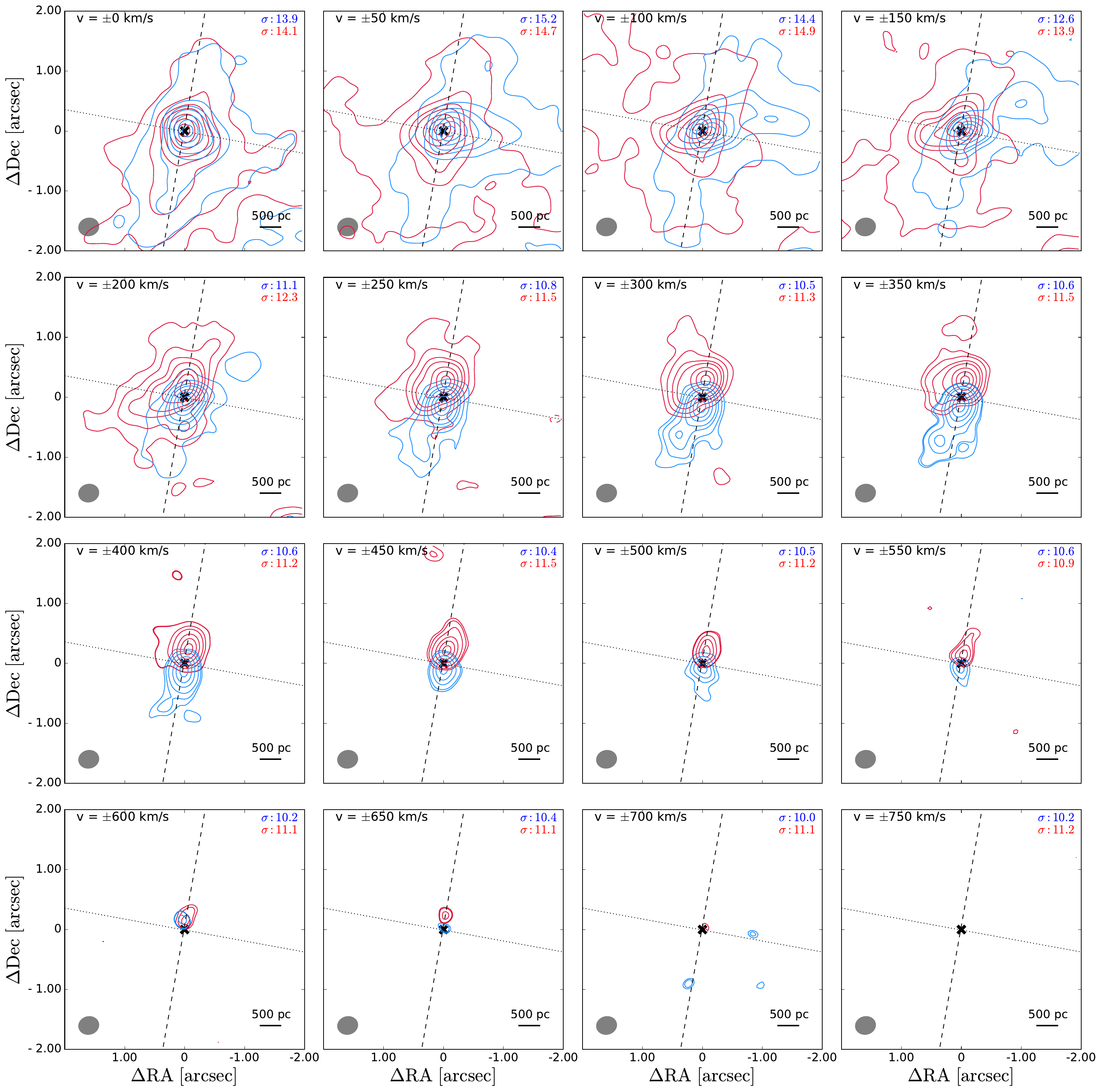} 
\caption{12112+0305 NE.} 
\end{figure*}

\begin{figure*} 
\centering 
\includegraphics[width=0.99\textwidth]{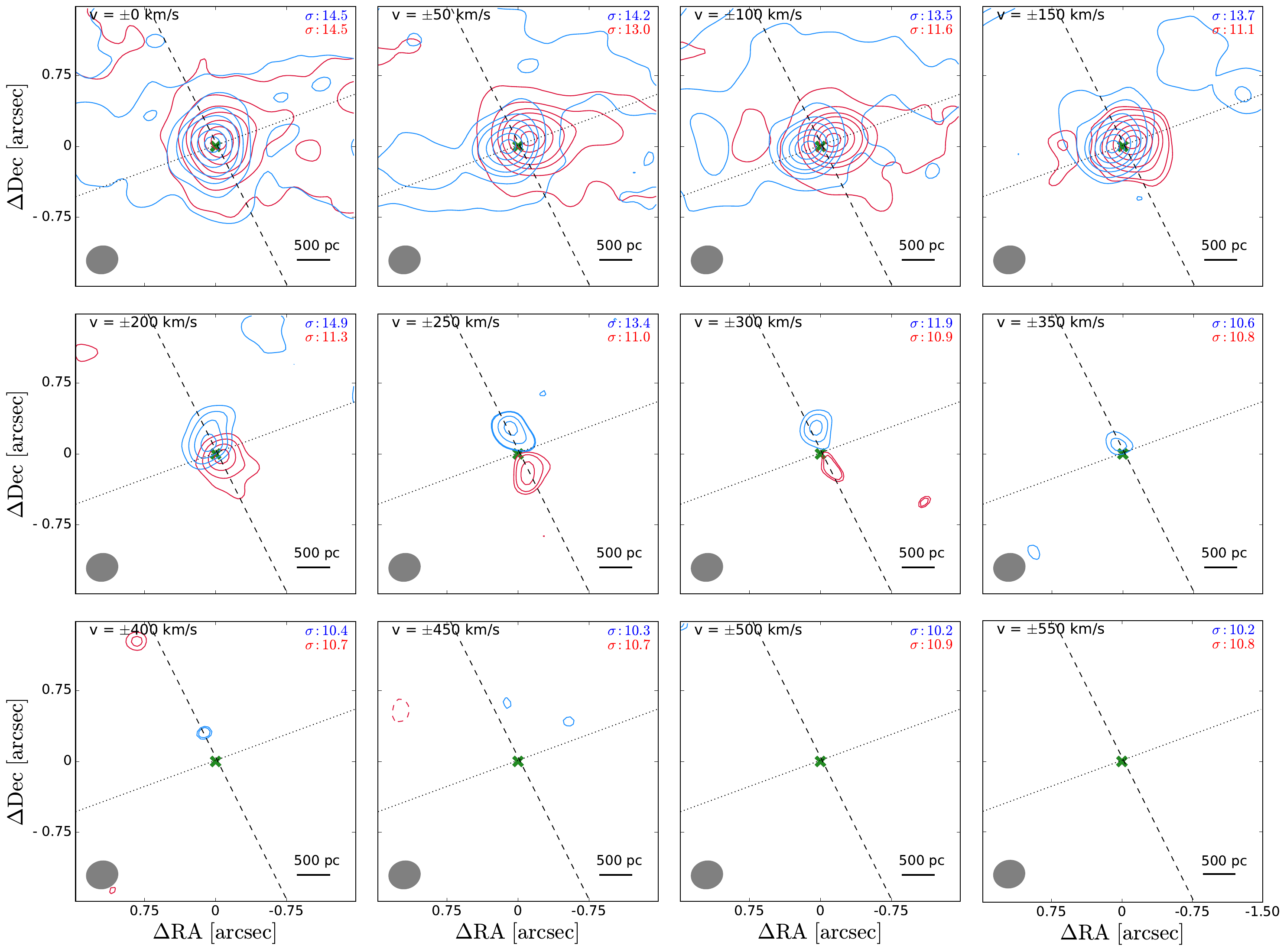} 
\caption{12112+0305 SW.} 
 \end{figure*}

\begin{figure*} 
\centering 
\includegraphics[width=0.99\textwidth]{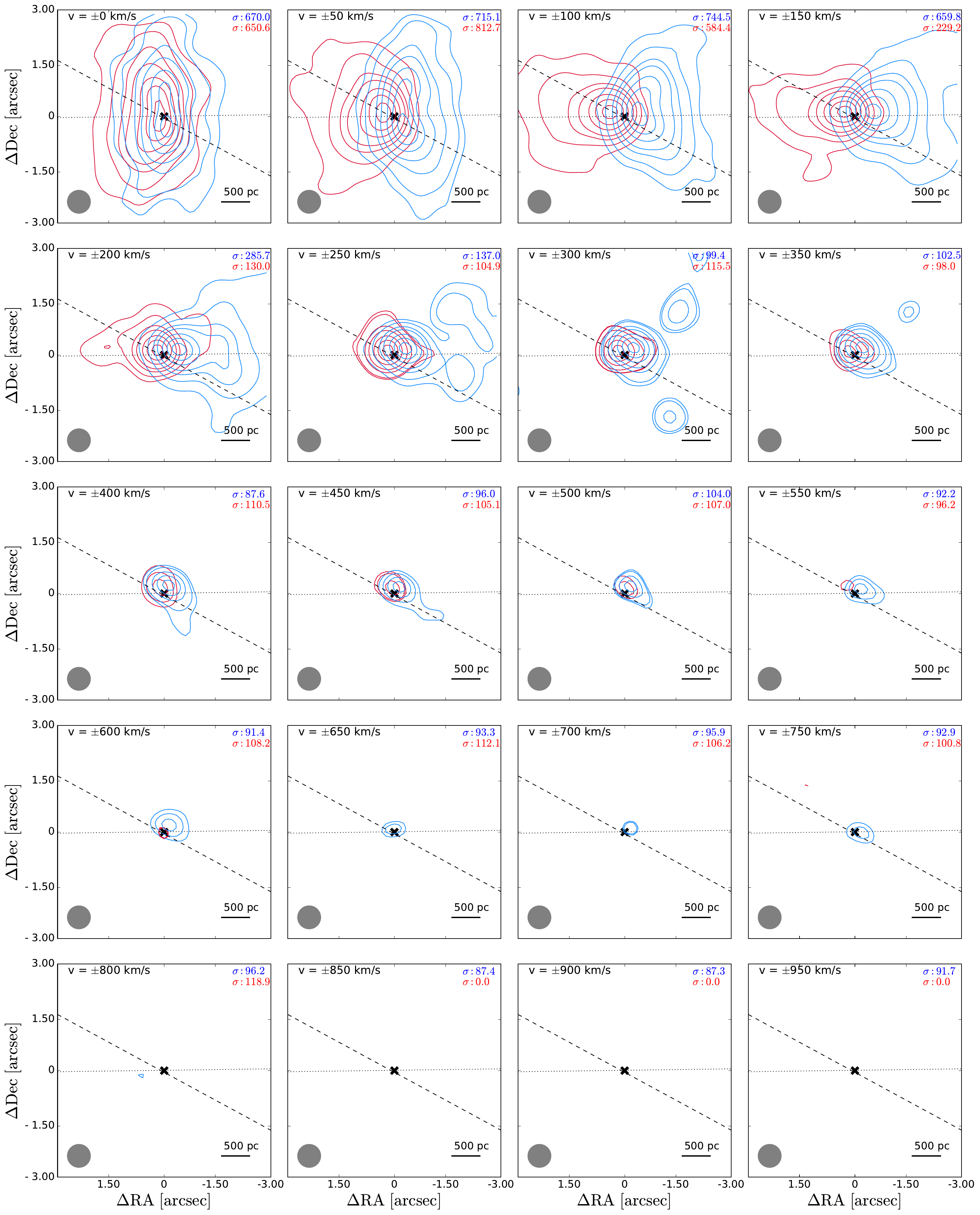} 
\caption{13120-5453.} 
\end{figure*}

\begin{figure*} 
\centering 
\includegraphics[width=0.99\textwidth]{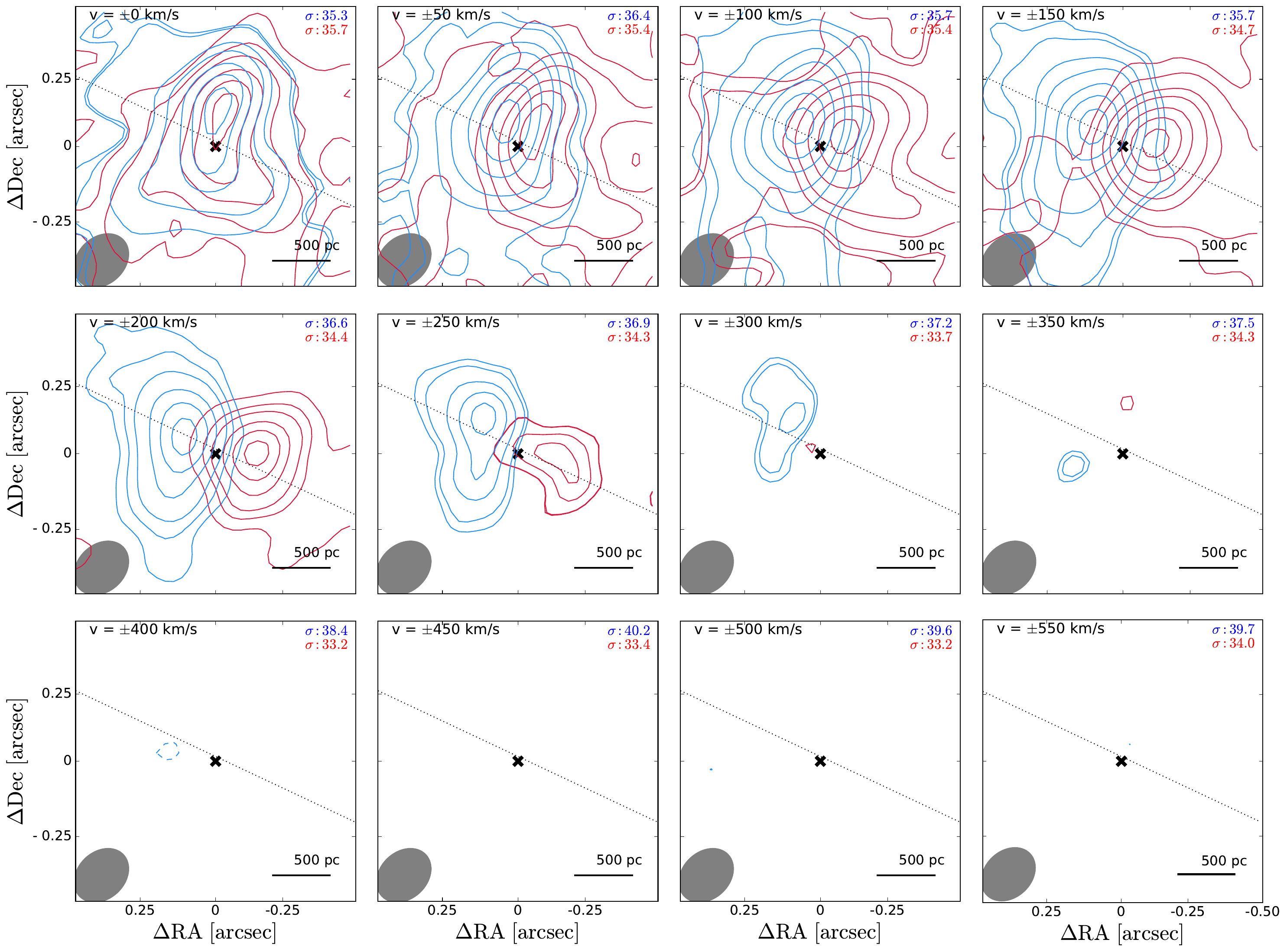} 
\caption{13451+1232 W.} 
\end{figure*}

\begin{figure*} 
\centering 
\includegraphics[width=0.99\textwidth]{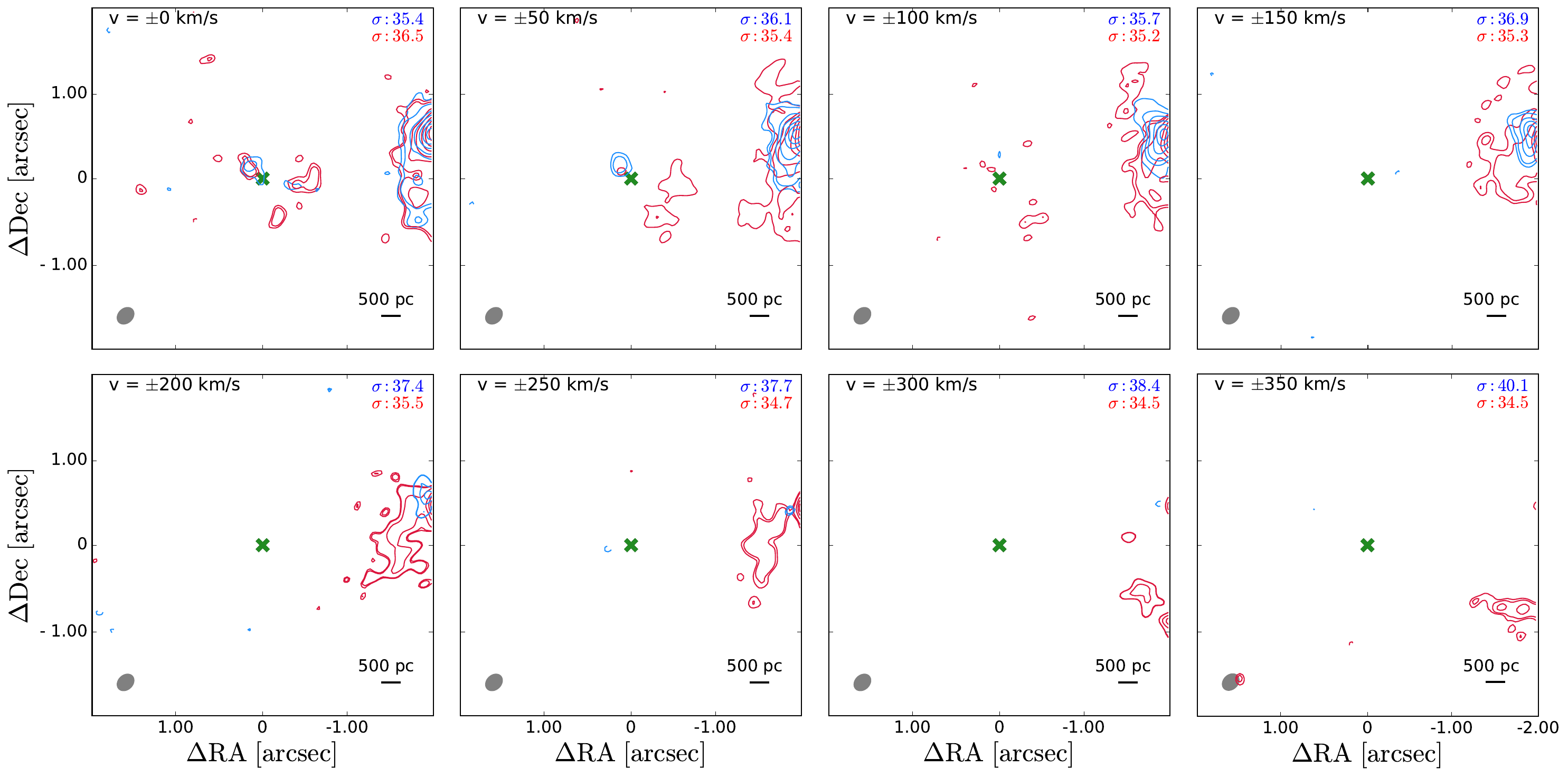} 
\caption{13451+1232 E.} 
 \end{figure*}

\begin{figure*} 
\centering 
\includegraphics[width=0.99\textwidth]{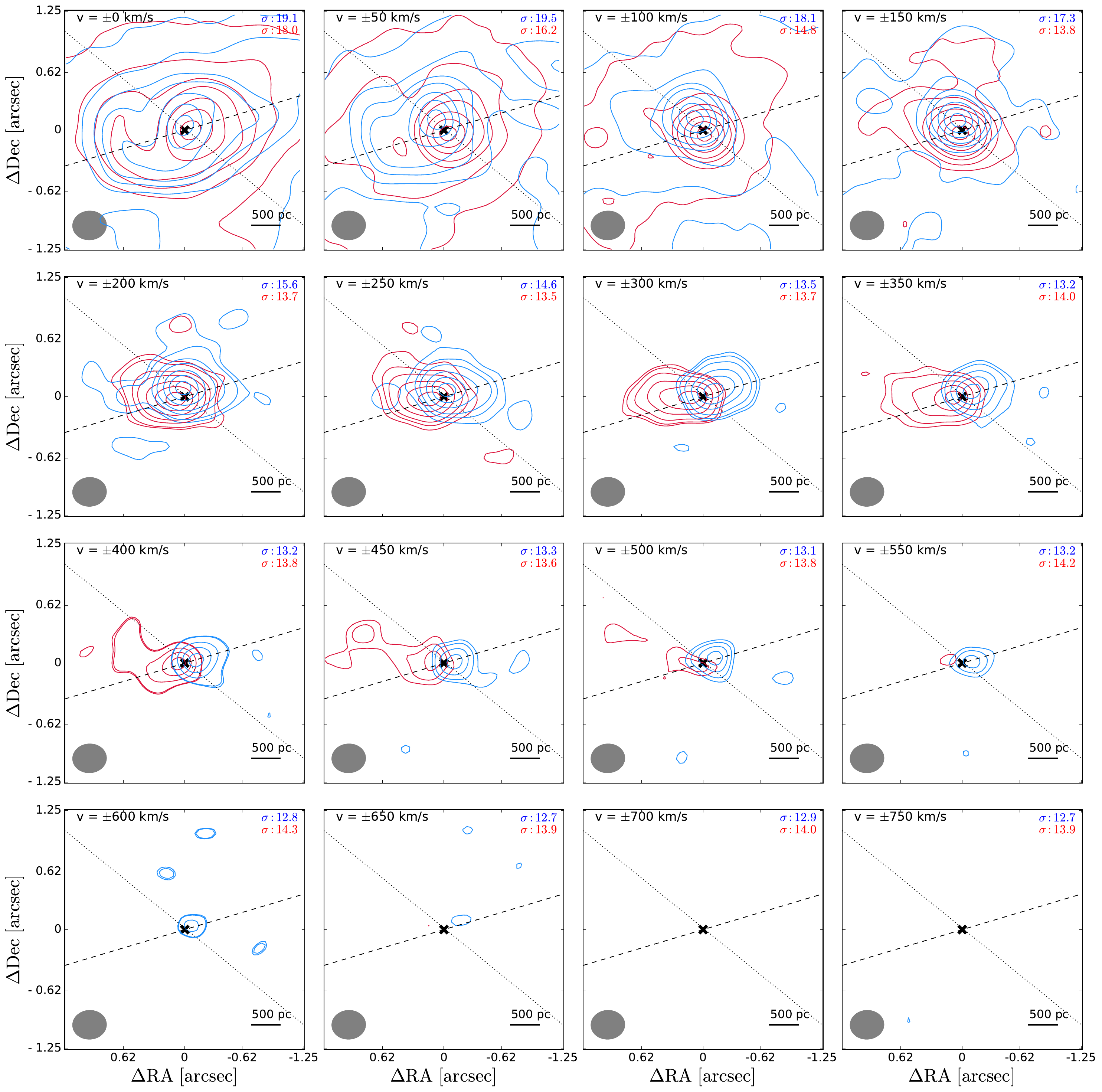} 
\caption{14348-1447 SW.} 
\end{figure*}

\begin{figure*} 
\centering 
\includegraphics[width=0.99\textwidth]{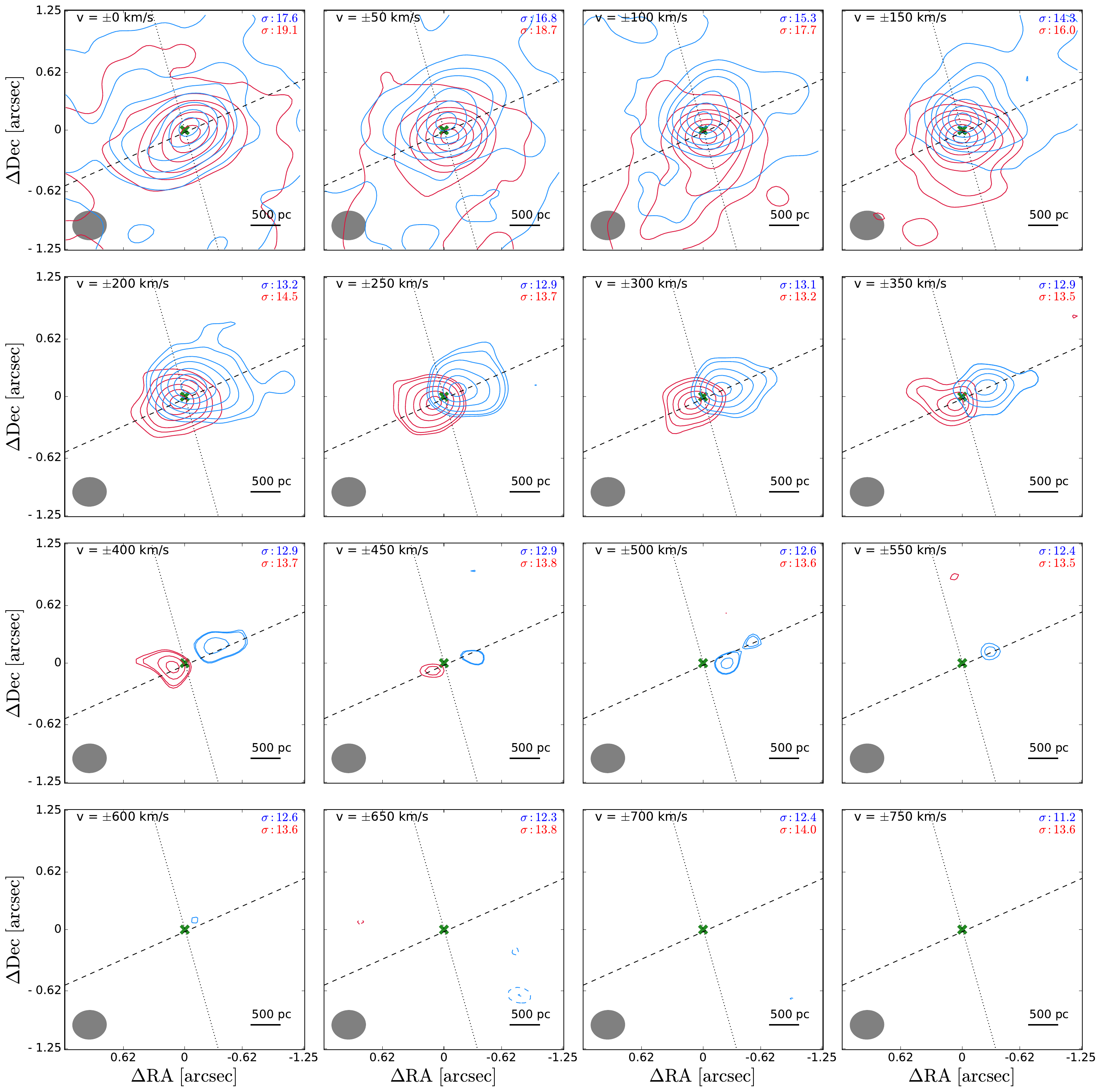} 
\caption{14348-1447 NE.} 
 \end{figure*}

\begin{figure*} 
\centering 
\includegraphics[width=0.99\textwidth]{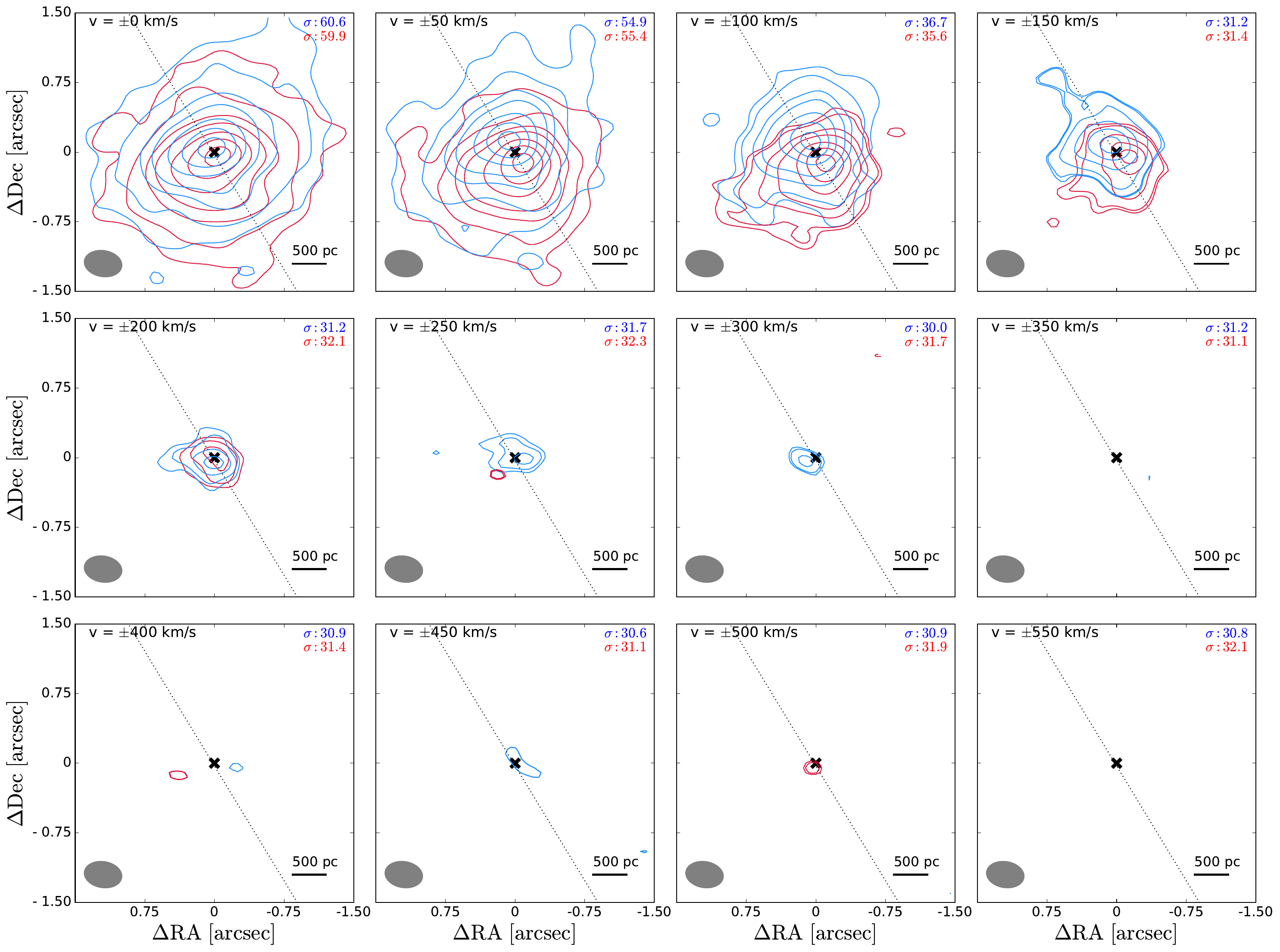} 
\caption{14378-3651.} 
\end{figure*}

\begin{figure*} 
\centering 
\includegraphics[width=0.99\textwidth]{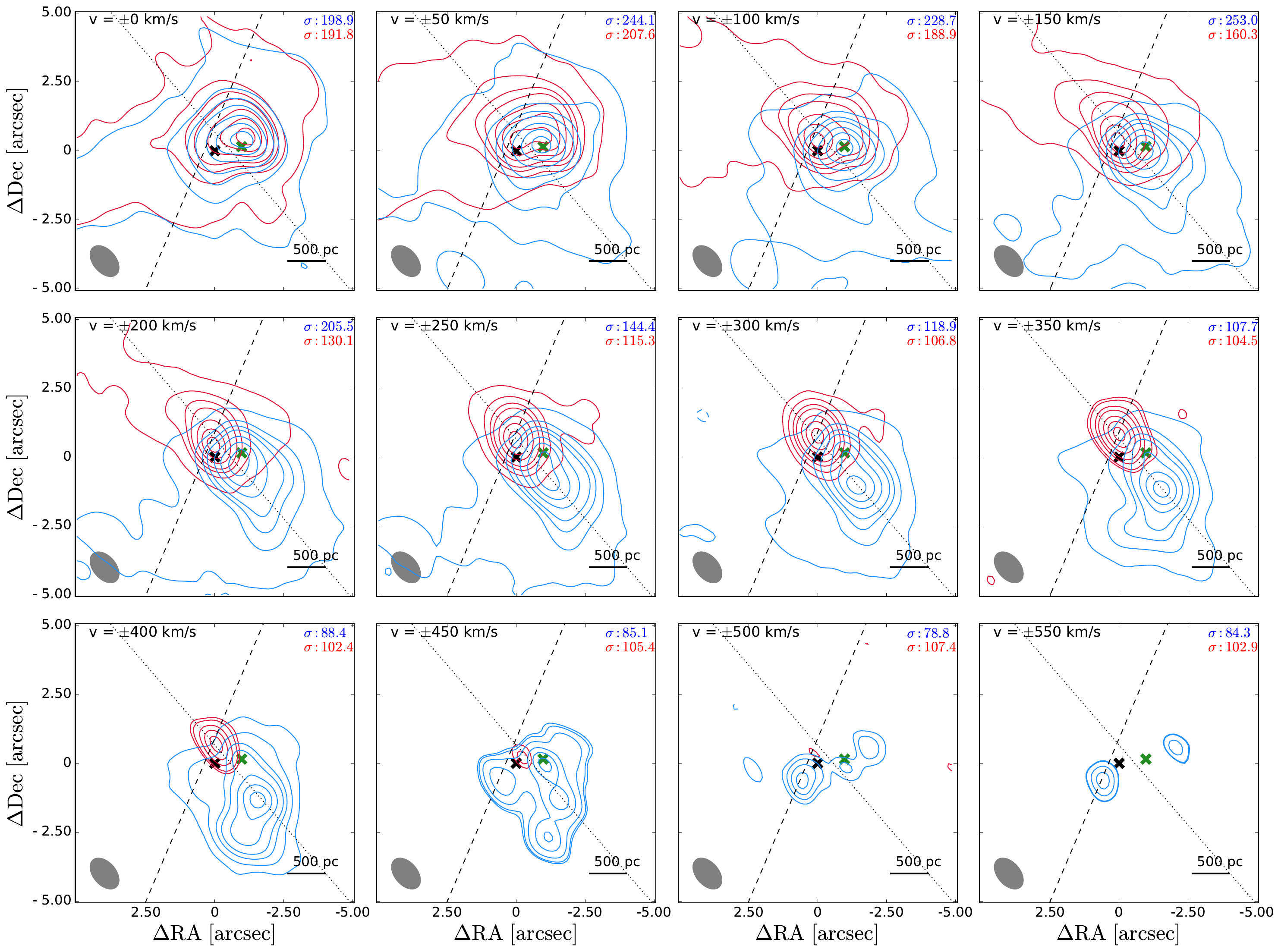} 
\caption{15327+2340.} 
\end{figure*}

\begin{figure*} 
\centering 
\includegraphics[width=0.99\textwidth]{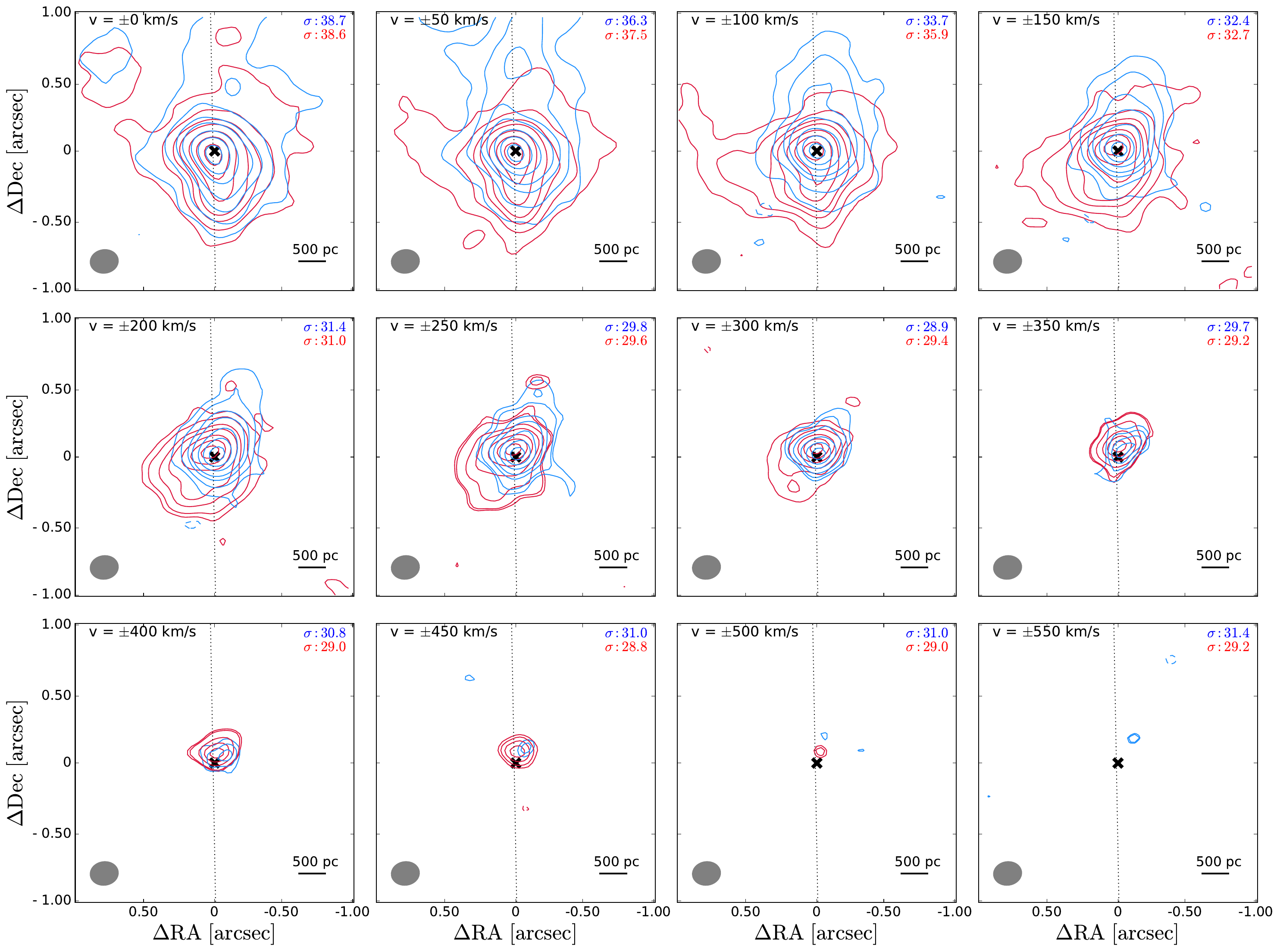} 
\caption{16090-0139.} 
\end{figure*} 
 
\begin{figure*} 
\centering 
\includegraphics[width=0.99\textwidth]{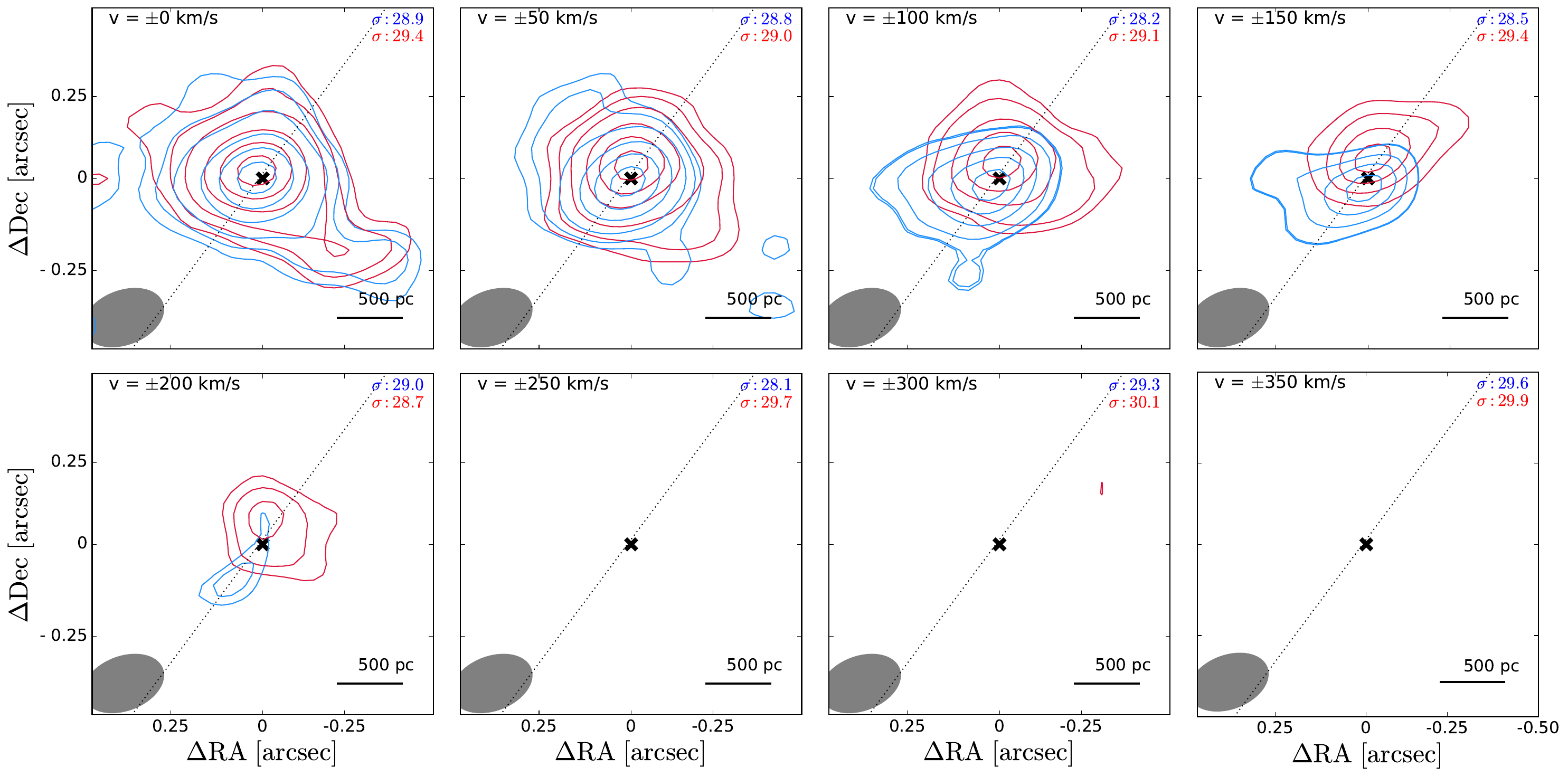} 
\caption{16156+0146 NW.} 
\end{figure*}

\begin{figure*} 
\centering 
\includegraphics[width=0.99\textwidth]{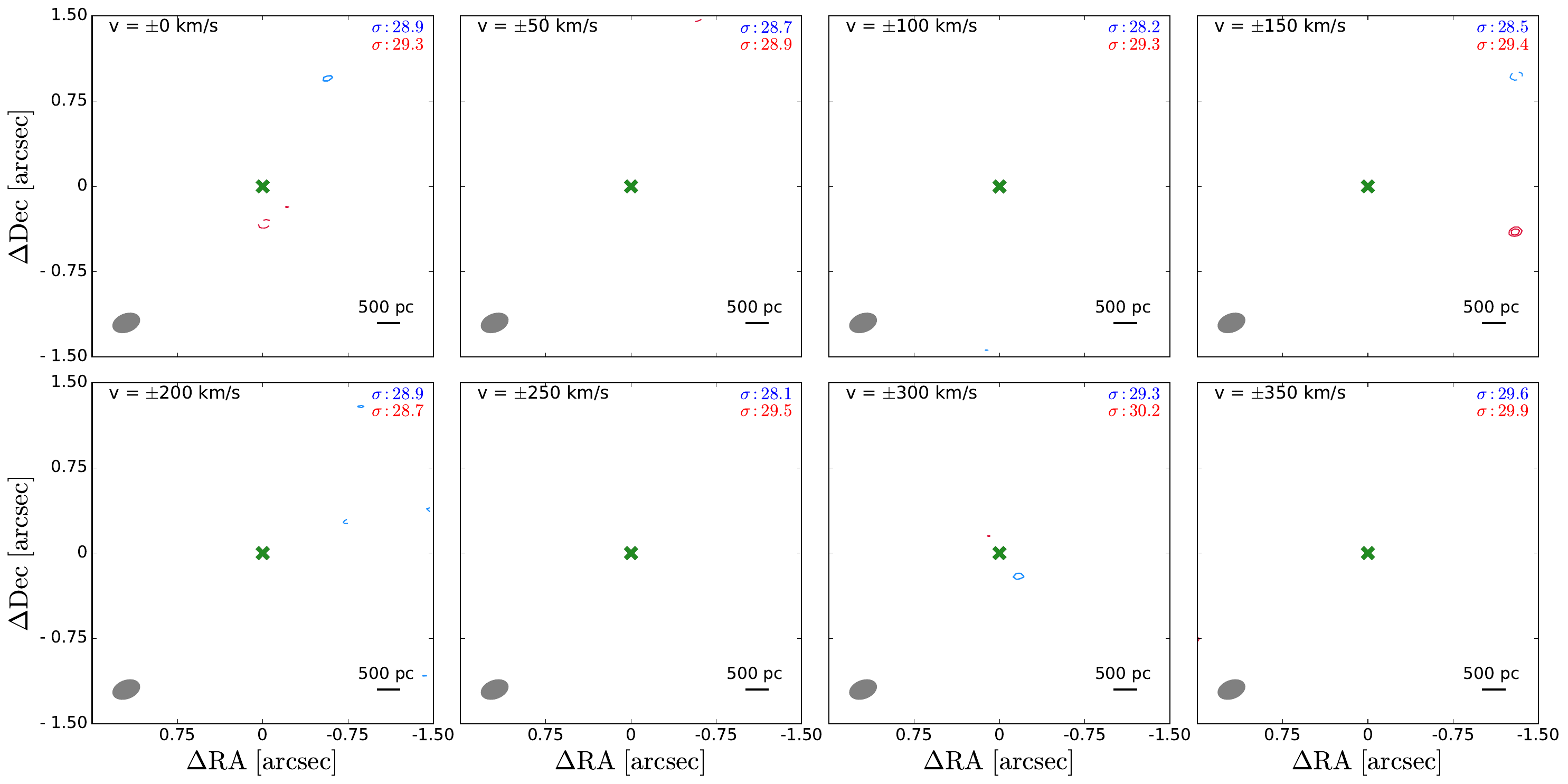} 
\caption{16156+0146 SE.} 
 \end{figure*} 
 
\begin{figure*} 
\centering 
\includegraphics[width=0.99\textwidth]{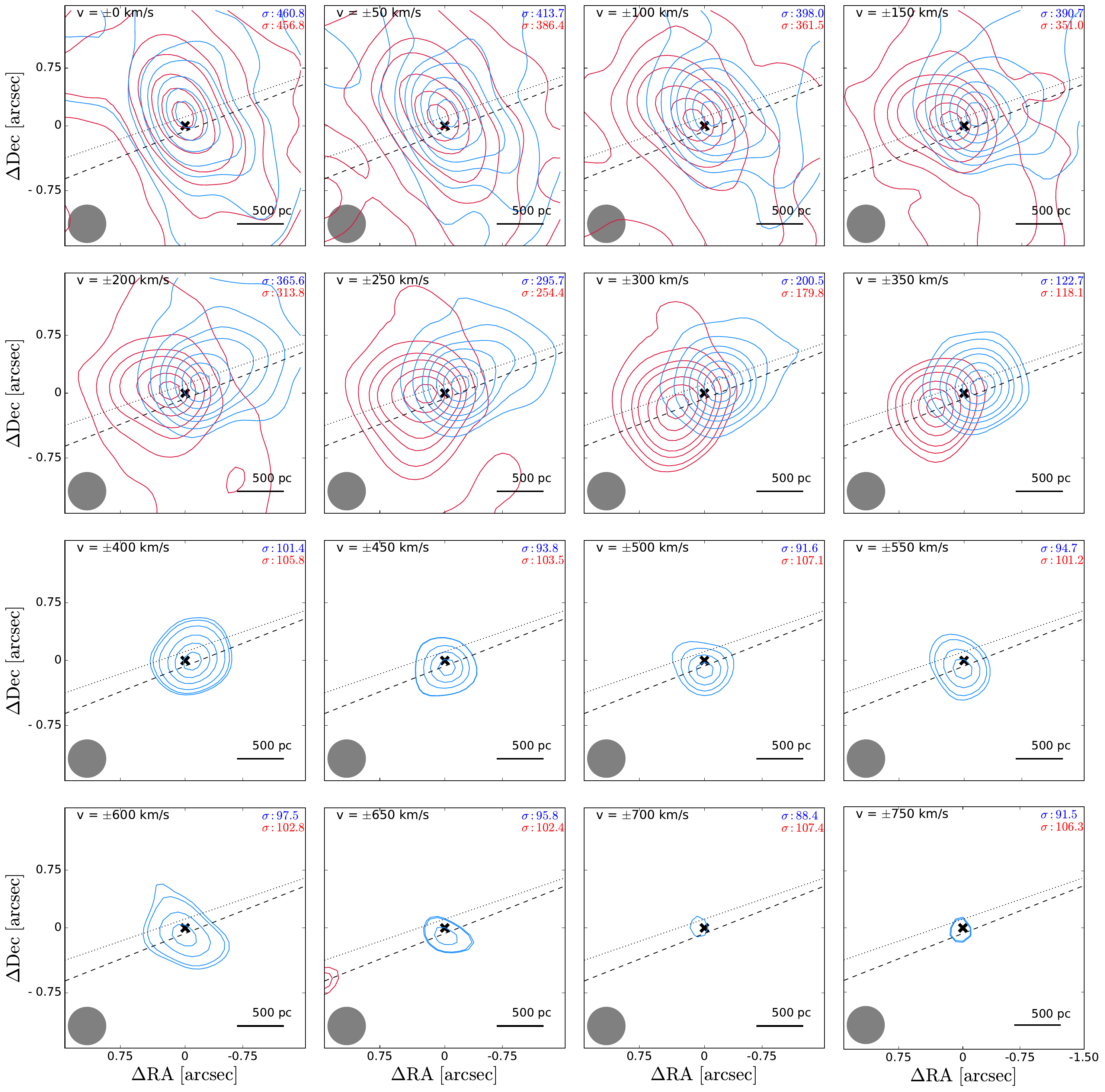} 
\caption{17208-0014.} 
\end{figure*}

\begin{figure*} 
\centering 
\includegraphics[width=0.99\textwidth]{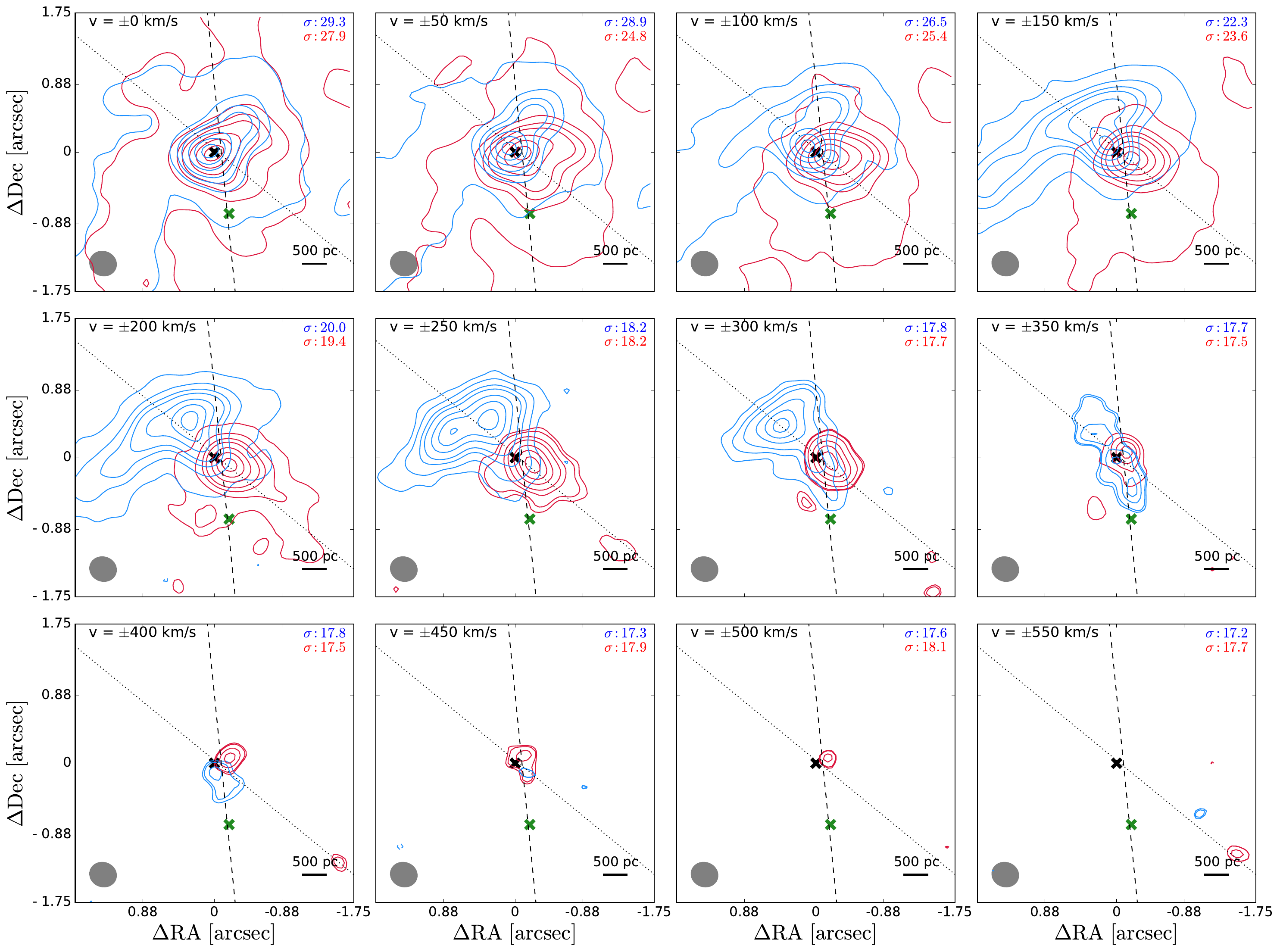} 
\caption{19297-0406 N.} 
\end{figure*}

\begin{figure*} 
\centering 
\includegraphics[width=0.99\textwidth]{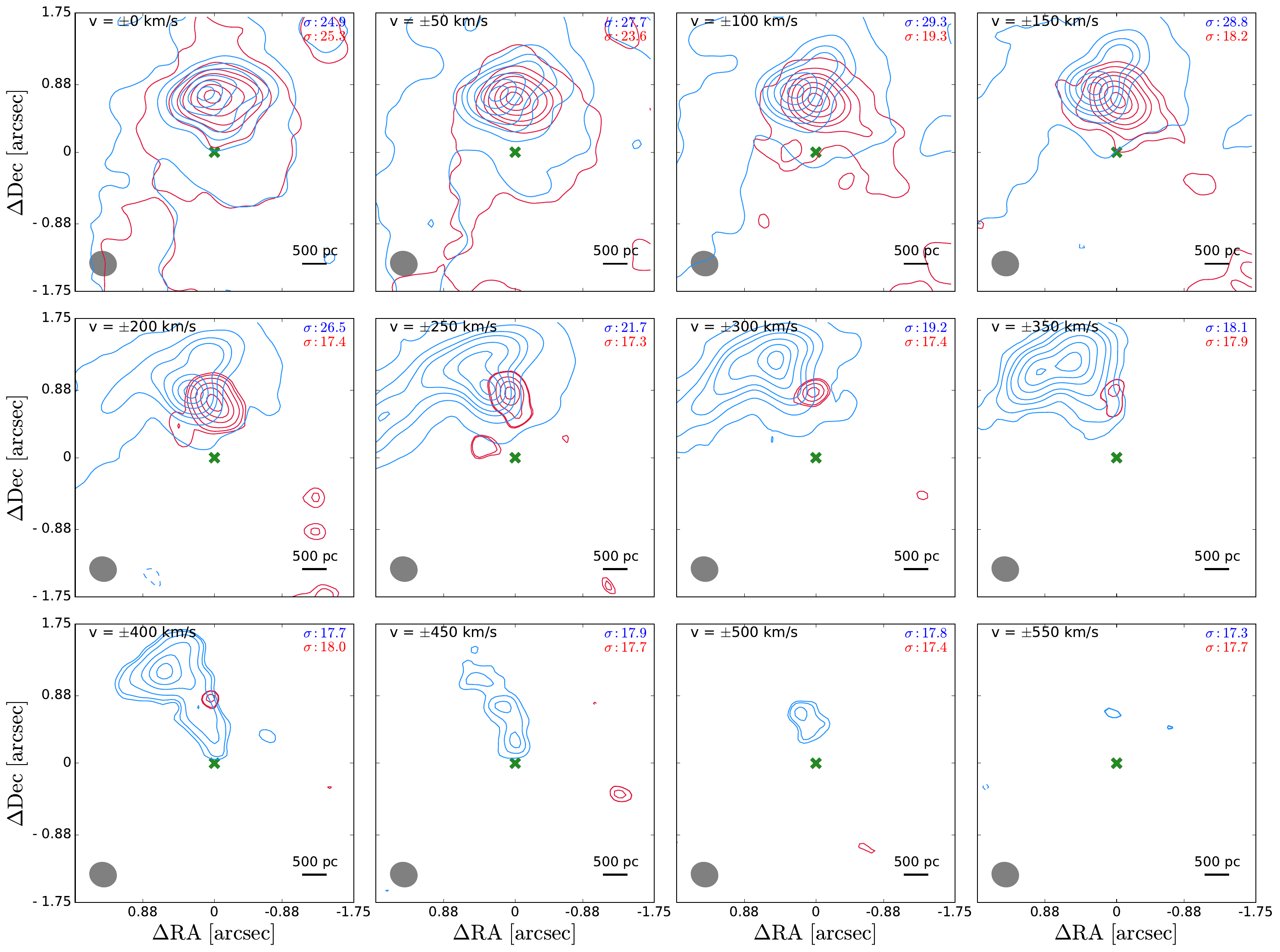} 
\caption{19297-0406 S.} 
 \end{figure*}

\begin{figure*} 
\centering 
\includegraphics[width=0.99\textwidth]{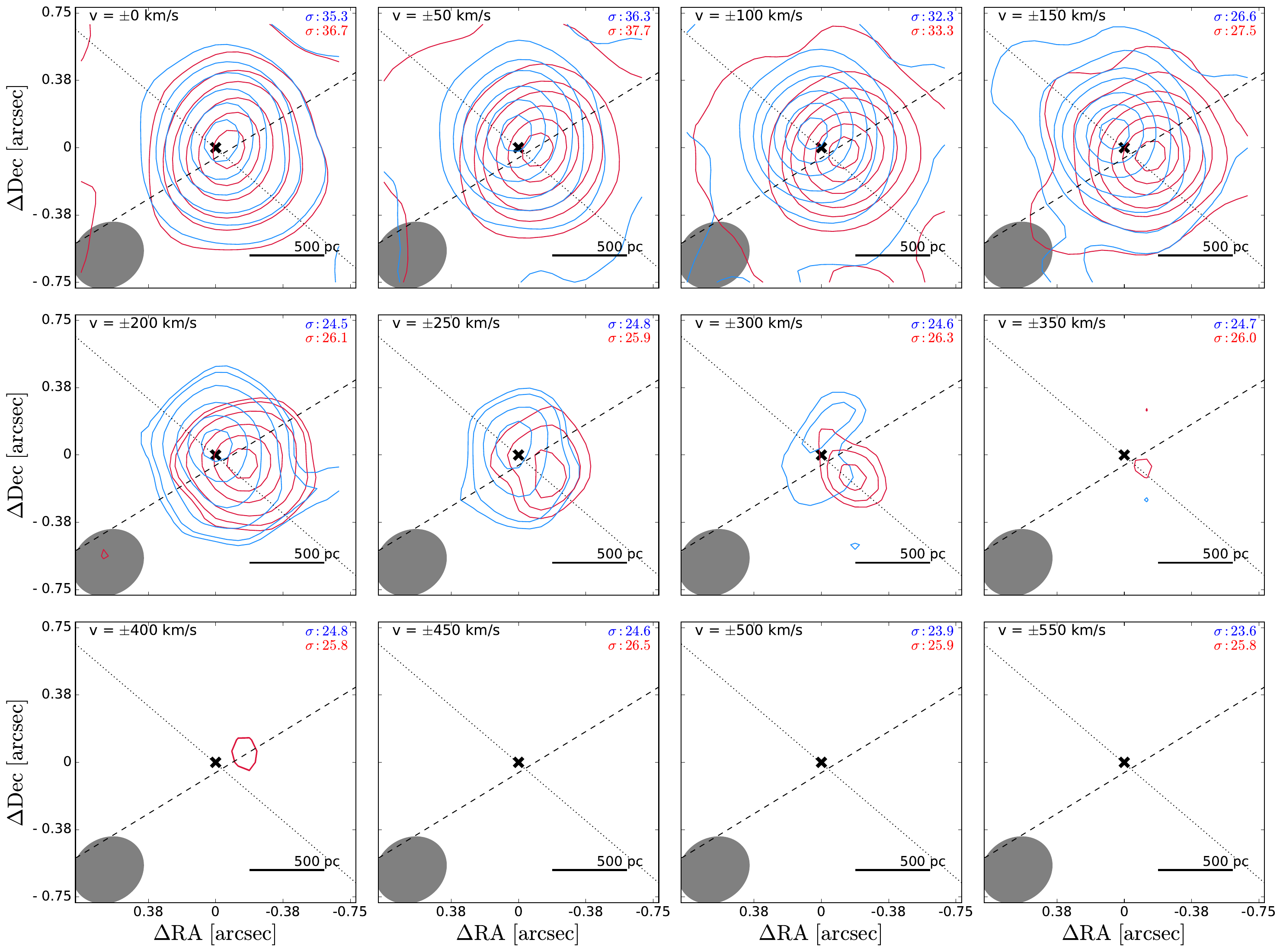} 
\caption{19542+1110.} 
\end{figure*}

\begin{figure*} 
\centering 
\includegraphics[width=0.99\textwidth]{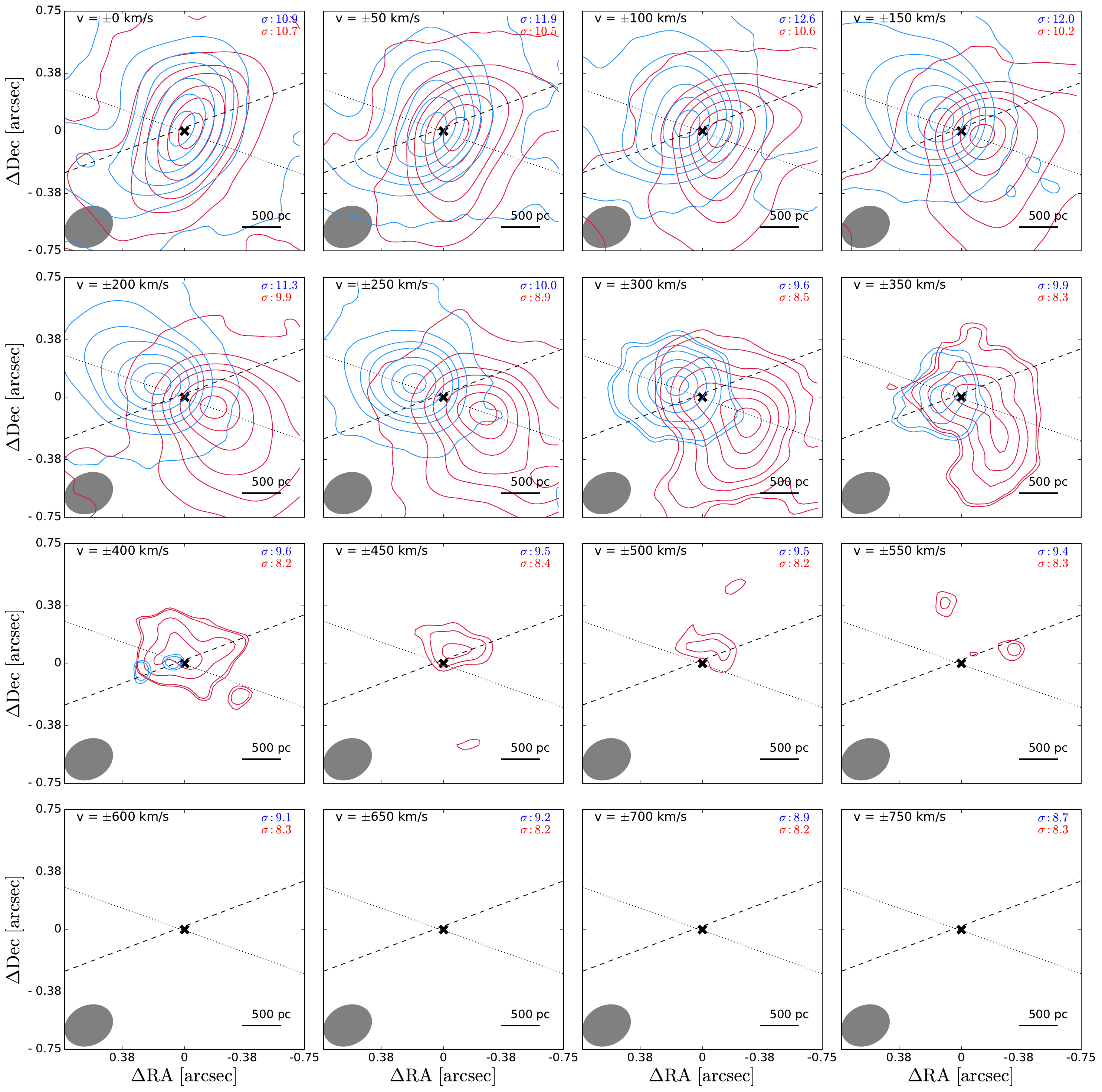} 
\caption{20087-0308.} 
\end{figure*} 
 
 \clearpage
\begin{figure*} 
\centering 
\includegraphics[width=0.99\textwidth]{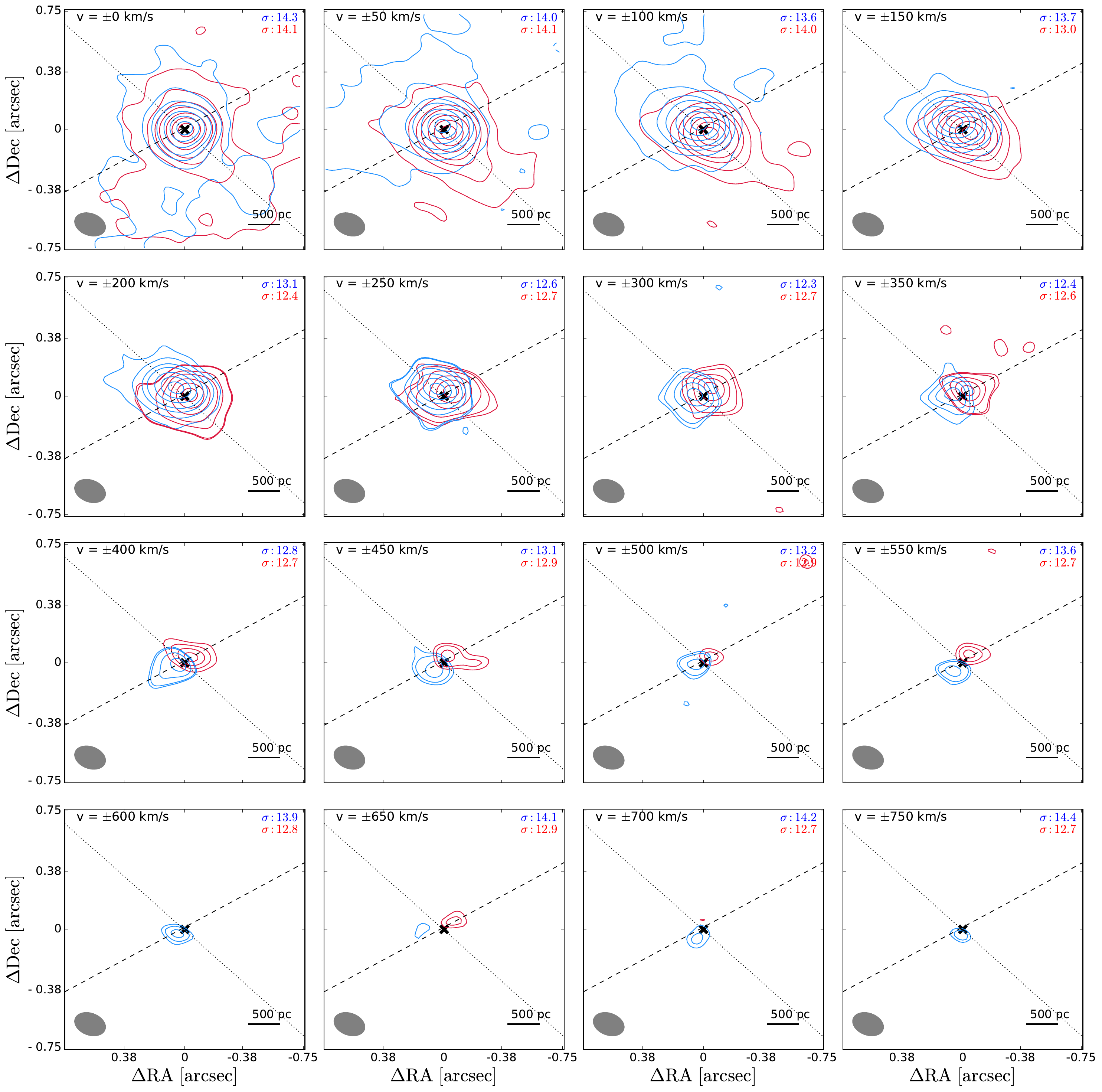} 
\caption{20100-4156 SE.} 
\end{figure*}

\begin{figure*} 
\centering 
\includegraphics[width=0.99\textwidth]{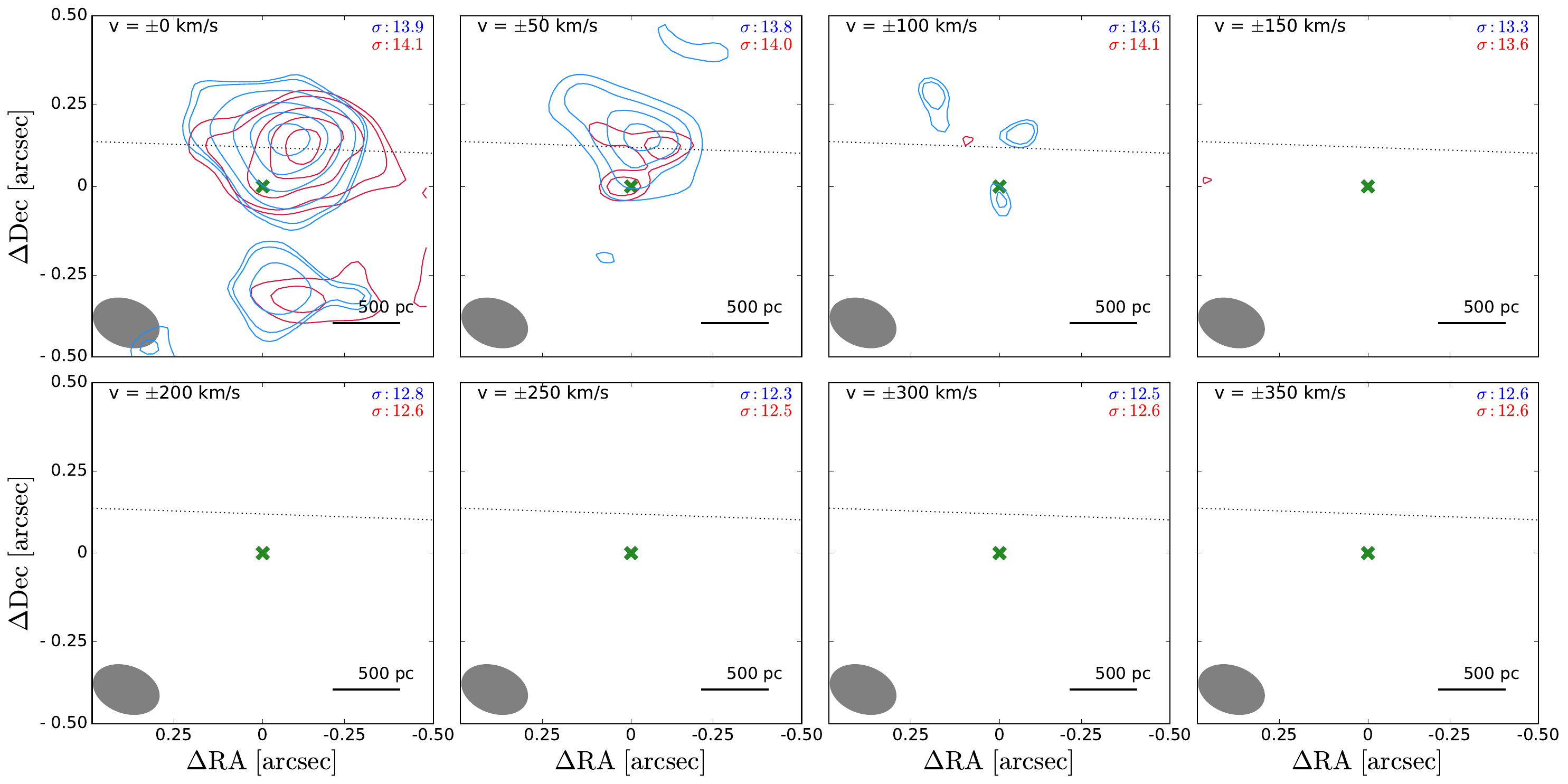} 
\caption{20100-4156 NW.} 
 \end{figure*}

\begin{figure*} 
\centering 
\includegraphics[width=0.99\textwidth]{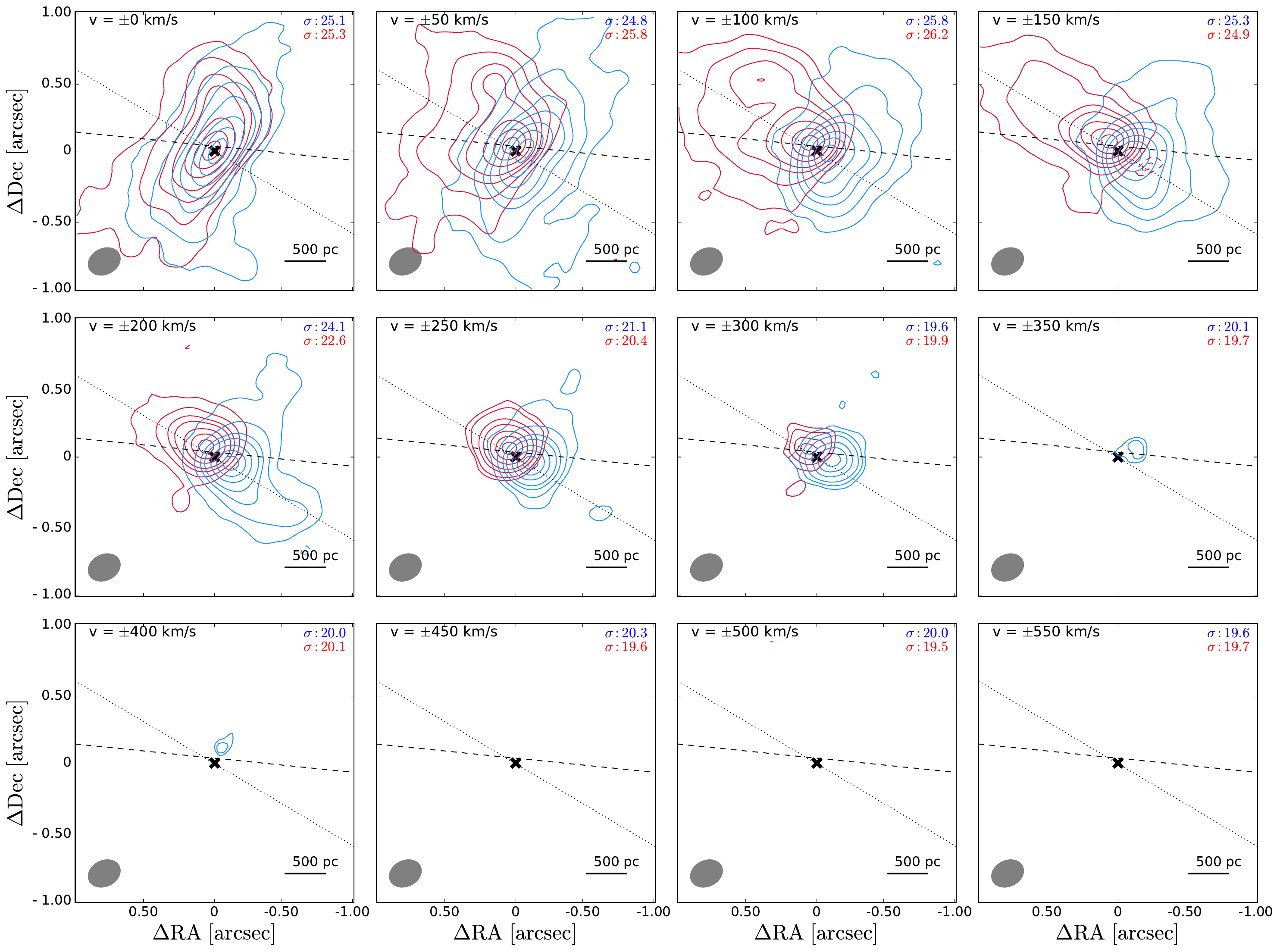} 
\caption{20414-1651.} 
\end{figure*}

\begin{figure*} 
\centering 
\includegraphics[width=0.99\textwidth]{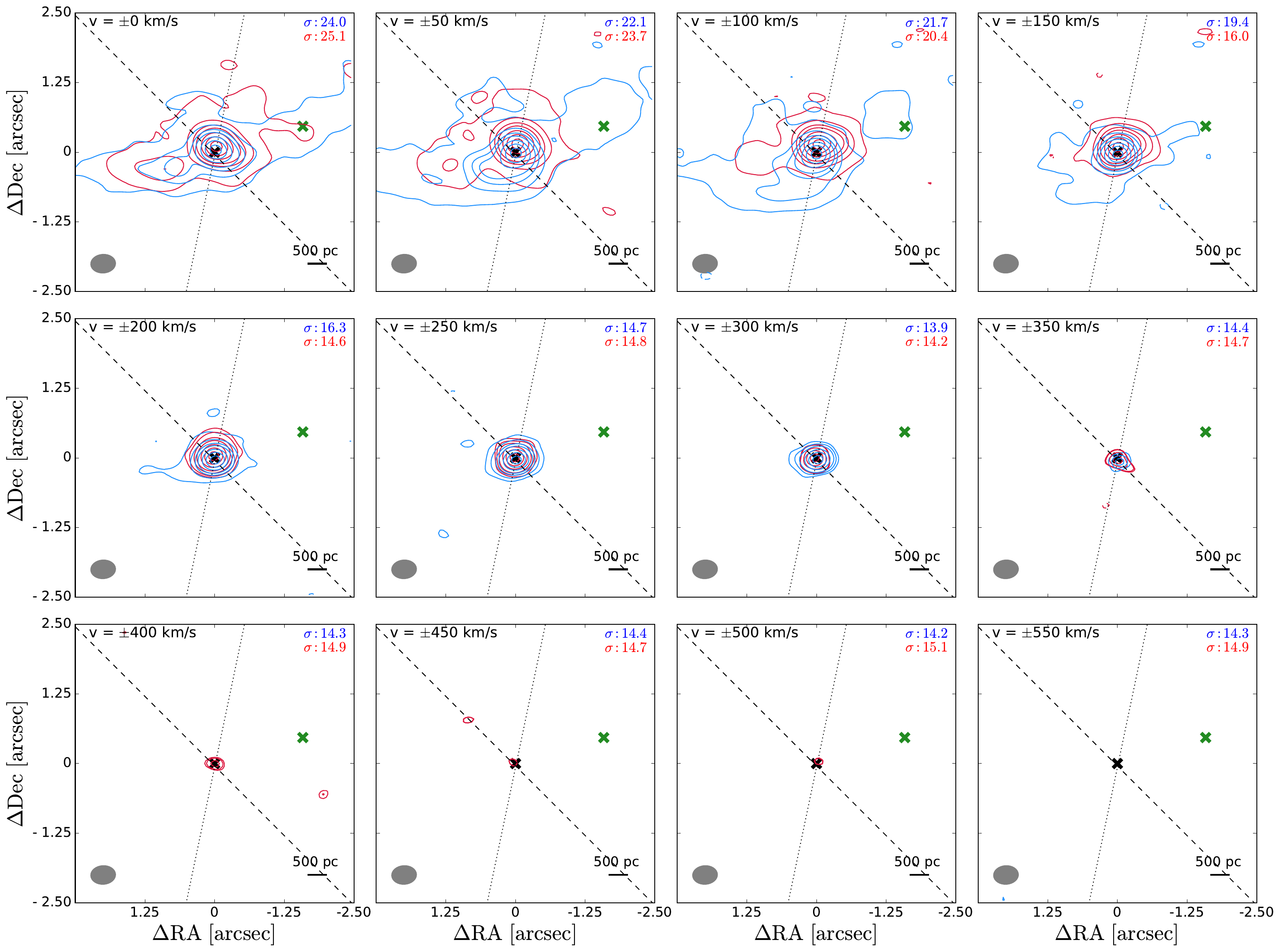} 
\caption{22491-1808 E.} 
\end{figure*}

\begin{figure*} 
\centering 
\includegraphics[width=0.99\textwidth]{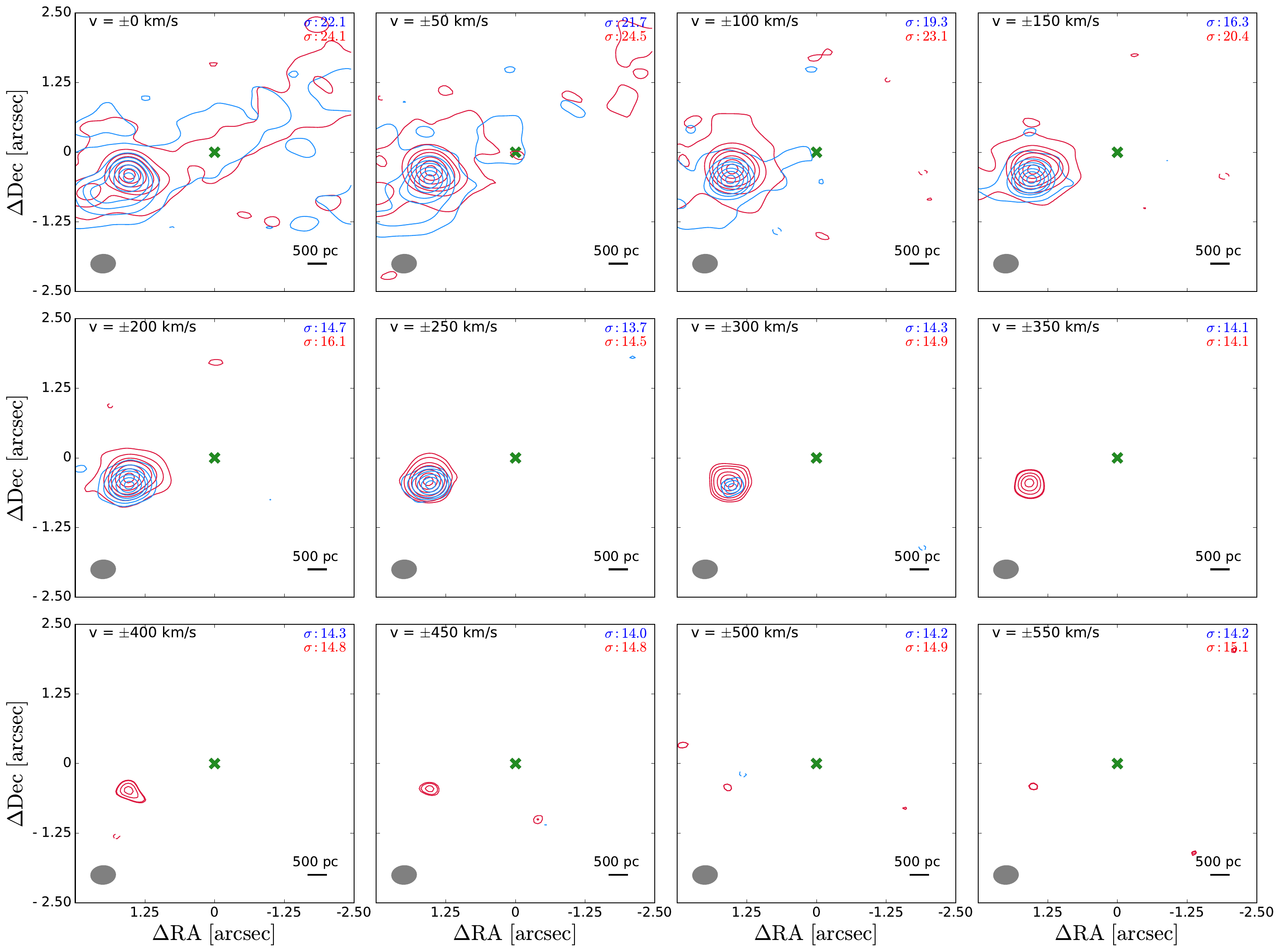} 
\caption{22491-1808 W.} 
 \end{figure*}

\end{appendix}
\end{document}